\documentclass[9pt,twocolumn,twoside,lineno]{pnas-new}
\pdfoutput=1

\templatetype{pnasresearcharticle} % Choose template 
\title{A Stochastic Heat Engine Based on Prandtl-Tomlinson Model}

\author[a,1]{Dongyang Zhao}

\affil[a]{Department of Mechanical Engineering, Tsinghua University, 100084 Beijing, China}
\leadauthor{Zhao} 

\significancestatement{In nanofriction, energy can be dissipated through a phenomenon called stick-slip, where work is done on the system in `stick' and then dissipated to the environment as heat in `slip'. On the contrary, heat engine can extract heat from the environment to the system and convert it into work output. To transform stick-slip into an energy-extracted process, we design a stochastic heat engine based on the Prandtl-Tomlinson (PT) model, which is fundamental in atomic stick-slip friction. Two mechanisms of work output are distinguished. One leads to a Carnot-like approximate efficiency and the other can increase the work output and also reduce it if excessive. Exotic nonlinear bifurcation as well as stochastic resonance of the PT model are analyzed preliminarily too.}

\authorcontributions{D.Y.Z. designed and performed research, analyzed data and wrote the paper.}
\authordeclaration{The authors declare no competing interest.}
\correspondingauthor{\textsuperscript{1}To whom correspondence should be addressed. E-mail: dy-zhao14@mails.tsinghua.edu.cn}

\keywords{Stochastic heat engine $|$ Prandtl-Tomlison Model $|$ Stick-slip $|$ Nonlinear Bifurcation $|$ Stochastic Resonance} 

\begin{abstract}
Stick-slip is a ubiquitous phenomenon in many scientific fields, such as earthquake and glacier dynamics, acoustics, cell biology, interface science and tribology. As a fundamental mechanism of energy dissipation in nanofriction, it can be interpreted by the Prandtl-Tomlinson (PT) model. In this paper we will show that aided by a specifically designed temperature field, stick-slip can be used to extract energy from the environment, i.e. forming a stochastic heat engine based on PT model (PTSHE). Utilizing Langevin dynamics simulation and the framework of stochastic thermodynamics, two mechanisms of work output, i.e. the potential mechanism and the thermolubricity mechanism, are distinguished. An approximate mean cycle work output limit based on the former one is derived, reminiscent of Carnot's limit. The latter one can make the mean cycle work output limit larger than that predicted by the former one while the excess of it can also lead to work output reduction. The mean cycle work curves with respect to the driving velocity is characteristic of PT model in both the PTSHE and nanofriction. The nonlinear bifurcation in zero temperature and the stochastic resonance in finite temperature of the PT model are analyzed preliminarily. With the corrugation number of the PT model increasing, the mean cycle work output limit first increases and then decreases. Besides stick-slip nanofriction and the PTSHE, the PT model is a promising system for studying nonlinear double- or multiple-well dynamics and is valuable to be explored further both theoretically and experimentally.
\end{abstract}

\begin{document}

\maketitle
\thispagestyle{firststyle}
\ifthenelse{\boolean{shortarticle}}{\ifthenelse{\boolean{singlecolumn}}{\abscontentformatted}{\abscontent}}{}

\begin{figure}[t]
\centering
\includegraphics[width=\linewidth]{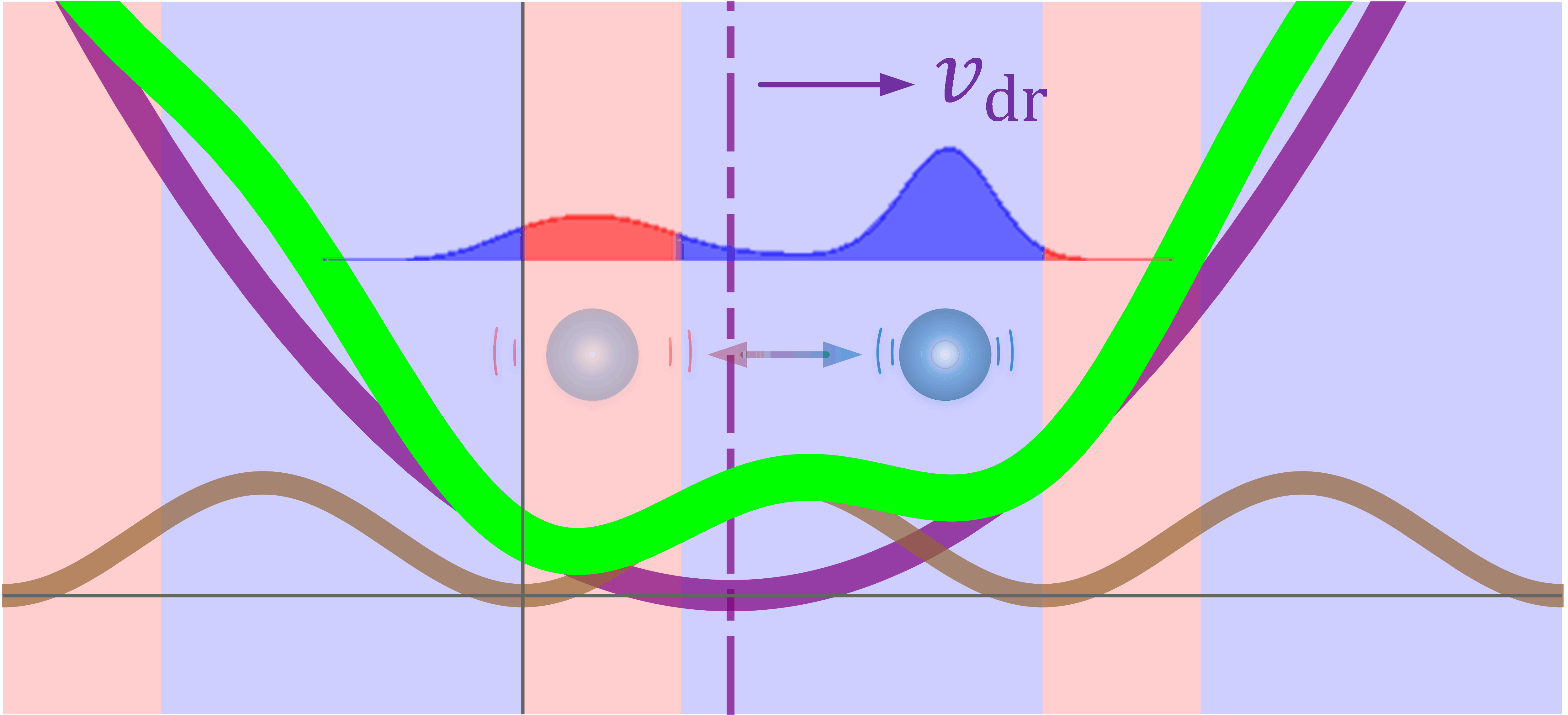}
\caption{Diagram of the stochastic heat engine based on PT model (PTSHE). The one-dimension PT model consists of three elements: the particle $m$ with time-varying position $x(t)$, the purple harmonic potential $V_{\rm h}=\frac12\kappa[x(t)-X(t)]^2$ with uniformly moving center $X(t)=v_{\rm dr}t$, and the brown lattice potential $V_{\rm l}=\frac{V_0}2[1-\cos(\frac{2\pi}ax(t))]$ with lattice period $a$. As the harmonic potential center moves forward, the shape of the green resultant potential $V=V_{\rm h}+V_{\rm l}$ changes and can cause stick-slip at zero temperature (SI Appendix Sec. 3). In the PTSHE, the geometry of the resultant potential curve couples with the duty ratio of the periodic temperature field, which modulates the thermal fluctuation of the particle. The particle oscillates around one of, or jumps back and forth between, the two local minimum points of $V$, with probability $p(x(t),t)$. We can see that although the left local minimum of $V$ is lower than the right one at the selected instant here, the particle distributes on the right more likely due to the specially designed temperature field, as is indicated by the distribution curve and the transparency of the particle. 
 	}
\label{fig:PTmodel}
\end{figure}

% If your first paragraph (i.e. with the \dropcap) contains a list environment (quote, quotation, theorem, definition, enumerate, itemize...), the line after the list may have some extra indentation. If this is the case, add \parshape=0 to the end of the list environment.
%\dropcap{S}tick-slip is an ubiquitous phenomenon in many scientific fields, such as earthquake and glacier dynamics \cite{NatureEarthquake,NatureStickslipIcestream,VolcanicSeismicity,StickslipPakistan,ScienceSSMechanismEarthquake,ScienceSSIce,SciencySSRockExperiment,ScienceSSEarthquakeHeatflow,ScienceSSSlowEarthquake,ScienceSSSorceParameterEarthquake}, acoustics \cite{StickslipLobsters,fletchermusicalinstmt}, cell biology \cite{SASSCelladhesion,PNASCelldynamics}, interface science \cite{PNASNanodroplet,FoamsStickSlip,Stickslipmercury,NatureStickslipLuminence,Resonantstickslip,ScienceSSBoundaryLub} and tribology \cite{Slip-stickFrictionNature,PNASSticksliplubricant,PNASArticularjoints}. 
\dropcap{F}riction converts mechanical work into heat dissipation irreversibly and cause the efficiency of heat engines to reduce below the Carnot limit. Due to the development of Friction Force Microscope (FFM), study of friction has advanced to the nanoscale regime \cite{KryPhysSF,VanColloquiumFriction}. The tip of the FFM driven by a unifromly moving support goes through the potential landscape of an atomically plain surface. At low temperature, due to the ups and downs of the surface potential, the tip sticks and slips, causing great energy dissipation, which can be modelled by the Plandtl-Tomlinson (PT) model \cite{Prandtl,Tomlinson} (Figure \ref{fig:PTmodel} and SI Appendix Sec. 3).
% \begin{figure}
% \includegraphics[width=8.6cm]{Figures/PTModel.pdf}%
% \caption{\label{fig:PTmodel}(Color Online)Diagrams of the PT model and the stochastic heat engine.
% 	(a)The PT model consists of three elements: the mass point $m$, the harmonic spring force $F_{\rm h}$ and the surface lattice force $F_{\rm l}$. 
% 	(b)The stochastic heat engine is constituted by the trapped ion friction emulator in spatial periodic heat bath.In the trapped ion friction emulator, the laser cooling particle serves as the mass point in the PT model; the Paul trap serves the harmonic potential $V_{\rm h}$; and the optical lattice serves the surface lattice potential $V_{\rm l}$. 
% 	(c)In the stochastic heat engine, the geometry of the resultant potential $V=V_{\rm h}+V_{\rm l}$ coupled with the duty ratio of the period of the temperature of the heat baths, which modulates the thermal fluctuation of the particle. If the particle is fluctuating around the potential local minimum behind the center of the harmonic trap, $x(t)<X(t)$, as is the case in the diagram, the harmonic trap pulls the particle and work is done on the particle. However, if the particle is fluctuating around the potential minimum ahead of the center of the harmonic trap, i.e. if the particle is fluctuating around the right local minimum point of the resultant potential $V$, $x(t)>X(t)$, the particle will pull the harmonic trap and work is done by the particle on the harmonic trap.} 
% \end{figure}
A trapped ion friction simulator has been theorectically proposed \cite{TIFENC,TIFEPRBTosatti,TrappedIonFKEPJBGarcia,TrappedIonPruttivarasinNJP} and then experimentally implemented \cite{ScienceTIFE} and utilized to study the velocity dependence of the friction force of a single ion driven through an optical lattice by a Paul trap \cite{NPVelocityTuning}, emulating the PT model. At high temperature or low velocity, thermal fluctuations of the ion (or the tip of FFM) make it jump over the middle energy barrier easily with a vanishing amount of energy loss, which is known as thermolubricity \cite{Thermolubricity}. 

Thermal fluctuation also plays an important role in a lot of microscale \cite{Stirling,BrownianCarnot} and nanoscale \cite{UnderdampedEngine,SingleatomScience,SingleIonPRL,SingleIonPRL2} heat engines. Using a single micrometer-sized colloidal particle as the working agent and a breathing laser trap, i.e. a parabola potential $V_{\rm h}'=\frac12\kappa(t) x(t)^2$ with controllable stiffness $\kappa(t)$ coupled with heat baths of relevant temperature protocols, a stochastic heat engine has been analyzed using stochastic thermodynamics \cite{SchmiedlStochaHeatEngine} and realized experimentally \cite{Stirling}. In the microscopic Stirling engine cycle in \cite{Stirling}, the harmonic potential expands [$\kappa(t)$ increases] in the high temperature stage and heat is absorbed while it shrinks [$\kappa(t)$ decreases] in the low temperature stage and work is outputted.

In analogy, in the PT model, the resultant potential superposed by the harmonic and the sinusoidal lattice potential has a local minimum which rises and uplifts the particle's potential energy in the stick stage when the harmonic potential moves forward [$X(t)$ increases], cf. SI Appendix Sec. 3. If the particle is immersed in high temperature during this stage and in low temperature in the rest of one period, i.e. coupling the resultant potential with a specific temperature field, energy may be extracted from the heat bath. In this paper, we will design such a heat engine, which will transform the energy-dissipated stick-slip process into an energy-extracted process. 

We will first describe the design of the stochastic heat engine based on PT model (PTSHE) and then analyze it in detail through Langevin dynamics simulation and the framework of stochastic thermodynamics in the Results section, the contents of which are indicated by the subheadings. In the Discussion section, we will go deep into the results further, including comparing the PTSHE with the analogous B{\"u}ttiker-Landauer heat engine as a complement to this introduction.

\begin{figure*}[t]
\centering
\includegraphics[width=0.4475\linewidth]{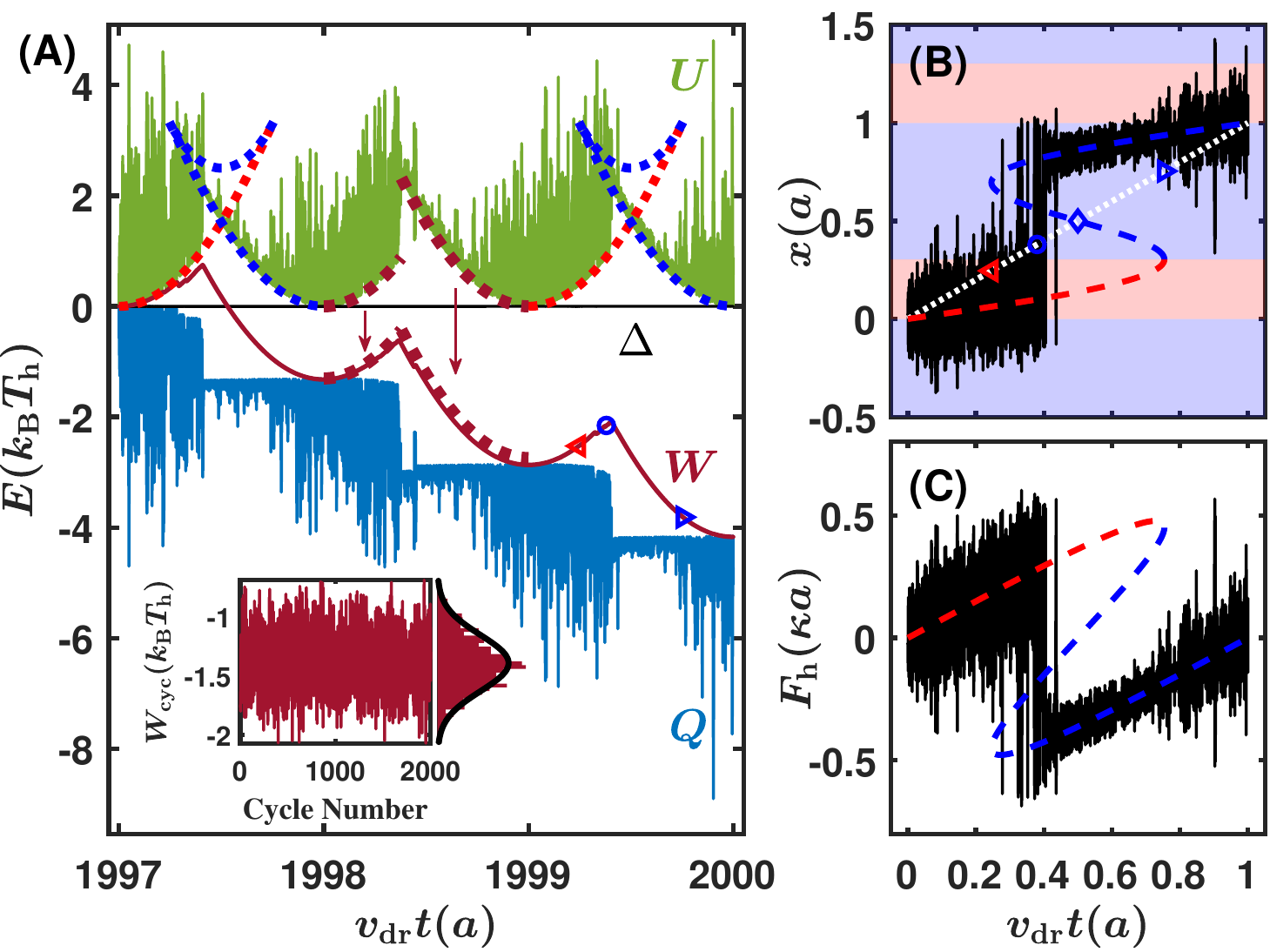}\includegraphics[width=0.545\linewidth]{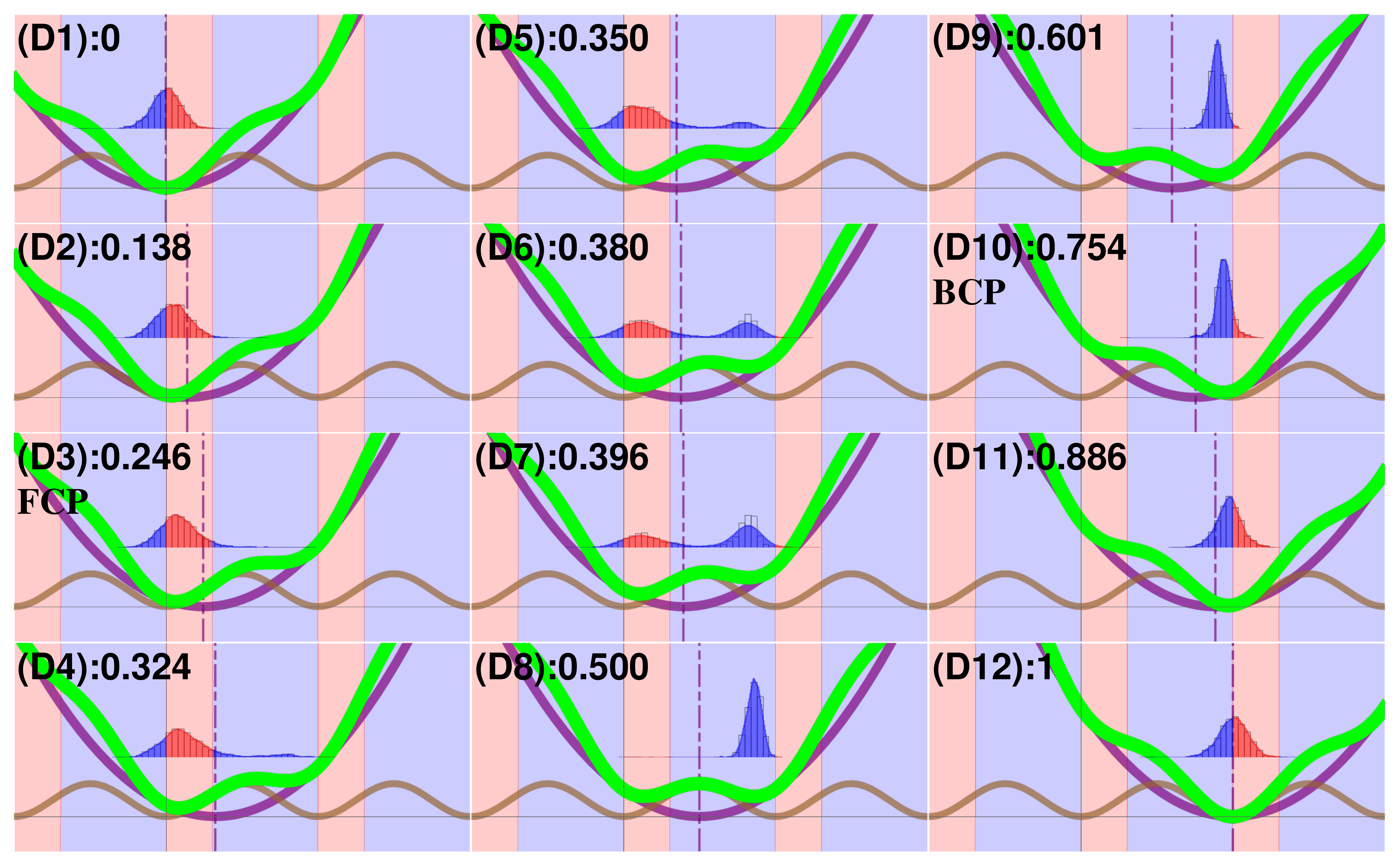}%
\caption{Simulation results of the PTSHE. (A) The internal energy $U$, work input $W$, heat to the heat bath $Q$ and the difference between them $\Delta=\Delta U+Q-W$ in the last three consecutive engine cycles among 2000 in total at $v_{\rm dr}=10^{-5}\rm m/s$. The work decreases to $-\infty$, i.e. work is outputted. The dotted red and blue curves are the resultant potential at the balanced points (SI Appendix Sec. 2) with the red ones in the hot zone and the blue ones in cold. The inset of (A) represents the cycle work of 2000 consecutive simulation cycles and its distribution. The $x-$coordinate $v_{\rm dr}t(a)$ is the nondimensional position of the driver center, which is proportional to time $t$. The dimension of the energy is $k_{\rm B}T_{\rm h}$ with $k_{\rm B}$ the Boltzmann's constant and $T_{\rm h}$ the high temperature. The dotted dark red curves indicate the shape similarity between the work curve and the lower bound of the internal energy. $\Delta\equiv0$ indicates the first law of thermodynamics (SI Appendix Sec. 5). (B) The displacement of the particle in the typical 2000th cycle with respect to the driver center's position, relative to the starting point of this cycle. The dotted white line represents the driver center's position with respect to itself.  %The dotted white line represents the corresponding position of the driver center at the same time. 
The dashed red and blue curves are the loci of the balanced points (SI Appendix Figure S3) in the hot and cold zones respectively. The red and blue zones represent the high and low temperatures repsectively. (C) Variation of the harmonic force in the 2000th engine cycle. The $x-$coordinate matches to that of (B). Time ratio of negative harmonic force in one cycle is larger than that of positive one, consistent with the work output. The dashed red and blue curves are the balanced harmonic force at the balanced point in hot and cold zones respectively (SI Appendix Figure S3). (D1)-(D12) Diagrams of the shapes of the resultant potential and the corresponding displacement distributions of the particle at different instants of the engine cycle. The displacement distribution at a certain instant relative to the latest cycle starting point (D1) is obtained from the total 2000 simulation cycles.
%the corresponding frequency distributions of the particle's displacement relative to the cycle starting point (D1) at different instants of the engine cycle. 
The decimal following the id of each frame is the nondimensional position of the driver center (or the nondimensional instant) relative to the cycle starting point (D1). The purple curve represents the moving harmonic potential and the brown one represents the lattice potential. The red and blue zones represent high and low temperatures respectively. In one engine cycle, the particle starts at the point when the harmonic potential local minimum overlaps with one of the lattice potential local minimum (D1). As the driver center goes forward, the shape of the resultant potential curve changes. After the appearance of the FCP (D3), there occurs a new local minimum on the right. In (D7), the particle distributes in the right well more likely, although the right local minimum is still higher than the left one. At the middle of the cycle (D8), the particle almost completely distributes in the right well and stays there until the end of one cycle (D12). Parameters: $\eta=3.0$, $\mu=4\times10^4\rm s^{-1}$, ${\it\Theta}_{\rm h,c}=0.4,0.04$ and others are in SI Appendix Sec. 8.H.}
\label{fig:CyclesTh04Tc004}
\end{figure*}

\section*{Results}

%Please note that whilst this template provides a preview of the typeset manuscript for submission, to help in this preparation, it will not necessarily be the final publication layout. For more detailed information please see the \href{https://www.pnas.org/page/authors/format}{PNAS Information for Authors}.
%
%If you have a question while using this template on Overleaf, please use the help menu (``?'') on the top bar to search for \href{https://www.overleaf.com/help}{help and tutorials}. You can also \href{https://www.overleaf.com/contact}{contact the Overleaf support team} at any time with specific questions about your manuscript or feedback on the template.

\subsection*{Engine Cycle and Temperature Protocol}
The diagram of our PTSHE is given in Figure \ref{fig:PTmodel}. As the center of the harmonic potential (driver center) goes forward to the right with uniform velocity $v_{\rm dr}$, the resultant potential,
\begin{eqnarray}
V(x(t),t)=\frac12\kappa[x(t)-v_{\rm dr}t]^2+\frac{V_0}2[1-\cos(\frac{2\pi}ax(t))],
\end{eqnarray}
changes periodically, as can be seen in Figure \ref{fig:CyclesTh04Tc004}(D1)-(D12), where we plot the resultant potential (the green curves) at different instants of one engine cycle. The nondimensional corrugation number $\eta=\frac{2\pi^2V_0}{\kappa a^2}$ determines the shape of the resultant potential. As $\eta$ ranges between $1$ and $4.6$, the resultant potential has at most two local minima \cite{ScienceTIFE,MultislipTIHE,CorrugationNumber}, cf. SI Appendix Sec. 4. 

In one engine cycle, the driver center travels from one lattice potential local minimum to the next one, with the temporal period $\frac{a}{v_{\rm dr}}$. The particle starts around the global minimum point of the resultant potential [Figure \ref{fig:CyclesTh04Tc004}(D1)] and is driven through high (red) and low (blue) temperature zones in succession. The resultant potential first has one global minimum [Figure \ref{fig:CyclesTh04Tc004}(D1)-(D3)], then two local minima separated by one local maximum [(D4)-(D9)], and then one global minimum again [(D10)-(D12)]. The local or global minimum points are stable balanced points where the harmonic force and the lattice force are equal while the local maximum point is unstable balanced point.

The spatially varying temperature of the heat bath couples with the resultant potential. In one spatial lattice period, the high temperature zone (red one) ranges from one lattice potential local minimum to the point where the resultant potential left local minimum and middle local maximum merge [backward critical point (BCP) in Figure \ref{fig:CyclesTh04Tc004}(D10)], i.e. as the left global or local minimum rises, it won't get out of the hot zone until it disappears. The low temperagure zone (blue one) ranges from the BCP to the next lattice potential local minimum point. The procedure to calculate the BCP as the boundary of high and low temperature zone is given in SI Appendix Sec. 2.

\subsection*{Four Stages of the Engine Cycle and the Potential Mechanism of Work Output}
We use Langevin dynamics simulation and the framework of stochastic thermodynamics (Methods and SI Appendix Sec. 5, 7 and 8) to analyze the the PTSHE. The solid dark red curve in Figure \ref{fig:CyclesTh04Tc004}(A) represents the work input by the driver at a low driving velocity $v_{\rm dr}=10^{-5}\rm m/s$ (i.e. the particle is at qusi-equilibrium state, see the next subsection). In each cycle, the work curve first increases and then decreases. The variation of work in the decreasing stage is greater than that in the increasing stage, i.e. outputted work is more than inputted work and in an entire cycle, work is outputted. The cycle work $W_{\rm cyc}$ is a stochastic variable and distributes approximately normally [inset of Figure \ref{fig:CyclesTh04Tc004}(A)] with the the mean value negative. Therefore, the PTSHE can indeed output work stably. As a comparison, we also simulate the cases of the homogeneous temperature heat bath with high and low temperature in SI Appendix Sec. 6.A and 6.B, where the mean cycle work is positive and on average work has to be inputted. The dimension of the work is $k_{\rm B}T_{\rm h}$, indicating that the PTSHE works in micro- or nanoscale where thermal fluctuations are significant \cite{PhysTodaySmallSystems,Stirling}.

Diagrams of the resultant potential energy curve and the particle's displacement distribution at 12 selected intants in one engine cycle are given in Figure \ref{fig:CyclesTh04Tc004}(D1)-(D12). The entire engine cycle can be divided into four stages. 
\subsubsection*{First stage} From the beginning (D1) to the appearance of the FCP [forward critical point, (D3)]. On the work curve of the 2000th cycle in Figure \ref{fig:CyclesTh04Tc004}(A), this stage ranges from the beginning to the red left triangle. In this stage, the resultant potential energy has only one global minimum moving forward and upward. The particle fluctuates around the minimum point [Figure \ref{fig:CyclesTh04Tc004}(B)] and the particle's displacement relative to the cycle starting point at a specific instant of the cycle has an approximate normal distribution centered at the minimum point [(D1)-(D3)]. On average the particle falls behind the driver center gradually and is pulled. Hence work is done on the particle so that the work curve goes up. In this stage, the harmonic force fluctuates around and rises along the balanced harmonic force curve [Figure \ref{fig:CyclesTh04Tc004}(C)], leading to the slope of the work curve increasing. As work is done on the particle, the resultant potential energy of the particle increases (explained below) and as the driver moves forward the particle fluctuates more in the hot zone so that its kinetic energy also increases. As a whole, the internal energy $U$ of the particle increases in this stage [Figure \ref{fig:CyclesTh04Tc004}(A)].

\subsubsection*{Second stage} From the appearance of the FCP to the cusp instant of the work curve. After the FCP appears, the resultant potential energy has two local minima and becomes double-well [(D4)-(D9)]. Because of the violent thermal fluctuation of the particle in the high temperature zone, the particle will sometimes cross over the middle energy barrier from the left well to the right one and the jumping back and forth event occurs more frequently as the driver moves forward [Figure \ref{fig:CyclesTh04Tc004}(B)]. In the right well, the particle is on average ahead of the driver center, i.e. $x>v_{\rm dr}t$, so the harmonic force $F_{\rm  h}=\kappa[v_{\rm dr}t-x(t)]<0$ [Figure \ref{fig:CyclesTh04Tc004}(C)] and the driver is driven by the particle and work is done by the particle to the driver. At the beginning of this stage, the backward energy barrier is low and it's easy for the particle to jump back to the left well and the particle's position still mainly distributes nearby the left balanced point [(D4)-(D5)]. Thereby, the net work input is still positive on average and the work curve continues to rise. As the driver goes forward, the forward barrier gets lower and the backward barrier gets higher, so the particle jumps forward over the barrier more easily and the right peak of its displacement distribution gets higher. In (D6), the two peaks has equal height approximately and afterwards the right peak is higher than the left one, although the left local minimum is still lower than the right one (D7).
%, \textcolor{red}{due to the low fluctuation of the particle in the right cold zone.} 
At some point the particle lanches over the barrier to the right and never jumps back. Before the jumping point, the harmonic force is on average positive while after that, the harmonic force is on average negative [Figure \ref{fig:CyclesTh04Tc004}(C)]. So the slope of the work curve is positive before and negative after the jumping point, causing the cusp. The blue circle in Figure \ref{fig:CyclesTh04Tc004}(A) and (B) represents the instant of (D6), around which the particle stays in the two wells equally likely. Before or after this instant, the jumping point (i.e. the instant of the cusp) of one cycle occurs stochasticly, causing the cycle work a stochastic variable, which will be discussed in detail below. And therefore the time durations of this and next stage are stochastic.
\subsubsection*{Third stage} From the cusp instant of the work curve to the appearance of the BCP [(D10) and the blue right tiangle on the work curve in Figure \ref{fig:CyclesTh04Tc004}(A)]. Because of the low temperature, the particle calms down in the right well. In (D8), while the forward and backward energy barriers are equal, the distribution of the particle's displacement is almost completely on the right cold zone. Pushed by the lattice force, the particle is ahead of the driver center [Figure \ref{fig:CyclesTh04Tc004}(B)] and the harmonic force is negative [Figure \ref{fig:CyclesTh04Tc004}(C)] so that work is done by the particle on the driver and is outputted. As the driver moves forward, the distance between it and the particle gets smaller [Figure \ref{fig:CyclesTh04Tc004}(B)] and the absolute harmonic force reduces along the balanced force curve [Figure \ref{fig:CyclesTh04Tc004}(C)]. Therefore the work curve descends with decreasing absolute slope. In the cold zone, as the potential well gets deeper and narrower, the distribution gets thinner [(D9)-(D10)] and the amplitude of the particle's thermal fluctuation gets smaller [Figure \ref{fig:CyclesTh04Tc004}(B)].
%because the amplitude of the particle's oscillation (and hence that of its thermal fluctuation) gets smaller [Figure \ref{fig:CyclesTh04Tc004}(B)]. 
The fluctuations of the internal energy $U$ and the heat to the environment $Q$ are also inhibited [Figure \ref{fig:CyclesTh04Tc004}(A)].
\subsubsection*{Fourth stage} From the disappearance of the BCP to the end of one cycle (D12). In this stage, there is again only one global minimum, which is now ahead of the driver center. Work is still outputted and the work curve continues to descend. The absolute slope of the work curve decreases as the driver center catches up with the particle gradually [Figure \ref{fig:CyclesTh04Tc004}(B)] and the absolute value of the average harmonic force reduces [Figure \ref{fig:CyclesTh04Tc004}(C)]. As the driver goes close to the end of the cycle, the particle frequently fluctuates to the hot zone of next cycle so that the amplitude of its thermal fluctuation increases gradually [Figure \ref{fig:CyclesTh04Tc004}(B)] and its displacement distribution diverges, although the potential well continues to get deeper and narrower. The fluctuation of the internal energy $U$ and the released heat $Q$ increases too [Figure \ref{fig:CyclesTh04Tc004}(A)].

%In Figure \ref{fig:CyclesTh04Tc004}(B), we can see that the black curve, which describes the position of the particle with respect to the driver center, hovers around the left local minimum of the resultant potential energy (the green dashed curve, SI Appendix) and is behind the driver center (the blue dotted line) at the beginning. So the friction force [Figure \ref{fig:CyclesTh04Tc004}(C)] is mainly positive and the work curve increases. However at around the blue circle it jumps back and forth between the two minima and then calms down at the other side and then hovers around the right local minimum until the end of one cycle. So the friction force mainly has minus sign and the work curve decreases. After averaging over one lattice period of the driver center, the mean value has minus sign and the work curve decreases. This is mean ``minus friction force'' in the sense of the lateral force, which is equivalent to work output. 

In Figure \ref{fig:CyclesTh04Tc004}(A), the dotted red and blue curves represent the resultant potential energy at the local extremum points (i.e. balanced points) with respect to the driver center's position [SI Appendix Figure S1(B)].
%, we can see that in one cycle a vertical line at a specific instant intersects with the red and blue dashed curves once and then three times and then once again, corresponding to the resultant potential energy's one global minimum at first and then two local minima and one local maximum and finally one global minimum again.
The left minimum branch (dotted red) and the right minimum branch (dotted blue) cross in the middle three-extremum zone. According to our design, the rising left minimum branch is immersed in high temperature, while the middle maximum branch and the setting right minimum branch is immersed in low temperature. The internal energy $U$ fluctuates, with the two local minimum branches as its lower bound, indicating that the particle's kinetic energy frequently relaxes to near zero and then regains, mainly resulting from the stochastic forces.

At the cusp instant of the work curve, the internal energy $U$ lanches from the left minimum branch to the right one like climbing a cliff, while the heat $Q$ falls off a cliff, indicating that heat is absorbed by the particle and transformed into its internal energy abruptly and won't be released back. % The reason of the cliff is the energy barrier. On the left and right of the cliff, heat absorbing (because of the fluctuating force) and releasing (because of the fluctuating force and the damping force) exist simultaneously. 
On the left of the energy barrier heat is absorbed from the high temperature heat bath and the kinetic energy of the particle increases, leading to its high fluctuating amplitude and its jumping over the barrier to the right, where the potential energy is high and the kinetic energy is low. After the particle calms down on the right, unlike the kinetic energy, the potential energy can't transform back to heat and will gradually transform into work output, in analogy to a falling body. The shape of the ascent and descent stages of the internal energy on the left and right minimum branches are similar to the respective ascent and descent segments of the work curve as indicated by the dotted dark red curves in Figure \ref{fig:CyclesTh04Tc004}(A), i.e. approximately the work input transforms into the potential energy of the particle on the ascent branch and the potential energy on the descent branch transforms into work output. A derivation to explain the similarity is given in SI Appendix Sec. 6.C. The net work output in one cycle is the difference between the work output in the descent segment and the work input in the ascent segment. Because of the symmetry of the left minimum branch and the right one, if the cusp is before the middle instant of the engine cycle (the blue diamond), the descent segment is larger than the the ascent segment in height and work is outputted as a whole, cf. Figure \ref{fig:Wcyc_Vdr}(A3). 

On the other hand, if the cusp is after the middle instant of the cycle as in the homogeous low temperature case (SI Appendix Sec. 6.A), the descent segment of the work curve is smaller than the ascent segment in height and work should be inputted. This is the stick-slip process, where the ascent segment is in the `stick' stage and the descent segment is in the `slip' stage. In the cliff, potential energy transforms into kinetic energy and then the kinetic energy is dissipated into heat because of damping. In our PTSHE, this process is reversed: heat is first absorbed and transforms into kinetic energy, and then kinetic energy transforms into potential energy.

At homogeneous high temperature (SI Appendix Sec. 6.B), due to such large stochastic forces, the particle almost cannot feel the middle energy barrier and jumps between the two wells frequently after the right local minimum appears and before the left local minimum disappears so that the positive and negative harmonic forces (and also the work input and output) cancel out a lot in an entire cycle. This phenomenon is thermolubricity \cite{Thermolubricity} and will be revisited below. %Nevertheless, there is still work input and the mean friction force is positive in consistency with the second law of thermodynamics.

% Because the particle sometimes stays on the right local minimum where the harmonic force is negative, the average harmonic force is less than that if it oscillates around the left one completely. So the slope of the work curve rises slower than the potential curve.

At this point we can summarize that the mechanism for the PTSHE to output work has two features: (1) the cusp point of the work curve is before the middle instant of one cycle; (2) the shape of the two segments of the work curve is similar to that of the two local minimum branches of the balanced resultant potential energy curve, cf. Figure \ref{fig:Wcyc_Vdr}(A3). The two features are realized by the special geometry of the balanced resultant potential energy and the spatially alternative high and low temperature field coupled with it. This mechanism will be referred to as potential mechanism. And there is still a second mechanism for work output related to the thermolubricity effect to be elucidated deeply below.

\subsection*{Mean Cycle Work with Respect to Driving Velocity at Different Parameters}

\begin{figure}[H]
\centering
\includegraphics[width=0.49\textwidth]{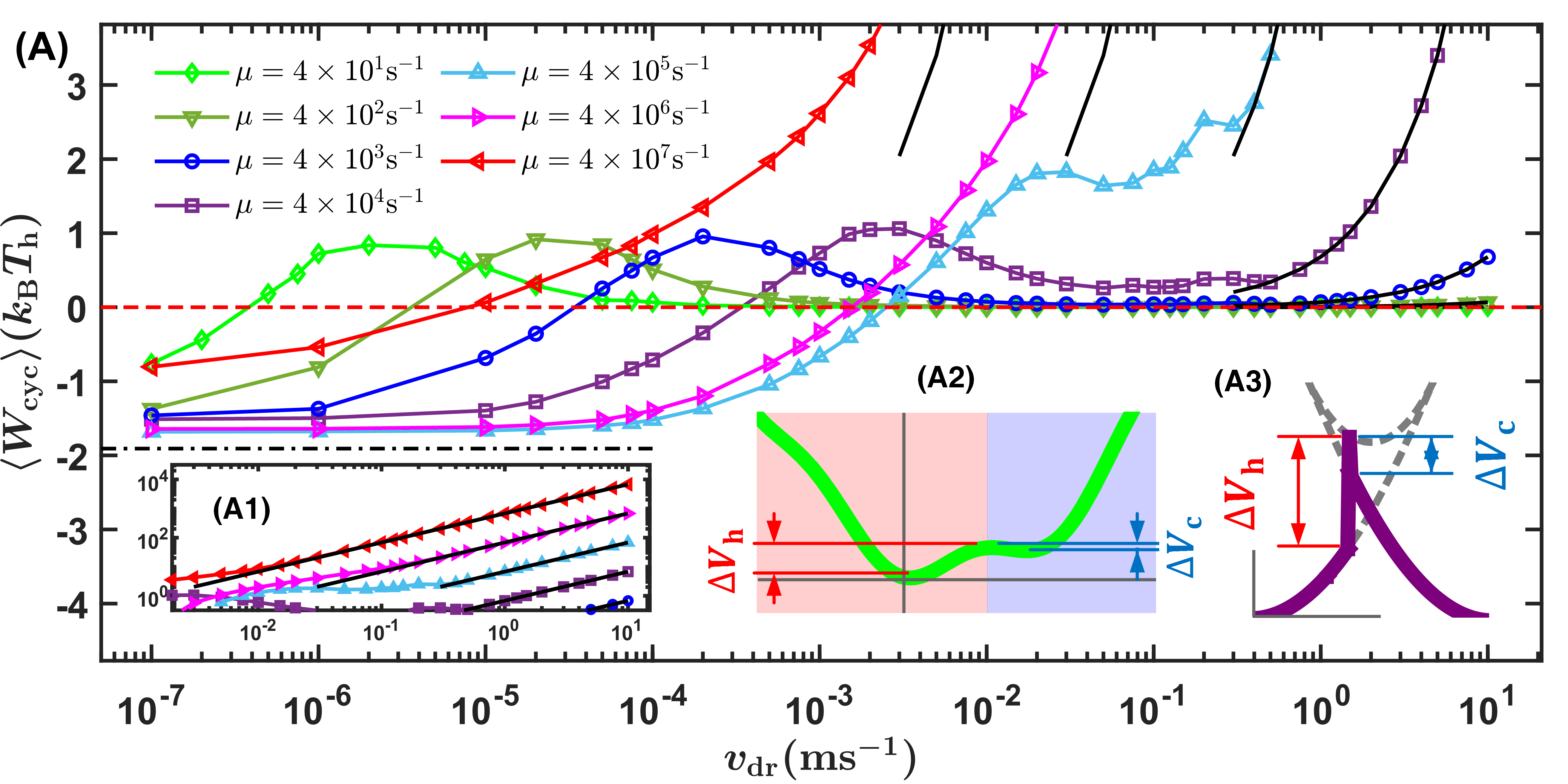}\\
\includegraphics[width=0.49\textwidth]{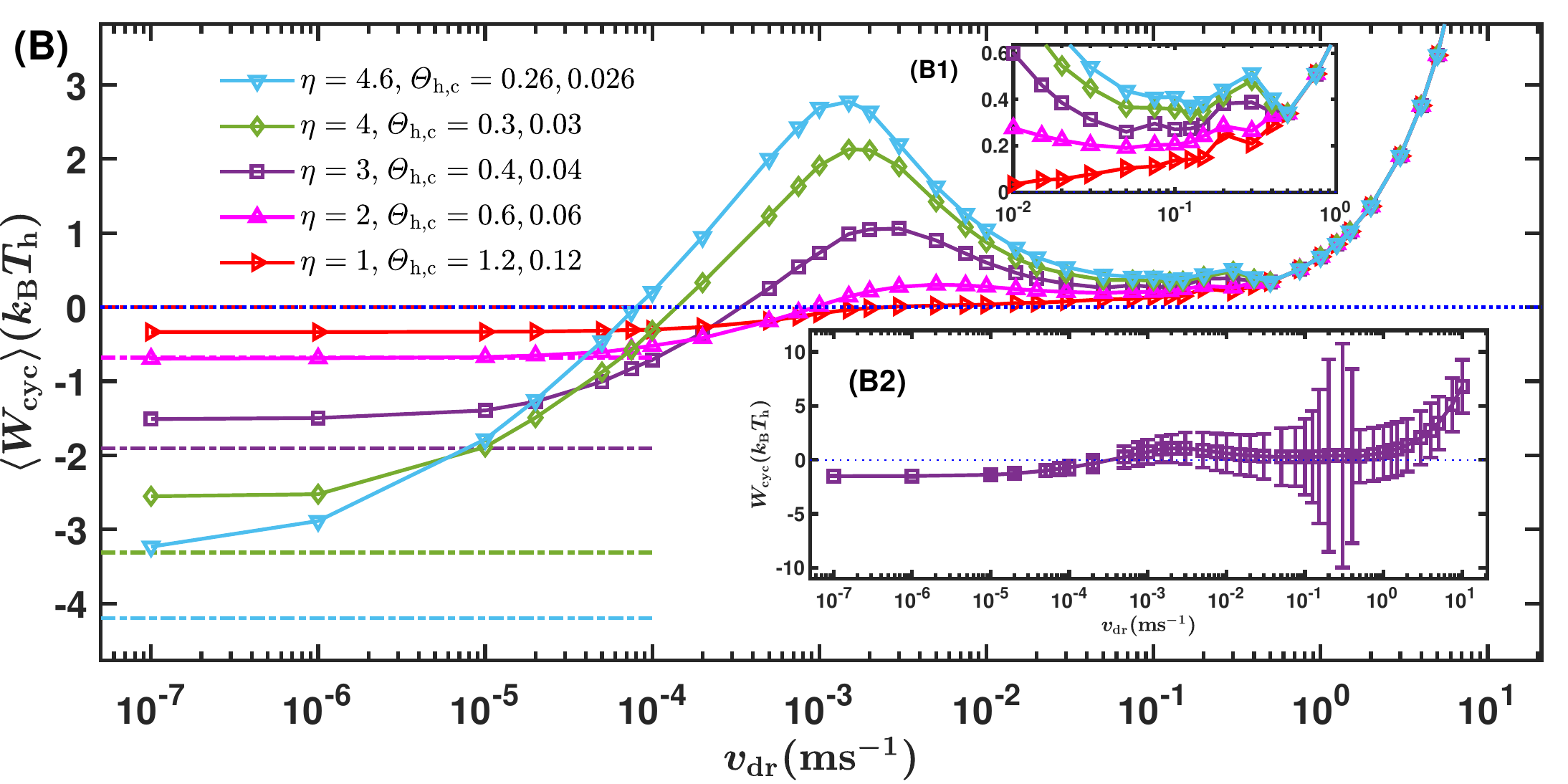}\\\includegraphics[width=0.49\textwidth]{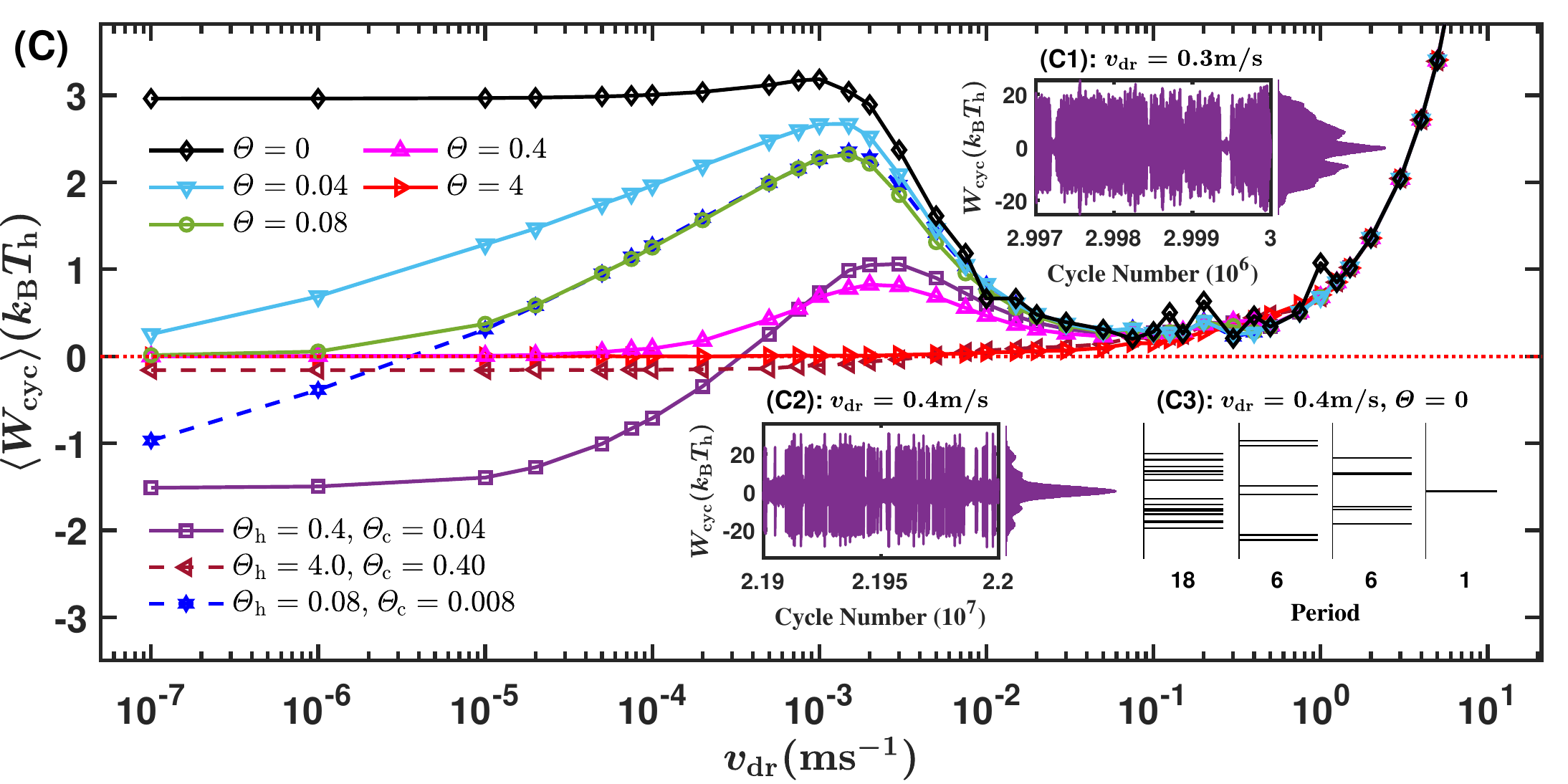}
\caption{The mean cycle work with respect to the driving velocity at different parameters. (A) The damping coefficient $\mu$ varies with $\eta=3.0,\ {\it\Theta}_{\rm h,c}=0.4,0.04$. The solid black curves on the right are the high velocity limit of $\langle W_{\rm cyc}\rangle$ (SI Appendix Sec. 6.D). The high-velocity parts of the curves outside the figure are plotted with log-log coordinate in the inset (A1). The dot-dashed line on the left is the equilibrium limit of the mean cycle work approximated based on the potential mechanism: $W_{\rm cyc,e.p.}$. The green curve in the inset (A2) represents the resultant potential curve at the mean cusp instant when the probability of the particle jumping forward over the middle energy barrier and that of it jumping backward are equal. The inset (A3) is the schematic describing how to calculate $W_{\rm cyc,e.p.}$. (B) $\eta$ varies with $\mu=4\times10^4\rm s^{-1}$. The absolute temperatures $T_{\rm h,c}$ of the heat bath are the same, while the nondimensional temperatures ${\it\Theta}_{\rm h,c}=T_{\rm h,c}/V_0$ are different for different $\eta$ because of the different $V_0$ at different $\eta$ with $\kappa$ keeping constant (SI Appendix Sec. 8.H). The dot-dashed lines on the left represent $W_{\rm cyc,e.p.}$ at different $\eta$ with the same color as the corresponding $\langle W_{\rm cyc}\rangle-v_{\rm dr}$ curve. The inset (B1) shows the detail of the curves at $v_{\rm dr}\in[10^{-2},10^0]\rm m/s$. The inset (B2) gives the standard deviations of $W_{\rm cyc}$ for the case of $\eta=3.0,\ {\it\Theta}_{\rm h,c}=0.4,0.04$. (C) The temperature field of the heat bath varies with $\eta=3.0,\ \mu=4\times10^4\rm s^{-1}$. The insets (C1) and (C2) plot the cycle work near the end of the simulation time range and its count distribution at $v_{\rm dr}=0.3\rm m/s$ and $0.4\rm m/s$ respectively 
%on the purple curve 
for the case of ${\it\Theta}_{\rm h,c}=0.4,0.04$ (SI Appendix Figure S40 in Sec. 6.I), and the inset (C3) represents the cycle work count distribution of the four solutions at $v_{\rm dr}=0.4\rm m/s$ and ${\it\Theta}=0$ with different periods, cf. SI Appendix Figure S42 in Sec. 6.J. The six subfigures in (C2) and (C3) share the same $W_{\rm cyc}$ axis on the left. The dimension of all the $\langle W_{\rm cyc}\rangle$ is $k_{\rm B}T_{\rm h}$, with $T_{\rm h}$ the high absolute temperature of the case of $\eta=3.0$, $\mu=4\times10^4s^{-1}$ and ${\it\Theta}_{\rm h,c}=0.4,0.04$ [the purple curve with squares in (A) ,(B) and (C)].}
\label{fig:Wcyc_Vdr}
\end{figure}

The mean cycle work 
\begin{equation}\label{eqn:wcyc}
\langle W_{\rm cyc}\rangle=-\langle\int_{t_0}^{t_0+\frac a{v_{\rm dr}}}\kappa[x(t)-v_{\rm dr}t]v_{\rm dr}\mathrm dt\rangle
\end{equation}
with respect to the driving velocity $v_{\rm dr}$ is plotted in Figure \ref{fig:Wcyc_Vdr} at different parameters.

In Figure \ref{fig:Wcyc_Vdr}(A), we change the damping coefficient $\mu$ from $0$ to $4\times10^7\rm s^{-1}$. When $\mu$ is as high as $4\times10^6\rm s^{-1}$, the $\langle W_{\rm cyc}\rangle-v_{\rm dr}$ curve is nearly exponential, indicating it's in the overdamped regime, and there is stable work output at low driving velocity $v_{\rm dr}$. In the underdamped regime where $\mu\lesssim 4\times10^5\rm s^{-1}$, there is stable work output at low $v_{\rm dr}$ too and the work output in both the overdamped and underdamped regimes seems approximately identical, cf. SI Appendix Sec. N.3.

In Figure \ref{fig:Wcyc_Vdr}(B), we can see that at low $v_{\rm dr}$, the mean cycle work has a limit which gets lower with $\eta$ increasing from $1$ to $4.6$, i.e. we can get more work output by increasing $V_0$ with constant $\kappa$ (SI Appendix, Sec. 8.H). This is the significance of the lattice potential in the PTSHE, cf. SI Appendix Sec. F.3.

In both Figure \ref{fig:Wcyc_Vdr}(A) and (B), we can see that work output occurs at low driving velocity. Therefore low driving velocity, i.e. qusi-equilibrium, is also a condition for stable work output. We can derive an approximation of the equilibrium limit of the mean cycle work based on the potential mechanism. Assume that the boundary of the high and low temperature zone is at the barrier peak point exactly [Figure \ref{fig:Wcyc_Vdr}(A2)]. At the mean cusp instant, the probability of the particle jumping forward over the middle energy barrier from the hot zone is equal to the probability of it jumping backward from the cold zone. Therefore at equilibrium the jumping forward and backward barriers $\Delta V_{\rm h,c}$ and the high and low temperatures $T_{\rm h,c}$ should satisfy 
$
\exp(-\frac{\Delta V_{\rm h}}{k_{\rm B}T_{\rm h}})=\exp(-\frac{\Delta V_{\rm c}}{k_{\rm B}T_{\rm c}})
$ (SI Appendix Sec. 6.E),
leading to 
\begin{equation}\label{eqn:carnotlikeformula}
\frac{\Delta V_{\rm h}}{\Delta V_{\rm c}}=\frac{T_{\rm h}}{T_{\rm c}},
\end{equation}
which is reminiscent of Carnot's formula if $\Delta V_{\rm h}$ and $\Delta V_{\rm c}$ are replaced by the absorbed and released heat $Q_{\rm h}$ and $Q_{\rm c}$ in the hot and cold heat bath respectively.
%It seems not as easy to obtain the efficiency of the PTSHE in the underdamped case as that in the overdamped case \cite{BLConstForcePRE,BLConstForcePAStochEnergetics,BrownianHeatEngineConstForceEPJB}, in that the particle oscillates violently and traverses through different temperature zones all the time so that it's not easy to calculate from the Langevin dynamics simution results the heat absorbed in the hot heat bath $Q_{\rm h}$ and that released in the cold one $Q_{\rm c}$, although not impossible. 
In line with the stochastic thermodynamics, the cycle heat absorbed from the hot heat bath should be calculated by $Q_{\rm h}=\int_{\rm hot\ zone}\text{\dj} Q=\int_{\rm hot\ zone}[m\mu\dot{x}(t)-\xi(t)]\mathrm dx$ [\cite{ConversionHeatWorkReviewPR} and SI Appendix Sec. 5], which is a little tough to calculate in that the particle fluctuates violently and traverses through different temperature zones all the time and the integral intervals are divided into many random non-overlapping subintervals. The same case occurs to the heat released to the cold heat bath $Q_{\rm c}$. To circumvent this difficulty, here we can approximate $Q_{\rm h,c}$ by $\Delta V_{\rm h,c}$ and then the Carnot-like limit efficiency for the PTSHE can be defined as $\Lambda=1-T_{\rm c}/T_{\rm h}$, which is an approximation \cite{EfficiencyBrownianHeatEngine}.
 
With Eq. \ref{eqn:carnotlikeformula} and the geometry of the balanced resultant potential curve, the appximate cusp instant can be calculalted. And considering the similarity of the shape of the work curve and that of the lower bound of the internal energy, the equilibrium limit of the mean cycle work output can be approximated by $W_{\rm cyc,e.p.}=\Delta V_{\rm h}-\Delta V_{\rm c}$, cf. Figure \ref{fig:Wcyc_Vdr}(A3). The subscript e. of $W_{\rm cyc,e.p.}$ represents equilibrium and p. represents potential indicating that $W_{\rm cyc,e.p.}$ is approximated based on the potential mechanism. A nonlinear equation system should be solved to calculate $W_{\rm cyc,e.p.}$ (SI Appendix Sec. 6.E) and the obtained value of $-W_{\rm cyc,e.p.}$ is represented by the dot-dashed lines on the left of Figure \ref{fig:Wcyc_Vdr}(A) and (B). As $W_{\rm cyc,e.p.}$ is determined by $T_{\rm h}/T_{\rm c}$ and $\eta$ and independent on $\mu$, there is only one $W_{\rm cyc,e.p.}$ in Figure \ref{fig:Wcyc_Vdr}(A) and five different $W_{\rm cyc,e.p.}$'s in Figure \ref{fig:Wcyc_Vdr}(B). For $\eta\gtrsim3.0$, $W_{\rm cyc,e.p.}$ is an upper bound, partly because the boundary of high and low temperature is designate on the left of the barrier peak, resulting in the particle frequently fluctuating to the low temperature zone and can't lanch to the right until the energy barrier is lower. If we set the boundary at the barrier peak point once it appears, the results will be closer to $W_{\rm cyc,e.p.}$. For $\eta\lesssim2.0$, the simulating results is larger than $W_{\rm cyc,e.p.}$, especially at $\eta=1.0$ where the approximate $W_{\rm cyc,e.p.}$ should be 0. The reason is that the other mechanism of work output takes effect and we will analyze it in the next subsection. 
%So the Carnot-like efficiency $\Lambda=1-T_{\rm c}/T_{\rm h}$ of our PTSHE is not always a lower bound. 
At homogeneous finite temperature $ W_{\rm cyc,e.p.}$ is equal to zero, which is accurate as shown in Figure \ref{fig:Wcyc_Vdr}(C).

\subsection*{The Second Mechanism of Work Output and the Consequence of Its Excess}
In Figure \ref{fig:Wcyc_Vdr}(B2), the standard deviations of $W_{\rm cyc}$ of the case of $\eta=3.0$, $\mu=4\times10^4\rm s^{-1}$, ${\it\Theta}_{\rm h,c}=0.4,0.04$ are plotted% in the inset (B2)
. We can see that at low $v_{\rm dr}$, the standard deviations of $W_{\rm cyc}$ are small, indicating that each cycle work is nearly the same. The reason is that at very low driving velocity 
%or very high temperature (so that the kinetic energy is much higher than the energy barrier) 
the particle has enough time to relax to equilibrium state at each instant, so that 
%the time average is nearly the same as the ensemble average, i.e. 
the time integral of each cycle $W_{\rm cyc}$ (time average) is equal to the ensemble average over all the cycles $\langle W_{\rm cyc}\rangle$, and thus the standard deviation of $W_{\rm cyc}$ is zero. Therefore the nonzero standard deviations of $W_{\rm cyc}$ at higher velocities indicate nonequilibrium. So instead of a nonequilibrium ensemble of engine cycles [Figure \ref{fig:CyclesTh04Tc004}(D)], in terms of the equilibrium limit of the mean cycle work, we can analyze one cycle at a low driving velocity at which the standard deviation of $W_{\rm cyc}$ is small enough, whose weakness, nonetheless, is that it's not easy to obtain the displacement distribution at certain instant due to the large number of simulation time steps in one single engine cycle.

\begin{SCfigure}[\sidecaptionrelwidth][tbhp]
\centering
\includegraphics[width=0.30\textwidth]{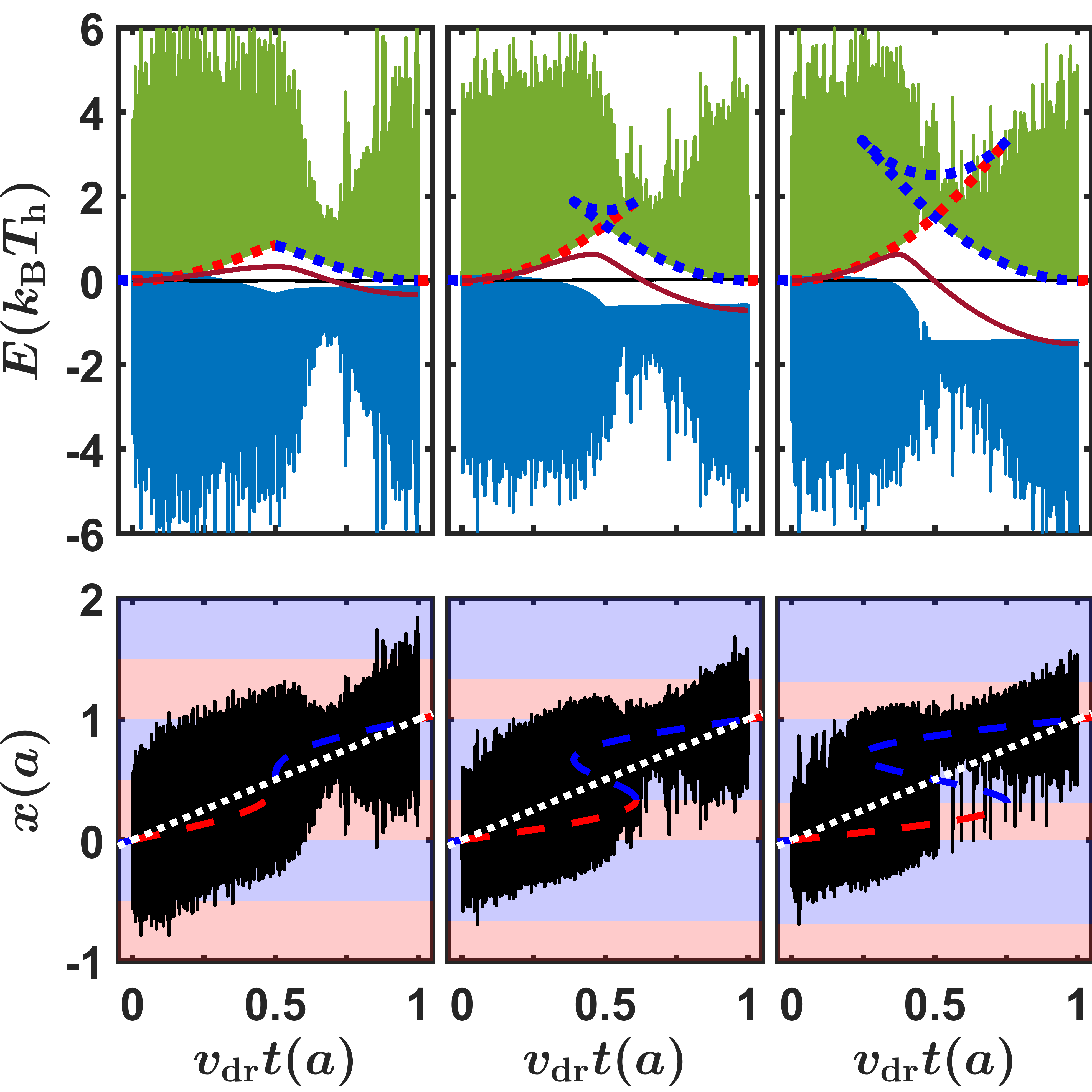}
 \caption{Energy and displacement of the particle at $v_{\rm dr}=10^{-7}\rm m/s$. From left to right $\eta=1,2$ and $3$ respectively. Other parameters are the same as those in Figure \ref{fig:Wcyc_Vdr}(B). Types and colors of the curves are the same as those in Figure \ref{fig:CyclesTh04Tc004}(A) and (B). $k_{\rm B}T_{\rm h}$ is equal to that in Figure \ref{fig:Wcyc_Vdr}. At low $\eta$, the low energy barrier leads to the particle's kinetic energy partly converted into its potential energy, so the work input is reduced before the cusp of the work curve.} 
 \label{fig:Eta1en2en3} 
 \end{SCfigure} 
 
In Figure \ref{fig:Eta1en2en3}, we plot the energy and displacement curves of the particle at $v_{\rm dr}=10^{-7}\rm m/s$ for the cases of $\eta=1,2$ and $3$ corresponding to those in Figure \ref{fig:Wcyc_Vdr}(B). For the middle $\eta=2$ case, we can see that the work curve before the cusp point is lower than the dotted red curve, different from that of the right $\eta=3$ case, so that the ascent segment of the work curve is shortened in height, resulting in the net cycle work output increasing. From SI Appendix Sec. 6.C we know that if the average position of the particle is on the balanced point exactly, the work curve will overlap with the dotted balanced resultant potential curves. However, from the bottom displacement curve, we can see that the average position of particle approaches the driver center (the white dotted curve) before the cusp instant. Although at the same absolute temperature $T_{\rm h}$ as the $\eta=3$ case, the nondimensional temperature ${\it\Theta}_{\rm h}$ of the $\eta=2$ case relative to the lattice potential amplitude $V_0$ is higher ($0.6$ v.s. $0.4$), resulting in the particle's easily crossing over the energy barrier, its tending to neglect the lattice potential and its distribution center approaching to the driver center (SI Appendix Sec. 6.F.1). Therefore the harmonic force and thus the slope of the work curve both reduce so that the work curve gets lower.

For the left $\eta=1$ case, although there is no energy barrier and only one global minimum first immersed in the hot and then in the cold zone, represented by the dotted red and blue curves (SI Appendix Sec. 6.F.2), the particle still fluctuates around the minimum point and the work curve is still similar to the dotted curves especially in the second half of the cycle. In the first half, the work curve is gradually lower than the dotted red curve in that at $\eta=1$ there is no energy barrier and the particle almost cannot feel the lattice potential so the particle again approaches to the driver center like in the $\eta=2$ case and the harmonic force from the driver to the particle gets smaller and the inputted work is reduced.

In the left top subfigure of Figure \ref{fig:Eta1en2en3}, we can see that the upper bound of the dark blue heat curve of the $\eta=1$ case decreases before and then increases a little after the middle of the cycle. In the hot zone heat is absorbed and transformed into kinetic energy, leading to the particle fluctuating violently approaching to the driver center where the resultant potential is high. So the kinetic energy is transformed into potential energy and the work input needed to increase the resultant potential energy is reduced. After the middle instant, the minimum point of the reslutant potential is ahead of the driver center and there is still a short period when the particle frequently traverses the hot zone and approaches to the driver center from ahead. Therefore the high kinetic energy becomes resistance and the work output tends to stay unchanged in a short period 
%so that as the particle goes deep into the cold zone, 
and both the potential and the kinetic energy reduce and transform into heat (SI Appendix Sec. 6.F.2). So after the middle of the cycle the upper bound of the heat curve rises a little, which is smaller than the decreased value before the middle instant. In an entire cycle heat absorbed is larger than heat released and the net heat absorbed is converted into the net work output.

So we can see that although there is no energy barrier in the $\eta=1$ case so that the equilibrium limit of the cycle work output $W_{\rm cyc,e.p.}$ approximated by the potential mechanism is zero, due to the reduction of the ascent segment of the work curve caused by the high temperature and with the descent segment changed a little, there is still a small quantity of work output.

% the upper bound of the heat curve decreases gradually and the work curve is lower than the dashed red curve.
In the $\eta=2$ and $3$ cases in Figure \ref{fig:Eta1en2en3}, we can also see that before the middle of the engine cycle, heat is absorbed and gradually transformed into the particle's potential energy, leading to the work input reduced. It's more distinct for the $\eta=2$ case because the high nondimensional temperature and low energy barrier lead to the particle crossing over the energy barrier easily and its fluctuating center is near the driver center. For the $\eta=3$ case, the nondimensional temperature is low and the energy barrier is high so it's harder for the particle to cross over the barrier. Because at such a low driving velocity the particle has enough time to traverse the middle states between the two local minimum so that the particle is nearly at equilibrium at each instant and the upper bound of the heat curve decreases gradually rather than suddenly like that in Figure \ref{fig:CyclesTh04Tc004}(B). Nonetheless, the particle still stays most of the time on the right after and left before the cusp instant and there is still a cusp on the work curve, although a little dull. And the work input reduction is not that significant.

In all the three cases in Figure \ref{fig:Eta1en2en3}, after the cusp instant, the particle is mainly in the cold zone and on average fluctuates around the right local minimum point so the work curve is nearly parallel to the dotted blue curve and work is done by the particle to the driver and is outputted.

So we can conclude the second mechanism for work output is that the particle's position distribution center approaches to the driver center, so that the harmonic force and thus the work input reduces. This mechanism is realized by the low energy barrier or equivalently the particle's high kinetic energy, resulting in the particle's tending to neglect the lattice potential energy.
%So we can conclude that the second mechanism for work output is that in the high temperature zone the low forward energy barrier or equivalently the particle's high kinetic energy makes it tend to neglect the lattice potential energy
%%cross over the middle energy barrier more easily 
%and fluctuate violently with its position distribution center approaching to the driver center, leading to the harmonic force and thus the work input reduced. 
In this mechanism heat is absorbed and first transformed into kinetic energy, and then the kinetic energy transforms into the potential energy so that the work input needed to increase the potential energy is reduced. This effect is similar to thermolubricity \cite{Thermolubricity} and will be referred to as thermolubricity mechanism.

The equilibrium limit of the mean cycle work output $W_{\rm cyc,e.p.}$ we derived in the previous subsection only considers the potential mechanism and it gives us a skeleton with an ascent segment and a descent segment [Figure \ref{fig:Wcyc_Vdr}(A3)]. The thermolubricity mechanism, on the other hand, reduces the ascent segment. If we keep the low temperature ${\it\Theta}_{\rm c}$ constant and increase the high temperature ${\it\Theta}_{\rm h}$, the cycle work output from the potential mechanism will increase because of the cusp instant moving to the left [Figure \ref{fig:Wcyc_Vdr}(A3)], and due to the thermolubricity mechanism the work input will be reduced further. However, we cannot increase ${\it\Theta}_{\rm h}$ without limit. 
\begin{SCfigure*}[\sidecaptionrelwidth][tbhp]
\centering
\includegraphics[width=0.595\textwidth]{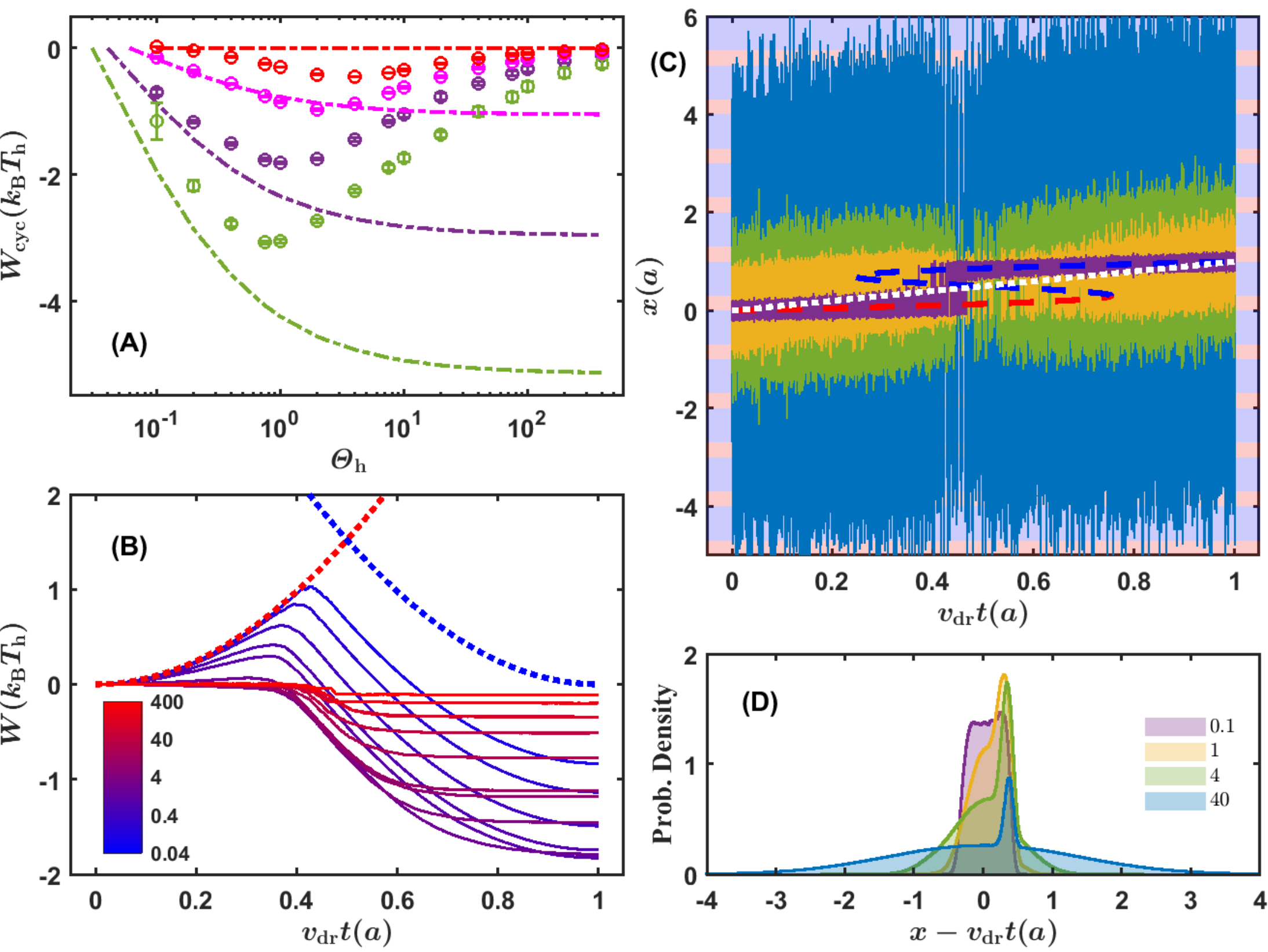}
\caption{The mean cycle work varies with the high temperature. (A) The cycle work $W_{\rm cyc}$ at $v_{\rm dr}=10^{-7}\rm m/s$ with respect to the nondimensional high temperature ${\it\Theta}_{\rm h}$ at different $\eta$. The circles with their errorbars represent the mean values and standard deviations averaged from 18 simulation cycles% of Langevin dynamics
. The dot-dashed curves represent the approximated equilibrium cycle work limit $-W_{\rm cyc,e.p.}$. From bottom to top, the curves and circles with the same color represent the cases of $\eta=4.0$, ${\it\Theta}_{\rm c}=0.03$; $\eta=3.0$, ${\it\Theta}_{\rm c}=0.04$; $\eta=2.0$, ${\it\Theta}_{\rm c}=0.06$ and $\eta=1.0$, ${\it\Theta}_{\rm c}=0.12$ respectively with $\mu=4\times10^{4}\rm s^{-1}$. The absolute low temperature $T_{\rm c}$ is the same for all the four cases [cf. Figure \ref{fig:Wcyc_Vdr}(B)]. (B) The work curves in one cycle corresponding to the purple circles of $\eta=3$, ${\it\Theta}_{\rm c}=0.04$ in (A), with the high temperature ${\it\Theta}_{\rm h}$ denoted by the gradually varying color of the curves. The dotted red and blue curves represent the resultant potential at the balanced points [cf. Figure \ref{fig:CyclesTh04Tc004}(A)]. In (A) and (B), $k_{\rm B}T_{\rm h}$ is equal to that in Figure \ref{fig:Wcyc_Vdr}. (C) The displacement of the particle in one cycle for the case of $\eta=3$, ${\it\Theta}_{\rm c}=0.04$ in (A) at four typical high temperatures: ${\it\Theta}_{\rm h}=0.1$(purple), $1$(dark yellow), $4$(dark green) and $40$(dark blue). The dotted white line denotes the position of the driver center. The dashed red and blue curves represent the position of the balanced points of the resultant potential [cf. Figure \ref{fig:CyclesTh04Tc004}(B)]. (D) The probability density distributions of the particle's relative displacement to the driver center $x-v_{\rm dr}t$ in one cycle at four typical ${\it\Theta}_{\rm h}$ for the case of $\eta=3$, ${\it\Theta}_{\rm c}=0.04$ in (A).} 
\label{fig:WvsT} 
\end{SCfigure*}
In Figure \ref{fig:WvsT}(A), %the circles with errorbar represent the simulation mean cycle work 
the circles represent the simulation mean cycle work at $v_{\rm dr}=10^{-7}\rm m/s$ varying with ${\it\Theta}_{\rm h}$. The absolute low temperature $T_{\rm c}$ of the four cases of $\eta=1,2,3$ and $4$ is the same. We can see that as %the nondimensional high temperature 
${\it\Theta}_{\rm h}$ increases, the mean cycle work at $v_{\rm dr}=10^{-7}\rm m/s$ first decreases and then increases for all the four cases. The dot-dashed curves represent %the equilibrium mean cycle work limit $W_{\rm cyc,e.p.}$ which takes only the potential mechanism into consideration and decreases to a lower limit with increasing ${\it\Theta}_{\rm h}$. 
the equilibrium mean cycle work limit $-W_{\rm cyc,e.p.}$ %considering only
approximated by the potential mechanism, which decreases to a lower limit with ${\it\Theta}_{\rm h}$ increasing (SI Appendix, Eq. 41). As we have explained, the offset of the descending stage of each set of circles from the corresponding dot-dashed curves of the same $\eta$ and ${\it\Theta}_{\rm c}$ results from the thermolubricity mechanism. 

We next explain that the ascending stage of each set of circles roots in the same mechanism. In Figure \ref{fig:WvsT}(B), the work curves in one cycle corresponding to the purple circles of $\eta=3$ in Figure \ref{fig:WvsT}(A) are plotted.
%except for the one of ${\it\Theta}_{\rm h}={\it\Theta}_{\rm c}=0.04$, which should overlap with the dashed red and blue curves and lead to vanishing mean cycle work. 
We can see that from ${\it\Theta}_{\rm h}=0.1$ to $1.0$, the cusps of the work curves move to the left because the ratio ${\it\Theta}_{\rm h}/{\it\Theta}_{\rm c}$ increases, cf. Figure \ref{fig:Wcyc_Vdr}(A3). At the same time the ascent segments of the work curve before the cusps gradually get lower and lower because of the thermolubricity mechanism. In this range of ${\it\Theta}_{\rm h}$, the descent segments of the work curves keep parallel to each other and also to the dotted blue curve. As a whole the cycle work output increases in this range of ${\it\Theta}_{\rm h}$. However, as ${\it\Theta}_{\rm h}\geq2$ the ascent segments of the work curves become nearly flat and the cusps almost disappear, in that the temperature is so high that the thermolubricity effect touches ground. Moreover, the descent segments are no longer parallel to the dotted blue curve and get flatter and flatter with ${\it\Theta}_{\rm h}$ increasing, leading to that the right ends of the work curves rise so that the cycle work output decreases. In Figure \ref{fig:WvsT}(C), we plot four typical displacement curves of ${\it\Theta}_{\rm h}=0.1,1,4$ and $40$ corresponding to the four respective purple circles at $\eta=3$ in Figure \ref{fig:WvsT}(A). We can see that when ${\it\Theta}_{\rm h}$ is high, the particle fluctuates violently with amplitute as high as several lattice periods as it is close to the end of the engine cycle because it frequently traverses through the high temperature zone in the next lattice period. So the particle's mean position almost overlaps with the driver center rather than the right local or global minimum point. The harmonic force then is reduced to near zero so that the absolute slope of the work curve decreases to near zero and it becomes flat at its right end. As ${\it\Theta}_{\rm h}$ goes higher, the range of the particle's no feeling of the lattice potential extends so that the residual descent segment of the work curve shrinks and the cycle work output decreases further. 

The quantity $x-v_{\rm dr}t$, which is the integrand of $W_{\rm cyc}$ in Eq. \ref{eqn:wcyc} neglecting a constant, can be regarded as the particle's displacement relative to the driver center. In Figure \ref{fig:WvsT}(D), we plot the probability density distributions of the $(x-v_{\rm dr}t)/a$ value in an engine cycle at $v_{\rm dr}=10^{-7}\rm m/s$ at four ${\it\Theta}_{\rm h}$. The average value of the distribution is proportional to the cycle work $W_{\rm cyc}$. %the area under which is proportional to the cycle work $W_{\rm cyc}$. 
At ${\it\Theta}_{\rm h}=0.1$, there is a little pingo on the right because the nonhomogeneous temperature causes the particle jumps forward to the right local minimum a little earlier so that in an entire cycle the particle tends to distribute a little more on the right of the driver center. From ${\it\Theta}_{\rm h}=0.1$ to ${\it\Theta}_{\rm h}=1$, the left plateau declines while the right hill rises so that the particle distributes more on the right of the driver center and the cycle work increases. When ${\it\Theta}_{\rm h}>1$, however, both sides of the distribution curve expand so that the right hill gets thin and short in that ${\it\Theta}_{\rm h}$ is so high that the work output part is diluted.

In Figure \ref{fig:WvsT}(A), all the standard deviations of the mean cycle work are small except the point at ${\it\Theta}_{\rm h}=0.1$ of the case of $\eta=4$, ${\it\Theta}_{\rm c}=0.03$ because the temperature is so low that the velocity $v_{\rm dr}=10^{-7}\rm m/s$ is not low enough for the particle to relax to equilibrium with such a high middle energy barrier. As we have stated above, the small standard deviations at low driving velocity indicate equilibrium.

In SI Appendix, Sec. 6.G, we give the results corresponding to those in Figure \ref{fig:WvsT}(B), (C) and (D) for the cases of $\eta=1$, $2$ and $4$ for reference. 
 
%  Therefore, there are two mechanisms of work output, from potential energy and from kinetic energy. At large $\eta$ in $[1,4.6]$, the former dominates and because of the high energy barrier, the latter is suppressed. Nonetheless, we can still see that in Figure \ref{fig:CyclesTh04Tc004}(A) before the cliffs of $Q$, there are slow descent stages, corrosponding to which the internal energy $U$ fluctuats violently and the work curve $W$ increases more slowly than the resultant potential. And that is rooting in the transformation from kinetic energy to work output. On the contrary, as $\eta$ gets lower in $[1,4.6]$, the energy barrier decreases and the latter plays a more and more important role while the former is suppressed.

%\begin{figure}[H]
%\centering
%% \includegraphics[width=0.3935\textwidth]{Figures/Wcyc_limit.pdf}
% \includegraphics[width=0.50\textwidth]{Figures/StickSlips.pdf}
% \caption{} 
% \label{fig:Wcyclimit} 
% \end{figure}

\subsection*{The Stall Regime of the PTSHE: Connection with Nanofriction}
As the driving velocity increases exceeding the zero points of the $\langle W_{\rm cyc}\rangle-v_{\rm dr}$ curves, the PTSHE stalls \cite{Stirling,SchmiedlStochaHeatEngine}, i.e. work has to be inputted to the particle by the driver. 
%We analyze this regime in detail and elucidate the connection between the PTSHE and nanofriction in SI Appendix, Sec. 6.H.
When the damping coefficient $\mu$ is small (underdamped), at the stall regime, the $\langle W_{\rm cyc}\rangle-v_{\rm dr}$ curve first increases to a plateau and then decreases to a valley and finally increases to $+\infty$ [Figure \ref{fig:Wcyc_Vdr}(A)]. The position of the plateau moves to the left being approximately proportional to the decreasing $\mu$ and the range of the valley expands both to the left and right. The plateau also rises and moves to the left as $\eta$ increases at the same absolute temperature [Figure \ref{fig:Wcyc_Vdr}(B)].  When $\mu$ is large enough (overdamped) the plateau and the valley both vanish. At very high driving velocity the mean cycle work increases linearly [Figure \ref{fig:Wcyc_Vdr}(A1)] because the damping force dominates and thus an analytical expression can be derived (SI Appendix Sec. 6.D) represented by the solid black curves in Figure \ref{fig:Wcyc_Vdr}(A). It is independent on $\eta$ and ${\it\Theta}_{\rm h,c}$ indicated by the coincident part of the $\langle W_{\rm cyc}\rangle-v_{\rm dr}$ curves in Figure \ref{fig:Wcyc_Vdr}(B) and (C).

%At very high velocity the mean cycle work increases exponentially because of the great damping force, which depends solely on the damping coefficient, as can be seen by the black curves in Figure \ref{fig:Wcyc_Vdr}(A) (SI Appendix, Sec. 6.D) and the coincidence of the $\langle W_{\rm cyc}\rangle-v_{\rm dr}$ curves in Figure \ref{fig:Wcyc_Vdr}(B) and (C).

In homogeneous (not very high) temperature, the $\langle W_{\rm cyc}\rangle-v_{\rm dr}$ curves in Figure \ref{fig:Wcyc_Vdr}(C) are characteristic in nanofriction \cite{NonMonoVelDependenceAF,VelFExcitation,PNASArticularjoints}. The mean cycle work $\langle W_{\rm cyc}\rangle$ is equal to the mean harmonic (minus friction) force $\langle \bar F_{\rm h}\rangle$ times the lattice period $a$. The $\langle W_{\rm cyc}\rangle-v_{\rm dr}$ curve goes to zero at low driving velocity in homogeneous finite temperature [Figure \ref{fig:Wcyc_Vdr}(C)] corresponding to the thermal drift regime of nanofriction \cite{KryPhysSF}. As the temperature reduces, the thermal drift regime shrinks to the left and disappears at zero temperature. Except for the negative mean cycle work, the $\langle W_{\rm cyc}\rangle-v_{\rm dr}$ curves of the PTSHE in Figure \ref{fig:Wcyc_Vdr}(A) and (B) are similar to those of homogeneous finite temperature heat baths in Figure \ref{fig:Wcyc_Vdr}(C). At the ascent stage after the zero point and before the plateau peak, stick-slip occurs so that work has to be inputted to compensate the energy dissipation. After the plateau peak, stick-slip is weakened because of the interference of the not completely attenuated oscillation from the last cycle \cite{VelocityWeakening,NPVelocityTuning}. We analyze the transition of the energy, displacement and harmonic force in one cycle with the driving velocity in SI Appendix, Sec. 6.H.

\subsection*{Bifurcation of Mean Cycle Work with Driving Velocity as Parameter at Zero Temperature}

\begin{figure}[H]
\centering
\includegraphics[width=0.48\textwidth]{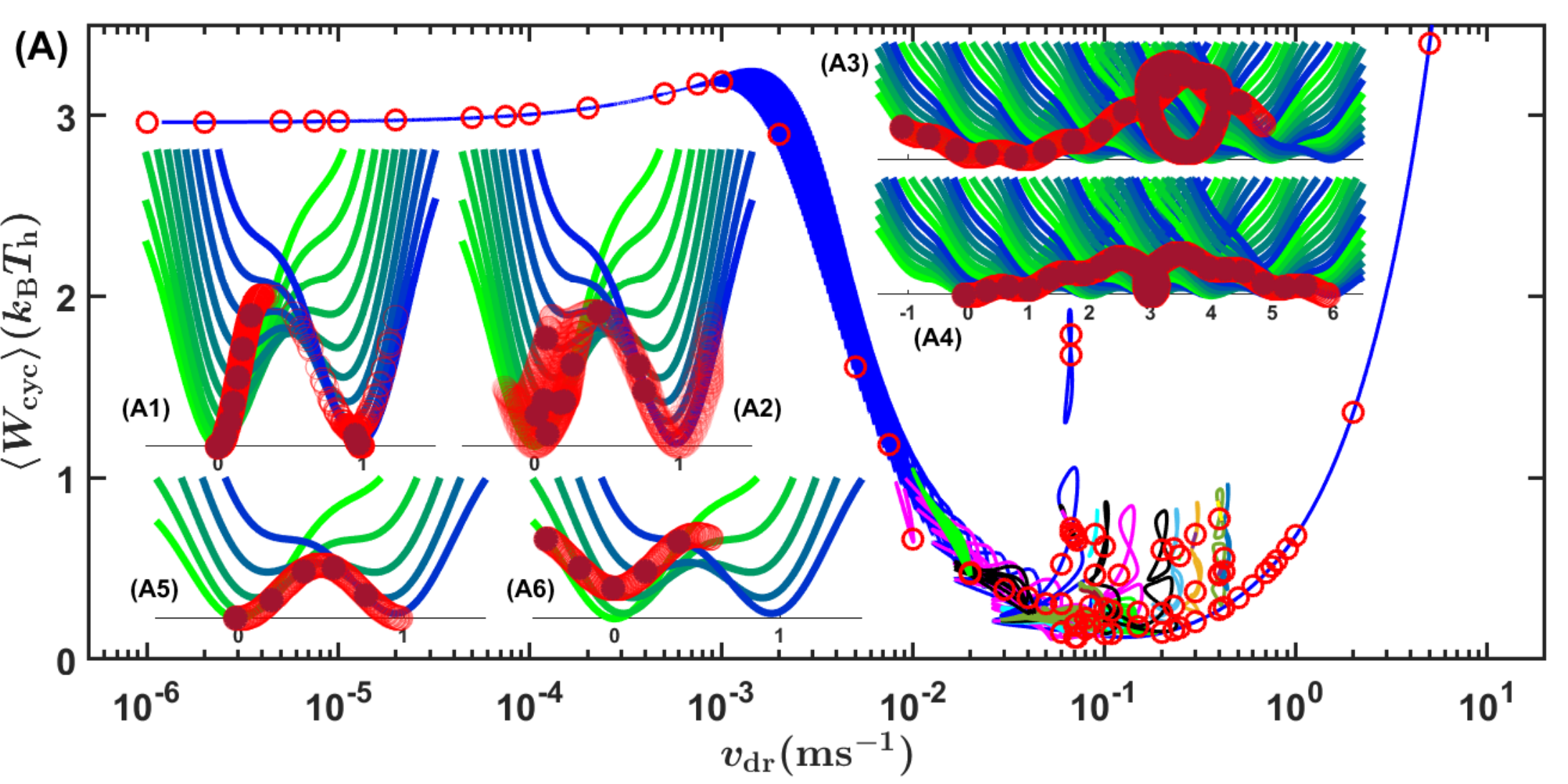}\\
\includegraphics[width=0.48\textwidth]{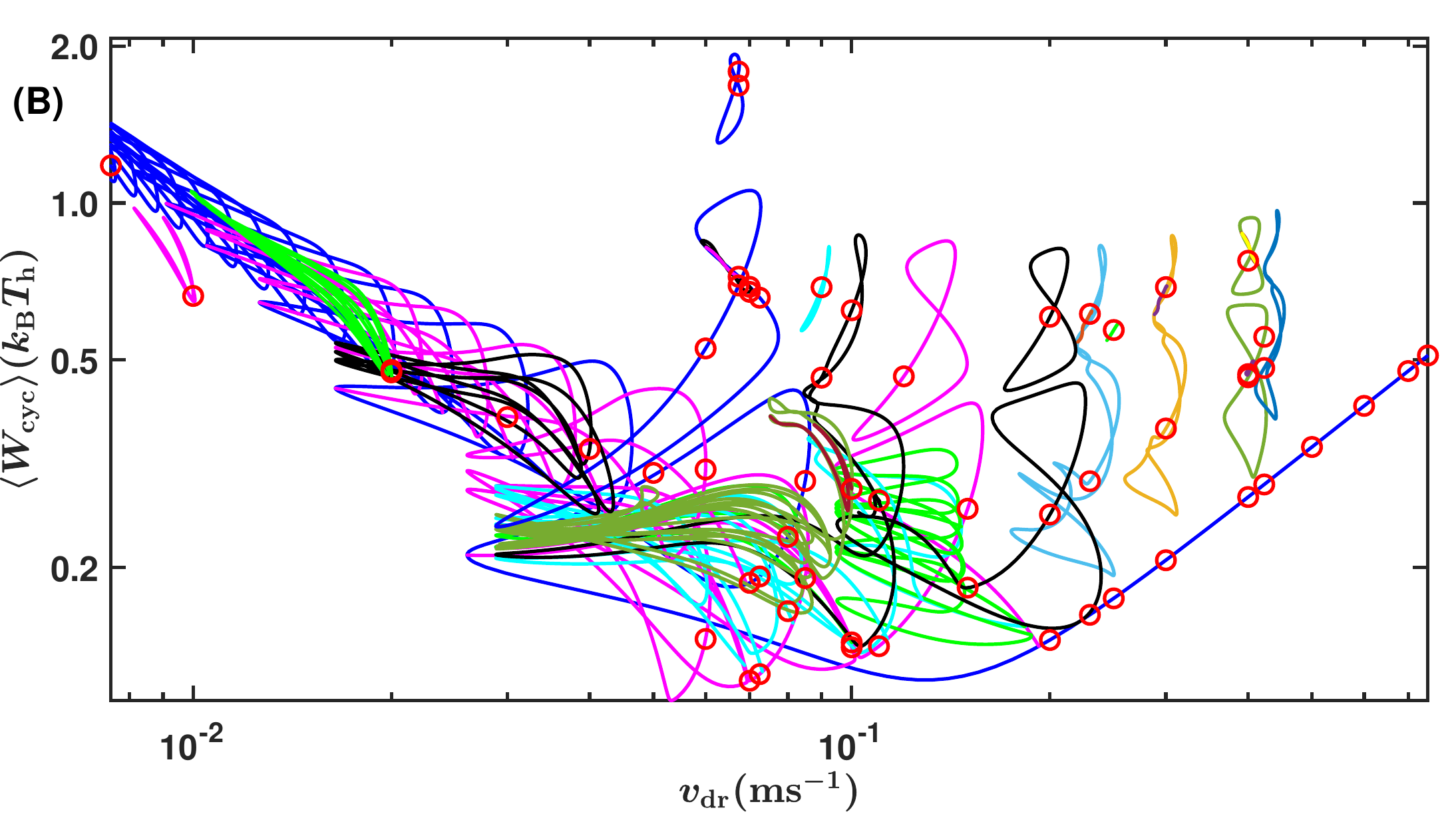}
\caption{The bifurcation diagram of $\langle W_{\rm cyc}\rangle$ with $v_{\rm dr}$ as parameter at zero temperature. In (A), the red circles are simulation results and the solid curves are results of continuation. Curves with different color have different cycle number period $P_{\rm c.n.}=Tv_{\rm dr}/a$. (B) is a zoomin of (A). The dimension $k_{\rm B}T_{\rm h}$ of $\langle W_{\rm cyc}\rangle$ is the same as that in Figure \ref{fig:Wcyc_Vdr}. More details of the upper one are given in SI Appendix Sec. 6.K.2. Here $\eta=3.0,\mu=4\times10^{-1}\rm s^{-1}$ and other parameters are given in SI Appendix Sec 8.H. The six insets in (A) are schematics of the particle's position (dark red balls) projected on the resultant potential curves (with color gradually varying from green to blue during one engine cycle) varying with time. (A1), (A2) and (A6) correspond to $v_{\rm dr}=10^{-5}\rm m/s$ ($P_{\rm c.n.}=1$, stick-slip), $0.003\rm m/s$ (no period, velocity weakening) and $10\rm m/s$ ($P_{\rm c.n.}=1$) respectively in one engine cycle, cf. SI Appendix Figure S25. (A3), (A4) and (A5) correspond to three solutions at $v_{\rm dr}=0.4\rm m/s$, with $P_{\rm c.n.}=6,\ 6$ and $1$ respectively in Figure \ref{fig:Wcyc_Vdr}(C3), cf. SI Appendix Sec. 6.J.} 
\label{fig:Bifurcation} 
\end{figure}
 
In Figure \ref{fig:Wcyc_Vdr}(B1) and (C) we find that the $\langle W_{\rm cyc}\rangle -v_{\rm dr}$ curves are not smooth in the velocity range $(7.5\times10^{-3},3)\rm m/s$ where there are some small peaks especially on the zero temperature curve. 
%and at these peaks the cycle work has no period [Figure \ref{fig:Wcyc_Vdr}(C1)] or period larger than one [Figure \ref{fig:Wcyc_Vdr}(C2)] rather than period one [Figure \ref{fig:Wcyc_Vdr}(C3)]. 
What's more, in this velocity regime, the standard deviations are large [Figure \ref{fig:Wcyc_Vdr}(B2)] and the count distributions of $W_{\rm cyc}$ have some peculiar peaks [Figure \ref{fig:Wcyc_Vdr}(C1) and (C2), and SI Appendix Sec. 6.I]. We then simulated more results for the zero temperature case (the differential equation of which is ordinary), and found that in this regime, the mean cycle work $\langle W_{\rm cyc}\rangle$ has multiple solutions with different limit cycles dependent on the initial conditions, as shown by more than one $\langle W_{\rm cyc}\rangle$ value (the red circles) at some $v_{\rm dr}$ in Figure \ref{fig:Bifurcation}, cf. SI Appendix Sec. 6.J. Different solution may have different period and solutions with the same period may lead to different $\langle W_{\rm cyc}\rangle$, e.g. the four solutions at $v_{\rm dr}=0.4\rm m/s$ shown in Figure \ref{fig:Wcyc_Vdr}(C3) and three of them in Figure \ref{fig:Bifurcation}(A3), (A4) and (A5). That's a kind of nonlinear bifurcation resulting from the sinusoidal lattice potential of the PT model. Through continuation with MatCont \cite{MatcontRef}, we can obtain the bifurcation diagram of $\langle W_{\rm cyc}\rangle$ with $v_{\rm dr}$ as the parameter in Figure \ref{fig:Bifurcation} (SI Appendix Sec. 6.K).

%At finite temperature, in the strong nonlinear regime it takes the particle a long time to traverse all the stable solutions at a specific velocity, so that we have simulate a longer time to obtain stable mean and standard deviation values.

We can see that at zero temperature, the $\langle W_{\rm cyc}\rangle-v_{\rm dr}$ curve shows strong nonlinearity. At low and high driving velocity ends there is only one solution with period $T=a/v_{\rm dr}$ for each $v_{\rm dr}$ [Figure \ref{fig:Bifurcation}(A1) and (A6)]. While $v_{\rm dr}$ is in $(8\times10^{-4},0.2)\rm m/s$, the blue backbone curve with period $T=a/v_{\rm dr}$ twines up from right to left, on which there are a lot of period doubling branches \cite{Seydelbook}, among which we plot only a small part (more detail is given in SI Appendix Figure S44). In the velocity range $(0.02,0.5)\rm m/s$, there're a lot of isolated loops with different cycle number period $P_{\rm c.n.}=Tv_{\rm dr}/a$ indicated by different color. And almost on all the isolated loops, there are period doublings, only several of which are ploted. From Figure \ref{fig:Bifurcation}(B), we can see that it's nearly chaotic when $v_{\rm dr}$ is in the interval $(0.025,0.09)\rm m/s$.

When the frequency of the driver center's moving over the lattice periods, i.e. $v_{\rm dr}/a$, equals from the frequency of the harmonic potential $\frac{\omega_0}{2\pi}$ to the maximum frequency of the resultant potential $\frac{\omega_0}{2\pi}\sqrt{1+\eta}$, the driving velocity $v_{\rm dr}$ ranges from $0.0673$ to $0.1347\rm m/s$. This range is contained in the strong nonlinear bifurcation regime characterized by many isolated loops. So the strong nonlinear bifurcation is related to resonance. In Figure \ref{fig:Bifurcation}, we can actually see at $v_{\rm dr}=0.0673$ two red circles much higher than others and a clear resonance peak nearby. Moreover, in Figure \ref{fig:Wcyc_Vdr}(B2), we can see that in this regime the standard deviation of $W_{\rm cyc}$ has a peak at nearby $v_{\rm dr}=0.3\rm m/s$, indicating that stochastic resonance occurs \cite{StochasticResonance}. Compared with the example of the double-well potential superposed with a periodic driving in \cite{StochasticResonance}, in the PT model the particle has more than two states to switch among. Here the different states correspond to different steady-state periodic solutions of the zero temperature Langevin equation, which can be represented by the bifurcation diagram Figure \ref{fig:Bifurcation}, rather than the local minima of the potential like that in \cite{StochasticResonance}.
%, resulting from the bifurcation of $\langle W_{\rm cyc}\rangle$ at zero temperature
In Figure \ref{fig:Bifurcation}(B) we can see that when there is no noise (${\it\Theta}=0$) at $v_{\rm dr}=0.4\rm m/s$ there are more than two states, four of which are plotted in Figure \ref{fig:Wcyc_Vdr}(C3) represented by the steady-state count distribution of $W_{\rm cyc}$ and three of which are plotted in Figure \ref{fig:Bifurcation}(A3), (A4) and (A5) represented by the particle's position projected on the resultant potential curve varying with time at steady state. At finite temperature (e.g. ${\it\Theta}_{\rm h,c}=0.4,0.04$), the particle can switch induced by noise among these states so that $W_{\rm cyc}$ has a peculiar distribution in Figure \ref{fig:Wcyc_Vdr}(C2). At different $v_{\rm dr}$ the states are different, so the distributions of $W_{\rm cyc}$ at different $v_{\rm dr}$'s are different (SI Appendix Sec. 6.I). When stochastic resonance occurs, the particle's hopping among these states synchronizes with the driver center's moving over the lattice periods, leading to the standard deviation of $W_{\rm cyc}$ maximizes (SI Appendix Sec. 6.I).

\subsection*{Mean Cycle Work at $\eta>4.6$}

\begin{figure}[H]
\centering
\includegraphics[width=0.485\textwidth]{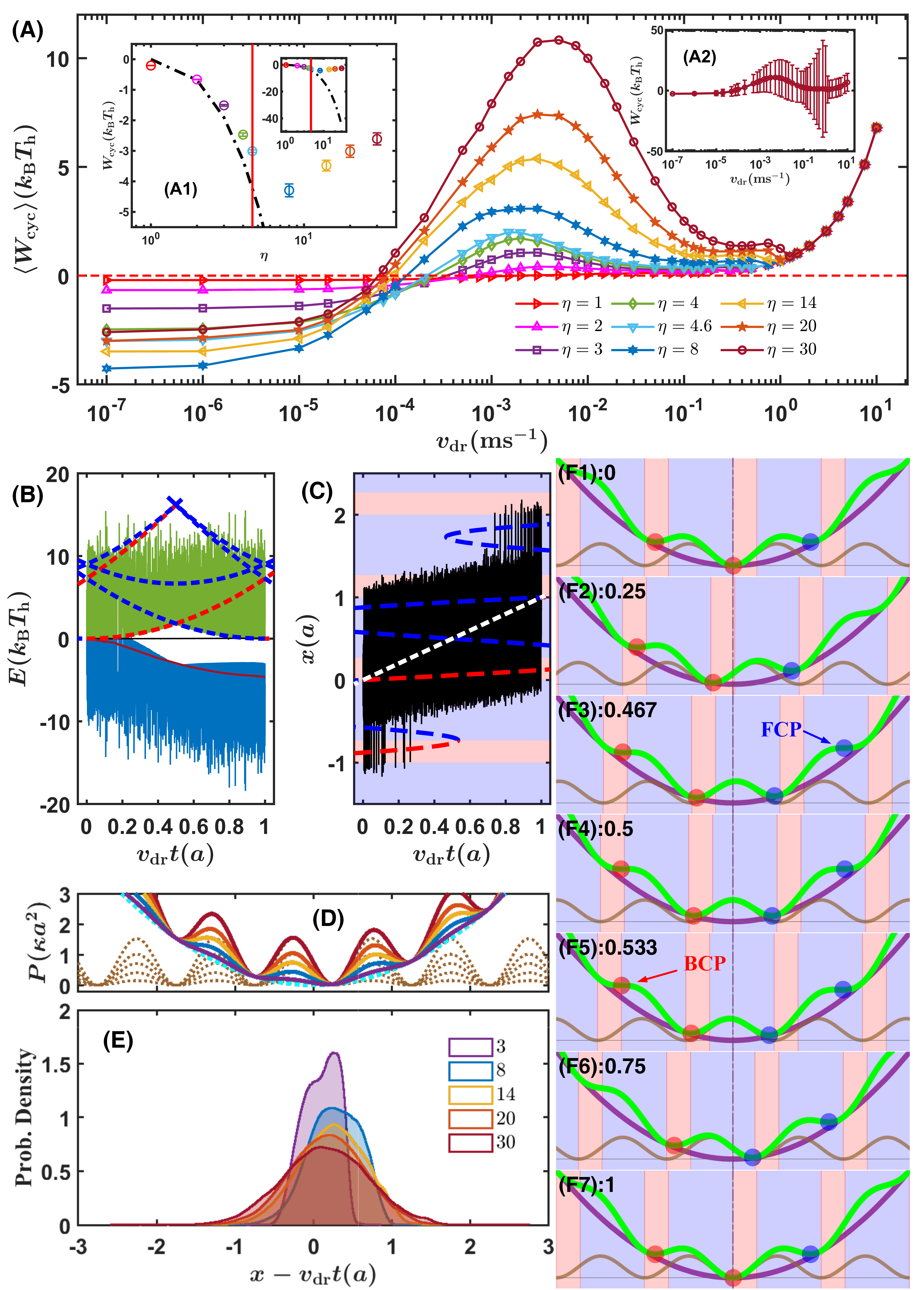}
\caption{Simulation results of the cases of $\eta>4.6$. (A) Mean cycle work at ${\it\Theta}_{\rm h,c}=0.4,0.04$, $\mu=4\times10^{4}\rm s^{-1}$ from $\eta=1$ to $\eta=30$. The circles with their errorbars in the inset (A1) are the mean values and standard deviations of $W_{\rm cyc}$ averaged from 18 simulation cycles at $v_{\rm dr}=10^{-7}\rm m/s$ with respect to $\eta$, and the colors of the circles are the same as the corresponding $\langle W_{\rm cyc}\rangle-v_{\rm dr}$ curves. As $\eta$ increases, the low velocity limit of $\langle W_{\rm cyc}\rangle$ first decreases and then increases. The black dot-dashed curve is the equilibrium cycle work approximated by the potential mechanism $-W_{\rm cyc,e.p.}$, which decreases to $-\infty$ with $\eta$ increasing [see the inset of (A1)]. The red vertical line represents the critical value $\eta=4.6033388487517$ (SI Appendix Sec. 4). The inset (A2) gives the standard deviations of $W_{\rm cyc}$ at $\eta=30$. (B) and (C) Energy and displacement curves at $\eta=8$, ${\it\Theta}_{\rm h,c}=0.4,0.04$ and $\mu=4\times10^{4}\rm s^{-1}$, curve colors and types are the same as those in Figure \ref{fig:CyclesTh04Tc004}(A) and (B). (D) The solid curves are the resultant potential curves at different $\eta$ denoted by different colors corresponding to those in (E). The cyan dotted curve represents the harmonic potential which is kept constant. The brown dotted curves represent the sinusoidal lattice potential leading to different $\eta$. (E) The probability density distribution of the particle's relative displacement to the driver center $x-v_{\rm dr}t$ %in one cycle 
at $v_{\rm dr}=10^{-7}\rm m/s$ and different $\eta$. %with the same colors as the corresponding circles at the same $\eta$ in (A1). 
Each distribution curve is obtained from three consecutive steady state engine cycles. (F1)-(F7) Diagrams of the resultant potential curves at different instants in one engine cycle of the $\eta=8$ case. Here we keep the harmonic potential stationary and move the lattice potential and the temperature field to the left uniformly, which is equivalent to keeping the latter two stationary and moving the former to the right uniformly like that in Figure \ref{fig:CyclesTh04Tc004}(D). The red balls denote the local minima in the hot zones and the blue ones denote those in the cold zones. The decimals following the id's are the relative instant to the starting of the cycle (F1).} 
\label{fig:HighEta} 
\end{figure}

In the above we only considered the cases of $\eta\leq4.6$, in which the resultant potential curves have at most two local minima. In this subsection, we will preliminarily consider the cases of $\eta>4.6$ with the resultant potential curves having more than two local minima (SI Appendix Sec. 4). In Figure \ref{fig:HighEta}(A), we plot the $\langle W_{\rm cyc}\rangle-v_{\rm dr}$ curves at different $\eta$ and ${\it\Theta}_{\rm h,c}=0.4,0.04$. 
%, $\mu=4\times10^{4}\rm s^{-1}$. 
We can see that while the $\langle W_{\rm cyc}\rangle-v_{\rm dr}$ curves have similar shapes at different $\eta$, the low velocity limits of $\langle W_{\rm cyc}\rangle$ are very different. In the inset (A1), the mean values and standard deviations of $W_{\rm cyc}$ at $v_{\rm dr}=10^{-7}\rm m/s$ are plotted with respect to $\eta$, from which we can see that the absolute value of the low velocity limit of $\langle W_{\rm cyc}\rangle$ first increases and then decreases with $\eta$ increasing. On the other hand, the approximate equilibrium cycle work $W_{\rm cyc,e.p.}$ represented by the dot-dashed black curve, however, decreases to $-\infty$ [%the inset of (A1), 
cf. SI Appendix Figure S55(d)]. We can also see that when $\eta$ is large the standard deviations are also large especially at $\eta>4.6$, indicating that equilibrium needs to be achieved at lower driving velocity than $v_{\rm dr}=10^{-7}\rm m/s$. Next we will give a preliminary explanation to these phenomena. 

In Figure \ref{fig:HighEta}(F1)-(F7), the resultant potential curves at $\eta=8$ are plotted at different instants in one engine cycle. Here instead of moving the harmonic potential uniformly to the right, we keep it stationary and move the temperature field and the sinusoidal potential to the left with the same magnitude of velocity, which will lead to the same results. It looks like a rack moving to the left and driving a gear. We can see that there are at most four local minima between the instant of the FCP (F3) and the instant of the BCP (F5), and three in the rest of one cycle. The red and blue balls denote the local minima in the hot and cold zones respectively. From the starting point of one cycle (F1) to the instant of the BCP (F5), the leftmost red ball, i.e. the leftmost local minimum, goes leftward and upward in the same hot zone and then disappears at the instant of the BCP. The middle red ball, i.e. the middle global minimum in (F1) goes leftward and upward in its right neighbor hot zone and becomes the leftmost red ball at the end of one cycle. So it takes more than one cycle for the middle red ball starting at the global minimum point to go through the hot zone to the BCP and then out of the hot zone, rather than within one cycle as $\eta\in(1,4.6]$. 

Besides, the particle now can stay in one of these wells or hop among them. For example, in (F2) or (F6), besides to the middle well, the particle can jump out of the leftmost well to the rightmost one. This phenomenon is similar to multislip \cite{MultislipTIHE} of nanofriction in homogeneous temperature and in Figure \ref{fig:HighEta}(C) we can see that at the beginning and end of one cycle when the forward and backward energy barrier from the middle well are low enough [(F1) and (F7)], the displacement of the particle covers all the three local minima, so that the particle will with high probability jump from the leftmost well to the rightmost one or vice versa.

%multislip will occur with high probability, especially when it's close to the end when the leftmost well is in a hot zone while the other two wells are both in cold zones (F6).
% and can actually be observed from the simulation results, in which the dashed red and blue curves represent the local minimum points of the wells in the hot and cold zones respectively, like that in Figure \ref{fig:CyclesTh04Tc004}. 
%Another reason for the particle's covering all the three local minima at the beginning and end of one cycle is nevertheless, the high temperature so that the particle tends to neglect the lattice potential and fluctuate around the driver center, i.e. the thermolubricity mechanism works, cf. SI Appendix Figure S45. 
% and the latter effect surpasses the former because the rightmost well is in the cold zone, so that the work output from the latter is larger than the work input from the former. So for the $\eta=8$ case, this effect increases the mean cycle work output. 

In Figure \ref{fig:HighEta}(B), we can see that the dotted curves of the resultant potential at the balanced points are more complex than those of the $\eta\leq4.6$ cases (Figure \ref{fig:Eta1en2en3}). An unbroken version of these curves is given in SI Appendix Figure S51. In the $\eta=8$ case, the potential mechanism still plays a leading role and we can see that both the right energy barrier and the right local minimum (blue ball) neighbor to the middle global minimum (red ball) in (F1) have already occurred before the beginning of the engine cycle, and the work curve in its middle stage is approximately parallel to the bottommost dotted blue curve, which is the mark of the potential mechanism similar to the $\eta\leq4.6$ cases in Figure \ref{fig:Eta1en2en3}.
%which represents the resultant potential energy transition from the rightmost local minimum in (F1) to the middle global minimum in (F7). 
%However, because the thermolubricity mechanism, which is now negative, and also the backward multislip effect, at the beginning the work curve descends slower than the corresponding dashed blue branch.
However, because at the beginning and end of one cycle the particle can hop among the wells due to the low energy barriers %and the high temperature zones nearby the global minimum point [Figure \ref{fig:HighEta}(C)] 
and tends to neglect the lattice potential, the beginning and end stage of the work curve descend slower than the bottommost dotted blue branch. 

Therefore, we can conclude that for the $\eta=8$ case, the potential mechanism results in its high mean cycle work output, which is declined a little by the particle's covering more wells at the beginning and end of one engine cycle. However, when $\eta$ increases further, the left and right wells neighbor to the middle one 
%tend to ``intrude'' more deeply into the current engine cycle 
become relatively more reachable for the particle (cf. SI Appendix Sec. L), resulting in that the particle spends a longer period remaining in the neighbor wells both ahead of and behind the global one, i.e. covering more wells.
% at the same ${\it\Theta}_{\rm h,c}$ the absolute temperature becomes higher,
%and  
%However, when $\eta$ increases further, there will be more local minima on the left and right [Figure \ref{fig:HighEta}(D)], from the middle, the particle can jump to its left first well and then from the left first well to the left second well, or it can jump from the middle well to the left second or third well directly and so on, and so can it jump to the right wells far away from the middle. 
%Therefore at high $\eta$, the particle tends to disperse to a wide range at the same ${\it\Theta}_{\rm h,c}=0.4,0.04$.
%so that the particle will be resisted by the many energy barriers. 
In Figure \ref{fig:HighEta}(E), we can see that when $\eta$ increases from $8$ to $30$, the probability density distribution of the particle's displacement relative to the driver center gets fat both to the left and right and the average value of the distribution, which is approximately proportional to the mean cycle work, gets diluted and approaches to zero. So at $\eta>8$ the particle's %remaining longer in the side (especially backward) 
covering more wells leads to the work output declined more, cf. SI Appendix Sec. L. 

%Therefore, we can conclude that for the $\eta=8$ case, the potential mechanism results in its high mean cycle work output, which is declined a little by the particle's hopping among the wells. 
%For the cases of $\eta>8$, the hopping-among-the-wells effect is more significant and causes the work output decreases, cf. SI Appendix Figure S46. 
%We must note that the negative thermolubricity effect may still play an important role because of the high absolute temperature resulting from the large $V_0$ of high $\eta$ at the same relative nondimensional temperature ${\it\Theta}_{\rm h,c}$. 
In the above analysis, we have kept the nondimensional temperatures ${\it\Theta}_{\rm h,c}$ constant with $\eta$ varying. When $\eta$ is higher, the absolute temperatures $T_{\rm h,c}$ are also higher due to the higher $V_0$. %So the negative thermolubricity mechanism could play an important role in the reduction of $\langle W_{\rm cyc}\rangle$ at high $\eta$. 
In terms of the particle's position distribution center approaching to the driver center and its tending to neglect the lattice potential, the high absolute temperatures, or rather thermolubricity, play an important role in the reduction of the equilibrium cycle work output at high $\eta$.
%We can check this by decreasing ${\it\Theta}_{\rm h,c}$ to see whether the equilibrium cycle work output limit increases and has a maximum with $\eta$ fixed at a high value. We can also keep the absolute temperatures $T_{\rm h,c}$ constant to see how the mean cycle work output varies with $\eta$.
%And we can presume that the at high $\eta$, the maximum work output with respective to the nondimensional temperature ${\it\Theta}_{\rm h,c}$ decreases with increasing $\eta$. 
To have a complete understanding about the effect of the increasing number of teeth of the resultant potential with increasing $\eta$ on the equilibrium limit of $\langle W_{\rm cyc}\rangle$, %more simulation should be done to 
we can keep the absolute temperatures $T_{\rm h,c}$ constant to see how $\langle W_{\rm cyc}\rangle $ varies with $\eta$ or keep $\eta$ constant at a high value to see how $\langle W_{\rm cyc}\rangle $ varies with ${\it\Theta}_{\rm h,c}$, at low driving velocity. 
However, because of the large number of time steps at high $\eta$, it's not easy by Langevin dynamics simulation to achieve a low enough driving velocity necessary for the particle to relax to qusiequilibrium in low %nondimensional temperatures 
${\it\Theta}_{\rm h,c}$, see the next subsection. The circles' large standard deviations at $\eta\geq8$ in Figure \ref{fig:HighEta}(A1) indicate $v_{\rm dr}=10^{-7}\rm m/s$ is not low enough. A qualitative analysis is given in SI Appendix, Sec. 6.L.
 %So our analysis for the high $\eta$ cases is preliminary. %Other candidate such as the Fokker-Planck method \cite{VANKAMPENFokkerPlanck}
Methods based on the Fokker-Planck or Kramers equation \cite{KramersEqBookRisken}
% and the transition state theory \cite{StochTDNanofriction} 
are promising to be used to investigate this problem further, with which other features such as 
%whether there is maximum with respect to ${\it\Theta}_{\rm h,c}$ at high $\eta$ and 
the increasing plateau peak value with the increasing $\eta$ may also be elucidated.
\section*{Discussion}
\subsection*{More on the Temperature}
In the following discussion, we return to the $\eta\leq4.6$ cases, unless otherwise specified. In the above we have focused mainly on the effect of the high temperature $\it\Theta_{\rm h}$. Nevertheless, attention should also be paid to the low temperature $\it\Theta_{\rm c}$, whose value we have chosen is kind of mediate. When the low temperature is high, such as the ${\it\Theta}_{\rm h,c}=4.0,0.4$ case in Figure \ref{fig:Wcyc_Vdr}(C), the mean cycle work output is much smaller than the ${\it\Theta}_{\rm h,c}=0.4,0.04$ case. That's because it's easy for the particle to cross over the barrier from both the high and low temperature zone at such high temperatures and the particle's fluctuation center is close to the driver center whether behind or ahead of it, i.e. the negative thermolubricity mechanism. Nevertheless, due to the potential mechanism, there is still a bit of net work output.
%In the low temperature zone, the fluctuation center is a little closer to the right local minimum and a little farther away from the driver center than that in the high temperature zone so that there is still a bit of net work output. 
Because in the PTSHE ${\it\Theta}_{\rm h}>{\it\Theta}_{\rm c}$, when ${\it\Theta}_{\rm c}$ gets extremely high, the result is similar to the homogeneous very high temperature case (SI Appendix Sec. 6.B) and the mean cycle work output will go to zero. 
%The key point of thermolubricity effect is that high temperature makes the particle's position distribution more normal and centered closer to the driver center. 

In Figure \ref{fig:Wcyc_Vdr}(C), the result of the ${\it\Theta}_{\rm h,c}=0.08,0.008$ case is also given. We can see that it's more difficult for this low temperature case to achieve equilibrium, in that the probabilities of jumping over the barrier, $\exp(-\frac{\Delta V_{\rm h,c}}{k_{\rm B}T_{\rm h,c}})$, are small in both the forward and the backward direction and the driving velocity should be small enough for the crossover event to occur considerable times in one cycle.
%[compare the cases of $1\leq\eta\leq4.6$ at the same absolute temperatures in Figure \ref{fig:Wcyc_Vdr}(B) and at the same nondimensional temperatures in Figure \ref{fig:HighEta}(A)]. \textcolor{red}{At the same absolute temperatures, the velocity range of the work output shrinks as $\eta$ increases.} 
It's not easy to use Langevin dynamics simulation to explore the cases at such low temperature and we may turn to the Fokker-Planck or Kramers equation method \cite{KramersEqBookRisken}. 

So to achieve distinct mean cycle work output, the low temperature should be lower than the lattice potential energy barrier $V_0$ enough to avoid the negative thermolubricity effect: ${\it\Theta}_{\rm c}=\frac{k_{\rm B}T_{\rm c}}{V_0}\ll1$.
% $k_{\rm B}T_{\rm c}\ll\Delta V_{\rm c}$ or approximately ${\it\Theta}_{\rm c}=\frac{k_{\rm B}T_{\rm c}}{V_0}\ll1$ [$\Delta V_{\rm h,c}$ is inherited from Figure \ref{fig:Wcyc_Vdr}(A3)]. 
And to avoid the driving velocity being too small, it should not be too low. ${\it\Theta}_{\rm c}\approx0.01$ is a good choice for the parameter ranges considered for the $\eta\leq4.6$ cases above. As for the high temperature, as we have elucidated, it should not be too high or too low either and ${\it\Theta}_{\rm h}\approx1$
%(or approximately $k_{\rm B}T_{\rm h}\approx\Delta V_{\rm h}$) 
is a good choice for the parameter ranges considered for the $\eta\leq4.6$ cases above.
%up to now (cf. the subsection Mean cycle work at $\eta>4.6$). 

Some more remarks are given in SI Appendix Sec. 6.M.

\subsection*{Practical Implementation of the PTSHE}
There are three systems promising for the practical realization of the PTSHE from atomic- to microscale: the trapped ion system like the trapped ion friction simulator \cite{PRLTrappedIonOpticalLattice,ScienceTIFE,NPVelocityTuning,MultislipTIHE}, the optical levitated nanosphere system in cavity \cite{CavityCoolNanospherePRL} and the colloidal microparticle system in liquid medium \cite{Stirling}. We give a preliminary discussion about some topics on the practical implementation of the PTSHE in SI Appendix Sec. 6.N.

\subsection*{Comparison with the B{\"u}ttiker-Landauer Heat Engine} 
The PTSHE is reminiscent of the Brownian motor or ratchet \cite{HanggiBrownianMotor,ReimannPeterReview,BrownianMotorPhysicsToday,ParticleSepReview} %,EntropicSplitterParticleSepPRL,EntropicTransportKineticScalingPRL,EntropicParticleTransportPeriodicChannelBiosystems,NanofluidicRockingBrownianMotorsScience
in which thermal fluctuations (or diffusion) play an important role and the heat bath is usually of constant temperature. It could be traced back to Feymann's ratchet and pawl \cite{Feynman}. In \cite{ButtikerTransport}, B{\"u}ttiker calculated the flux of an overdamped particle transmitting through a periodic potential with a spatially varying diffusion coefficient at the same period, which can be realized by a spatially varying temperature field. He showed that when the diffusion coefficient is not in phase with the potential, the particle can achieve directional movement. This mechanism is known as the B{\"u}ttiker-Landauer ratchet \cite{LandauerTransport,KampenNonuniformTemp,SoftHeatengineReview,ConversionHeatWorkReviewPR}. With a load applied on the particle, the ratchet becomes a heat engine \cite{OptPotentialTempRatchetPRE}. Usually the load force is predetermined and constant \cite{EfficiencyBrownianHeatEngine,OptPotentialTempRatchetPRE,BLConstForcePRE,BLConstForcePAStochEnergetics,BrownianHeatEngineConstForceEPJB,BHEConstForceEPJBVaryingPotential,TightCouplingHeatBrownianPRE}.
Or it can also vary with time as in \cite{TimevaryingforcePNAS,TimevaryingforcePhysicaA}, although the temperature is constant in the two works.
%,BLPRERef,BLRef2,BLRef3 
In our PTSHE, besides the periodic sinusoidal lattice potential and the alternative high and low temperature field, an uniformly moving harmonic potential is added, resulting in that the load, i.e. the harmonic force, is stochasticly varying. This property is characteristic of the PT model. And we have shown that with the addition of such kind of an external load, the limit extractable work increases with the corrugation number $\eta$ when $\eta$ is not very large. Besides, with a uniform driving velocity $v_{\rm dr}$, we can define the engine cycle conveniently by identifying it as the process for the driver center moving over one lattice period, so that we can discuss the cycle work $W_{\rm cyc}$. On the other hand, the driving velocity $v_{\rm dr}$ of the harmonic potential should be uniform, so feedback control may be needed in practical implementation. Instead of moving the harmonic potential, we can try to move uniformly the sinusoidal potential and the temperature field, i.e. making two standing waves into two travelling waves, cf. Figure \ref{fig:HighEta}(F1)-(F7).
%depending on the specific implementation strategy.
Nonuniform driving velocity of the harmonic potential, predetermined or stochastic, should be okay and may lead to better performance.

% Further comparison between the B{\"u}ttiker-Landauer heat engine and our PTSHE can be made.

%In \cite{FKmodelBLJSM}, a one dimensional Frenkel-Kontorova model of interacting particles moving over a periodic substrate, which is another frequently-used model in nanofriction, in spatially varying temperature profile is analyzed. Without load, a center-of-mass velocity can be developed when the substrate and the temperature profile are spatially asymmetric. This model can also be generalized by adding a harmonic potential \cite{ScienceTIFE,FKTIHENJP} as the PTSHE. We also have noted that in \cite{MultipleInteractingThermalRatchetEPL}, directed transport is found of a one-dimensional hard-rod fluid undergoing collisions in spatially inhomogeneous heat bath and periodic potential with an appropriate relative phase shift between them. This system can also be seen as a generalization of the B{\"u}ttiker-Landauer ratchet from a single particle to many interacting particles \cite{ButtikerTransport} and we can introduce a moving harmonic potential to this system too.

Based on the principles of stable work output, there is potential to design other stochastic heat engines of this type with a resultant potential energy curve of other shape and an appropriate temperature field. An asymmetric lattice potential composed of two or more harmonics superposed with a harmonic potential may be okay, and we can consider the optimum shape of the asymmetric lattice potential. Optimization of the driving velocity, damping coefficient, temperature profile, the shape of the resultant potential energy and other parameters of the PTSHE is an interesting topic.

\subsection*{The Nonlinearity of the PT Model}
%When the frequency of the driver center moving over the lattice periods, i.e. $v_{\rm dr}/a$, euquals from the frequency  of the harmonic potential $\omega_0$ to the maximum frequency of the resultant potential $\omega_0\sqrt{1+\eta}$, the driving velocity $v_{\rm dr}$ ranges from $0.0673$ to $0.2964$. This range is contained in that of the strong nonlinear bifurcation regime. So we can speculate that the nonlinear bifurcation regime partly results from resonance, which is interesting to explore further.

%The heat bath with spatially alternative high ang low temperature is not only difficult to realize in experiment but also not easy to define. Here, 
%\subsection*{Nonlinearity of the PT Model}
We have seen that the nonlinearity of the PT model plays an essential role both in the design of the PTSHE and in the occurance of stick-slip.
% The resultant potential energy should have multiple extrema for the particle to jump from the low energy minimum over the barrier to the higher one or the other way around. 
 The nonlinearity of stick-slip friction has been discussed in \cite{NatureNonlinearFriction,FrictionNonlinearDynamics,PRLSSChaos,PREChaos}.  In this paper we have given a preliminary analysis of the bifurcation of the mean cycle work of the PT model with the driving velocity as parameter at zero temperature. At finite temperature, the nonlinear bifurcation can be reflected by the stochastic resonance represented by the standard deviations of $W_{\rm cyc}$ as in Figure \ref{fig:Wcyc_Vdr}(B2), Figure \ref{fig:HighEta}(A2) and SI Appendix Figure S18. 
%It's promising to design nonlinear sensors taking advantage of the variation of the standard deviations of the cycle work with respect to the driving velocity. 
In homogeneous temperature (SI Appendix Figure S18), this variation can be experimentally validated through levitating a nanosphere by a sinusoidal optical lattice superposed by a harmonic Paul trap as in \cite{CavityCoolNanospherePRL} or through the trapped ion system \cite{PRLTrappedIonOpticalLattice,ScienceTIFE,NPVelocityTuning,MultislipTIHE} which we have actually adopted as the model system in this paper (SI Appendix Sec. 8.H).
%It not only is  noteworthy in nanofriction but also may inspire new nonlinear nanoscale sensors \cite{NCNonlinearSensor,NMNonlinearSwitching,NLNonlinearNanoresonator} based on the PT model. 

%In \cite{CavityOMOptLNPNAS}, D. Chang et al. proposed that by introducing specifically tailored optical potentials, nonlinear motion should be possible to be produced. 
%The bifurcation diagram we have derived is at zero temperature. 
%At finite temperature, however, it's not easy to obtain such kind of diagram unless the temperature is so low, which is in high demand of the cooling strategy. 
%Even so, the nonlinear bifurcation can still be reflected by the standard deviations of the $W_{\rm cyc}-v_{\rm ddr}$ curves as that in Figure \ref{fig:Wcyc_Vdr}(B2) and Figure \ref{fig:HighEta}(A2), in which the standard deviations have a maximum, which may result from stochastic resonance \cite{StochasticBistableDynamicsNC} and we will explore further in the future. 
%It's promising to design nonlinear sensors taking advantage of the maximum of the standard deviations of the cycle work with respect to the driving velocity. 
%At homogeneous temperature, the simulation results can be experimentally validated through levitating a nanosphere by a sinusoidal optical lattice superposed by a harmonic Paul trap as in \cite{CavityCoolNanospherePRL} or through the trapped ion system \cite{PRLTrappedIonOpticalLattice,ScienceTIFE,NPVelocityTuning,MultislipTIHE}, which we have actually adopted as the model system in this paper.

The stochastic resonance peak's position is not as easy to calculate %as the case of the double-well potential superposed with a periodic driving force in \cite{StochasticResonance} 
and should be treated in the future. And the stochastic resonance of the PT model system should be investigated further to explore its potential to apply on noise-induced amplification schemes \cite{StochasticBistableDynamicsNC,StochasticSwitchingCantilever} for weak signal detection. 
%We also find that when projected to the $\langle W_{\rm cyc}\rangle -v_{\rm dr}$ space, some unclosed curves become closed. 
The stability of the bifurcation diagram in Figure \ref{fig:Bifurcation} hasn't been considered. The nonlinearity of the ordinary differential equation based on PT model is peculiar and there are still a lot of its features to be explored (SI Appendix Sec. 6.K.3).

\subsection*{Conclusion and Outlook}
To conclude, we designed a stochastic heat engine based on the Prandtl-Tomlinson model. With the Langevin dynamics simulation and the framework of stochastic thermodynamics we obtained the particle's phase trajectories and the thermodynamic quantities, from which we analyzed two mechanisms of work output of the PTSHE. Based on the potential mechanism, we obtained an approximation of the equilibrium cycle work output and an approximate Carnot-like efficiency of the PTSHE. The thermolubricity mechanism makes the equilibrium cycle work output greater than the approximation based on the potential mechanism and can also lead to the work output reduction if the temperature is excessively high. Mean cycle work is analyzed over a wide range of velocity, giving us a complete picture of the work output regime of the PTSHE as well as the work input regime when it stalls. The latter is similar to that in the nanofriction and we have compared the PTSHE with the homogeneous temperature cases in nanofriction.
%through typical quantities during one cycle. 
In the middle velocity regime of the mean cycle work, the strong and peculiar nonlinear bifurcation at zero temperature and the stochastic resonance at finite temperature of the mean cycle work with respect to the driving velocity are analyzed preliminarily, both of which are still needed to be explored further. When $\eta>4.6$, the PTSHE has some different features from the $\eta\leq4.6$ cases and we found that the work output first increases and then decreases with $\eta$ at the same ${\it\Theta}_{\rm h,c}$ and a preliminary analysis is given which still needs to be investigated further.

%The ranges of and mutual relations between the temperature and other quanties of the PTSHE are discussed in detail with emphasis on the practical realizability. 

 Combining the stick-slip process and the heat engine, the PTSHE can serve as a thermolubricity strategy and is a candidate system to study the thermodynamics in nanofriction \cite{FrictionNegativeworktail,FrictionNonequilibrium,StochTDNanofriction}.  The double-well potential is common in many laser trap experiments such as that in verification of the Landauer's limit by a single colloidal particle trapped in a modulated double-well potential \cite{LandauerLimNature} and that in the direct measurement of Kramers turnover with a optically levitated nanoparticle \cite{KramersTurnoverNatureNano}. Implementation of the PT model at nanoscale through trapped ion system or optically levitated nanoparticle system or at microscale through laser trapped colloidal particle system gives a candidate scheme for studies relevant to the bistable dynamics \cite{StochasticBistableDynamicsNC} or multistable dynamics (when $\eta>4.6$). In the micro- or nanoscale, the thermodynamics quantities are stochastic variables due to the nonneglected thermal fluctuations and the framework of stochastic thermodynamics  \cite{SeifertSTDReview,SekimotoFirstlaw1,SekimotoFirstlaw12} is an important tool to analyze the experimental results \cite{LandauerLimNature,FirstLawPRL,Stirling,BrownianCarnot}.
The PTSHE can be analyzed further by the Fokker-Planck or Kramers equation method \cite{KramersEqBookRisken}, with which the accurate efficiency, the very low temperature case and the high $\eta$ cases may be treated relatively easily and other issues such as the entropy product, the fluctuation theorems \cite{SeifertSTDReview} and the power of the PTSHE can also be investigated.
%And we can also speculate that in biosystems, some macromolecues or cells may be driven aided by the temperature gradients caused by some nonhomogeneous chemical field in a similar way \cite{KinesinProtein1,KinesinProtein2,BrownianRatchetMolecularMotors,BrownianMotorPhysicsToday}. 
% , as in \cite{ButtikerTransport,LandauerTransport,KampenNonuniformTemp}, where the Fokker-Planck method is used to calculate the probability of the particle at different local stable states and the fluxes between them at the existence of temperature [or diffusion coefficient \cite{ButtikerTransport}] nonuniformity.
The alternative high and low temperature field not only has promising appllication but also draws forth many interesting physics and is worth being studied. In this paper we consider the PTSHE as a classical system and the effect of quantum tunneling through the barriers should be considered when implemented with, for instance, the trapped ion system at very low temperature.

\matmethods{\label{sec:methods}The dynamics of the particle in the harmonic and sinusoidal lattice potential and the alternative high and low temperature field can be described by the Langevin equation \cite{KryPhysSF,VanColloquiumFriction,DongAnalytical,MuserVelocity,WangZijianEnergy}
\begin{eqnarray}\label{LEqn}
m\ddot{x}(t)=-m\mu\dot{x}(t)-\frac{\partial V(x(t),X(t))}{\partial x}+\xi(t),
\end{eqnarray}
where the left hand side (LHS) is the inertial force. The first term on the right hand side (RHS) is the damping force term which is proportional to the particle's velocity $\dot{x}(t)$ with $\mu$ the damping coefficient; the second term on the RHS is the conservative force exerting by the potential field of the PT model, $\frac{\partial V(x(t),X(t))}{\partial x}=F_{\rm h}+F_{\rm l}=\kappa[x(t)-v_{\rm dr}t]+\frac\pi aV_0\sin(\frac{2\pi}ax(t))$; the third term on the RHS is the fluctuating force $\xi(t)=\Gamma\zeta(t)$ satisfying the fluctuation-dissipation theorem $\Gamma=\sqrt{2m\mu k_BT(x)}$ with a zero mean Gaussian white noise whose covariance $\langle\zeta(t)\zeta'(t)\rangle=\delta(t-t')$, where $\delta(t)$ is the Dirac delta function. We integrate the Langevin equation with the 4th order stochastic Runge-Kutta method and the detail is given in SI Appendix, Sec. 8, where the numerical method is evaluated carefully to gurantee that it's reasonable and practical. % For conveinence, we use the parameters in the experiment of the trapped ion friction emulator\cite{RN8469,RN8483,RN8471} and we will comment on the generalization of our results to larger scale and higher temperature at the end of the letter.

After integration of the Langevin equation, we obtain the phase trajectories of the particle $(x(t),\dot{x}(t))$. %with the corresponding velocities $\dot{x}(t)$. 
And we use the framework of stochastic thermodynamics \cite{SeifertSTDReview,SekimotoFirstlaw1,SekimotoFirstlaw12} to extract %the thermodynamic quantities by
the change of the particle's internal energy $\Delta U$, the heat from the particle to the heat bath $Q$ and the work done on the particle $W$ by
\begin{equation}
\begin{aligned}
\Delta U&=U(t)-U(t_0),\\
Q&=-\int_{t_0}^{t}\left\{\kappa[x(t)-v_{\rm dr}t]+\frac\pi aV_0\sin[\frac{2\pi}ax(t)]\right\}\mathrm dx\\
&\qquad-\left[\frac12m\dot{x}(t)^2-\frac12m\dot{x}(t_0)^2\right],\\
W&=-\int_{t_0}^t\kappa\left[x(t)-v_{\rm dr}t\right]v_{\rm dr}\mathrm dt.
\end{aligned}
\end{equation}
%where $\Delta U$ is the change of the particle's internal energy, $Q$ is the heat from the particle to the heat bath and $W$ is the work done on the particle. 
The details of the derivation are given in SI Appendix, Sec. 5. 
}

\showmatmethods{} % Display the Materials and Methods section

\acknow{I'm deeply indebted to Dr. Dorian Gangloff (Cambridge University, UK) and Prof. Vladan Vuleti{\'c} (Massachusetts Institute of Technology, USA) for their answering my so many questions about \cite{NPVelocityTuning,ScienceTIFE} and their kind offer of the original experimental data and codes of Figure 2 in \cite{NPVelocityTuning}. I thank Prof. Tian Yu for helpful discussion. I also acknowledge Prof. Kim Kihwan for his comments about the experimental realization and Prof. Ou-Yang Zhong-Can and Prof. Tu Zhan-Chun for their kind feedback of the manuscript. I thank Prof. Guo Fei for his reading and advice on the manuscript and he and Prof. Liu Xiangfeng's support of this research. This work is partially supported by the National Natural Science Foundation of China (Grant No.U1937602).}

\showacknow{} % Display the acknowledgments section

% Bibliography
\bibliography{ms}

\end{document}

%% Comment out or remove this line before generating final copy for submission; this will also remove the warning re: "Consecutive odd pages found".
%\instructionspage  

\maketitle

%% Adds the main heading for the SI text. Comment out this line if you do not have any supporting information text.
\SItext
This supporting information appendix is a little long for the purpose of describing the mathematical models in detail and pedagogically. Although we have tried our best to make this paper self-consistent, there are still aspects we cannot include. If there are anywhere unclear or any mistakes, the author is grateful for the readers to contact him freely. 

Many derivations and results in this supporting information appendix are not original and must have been derived or obtained before, although we have tried our best to cite the relevant works. If we didn't cite what should be cited, the originality should be credited to the original authors.

This SI Appendix has 8 sections. In Sec. \ref{sec:refabs} we give some references for the Abstract of the main text for the readers to have a knowledge of the ubiquity of the stick-slip phenomenon in many scientific fields.

In Sec. \ref{CriticalPtsofTempZoneSec}, we show how to determine the boundaries of the hot and the cold zones of the PTSHE. 

In Sec. \ref{sec:georoot}, we introduce the geometric root of the stick-slip phenomenon based on the PT model with different quantities. Curves of these quantities at the balanced points with respect to the driver center's position are plotted and are used in analyze the mechanisms of stable work output of the PTSHE in the main text. %The key to the potential mechanism of work output of the PTSHE is also discribed in this section.

In Sec. \ref{sec:critcalvalueeta}, we give a brief derivation of the critical values of $\eta$. In the main text, we mainly focus on the range $[1,4.6]$ and the higher order behavior of the PTSHE when $\eta>4.6$ is also discussed preliminarily.

In Sec. \ref{sec:firstlaw}, the first law of thermodynamics is derived from the Langevin equation utilizing the framework of stochastic thermodynamics. The formula of calculating the internal energy, released heat and inputted work are derived, which are mentioned in the Methods section of the main text.

In Sec. \ref{sec:apndRD}, we gather all the appendices for the Results and Discussion section of the main text.

In Sec. \ref{FluctDissTheoSec}, we give a brief derivation of a form of fluctuation-dissipation theorem based on the linear Langevin equation for the harmonic oscillator, which is equivalent to the equipartition theorem. And we have assumed that it is also satisfied by our nonlinear Langevin Eq. 4 in the main text. We didn't find a detailed derivation in the textbooks accessible to us so we derive it ourselves here for self-consistency of the paper. It is also used for evaluation of the SRK4 method in the next section.

In Sec. \ref{Langevindynamicssimulation}, we introduce the Langevin dynamics simulation in detail. Because Langevin simulation is the main methods of this paper, we describe as elaborately as we could. We evaluate the SRK4 method carefully to make sure that it satisfys the equipartition theorem with enough accuracy for the case of the linear Langevin equation, that it performs as well as another nonlinear method with the appropriately chosen time stepsize, that it satisfys the first law of thermodynamics with low accumulative error and that it can duplicate the experimental results without fitting parameters.

The last two sections are a little long so we arrange them at the end of this SI Appendix. They can also be placed above Sec. \ref{sec:apndRD}.

For the use of this SI Appendix, we suggest the readers first read the main text subsection by subsection and glance over the corresponding cited parts of this SI Appendix without thorough reading to have a first impression. After finishing reading at least one subsection of the main text, the readers can read through the corresponding SI Appendix parts if necessary. As for a specific subsection in this SI Appendix, we suggest the readers first read through the entire subsection including all the figures without caring too much about the incomprehensions encountered, which may be figured out after finishing reading the whole subsection.

\tableofcontents

%\subsection*{Subhead}
%Type or paste text here. This should be additional explanatory text such as an extended technical description of results, full details of mathematical models, etc.   
\section{References for the abstract of the main text}
\label{sec:refabs}
In the abstract we mention that stick-slip is an ubiquitous phenomenon in many scientific fields and here are some references: earthquake and glacier dynamics \cite{NatureEarthquake,NatureStickslipIcestream,VolcanicSeismicity,StickslipPakistan,ScienceSSMechanismEarthquake,ScienceSSIce,SciencySSRockExperiment,ScienceSSEarthquakeHeatflow,ScienceSSSlowEarthquake,ScienceSSSorceParameterEarthquake}, acoustics \cite{StickslipLobsters,fletchermusicalinstmt}, cell biology \cite{SASSCelladhesion,PNASCelldynamics}, interface science \cite{PNASNanodroplet,FoamsStickSlip,Stickslipmercury,NatureStickslipLuminence,Resonantstickslip,ScienceSSBoundaryLub} and tribology \cite{Slip-stickFrictionNature,PNASSticksliplubricant,PNASArticularjoints}. 
\section{Determination of the boundaries of different temperature zones}
\label{CriticalPtsofTempZoneSec}
We first nondimensionalize the resultant potential
\begin{equation}\label{ResPotential}
V(x(t),X(t))=\frac12\kappa[x(t)-X(t)]^2+\frac{V_0}{2}[1-\cos(\frac{2\pi}ax(t))]
\end{equation}
to 
\begin{equation}\label{Vtau}
\tilde{V}(z(\tau),\tilde{X}(\tau))=\frac{V(x(t))}{\kappa a^2/(4\pi^2)}=\frac12[z(\tau)-\tilde{X}(\tau)]^2+\eta[1-\cos(z(\tau))],
\end{equation}
where $z(\tau)=\frac{2\pi}ax(t)$, $\tau=\frac{\omega_0}{2\pi}t$, $\tilde{X}(\tau)=\frac{2\pi}aX(t)$, $\eta=\frac{2\pi^2V_0}{\kappa a^2}$ and $X(t)=v_{\rm dr}t$. Here $\omega_0=\sqrt{\kappa/a}$ is the intrinsic frequency of the parabola potential $V_{\rm h}$.

\begin{figure}[H]
\centering
\includegraphics[width=\textwidth]{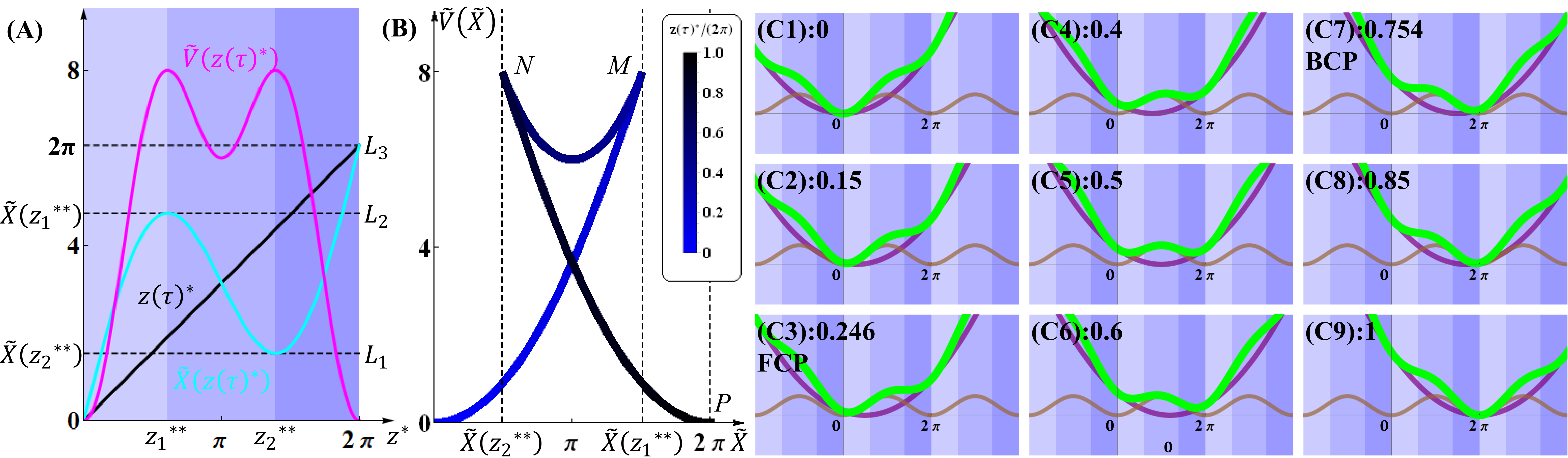}
\caption{Determination of the boundaries of different temperature zones. (A) The resultant potential $\tilde V(z(\tau)^*)$ at the balanced point $z(\tau)^*$ and the corresponding change of the driver center $\tilde{X}(z(\tau)^*)$ in one period with respect to $z(\tau)^*$. Two critical points $z_1^{**}$ (BCP) and $z_2^{**}$ (FCP) separate one spatial lattice period into three zones.
% , where there is one global or local minimum of the resultant potential energy $\tilde{V}$, zero or one local maximum and one global or local minimum again from the left zone to the right one respectively. 
Two lines $L_1:\tilde{X}(z(\tau)^*)=\tilde{X}(z_2^{**})$ and $L_2:\tilde{X}(z(\tau)^*)=\tilde{X}(z_1^{**})$ separate one temporal period, in which the driver center $\tilde{X}(z(\tau)^*)$ goes over one lattice period, into three parts. Between the $x-$axis and $L_1$, there is one solution for a horizontal line intersecting the $\tilde{X}(z(\tau)^*)-z(\tau)^*$ curve, corresponding to the only one global minimum in (C1)-(C2). Between $L_1$ and $L_2$, there are three solutions for a horizontal line intersecting the $\tilde{X}(z(\tau)^*)-z(\tau)^*$ curve, corresponding to the two local minima, and one local maximum in between in (C4)-(C6). Between $L_2$ and $L_3:\tilde{X}(z(\tau)^*)=2\pi$, there is one solution again for a horizontal line intersecting the $\tilde{X}(z(\tau)^*)-z(\tau)^*$ curve, corresponding to the only one global minimum in (C8)-(C9). (B) The $\tilde{V}(z(\tau)^*,\tilde{X}(z(\tau)^*))-\tilde{X}(z(\tau)^*)$ curve with $z(\tau)^*$ as parameter. $\wideparen{OM}$ represents the left (first global and then local) minimum's resultant potential rising with $\tilde{X}$ from the starting instant of one temporal period (C1) to the appearance of the BCP (C7). $\wideparen{NM}$ represents the middle maximum's resultant potential changing with $\tilde{X}$ from the disappearance of the FCP to the appearance of the BCP [(C3) to (C7)], with $N$ corresponding to the instant of the FCP (C3) and $M$ the instant of the BCP (C7). $\wideparen{NP}$ represents the right (first local and then global) minimum's resultant potential setting with $\tilde{X}$ from the disappearance of the FCP (C3) to the end of the same temporal period (C9). In the interval $(\tilde{X}(z_2^{**}),\tilde{X}(z_1^{**}))$, the three extrema coexist, which is more clear in (C4) to (C6). It is crucial to note that on $\wideparen{OM}$, the left global or local minimum is always behind the driver center while on $\wideparen{NP}$, the right local or global minimum is always ahead of the driver center. In our PTSHE, the particle will stay around the minimum points (stable balanced points) of the resultant potential, and if it is around the left one, the driver pulls the particle and work is inputed to the particle, while if it is around the right one, the particle pulls the driver and work is outputed from the particle to the driver. 
 (C1)-(C9) The key frames of the potential cruves at 9 selected instants during one temporal period.
%time (which is proportional to the position of the driver center $\tilde{X}(\tau)=\tilde{v}\tau$, where $\tilde v=\frac{2\pi}a\frac{2\pi}{\omega_0}v_{\rm dr}=\frac{4\pi^2}{a\omega_0}v_{\rm dr}$ is the nondimensional driving velocity) in one period. 
The purple, brown and green curves represent the moving harmonic potential, the lattice potential and the resultant potential respectively. The decimal following each label represents the relative position of the driver center $X(t)/a={v}_{\rm dr}t/a$ to its starting point in (C1), which is proportional to the time $t$.
%  different instants in one period $T=\frac a{v_{\rm dr}}$, i.e. $\frac tT=\frac\tau{2\pi}$, relative to the starting of the specific temporal period. It corresponds to different positions of the driver center $\tilde{X}(\tau)=\tilde{v}_{\rm dr}\tau$, which goes forward to the right with uniform velocity $v_{\rm dr}$.
}
\label{DeterminationCritPts}
\end{figure}

At a specific driver center position $\tilde{X}(\tau))$, set the first partial derivative of $\tilde{V}(z(\tau),\tilde{X}(\tau))$ with respect to $z$ to be zero,
\begin{eqnarray}
\frac{\partial \tilde{V}(z(\tau),\tilde{X}(\tau))}{\partial z}=[z(\tau)-\tilde{X}(\tau)]+\eta\sin(z(\tau))=0,
\end{eqnarray}
yielding
\begin{equation}\label{Xtau}
\tilde{X}(\tau)=z(\tau)+\eta\sin(z(\tau)),
\end{equation}
solving which we can obtain the local minima and maxima, i.e. the stable and unstable balanced points, of $\tilde V(z(\tau),\tilde{X}(\tau))$: $z(\tau)^*$, satisfying
\begin{equation}\label{Xtau2}
\tilde{X}(z(\tau)^*)=z(\tau)^*+\eta\sin(z(\tau)^*).
\end{equation}
Substitute Eq. \ref{Xtau2} into Eq. \ref{Vtau}, we will get
\begin{equation}\label{Vtau2}
\tilde{V}(z(\tau)^*)=\frac12\eta^2\sin^2(z(\tau)^*)+\eta[1-\cos(z(\tau)^*)].
\end{equation}
%which describes the locus of the extremum (balanced) points $(z(\tau)^*,\tilde{V}(z(\tau)^*))$. 
%Here we distinguish the extremum points with $*$ from other points. 
The meaning of $\tilde{V}(z(\tau)^*)$ is that at $z(\tau)^*$, the nondimensional resultant potential $\tilde{V}(z(\tau),\tilde{X}(z(\tau)^*))$ achieves its extremum $\tilde{V}(z(\tau)^*)$ and the driver center is at $\tilde{X}(z(\tau)^*)$,
%Eq. \ref{Xtau} can be expressed with respect to $z(\tau)^*)$ as 
%\begin{equation}\label{Xtau2}
%\tilde{X}(z(\tau)^*)=z(\tau)^*+\eta\sin(z(\tau)^*).
%\end{equation}
cf. Figure \ref{DeterminationCritPts}(A), in which we plot $\tilde{V}(z(\tau)^*)$, $\tilde{X}(z(\tau)^*)$ and $z(\tau)^*$ with respect to $z(\tau)^*$ in one period $[0,2\pi)$. There are three zones separated by the two maximum points of $\tilde{V}(z(\tau)^*)$. Let the second partial derivative of $\tilde{V}(z(\tau),X(\tau))$ with respect to $z$ at $z(\tau)^*$ equal zero,
\begin{eqnarray}
\frac{\partial^2\tilde{V}(z(\tau)^*,\tilde{X}(z(\tau)^*))}{\partial z^2}=1+\eta\cos(z(\tau)^*)=0,
\end{eqnarray}
we get the critical maximum points separating the zones,
\begin{equation}\label{CritPts}
z_1^{**}=2n\pi+\arccos(-\frac1\eta),
z_2^{**}=2n\pi-\arccos(-\frac1\eta),n\in\mathbb Z.
\end{equation}
In the period $[0,2\pi)$, the two critical points are $z^{**}_1=\arccos(-\frac1\eta)$ and 
$z^{**}_2=2\pi-\arccos(-\frac1\eta)$. We will refer to $z^{**}_1$ and $z^{**}_2$ as the backward critical point (BCP) and the forward critical point (FCP) respectively.

In Figure \ref{DeterminationCritPts}(C1)-(C9), we plot the potential curves at 9 selected instants during one temporal period during which the driver center goes over a lattice period. The FCP and the BCP are both inflection and stationary points of the resultant potential curve $\tilde{V}(z(\tau),X(\tau))$. From the starting instant of one temporal period [Figure \ref{DeterminationCritPts}(C1)] to the appearance of FCP [Figure \ref{DeterminationCritPts}(C3)], the resultant potential energy curve has only one global minimum in the left zone. There are no other extrema in the other two zones of one spatial lattice period. From the disappearance of the FCP to the appearance of the BCP [Figure \ref{DeterminationCritPts}(C7)], there are two local minima and one local maxima in the left, the right and the middle zone of one spatial lattice period, respectively. After the disappearance of the BCP [Figure \ref{DeterminationCritPts}(C7)] till the end of one temporal period [Figure \ref{DeterminationCritPts}(C9)], there is only one global minimum again in the right zone. So in one temporal period, the left zone covers the left local or global minimum of $\tilde{V}(z(\tau),X(\tau))$, the middle zone covers the local maximum of $\tilde{V}(z(\tau),X(\tau))$, and the right zone covers the right local or global minimum of $\tilde{V}(z(\tau),X(\tau))$. The FCP appears and divides into the right local minimum and the local maximum and from then on there are three extrema, until the left local minimum and the local maximum merges into the BCP that then disappears. 

From the starting instant [Figure \ref{DeterminationCritPts}(C1)] to the BCP [Figure \ref{DeterminationCritPts}(C7)], the global or left local minimum rises from zero to its highest value before it merges into the BCP and disappears, and during this process it stays in the left zone $[0,z^{**}_1)$ in the temporal interval $[0,z^{**}_2/\tilde{v}_{\rm dr})$. We set the spacial intervals $[2n\pi,z^{**}_1+2n\pi),n\in\mathbb Z$ in the high temperature $T_{\rm h}$ and the rest of the spatial lattice periods $[z^{**}_1+2n\pi,2\pi),n\in\mathbb Z$ in the low temperature $T_{\rm c}$.
 
In Figure \ref{DeterminationCritPts}(A), we also plot $\tilde{X}(z(\tau)^*)$ with respect to $z(\tau)^*$, which has one local minimum and one local maximum. The extremum points can be obtained by setting the partial derivative of Eq. \ref{Xtau2} to be zero,
\begin{equation}
\frac{\partial\tilde{X}(z(\tau)^*)}{\partial z^*}=1+\eta\cos(z(\tau)^*)=0,
\end{equation}
leading to the same values as Eq. \ref{CritPts}. The $\tilde{X}(z(\tau)^*)$ is ploted with respect to the stationary point $z(\tau)^*$ at which the resultant potential achieves its extremum. In the left zone, at each point $z(\tau)^*$, the resultant potential can only has the global or the left local minimum. In the middle zone, at each point $z(\tau)^*$, the resultant potential can only has the local maximum. While in the right zone, at each point $z(\tau)^*$, the resultant potential can only has the global or the right local minimum. In the middle zone $\tilde{X}(z(\tau)^*)$ goes down as $z(\tau)^*$ increases because in this zone the resultant potential energy has only one local maximum moving backwards to the left as the driver goes forward [Figure \ref{DeterminationCritPts}(C3)-(C7)]. The two extremum points of $\tilde{X}(z(\tau)^*)$ divide one temporal period into three parts too. When $\tilde{X}(z(\tau)^*)\in[0,\tilde{X}(z_2^{**}))$, i.e. between the $x-$axis and $L_1$, there is only one global minimum [Figure \ref{DeterminationCritPts}(C1)-(C2)]. When $\tilde{X}(z(\tau)^*)\in(\tilde{X}(z_2^{**}),\tilde{X}(z_1^{**}))$, i.e. between the $L_1$ and $L_2$, there are three extrema, i.e. the left local minimum, the middle local maximum and the right local minimum [Figure \ref{DeterminationCritPts}(C4)-(C6)]. It's clear that there are three intersection points of the $\tilde{X}(z(\tau)^*)-z(\tau)^*$ curve with a horizontal line in this interval, distributed in the three zones separated by the FCP ($z_2^{**}$) and BCP ($z_1^{**}$). When $\tilde{X}(z(\tau)^*)\in(\tilde{X}(z_1^{**}),2\pi)$, i.e. between $L_2$ and $L_3$, there is only one global minimum again [Figure \ref{DeterminationCritPts}(C8)-(C9)].

From Eq. \ref{Vtau2} and Eq. \ref{Xtau2}, we can treat $\tilde{V}(z(\tau)^*)$ as a function of $\tilde{X}(z(\tau)^*)$ with $z(\tau)^*$ as parametar,
\begin{equation}\label{eq:latticepotentialstickslip}
\begin{cases}
\tilde{X}(z(\tau)^*)=z(\tau)^*+\eta\sin(z(\tau)^*),\\
\tilde{V}(z(\tau)^*)=\frac12\eta^2\sin^2(z(\tau)^*)+\eta[1-\cos(z(\tau)^*)],
\end{cases}
\end{equation}
whose curve is plotted in Figure \ref{DeterminationCritPts}(B).  In Figure \ref{DeterminationCritPts}(B), it's clearer that as the driver moves from left to right in one period, there is one global minimum in the interval $[0,\tilde{X}(z_2^{**}))$, then three local extrema in $(\tilde{X}(z_2^{**}),\tilde{X}(z_1^{**}))$ and then one global minimum again in $(\tilde{X}(z_1^{**}),2\pi)$. In Figure \ref{DeterminationCritPts}(B), the singular points $M(\tilde{X}(z_1^{**}),\tilde{V}(z_1^{**}))$ and $N(\tilde{X}(z_2^{**}),\tilde{V}(z_2^{**}))$ correspond to the instants when (and also the position of the driver center where) the BCP and the FCP appear respectively. The curve $\wideparen{OM}$ represents the rising of the left minimum with respect to $\tilde{X}(z(\tau)^*)$, which corresponds to Figure \ref{DeterminationCritPts}(C1)-(C7). At $M$, the left local minimum merges with the middle local maximum and both of them then disappear. In the main text, we refer to the curve $\wideparen{OM}$ as the left minimum branch and is colored in red because this branch is immersed in the hot zone. The right minimum branch $\wideparen{NP}$ and the middle maximum branch $\wideparen{MN}$ are immersed in the cold zone and are colored in blue.

\section{The geometric explanation of the stick-slip phenomenon based on PT model}
\label{sec:georoot}

\begin{figure}[H]
\centering
\includegraphics[width=\textwidth]{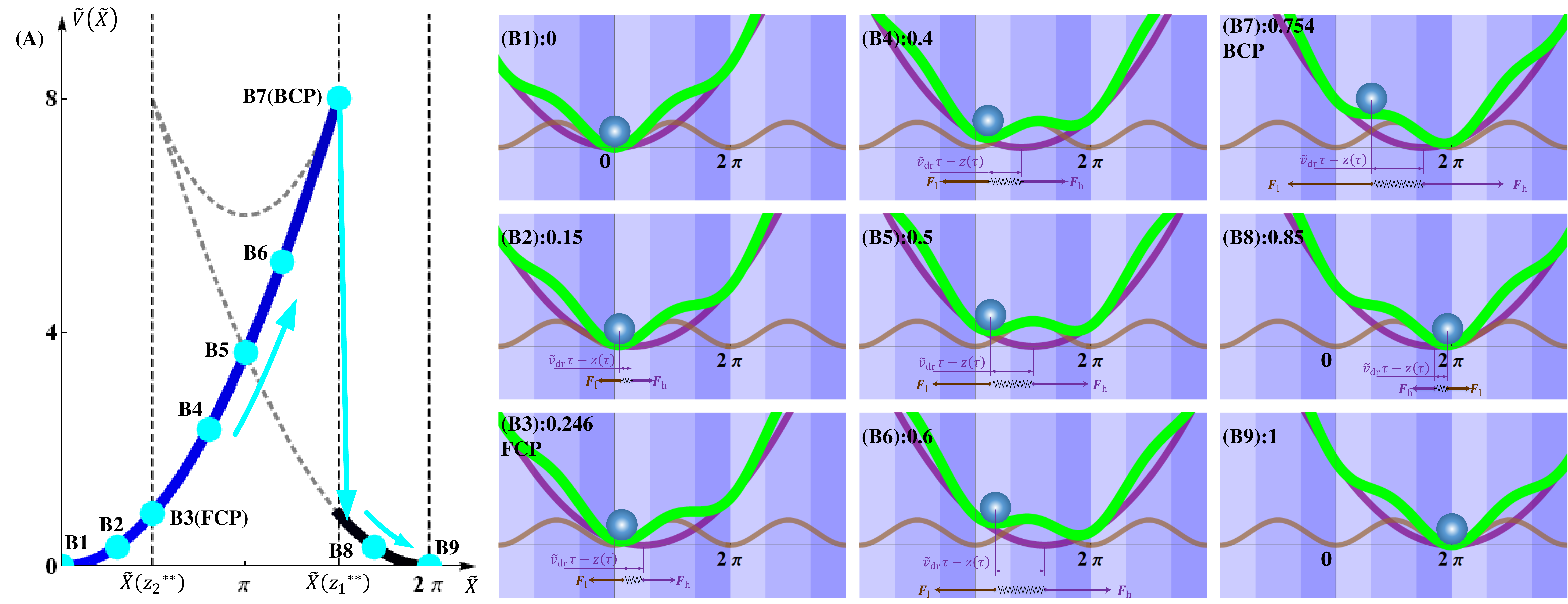}
\caption{The stick-slip process interpreted by the PT model. The resultant potential energy change of the particle in the stick-slip process is represented by the solid curves in (A) with the blue balls B1-B9 corresponding to the frames in (B1)-(B9). The dashed gray backbone curve is inherited from Figure \ref{DeterminationCritPts}(B). As the particle is driven to the right by the driver, the potential energy of it rises from minimum (B1) to maximum at the BCP (B7), which is the stick process. After the BCP disappears, the particle is exerted by a resultant force pointing to the right and accelerates (slips) to the right minimum which is now global. And then the particle is pushed by the lattice force and drives the driver to the right until the end of one period (B9), where the driver center and the particle overlap. Work is done on the particle from the driver before the instant of the BCP, while work is outputed from the particle to the driver after the particle has slipped to the right global minimum. Here the particle is assumed to be acted on by an infinitely large damping force besides the harmonic force and the lattice force at zero temperature so that it tends to stay at the local minimum (stable balanced) point all the time and slips exactly at the instant of the BCP.}
\label{VXForwardCycle}
\end{figure}

In Figure \ref{VXForwardCycle}(B1)-(B9) there is a particle driven by the driver starting at the global minimum of the resultant potential when the harmonic potential minimum overlaps with one of the lattice potential minima [Figure \ref{VXForwardCycle}(B1)]. The resultant potential energy of the particle is represented by the solid curves in Figure \ref{VXForwardCycle}(A) with the blue balls B1-B9 corresponding to the frames in (B1)-(B9). Assuming the particle is exerted by an infinitely large damping force besides the harmonic force $F_{\rm h}$ from the driver and the lattice force $F_{\rm l}$ at zero temperature so that there is no relaxing process and it tends to stay in balance all the time, i.e. this is a qusistatic model. Then the particle will stay on the left (at first global and then local) minimum following the driver. The potential energy of the particle rises along the curve $\wideparen{OM}$ to the maximum value $\tilde{V}(z_1^{**})$ at the BCP $z(\tau)=z_1^{**}$ in Figure \ref{VXForwardCycle}(A) and this is the stick process. At the BCP the left local minimum merges with the middle local maximum and then both of them disappear and the harmonic force $F_{\rm h}$ and the lattice force $F_{\rm l}$ can't continue to balance with each other because $\frac{\partial V}{\partial x}=F_{\rm h}+F_{\rm l}$ can't any longer equal $0$. The particle will then be exerted by the resultant force $\frac{\partial V}{\partial x}=F_{\rm h}+F_{\rm l}>0$ pointing to the right and accelerate (slip) until $\frac{\partial V}{\partial x}=F_{\rm h}+F_{\rm l}=0$ again at the right minimum which is now the only global minimum. Then the particle remains at this global minimum point until the end of the period when the harmonic potential minimum overlaps with the next lattice potential minimum. This is the geometric explanation of the stick-slip process \cite{KryPhysSF,VanColloquiumFriction,DongAnalytical,MuserVelocity,WangZijianEnergy}.

Note that after the particle slips and stays in the right global minimum, the particle is on the right of (ahead of) the driver center and $F_{\rm l}>0$, i.e. pointing to the right, and $F_{\rm h}<0$, i.e. pointing to the left, so that the particle is pushed by the lattice force $F_{\rm l}$ and the particle drives the driver forward,
%with velocity $v_{\rm dr}$, 
i.e. work is done by the particle on the driver. That's the key point for the work output of our PTSHE.

\begin{figure}[H]
\centering
\includegraphics[width=0.8\textwidth]{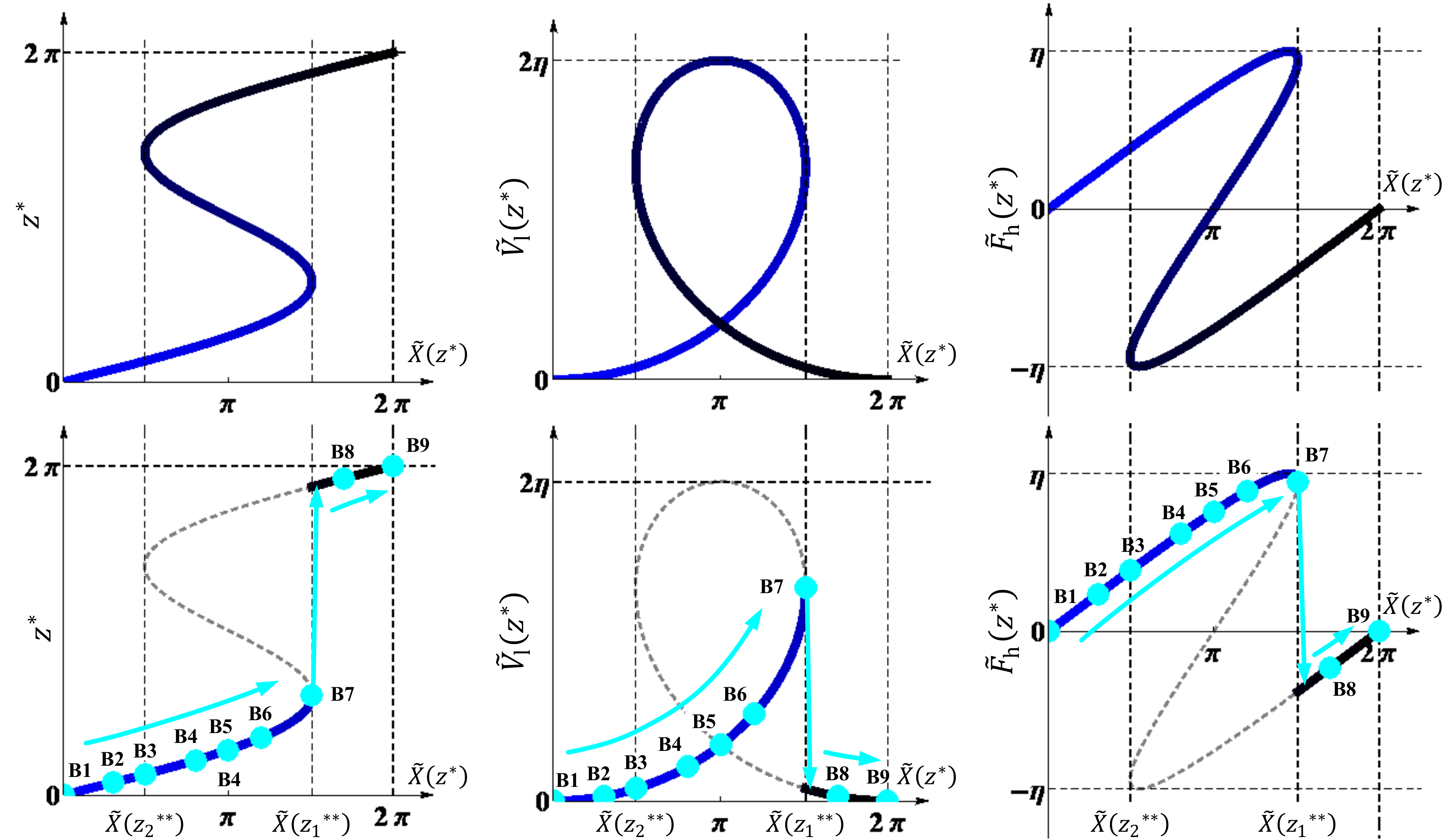}
\caption{Quantities at the balanced points of the PT model and the stick-slip process represented by them. The upper three subfigures are the balanced point position $z^*$, the lattice potential $\tilde V_{\rm l}(z^*)$ at the balanced point and the harmonic force $\tilde F_{\rm h}(z^*)$ at the balanced point with respect to the position of the driver center $\tilde X(z^*)$ at the balanced point from the left to the right respectively. The bottom three subfigures are the stick-slip process represented by the upper three quantities respectively at infinitely large damping force and zero temperature as in Figure \ref{VXForwardCycle}(A). The blue balls B1-B9 in the bottom three subfigures correspond to the 9 frames in Figure \ref{VXForwardCycle}(B1)-(B9).}
\label{fig:StickSlips}
\end{figure}

In Figure \ref{fig:StickSlips}, we plot three more quantities with respect to the driver center's position, which we will use in the main text and below. The upper left one is the nondimensional balanced point position $z(\tau)^*$, i.e. the position of the resultant potential extrema. It can be expressed by the parametric equation
\begin{equation}
\begin{cases}
\tilde X(z(\tau)^*)=z(\tau)^*+\eta\sin(z(\tau)^*),\\
z(\tau)^*=z(\tau)^*,
\end{cases}
\end{equation}
with $z(\tau)^*$ as the parameter. This curve is the same as the cyan curve in Figure \ref{DeterminationCritPts}(A) except for the exchange of the $x$- and $y$- coordinates.

The upper middle one is the nondimensional lattice potential $\tilde V_{\rm l}(z(\tau)^*)$ at the balanced point $z(\tau)^*$. It can be expressed by the parametric equation
\begin{equation}
\begin{cases}
\tilde X(z(\tau)^*)=z(\tau)^*+\eta\sin(z(\tau)^*),\\
\tilde V_{\rm l}(z(\tau)^*)=\eta[1-\cos(z(\tau)^*)],
\end{cases}
\end{equation}
with $z(\tau)^*$ as the parameter.

The upper right one is the nondimentional harmonic force 
\begin{equation}\tilde F_{\rm h}(z(\tau)^*)=\frac{F_{\rm h}(x(t))}{\kappa a/(2\pi)}=\frac{\kappa[X(t)-x(t)]}{\kappa a/(2\pi)}
\end{equation} 
at the balanced point $z(\tau)^*$. The paramteric equation for it is
\begin{equation}
\begin{cases}
\tilde X(z(\tau)^*)=z(\tau)^*+\eta\sin(z(\tau)^*),\\
\tilde F_{\rm h}(z(\tau)^*)=\tilde X(z(\tau)^*)-z(\tau)^*=\eta\sin(z(\tau)^*),
\end{cases}
\end{equation}
with $z(\tau)^*$ as the parameter.

The lower three schemetics are the stick-slip process represented by the upper three quantities. We can see that all of the four quanties [$z^*$, $\tilde V_{\rm l}(z^*)$, $\tilde F_{\rm h}(z^*)$ in Figure \ref{fig:StickSlips} and $\tilde V(z^*)$ in Figure \ref{VXForwardCycle}] slips at the same driver center position $\tilde X(z_1^{**})$.
\section{The derivation of the critical values of $\eta$}
\label{sec:critcalvalueeta}

\begin{figure}[H]
\centering
\includegraphics[width=0.8\textwidth]{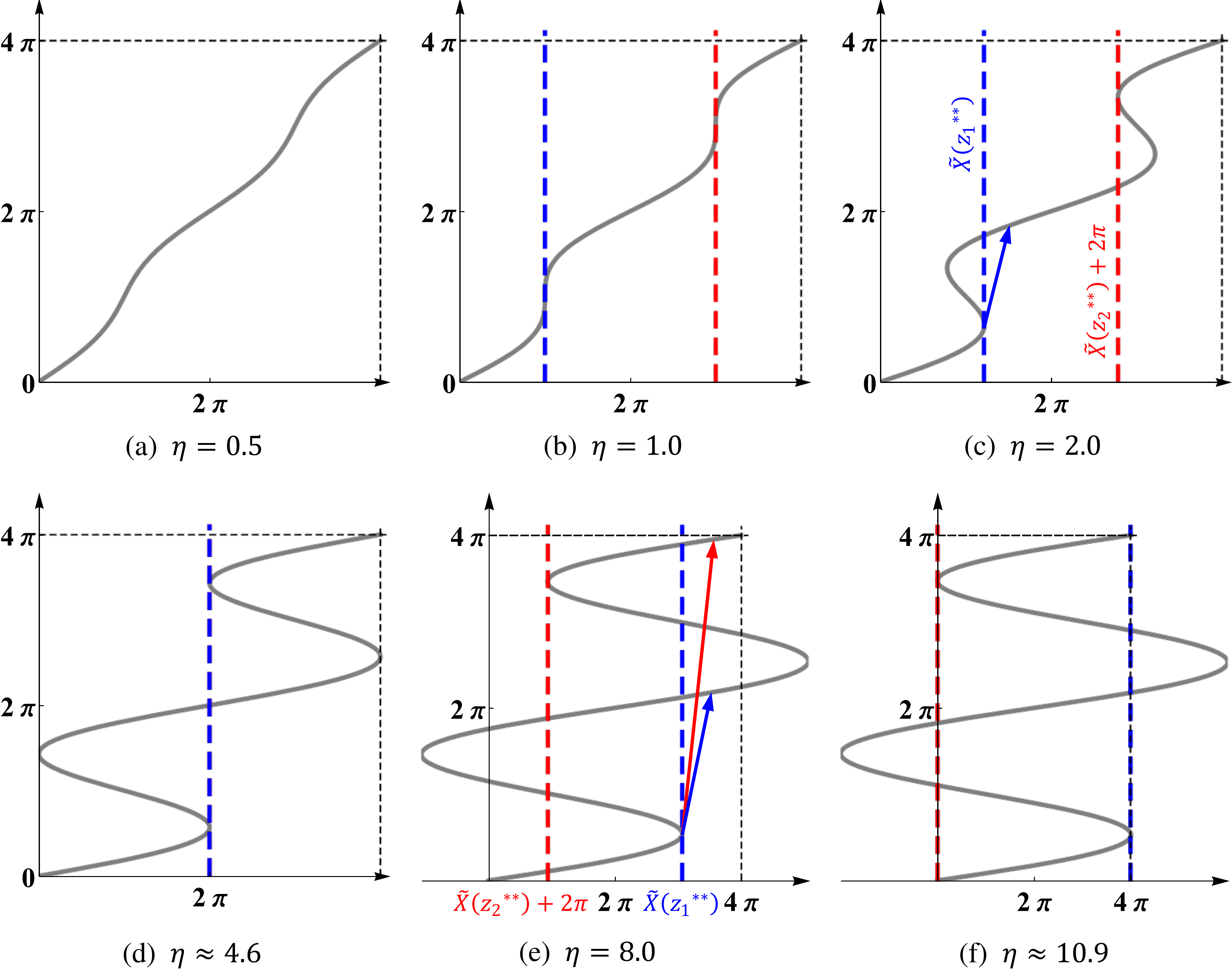}
\caption{Schematic of calculating the critical value of $\eta$. All the x-coordinates are $\tilde X(z^*)$ and all the y-coordinates are $z*$. The blue dashed line corresponds to the driver center position at the BCP of the first cycle $\tilde X(z_1^{**})$ and the red one corresponds to that at the FCP of the second cycle $\tilde X(z_2^{**})+2\pi$. At the first critical value $\eta=1.0$ (b), the two dashed lines occur. At the second critical value $\eta\approx4.6$ (d), the two dashed lines overlap at $2\pi$. At the third critical value $\eta\approx10.9$ (f), $\tilde X(z_1^{**})=4\pi$ and $\tilde X(z_2^{**})+2\pi=0$. The arrows in (c) and (e) represent the slip event and when $\eta=8$, there are two possible landing points for the particle at the slip instant.}
\label{fig:CritEta}
\end{figure}

In the main text, we mention that our attention is mainly payed to $\eta$ in the range of $[1,4.6]$, where the resultant potential energy has at most two maxima. The critical valued of $\eta$ can be obtained by the method used in \cite{CorrugationNumber}. Here we give a geometric derivation. We take the $z^*-\tilde X(z^*)$ curve as an example. In Figure \ref{fig:CritEta}, we can see that when $\eta=0.5$, the $z^*-\tilde X(z^*)$ curve is monotonous while when $\eta=2.0$, it becomes nonmonotonous and there is a fold in one cycle with two points, i.e. the FCP $(\tilde X(z_2^{**}),z_2^{**})$ and the BCP $(\tilde X(z_1^{**}),z_1^{**})$, having infinite slope, both of which are symmetric about the middle of the cycle. Therefore there is a critical value of $\eta$ between $0.5$ and $2$, at which a point with infinite slope occurs, i.e. at this point
\begin{equation}\label{eta1}
\frac{\mathrm d\tilde X(z^*)}{\mathrm d z^*}=1+\eta_1\cos(z^*)=0.
\end{equation}
From symmetry we know that this point is $z^*=\pi$, so that $\eta_1=1$.

At $\eta=2.0$, $\tilde X(z_1^{**})<2\pi$, the stick-slip event has only one landing point at the instant of the BCP as indicated by the blue arrow in Figure \ref{fig:CritEta}(c). While at $\eta=8.0$, $\tilde X(z_1^{**})>2\pi$, the stick-slip event can have two landing points at the instant of the BCP as indicated by the blue and red arrows in Figure \ref{fig:CritEta}(e). Therefore, at the second critical value of $\eta$, 
\begin{equation}
\tilde X(z_1^{**})=2\pi=z_1^{**}+\eta_2\sin(z_1^{**})=\arccos(-\frac1{\eta_2})+\eta_2\sin(\arccos(-\frac1{\eta_2})),
\end{equation}
where Eq. \ref{CritPts} is utilized. Solve this equation with a nolinear solver, we can obtain the second critical value $\eta_2=4.6033388487517$. By the same reasoning we can obtain the other critical values of $\eta$ by solving the nonlinear equations
\begin{equation}
\tilde X(z_1^{**})=z_1^{**}+\eta_{n+1}\sin(z_1^{**})=\arccos(-\frac1{\eta_{n+1}})+\eta_{n+1}\sin(\arccos(-\frac1{\eta_{n+1}}))=2\pi n,\ n=2,3,4,\cdots
\end{equation}
The third critical value $\eta_3=10.949879869826264$, and the fourth critical value $\eta_4=17.24976556755863$.

When we set $n=0.5$ in this equation, i.e. $\tilde X(z_1^{**})=\pi$, we will obtain the same result as Eq. \ref{eta1}.

We can also obtain the same results utilizing the FCP (the red dashed lines) by solving the equations 
\begin{equation}
\tilde X(z_2^{**})+2\pi=z_2^{**}+\eta_{n+1}\sin(z_2^{**})=2\pi-\arccos(-\frac1{\eta_{n+1}})+\eta_{n+1}\sin(2\pi-\arccos(-\frac1{\eta_{n+1}}))=2\pi-2\pi(n-1),\ n=0.5,1,2,\cdots
\end{equation}

\section{The first law of thermodynamics derived from the Langevin equation}
\label{sec:firstlaw}
We use the framework of stochastic thermodynamics \cite{SeifertSTDReview,SekimotoFirstlaw1,SekimotoFirstlaw12,Stirling,BrownianCarnot} to calculate the thermodynamic quantities of the particle from its displacement and velocity obtained from the Langevin dynamics simulation. Mutiplying both sides of Eq. 4 in the main text with $\mathrm dx$ and then adding $\frac{\partial V(x(t),X(t))}{\partial X}\mathrm dX$ to both sides yield
\begin{equation}
m\ddot{x}(t)\mathrm dx+\frac{\partial V(x(t),X(t))}{\partial x}\mathrm dx+\frac{\partial V(x(t),X(t))}{\partial X}\mathrm dX\nonumber=[-m\mu\dot{x}(t)+\xi(t)]\mathrm dx+\frac{\partial V(x(t),X(t))}{\partial X}\mathrm dX,
\end{equation}
which can be reorganized into
\begin{equation}
\mathrm d[\frac12m\dot{x}(t)^2+V(x(t),X(t))]+[m\mu\dot{x}(t)-\xi(t)]\mathrm dx\nonumber=\frac{\partial V(x(t),X(t))}{\partial X}\mathrm dX.
\end{equation}
In accordance with the first law of thermodynamics, the first term on the LHS can be identified as the differential internal energy
of the particle $\mathrm dU$, with $U(t)=\frac12m\dot{x}(t)^2+V(x(t),X(t))$ and the kinetic energy of the particle $K(t)=\frac12m\dot{x}(t)^2$. $V(x(t),X(t))$ is still the particle's resultant potential energy. The second term on the LHS can be identified as the differential heat transferred to the heat bath from the particle $\text{\dj} Q=[m\mu\dot{x}(t)-\xi(t)]\mathrm dx$. And the RHS term can be identified as the differential work done by the driver to the particle $\text{\dj}W=\frac{\partial V(x(t),X(t))}{\partial X}\mathrm dX=-\kappa[x(t)-X(t)]\mathrm dX$. Then we can obtain the change of the internal energy $\Delta U$, the heat to the heat baths $Q$ and the work input $W$ by integration:
\begin{eqnarray}
\Delta U&=&\int_{U(t_0)}^{U(t)}\mathrm dU=U(t)-U(t_0),\\
Q&=&\int_{x(t_0)}^{x(t)}\text{\dj}Q=\int_{x(t_0)}^{x(t)}\left[m\mu\dot{x}(t)-\xi(t)\right]\mathrm dx\notag\\
&=&\int_{x(t_0)}^{x(t)}\left[-\frac{\partial V(x(t),X(t))}{\partial x}-m\ddot{x}(t)\right]\mathrm dx\notag\\
&=&-\int_{t_0}^{t}\left\{\kappa[x(t)-v_{\rm dr}t]+\frac\pi aV_0\sin[\frac{2\pi}ax(t)]\right\}\mathrm dx-\int_{t_0}^{t}\mathrm d\left[\frac12m\dot{x}(t)^2\right],\notag\\
&=&-\int_{t_0}^{t}\left\{\kappa[x(t)-v_{\rm dr}t]+\frac\pi aV_0\sin[\frac{2\pi}ax(t)]\right\}\mathrm dx-\left[\frac12m\dot{x}(t)^2-\frac12m\dot{x}(t_0)^2\right],\label{QInt}\\
W&=&\int_{x(t_0)}^{x(t)}\text{\dj}W=\int_{X(t_0)}^{X(t)}-\kappa\left[x(t)-X(t)\right]\mathrm dX\notag\\
&=&-\int_{t_0}^t\kappa\left[x(t)-v_{\rm dr}t\right]v_{\rm dr}\mathrm dt.
\end{eqnarray}
The third equality of Eq. \ref{QInt} results from Langevin Eq. 4 in the main text. The mid-point rule, i.e. the Stratonovich rule, should be used to discretize the integral in Eq. \ref{QInt} \cite{SeifertSTDReview,BrownianCarnot}. A brief derivation is given in Sec. \ref{Langevindynamicssimulation}.\ref{midpointrule}.

\section{Appendices for the Results and the Discussion section of the main text}
\label{sec:apndRD}
\subsection{The homogeneous low temperature heat bath: stick-slip}

%For the homogeneous lower temperature case, we can see that the work curve in Figure \ref{CyclesTc004} goes to $+\infty$, i.e. work is inputted. 

\begin{figure}[H]
\centering
\includegraphics[width=7.7cm]{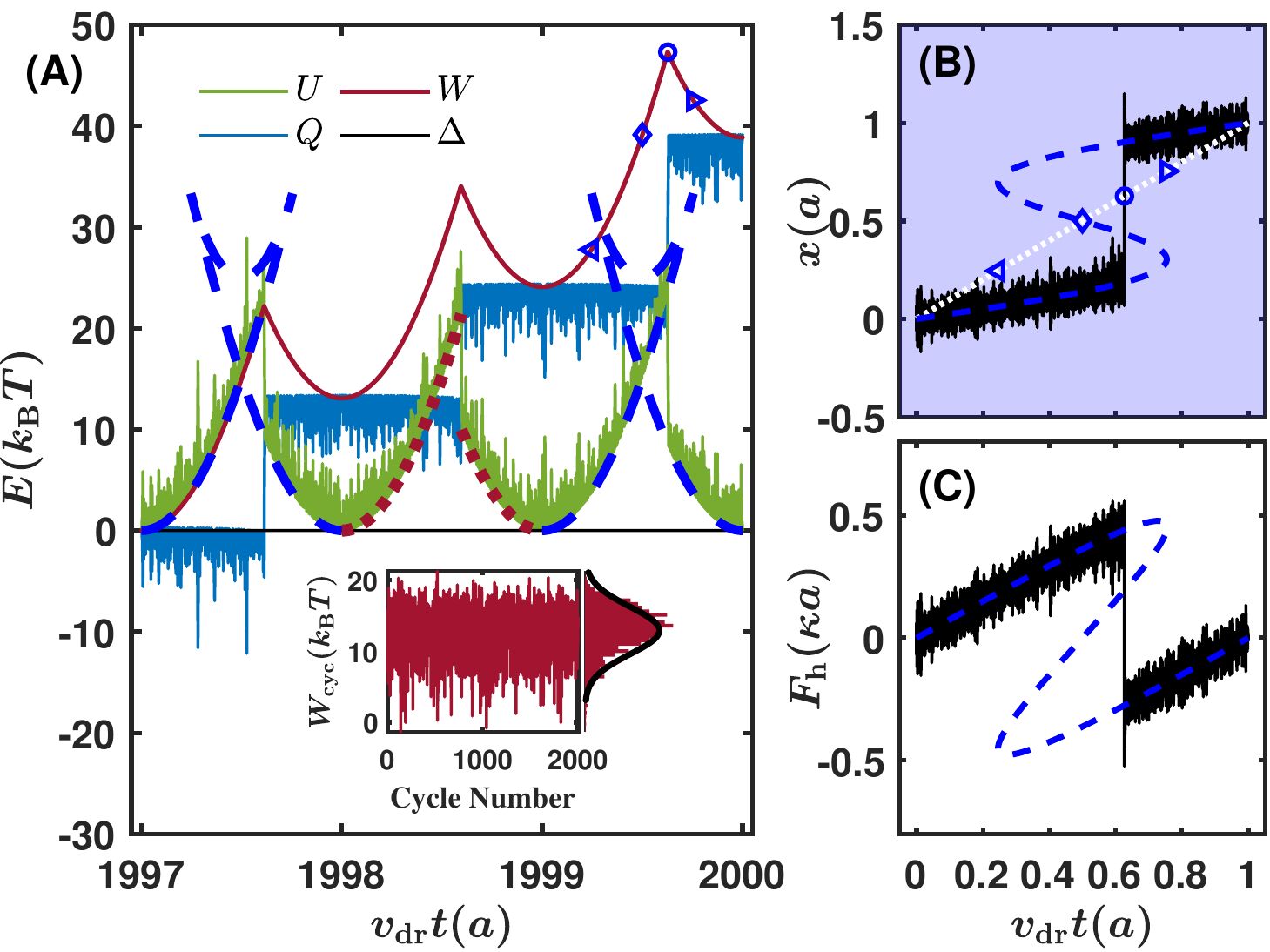}
\includegraphics[width=9.5cm]{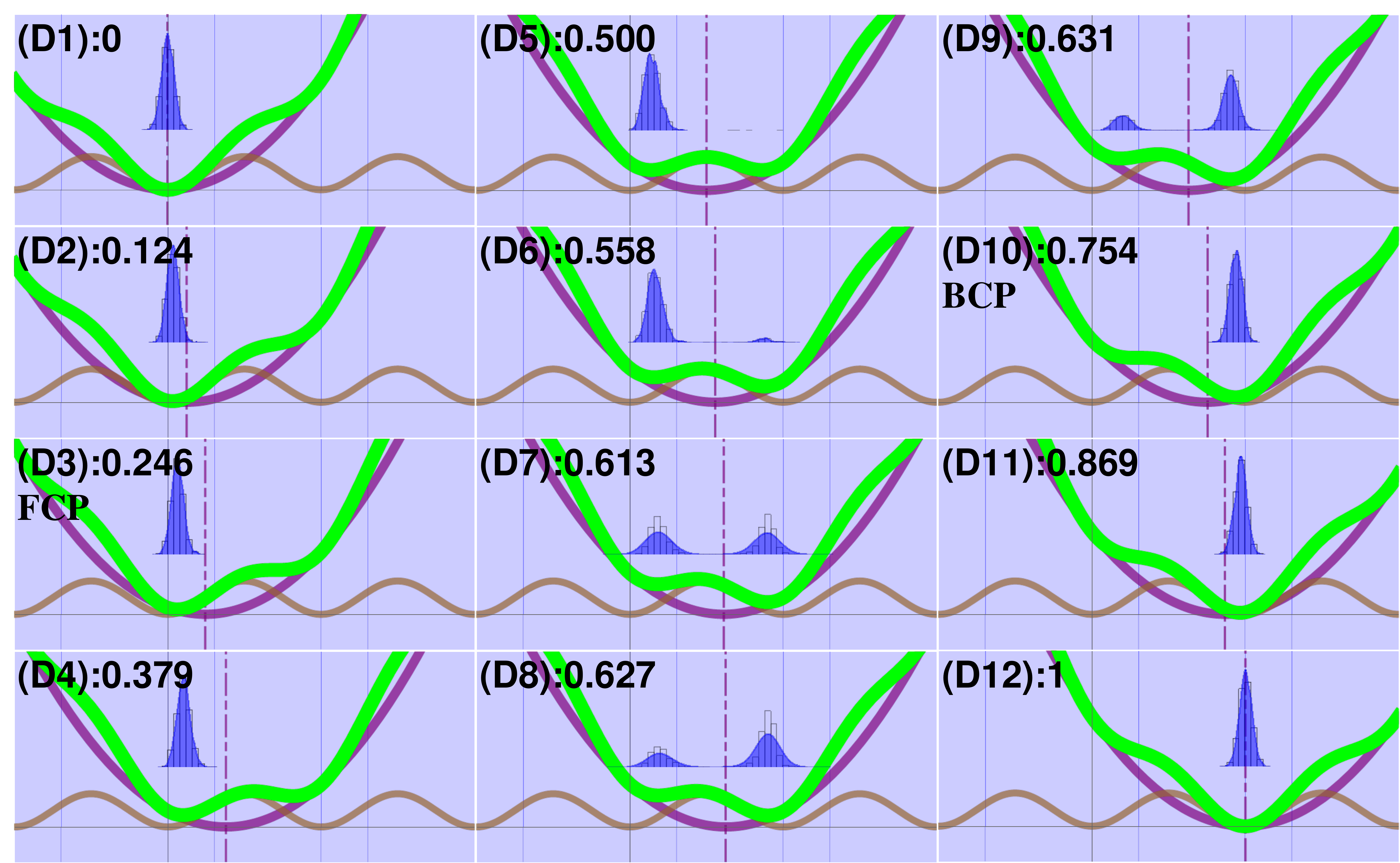}
\caption{Simulation results for the particle immersed in the homogeneous low temperature ${\it\Theta}=0.04$. (A) The internal energy $U$, work input $W$, heat to the heat bath $Q$ and the difference between them $\Delta=\Delta U+Q-W$ in the last three consecutive low temperature isothermal cycles among 2000 in total at $v_{\rm dr}=10^{-5}\rm m/s$. The work increases to $+\infty$, i.e. work is inputted. The dashed blue curves are the resultant potential at the balanced points (Sec. \ref{CriticalPtsofTempZoneSec}). The inset of (A) represents the cycle work in 2000 consecutive cycles and its distribution. The $x-$coordinate $v_{\rm dr}t(a)$ is the nondimensional position of the driver center, which is proportional to time $t$. The dimension of the energy is $k_{\rm B}T$ with $k_{\rm B}$ the Boltzmann's constant and $T(=\Theta V_0)$ the absolute temperature. The dotted dark red curves indicate the lower bound of the internal energy. $\Delta\equiv0$ indicates the first law of thermodynamics (Sec. \ref{sec:firstlaw}). (B) The displacement of the particle in the typical 2000th homogeneous low temperature isothermal cycle with respect to the driver center's position, relative to the starting point of the cycle. The dotted white line represents the driver center's position with respect to itself. The dashed blue curve is the loci of the balanced point (Figure \ref{fig:StickSlips}). The blue background represents the homogeneous low temperature. (C) The variation of harmonic force in the 2000th cycle. The $x-$coordinate matches that of (B).  Time ratio of minus harmonic force in one cycle is smaller than that of plus one, consistent with the work input. The dashed blue curve is the balanced harmonic force at the balanced point (Figure \ref{fig:StickSlips}). (D1)-(D12) Diagrams of the shapes of the resultant potential and the corresponding displacement distributions of the particle at different instants of the homogeneous low temperature isothermal cycle. The displacement distribution at a certain instant relative to the latest cycle starting point is obtained from the total 2000 simulation cycles. The decimal following the id of each frame is the nondimensional position of the driver center (or the nondimensional instant) relative to the cycle starting point (D1). The purple curve represents the moving harmonic potential and the brown one represents the lattice potential. The blue background represents the homogeneous low temperature. In (D6)-(D9), although the particle's displacement distribution obtained from the total 2000 simulation cycles disperses around both of the two local minimum points, in terms of a specific cycle, the particle only stays around one of the two local minimum points, which can be seen in (B) where the jumping-back-and-forth events between the two local minimum points are rare. This is an indicator of nonequilibrium in such a low temperature at such a relatively high velocity (see the text). Such a low temperature leads to the low fluctuation of the particle so that it hardly jumps over the energy barrier until the forward barrier is low enough. Therefore, in (D4) and (D5), although there are two resultant potential local minima, the particle's position nearly all distributes nearby the left local minimum point. And in (D6), even though the right local minimum is lower than the left one, the particle still tends to distribute around the left one. The relatively high driving velocity leads to the particle unable to arrive at equilibrium state. For instance, in (D7) even though the right local minimum is lower than the left one, the particle distributes around the left and right local minimum points approximately equally likely rather than distributing more around the right one. After (D7), the forward energy barrier is low enough for the particle to cross over at this temperature %while the backward energy barrier is high and the jumping backward event is still rarer, so the particle tends to stay around the right local minimum and the distribution of the particle's position is approximately proportional to the relative depth of the two local minima until the appearance of the BCP (D10), where the left local minimum disappears.
so the particle has a higher probability to stay around the right local minimum point until the appearance of the BCP (D10), where the left local minimum disappears. Parameters: $v_{\rm dr}=10^{-5}\rm m/s$, $\eta=3.0$, $\mu=4\times10^4\rm s^{-1}$, ${\it\Theta}=0.04$ and others are in Sec. \ref{Langevindynamicssimulation}.\ref{ParametersUsed}.}
\label{CyclesTc004}
\end{figure}

When the temperature is low (${\it\Theta}=0.04$), the fluctuation of the particle is small. In Figure \ref{CyclesTc004}, we plot the simulation results for the particle immersed in the homogeneous low temperature ${\it\Theta}=0.04$ at $v_{\rm dr}=10^{-5}\rm m/s$. We can also divide one low temperature isothermal cycle into four stages as we have done for the PTSHE in the main text. 

The first stage ranges from the beginning of one cycle (D1) to the instant of the FCP [(D3) and the left blue triangle in (A) and (B)]. In this stage there is only one global minimum on the resultant potential curve gradually falling behind the driver center and work is inputted. 

The second stage ranges from the instant of the FCP to the cusp instant of the work curve. In this stage there coexist two local minima and an energy barrier between them. The particle is still behind the driver center and work is still inputted.

The third stage ranges from the cusp instant to the instant of the BCP [(D10) and the right blue triangle in (A) and (B)]. In this stage there are still two local minima and the particle has jumped over the barrier and fluctuates around the right local minimum point. The particle is ahead of the driver center so that work is outputted. 

The fourth stage ranges from the instant of the BCP to the end of one cycle (D12). In this stage, there is again only one global minimum and the particle is ahead of the driver center so that work is again outputted. 

When the driving velocity is very low so that at each instant the system is at equilibrium state, there will be enough time for the event of jumping back and forth over the middle energy barrier to occur and the displacement distribution of the particle will be proportional to $\exp(-\frac{V}{k_{\rm B}T})$. Then the cusp will be in the middle of one cycle exactly and no longer that sharp and the cycle work will be zero. In Figure 3(C) in the main text, we can actually see that the ${\it\Theta}=0.04$ curve goes to zero as $v_{\rm dr}\rightarrow0$. Nonetheless, we can see from this curve that at $v_{\rm dr}=10^{-5}\rm m/s$, the cycle work input is much greater than zero, indicating that the particle cannot achieve equilibrium state at such a relatively high driving velocity in such a low temperature. In other words, due to the small fluctuation in the low temperature and the relatively high driving velocity,
%(so the particle is in nonequilibrium state, cf. the last but one paragraph in this subsection), 
the particle won't be able to go over the high forward energy barrier until it is low enough and it rarely jumps back and forth between the two wells. The latter can be seen in Figure \ref{CyclesTc004}(B) where the lanching from the left minimum point to the right one nearby the blue circle ($v_{\rm dr}t=0.613a$) is sharp and there is nearly no jumping back and forth, while the former can be concluded from that in all the 2000 simulation cycles the particle almost never crosses over the energy barrier before the middle of one cycle, cf. Figure \ref{CyclesTc004}(D5), 
%It is only when the driver moves a little farther ahead of the middle of one cycle that the particle will begin to go over the energy barrier. 
and not until $v_{\rm dr}t/a=0.613$ [Figure \ref{CyclesTc004}(D7)] will the particle distribute approximately equally likely around either (not both) of the two local minima. % (almost without jumping back and forth between the two wells).
%(because of nonequilibrium and the low thermal fluctuation, cf. the last but one paragraph in this subsection). 
 %After jumping over the barrier from left, the particle's high kinetic energy transformed from the potential energy is dissipated quickly because of damping. 
%Once the particle jumps over the energy barrier to the right, it almost never jumps back, in that the fluctuation is very small at such low temperature, which is different from the PTSHE in the main text.
As in the PTSHE, the jumping instant is stochastic and ranges approximately from the the middle of one cycle [Figure \ref{CyclesTc004}(D5)] to before the BCP [Figure \ref{CyclesTc004}(D10)].

At the instant of the cusp, when the particle slips, the internal energy $U$ falls off a cliff while the released heat $Q$ lanches over a cliff, so that the high potential energy around the left local minimum transforms into the low potential energy around the right local minimum and kinetic energy. Then most of the kinetic energy dissipates into heat quickly because of damping. 

We can see that a main difference between the PTSHE and the homogeneous low temperature heat bath case is whether the instant of the cusp is before or after the middle of one cycle.

As in such low temperature and at such a relatively high velocity the jumping-back-and-forth events are rare and the particle's distribution nearly always centered at one of the local (or at the global) minimum point(s), the shape of the work curves is nearly the same as that of the resultant potential energy local minimum branches, cf. Sec. \ref{sec:apndRD}.\ref{SimilaritybtwWorkPotential}.

 \subsection{The homogeneous high temperature heat bath: thermolubricity}

\begin{figure}[H]
\centering
\includegraphics[width=7.7cm]{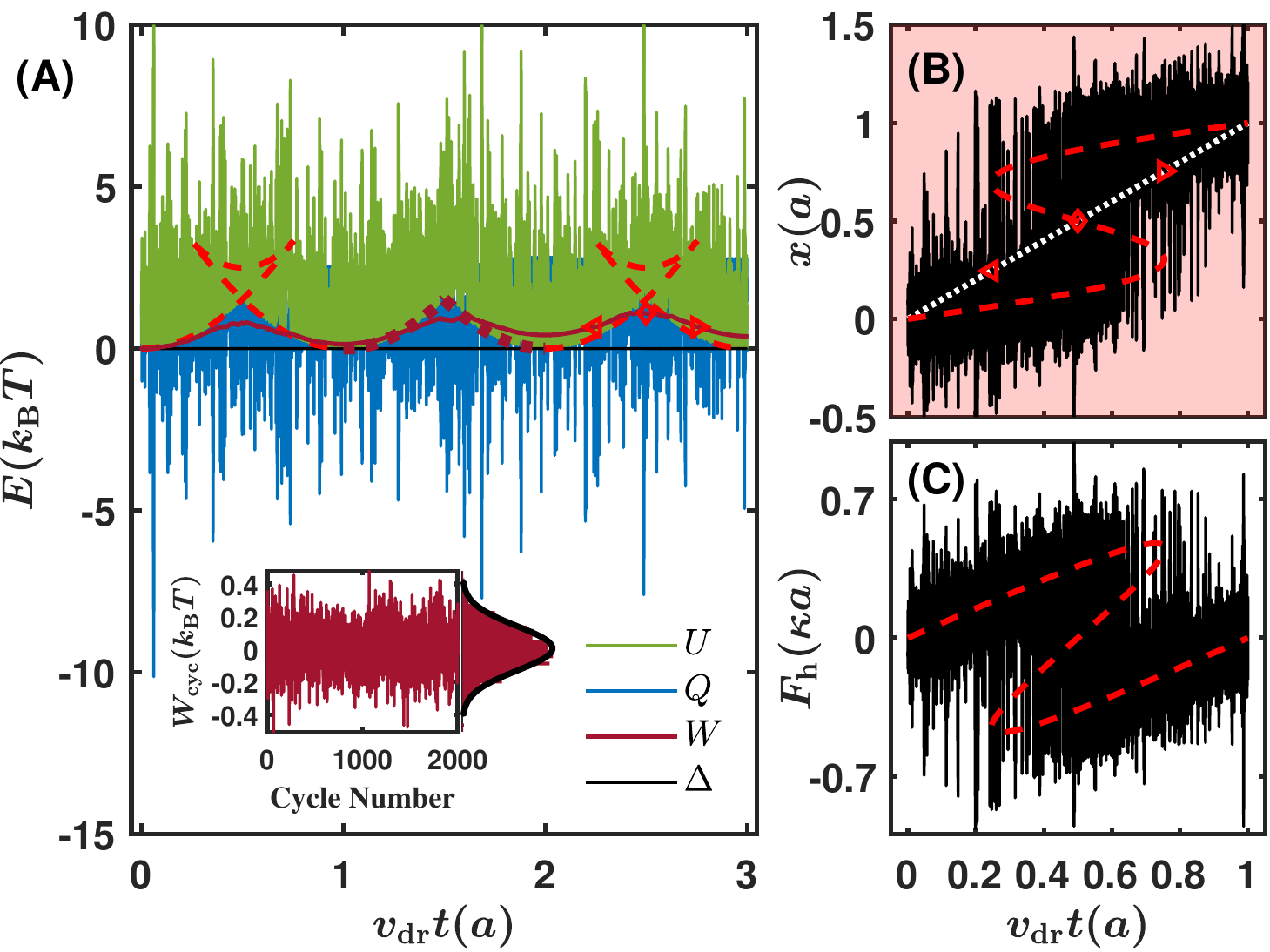}
\includegraphics[width=9.5cm]{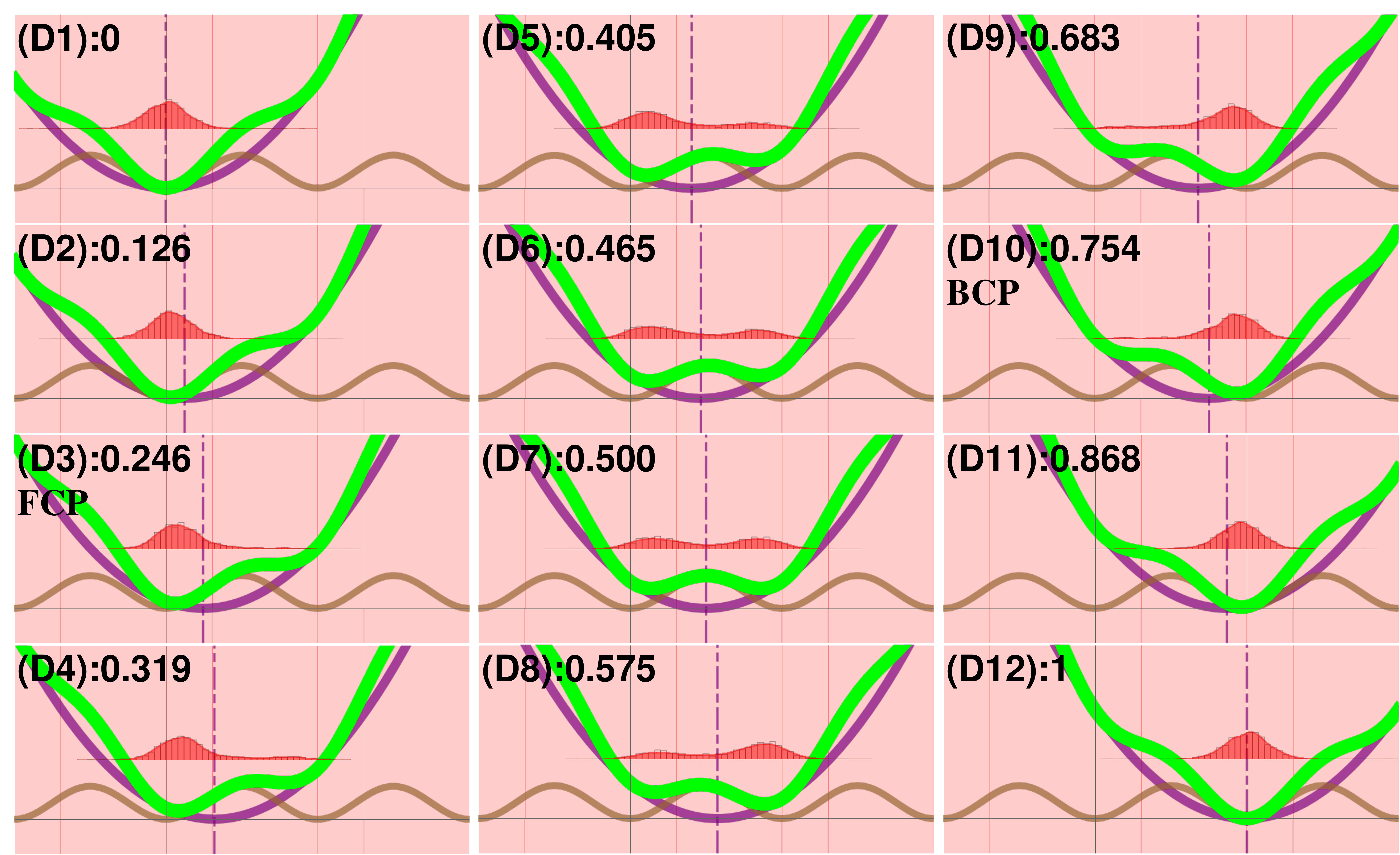}
\caption{Simulation results for the particle immersed in the homogeneous high temperature ${\it\Theta}=0.4$. (A) The internal energy $U$, work input $W$, heat to the heat bath $Q$ and the difference between them $\Delta=\Delta U+Q-W$ in the last three consecutive high temperature isothermal cycles among 2000 in total at $v_{\rm dr}=10^{-5}\rm m/s$. The dashed red curves are the resultant potential at the balanced points (Sec. \ref{CriticalPtsofTempZoneSec}). The inset of (A) represents the cycle work in 2000 consecutive cycles and its distribution. The mean cycle work is approximately zero. The $x-$coordinate $v_{\rm dr}t(a)$ is the nondimensional position of the driver center, which is proportional to time $t$. The dimension of the energy is $k_{\rm B}T$ with $k_{\rm B}$ the Boltzmann's constant and $T(=\Theta V_0)$ the absolute temperature. The dotted dark red curves indicate the lower bound of the internal energy. $\Delta\equiv0$ indicates the first law of thermodynamics (Sec. \ref{sec:firstlaw}). (B) The displacement of the particle in the typical 2000th homogeneous high temperature isothermal cycle with respect to the driver center's position, relative to the starting point of the same cycle. The dotted white line represents driver center's position with respect to itself. The dashed red curve is the loci of the balanced point (Figure \ref{fig:StickSlips}). The red background represents the homogeneous high temperature.
%If the the driver center is ahead of the ion, i.e. $v_{\rm dr}t>x$, then the friction force is positive and work is inputted. On the other hand, if the ion jumps forward and is ahead of the driver center, then $v_{\rm dr}t<x$ and the friction force is negtive and work is outputted.
Because of the violent fluctuation of the particle in such a high temperature, once getting through the FCP (the left triangle) the particle's shuttling between the two resultant potential local minimum points begins and then becomes even intense because of the forward energy barrier getting lower with the driver center moving forward. Nearby the middle instant of one cycle, the shuttling reaches its peak and then it recedes because of the rising again of the backward energy barrier until the instant of the BCP (the right triangle). (C) The variation of harmonic force in the 2000th cycle. The $x-$coordinate matches that of (B). Ratio of time of minus harmonic force in one cycle is approximately equal to that of the plus one, consistent with the near-zero mean cycle work. The dashed red curve is the balanced harmonic force at the balanced point (Figure \ref{fig:StickSlips}). The amplitude of the oscillation of the harmonic force is larger than that of the PTSHE and that of the homogeneous low temperature case. And it is even doubled around the middle instant of one cycle because of the violent shuttling of the particle between the two local minimum points. 
%The ratio of time of plus harmonic force is nearly equal to the minus harmonic force, in consistency with the near-zero mean cycle work. 
(D1)-(D12) Diagrams of the shapes of the resultant potential and the corresponding displacement distributions of the particle at different instants of the homogeneous high temperature isothermal cycle. The displacement distribution at a certain instant relative to the latest cycle starting point is obtained from the total 2000 simulation cycles. The decimal following the id of each frame is the nondimensional position of the driver center (or the nondimensional instant) relative to the starting point of the cycle (D1). The purple curve represents the moving harmonic potential and the brown one represents the lattice potential. The red background represents the homogeneous high temperature. In the homogeneous high temperature isothermal cycle case the shuttling process begins almost immediately the right potential local minimum appears at the instant of the FCP and won't end until the left potential local minimum disappears at the instant of the BCP. When the particle fluctuates around the left local minimum point, it is on average behind the driver center and the driver pulls it and does work on it, while when it oscillates around the right local minimum point, it is on average ahead of the dirver center so that it pulls the driver and outputs work. Shortly after the appearance of the right local minimum, the particle is in high probability on the left local minimum and there is work input to the particle as a whole. Whereas, as the driver is moving forward to the right, the left local minimum is going up and the right local minimum is going down. So the particle is more and more likely to stay around the right local minimum point and less and less likely to stay around the left local minimum point [(D4)-(D6)]. As a result, the work consumed by the particle gradually gets less and that outputted by it gradually gets more, which leads to descending slope of the work curve in (A). When the work input and output of the particle are balanced, the slope of the work curve will be zero, which won't happen until the driver moves close to the middle of one cycle ($v_{\rm dr}t=\frac12a$) when the two potential local minima are equal (D7). As the driver continues to move forward and the local minimum on the left goes up while the right one goes down, the particle oscillates around the right local minimum point with higher probability and is more likely to pull the driver. Thus the work done by the particle to the driver is more than that done by the driver to the particle and the particle outputs work in total. The work output of the particle increases and the input decreases due to the lowering of the right resultant potential local minimum and the highering of the left one, leading to the absolute slope of the work curve increasing until around the instant of the BCP. Parameters: $v_{\rm dr}=10^{-5}\rm m/s$, $\eta=3.0$, $\mu=4\times10^4\rm s^{-1}$, ${\it\Theta}=0.4$ and others are given in Sec. \ref{Langevindynamicssimulation}.\ref{ParametersUsed}.} 
\label{CyclesTh04}
\end{figure}

When the temperature is homogeneous and high (${\it\Theta}=0.4$) and the driving velocity is low ($v_{\rm dr}=10^{-5}\rm{m/s}$), the particle tends to be 
%nearly alwarys
at thermal equilibrium state at each instant, because in such a high temperature the particle's acute fluctuation leads to its easily crossing over the energy barrier to achieve equilibrium quickly, which is quicker than the driver's moving forward at such a relatively low driving velocity.
%the particle has high kinetic energy and can return from out of equilibrium to equilibrium quickly, which is quicker than the driver's moving forward at such a relatively low driving velocity. The particle's acute fluctuation leads to its easily crossing over the energy barrier to quickly achieve equilibrium. 
So in the homogeneous high temperature ${\it\Theta}=0.4$ case, $v_{\rm dr}=10^{-5}\rm{m/s}$ is a relatively low driving velocity rather than a high one compared with the homogeneous low temperature ${\it\Theta}=0.04$ case in the last subsection.
%Even if there is an energy barrier, the equilibrium can be achieved quickly due to the the particle's acute fluctuation leading to its easily crossing over. 
As we have mentioned in the last subsection, at equilibrium the particle's displacement obeys the Boltzmann distribution and is proportional to $\exp(-\frac{V}{k_{\rm B}T})$ which can be verified in Figure \ref{CyclesTh04}(D1)-(D12). 
%The ${\it\Theta}=0.4$ case has the same results as the ${\it\Theta}=0.04$ case if the driving velocities of both cases are low enough for the particle to relax to equilibrium at each instant, i.e. the low velocity limits of both cases are the same. 
As $v_{\rm dr}\rightarrow0$, the particle is at equilibrium at each instant so that the isothermal cycle is reversible and the work input in the forward direction is the same as the work output in the reversed direction. And because of the symmetry of the forward and the reversed cycle, the work curve in one cycle is symmetric about the middle so that the low velocity limit of the mean cycle work %$\langle W_{\rm cyc,limit}\rangle$ 
is zero as can be verified in Figure 3(C) in the main text. 

In fact $v_{\rm dr}=10^{-5}\rm{m/s}$ is still not low enough so that the standard deviations of $W_{\rm cyc}$ is not very small as can be seen in the inset of Figure \ref{CyclesTh04}(A) and Figure \ref{fig:Wcyc_T004T04_Error}. When the temperature is high, the mean cycle work and thus the mean friction force can achieve near-zero at a relatively low driving velocity. This effect is thermolubricity \cite{Thermolubricity}.

When the driving velocity goes to zero, the isothermal cycle is reversible and the work input is zero. On the other hand, when the driving velocity is finite, the isothermal cycle is irreversible so that the work input is nozero. So the second law of thermodynamics is obeyed.

At low enough driving velocity so that the particle is nearly at equilibrium at each instant, the homogeneous high temperature isothermal cycle can also be divided into four stages. 

The first stage ranges from the starting point of one cycle [Figure \ref{CyclesTh04}(D1)] to the instant of the FCP [Figure \ref{CyclesTh04}(D3) and the left red triangle in Figure \ref{CyclesTh04}(A) and (B)]. In this stage there is only one global minimum on the resultant potential curve and on average the driver center is ahead of the particle so that work is done by the driver to the particle. The mean value point of the particle's displacement distribution, which is approximately Boltzmann's distribution and proportional to $\exp(-\frac{V}{k_{\rm B}T})$, gradually gets nearer to the driver center and farther away from the global minimum point so that the work curve gradually gets lower than the dashed red balanced resultant potential curve, cf. Sec. \ref{sec:apndRD}.\ref{SimilaritybtwWorkPotential}.

The second stage ranges from the instant of the FCP to the maximum instant of the work curve of one cycle, which approximately coincides with the middle instant of one cycle [Figure \ref{CyclesTh04}(D7) and the red diamond in Figure \ref{CyclesTh04}(A) and (B)]. In this stage there are two local minima and an energy barrier between them. The particle's displacement distribution has two peaks and the left one is higher than the right one so that the mean displacement is between the left local minimum point and the driver center and gradually approaches the driver center as the left local minimum gets higher and the right one gets lower, with the driver moving forward. At the middle instant the two local minima are equal in height so that the two peaks of the displacement distribution are of the same height and the mean displacement are around the driver center. Afterwards the mean displacement is ahead of the driver center. Therefore, before the middle of one cycle, work is inputted and after the middle of one cycle, work is outputted so that there is a maximum on the work curve nearby the middle instant. When the driving velocity goes to zero, the maximum point will be at the middle instant of the cycle exactly.

The third stage ranges from the maximum instant of the work curve of one cycle to the instant of the BCP [Figure \ref{CyclesTh04}(D10)]. In this stage, there are still two local minima and one energy barrier between them. The displacement distribution has two peaks,  the left of which gradually gets lower than the right one until it vanishes at the BCP instant. The mean displacement is ahead of and gradually gets farther away from the driver center so that the work curve goes down and its absolute slope gets larger.

The fourth stage ranges from the instant of the BCP to the end of one cycle [Figure \ref{CyclesTh04}(D12)]. In this stage, there is again one global minimum and one displacement distribution peak. The mean displacement is between the driver center and the global minimum point so that work is still outputted. As the driver center gradually catches up with the global minimum point and also the mean displacement point (at the end of one cycle the driver center, the global minimum point and the mean displacement point are coincident), the absolute slope of the work output curve decreases to zero.

We have to note that the analysis above and in Figure \ref{CyclesTh04} is qualitative and not very rigorous. It's an interesting problem to find whether the inflection points of the work curve are at the FCP and the BCP instant and the equilibrium mean displacement value at each instant could also be figured out.

\subsection{The similarity of the shape of the work curve and that of the lower bound of the internal energy} 
\label{SimilaritybtwWorkPotential}
The lower bound of the internal energy is on the resultant potential energy minimum branches indicating that the particle frequently relaxes to one of the stable balanced points due to the damping force and because of the stochastic force, the particle can regain kinetic energy and the internal energy will again exceed the local minimum resultant potential energy.
%because if the kinetic energy of the particle relaxes to zero, the particle will with high probability stay at the local minimum where the harmonic force and the lattice force are balanced.
%because the resultant potential energy minimum at each instant has the amount of potential energy that can't transform into kinetic energy. 
%We will derive that the inputted work is approximately equal to the increment of the resultant potential energy and the outputted work is approximately equal to the decrement of the resultant potential energy.

At low velocity, the particle fluctuates around a stable balanced point which is approximately the average position of the particle. At the balanced point $z(\tau)^*$ the driver center's position $\tilde X(z(\tau)^*)$ satisfys Eq. \ref{Xtau2}. Therefore at low velocity the inputted work can be approximated by
\begin{equation}
\begin{aligned}
W(\tau)&\approx\int_{t_0}^t\kappa[X(x(t)^*)-x(t)^*]\mathrm dX(x(t)^*)\\
&=\int_{\tau_0}^\tau\kappa a^2[\tilde X(z(\tau)^*)-z(\tau)^*]\mathrm d\tilde X(z(\tau)^*)\\
&=\int_{\tau_0}^\tau\kappa a^2\eta\sin(z(\tau)^*)\mathrm d[z(\tau)^*+\eta\sin(z(\tau)^*)]\\
&=\kappa a^2\eta\left[-\cos(z(\tau)^*)+\cos(z(\tau_0)^*)\right]+\kappa a^2\eta^2\frac12\left[\sin^2(z(\tau)^*)-\sin^2(z(\tau_0)^*)\right]\\
&=\kappa a^2\eta\left[-\cos(z(\tau)^*)+\cos(z(\tau_0)^*)\right]+\kappa a^2\eta^2\frac12\left\{\left[\frac1\eta(\tilde X(z(\tau)^*)-z(\tau)^*)\right]^2-\sin^2(z(\tau_0)^*)\right\}\\
&=\frac12\kappa a^2\left[\tilde X(z(\tau)^*)-z(\tau)^*\right]^2+\kappa a^2\eta\left[1-\cos(z(\tau)^*)\right]+\kappa a^2\eta\left[-1+\cos(z(\tau_0)^*)\right]-\frac12\kappa a^2\eta^2\sin^2(z(\tau_0)^*)\\
&=\kappa a^2\tilde V(z(\tau)^*,\tilde X(z(\tau)^*))+\kappa a^2\eta\left[-1+\cos(z(\tau_0)^*)\right]-\frac12\kappa a^2\eta^2\sin^2(z(\tau_0)^*)\\
&=C_1\tilde V(z(\tau)^*,\tilde X(z(\tau)^*))+C_2,
\end{aligned}
\end{equation}
where $C_1$ and $C_2$ are constants. Therefore the shape of the work curve and that of the resultant potential energy local minmum branch are similar.

Intuitively, when the driver center moves forward ahead of the particle, the particle fluctuates around the left local minimum point of the resultant potential energy which goes forward and upward, so the work done on the particle is transformed into its potential energy. On the other hand, when the driver center moves forward behind the particle, the particle fluctuates around the right local minimum point of the resultant potential energy which goes forward and downward so the potential energy is transformed into work output. The latter is analogous to the free falling body transforming its gravity potential energy into kinetic energy which can be utilized to do work, and of course here the particle is not free.
%At the ascent segment $\cos(z(\tau_0)^*)=1,\sin^2(z(\tau_0)^*=0$, so that 
%\begin{equation}
%\begin{aligned}
%W(\tau)=\kappa a^2\tilde V(z(\tau)^*,\tilde X(\tau)^*).
%\end{aligned}
%\end{equation}
%At the descent segment, the starting instant $\tau_0$ is at the cusp: $\tau_0=\tau_{
%\rm cusp}$, and $W(\tau_{\rm cusp})=\kappa a^2\tilde V(z(\tau_{\rm cusp}),\tilde X(\tau_{\rm cusp}))$ so that
%\begin{equation}
%\begin{aligned}
%W(\tau)=\kappa a^2\tilde V(z(\tau)^*,\tilde X(\tau)^*).
%\end{aligned}
%\end{equation}
\subsection{The high velocity limit of the mean cycle work}
\label{sec:hvlim}
Here we give a derivation of the high velocity limit of the mean cycle work. The nonlinear Langevin equation 4 in the main text can be reformulated into 
\begin{equation}\label{eqn:NLhvlim}
m\ddot{x}(t)+m\mu\dot{x}(t)+\kappa\left[x(t)-v_{\rm dr}t\right]+\frac\pi aV_0\sin\left(\frac{2\pi}ax(t)\right)-\xi(t)=0.
\end{equation}
At very high driving velocity $v_{\rm dr}$, the damping force $m\mu\dot x(t)$ dominates on the LHS, so that we can omit the inertia force $m\ddot x(t)$, the lattice force $\frac\pi aV_0\sin\left(\frac{2\pi}ax(t)\right)$ and the fluctuation force $\xi(t)$. Then, Eq. \ref{eqn:NLhvlim} can be approximated by
\begin{equation}
m\mu\dot{x}(t)+\kappa\left[x(t)-v_{\rm dr}t\right]=0,
\end{equation}
i.e. 
\begin{equation}
-\kappa\left[x(t)-v_{\rm dr}t\right]=m\mu\dot{x}(t).
\end{equation}
Then we can calculate the work through
\begin{equation}
W=-\int_{t_0}^t\kappa\left[x(t)-v_{\rm dr}t\right]v_{\rm dr}\mathrm dt=\int_{t_0}^tm\mu\dot{x}(t)v_{\rm dr}\mathrm dt=m\mu v_{\rm dr}\int_{t_0}^t\mathrm dx(t)=m\mu v_{\rm dr}[x(t)-x(t_0)].
\end{equation}
And the mean cycle work at high velocity can be calculated by
\begin{equation}
\langle W_{\rm cyc,hvlim}\rangle=m\mu v_{\rm dr}[\langle x(t_0+\frac{a}{v_{\rm dr}})\rangle-x(t_0)]=m\mu v_{\rm dr}a,
\end{equation}
which is linear with $v_{\rm dr}$. 

Note that at the high driving velocity limit, the time range of one engine cycle is small so that we can neglect the fluctuation force $\xi(t)$, which has a low probability to achieve very high value in such a short period. However, at the low driving velocity limit, the time range of one engine cycle is large and the fluctuation force $\xi(t)$ has a high probability to achieve very high value. Then we cannot omit $\xi(t)$. When the damping coefficient $\mu$ is large, we can still omit the inertia force $m\ddot x(t)$ and obtain the overdamped Langevin equation
\begin{equation}
m\mu\dot{x}(t)+\kappa\left[x(t)-v_{\rm dr}t\right]+\frac\pi aV_0\sin\left(\frac{2\pi}ax(t)\right)-\xi(t)=0.
\end{equation}
%\subsection{The calculation of the theoretical low velocity limit of mean cycle work}
\subsection{The equilibrium limit of the mean cycle work approximated by the potential mechanism}
\label{sec:wcyceploweta}
\begin{figure}[H]
 \centering
\includegraphics[width=0.5\textwidth]{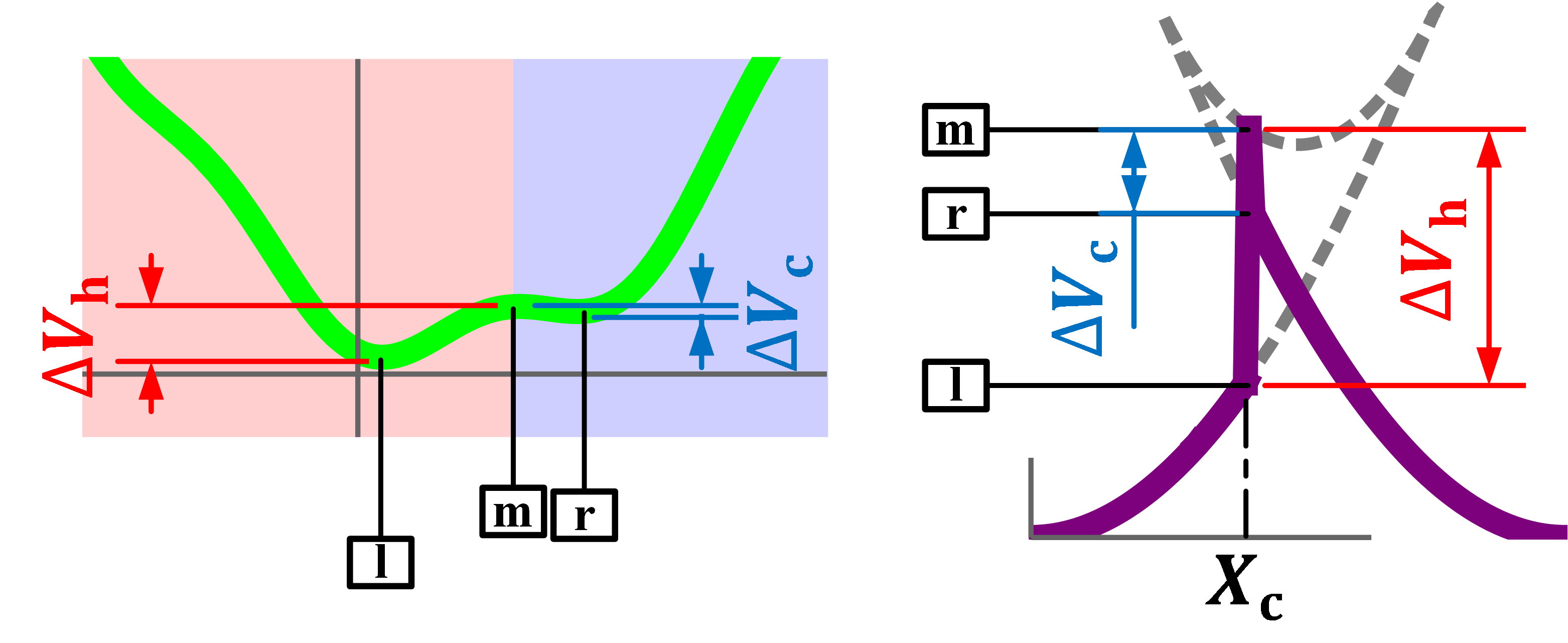}
\caption{The schematic of calculating the equilibrium limit of the mean cycle work $W_{\rm cyc,e.p.}$. In the left subfigure, the green curve represents the resultant potential at the cusp instant when the driver center is at $X_{\rm c}$. $\rm l,m,r$ represents the left local minimum, the middle local maximum and the right local minimum respectively. The red and blue backgrounds represent the high and low temperature heat baths respectively. In the right subfigure, the dashed gray curve represents the resultant potential at its local extremum points. The solid purple curve represents the resultant potential energy at the local minimum points passed through by the particle, with the vertical
%dot-dashed 
line at $X_{\rm c}$ representing %the average driver center position where 
the particle jumping from the hot zone to the cold one.}
\label{Carnot}
 \end{figure}

At equilibrium in the classical limit, we assume the probability distribution of the particle's energy $E$ relative to its resultant potential energy local minimum
%nearby the local minimum points of the resultant potential energy 
has a Boltzmann distribution \cite{bylinskiiphdthesisEnergyDist}
\begin{equation}
p_{\rm E}=\frac1{k_{\rm B}T}\exp(-\frac{E}{k_{\rm B}T}).
\end{equation}
We also assume that on the left of the energy barrier peak point [Figure 3(A2) in the main text and Figure \ref{Carnot}], the heat bath is of high temperature and on the right of the energy barrier peak point, the heat bath is of low temperature, which is different from the periodic alternative high and low temperature heat bath of the PTSHE. In the hot zone, the probability of the particle jumping forward over the energy barrier can be approximated \cite{bylinskiiphdthesisEnergyDist} by
\begin{equation}
\begin{aligned}
P_{\rm h}(E>\Delta V_{\rm h})&=\int_{\Delta V_{\rm h}}^{+\infty}p_{\rm E}\mathrm dE\\
&=\int_{\Delta V_{\rm h}}^{+\infty}\frac1{k_{\rm B}T_{\rm h}}\exp(-\frac{E}{k_{\rm B}T_{\rm h}})\mathrm dE\\
&=\exp(-\frac{\Delta V_{\rm h}}{k_{\rm B}T_{\rm h}}),
\end{aligned}
\end{equation}
and that of the particle jumping backward over the energy barrier from the cold zone to the hot one can be approximated by
\begin{equation}
\begin{aligned}
P_{\rm c}(E>\Delta V_{\rm c})&=\int_{\Delta V_{\rm c}}^{+\infty}p_{\rm E}\mathrm dE\\
&=\int_{\Delta V_{\rm c}}^{+\infty}\frac1{k_{\rm B}T_{\rm c}}\exp(-\frac{E}{k_{\rm B}T_{\rm c}})\mathrm dE\\
&=\exp(-\frac{\Delta V_{\rm c}}{k_{\rm B}T_{\rm c}}).
\end{aligned}
\end{equation}
At the average cusp instant of the work curve,
%[cf. Figure 2(A) in the main text], 
the probabilities of the particle jumping forward and backward over the energy barrier satisfy
\begin{equation}
P_{\rm h}(E>\Delta V_{\rm h})=P_{\rm c}(E>\Delta V_{\rm c}),
\end{equation}
i.e.
\begin{equation}
\exp(-\frac{\Delta V_{\rm h}}{k_{\rm B}T_{\rm h}})=\exp(-\frac{\Delta V_{\rm c}}{k_{\rm B}T_{\rm c}}),
\end{equation}
leading to Eq. 3 in the main text.
The equilibrium limit of the mean cycle work 
\begin{equation}
\label{eqn:eq3maintext}
W_{\rm cyc,e.p.}=\Delta V_{\rm h}-\Delta V_{\rm c}.
\end{equation}

It is shown in the inset of Figure 3(A2) and (A3) in the main text and Figure \ref{Carnot} that at the average cusp instant when the driver center is at $\tilde X_{\rm c}$ the two local minima points $z_l,\ z_r$ and the local maximum point $z_m$ satisfy
\begin{equation}\label{eqn:zlXcconstraint}
\begin{aligned}
\frac{\partial \tilde V}{\partial z}(z_l)&=z_l-\tilde X_{\rm c}+\eta\sin(z_l)=0,\\
\frac{\partial \tilde V}{\partial z}(z_m)&=z_m-\tilde X_{\rm c}+\eta\sin(z_m)=0,\\
\frac{\partial \tilde V}{\partial z}(z_r)&=z_r-\tilde X_{\rm c}+\eta\sin(z_r)=0,
\end{aligned}
\end{equation}
(c.f. Eq. \ref{Vtau}) and the nondimentional $\Delta\tilde V_{\rm h}$ and $\Delta\tilde V_{\rm c}$ can be calculated by
\begin{equation}
\begin{aligned}
\Delta \tilde V_{\rm h}&
=\tilde V(z_m)-\tilde V(z_l),\\
\Delta \tilde V_{\rm c}&
=\tilde V(z_m)-\tilde V(z_r).
\end{aligned}
\end{equation}
Considering Eq. 3 in the main text (or Eq. \ref{eqn:eq3maintext}), we obtain the equation system 
\begin{equation}
\begin{cases}
\tilde V(z_m)-\tilde V(z_l)=\frac{{\it\Theta}_{\rm h}}{{\it\Theta}_{\rm c}}[\tilde V(z_m)-\tilde V(z_r)],\\
z_l-\tilde X_{\rm c}+\eta\sin(z_l)=0,\\
z_m-\tilde X_{\rm c}+\eta\sin(z_m)=0,\\
z_r-\tilde X_{\rm c}+\eta\sin(z_r)=0,
\end{cases}
\end{equation}
which can be solved by a nonlinear solver with the ranges of $z_l$, $z_m$ and $z_r$ being
\begin{equation}
\begin{cases}
z_l&\in(0,\arccos(-1/\eta)),\\
z_m&\in(\arccos(-1/\eta),2\pi-\arccos(-1/\eta)),\\
z_r&\in(2\pi-\arccos(-1/\eta),2\pi),
\end{cases}
\end{equation}
c.f. Sec. \ref{CriticalPtsofTempZoneSec}.

Therefore we can get $z_l$, $z_r$ and $z_m$ and then $\Delta \tilde V_{\rm h}$ and $\Delta \tilde V_{\rm c}$. So the equilibrium limit of the mean cycle work approximated by the potential mechanism
\begin{equation}\label{worklimit_tight}
W_{\rm cyc,e.p.}=\frac{m\omega_0^2a^2}{4\pi^2}(\Delta \tilde V_{\rm h}-\Delta \tilde V_{\rm c}).
\end{equation}

As the ratio $\frac{T_{\rm h}}{T_{\rm c}}=\frac{{\it\Theta}_{\rm h}}{{\it\Theta}_{\rm c}}\rightarrow+\infty$ so that $\frac{\Delta V_{\rm h}}{\Delta V_{\rm c}}\rightarrow+\infty$ (the main text Eq. 3), we can see from Figure \ref{Carnot} that the jumping point $X_{\rm c}\rightarrow\tilde X(z_2^{**})$, cf. Figure \ref{DeterminationCritPts}(B). Then both $z_r$ and $z_m$ go to $z_2^{**}=2\pi-\arccos(-\frac1\eta)$ (Eq. \ref{CritPts}) and in the limit $z_l$ satisfies
\begin{equation}
z_l-\tilde X(z_2^{**})+\eta\sin(z_l)=0,
\end{equation}
with $\tilde X(z_2^{**})=z_2^{**}+\eta\sin(z_2^{**})$ and $z_l\in(0,\arccos(-1/\eta))$. Solve this equation with a nonlinear solver we can get the limit of $z_l$, and then we can obtain the limit of $W_{\rm cyc,e.p.}$ as $\frac{{\it\Theta}_{\rm h}}{{\it\Theta}_{\rm c}}\rightarrow+\infty$ by Eq. \ref{worklimit_tight}. This limit is an upper bound of $W_{\rm cyc,e.p.}$.

At $\eta=2$, $3$ and $4$, as $\frac{{\it\Theta}_{\rm h}}{{\it\Theta}_{\rm c}}\rightarrow+\infty$,
\begin{equation}
-W_{\rm cyc,e.p.}\rightarrow-1.05k_{\rm B}T_{\rm h},\ -2.96k_{\rm B}T_{\rm h}\ \text{and}\ -5.15k_{\rm B}T_{\rm h},
\end{equation} 
respectively with $k_{\rm B}T_{\rm h}=0.4\frac{3m\omega_0^2a^2}{2\pi^2}$ the same as that of the main text Figure 5(A) and the three limits are what the three corresponding dot-dashed curves in the main text Figure 5(A) go to as ${\it\Theta}_{\rm h}\rightarrow+\infty$. At $\eta=1$, $-W_{\rm cyc,e.p.}\equiv0$.

\subsection{The reason for the higher work output at $\eta=2$ and $\eta=1$ than the potential mechanism approximation}
%In Figure \ref{SimuResultEta2}(A), we can see that for the $\eta=2$ case, in the typical 2000th engine cycle, the heat $Q$ absorbed before the the cliff instant is of a large amount and larger than that absorbed at the cliff instant. Because of the relatively flat resultant potential energy at $\eta=2$, the particle oscillates in a large spatial range even before the FCP occurs [Figure \ref{SimuResultEta2}(B)] and frequently goes ahead of the driver center. So besides the amount of work output transformed from the potential energy, which is mainly converted from the heat absorbed at the cliff instant, a large amount of kinetic energy, which is converted from the heat absorbed in the hot zone, is transformed into the particle's potential energy resulting in the work input being reduced, cf. the main text. And this amount of reduced work input is relatively more significant than that of the high $\eta$ case at the same absolute temperature, where the particle's crossing over the middle energy barrier happends less frequently bescause of the high energy barrier.
 
%Therefore, different from the high $\eta$ cases, the flat resultant potential energy leading to more kinetic energy transforming into work output. And that's the mechanism of the higher mean cycle work output limit than that calculated by Eq. (\ref{worklimit_tight}) at the low $\eta$'s. 

\subsubsection{The case of $\eta=2$}
\begin{figure}[H]
\centering
\includegraphics[width=0.44\textwidth]{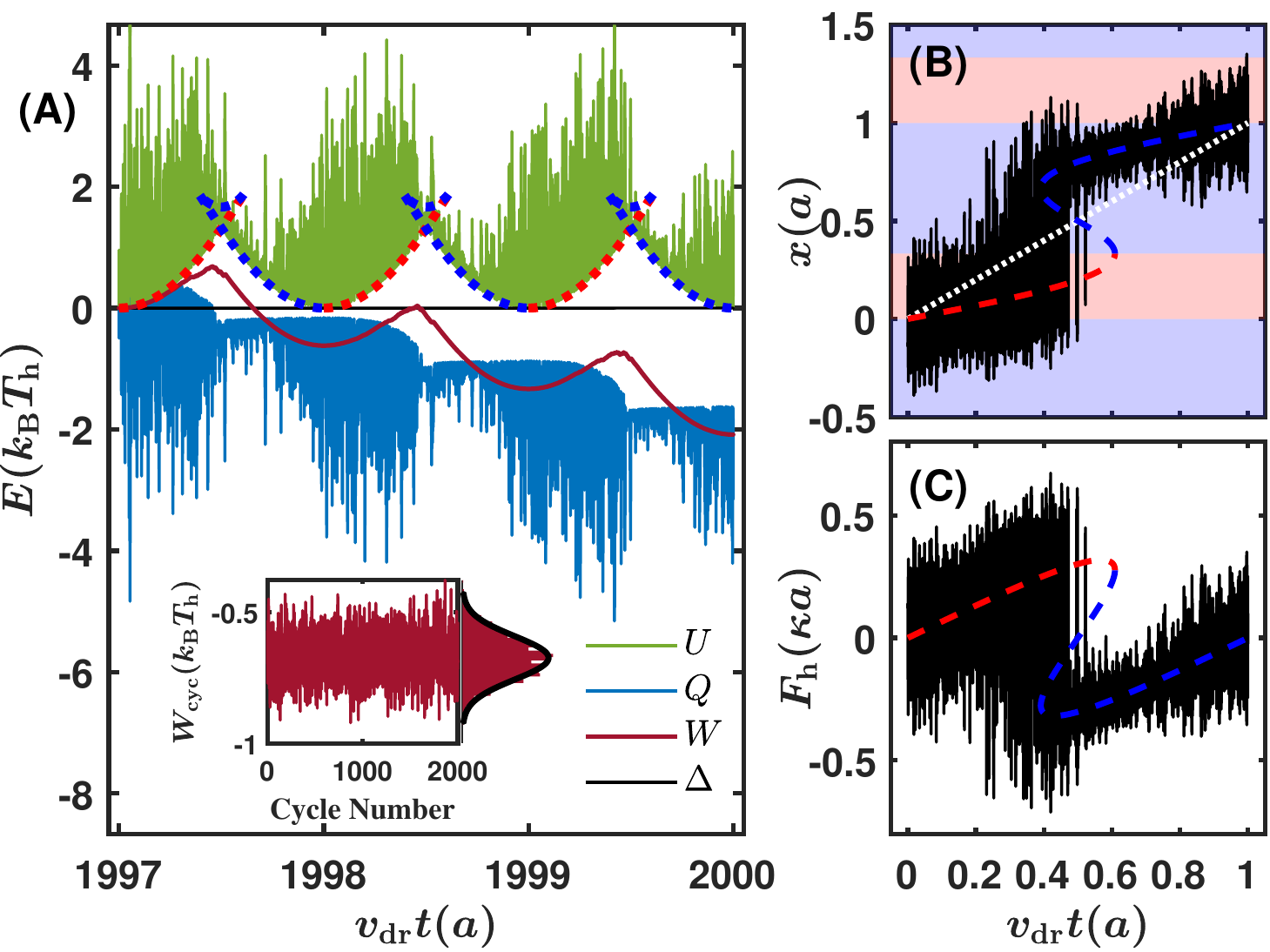}
\includegraphics[width=0.54\textwidth]{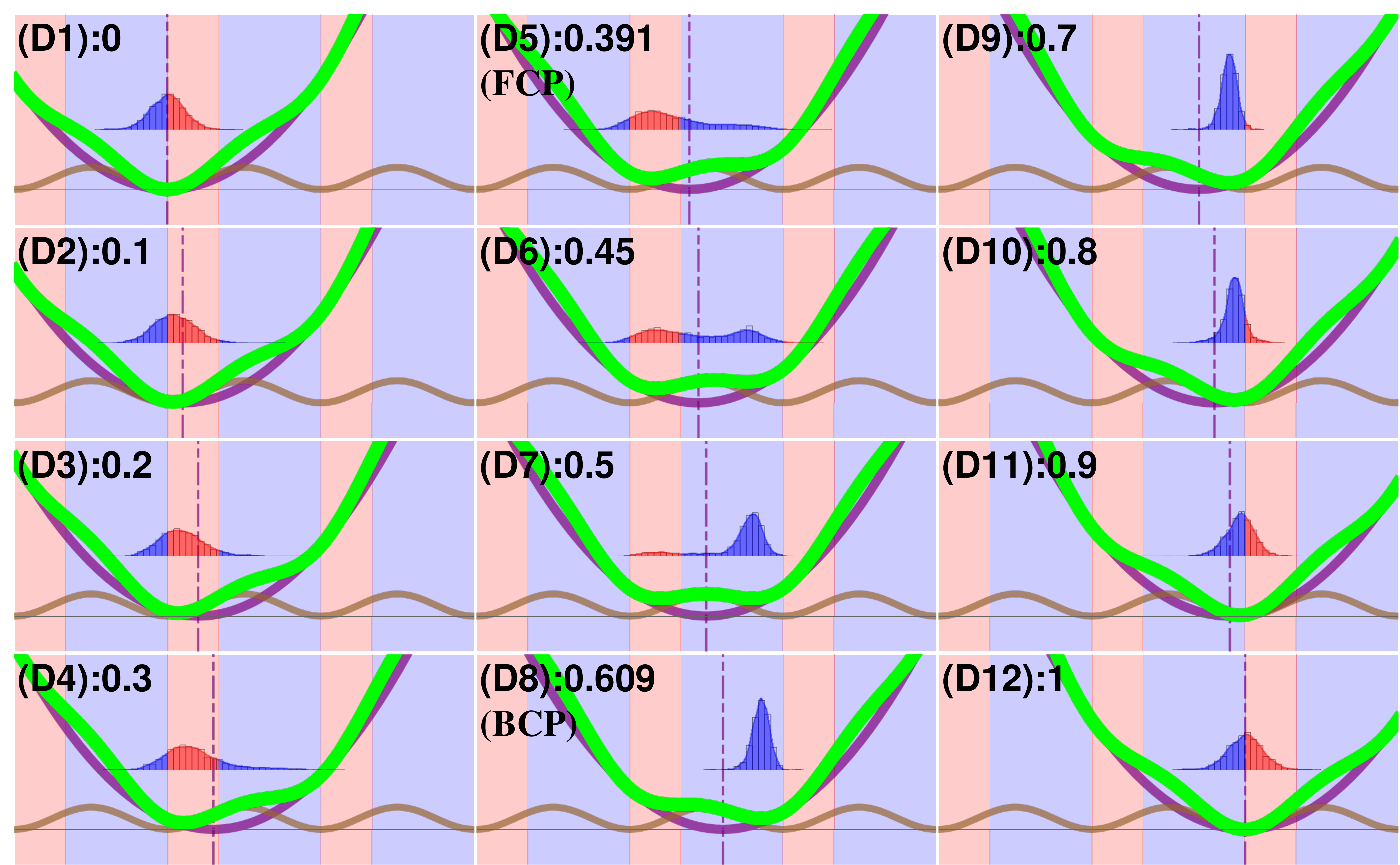}
\caption{The simulation results of the case of $\eta=2,\ \mu=4\times10^4\rm s^{-1},\ {\it\Theta}_{\rm h,c}=0.6,0.06$ at $v_{\rm dr}=10^{-5}\rm m/s$. The layout of this figure is similar to that of the main text Figure 2. The colors and types of the curves are the same as those in the main text Figure 2. The temperature $T_{\rm h}$ in the dimension $k_{\rm B}T_{\rm h}$ is the same as that
% of the $\eta=3$ case
 in the main text Figure 3. In the typical 2000th engine cycle, the heat $Q$ absorbed before the the cliff instant is of a large amount and larger than that absorbed at the cliff instant, cf. (A). Because of the relatively flat resultant potential resulting from the relatively flat lattice potential, the particle oscillates in a large spatial range even before the FCP occurs and frequently goes ahead of the driver center [(B) and (D3)-(D5)]. So besides the amount of work output transformed from the potential energy, which is mainly converted from the heat absorbed at the cliff instant, a large amount of kinetic energy, which is converted from the heat absorbed in the hot zone before the cliff instant, is transformed into the particle's potential energy, resulting in the work input being reduced, cf. the main text. And this amount of reduced work input is relatively more significant than that of the high $\eta$ cases at the same absolute temperature, where the particle's crossing over the middle energy barrier happends less frequently bescause of the higher energy barrier. Other parameters are given in Sec. \ref{Langevindynamicssimulation}.\ref{ParametersUsed}.}
\label{SimuResultEta2}
\end{figure}

In Figure 4 of the main text, we plot the energy and displacement curves of the particle at $v_{\rm dr}=10^{-7}\rm m/s$ for the case of $\eta=2$. Comparing with the results at $v_{\rm dr}=10^{-7}\rm m/s$, we can see that the nonequilibrium features are evident at $v_{\rm dr}=10^{-5}\rm m/s$ in Figure \ref{SimuResultEta2}. The cusp of the work curve is sharper and there is a short cliff on the heat curve [Figure \ref{SimuResultEta2}(A)] and also on the internal energy [Figure \ref{SimuResultEta2}(A)], the displacement [Figure \ref{SimuResultEta2}(B)] and the harmonic force [Figure \ref{SimuResultEta2}(C)] curves. The standard deviation is finite (although not very large) as can be seen in the inset of Figure \ref{SimuResultEta2}(A). 

As we have mentioned in the main text, the advantage of the nonequilibrium (or qusi-equilibrium) results at $v_{\rm dr}=10^{-5}\rm m/s$ is that we can get the displacement distribution of the particle benefited from the less simulation time steps in one cycle%the smaller computation cost
. The displacement distribution at $v_{\rm dr}=10^{-5}\rm m/s$ in Figure \ref{SimuResultEta2}(D1)-(D12) can be treated as an approximation of the equilibrium displacement distribution. We can see that before the mean cusp instant [(D1)-(D6)] the particle's displacement distribution is divergent and a large part is ahead of the driver center, resulting in that the average position of the particle is farther away from the left global or local minimum point and nearer to the driver center so that the work curve gets lower than dotted red curve in (A), cf. Sec. \ref{SimilaritybtwWorkPotential}. On the other hand, after the cusp instant, the particle's displacement distribution is convergent and the average displacement is close to the right local or global minimum point [(D8)-(D10)]. So the work curve is similar to the dotted blue curve in (A).

\subsubsection{The case of $\eta=1$}
\label{sec:caseeta1}
%In the main text, we find that at $\eta=1$ and $\eta=2$ the limit of the mean cycle work is larger than that calculated by Eq. (\ref{worklimit_tight}). Here we analyze the reason. 

\begin{figure}[H]
\centering
\includegraphics[width=0.44\textwidth]{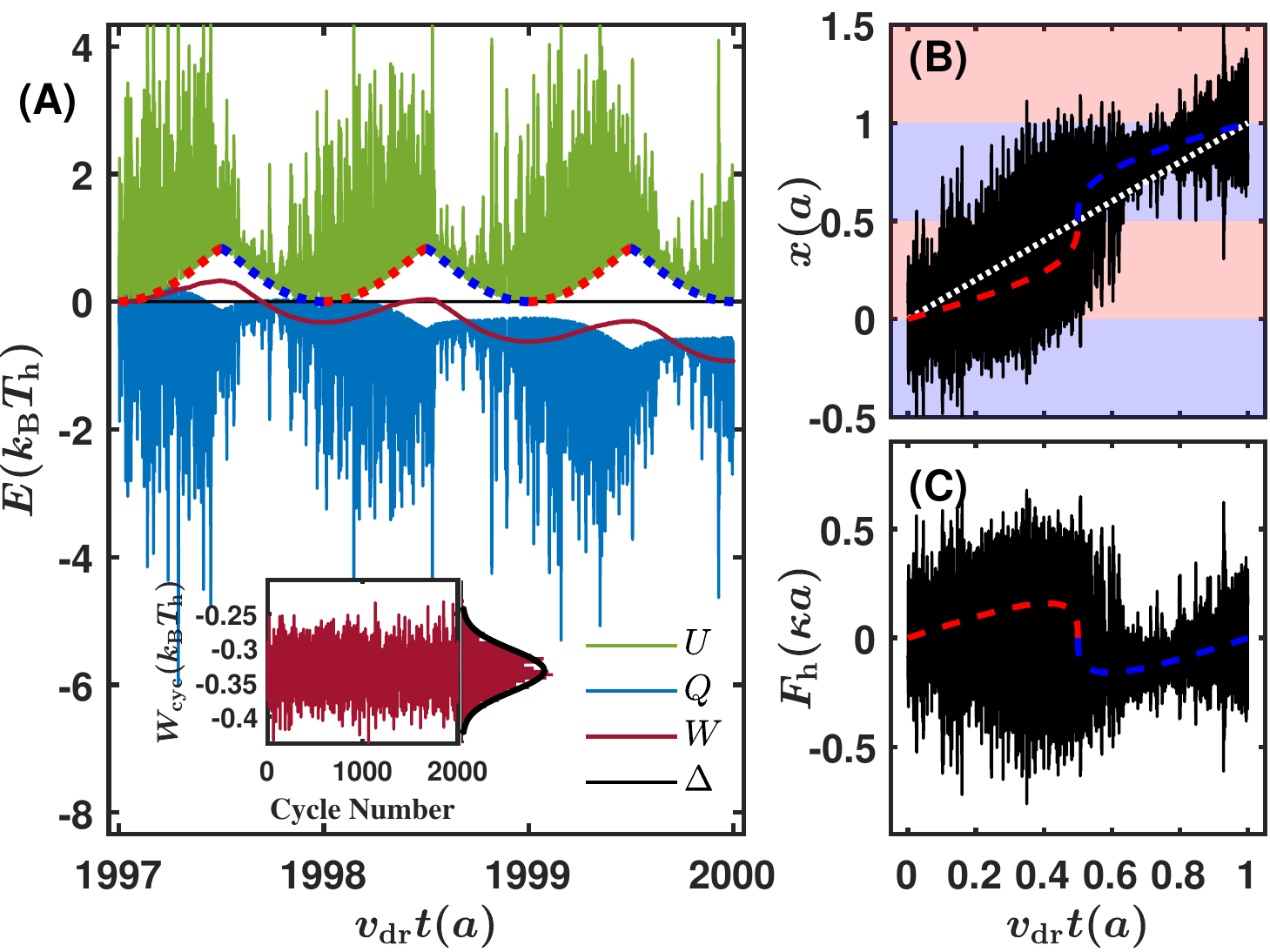}
\includegraphics[width=0.54\textwidth]{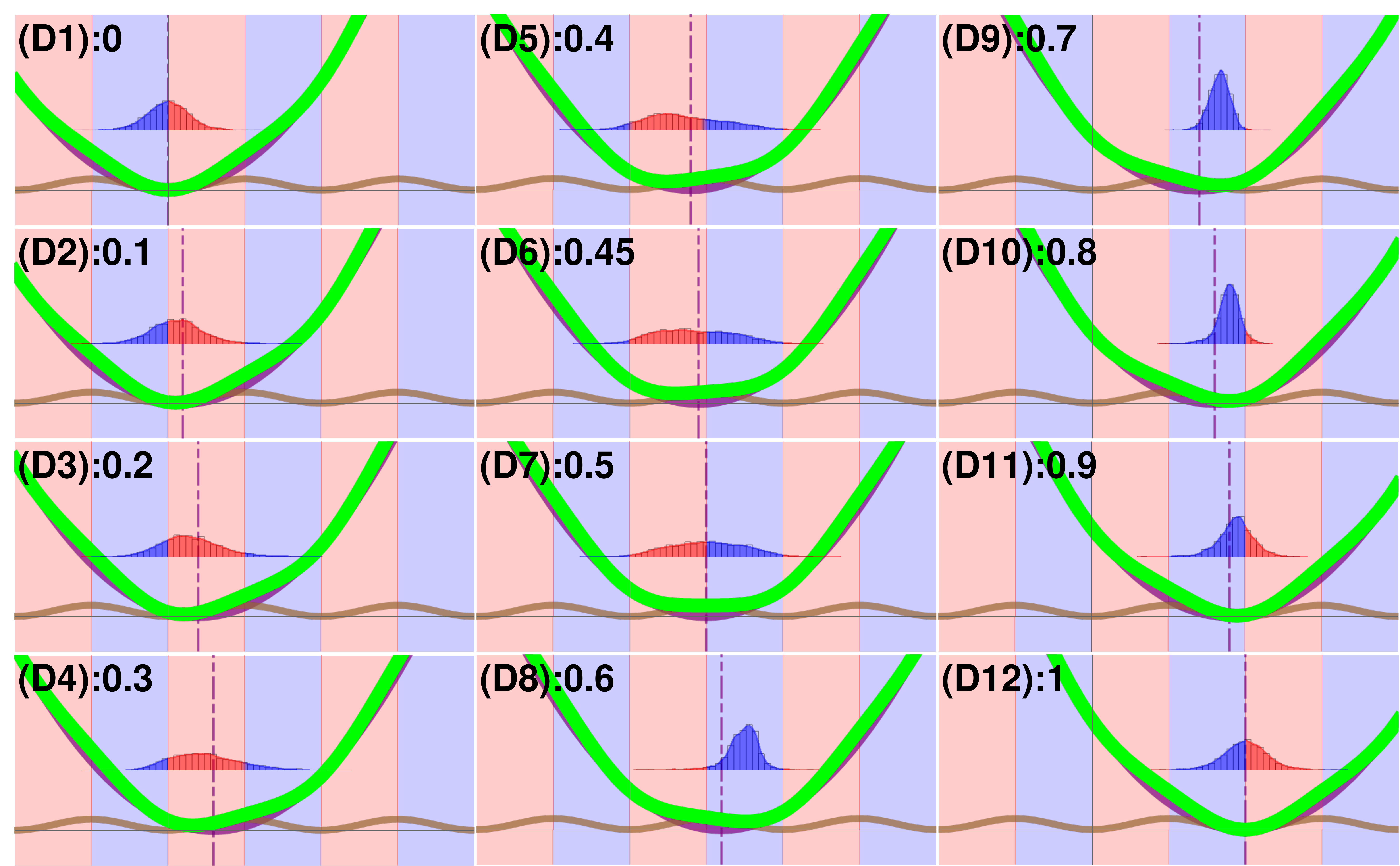}
\caption{The simulation results of the case of $\eta=1$, $\ \mu=4\times10^4\rm s^{-1}$, ${\it\Theta}_{\rm h,c}=1.2,0.12$ at $v_{\rm dr}=10^{-5}\rm m/s$. The layout of this figure is similar to that of the main text Figure 2. The colors and types of the curves are the same as those in the main text Figure 2. The temperature $T_{\rm h}$ in the dimesion $k_{\rm B}T_{\rm h}$is the same as that in the main text Figure 3. Other parameters are given in Sec. \ref{Langevindynamicssimulation}.\ref{ParametersUsed}.}
\label{SimuResultEta1}
\end{figure}

In Figure \ref{SimuResultEta1}(D1)-(D12), we can see the resultant potential at $\eta=1$ has only one global minimum and no energy barrier. At the starting point of one cycle (D1), the resultant potential is symmetric and then becomes asymmetric. Before the middle of one cycle (D7), the resultant potential is steeper on the left and flatter on the right of the global minimum and as the driver goes forward, the bottom of the resultant potential curve gets flatter and flatter. Therefore in the hot zone during the first half of one cycle, the particle distributes more and more dispersedly. 
%In Figure \ref{SimuResultEta1}(B), we can see that in the typical 2000th engine cycle, before $v_{\rm dr}t/a=0.5$, the particle oscillates to the right frequently after $v_{\rm dr}t/a\approx0.1$ and distributes beyond a full spatial period from $v_{\rm dr}t/a\approx0.3$ to $v_{\rm dr}t/a\approx0.5$. Therefore, there is a large part of time in the first half of one cycle when the particle is ahead of the driver center. In Figure \ref{SimuResultEta1}(A), we can see that the heat $Q$ decreases a lot before the middle of one cycle and the work curve increases slower than the lower bound of the internal energy curve, i.e. heat is absorbed from the heat bath and transformed into work.
After $v_{\rm dr}t/a=0.5$, 
%although it dosen't calm down immediately and oscillates backward behind the driver center [Figure \ref{SimuResultEta1}(B)], 
the distribution of the particle's position shrinks quickly and at $v_{\rm dr}t/a=0.6$ (D8) it distributes around the global minimum in the cold zone and is almost all ahead of the driver center. When approaching to the end of one cycle, the particle's displacement distribution expands again because of its frequently traversing to the hot zone in the next lattice period. 
%So after the middle of one cycle, work is first inputted and heat is released for a short time and then heat is no long release much and potential energy is transformed into work output notably [Figure \ref{SimuResultEta1}(A)]. 

%Before around the middle of the typical 2000th cycle, heat is absorbed by the particle and transformed into kinetic energy and potential energy. 
%As the resultant potential energy is flat, the kinetic energy is large and the potential is small. The particle oscillates with large amplitude and goes ahead of the driver center frequently so that a large amount of kinetic energy transforms into work output. 
%After the middle, most of the particle's kinetic energy first damped into heat and then the particle's potential energy transformes into work output.

In Figure \ref{SimuResultEta1}(B), we can see that after the middle of one cycle, the fluctuation amplitude of the particle reduces due to the damping force and also the small stochastic force in the cold zone, which can only maintain a small fluctuation amplitude of the particle. From (D7) to (D8), we can also see that after the middle instant the particle's displacement distribution shrinks. Small fluctuation and convergent displacement distribution indicate small kinetic energy and the amount of declined kinetic energy is dissipated into heat. So the upper bound of the heat curve ascends. On the other hand, shortly after the middle instant, the particle's high kinetic energy resulting from its frequently traversing backwards through the hot zone leads to its average displacement point nearly overlaping with the dirver center so that the work curve stays almost unchanged. In this period, the potential energy of the particle decreases. As the work output is nearly zero, the reduced potential energy is transformed into kinetic energy and the kinetic energy is transformed into heat, which also contributes to the ascent of the upper bound of the heat curve. As the driver continues to move forward, the particle goes deep into the cold zone and its displacement distribution shrinks with the average value point catches up with the global minimum point (D8). So the work curve gradually decreases and its slope increases and becomes similar to that of the dotted blue balanced resultant potential curve. Afterwards, the upper bound of the heat curve stays constant and the kinetic energy and the heat transforms into each other interactively. The work curve is parallel to the balanced resultant potential curve so the potential energy gradually transforms into work output.

\begin{figure}[H]
\centering
\includegraphics[width=0.5\textwidth]{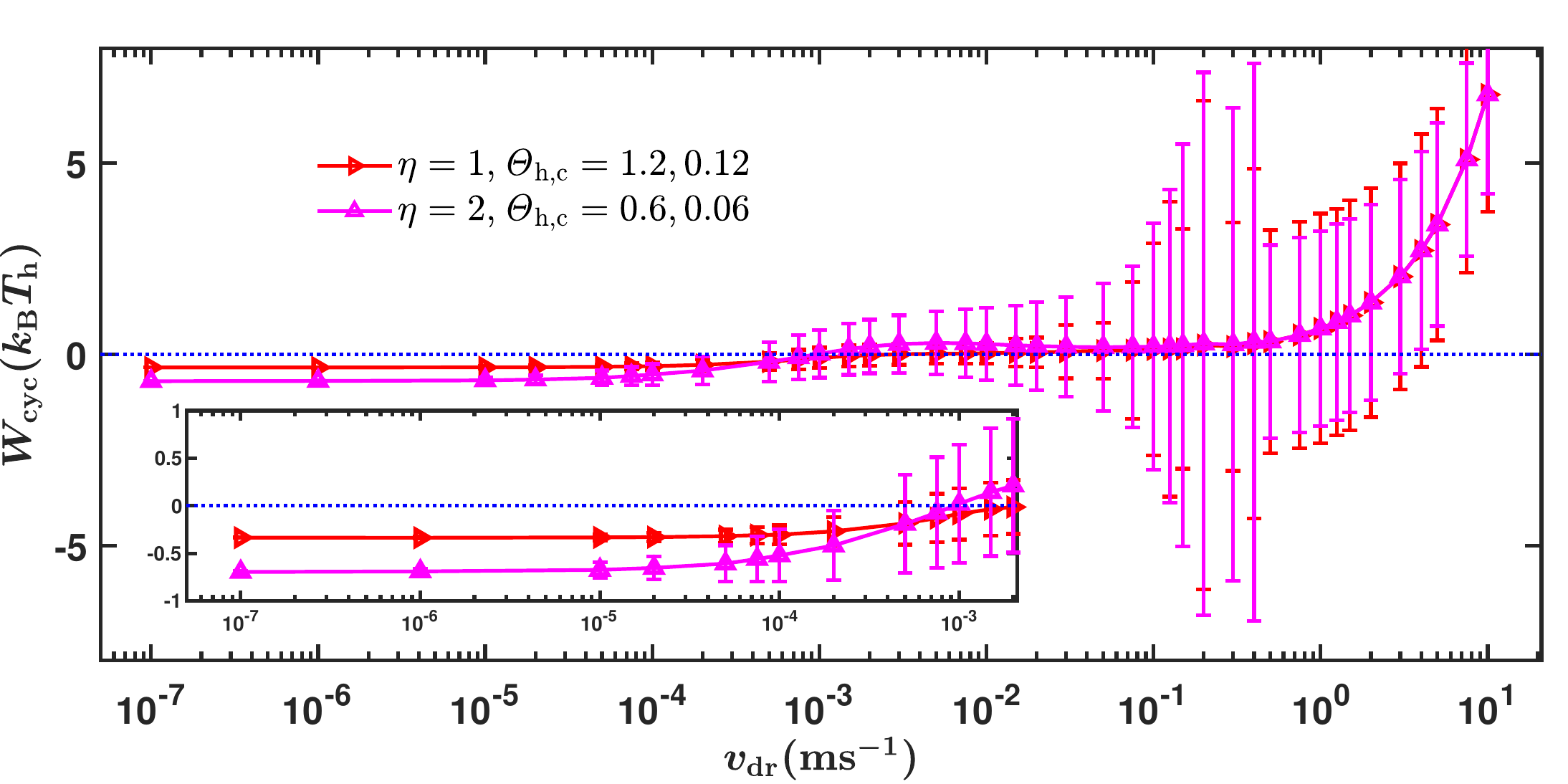}
\caption{The standard deviations of $W_{\rm cyc}$ of the cases of $\eta=1$ and $\eta=2$ in Figure 3(B) in the main text. Because of the higher ${\it\Theta}_{\rm h,c}$ in the $\eta=1$ case and there being no energy barrier resulting in that the particle relaxing to equilibrium more quickly, the standard deviations on the low velocity end are smaller than those of the $\eta=2$ case. On the high driving velocity end, however, the higher ${\it\Theta}_{\rm h,c}$ and no energy barrier make the particle fluctuates in a larger range so that the standard deviations of $W_{\rm cyc}$ are larger in the $\eta=1$ case. Another special characteristic of the standard deviations at such low $\eta$ is that the standard deviation has more than one peaks varying with the driving velocity. We can see that there are at least 3 peaks in the $\eta=1$ case and 2 peaks in the $\eta=2$ case.}
\label{Wcyc_Eta1_2_Error}
\end{figure}

In such high relative temperatures ${\it\Theta}_{\rm h,c}=1.2,0.12$ and because there is no energy barrier in the $\eta=1$ case, the nonequilibrium feature at $v_{\rm dr}=10^{-5}\rm m/s$ is not as distinct as the case of $\eta=2$. The standard deviation is small as can be seen in the insets of Figure \ref{SimuResultEta1}(A) and Figure \ref{Wcyc_Eta1_2_Error}. Nevertheless we can still see the difference between the results at $v_{\rm dr}=10^{-7}\rm m/s$ (Figure 4 in the main text) and that at $v_{\rm dr}=10^{-5}\rm m/s$ (Figure \ref{SimuResultEta1}), especially for the displacement curve. Because of the low driving velocity, the particle has enough time to traverse more states and seems to distributes widely and centered at the dotted white line particularly at the beginning and end of the cycle in the bottom left subfigure of the main text Figure 4. Nonetheless, from the work curve we know that the mean displacement point actually deviates from the driver center. Otherwise the slope of the work curve will be zero. When the driving velocity goes to zero, we can imagine that the displacement of the particle should cover the full range of the interval $(-\infty,+\infty)$. So it's necessary to obtain the displacement distribution through e.g. the Fokker-Planck or Kramers equation method \cite{KramersEqBookRisken} which may avoid the excessive computing resource and time cost of the Langevin dynamics simulation.

\subsubsection{The simulation results at $\eta<1$}

\begin{figure}[H]
\centering
\includegraphics[width=0.49\textwidth]{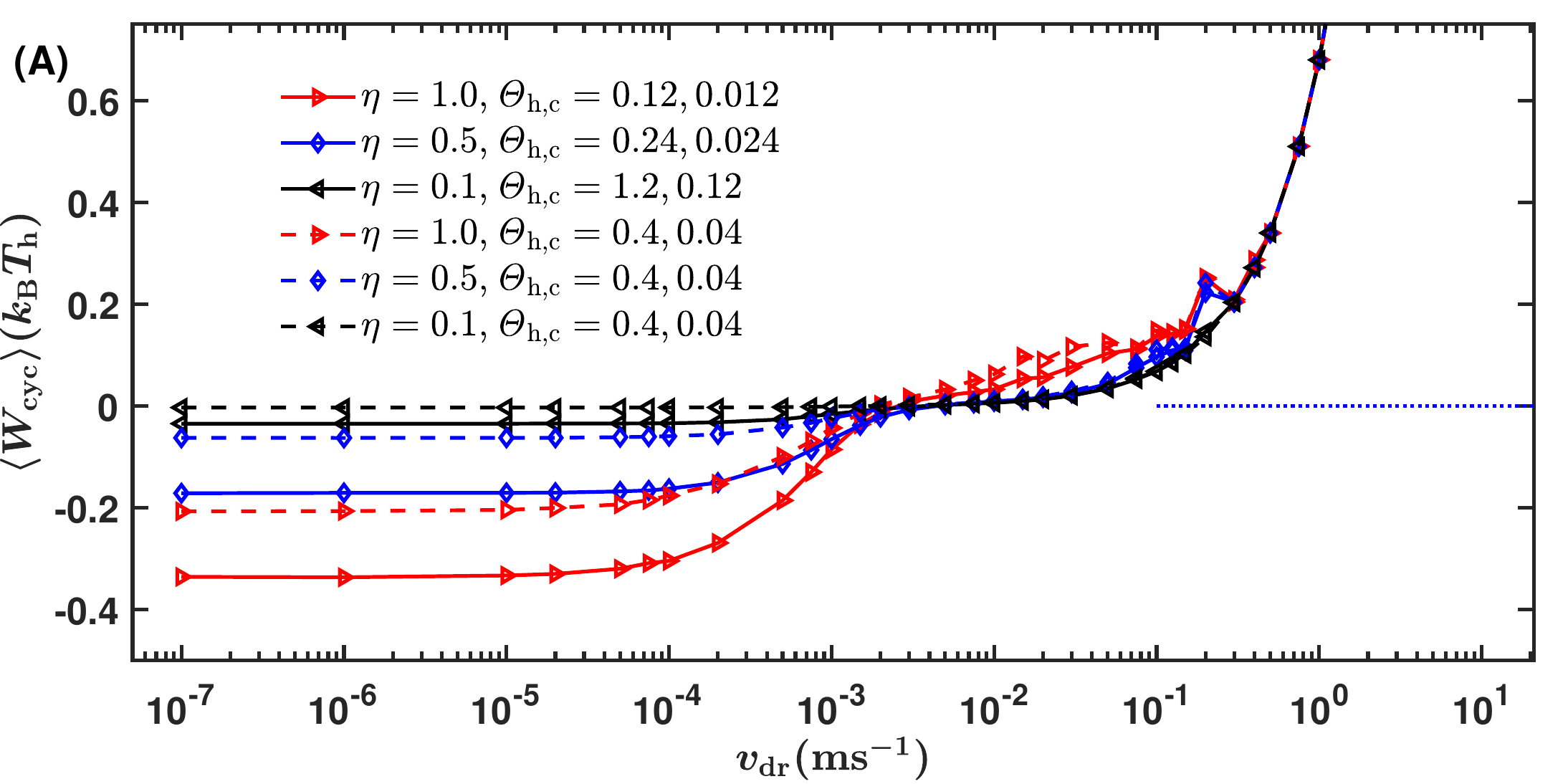}
\end{figure}
\begin{figure}[H]
\includegraphics[width=0.49\textwidth]{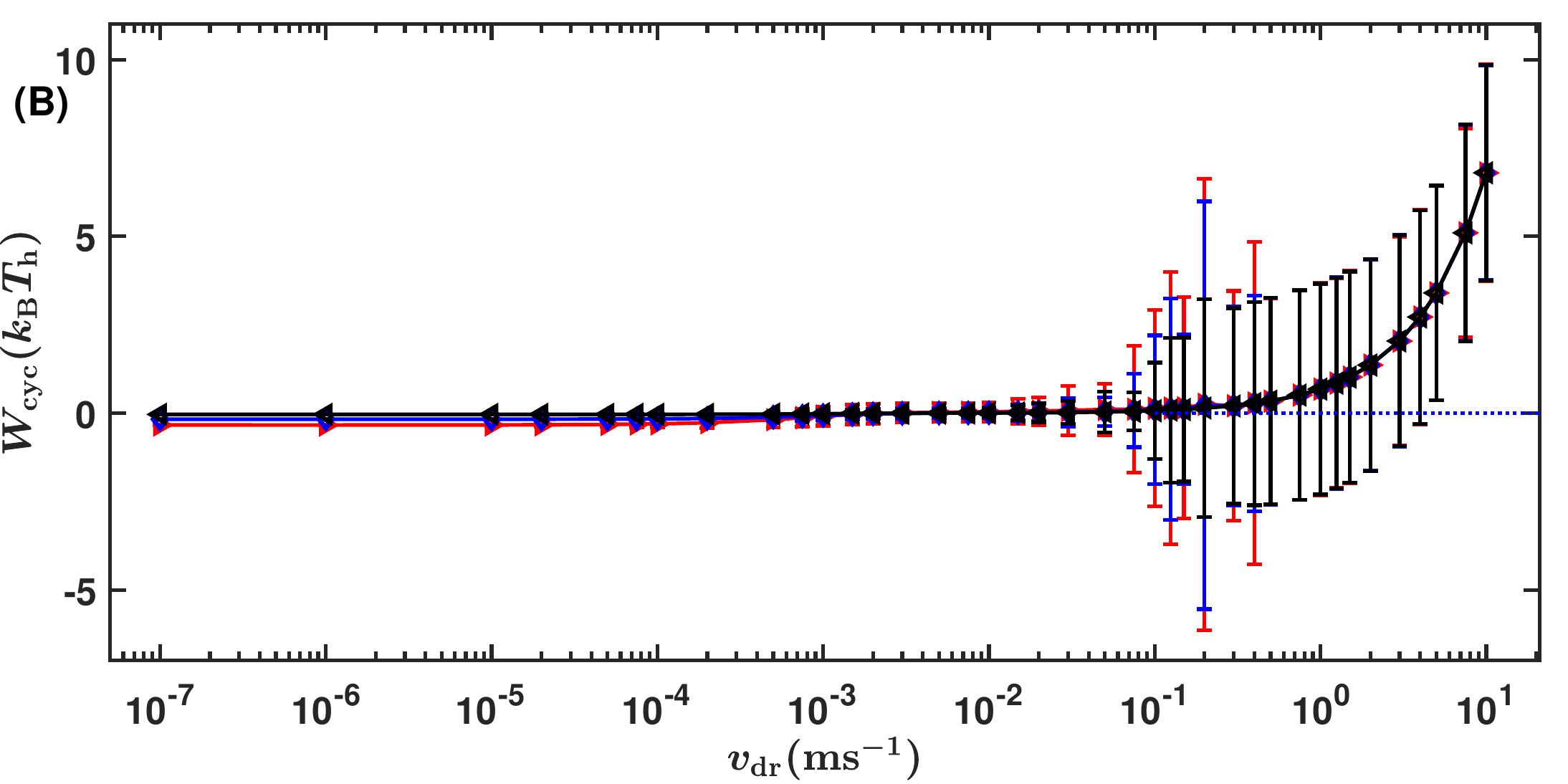}
\includegraphics[width=0.49\textwidth]{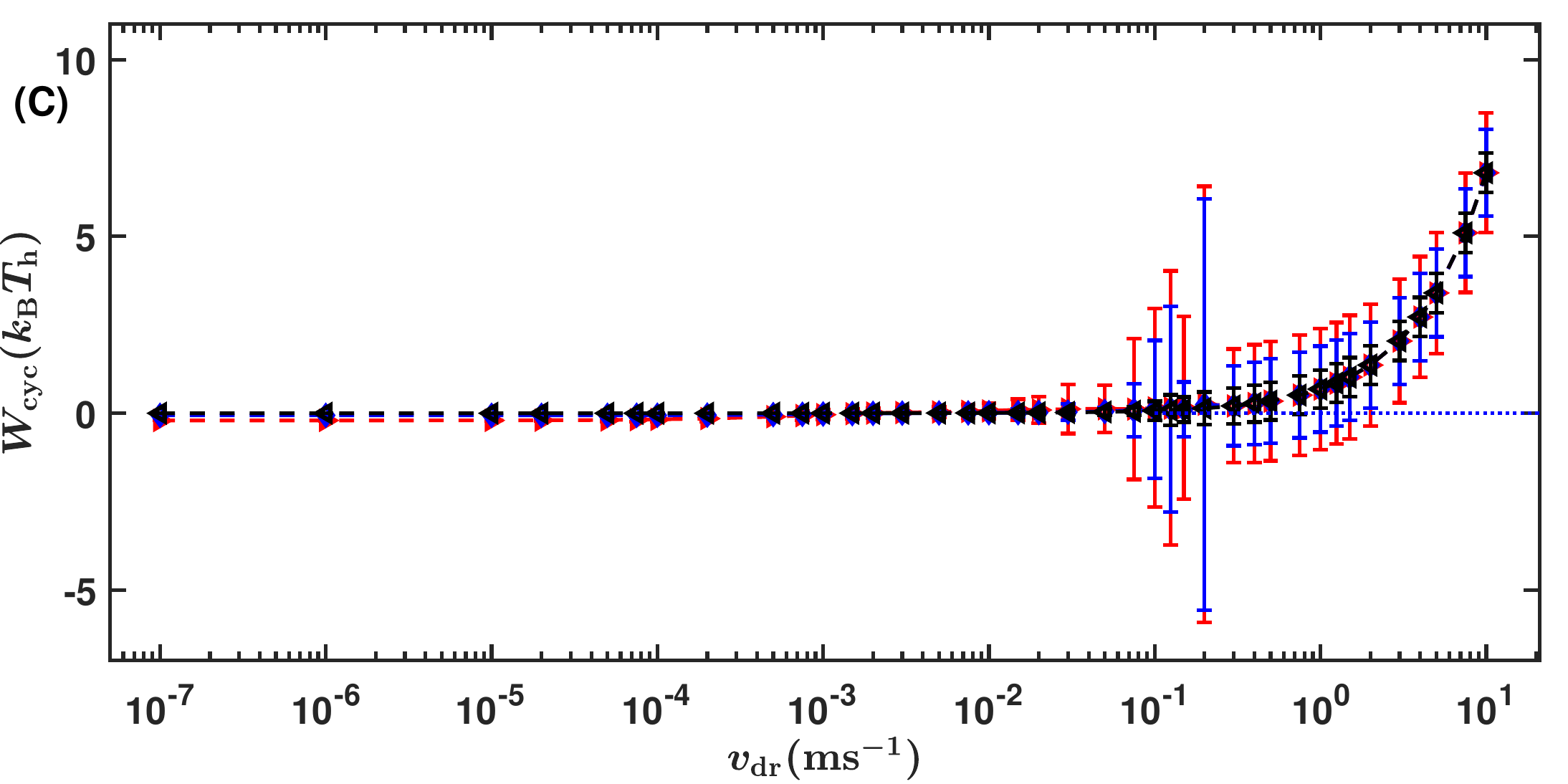}
\caption{Simulation results at $\eta\leq1$. Here $\mu=4\times10^4\rm s^{-1}$ and three of the six cases are of the same absolute high and low temperature $T_{\rm h,c}$ and the other three are of the same nondimensional high and low temperature ${\it\Theta}_{\rm h,c}$. The boundaries of the high and low temperature zones are at the local minimum and maximum points of the lattice potential, i.e. the driver center crosses the boundaries at the beginning, end and middle instants of one engine cycle 
%the high and low temperature zones gone through by the driver center within one engine cycle is bounded at the cycle middle instant 
like that in the case of $\eta=1$ in Figure \ref{SimuResultEta1}. The $\langle W_{\rm cyc}\rangle-v_{\rm dr}$ curves are plotted in (A) and the standard deviations of $W_{\rm cyc}$ are given in (B) and (C). We can see that with $\eta$ decreasing, the equilibrium work output decreases too, indicating the necessity of the sinusoidal lattice potential for the PTSHE to output work. In (B) and (C), we can speculate that for the cases of $\eta\leq1$ at the high driving velocity end, the standard deviation of $W_{\rm cyc}$ is determined by the absolute temperature rather than the nondimensional one, which is not true for the $\eta>1$ cases, cf. Figure \ref{Wcyc_Eta1_2_Error} where the absolute temperature for the $\eta=1$ and $\eta=2$ cases is the same while the standard deviations at the high driving velocity end are different. We can also see that the standard deviation has more than one peaks in the resonance regime at $\eta<1$ similar to the cases of $\eta=1$ and $2$ in Figure \ref{Wcyc_Eta1_2_Error}. Other parameters are given in Sec. \ref{Langevindynamicssimulation}.\ref{ParametersUsed}.}
\label{fig:Wcyc_Vdr_sprlub}
\end{figure}

In Figure \ref{fig:Wcyc_Vdr_sprlub}, we plot the simulation results at $\eta\leq1$. Three of them are of the same absolute high and low temperatures while the other three are of the same nondimensional high and low temperatures. For all of these cases, the high and low temperature boundaries are at the local maximum and minimum points of the lattice potential, i.e. the driver center crosses the boundaries at the beginning, end and middle instants of one engine cycle, like that in the case of $\eta=1$ in Figure \ref{SimuResultEta1}(D1)-(D12). We can see that as $\eta$ decreases, the mean cycle work output at the low driving velocity limit decreases too. Because we keep $\omega_0$ constant, the intensity of the lattice potential $V_0$ decreases proportionally with $\eta$. With $V_0$ decreasing, the resultant potential approaches to the parabola harmonic potential, which is symmetric about its center (i.e. the driver center) during the entire engine cycle. In a symmetric parabola resultant potential, the particle will oscillate around the driver center and distribute ahead of and behind the driver center symmetrically about the middle instant of its going through each high or low temperature zone, from entering to exiting it, %equally likely on average during one entire engine cycle in spite of the alternative high and low temperature field. 
%Actually, after going through each high or low temperature zone, from entering to exiting it,
%temperature zone in either homogeneous high or low temperature
so that the particle will neither output nor input work in total at equilibrium in each homogeneous temperature zone. % due to the symmetry of the parabola potential. 
Therefore the cycle work output at equilibrium approaches to zero as $\eta$ goes to zero. And this conclusion can be generalized to an alternative high and low temperature field with 
%its boundary deviating from the middle point of one cycle, i.e. the high and low temperature zone are of different length.
the high and low temperature zones of different length and with an arbitrary phase shift. Therefore it's necessary to have a finite lattice potential for the PTSHE to output work at low driving velocity.

\subsection{The simulation results with the high temperature ${\it\Theta}_{\rm h}$ varying for the $\eta=1,2,4$ cases}

\begin{figure}[H]
\centering
\begin{minipage}{0.329\textwidth}
\centerline{
\includegraphics[width=\textwidth]{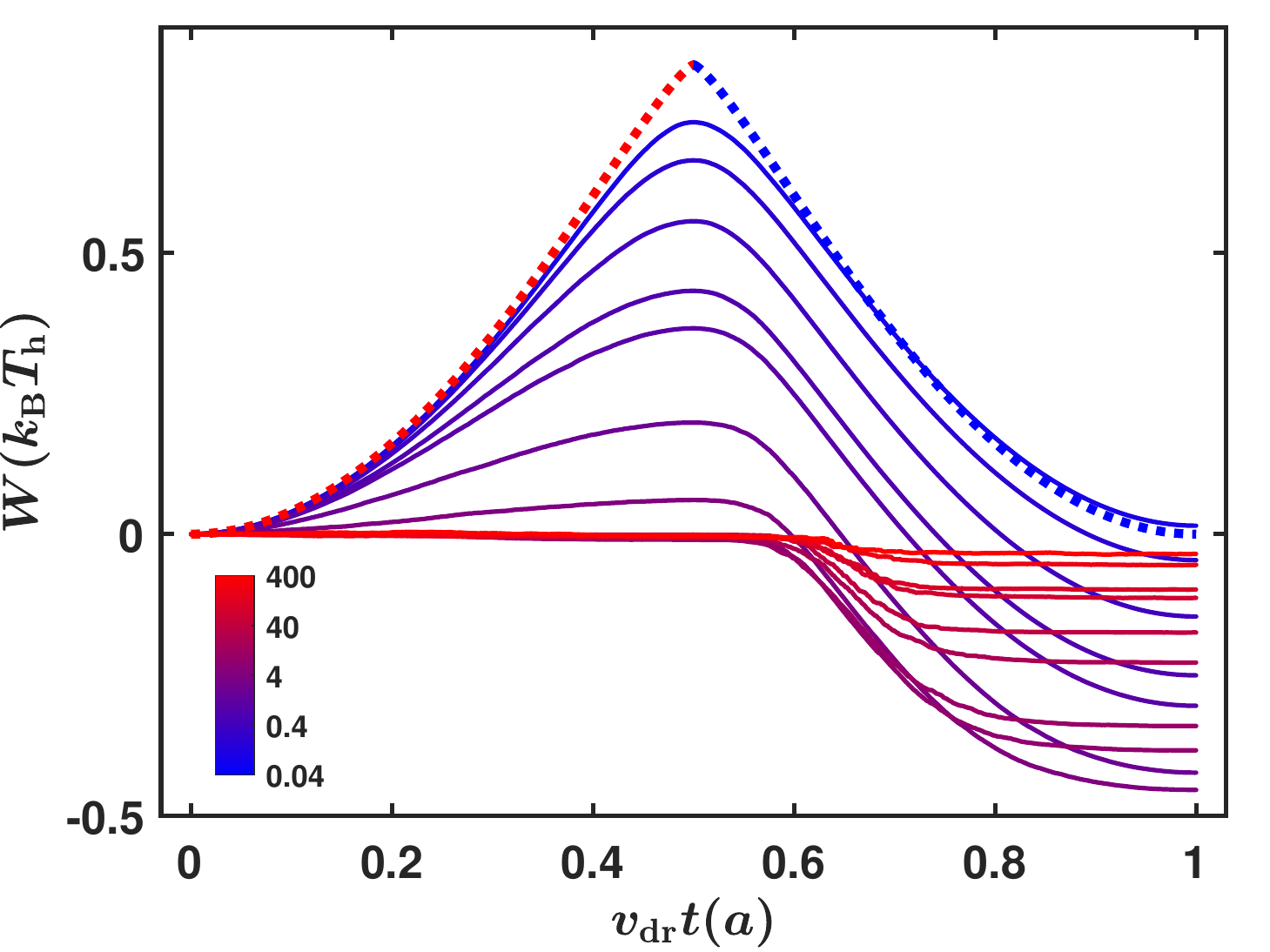}}
\centerline{(a)\ $\eta=1$}
\end{minipage}
\centering
\begin{minipage}{0.329\textwidth}
\centerline{
\includegraphics[width=\textwidth]{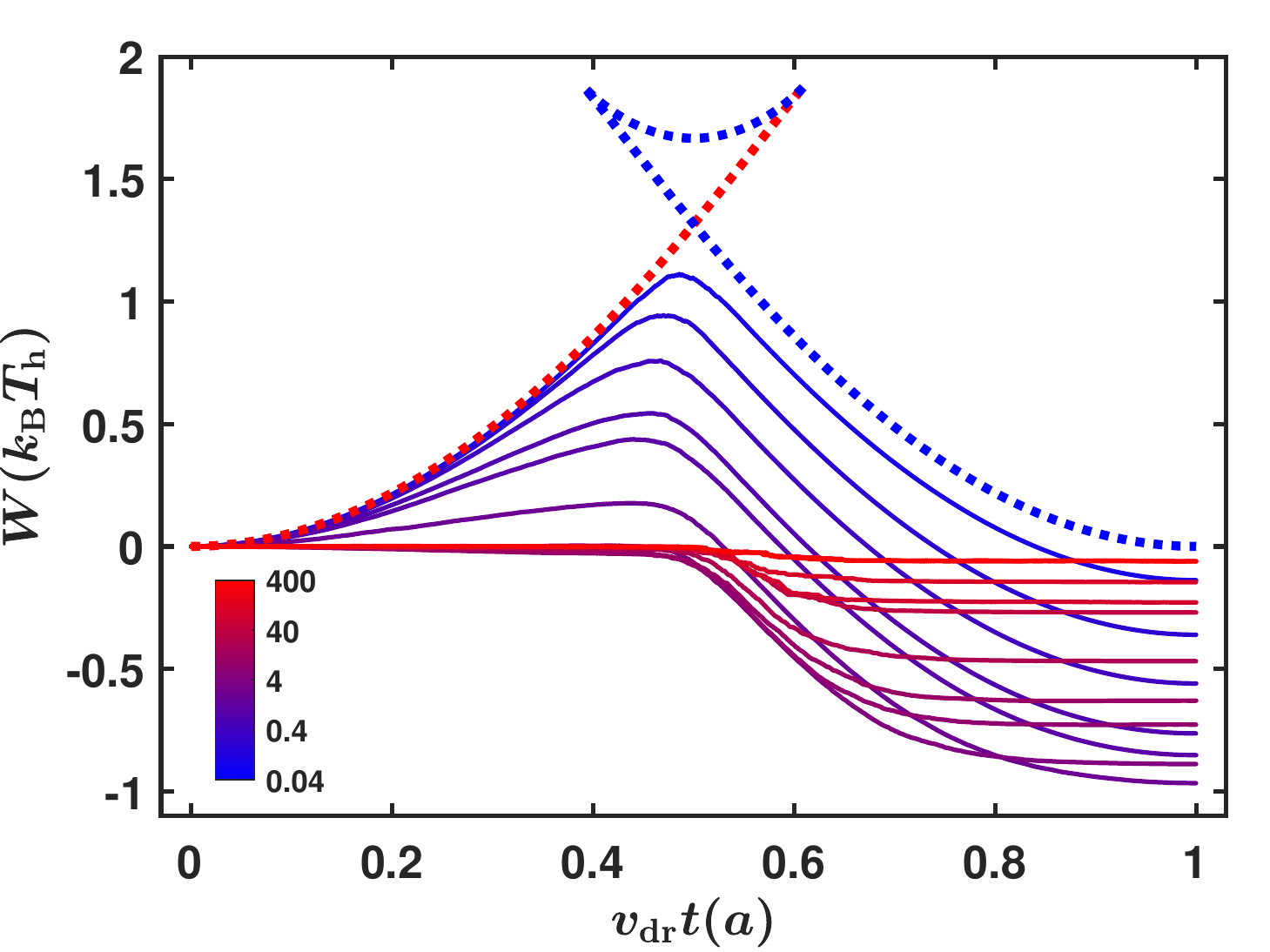}}
\centerline{(b)\ $\eta=2$}
\end{minipage}
\centering
\begin{minipage}{0.329\textwidth}
\centerline{
\includegraphics[width=\textwidth]{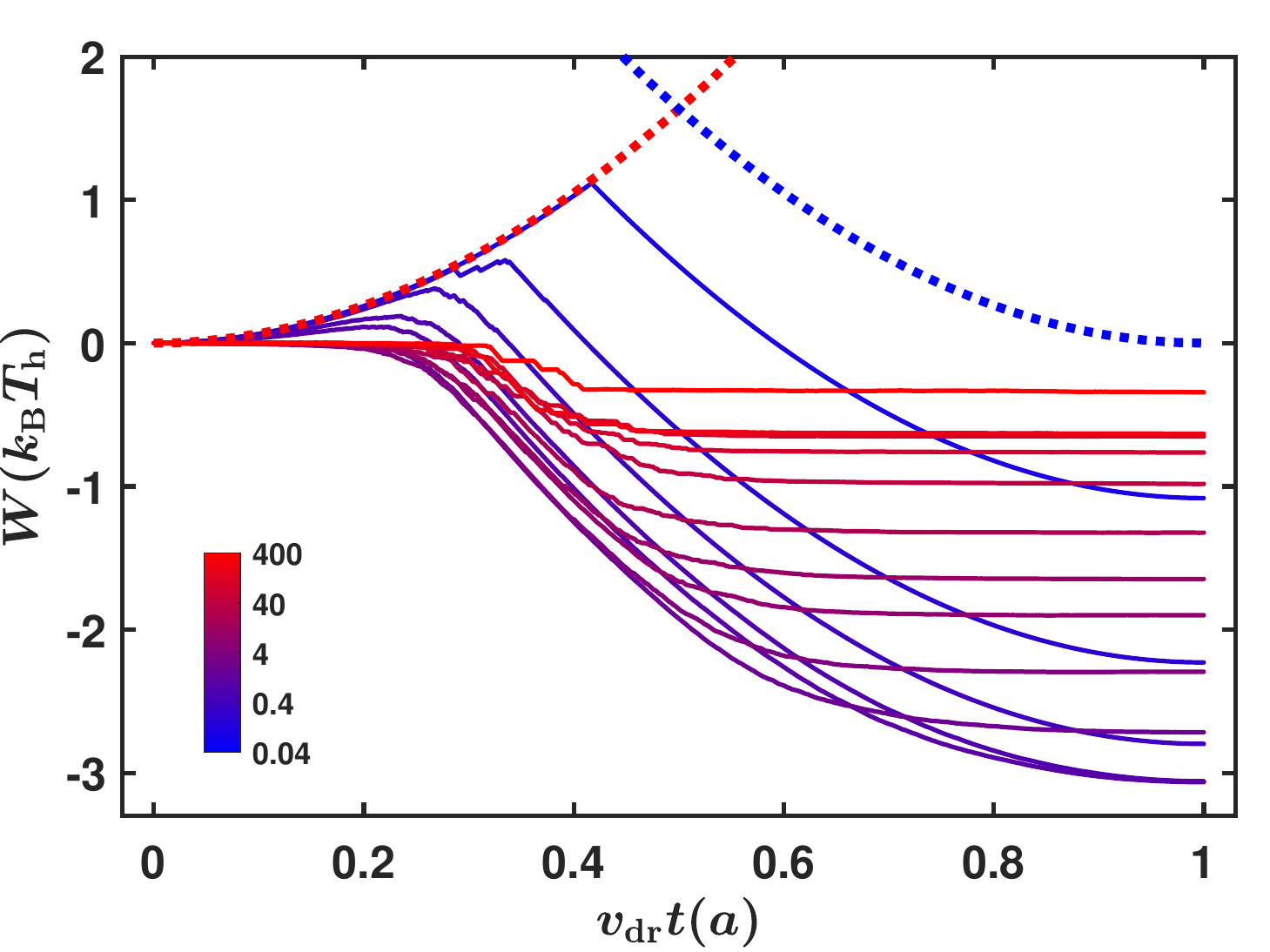}}
\centerline{(c)\ $\eta=4$}
\end{minipage}

\caption{The work curves in one cycle corresponding to the circles of the $\eta=1$, $2$ and $4$ cases in the main text Figure 5(A), with the high temperature ${\it\Theta}_{\rm h}$ denoted by the gradually varying color of the curves. The dot-dashed curves represent the resultant potential at the balanced points. Here $k_{\rm B}T_{\rm h}$ is equal to that of the $\eta=3$ case in the main text Figure 5.}
\label{fig:Eta124WorkCurves}
\end{figure}

In Figure \ref{fig:Eta124WorkCurves}, we plot the work curves in one cycle corresponding to the $\eta=1,2,4$ cases in the main text Figure 5(A). For the three cases, the shape of the work curve has a varying trend with respect to ${\rm\Theta}_{\rm h}$ similar to that of the $\eta=3$ case in the main text Figure 5(B). 
\\
\indent For the $\eta=1$ case, because there is no energy barrier so that the potential mechanism has no effect, the cusp (if there is one) is at the middle instant of the engine cycle exactly. And after the middle instant, the work curve tends to keep nearly unchanged for a short period due to the thermolubricity effect, cf. Sec. \ref{sec:apndRD}.\ref{sec:caseeta1}. The upmost curve represents the case of ${\it\Theta}_{\rm h}=0.04<{\it\Theta}_{\rm c}=0.12$, from which we can see that in this case the cycle work is positive, i.e. work should be inputted.

\begin{figure}[H]
\centering
\begin{minipage}{0.329\textwidth}
\centerline{
\includegraphics[width=\textwidth]{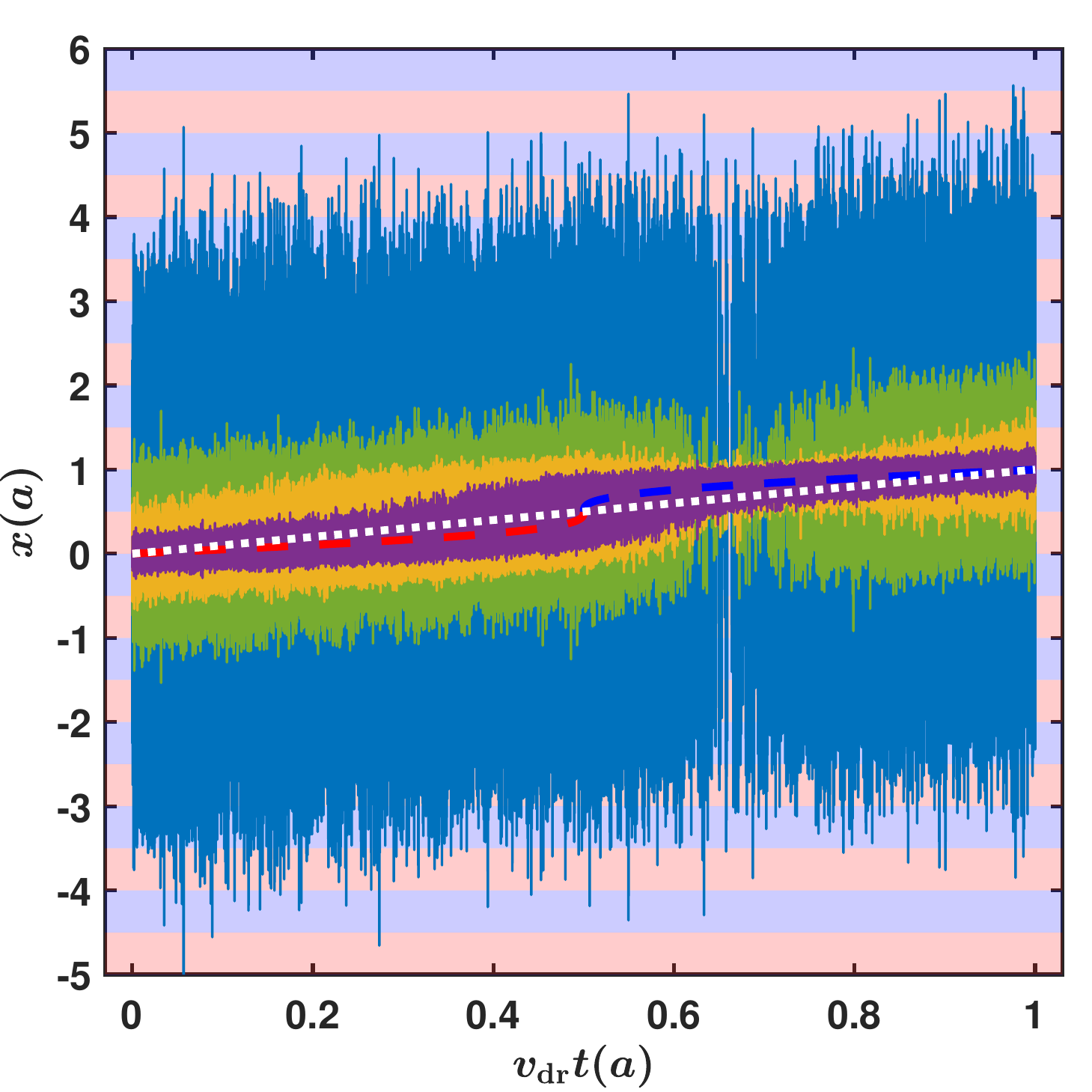}}
\centerline{(a)\ $\eta=1$.
%The dark blue, dark green, dark yellow and purple curves represent the cases of ${\it\Theta}_{\rm h}=40,4,1,0.2$ respectively.
}
\end{minipage}
\centering
\begin{minipage}{0.329\textwidth}
\centerline{
\includegraphics[width=\textwidth]{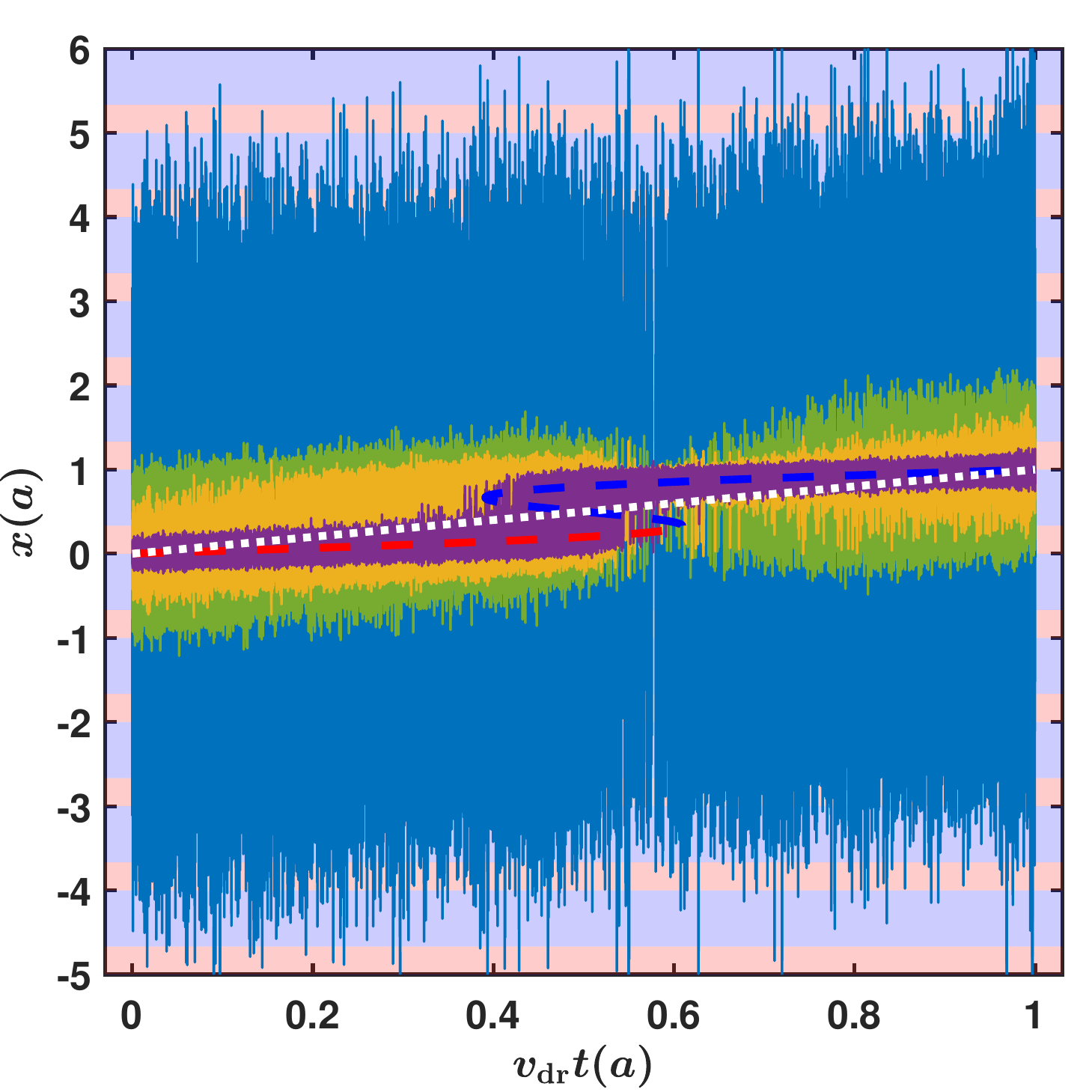}}
\centerline{(b)\ $\eta=2$.
%The dark blue, dark green, dark yellow and purple curves represent the cases of ${\it\Theta}_{\rm h}=40,2,0.75,0.1$ respectively.
}
\end{minipage}
\centering
\begin{minipage}{0.329\textwidth}
\centerline{
\includegraphics[width=\textwidth]{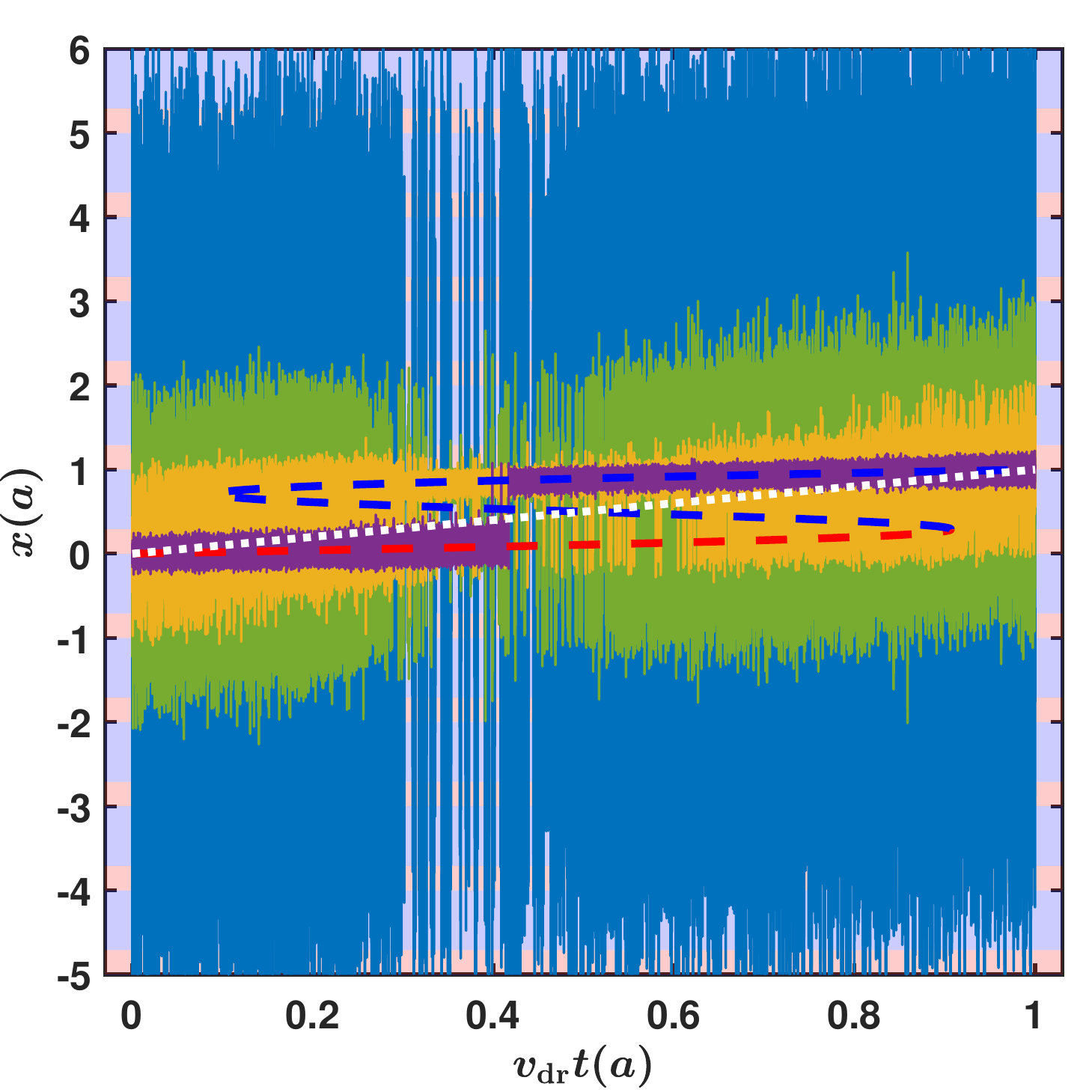}}
\centerline{(c)\ $\eta=4$.
%The dark blue, dark green, dark yellow and purple curves represent the cases of ${\it\Theta}_{\rm h}=40,4,0.75,0.1$ respectively.
}
\end{minipage}
\caption{The displacement of the particle in one cycle for the cases of $\eta=1$, $2$ and $4$ in the main text Figure 5(A) at four typical ${\it\Theta}_{\rm h}$'s. The dark blue, dark green, dark yellow and purple curves represent the cases of ${\it\Theta}_{\rm h}=40,4,1,0.2$ respectively in (a); $40,2,0.75,0.1$ respectively in (b) and $40,4,0.75,0.1$ respectively in (c). The dotted white line denotes the position of the driver center. The dashed red and blue curves represent the position of the balanced point of the resultant potential.}
\label{fig:Eta124XMultiTh}
\end{figure}

For the $\eta=2$ case, due to the low energy barrier, the cusp (if there is one) shifts to the left more mildly that the $\eta=3$ case.
\\
\indent For the $\eta=4$ case, the energy barrier is high and the cusp (if there is one) shifts to the left more violently than the $\eta=3$ case and the work curve is less smooth. When ${\it\Theta}_{\rm h}$ is small, the cusp is sharp because of nonequilibrium, cf. Figure \ref{fig:Eta124XMultiTh}(c) and the work curve nearly overlaps with the dotted red curve before the cusp instant, e.g. the upmost ${\it\Theta}_{\rm h}=0.04$ curve.

In Figure \ref{fig:Eta124XMultiTh}, we plot the displacement of the particle in one cycle for the cases of $\eta=1$, $2$ and $4$ in the main text Figure 5(A) at four typical ${\it\Theta}_{\rm h}$'s. Because of the higher $V_0$ value at higher $\eta$, the absolute temperature corresponding to ${\it\Theta}_{\rm h}=40$ is higher at $\eta=4$ than that at $\eta=2$ and $\eta=1$, so that the dark blue curve's range expands when $\eta$ increases from $1$ to $4$.

\begin{figure}[H]
\centering
\begin{minipage}{0.329\textwidth}
\centerline{
\includegraphics[width=\textwidth]{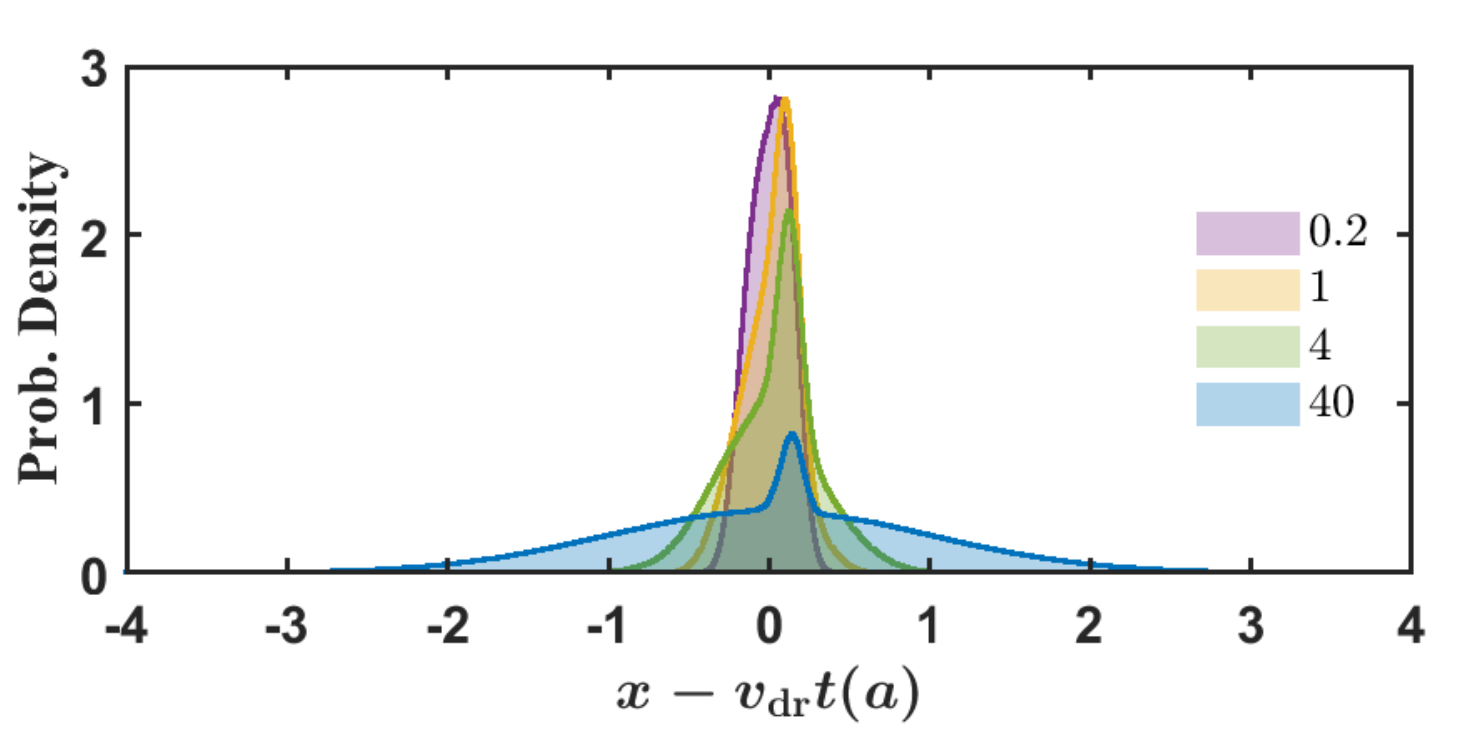}}
\centerline{(a)\ $\eta=1$}
\end{minipage}
\centering
\begin{minipage}{0.329\textwidth}
\centerline{
\includegraphics[width=\textwidth]{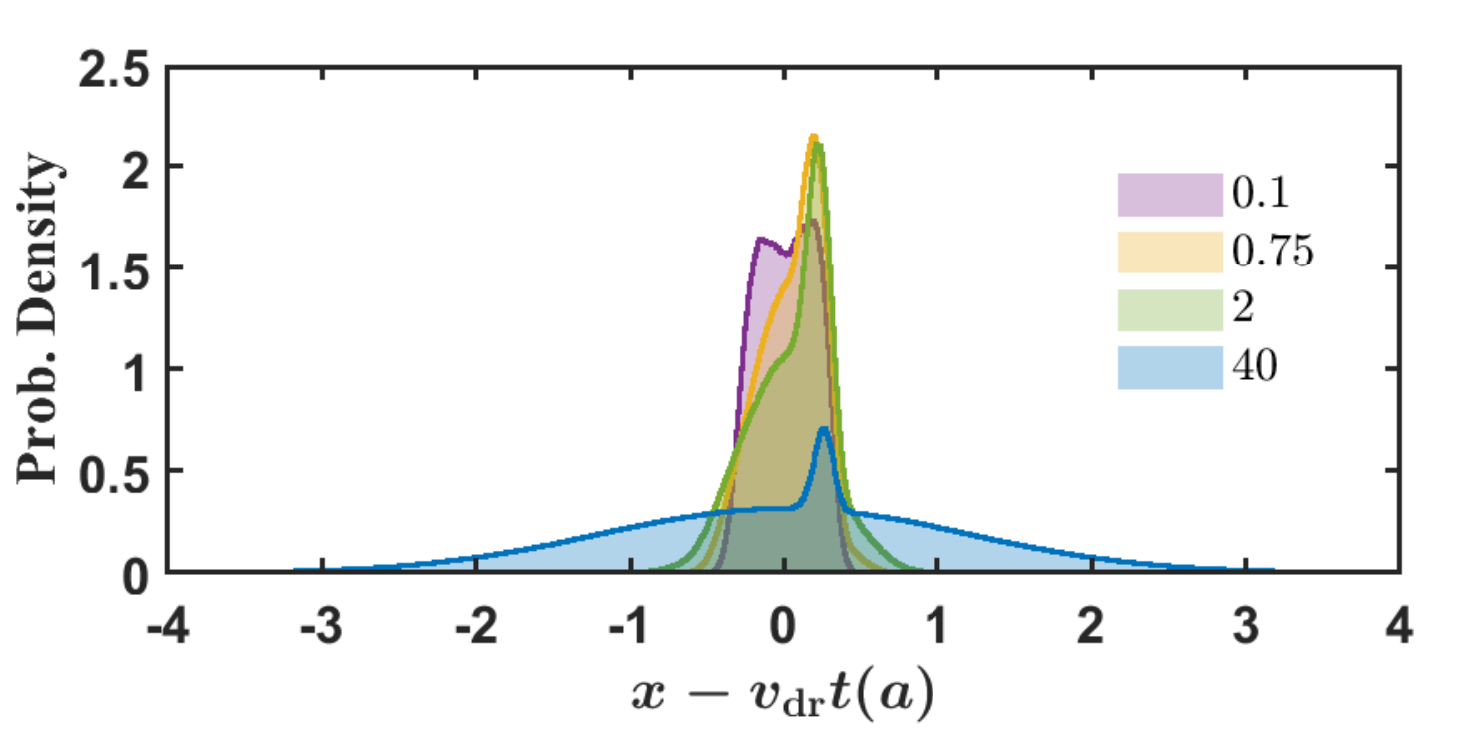}}
\centerline{(b)\ $\eta=2$}
\end{minipage}
\centering
\begin{minipage}{0.329\textwidth}
\centerline{
\includegraphics[width=\textwidth]{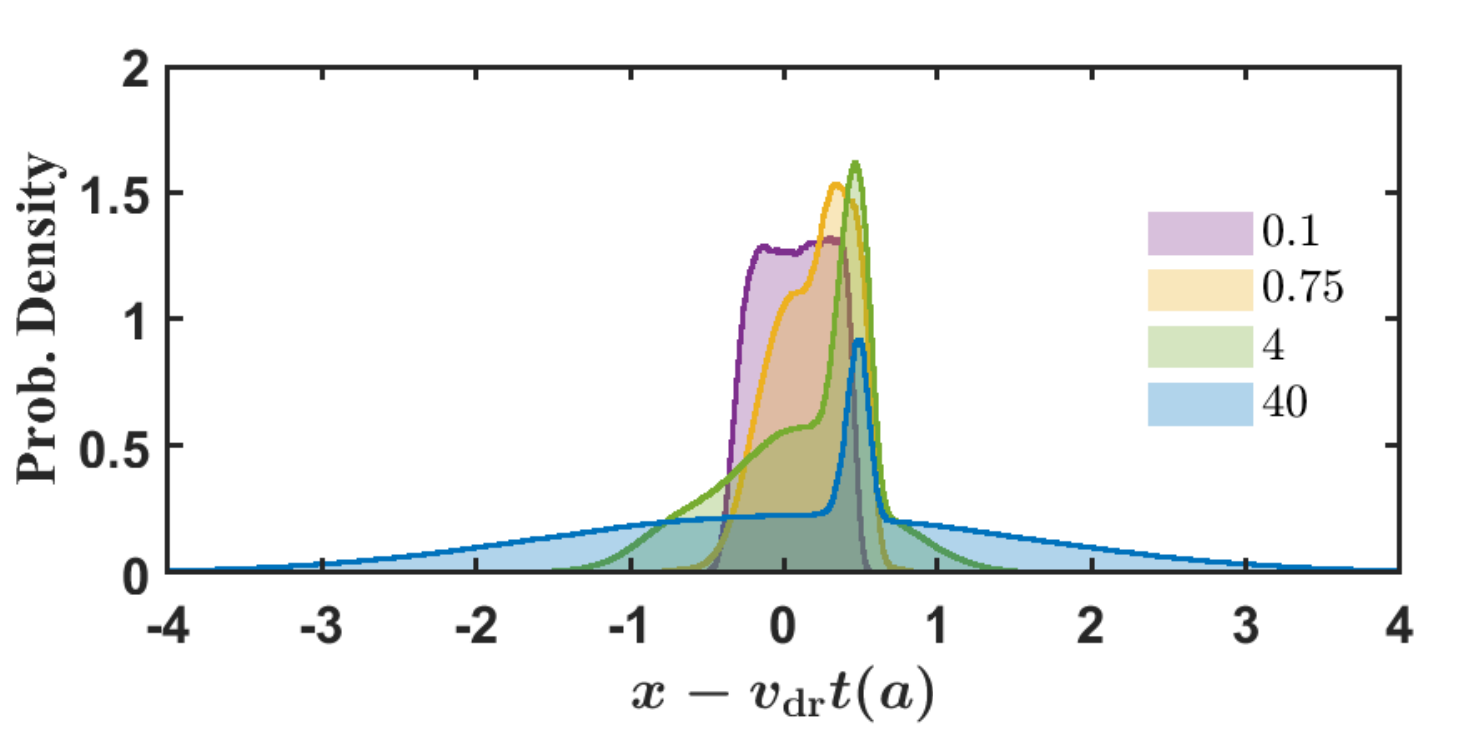}}
\centerline{(c)\ $\eta=4$}
\end{minipage}

\caption{The probability density distributions of the particle's relative displacement to the driver center $x-v_{\rm dr}t$ in one cycle at four typical ${\it\Theta}_{\rm h}$'s for the cases of $\eta=1$, $2$, and $4$ in the main text Figure 5(A).}
\label{fig:Eta124RXDistMultiTh}
\end{figure}

In Figure \ref{fig:Eta124RXDistMultiTh}, we plot the probability density distributions of the particle's relative displacement to the driver center $x-v_{\rm dr}t$ in one cycle at four typical ${\it\Theta}_{\rm h}$'s for the cases of $\eta=1$, $2$ and $4$ in the main text Figure 5(A). For all the four cases including the $\eta=3$ case in the main text, as the high temperature ${\it\Theta}_{\rm h}$ increases a peak extrudes on the right and with $\eta$ increasing, the peak's position shifts to the right so that the average value of the distribution and thus the cycle work increases.

\subsection{Transition of the energy, displacement and harmonic force during one cycle with the driving velocity}
\label{sec:transitionEDFwdv}
\begin{figure}[H]
\centering
 \includegraphics[width=\textwidth]{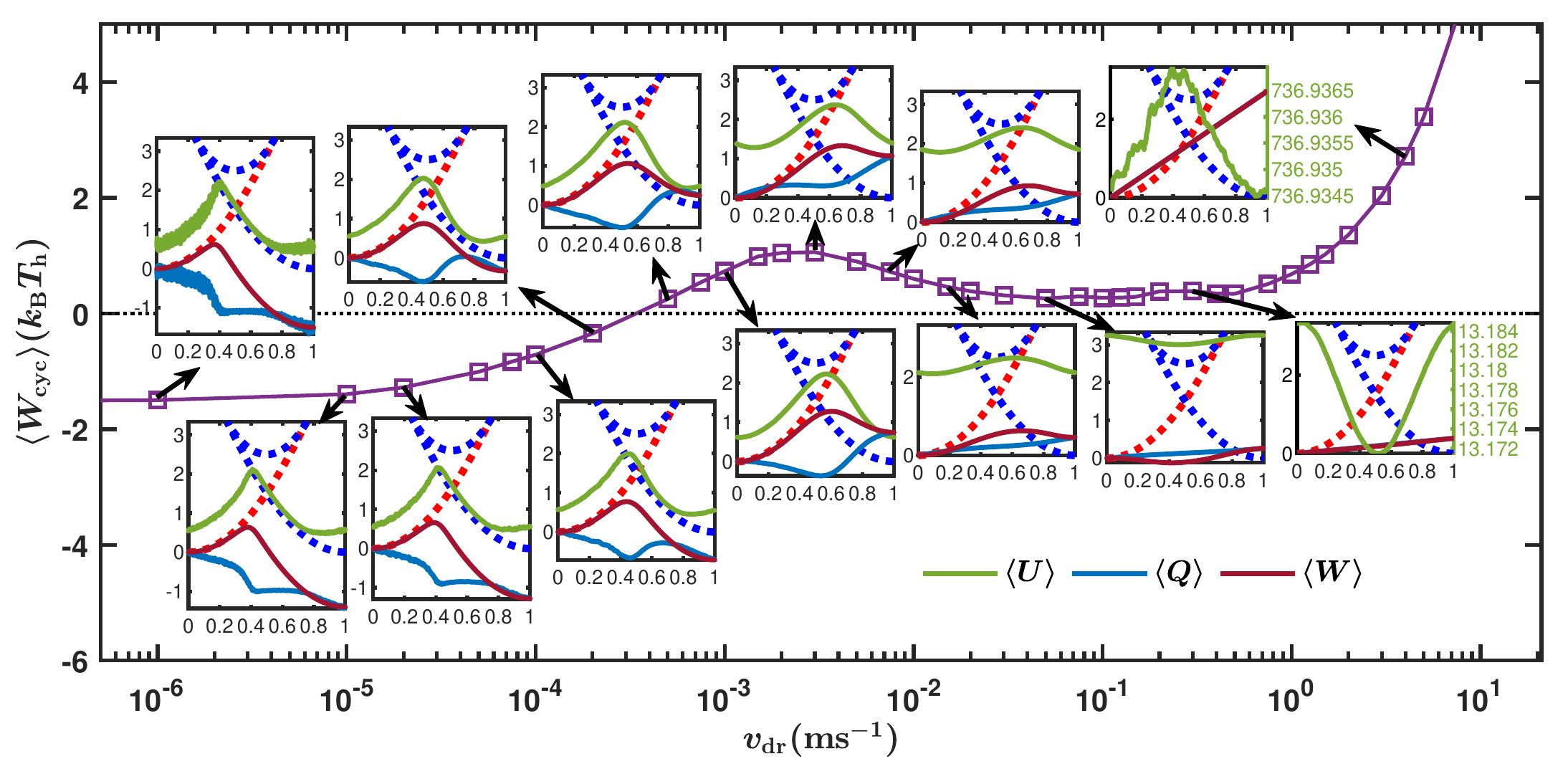}
 \caption{The transition of the energy curves with the driving velocity. The backbone purple $\langle W_{\rm cyc}\rangle-v_{\rm dr}$ curve is the same as the purple one in Figure 3(B) in the main text. The insets give the mean internal energy $\langle U\rangle$, the mean heat to the heat bath $\langle Q\rangle$ and the mean work input to the particle $\langle W\rangle$ during one cycle %averaged over a certain number of simulation engine cycles 
 at the corresponding driving velocities marked by the arrows. These insets are chosen from Figure \ref{fig:StickSlipsSingleTh04Tc004}.} 
 \label{fig:StickSlipsTransition} 
 \end{figure}

We can map the energy curves' transition with the driving velocity to the change of the $\langle W_{\rm cyc}\rangle-v_{\rm dr}$ curve, which is shown in Figure \ref{fig:StickSlipsTransition}. We can see that on the left end of the $\langle W_{\rm cyc}\rangle-v_{\rm dr}$ curve where work is outputted, the cusp of the mean work curve (here the mean internal energy curve also has a cusp) during one engine cycle is before the middle instant of the cycle. After the engine stalls the cusp of the mean internal energy curve surpasses the cycle middle instant to the right and the mean cycle work input $\langle W_{\rm cyc}\rangle$ increases until an extremum, and this is the stick-slip regime. 

\begin{figure}[H]
\centering
 \includegraphics[width=\textwidth]{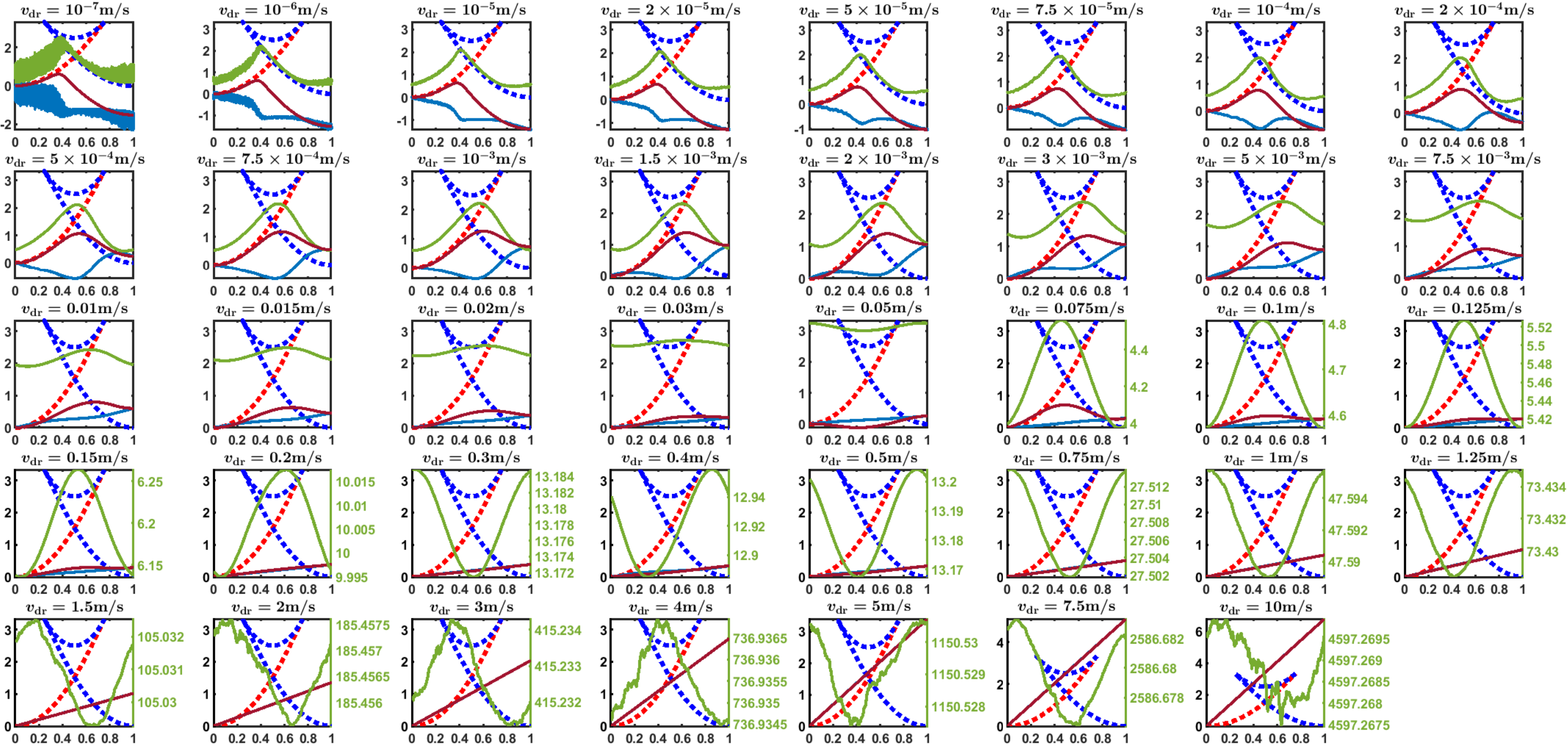}
 \caption{The mean internal energy $\langle U\rangle$ (dark green), the mean heat to the heat bath $\langle Q\rangle$ (dark blue) and the mean work input to the system $\langle W\rangle$ (dark red) during one cycle % averaged over the ensemble of the simulation engine cycles
at different driving velocities of the case of $\eta=3$, $\mu=4\times10^4\rm s^{-1}$ and ${\it\Theta}_{\rm h,c}=0.4,0.04$. All the $x$-coordinates are the driver center's nondimensional position $v_{\rm dr}t/a$ relative to the latest cycle starting point, i.e. the same as those in the main text Figure 2(B) and (C). All the $y$-coordinates and the dotted red and blue curves are the same as those in the main text Figure 2(A). Other parameters are given in Sec. \ref{Langevindynamicssimulation}.\ref{ParametersUsed}. The numbers of simulation cycles used to calculate the three mean quantity curves at each driving velocity are given in Table \ref{tab:Numberofcycles}. At each driving velocity, all of the simulation cycles are computed sequentially with the initial values of the first simulation cycle inherited from the end values of the last simulation cycle used to calculate the corresponding point at the same $v_{\rm dr}$ on the purple $\langle W_{\rm cyc}\rangle-v_{\rm dr}$ curve of the $\eta=3,{\it\Theta}_{\rm h,c}=0.4,0.04$ case in the main text Figure 3(B), so that we can make sure that the steady state has already been achieved and we don't need to exclude a certain number of cycles at the beginning transient process as we have done when calculating the mean values and standard deviations of $W_{\rm cyc}$ and obtaining the count distributions of $W_{\rm cyc}$ in Figure \ref{WcycDist}, cf. Sec. \ref{Langevindynamicssimulation}.\ref{ParametersUsed}.
\\From $v_{\rm dr}=10^{-7}\rm m/s$ to $7.5\times10^{-5}\rm m/s$, we can see that the mean internal energy curve before its cusp is higher than the dotted red balanced resultant potential curve because of the particle's high kinetic energy in the high temperature zone and so is the part of this curve near the end of the engine cycle because of the particle's frequently traversing to the hot zone in the next lattice period, compared with the segment on the right of the cusp where it is close to the dotted blue curve. At and after $v_{\rm dr}=10^{-4}\rm m/s$ the mean internal energy curve on the right of the cusp is uplifted from left to right gradually due to the kinetic energy from the oscillating relaxation process after the slipping event, which gets longer relative to the driver center's position with $v_{\rm dr}$ increases. And at around $v_{\rm dr}=10^{-3}\rm m/s$ the mean internal energy curve at the beginning of the cycle begins to be uplifted from the left end because of the residual kinetic energy from the last cycle's slipping event. At the end of one cycle this curve is also uplifted because of the periodicity so the entire mean internal energy curve begins to be uplifted. %At the end of one cycle this curve is also uplifted because of the kinetic energy residual kinetic from the slipping event in the same cycle. 
At and after $v_{\rm dr}=0.075\rm m/s$, the mean internal energy curve is higher than the entire dotted balanced resultant potential curve, so we put its $y$-axis to the right seperately, and we can see that its shape varies and it continues to be entirely uplifted gradually with the driving velocity increasing. When the driving velocity $v_{\rm dr}$ is very high, the mean internal energy curve stays nearly constant and the mean work and heat curves are nearly linear whose slopes increase with the increasing $v_{\rm dr}$.}
 \label{fig:StickSlipsSingleTh04Tc004} 
 \end{figure}

\begin{figure}[H]
\centering
 \includegraphics[width=\textwidth]{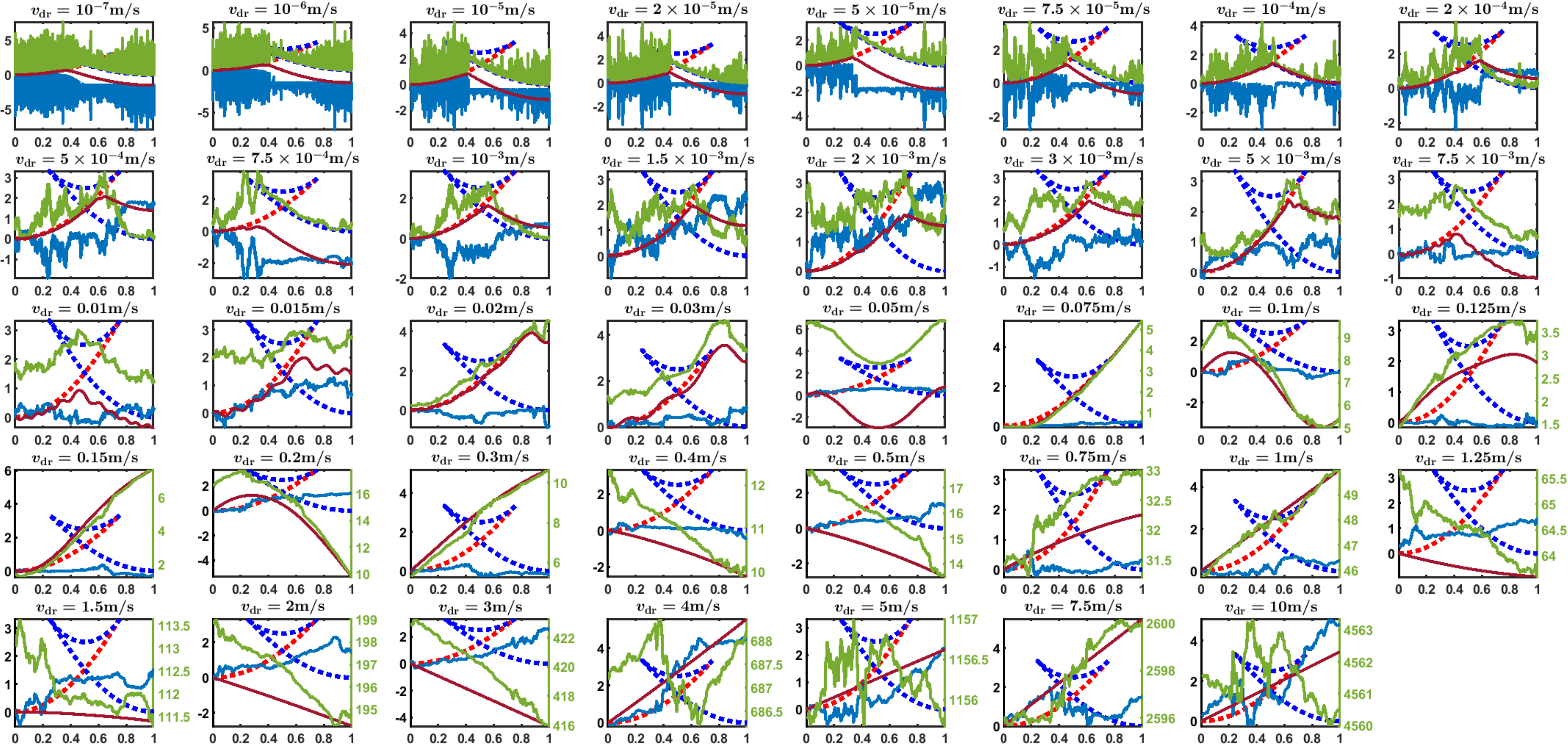}
 \caption{The internal energy $U$ (dark green), the heat to the heat bath $Q$ (dark blue) and the work input to the system $W$ (dark red) during one single cycle at different driving velocities of the case of $\eta=3$, $\mu=4\times10^4\rm s^{-1}$ and ${\it\Theta}_{\rm h,c}=0.4,0.04$. All the $x$-coordinates are the driver center's nondimensional position $v_{\rm dr}t/a$ relative to the latest cycle starting point, i.e. the same as those in the main text Figure 2(B) and (C). All the $y$-coordinates and the dotted red and blue curves are the same as those in the main text Figure 2(A). Other parameters are given in Sec. \ref{Langevindynamicssimulation}.\ref{ParametersUsed}. At each driving velocity, the simulation cycle is computed with the initial values inherited from the end values of the last simulation cycle used to calculate the corresponding point at the same $v_{\rm dr}$ on the purple $\langle W_{\rm cyc}\rangle-v_{\rm dr}$ curve of the $\eta=3,{\it\Theta}_{\rm h,c}=0.4,0.04$ case in the main text Figure 3(B), so that we can make sure that the steady state has already been achieved.
\\We can see that even without average, the work curve is smooth at all the driving velocities. From $v_{\rm dr}=10^{-7}\rm m/s$ to $2\times10^{-4}\rm m/s$, the work curves all have a cusp, while after average the cusp vanishes, cf. Figure \ref{fig:StickSlipsSingleTh04Tc004}. The stochastic feature of the one single cycle trajetories is clear compared with the mean curves in Figure \ref{fig:StickSlipsSingleTh04Tc004}.} 
 \label{fig:StickSlipsSingleOnecycleTh04Tc004} 
 \end{figure}

\begin{figure}[H]
\centering
\includegraphics[width=\textwidth]{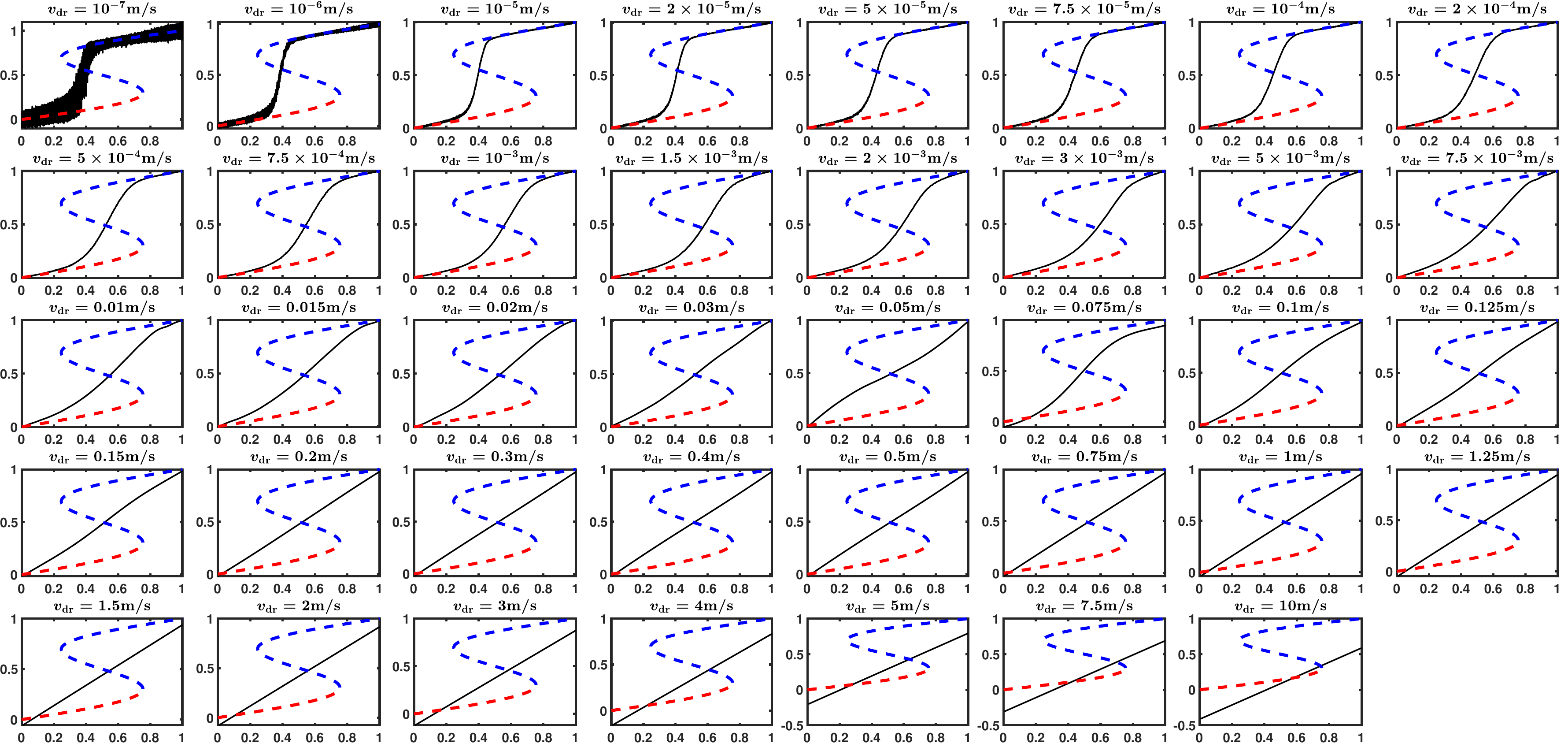}
\caption{The mean displacement of the particle $\langle x\rangle$ during one cycle at different driving velocities of the case of $\eta=3$, $\mu=4\times10^4\rm s^{-1}$ and ${\it\Theta}_{\rm h,c}=0.4,0.04$. All the $x$-coordinates are the driver center's nondimensional position $v_{\rm dr}t/a$ relative to the latest cycle starting point, i.e. the same as that in the main text Figure 2(B). All the $y$-coordinates and the dashed red and blue curves are the same as those in the main text Figure 2(B). Other parameters are given in Sec. \ref{Langevindynamicssimulation}.\ref{ParametersUsed}. The number of simulation cycles at each driving velocity is given in Table \ref{tab:Numberofcycles}. At each driving velocity, all of the simulation cycles are computed sequentially with the initial values of the first simulation cycle inherited from the end values of the last simulation cycle used to calculate the corresponding point at the same $v_{\rm dr}$ on the purple $\langle W_{\rm cyc}\rangle-v_{\rm dr}$ curve of the $\eta=3,{\it\Theta}_{\rm h,c}=0.4,0.04$ case in the main text Figure 3(B), so that we can make sure that the steady state has already been achieved and we don't need to exclude a certain number of cycles at the beginning transient process as we have done when calculating the mean values and standard deviations of $W_{\rm cyc}$ and obtainning the count distributions of $W_{\rm cyc}$ in Figure \ref{WcycDist}, cf. Sec. \ref{Langevindynamicssimulation}.\ref{ParametersUsed}.
\\We can see that average leads to the jumping cliff of the one single cycle displacement curve in Figure \ref{fig:StickSlipsSingleOnecycleXTh04Tc004} smoothened. When the driving velocity is very high, the mean displacement curve falls behind the dashed red and blue curves (so that the particle falls behind the driver center) a lot. On the other hand, at not very high driving velocities, the starting and end points of the mean displacement curves during one cycle overlap with the starting and end positions of the driver center in one cycle.} 
\label{fig:StickSlipsSingleXTh04Tc004} 
\end{figure}

\begin{figure}[H]
\centering
\includegraphics[width=\textwidth]{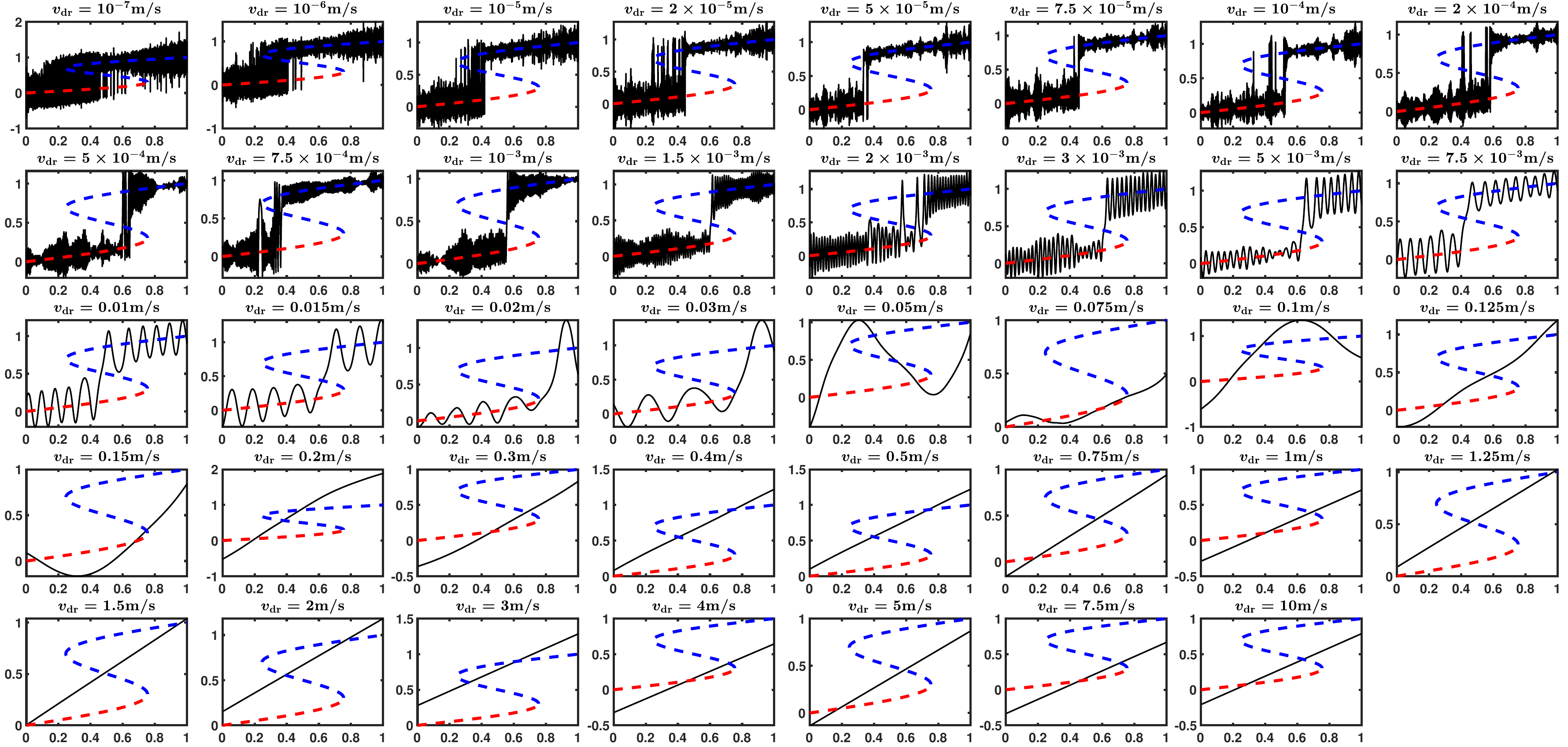}
\caption{The displacement of the particle $x$ during one single cycle at different driving velocities of the case of $\eta=3$, $\mu=4\times10^4\rm s^{-1}$ and ${\it\Theta}_{\rm h,c}=0.4,0.04$. All the $x$-coordinates are the driver center's nondimensional position $v_{\rm dr}t/a$ relative to the latest cycle starting point, i.e. the same as that in the main text Figure 2(B). All the $y$-coordinates and the dashed red and blue curves are the same as those in the main text Figure 2(B). Other parameters are given in Sec. \ref{Langevindynamicssimulation}.\ref{ParametersUsed}. At each driving velocity, the simulation cycle is computed with the initial values inherited from the end values of the last simulation cycle used to calculate the corresponding point at the same $v_{\rm dr}$ on the purple $\langle W_{\rm cyc}\rangle-v_{\rm dr}$ curve of the $\eta=3,{\it\Theta}_{\rm h,c}=0.4,0.04$ case in the main text Figure 3(B), so that we can make sure that the steady state has already been achieved.
\\As the driving velocity increases, the number of oscillations of the particle during one cycle are reduced. So the waveform of the displacement curve is gradually clearer with the increasing driving velocity $v_{\rm dr}$. The waves are modulated by the stochastic force so that their amplitudes are stochastically changing. In the resonance regime the particle's oscillation period is approximately equal to the cycle period so that there is approximately one wave in one cycle, which is affected by the stochastic force and irregular. At even higher driving velocity, the particle is far behind the driver center and the displacement curve is close to linear, cf. Figure \ref{fig:StickSlipsSingleOnecycleXT0}. Due to the stochastic force, the particle's one single cycle displacement curve is random and doesn't overlap with the corresponding mean displacement curve in Figure \ref{fig:StickSlipsSingleXTh04Tc004}, even at the high driving velocity end.} 
\label{fig:StickSlipsSingleOnecycleXTh04Tc004} 
\end{figure}

\begin{figure}[H]
\centering
 \includegraphics[width=\textwidth]{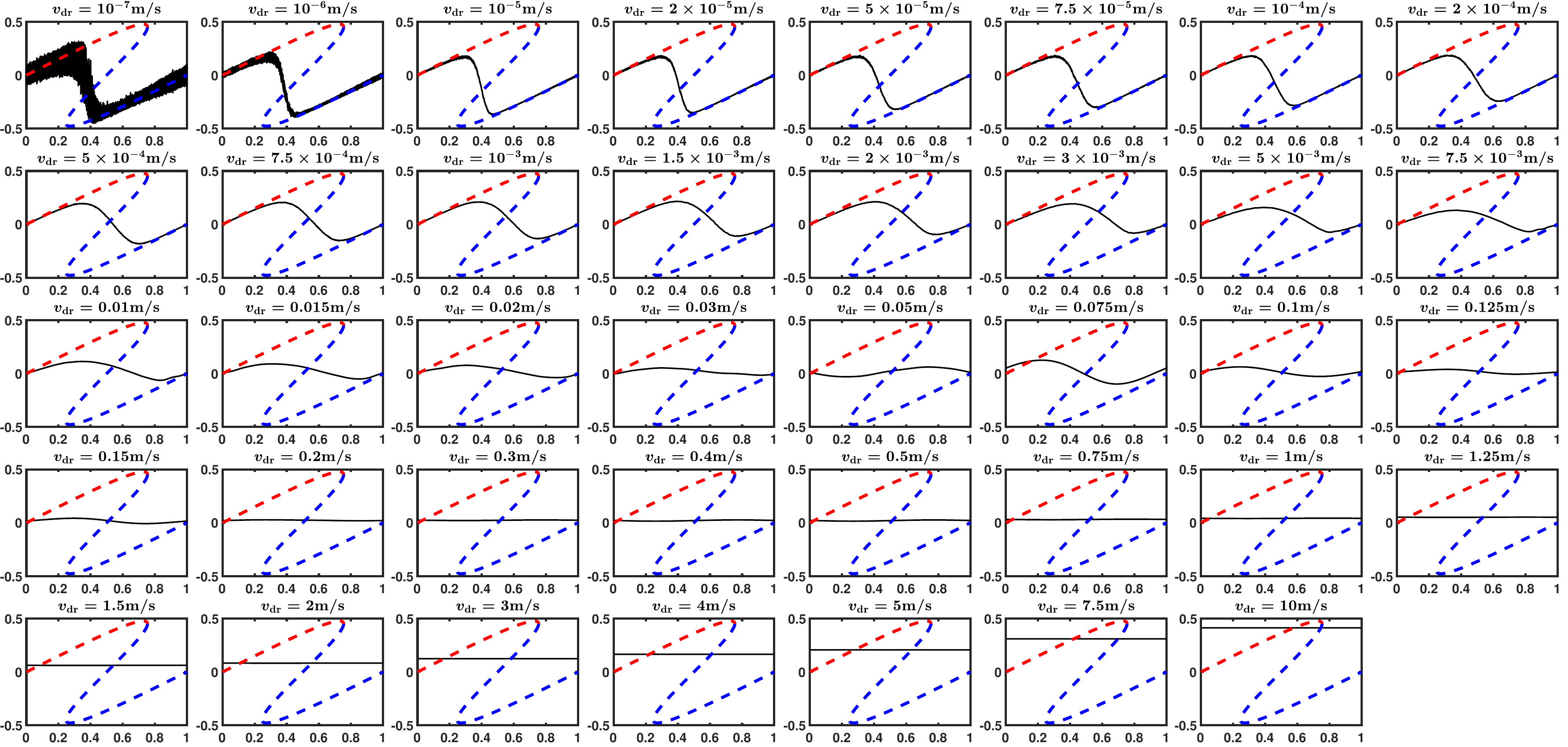}
 \caption{The mean harmonic force $\langle F_{\rm h}\rangle$ during one cycle at different driving velocities of the case of $\eta=3$, $\mu=4\times10^4\rm s^{-1}$ and ${\it\Theta}_{\rm h,c}=0.4,0.04$. All the $x$-coordinates are the driver center's nondimensional position $v_{\rm dr}t/a$ relative to the latest cycle starting point, i.e. the same as that in the main text Figure 2(C). All the $y$-coordinates and the dashed red and blue curves are the same as those in the main text Figure 2(C). Other parameters are given in Sec. \ref{Langevindynamicssimulation}.\ref{ParametersUsed}. The number of simulation cycles at each driving velocity is given in Table \ref{tab:Numberofcycles}. At each driving velocity, all of the simulation cycles are computed sequentially with the initial values of the first simulation cycle inherited from the end values of the last simulation cycle used to calculate the corresponding point at the same $v_{\rm dr}$ on the purple $\langle W_{\rm cyc}\rangle-v_{\rm dr}$ curve of the $\eta=3,{\it\Theta}_{\rm h,c}=0.4,0.04$ case in the main text Figure 3(B), so that we can make sure that the steady state has already been achieved and we don't need to exclude a certain number of cycles at the beginning transient process as we have done when calculating the mean values and standard deviations of $W_{\rm cyc}$ and obtaining the count distributions of $W_{\rm cyc}$ in Figure \ref{WcycDist}, cf. Sec. \ref{Langevindynamicssimulation}.\ref{ParametersUsed}.
\\Like the mean displacement curves in Figure \ref{fig:StickSlipsSingleXTh04Tc004}, the mean harmonic force curve overlaps with the dashed balanced force curves at the beginning and end stages of the engine cycle when the driving velocity is not very high. As the driving velocity increases, the waveform of the mean harmonic force curve changes and its amplitude declines. When $v_{\rm dr}$ is very high, the mean harmonic force stays nearly constant during one cycle, which is consistent with the approximate parallelism between the corresponding mean displacement curve and the driver center line [not plottd, cf. the main text Figure 2(B)] in Figure \ref{fig:StickSlipsSingleXTh04Tc004}. In this high driving velocity regime, the nearly constant mean harmonic force increases to balance the high damping force, cf. Figure \ref{fig:StickSlipsSingleOnecycleXT0}.
} 
 \label{fig:StickSlipsSingleFhTh04Tc004} 
 \end{figure}

\begin{figure}[H]
\centering
 \includegraphics[width=\textwidth]{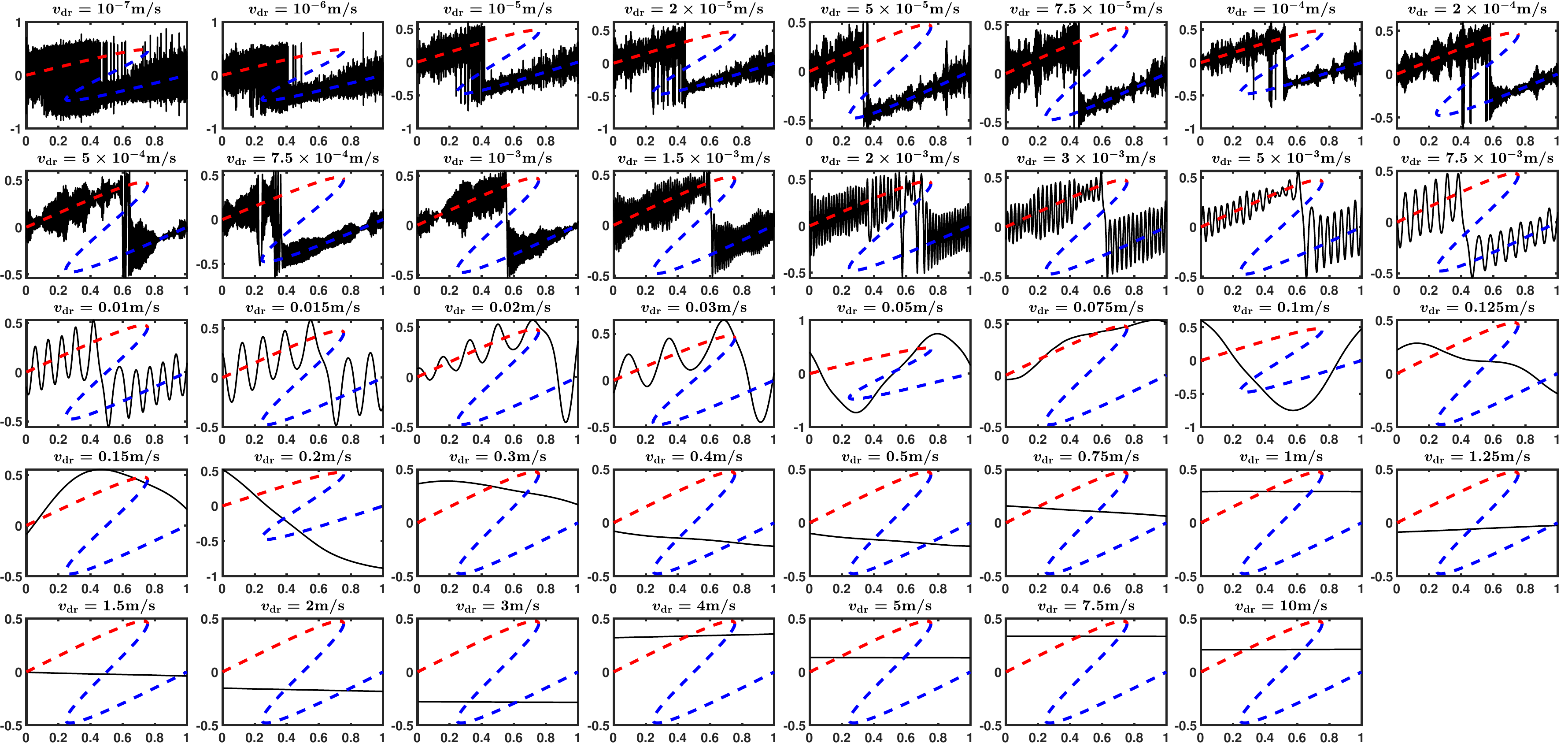}
 \caption{The harmonic force $F_{\rm h}$ during one single cycle at different driving velocities of the case of $\eta=3$, $\mu=4\times10^4\rm s^{-1}$ and ${\it\Theta}_{\rm h,c}=0.4,0.04$. All the $x$-coordinates are the driver center's nondimensional position $v_{\rm dr}t/a$ relative to the latest cycle starting point, i.e. the same as that in the main text Figure 2(C). All the $y$-coordinates and the dashed red and blue curves are the same as those in the main text Figure 2(C). Other parameters are given in Sec. \ref{Langevindynamicssimulation}.\ref{ParametersUsed}. At each driving velocity, the simulation cycle is computed with the initial values inherited from the end values of the last simulation cycle used to calculate the corresponding point at the same $v_{\rm dr}$ on the purple $\langle W_{\rm cyc}\rangle-v_{\rm dr}$ curve of the $\eta=3,{\it\Theta}_{\rm h,c}=0.4,0.04$  case in the main text Figure 3(B), so that we can make sure that the steady state has already been achieved.
\\The waveform of the one single cycle harmonic force curve gradually becomes clear and then becomes irregular in the resonance regime and finally becomes flat, consistent with the corresponding one single cycle displacement curve in Figure \ref{fig:StickSlipsSingleOnecycleXTh04Tc004}. Its amplitude is also modulated by the stochastic force at low driving velocities. At very high driving velocity, the nearly constant harmonic force in one cycle is random due to the stochastic force and not equal to the corresponding nearly constant mean harmonic force in Figure \ref{fig:StickSlipsSingleFhTh04Tc004}.
} 
 \label{fig:StickSlipsSingleOnecycleFhTh04Tc004} 
 \end{figure}

\begin{figure}[H]
\centering
 \includegraphics[width=\textwidth]{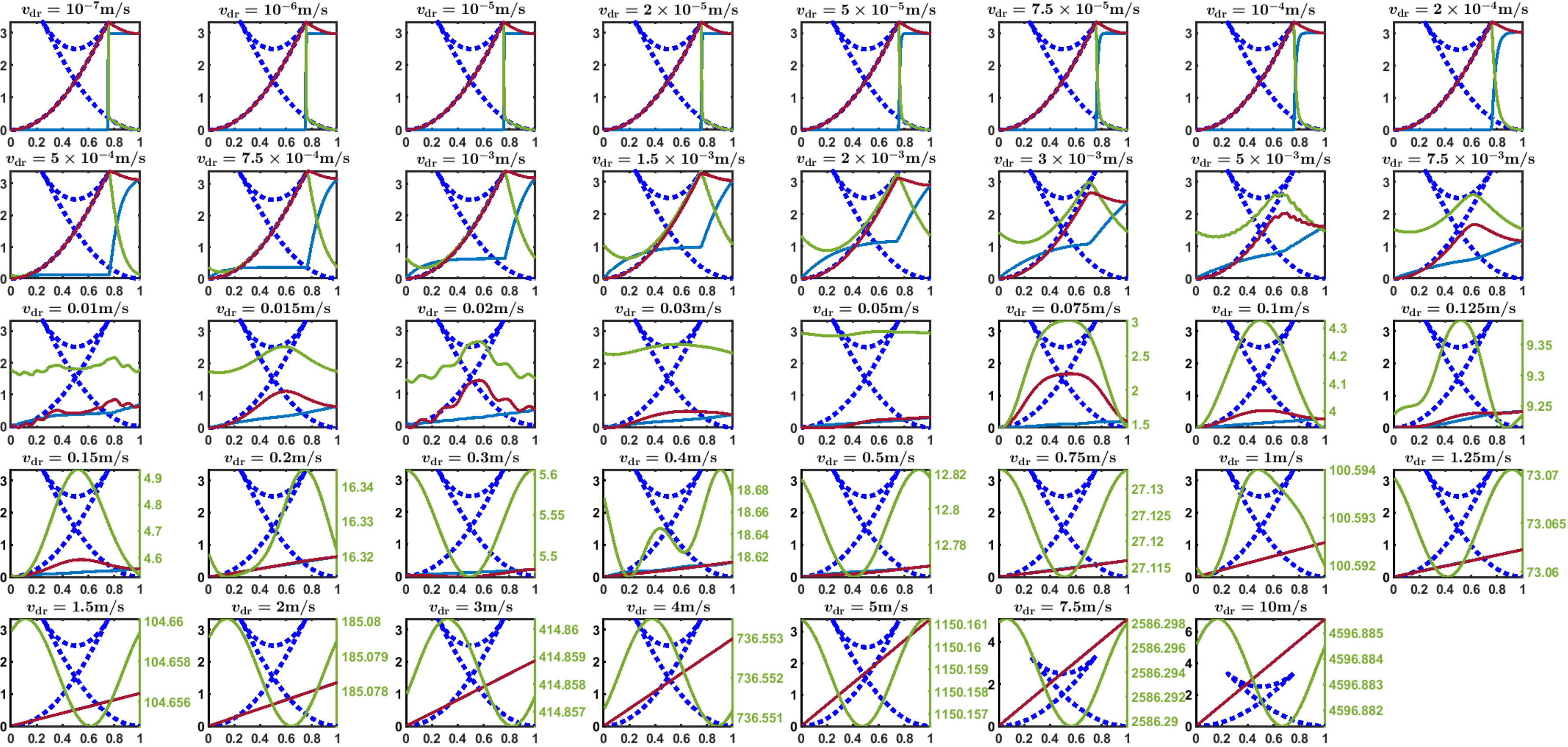}
 \caption{The mean internal energy $\langle U\rangle$ (dark green), the mean heat to the heat bath $\langle Q\rangle$ (dark blue) and the mean work input to the system $\langle W\rangle$ (dark red) during one cycle at different driving velocities of the case of $\eta=3$, $\mu=4\times10^4\rm s^{-1}$ and ${\it\Theta}=0$. All the $x$-coordinates are the driver center's nondimensional position $v_{\rm dr}t/a$ relative to the latest cycle starting point, i.e. the same as those in Figure \ref{CyclesTc004}(B) and (C). All the $y$-coordinates and the dotted blue curves are the same as those in Figure \ref{CyclesTc004}(A). Other parameters are given in Sec. \ref{Langevindynamicssimulation}.\ref{ParametersUsed}. The number of simulation cycles at each driving velocity is given in Table \ref{tab:Numberofcycles}. At each driving velocity, all of the simulation cycles are computed sequentially with the initial values of the first simulation cycle inherited from the end values of the last simulation cycle used to calculate the corresponding point at the same $v_{\rm dr}$ on the black $\langle W_{\rm cyc}\rangle-v_{\rm dr}$ curve of the ${\it\Theta}=0$ case in the main text Figure 3(C), so that we can make sure that the steady state has already been achieved and we don't need to exclude a certain number of cycles at the beginning transient process as we have done when calculating the mean values and standard deviations of $W_{\rm cyc}$ and obtaining the count distributions of $W_{\rm cyc}$ in Figure \ref{WcycDist}, cf. Sec. \ref{Langevindynamicssimulation}.\ref{ParametersUsed}.
\\In the first row of the subfigures, i.e. from $v_{\rm dr}=10^{-7}\rm m/s$ to $v_{\rm dr}=2\times10^{-4}\rm m/s$, the mean internal energy curve after the cusp (slip) instant is gradually uplifted from the dashed blue curve because of the oscillating relaxation process relative to the driver center gets longer with the driving velocity increasing, cf. the text at the beginning of this subsection. At and after $v_{\rm dr}=5\times10^{-4}\rm m/s$ the cycle starting (and also the cycle end, due to the periodicity) point of the mean internal energy curve is also gradually uplifted from the dotted blue curve due to the residual kinetic energy from the last cycle's slip. At and after $v_{\rm dr}=2\times10^{-3}$, the cusp of the mean internal energy curve begins to descend and move to the left, i.e. the residual kinetic energy from slipping in the last cycle arrives. At and after $v_{\rm dr}=0.1\rm m/s$, the mean internal energy curve is higher than the entire dashed blue balanced resultant potential curve and is entirely uplifted gradually, so we put its $y$-axis to the right separately. When the mean internal energy curve is on the dashed blue balanced resultant potential curve, the mean heat curve keeps constant, while when the mean internal energy curve is uplifted from the dashed curve, the mean heat curve becomes exponential correspondingly, in that the residual kinetic energy from slipping is partly dissipated into heat during the longer relaxation process relative to the driver center. From $v_{\rm dr}=10^{-7}\rm m/s$ to $2\times10^{-4}\rm m/s$ (subfigures in the first row), the mean work curve nearly unchanged, consistent with the nearly constant work input in this velocity range on the $\langle W_{\rm cyc}\rangle-v_{\rm dr}$ curve of the ${\it\Theta}=0$ case in the main text Figure 3(C). From $v_{\rm dr}=5\times10^{-4}\rm m/s$ to $10^{-3}\rm m/s$, the descent segment of the mean work curve is shortened in height because the longer oscillating relaxation process after slipping leads to the effective temperature increasing, which is similar to the thermolubricity mechanism. So there is a short elevation before the plateau peak point ($v_{\rm dr}=10^{-3}\rm m/s$) on the $\langle W_{\rm cyc}\rangle-v_{\rm dr}$ curve of the ${\it\Theta}=0$ case in the main text Figure 3(C). At and after $v_{\rm dr}=1.5\times10^{-3}\rm m/s$, the ascent segment of the mean work curve is also shortened in height, resulting in the decreasing of the mean cycle work in the velocity weakenning regime, which we have explained in the text at the beginning of this subsection. In the resonance regime, the three curves are all irregular. At very high driving velocities, the mean work and heat curves overlap with each other and the mean internal energy keeps nearly constant during one cycle, cf. Figure \ref{fig:StickSlipsSingleOnecycleXT0}.
} 
 \label{fig:StickSlipsSingleT0} 
 \end{figure}

\begin{figure}[H]
\centering
 \includegraphics[width=\textwidth]{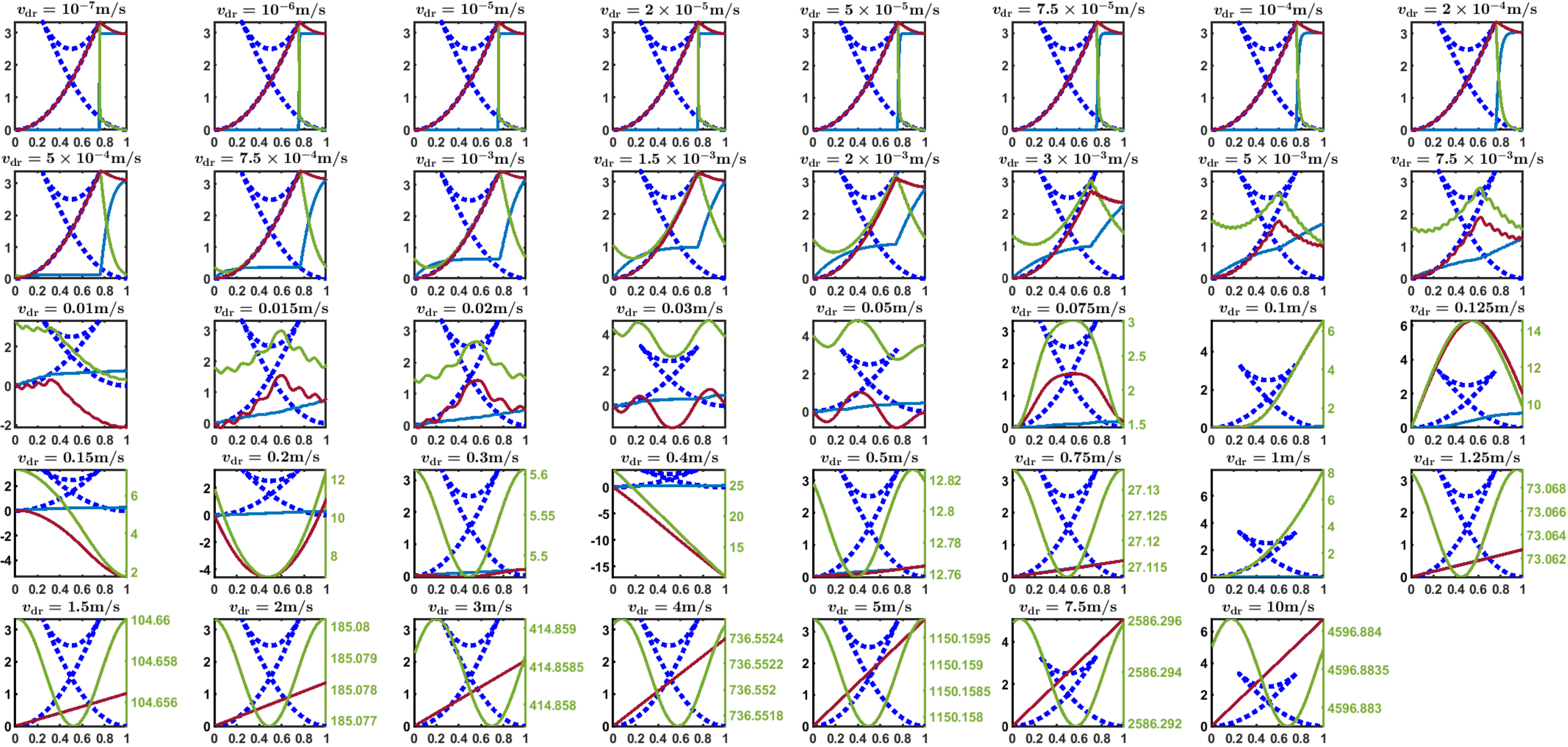}
 \caption{The internal energy $U$ (dark green), the heat to the heat bath $Q$ (dark blue) and the work input to the system $W$ (dark red) during one single cycle at different driving velocities of the case of $\eta=3$, $\mu=4\times10^4\rm s^{-1}$ and ${\it\Theta}=0$. All the $x$-coordinates are the driver center's nondimensional position $v_{\rm dr}t/a$ relative to the latest cycle starting point, i.e. the same as those in Figure \ref{CyclesTc004}(B) and (C). All the $y$-coordinates and the dotted blue curves are the same as those in Figure \ref{CyclesTc004}(A). Other parameters are given in Sec. \ref{Langevindynamicssimulation}.\ref{ParametersUsed}. At each driving velocity, the simulation cycle is computed with the initial values inherited from the end values of the last simulation cycle used to calculate the corresponding point at the same $v_{\rm dr}$ on the black $\langle W_{\rm cyc}\rangle-v_{\rm dr}$ curve of the ${\it\Theta}=0$ case in the main text Figure 3(C), so that we can make sure that the steady state has already been achieved.
\\The starting and end points of the one single cycle internal energy curve should have the same value if the particle's movement is periodic with as period the time of the driver center going over one cycle. If they are not the same, the particle's movement may not be periodic or have a period of the time of the driver center going over more than one lattice period. The latter is common in the velocity weakening regime and the resonance regime, cf. the main text subsection ``Bifurcation of Mean Cycle Work with Driving Velocity as Parameter at Zero Temperature''.
} 
 \label{fig:StickSlipsSingleOnecycleT0} 
 \end{figure}

\begin{figure}[H]
\centering
 \includegraphics[width=\textwidth]{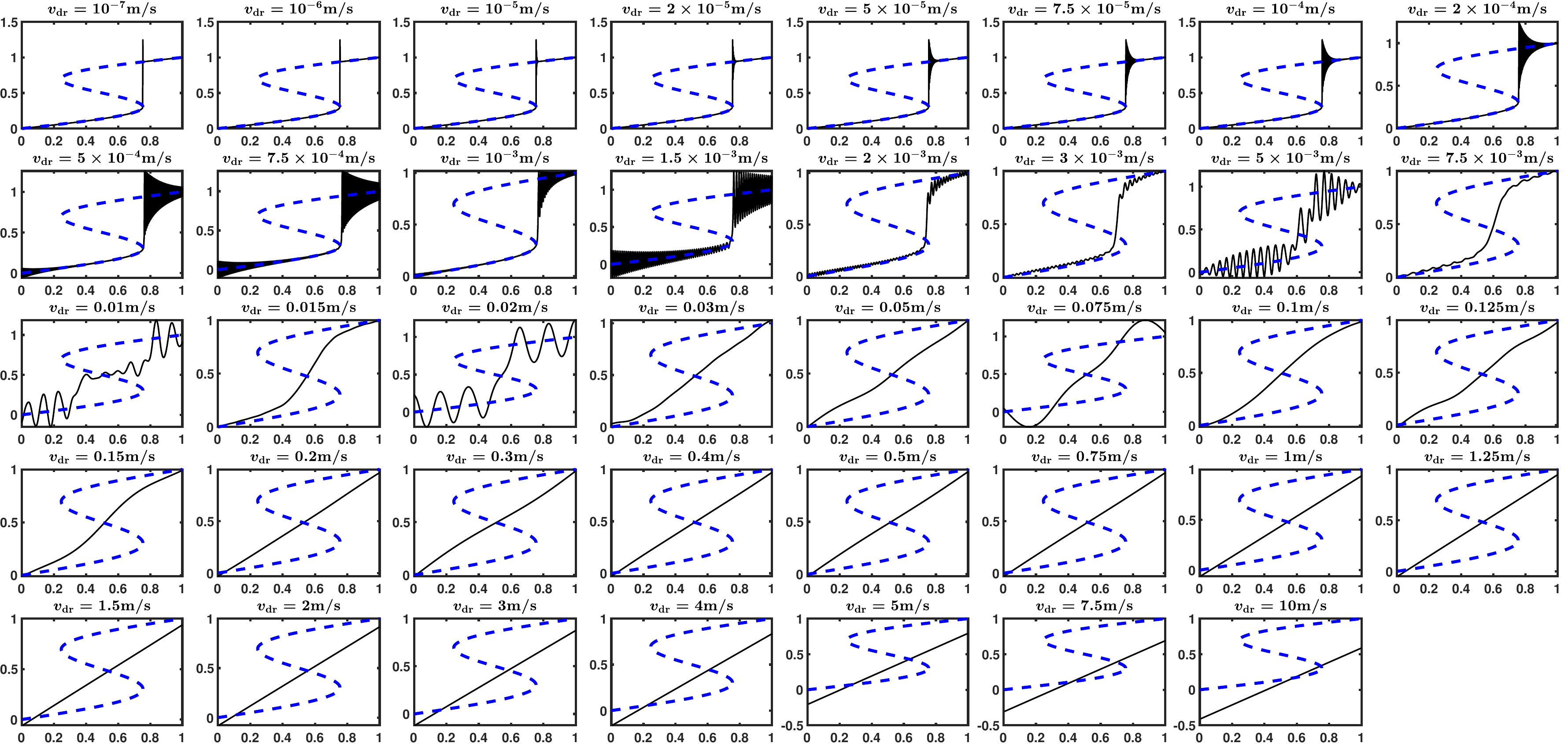}
 \caption{The mean displacement of the particle $\langle x\rangle$ during one cycle at different driving velocities of the case of $\eta=3$, $\mu=4\times10^4\rm s^{-1}$ and ${\it\Theta}=0$. All the $x$-coordinates are the driver center's nondimensional position $v_{\rm dr}t/a$ relative to the latest cycle starting point, i.e. the same as that in Figure \ref{CyclesTc004}(B). All the $y$-coordinates and the dashed blue curves are the same as those in Figure \ref{CyclesTc004}(B). Other parameters are given in Sec. \ref{Langevindynamicssimulation}.\ref{ParametersUsed}. The number of simulation cycles at each driving velocity is given in Table \ref{tab:Numberofcycles}. At each driving velocity, all of the simulation cycles are computed sequentially with the initial values of the first simulation cycle inherited from the end values of the last simulation cycle used to calculate the corresponding point at the same $v_{\rm dr}$ on the black $\langle W_{\rm cyc}\rangle-v_{\rm dr}$ curve of the ${\it\Theta}=0$ case in the main text Figure 3(C), so that we can make sure that the steady state has already been achieved and we don't need to exclude a certain number of cycles at the beginning transient process as we have done when calculating the mean values and standard deviations of $W_{\rm cyc}$ and obtaining the count distributions of $W_{\rm cyc}$ in Figure \ref{WcycDist}, cf. Sec. \ref{Langevindynamicssimulation}.\ref{ParametersUsed}.
\\If the mean displacement curve during one cycle is different from the corresponding one single cycle displacement curve in Figure \ref{fig:StickSlipsSingleOnecycleXT0} at the same driving velocity, then the particle's movement is not periodic or is periodic with as period the time of the driver center moving over more than one lattice period.
} 
 \label{fig:StickSlipsSingleXT0} 
 \end{figure}

\begin{figure}[H]
\centering
 \includegraphics[width=\textwidth]{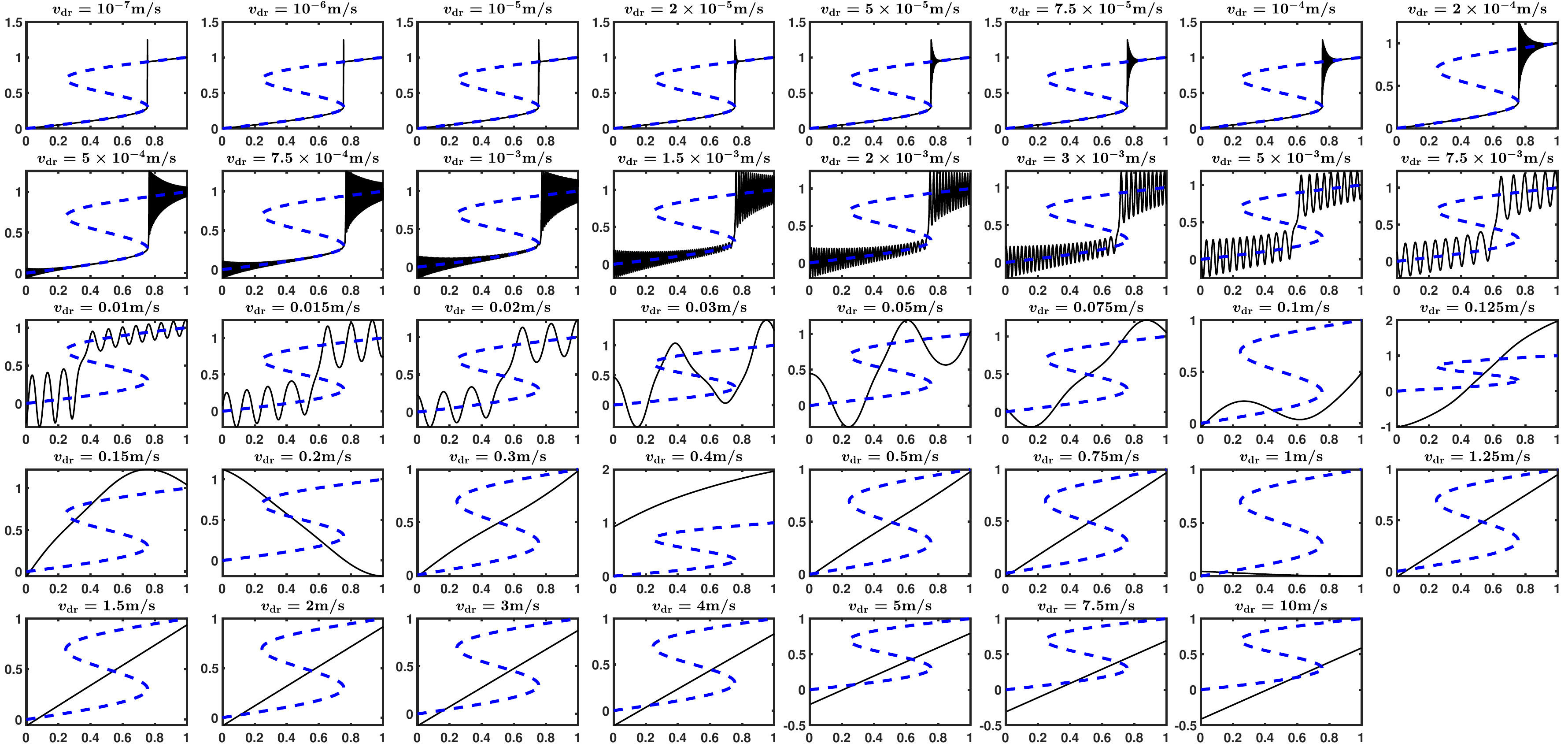}
 \caption{The displacement $x$ during one single cycle at different driving velocities of the case of $\eta=3$, $\mu=4\times10^4\rm s^{-1}$ and ${\it\Theta}=0$. All the $x$-coordinates are the driver center's nondimensional position $v_{\rm dr}t/a$ relative to the latest cycle starting point, i.e. the same as that in Figure \ref{CyclesTc004}(B). All the $y$-coordinates and the dashed blue curves are the same as those in Figure \ref{CyclesTc004}(B). Other parameters are given in Sec. \ref{Langevindynamicssimulation}.\ref{ParametersUsed}. At each driving velocity, the simulation cycle is computed with the initial values inherited from the end values of the last simulation cycle used to calculate the corresponding point at the same $v_{\rm dr}$ on the black $\langle W_{\rm cyc}\rangle-v_{\rm dr}$ curve of the ${\it\Theta}=0$ case in the main text Figure 3(C), so that we can make sure that the steady state has already been achieved.
\\In the text at the beginning of this subsection, we have analyzed the stick-slip regime and the velocity weakening regime. At $v_{\rm dr}=10^{-5}\rm m/s$, stick-slip occurs and we have projected the position of the particle on the resultant potential curves varying with time in the main text Figure 6(A1) which can be compared with the one single cycle displacement curve at the same driving velocity $v_{\rm dr}=10^{-5}\rm m/s$ in the subfigure at (row,column)=(1,3). The velocity weakening case at $v_{\rm dr}=0.003\rm m/s$ is plotted in the main text Figure 6(A2), which can also be compared with the corresponding one single cycle displacement curve in the subfigure at (row,column)=(2,6). At $v_{\rm dr}=0.4\rm m/s$ in the resonance regime, three solutions are plotted in the main text Figure 6(A3), (A4) and (A5), with the cycle number period $P_{\rm c.n.}=6$, $6$ and $1$ (Eq. \ref{eqn:cyclenumperiod}) respectively. In the corresponding subfigure at the same driving velocity at (row,column)=(4,4), this specific one single cycle displacement curve has a cycle number period $P_{\rm c.n.}>1$. At $v_{\rm dr}=10\rm m/s$, we can see that the particle's position is far behind the driver center in the main text Figure 6(A6) and in the subfigure (row,column)=(5,7). At such a high driving velocity, the damping force $\mu m\dot x$ is so large that the harmonic force has to be large enough to balance it so that the particle's displacement from the driver center $x-v_{\rm dr}t$ should be large. Because the sinusoidal lattice force's amplitude is constant, its effect compared with such a large harmonic force is small. Therefore, the particle's acceleration is small and its velocity is nearly constant so that the displacement curve is approximately linear at the high driving velocity end.
} 
 \label{fig:StickSlipsSingleOnecycleXT0} 
 \end{figure}

\begin{figure}[H]
\centering
 \includegraphics[width=\textwidth]{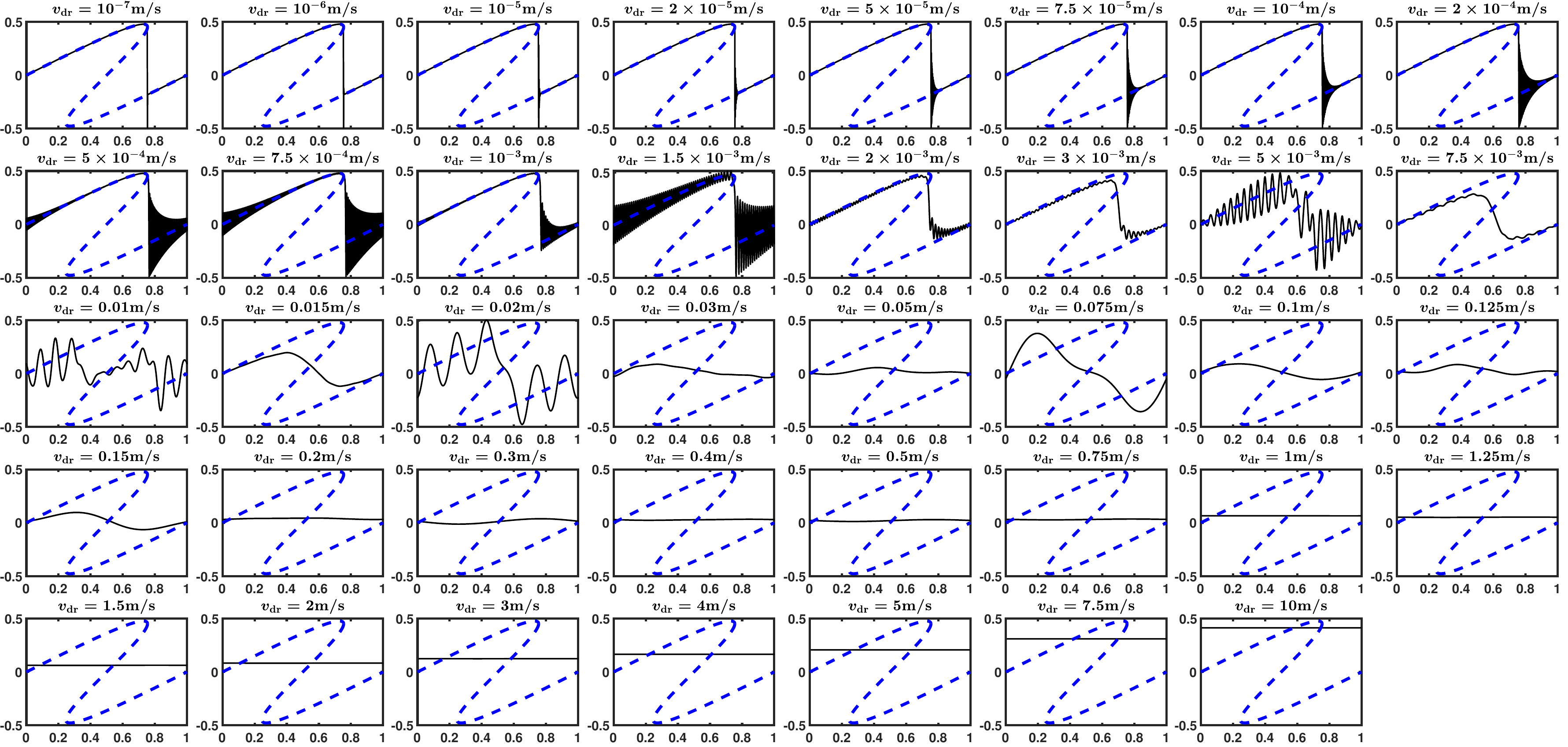}
 \caption{The mean harmonic force $\langle F_{\rm h}\rangle$ during one cycle at different driving velocities of the case of $\eta=3$, $\mu=4\times10^4\rm s^{-1}$ and ${\it\Theta}=0$. All the $x$-coordinates are the driver center's nondimensional position $v_{\rm dr}t/a$ relative to the latest cycle starting point, i.e. the same as that in Figure \ref{CyclesTc004}(C). All the $y$-coordinates and the dashed blue curves are the same as those in Figure \ref{CyclesTc004}(C). Other parameters are given in Sec. \ref{Langevindynamicssimulation}.\ref{ParametersUsed}. The number of simulation cycles at each driving velocity is given in Table \ref{tab:Numberofcycles}. At each driving velocity, all of the simulation cycles are computed sequentially with the initial values of the first simulation cycle inherited from the end values of the last simulation cycle used to calculate the corresponding point at the same $v_{\rm dr}$ on the black $\langle W_{\rm cyc}\rangle-v_{\rm dr}$ curve of the ${\it\Theta}=0$ case in the main text Figure 3(C), so that we can make sure that the steady state has already been achieved and we don't need to exclude a certain number of cycles at the beginning transient process as we have done when calculating the mean values and standard deviations of $W_{\rm cyc}$ and obtaining the count distributions of $W_{\rm cyc}$ in Figure \ref{WcycDist}, cf. Sec. \ref{Langevindynamicssimulation}.\ref{ParametersUsed}.
\\In homogeneous zero temperature, the mean harmonic force curves during one cycle are not smooth at low driving velocities, in that the oscillating relaxation process is the same in each single cycle and will stay the same after average, rather than being smoothened after average at some of the mediate driving velocities at which the particle's movement has no period or has a period of the time of the driver center moving over more than one lattice period, e.g. at $v_{\rm dr}=2\times10^{-3}\rm m/s$, $3\times10^{-3}\rm m/s$, $5\times10^{-3}\rm m/s$, $7.5\times10^{-3}\rm m/s$, etc. The mean harmonic force curves are also smoothened after average in the finite temperature cases as can be seen in Figure \ref{fig:StickSlipsSingleFhTh04Tc004}, \ref{fig:StickSlipsSingleFhT04} and \ref{fig:StickSlipsSingleFhT004}.
} 
 \label{fig:StickSlipsSingleFhT0} 
 \end{figure}

\begin{figure}[H]
\centering
 \includegraphics[width=\textwidth]{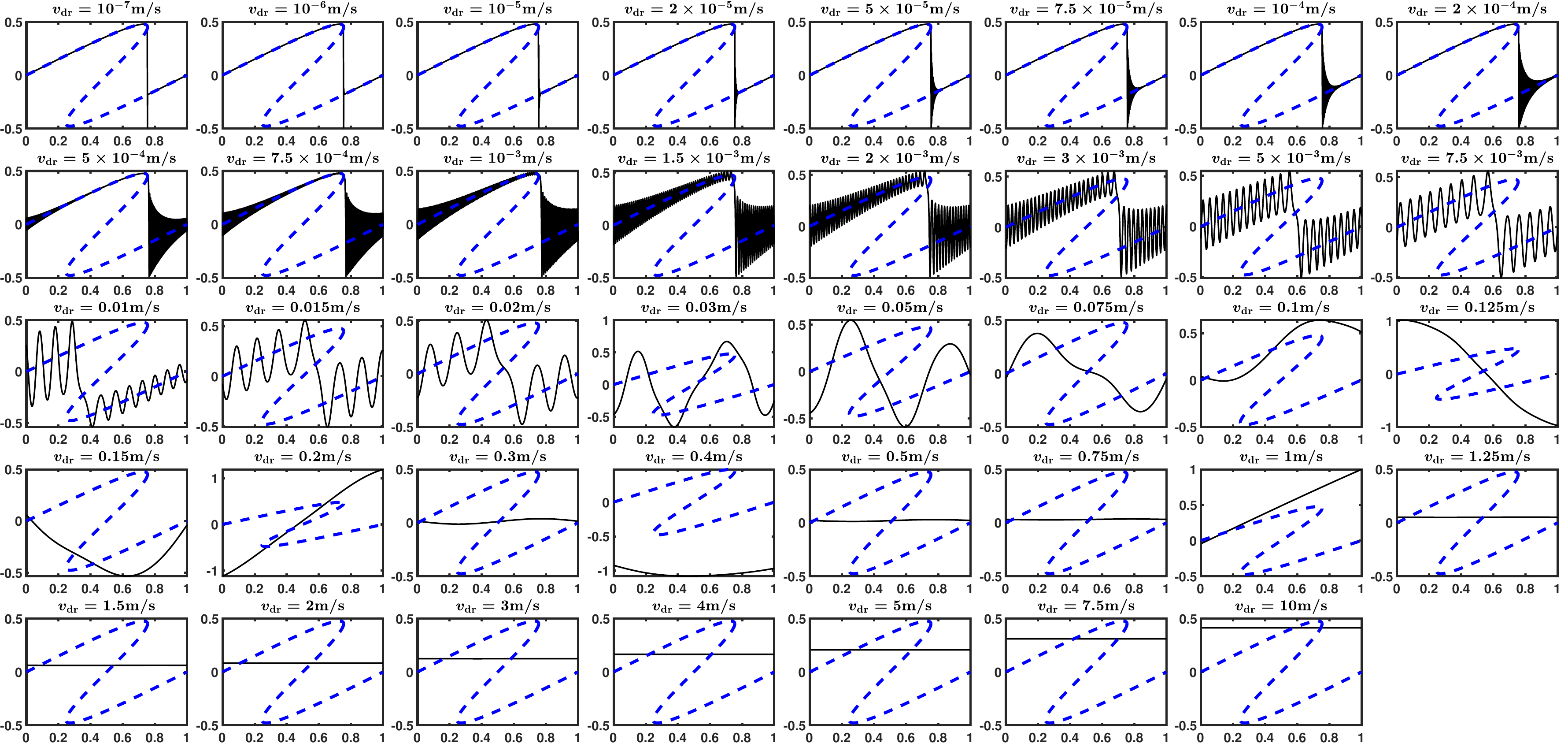}
 \caption{The harmonic force $F_{\rm h}$ during one single cycle at different driving velocities of the case of $\eta=3$, $\mu=4\times10^4\rm s^{-1}$ and ${\it\Theta}=0$. All the $x$-coordinates are the driver center's nondimensional position $v_{\rm dr}t/a$ relative to the latest cycle starting point, i.e. the same as that in Figure \ref{CyclesTc004}(C). All the $y$-coordinates and the dashed blue curves are the same as those in Figure \ref{CyclesTc004}(C). Other parameters are given in Sec. \ref{Langevindynamicssimulation}.\ref{ParametersUsed}. At each driving velocity, the simulation cycle is computed with the initial values inherited from the end values of the last simulation cycle used to calculate the corresponding point at the same $v_{\rm dr}$ on the black $\langle W_{\rm cyc}\rangle-v_{\rm dr}$ curve of the ${\it\Theta}=0$ case in the main text Figure 3(C), so that we can make sure that the steady state has already been achieved.
\\We can see that when the driving velocity is so high, the harmonic force is very large and keeps nearly constant, cf. Figure \ref{fig:StickSlipsSingleOnecycleXT0}.
} 
 \label{fig:StickSlipsSingleOnecycleFhT0} 
 \end{figure}

\begin{figure}[H]
\centering
 \includegraphics[width=\textwidth]{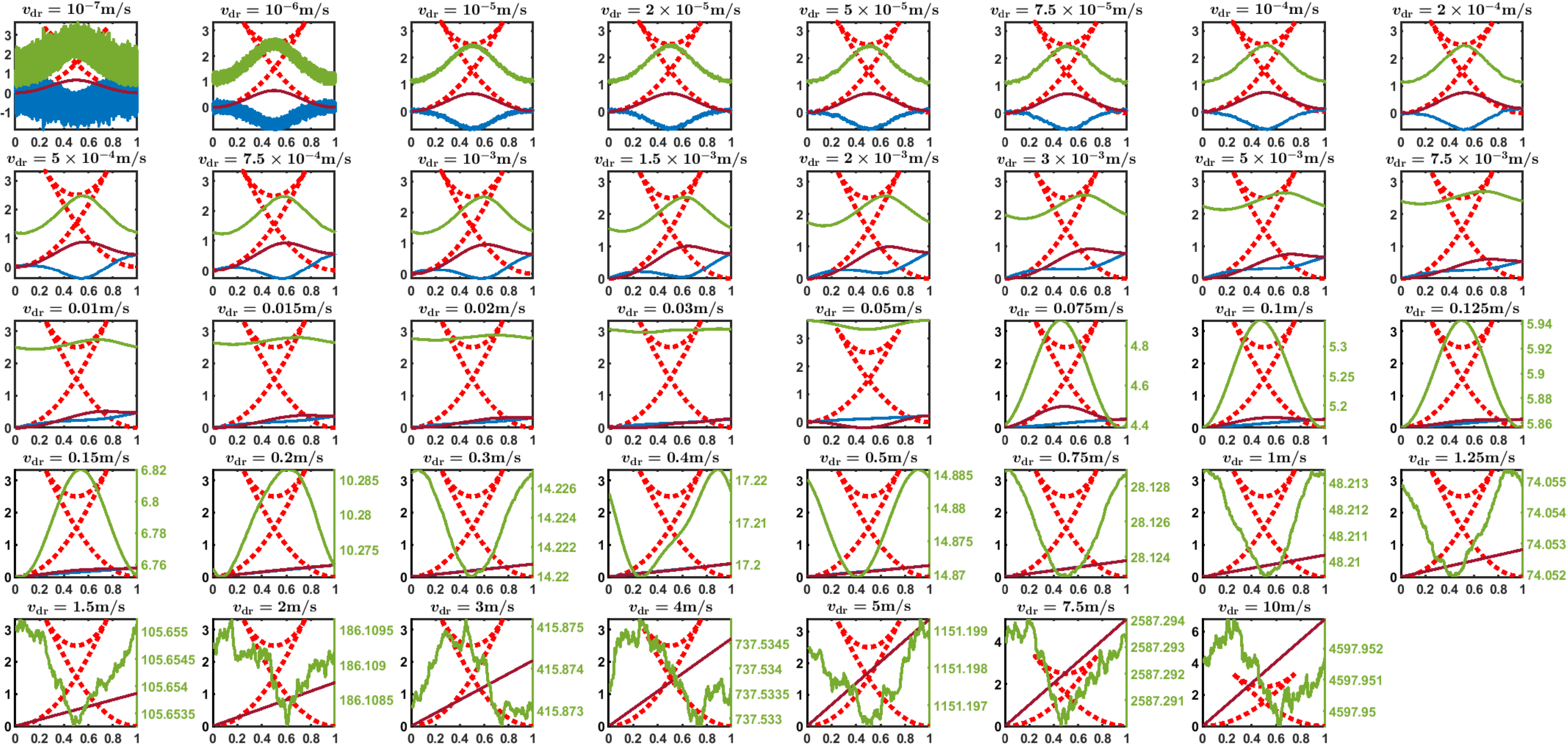}
 \caption{The mean internal energy $\langle U\rangle$ (dark green), the mean heat to the heat bath $\langle Q\rangle$ (dark blue) and the mean work input to the system $\langle W\rangle$ (dark red) during one cycle at different driving velocities of the case of $\eta=3$, $\mu=4\times10^4\rm s^{-1}$ and ${\it\Theta}=0.4$. All the $x$-coordinates are the driver center's nondimensional position $v_{\rm dr}t/a$ relative to the latest cycle starting point, i.e. the same as those in Figure \ref{CyclesTh04}(B) and (C). All the $y$-coordinates and the dotted red curves are the same as those in Figure \ref{CyclesTh04}(A). Other parameters are given in Sec. \ref{Langevindynamicssimulation}.\ref{ParametersUsed}. The number of simulation cycles at each driving velocity is given in Table \ref{tab:Numberofcycles}. At each driving velocity, all of the simulation cycles are computed sequentially with the initial values of the first simulation cycle inherited from the end values of the last simulation cycle used to calculate the corresponding point at the same $v_{\rm dr}$ on the pink $\langle W_{\rm cyc}\rangle-v_{\rm dr}$ curve of the ${\it\Theta}=0.4$ case in the main text Figure 3(C), so that we can make sure that the steady state has already been achieved and we don't need to exclude a certain number of cycles at the beginning transient process as we have done when calculating the mean values and standard deviations of $W_{\rm cyc}$ and obtaining the count distributions of $W_{\rm cyc}$ in Figure \ref{WcycDist}, cf. Sec. \ref{Langevindynamicssimulation}.\ref{ParametersUsed}.
\\At such a high temperature, the particle is easy to achieve equilibrium at each instant at low enough driving velocity. Because the temperature is homogeneous, the mean internal energy, heat and work curves are nearly all symmetric about the middle of one cycle when the driving velocity is low, such as those in the first 4 subfigures. At $v_{\rm dr}=10^{-7}\rm m/s$ and $10^{-6}\rm m/s$, the curves are noisy because of the small sample number of simulation cycles.
} 
 \label{fig:StickSlipsSingleT04} 
 \end{figure}

\begin{figure}[H]
\centering
 \includegraphics[width=\textwidth]{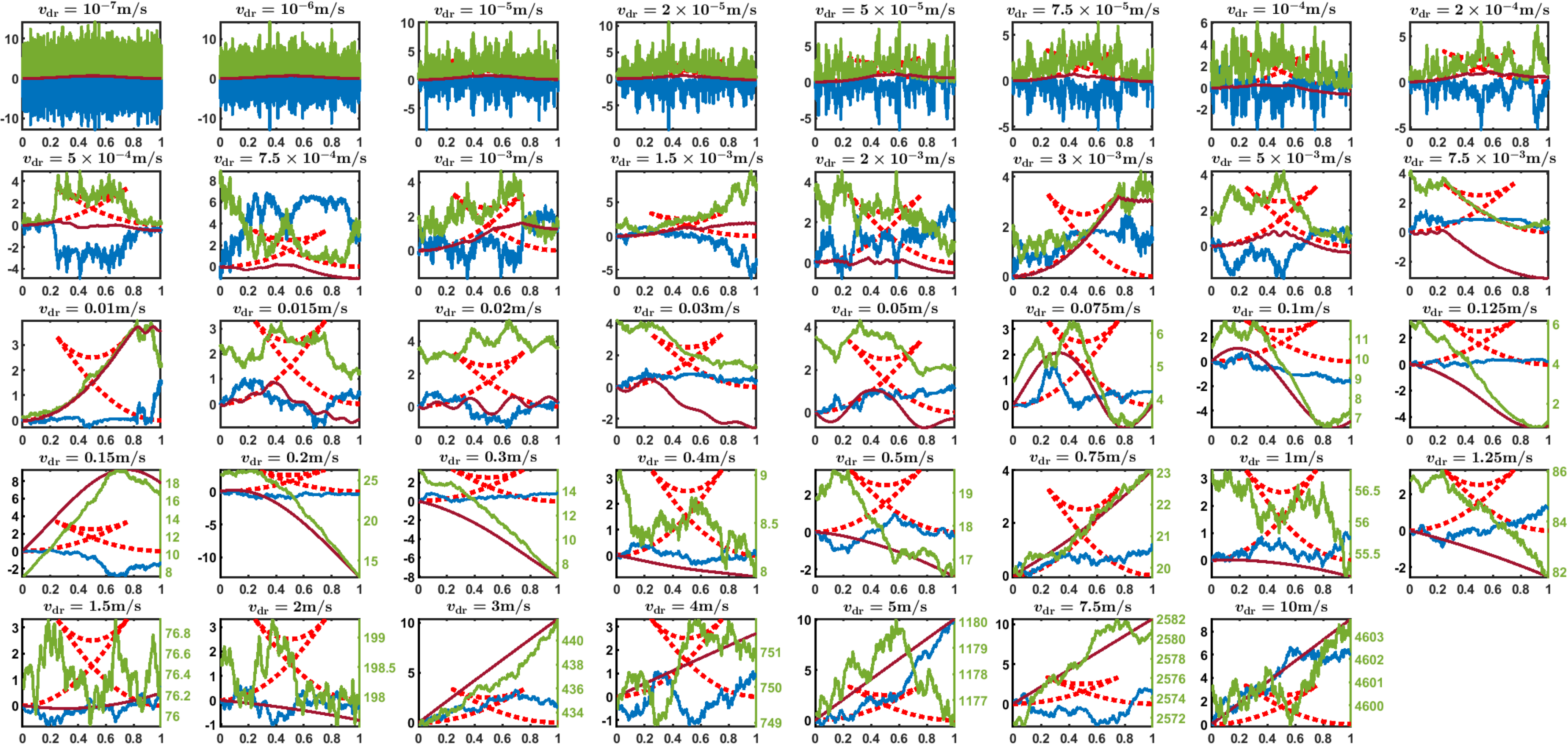}
 \caption{The internal energy $U$ (dark green), the heat to the heat bath $Q$ (dark blue) and the work input to the system $W$ (dark red) during one single cycle at different driving velocities of the case of $\eta=3$, $\mu=4\times10^4\rm s^{-1}$ and ${\it\Theta}=0.4$. All the $x$-coordinates are the driver center's nondimensional position $v_{\rm dr}t/a$ relative to the latest cycle starting point, i.e. the same as those in Figure \ref{CyclesTh04}(B) and (C). All the $y$-coordinates and the dotted red curves are the same as those in Figure \ref{CyclesTh04}(A). Other parameters are given in Sec. \ref{Langevindynamicssimulation}.\ref{ParametersUsed}. At each driving velocity, the simulation cycle is computed with the initial values inherited from the end values of the last simulation cycle used to calculate the corresponding point at the same $v_{\rm dr}$ on the pink $W_{\rm cyc}-v_{\rm dr}$ curve of the ${\it\Theta}_{\rm h,c}=0.4$ case in the main text Figure 3(C), so that we can make sure that the steady state has already been achieved.
\\At such a high temperature, the stochastic force is large so that the internal energy, heat and work curves during one single cycle are very different from the mean curves in Figure \ref{fig:StickSlipsSingleT04}. Although the stochastic force is of high amplitude in such a high temperature, the work curve is still very smooth relative to the internal erergy and the heat curves.
} 
 \label{fig:StickSlipsSingleOnecycleT04} 
 \end{figure}

\begin{figure}[H]
\centering
 \includegraphics[width=\textwidth]{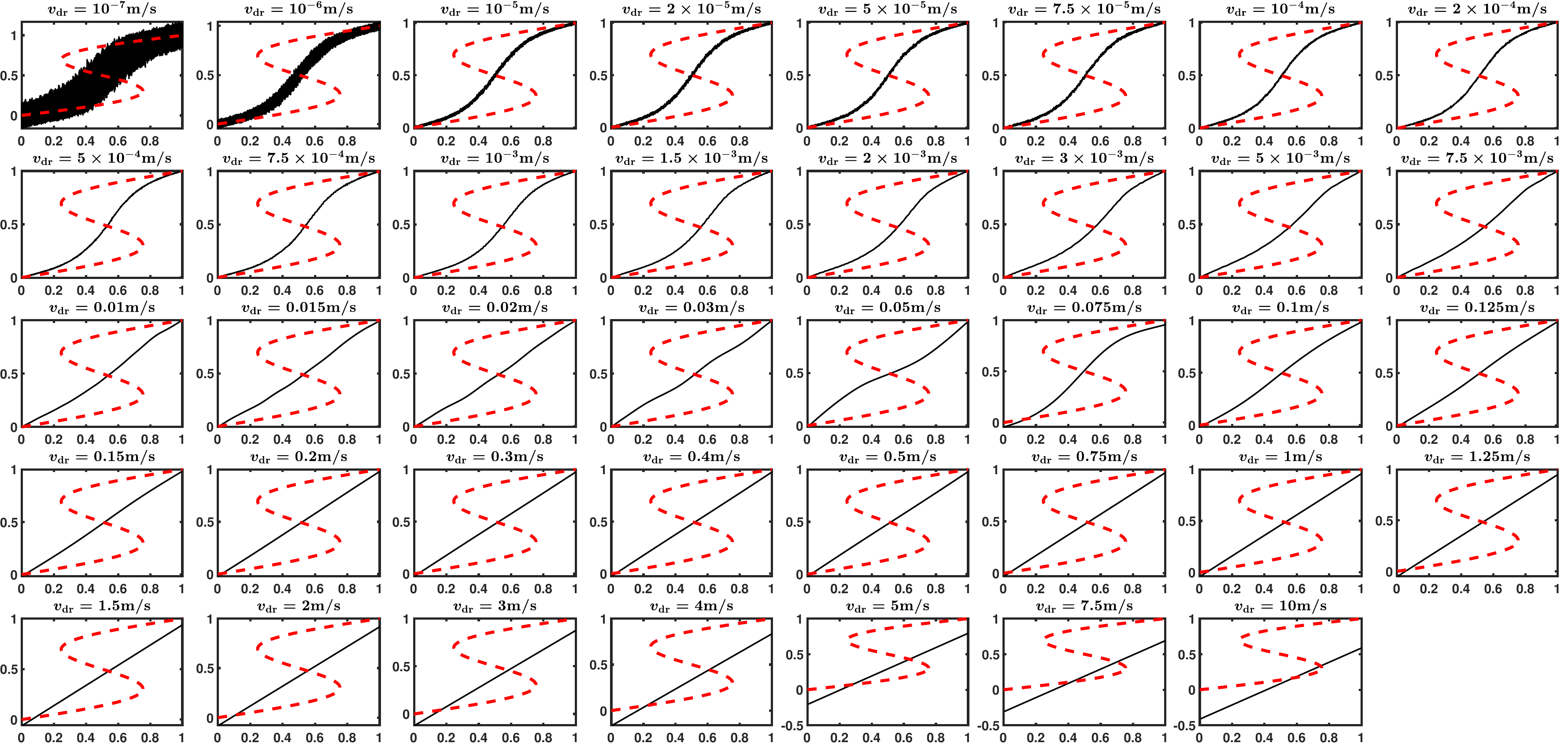}
 \caption{The mean displacement of the particle $\langle x\rangle$ during one cycle at different driving velocities of the case of $\eta=3$, $\mu=4\times10^4\rm s^{-1}$ and ${\it\Theta}=0.4$. All the $x$-coordinates are the driver center's nondimensional position $v_{\rm dr}t/a$ relative to the latest cycle starting point, i.e. the same as that in Figure \ref{CyclesTh04}(B). All the $y$-coordinates and the dashed red curves are the same as those in Figure \ref{CyclesTh04}(B). Other parameters are given in Sec. \ref{Langevindynamicssimulation}.\ref{ParametersUsed}. The number of simulation cycles at each driving velocity is given in Table \ref{tab:Numberofcycles}. At each driving velocity, all of the simulation cycles are computed sequentially with the initial values of the first simulation cycle inherited from the end values of the last simulation cycle used to calculate the corresponding point at the same $v_{\rm dr}$ on the pink $\langle W_{\rm cyc}\rangle-v_{\rm dr}$ curve of the ${\it\Theta}=0.4$ case in the main text Figure 3(C), so that we can make sure that the steady state has already been achieved and we don't need to exclude a certain number of cycles at the beginning transient process as we have done when calculating the mean values and standard deviations of $W_{\rm cyc}$ and obtaining the count distributions of $W_{\rm cyc}$ in Figure \ref{WcycDist}, cf. Sec. \ref{Langevindynamicssimulation}.\ref{ParametersUsed}.} 
 \label{fig:StickSlipsSingleXT04} 
 \end{figure}

\begin{figure}[H]
\centering
 \includegraphics[width=\textwidth]{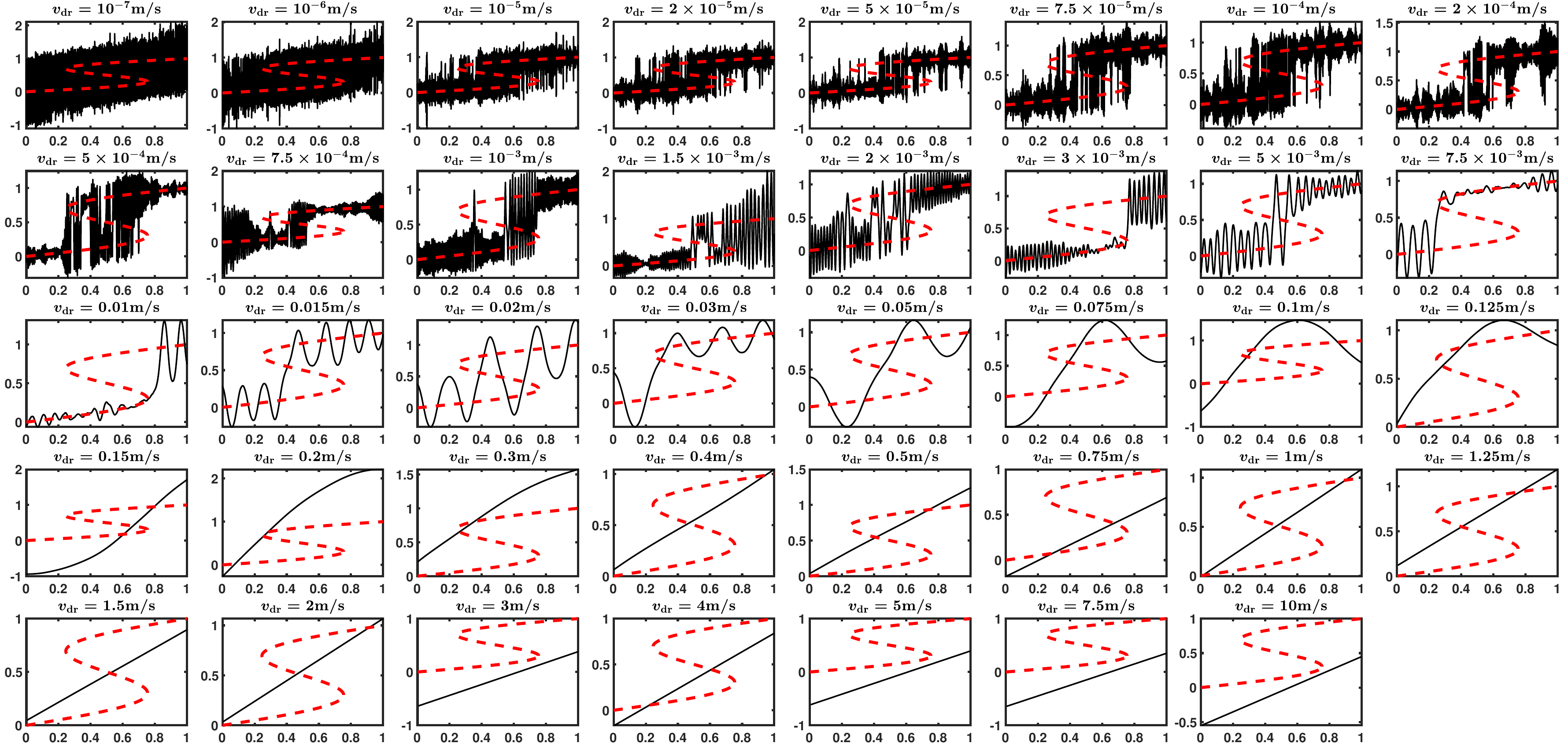}
 \caption{The displacement $x$ during one single cycle at different driving velocities of the case of $\eta=3$, $\mu=4\times10^4\rm s^{-1}$ and ${\it\Theta}=0.4$. All the $x$-coordinates are the driver center's nondimensional position $v_{\rm dr}t/a$ relative to the latest cycle starting point, i.e. the same as that in Figure \ref{CyclesTh04}(B). All the $y$-coordinates and the dashed red curves are the same as those in Figure \ref{CyclesTh04}(B). Other parameters are given in Sec. \ref{Langevindynamicssimulation}.\ref{ParametersUsed}. At each driving velocity, the simulation cycle is computed with the initial values inherited from the end values of the last simulation cycle used to calculate the corresponding point at the same $v_{\rm dr}$ on the pink $\langle W_{\rm cyc}\rangle-v_{\rm dr}$ curve of the ${\it\Theta}=0.4$ case in the main text Figure 3(C), so that we can make sure that the steady state has already been achieved.} 
 \label{fig:StickSlipsSingleOnecycleXT04} 
 \end{figure}

\begin{figure}[H]
\centering
 \includegraphics[width=\textwidth]{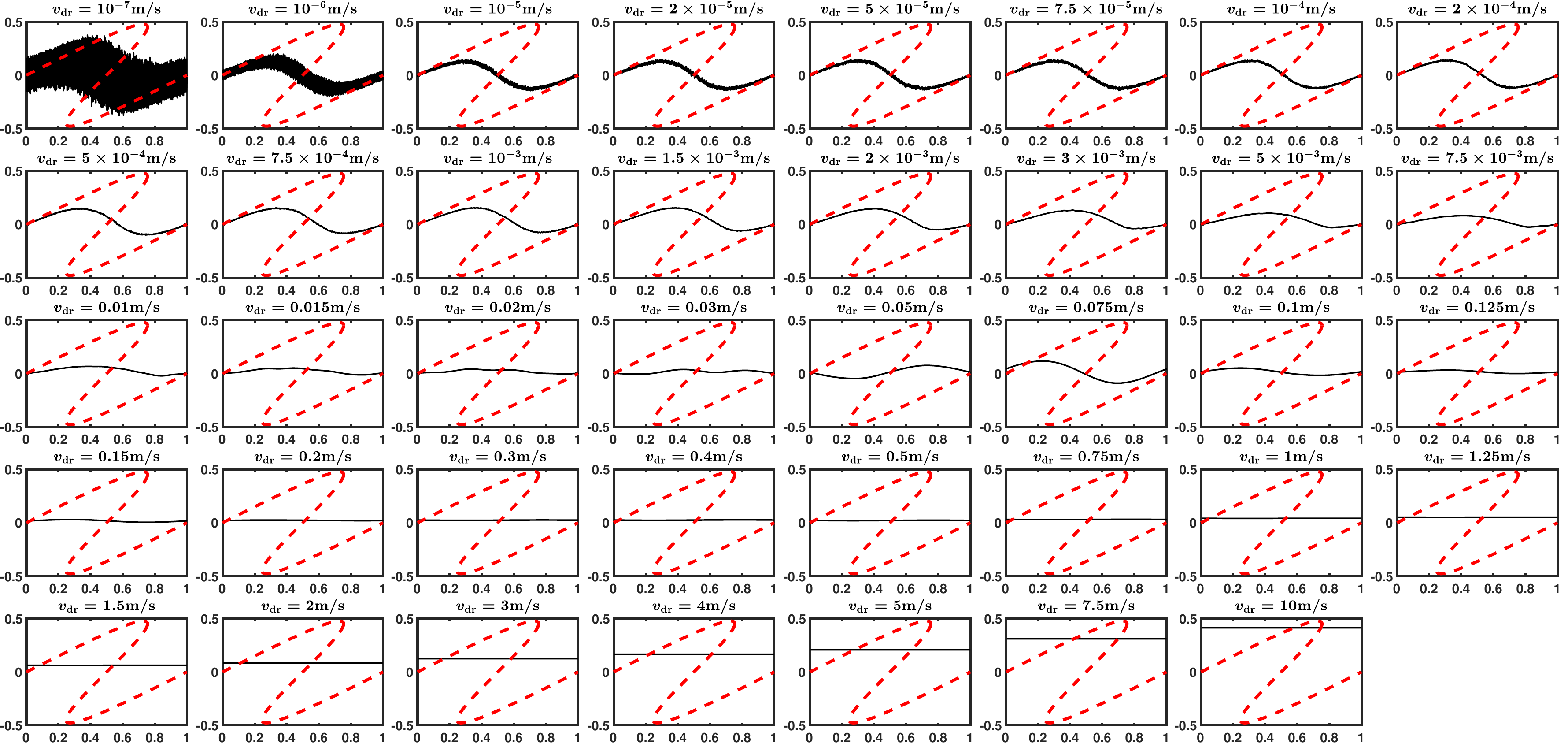}
 \caption{The mean harmonic force $\langle F_{\rm h}\rangle$ during one cycle at different driving velocities of the case of $\eta=3$, $\mu=4\times10^4\rm s^{-1}$ and ${\it\Theta}=0.4$. All the $x$-coordinates are the driver center's nondimensional position $v_{\rm dr}t/a$ relative to the latest cycle starting point, i.e. the same as that in Figure \ref{CyclesTh04}(C). All the $y$-coordinates and the dashed red curves are the same as those in Figure \ref{CyclesTh04}(C). Other parameters are given in Sec. \ref{Langevindynamicssimulation}.\ref{ParametersUsed}. The number of simulation cycles at each driving velocity is given in Table \ref{tab:Numberofcycles}. At each driving velocity, all of the simulation cycles are computed sequentially with the initial values of the first simulation cycle inherited from the end values of the last simulation cycle used to calculate the corresponding point at the same $v_{\rm dr}$ on the pink $\langle W_{\rm cyc}\rangle-v_{\rm dr}$ curve of the ${\it\Theta}=0.4$ case in the main text Figure 3(C), so that we can make sure that the steady state has already been achieved and we don't need to exclude a certain number of cycles at the beginning transient process as we have done when calculating the mean values and standard deviations of $W_{\rm cyc}$ and obtaining the count distributions of $W_{\rm cyc}$ in Figure \ref{WcycDist}, cf. Sec. \ref{Langevindynamicssimulation}.\ref{ParametersUsed}.} 
 \label{fig:StickSlipsSingleFhT04} 
 \end{figure}

\begin{figure}[H]
\centering
 \includegraphics[width=\textwidth]{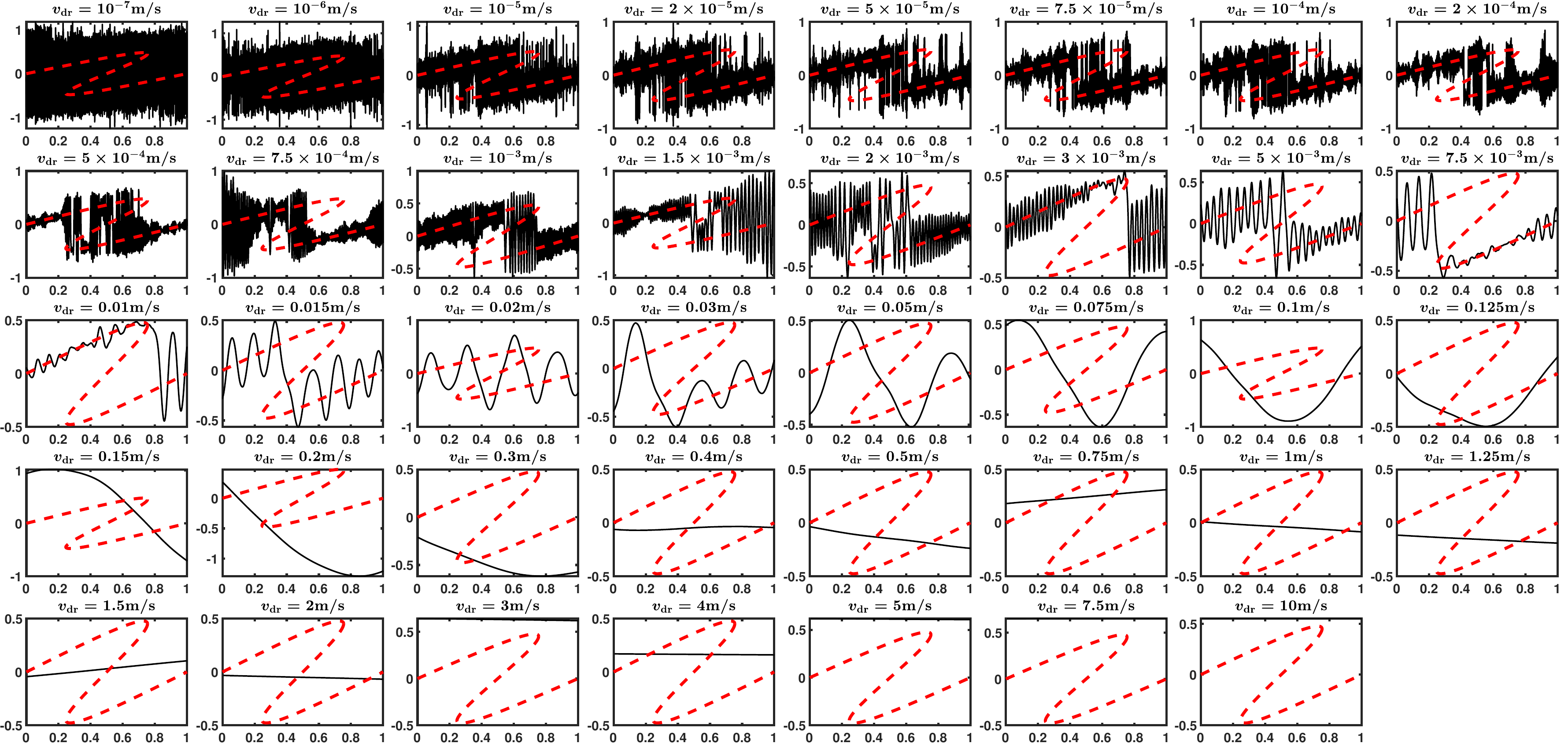}
 \caption{The harmonic force $F_{\rm h}$ during one single cycle at different driving velocities of the case of $\eta=3$, $\mu=4\times10^4\rm s^{-1}$ and ${\it\Theta}=0.4$. All the $x$-coordinates are the driver center's nondimensional position $v_{\rm dr}t/a$ relative to the latest cycle starting point, i.e. the same as that in Figure \ref{CyclesTh04}(C). All the $y$-coordinates and the dashed red curves are the same as those in Figure \ref{CyclesTh04}(C). Other parameters are given in Sec. \ref{Langevindynamicssimulation}.\ref{ParametersUsed}. At each driving velocity, the simulation cycle is computed with the initial values inherited from the end values of the last simulation cycle used to calculate the corresponding point at the same $v_{\rm dr}$ on the pink $\langle W_{\rm cyc}\rangle-v_{\rm dr}$ curve of the ${\it\Theta}=0.4$ case in the main text Figure 3(C), so that we can make sure that the steady state has already been achieved.} 
 \label{fig:StickSlipsSingleOnecycleFhT04} 
 \end{figure}

\begin{figure}[H]
\centering
 \includegraphics[width=\textwidth]{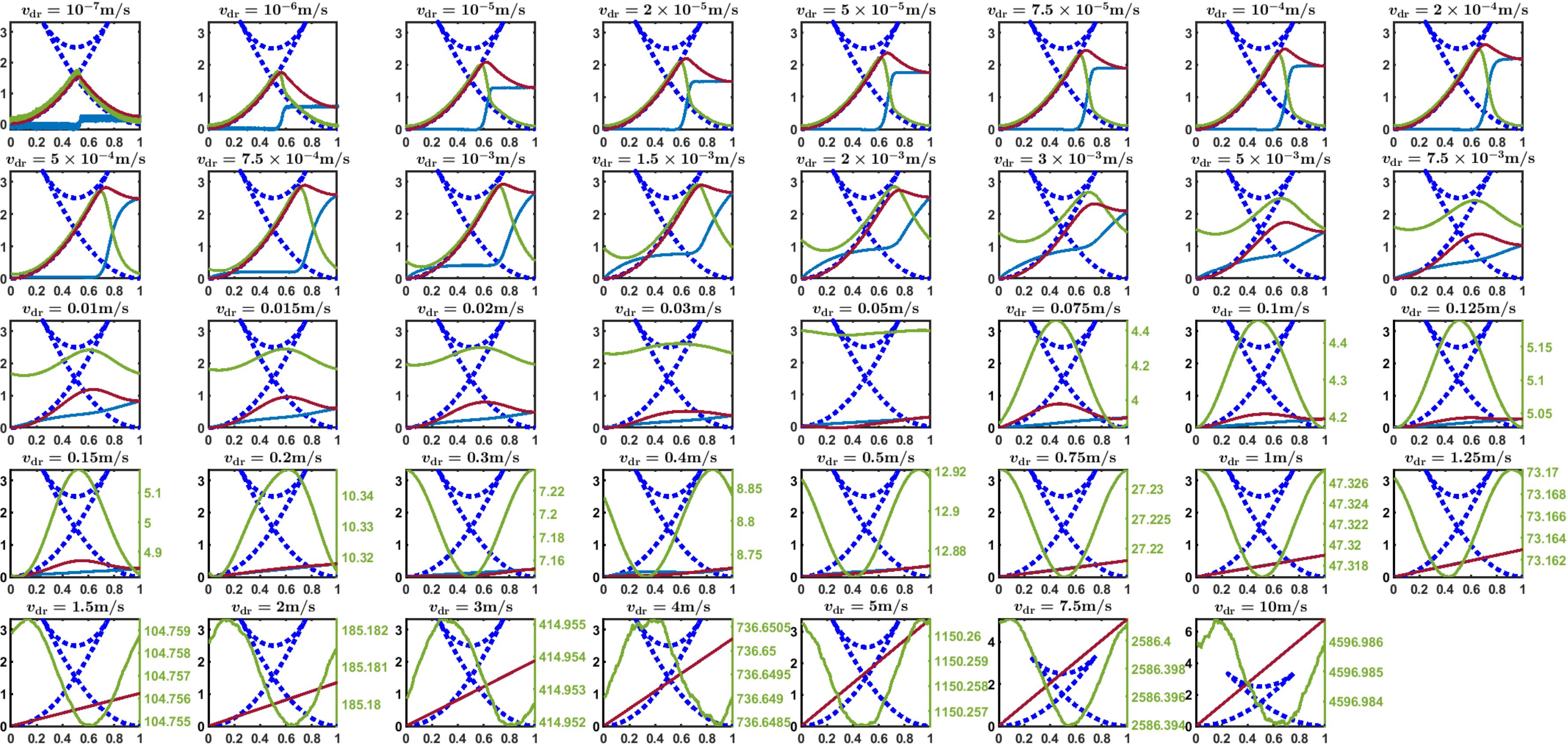}
 \caption{The mean internal energy $\langle U\rangle$ (dark green), the mean heat to the heat bath $\langle Q\rangle$ (dark blue) and the mean work input to the system $\langle W\rangle$ (dark red) during one cycle at different driving velocities of the case of $\eta=3$, $\mu=4\times10^4\rm s^{-1}$ and ${\it\Theta}=0.04$. All the $x$-coordinates are the driver center's nondimensional position $v_{\rm dr}t/a$ relative to the latest cycle starting point, i.e. the same as those in Figure \ref{CyclesTc004}(B) and (C). All the $y$-coordinates and the dotted blue curves are the same as those in Figure \ref{CyclesTc004}(A). Other parameters are given in Sec. \ref{Langevindynamicssimulation}.\ref{ParametersUsed}. The number of simulation cycles at each driving velocity is given in Table \ref{tab:Numberofcycles}. At each driving velocity, all of the simulation cycles are computed sequentially with the initial values of the first simulation cycle inherited from the end values of the last simulation cycle used to calculate the corresponding point at the same $v_{\rm dr}$ on the cyan $\langle W_{\rm cyc}\rangle-v_{\rm dr}$ curve of the ${\it\Theta}=0.04$ case in the main text Figure 3(C), so that we can make sure that the steady state has already been achieved and we don't need to exclude a certain number of cycles at the beginning transient process as we have done when calculating the mean values and standard deviations of $W_{\rm cyc}$ and obtaining the count distributions of $W_{\rm cyc}$ in Figure \ref{WcycDist}, cf. Sec. \ref{Langevindynamicssimulation}.\ref{ParametersUsed}.
\\Because ${\it\Theta}=0.04$ is not a very high temperature, the mean internal energy, heat, and work curves show some similarity to those of the homogeneous zero temperature case in Figure \ref{fig:StickSlipsSingleT0}. The main difference is that the cusp of the mean internal energy curve moves to the left at low driving velocity because of the finite temperature leading to the particle crossing over the middle energy barrier before the instant of the BCP. Comparing with Figure \ref{fig:StickSlipsSingleOnecycleT004}, we can see that the cliffs of the internal energy and heat curves at low driving velocity are no longer that steep after average because of the stochastic instant of the slipping at finite temperature.
} 
 \label{fig:StickSlipsSingleT004} 
 \end{figure}

\begin{figure}[H]
\centering
 \includegraphics[width=\textwidth]{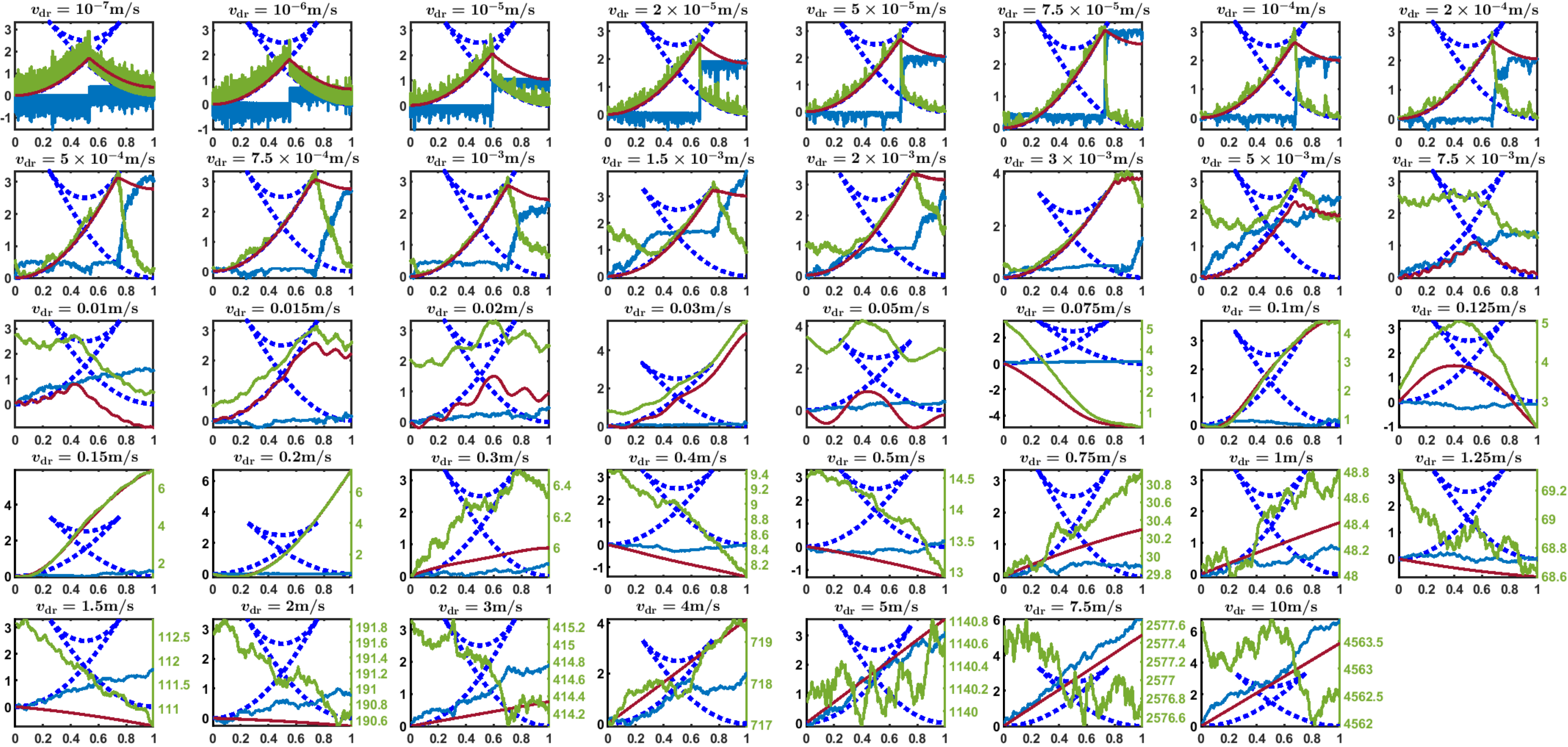}
 \caption{The internal energy $U$ (dark green), the heat to the heat bath $Q$ (dark blue) and the work input to the system $W$ (dark red) during one single cycle at different driving velocities of the case of $\eta=3$, $\mu=4\times10^4\rm s^{-1}$ and ${\it\Theta}=0.04$. All the $x$-coordinates are the driver center's nondimensional position $v_{\rm dr}t/a$ relative to the latest cycle starting point, i.e. the same as those in Figure \ref{CyclesTc004}(B) and (C). All the $y$-coordinates and the dotted blue curves are the same as those in Figure \ref{CyclesTc004}(A). Other parameters are given in Sec. \ref{Langevindynamicssimulation}.\ref{ParametersUsed}. At each driving velocity, the simulation cycle is computed with the initial values inherited from the end values of the last simulation cycle used to calculate the corresponding point at the same $v_{\rm dr}$ on the cyan $\langle W_{\rm cyc}\rangle-v_{\rm dr}$ curve of the ${\it\Theta}=0.04$ case in the main text Figure 3(C), so that we can make sure that the steady state has already been achieved.
\\%For this homogeneous low temperature case, we can see that the cliff of the one single cycle internal energy curve is steep at very low driving velocity because of the oscillating relaxation process is short relative to the driver center as can be seen in Figure \ref{fig:StickSlipsSingleOnecycleXT004}. With the driving velocity increasing, the cliff is less and less steep. 
Although the temperature is not very high, we can see that the stochastic feature of the one single cycle curves is still very distinct especially when the driving velocity is high so that the particle is at nonequilibrium, compared with the mean curves in Figure \ref{fig:StickSlipsSingleT004}. Therefore at finite temperature, we'd better analyze the mean results to eliminate stochasticity.% except in the low driving velocity range, where the particle relatively has enough time to achieve equilibrium.
} 
 \label{fig:StickSlipsSingleOnecycleT004} 
 \end{figure}

\begin{figure}[H]
\centering
 \includegraphics[width=\textwidth]{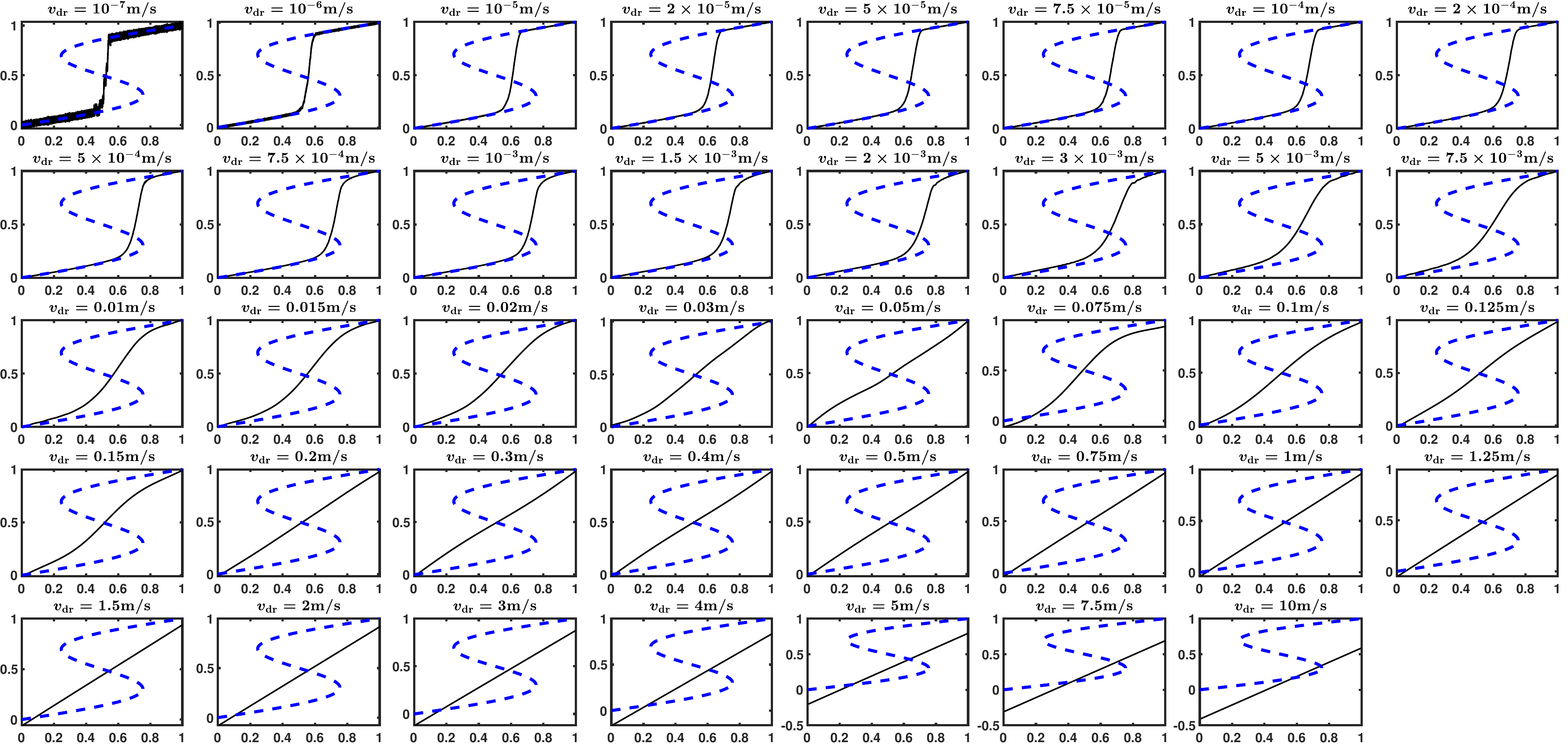}
 \caption{The mean displacement of the particle $\langle x\rangle$ during one cycle at different driving velocities of the case of $\eta=3$, $\mu=4\times10^4\rm s^{-1}$ and ${\it\Theta}=0.04$. All the $x$-coordinates are the driver center's nondimensional position $v_{\rm dr}t/a$ relative to the latest cycle starting point, i.e. the same as that in Figure \ref{CyclesTc004}(B). All the $y$-coordinates and the dashed blue curves are the same as those in Figure \ref{CyclesTc004}(B). Other parameters are given in Sec. \ref{Langevindynamicssimulation}.\ref{ParametersUsed}. The number of simulation cycles at each driving velocity is given in Table \ref{tab:Numberofcycles}. At each driving velocity, all of the simulation cycles are computed sequentially with the initial values of the first simulation cycle inherited from the end values of the last simulation cycle used to calculate the corresponding point at the same $v_{\rm dr}$ on the cyan $\langle W_{\rm cyc}\rangle-v_{\rm dr}$ curve of the ${\it\Theta}=0.04$ case in the main text Figure 3(C), so that we can make sure that the steady state has already been achieved and we don't need to exclude a certain number of cycles at the beginning transient process as we have done when calculating the mean values and standard deviations of $W_{\rm cyc}$ and obtaining the count distributions of $W_{\rm cyc}$ in Figure \ref{WcycDist}, cf. Sec. \ref{Langevindynamicssimulation}.\ref{ParametersUsed}.} 
 \label{fig:StickSlipsSingleXT004} 
 \end{figure}

\begin{figure}[H]
\centering
 \includegraphics[width=\textwidth]{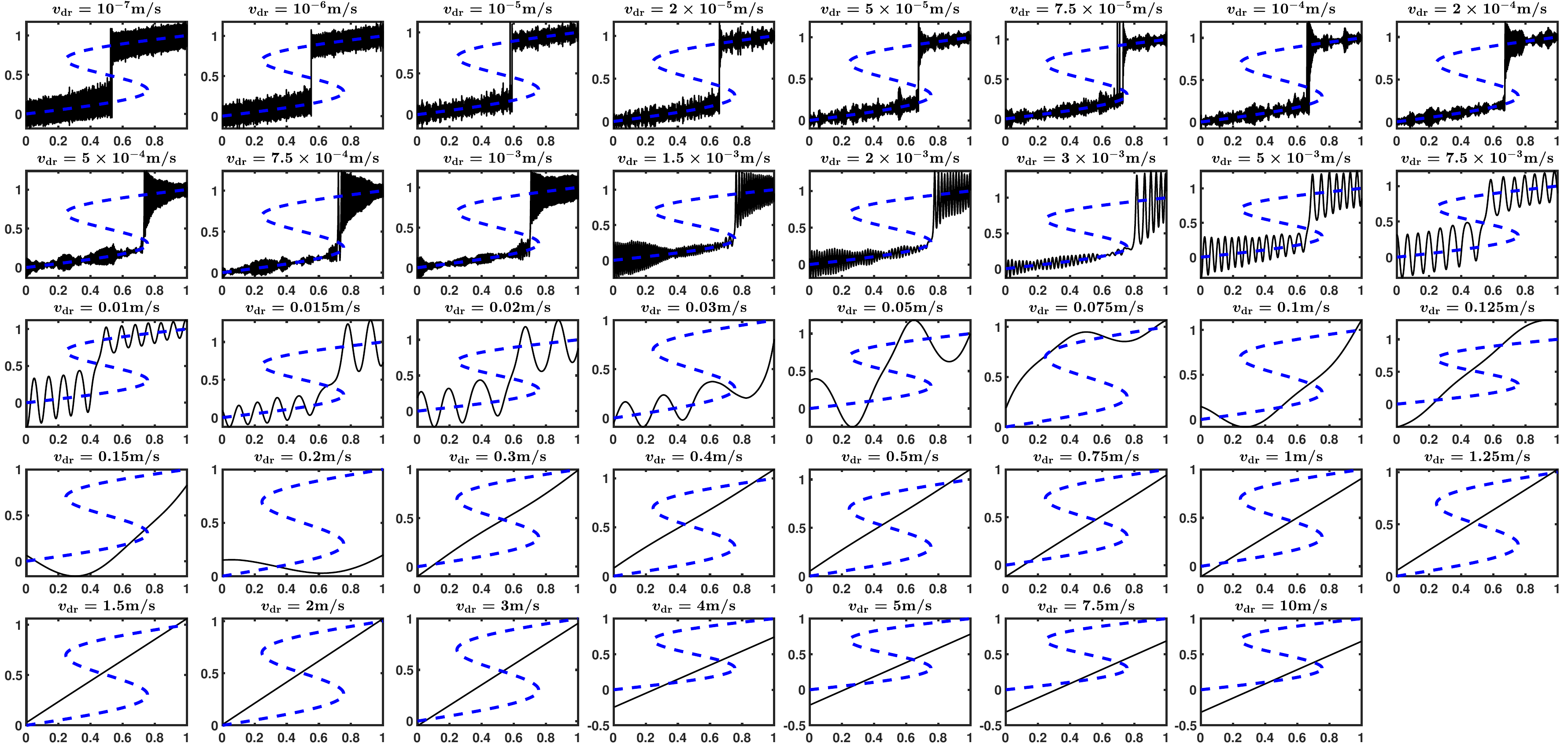}
 \caption{The displacement $x$ during one single cycle at different driving velocities of the case of $\eta=3$, $\mu=4\times10^4\rm s^{-1}$ and ${\it\Theta}=0.04$. All the $x$-coordinates are the driver center's nondimensional position $v_{\rm dr}t/a$ relative to the latest cycle starting point, i.e. the same as that in Figure \ref{CyclesTc004}(B). All the $y$-coordinates and the dashed blue curves are the same as those in Figure \ref{CyclesTc004}(B). Other parameters are given in Sec. \ref{Langevindynamicssimulation}.\ref{ParametersUsed}. At each driving velocity, the simulation cycle is computed with the initial values inherited from the end values of the last simulation cycle used to calculate the corresponding point at the same $v_{\rm dr}$ on the cyan $\langle W_{\rm cyc}\rangle-v_{\rm dr}$ curve of the ${\it\Theta}=0.04$ case in the main text Figure 3(C), so that we can make sure that the steady state has already been achieved.} 
 \label{fig:StickSlipsSingleOnecycleXT004} 
 \end{figure}

\begin{figure}[H]
\centering
 \includegraphics[width=\textwidth]{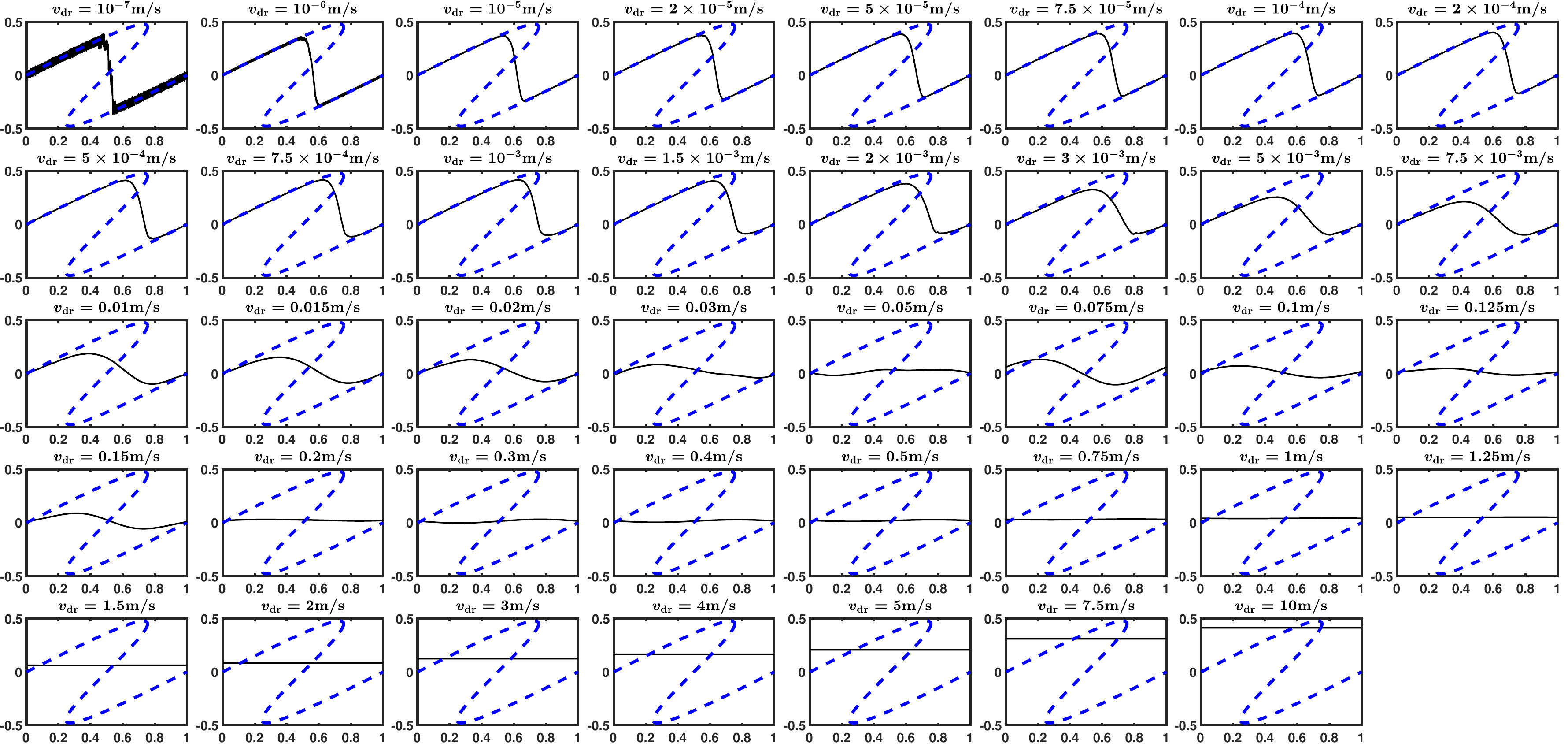}
 \caption{The mean harmonic force $\langle F_{\rm h}\rangle$ during one cycle at different driving velocities of the case of $\eta=3$, $\mu=4\times10^4\rm s^{-1}$ and ${\it\Theta}=0.04$. All the $x$-coordinates are the driver center's nondimensional position $v_{\rm dr}t/a$ relative to the latest cycle starting point, i.e. the same as that in Figure \ref{CyclesTc004}(C). All the $y$-coordinates and the dashed blue curves are the same as those in Figure \ref{CyclesTc004}(C). Other parameters are given in Sec. \ref{Langevindynamicssimulation}.\ref{ParametersUsed}. The number of simulation cycles at each driving velocity is given in Table \ref{tab:Numberofcycles}. At each driving velocity, all of the simulation cycles are computed sequentially with the initial values of the first simulation cycle inherited from the end values of the last simulation cycle used to calculate the corresponding point at the same $v_{\rm dr}$ on the cyan $\langle W_{\rm cyc}\rangle-v_{\rm dr}$ curve of the ${\it\Theta}=0.04$ case in the main text Figure 3(C), so that we can make sure that the steady state has already been achieved and we don't need to exclude a certain number of cycles at the beginning transient process as we have done when calculating the mean values and standard deviations of $W_{\rm cyc}$ and obtaining the count distributions of $W_{\rm cyc}$ in Figure \ref{WcycDist}, cf. Sec. \ref{Langevindynamicssimulation}.\ref{ParametersUsed}.} 
 \label{fig:StickSlipsSingleFhT004} 
 \end{figure}

\begin{figure}[H]
\centering
 \includegraphics[width=\textwidth]{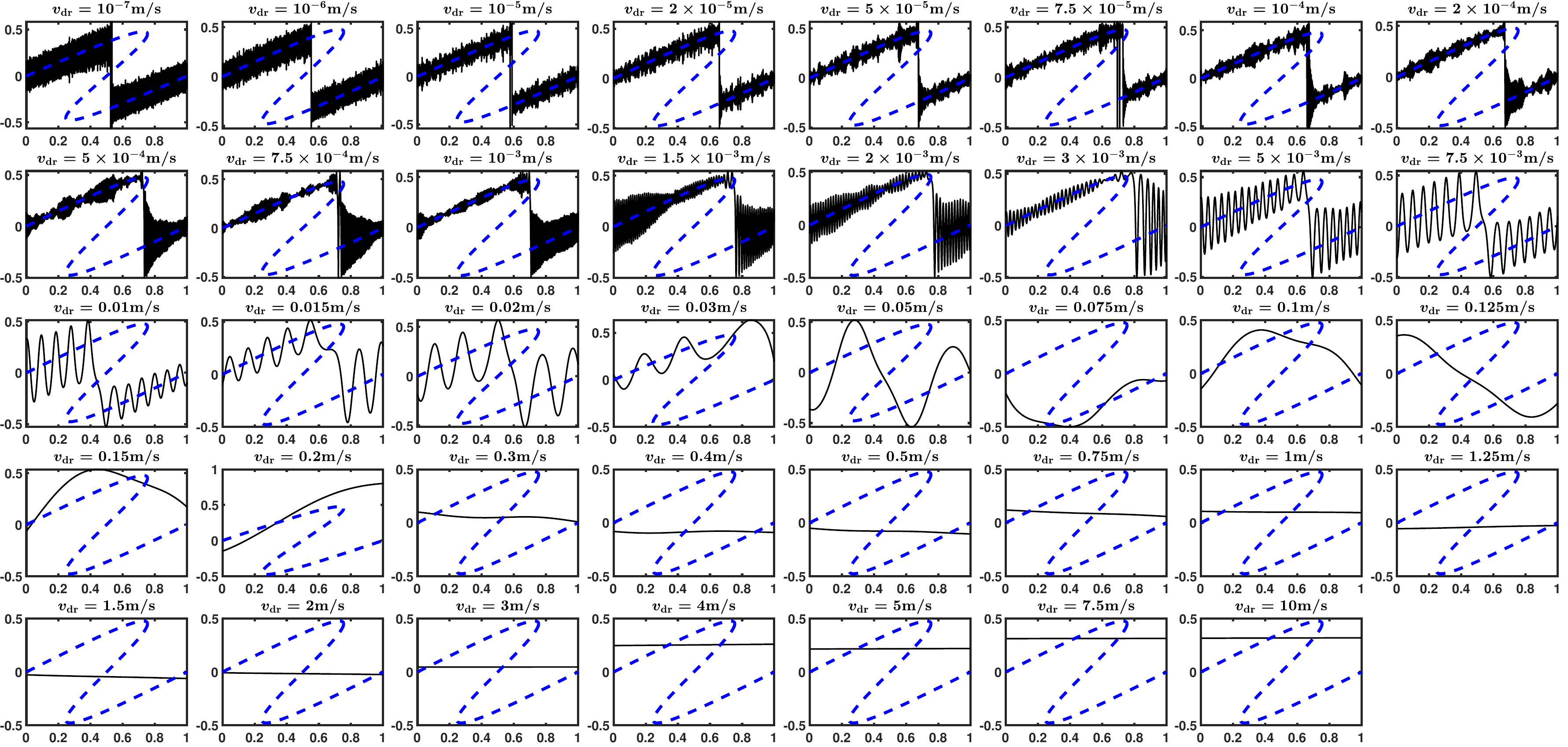}
 \caption{The harmonic force $F_{\rm h}$ during one single cycle at different driving velocities of the case of $\eta=3$, $\mu=4\times10^4\rm s^{-1}$ and ${\it\Theta}=0.04$. All the $x$-coordinates are the driver center's nondimensional position $v_{\rm dr}t/a$ relative to the latest cycle starting point, i.e. the same as that in Figure \ref{CyclesTc004}(C). All the $y$-coordinates and the dashed blue curves are the same as those in Figure \ref{CyclesTc004}(C). Other parameters are given in Sec. \ref{Langevindynamicssimulation}.\ref{ParametersUsed}. At each driving velocity, the simulation cycle is computed with the initial values inherited from the end values of the last simulation cycle used to calculate the corresponding point at the same $v_{\rm dr}$ on the cyan $\langle W_{\rm cyc}\rangle-v_{\rm dr}$ curve of the ${\it\Theta}=0.04$ case in the main text Figure 3(C), so that we can make sure that the steady state has already been achieved.} 
 \label{fig:StickSlipsSingleOnecycleFhT004} 
 \end{figure}

The insets of Figure \ref{fig:StickSlipsTransition} are chosen from Figure \ref{fig:StickSlipsSingleTh04Tc004}, in which the mean internal energy, mean heat and mean work curves at all the driving velocities of the data points on the $\langle W_{\rm cyc}\rangle-v_{\rm dr}$ curve (Table \ref{tab:Numberofcycles}) are plotted. The detail of how to calculate the mean values are given in the caption of this figure. For comparison, in Figure \ref{fig:StickSlipsSingleOnecycleTh04Tc004}, we plot the one single cycle internal energy, heat and work curves at all the simulation driving velocities correspondingly. And we also plot the mean displacement curves (Figure \ref{fig:StickSlipsSingleXTh04Tc004}), the one single cycle displacement curves (Figure \ref{fig:StickSlipsSingleOnecycleXTh04Tc004}), the mean harmonic force curves (Figure \ref{fig:StickSlipsSingleFhTh04Tc004}) and the one single cycle harmonic force curves (Figure \ref{fig:StickSlipsSingleOnecycleFhTh04Tc004}) at all the corresponding simulation driving velocities. These 6 figures are of the case of $\eta=3.0$, $\mu=4\times10^{4}\rm s^{-1}$ and ${\it\Theta}_{\rm h,c}=0.4,0.04$ corresponding to the purple $\langle W_{\rm cyc}\rangle-v_{\rm dr}$ curve in the main text Figure 3(B). We also give the similar $3\times6=18$ figures for the $3$ homogeneous temperature cases: $\eta=3.0$, $\mu=4\times10^{4}\rm s^{-1}$ and ${\it\Theta}=0$ [Figure \ref{fig:StickSlipsSingleT0}, \ref{fig:StickSlipsSingleOnecycleT0}, \ref{fig:StickSlipsSingleXT0}, \ref{fig:StickSlipsSingleOnecycleXT0} \cite{DisplacementCurveT0}, \ref{fig:StickSlipsSingleFhT0} and \ref{fig:StickSlipsSingleOnecycleFhT0}] corresponding to the black curve in the main text Figure 3(C); $\eta=3.0$, $\mu=4\times10^{4}\rm s^{-1}$ and ${\it\Theta}=0.4$ (Figure \ref{fig:StickSlipsSingleT04}, \ref{fig:StickSlipsSingleOnecycleT04}, \ref{fig:StickSlipsSingleXT04}, \ref{fig:StickSlipsSingleOnecycleXT04}, \ref{fig:StickSlipsSingleFhT04} and \ref{fig:StickSlipsSingleOnecycleFhT04}) corresponding to the pink curve in the main text Figure 3(C); $\eta=3.0$, $\mu=4\times10^{4}\rm s^{-1}$ and ${\it\Theta}=0.04$ (Figure \ref{fig:StickSlipsSingleT004}, \ref{fig:StickSlipsSingleOnecycleT004}, \ref{fig:StickSlipsSingleXT004}, \ref{fig:StickSlipsSingleOnecycleXT004}, \ref{fig:StickSlipsSingleFhT004} and \ref{fig:StickSlipsSingleOnecycleFhT004}) corresponding to the cyan curve in the main text Figure 3(C). From these figures, we can have an overall perspective about the transition of the energy, displacement and harmonic force during one cycle with the driving velocity and make it more clear about the reason for the ups and downs, i.e. different regimes such as the stick-slip regime mentioned above and the velocity weakening and resonance regimes explained below, of the $\langle W_{\rm cyc}\rangle-v_{\rm dr}$ curves. %A lot of information can be obtained from the energy, displacement and harmonic force curves at different driving velocities in Figure \ref{fig:StickSlipsSingleTh04Tc004}-\ref{fig:StickSlipsSingleFhT004}, and we mention some of them in the legends of these figures. Besides the ${\it\Theta}_{\rm h,c}=0.4,0.04$, ${\it\Theta}=0$ and ${\it\Theta}=0.04$ cases, we also plot the ${\it\Theta}=0.4$ case for comparison.
A lot of information can be obtained from these figures, and we will mention some in the following paragraphs and some others in the captions of these figures. %Besides the ${\it\Theta}_{\rm h,c}=0.4,0.04$, ${\it\Theta}=0$ and ${\it\Theta}=0.04$ cases, we also plot the ${\it\Theta}=0.4$ case for comparison.

It's not easy to recognize the oscillating relaxation process after slipping from the one single cycle displacement or harmonic force curves in Figure \ref{fig:StickSlipsSingleOnecycleXTh04Tc004} and \ref{fig:StickSlipsSingleOnecycleFhTh04Tc004} because of the stochastic feature of the single trajectories during one cycle resulting from the finite temperature. %${\it\Theta}_{\rm c}$.
We can see that the oscillating relaxation processes at finite temperature is modulated by the stochastic force in Figure \ref{fig:StickSlipsSingleOnecycleXTh04Tc004}, \ref{fig:StickSlipsSingleOnecycleFhTh04Tc004}; Figure \ref{fig:StickSlipsSingleOnecycleXT04}, \ref{fig:StickSlipsSingleOnecycleFhT04} and Figure \ref{fig:StickSlipsSingleOnecycleXT004}, \ref{fig:StickSlipsSingleOnecycleFhT004}. The oscillating relaxation process is more distinct in the homogeneous zero temperature case (${\it\Theta}=0$, Figure \ref{fig:StickSlipsSingleOnecycleXT0} and \ref{fig:StickSlipsSingleOnecycleFhT0}).
% and after being averaged, the relaxation porcess becomes exponential (cf. Figure \ref{fig:StickSlipsSingleTh04Tc004}, \ref{fig:StickSlipsSingleXTh04Tc004}, \ref{fig:StickSlipsSingleFhTh04Tc004} and Figure \ref{fig:StickSlipsSingleT004}, \ref{fig:StickSlipsSingleXT004}, \ref{fig:StickSlipsSingleFhT004}).

From the one single cycle displacement curve of the ${\it\Theta}=0$ case in Figure \ref{fig:StickSlipsSingleOnecycleXT0}, we can see that as the driving velocity increases before $v_{\rm dr}=10^{-3}\rm m/s$, i.e. the black $\langle W_{\rm cyc}\rangle-v_{\rm dr}$ curve's plateau peak point in the main text Figure 3(C), the particle's oscillating relaxation process after slipping gets longer {\it relative to the driver center} and gradually extends to next cycle. And at $v_{\rm dr}=10^{-3}\rm m/s$ the relative oscillating relaxation process is long enough to arrive at the slip instant of the next cycle and with $v_{\rm dr}$ increasing further the slip instant advances to the left. (We want to emphasize that the $x$-coordinate is the driver center's nondimensional position $v_{\rm dr}t/a$, so the displacement curves are relative to the driver center and with the driving velocity $v_{\rm dr}$ increasing, the oscillating relaxation process gets longer relative to the driver center's position.)
%After the plateau peak the relaxation process of last cycle excesses the slip instant and interferes with the new relaxation porcess. 
The residual kinetic energy from the last cycle is equivalent to a finite temperature heat bath and as the driving velocity increases, the increasing residual kinetic energy from the last cycle increases the effective temperature in the stick process so the particle tends to cross over the middle energy barrier earlier, i.e. slip earlier, resulting in the work input reduced, which is similar to the potential mechanism except that the slip instant is after the middle instant of the cycle. At the same time, the higher effective temperature also leads to the work input reduced further because of the thermolubricity mechanism. %The increasing effective temperature due to the increasing $v_{\rm dr}$ makes the cusp become dull. 
On the other hand, the kinetic energy from slipping also results in the increasing effective temperature immediately after slipping so that the work output after the slip instant is also reduced a little due to the negative thermolubricity mechanism, causing the net cycle work input to recover a little from the reduction before the slip instant. Nonetheless, as a whole $\langle W_{\rm cyc}\rangle$ reduces, and after the plateau peak point of the $\langle W_{\rm cyc}\rangle-v_{\rm dr}$ curve,
%the particle cannot relax to equilibrium in one engine cycle and the residual kinetic energy from the last cycle make the particle thermolubricted before the cusp and the input work is reduced. So 
the mean cycle work curve decreases with the driving velocity increasing. This velocity range is the so called velocity weakening regime \cite{VelocityWeakening,NPVelocityTuning}. 
%
%Incidentally, in Figure \ref{fig:StickSlipsSingleOnecycleXT0} we can see that the increasing effective temperature due to the increasing $v_{\rm dr}$ makes the cusp more and more dull.
\\
\indent As the driving velocity continues to increase so that $v_{\rm dr}/a$ is comparative to the frequency of the particle oscillating around the resultant potential local or global minimum points, the energy, displacement and harmonic force curves are odd (cf. Figure \ref{fig:StickSlipsSingleTh04Tc004}-\ref{fig:StickSlipsSingleFhT004}) due to the nonlinearity of the PT model, cf. the main text subsection ``Bifurcation of Mean Cycle Work with Driving Velocity as Parameter at Zero Temperature''. In this regime resonance occurs and we will refer to it as the resonance regime. After the resonance regime the mean cycle work curve increases linearly due to the so large damping force, cf. the main text Figure 3(A) and Sec. \ref{sec:apndRD}.\ref{sec:hvlim}.

% \begin{figure}[H]
% \centering
%\includegraphics[width=\textwidth]{SFigures/StickSlips_single_Th04Tc004.pdf}\\
%\includegraphics[width=\textwidth]{SFigures/StickSlips_single_onecycle_Th04Tc004.pdf}
%\caption{}
%\label{fig:StickSlipsEnergy}
% \end{figure}
% \begin{figure}[H]
% \centering
%\includegraphics[width=\textwidth]{SFigures/StickSlips_single_x_Th04Tc004.pdf}\\
%\includegraphics[width=\textwidth]{SFigures/StickSlips_single_onecycle_x_Th04Tc004.pdf}
%\caption{}
%\label{fig:StickSlipsx}
% \end{figure}
% \begin{figure}[H]
% \centering
%\includegraphics[width=\textwidth]{SFigures/StickSlips_single_Fh_Th04Tc004.pdf}\\
%\includegraphics[width=\textwidth]{SFigures/StickSlips_single_onecycle_Fh_Th04Tc004.pdf}
%\caption{}
%\label{fig:StickSlipsFh}
% \end{figure}

\subsection{The count distribution of $W_{\rm cyc}$ and its standard deviation}
\label{sec:WcycDist}
 \begin{figure}[H]
 \centering
\includegraphics[width=\textwidth]{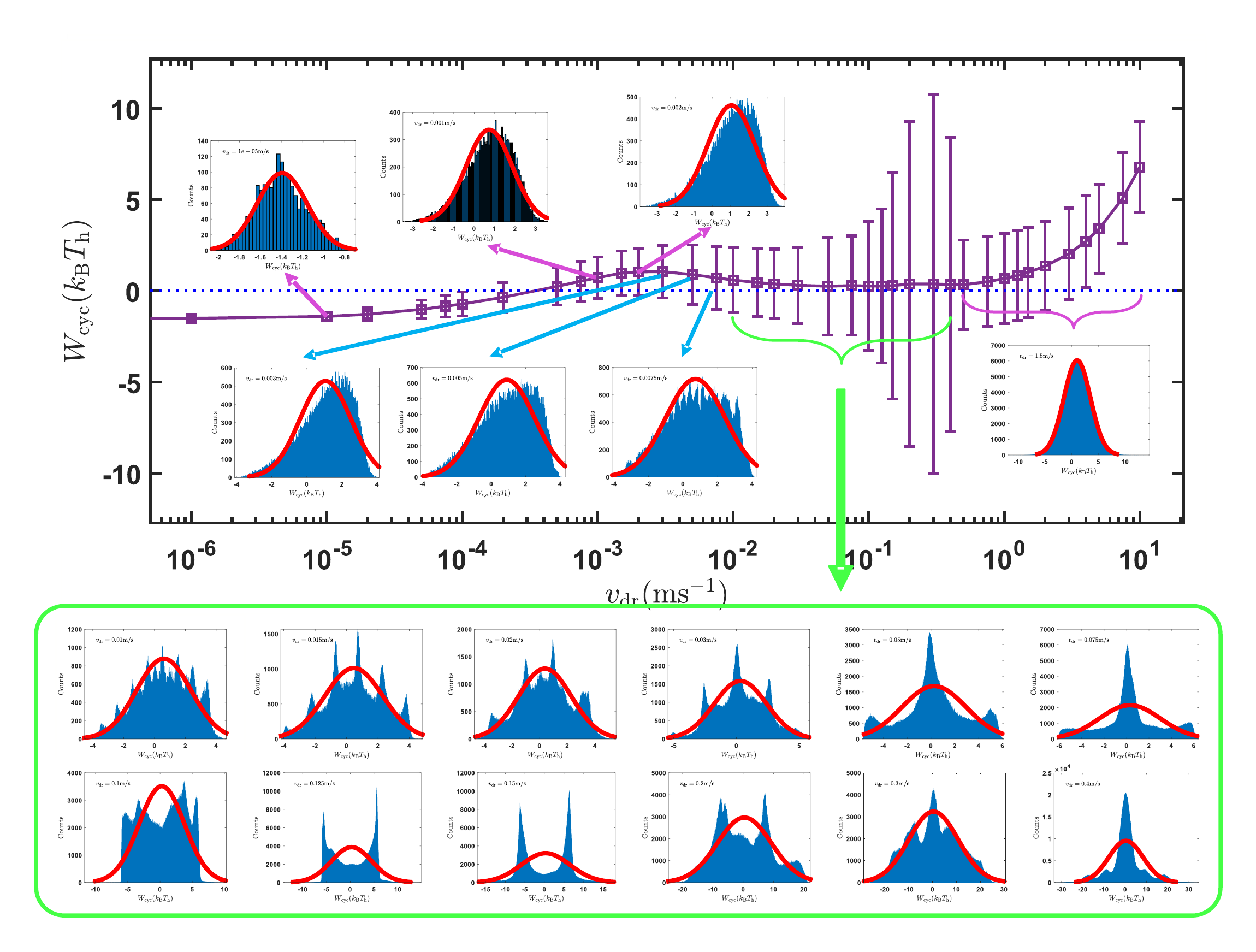}
\caption{The count distributions of $W_{\rm cyc}$ at different $v_{\rm dr}$'s of the case of $\eta=3$, $\mu=4\times10^4\rm s^{-1}$ and ${\it\Theta}_{\rm h,c}=0.4,0.04$ in the main text Figure 3(B). The red curve is a normal distribution fit. From the low driving velocity end to the high driving velocity limit, the distribution of $W_{\rm cyc}$ is first normal and then has several peaks and finally becomes normal again. When $v_{\rm dr}\in(10^{-2},4\times10^{-1})\rm m/s$, the distributions of $W_{\rm cyc}$ at different $v_{\rm dr}$'s are exotic. The sample of simulation engine cycles to obtain the count distribution of $W_{\rm cyc}$ at each driving velocity is the same as the sample to calculate the mean values and standard deviations of $W_{\rm cyc}$ at the same driving velocity, cf. Sec. \ref{Langevindynamicssimulation}.\ref{ParametersUsed}. Incidentally, the count distribution of $W_{\rm cyc}$ in the main text Figure 3(C1) and (C2) are inherited from here.}
\label{WcycDist}
 \end{figure}
 
In Figure \ref{WcycDist}, we plot the count distributions of $W_{\rm cyc}$ at different driving velocity $v_{\rm dr}$'s. In the high and low driving velocity end, the distributions are normal while in the middle driving velocity range the distributions are no longer normal and have several peaks especially in $(10^{-2},4\times10^{-1})\rm m/s$, indicating the periodicity and multiplicity of $W_{\rm cyc}$, cf. the main text Figure 3(C2) and (C3) and Sec. \ref{sec:apndRD}.\ref{sec:mulsolT0}. 
%\subsection{The standard deviations of $W_{\rm cyc}$ in the homogeneous high and low temperature cases}

\begin{figure}[H]
\centering
\includegraphics[width=0.5\textwidth]{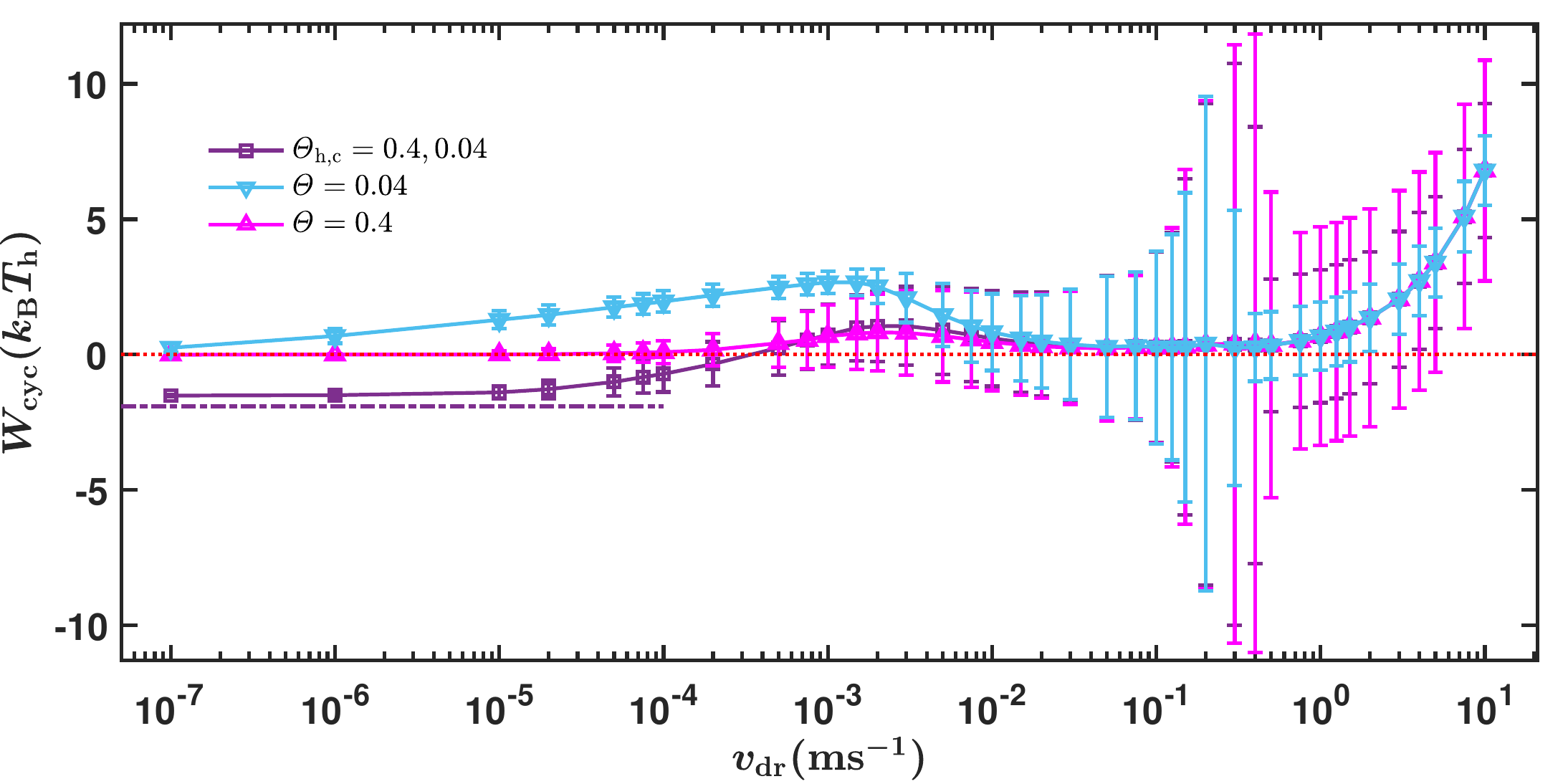}
\caption{The standard deviations of $W_{\rm cyc}$ in the homogeneous high (${\it\Theta}=0.4$) and low (${\it\Theta}=0.04$) temperature case. The ${\it\Theta}_{\rm h,c}=0.4,0.04$ case is plotted for comparison. The parameters are the same as those of the three corresponding cases in the main text Figure 3(C). We can see that in all the three cases, the standard deviation of $W_{\rm cyc}$ has a peak at a certain $v_{\rm dr}$ dependent on the specific temperature field. The nearly constant standard deviation at the high velocity end also depends on the temperature.} 
\label{fig:Wcyc_T004T04_Error}
\end{figure}

In Figure \ref{fig:Wcyc_T004T04_Error}, the standard deviations of $W_{\rm cyc}$ in the homogeneous high (${\it\Theta}=0.4$) and low (${\it\Theta}=0.04$) temperature cases are plotted with the ${\it\Theta}_{\rm h,c}=0.4,0.04$ case for comparison. We can see that at very low driving velocity, the standard deviations of the three cases are all very small, indicating the equilibrium at each instant and thus the reversibility of the cycles. 

With the driving velocity increasing, the standard deviation gradually goes up to a maximum when stochastic resonance occurs \cite{StochasticResonance}. After the maximum, the standard deviation goes down to a constant although the driving velocity is still increasing to the infinity. At the high driving velocity end, the stochastic noise (which is Gaussian white noise) is of larger time scale than the driver center (and the particle) moving over a lattice period, resulting in that the cycle work $W_{\rm cyc}$ has a normal distribution with constant standard deviations at increasing driving velocities. On the other hand, at the low driving velocity end, the stochastic noise is of smaller time scale than the driver center (and the particle) moving over a lattice period, so that at each instant of one cycle the particle relaxes to qusi-equilibrium and the ensemble average of the cycles, i.e. $\langle W_{\rm cyc}\rangle$, is nearly identical to the time average of one cycle, i.e. $W_{\rm cyc}$, cf. the main text subsection ``The Second Mechanism of Work Output and the Consequence of Its Excess''. As $v_{\rm dr}\rightarrow0$, $W_{\rm cyc}\rightarrow\langle W_{\rm cyc}\rangle$ and the standard deviation goes to zero, so the distribution of $W_{\rm cyc}$ goes to an one-point distribution. In the mediate driving velocity range, the time scale of the stochastic noise is of the same order as the time scale for the driver center (and the particle) to move over one lattice period so that the particle's movement synchronizes with the stochastic noise and stochastic resonance occurs at a certain $v_{\rm dr}$, leading to the standard deviation of $W_{\rm cyc}$ maximizing. The resonant driving velocity is affected by the temperature field and 
%it is noteworthy to investigate the stochastic reaonance of the PT model system in depth.
it is a noteworthy problem to seek for an analytical solution for the standard deviation of $W_{\rm cyc}$ varying with $v_{\rm dr}$.

% This mechanism can be used to design new kind of velocity sensors based and is worth being explored further.

\subsection{The multiple solutions of the zero temperature Langevin equation}
\label{sec:mulsolT0}

\begin{figure}[H]
\centering
 \begin{minipage}{0.329\textwidth}
 \centerline{
\includegraphics[width=\textwidth]{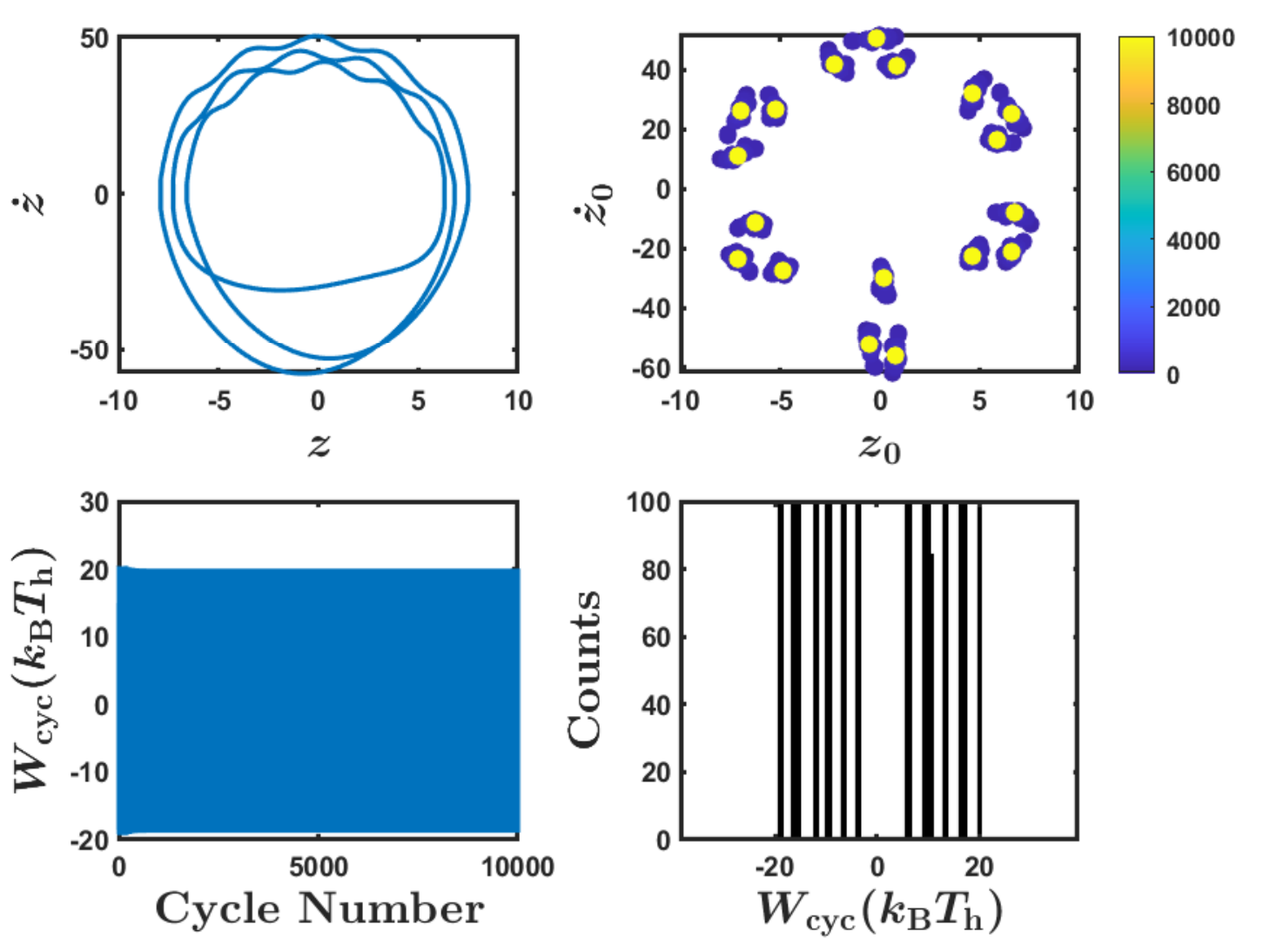}
 }
 \centerline{(a) $z(0)=-8,\dot z(0)=10$}
 \end{minipage}
 \begin{minipage}{0.329\textwidth}
 \centerline{
\includegraphics[width=\textwidth]{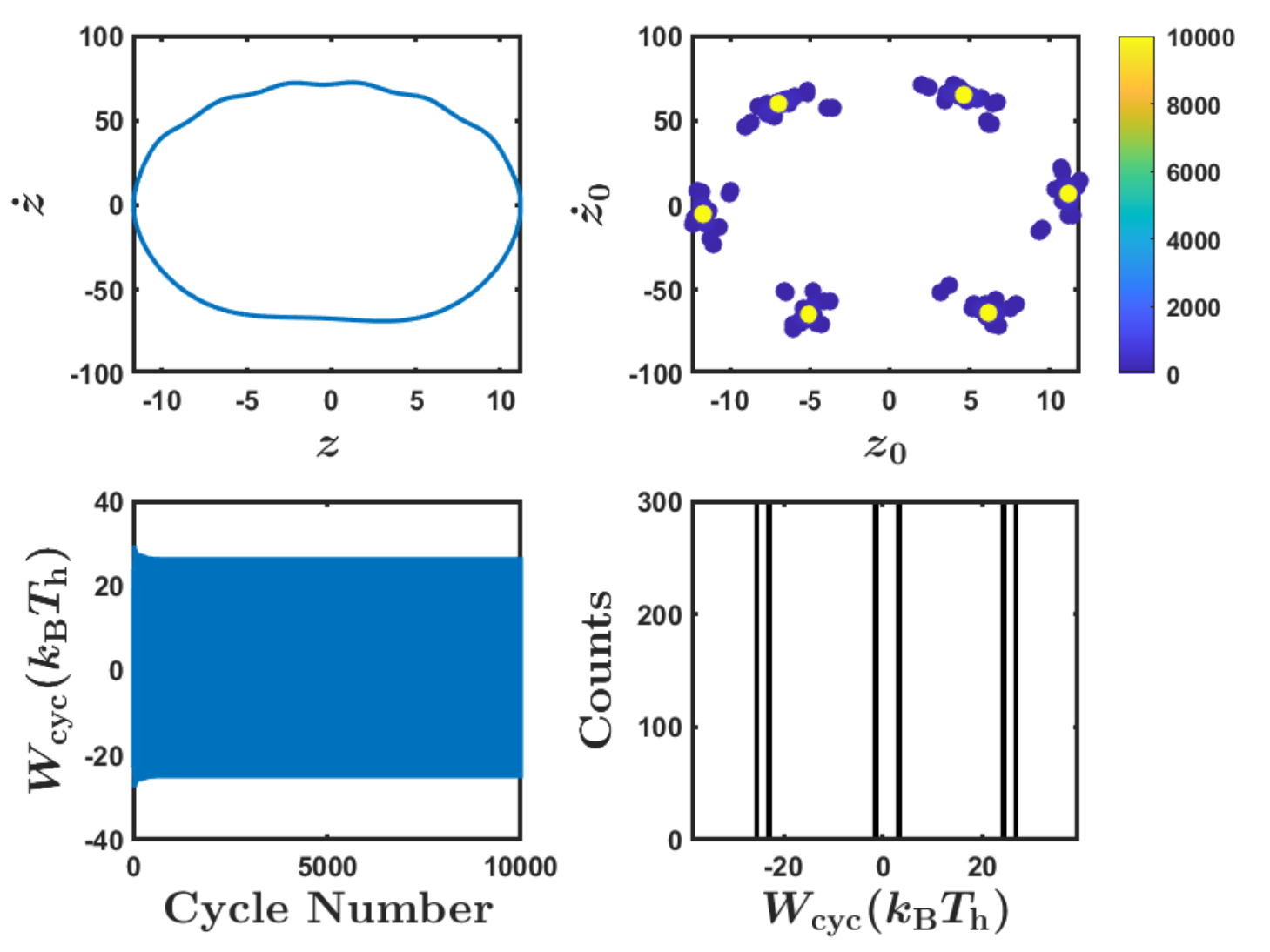}
 }
 \centerline{(b) $z(0)=-10,\dot z(0)=9$}
 \end{minipage}
 \begin{minipage}{0.329\textwidth}
 \centerline{
\includegraphics[width=\textwidth]{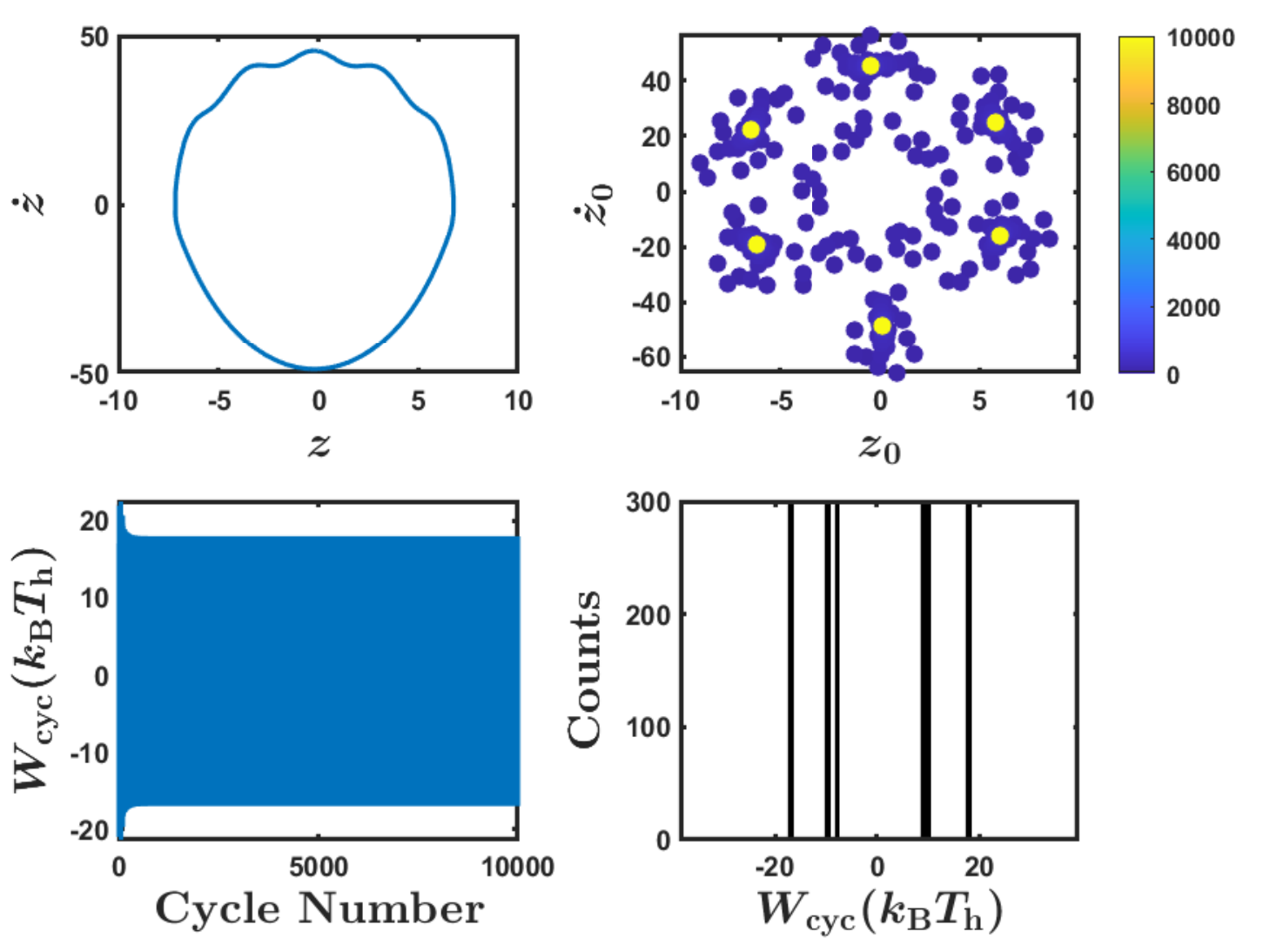}
 }
 \centerline{(c) $z(0)=-9,\dot z(0)=10$}
 \end{minipage}\\
 \begin{minipage}{0.329\textwidth}
 \centerline{
\includegraphics[width=\textwidth]{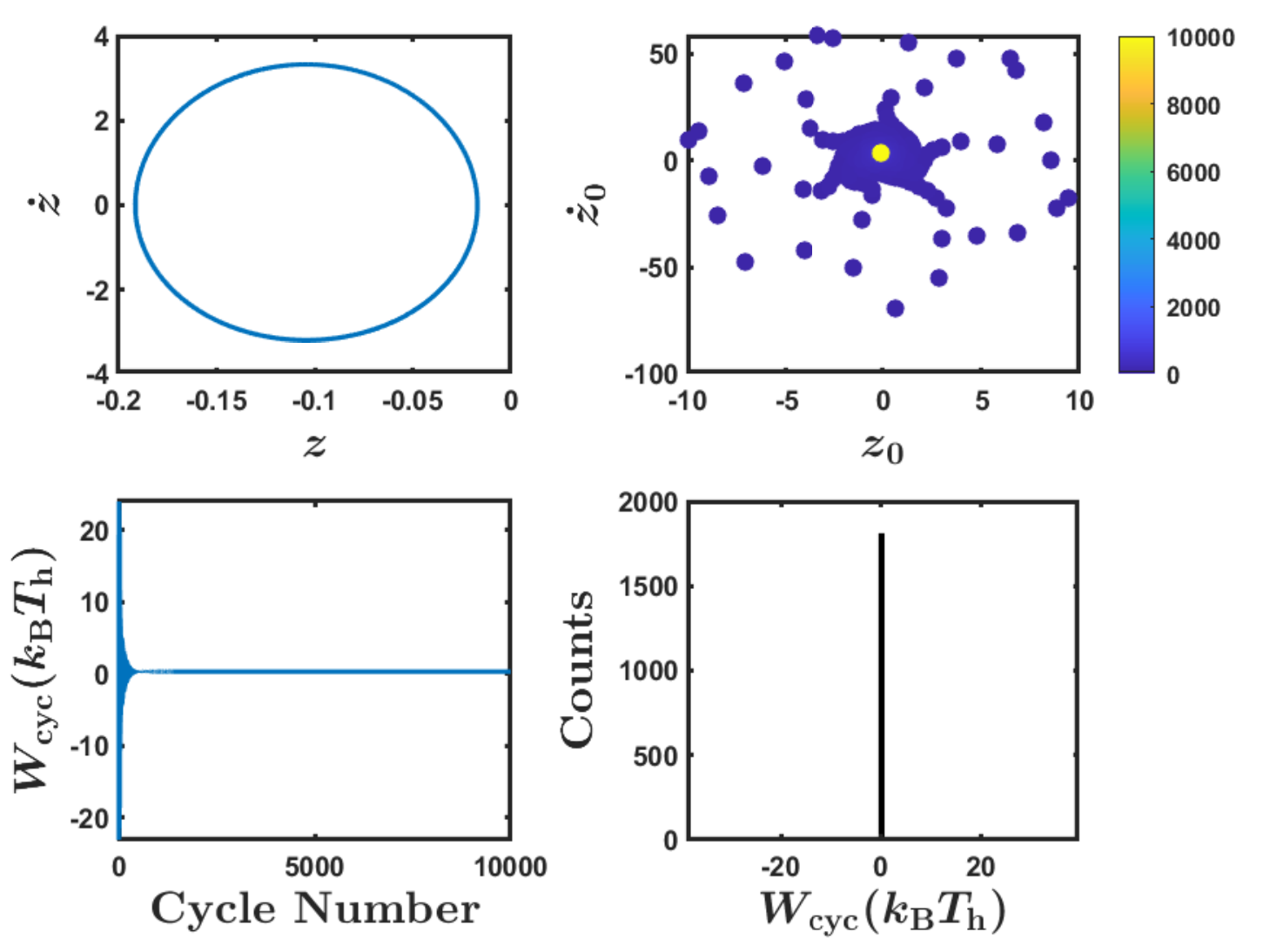}
 }
 \centerline{(d) $z(0)=-10,\dot z(0)=10$}
 \end{minipage}
 \begin{minipage}{0.329\textwidth}
 \centerline{
\includegraphics[width=\textwidth]{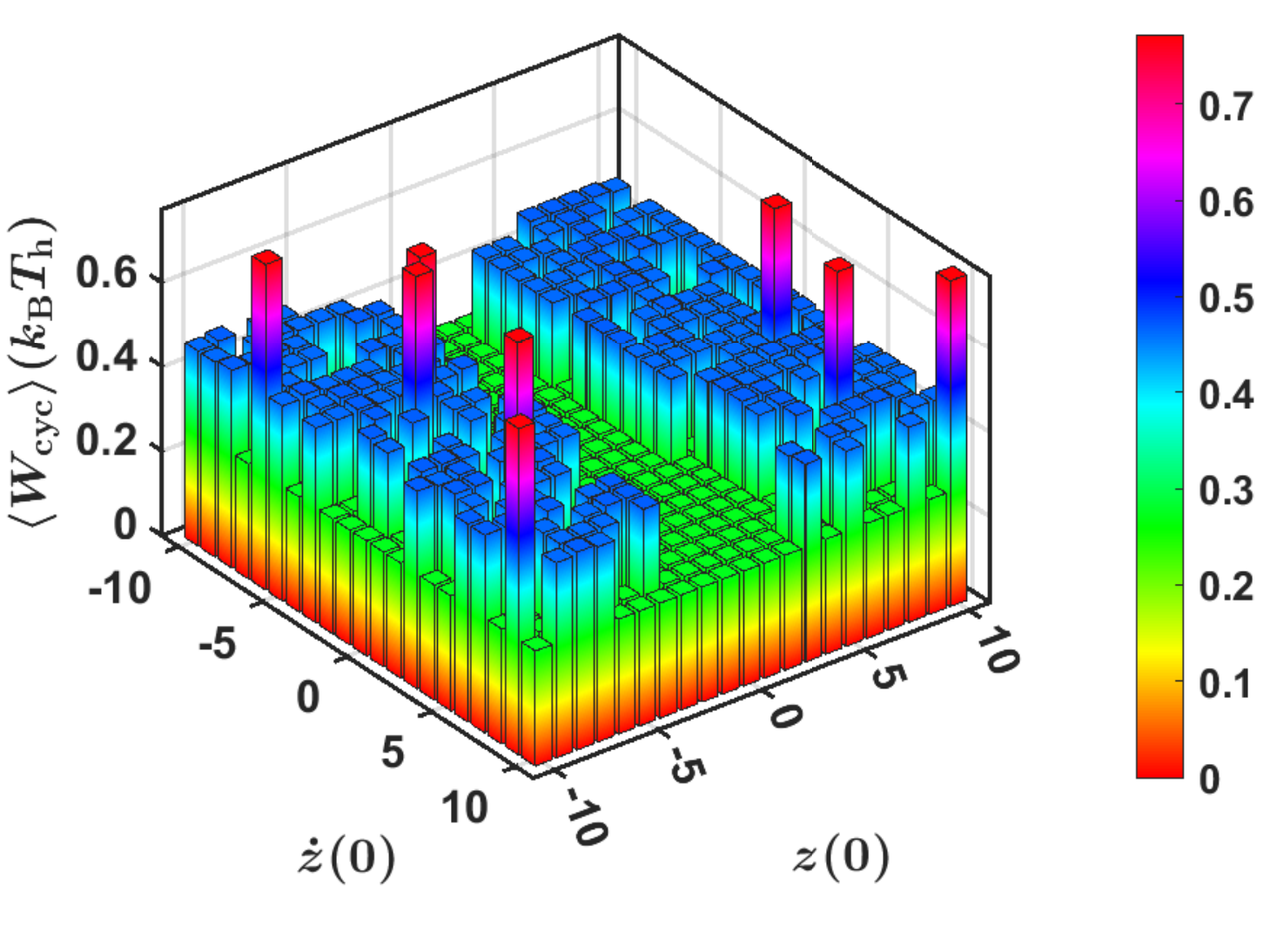}
 }
 \centerline{(e)}
 \end{minipage}
 \begin{minipage}{0.329\textwidth}
 \centerline{
\includegraphics[width=\textwidth]{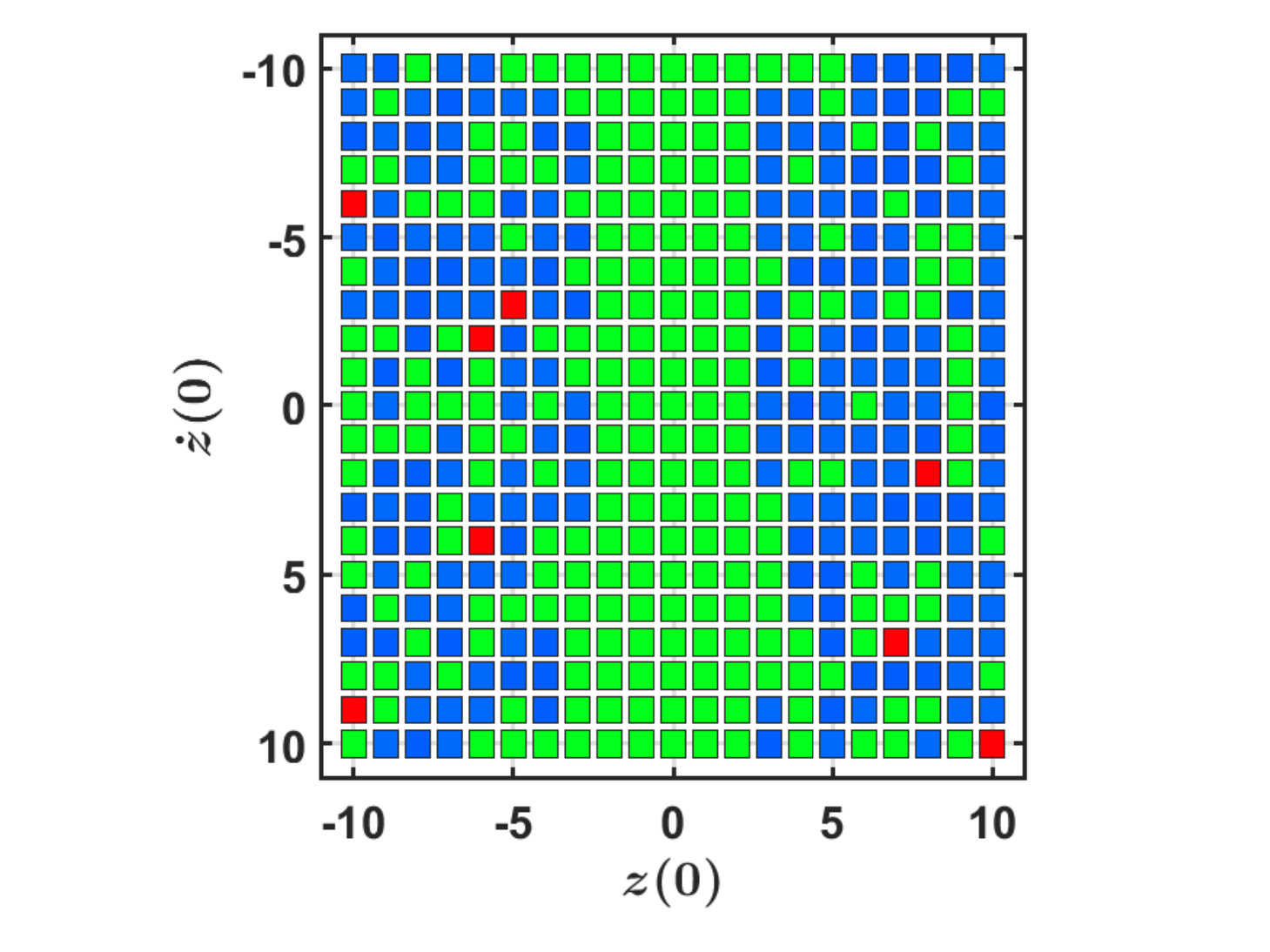}
 }
 \centerline{(f)}
 \end{minipage}

\caption{Different solutions of the zero temperature Langevin equation at $v_{\rm dr}=0.4\rm m/s$, $\eta=3.0$ and $\mu=4\times10^{4}\rm s^{-1}$. (a), (b), (c) and (d) Four different solutions of different initial conditions. In each of the four subfigures, the top left subgraph is the limit cycle on the $z-\dot z$ phase plane. The bottom left subgraph is the cycle work from the first to the last simulation cycle during the simulation time range. The bottom right subgraph is the count distribution of $W_{\rm cyc}$ of the last 1800 simulation cycles. The top right subgraph is the Poincare (or stroboscopic) map \cite{Seydel2010Ch7Stability} sampled at the starting point of each cycle, i.e. the set of the intial point $(z_0,\dot z_0)$ of each cycle with the cycle number indicated by the color of the points ranging from blue at the beginning to yellow at the end. (e) The mean cycle work $\langle W_{\rm cyc}\rangle$ corresponding to different initial conditions $(z(0),\dot z(0))$, which is different from the initial point $(z_0,\dot z_0)$ of each cycle. (f) The airview of (e). Other parameters are given in Sec. \ref{Langevindynamicssimulation}.\ref{ParametersUsed}. The main text Figure 3(C3) is inherited from (a), (b), (c) and (d) in this figure.}
\label{multiplesolutions0p4omgorgn}
\end{figure}

The nondimensional Langevin equation at zero temperature is an ordinary differential equation 
\begin{equation}
\ddot{z}+\beta\eta\dot z+4\pi^2(z-\tilde v\tau)-4\pi^2\eta\sin z=0,
\end{equation}
and is equivalent to the system
\begin{equation}\label{ODE}
\begin{cases}
\frac{\mathrm dz}{\mathrm d\tau}=\dot z,\\
\frac{\mathrm d\dot z}{\mathrm d\tau}=-\beta\eta\dot z-4\pi^2(z-\tilde{v}\tau)+4\pi^2\eta\sin z.
\end{cases}
\end{equation}
Substitute $y_1=z-\tilde{v}\tau,\ y_2=\dot z$ into Eq. \ref{ODE}, we obtain 
\begin{equation}\label{ODE_variablechanged}
\begin{cases}
\frac{\mathrm dy_1}{\mathrm d\tau}=y_2,\\
\frac{\mathrm dy_2}{\mathrm d\tau}=-\beta\eta y_2-4\pi^2y_1+4\pi^2\eta\sin(y_1+\tilde{v}\tau).
\end{cases}
\end{equation}
It's clearer now that this equation system has periodic solution from the last term on the right hand side of the second equation. Using the standard integrator ode113 in Matlab, which is of lower cost and higher precision than ode45 based on an explicit Runge-Kutta (4,5) formula \cite{MatlabODE}, we can obtain these periodic solutions. 

In Figure \ref{multiplesolutions0p4omgorgn}, four solutions at $v_{\rm dr}=0.4\rm m/s$ are plotted and we can see that different initial conditions lead to different limit cycles with different periods and different mean cycle work $\langle W_{\rm cyc}\rangle$. Solutions at other driving velocities are represented by the red circles in Figure \ref{BifurcationFigures} and the main text Figure 6. The mean cycle work $\langle W_{\rm cyc}\rangle$ is calculated from the last cycles of a number of multiple $P_{\rm c.n.}$'s (Eq. \ref{eqn:cyclenumperiod}) if the steady state solution is periodic. If the steady state solution is nonperiodic (including the cases in which we haven't achieved the steady state solution after a too long simulation time), the $\langle W_{\rm cyc}\rangle$ is calculated from the last cycles of a presumably adequate number.

\begin{figure}[H]
\centering
 \begin{minipage}{0.329\textwidth}
 \centerline{
\includegraphics[width=\textwidth]{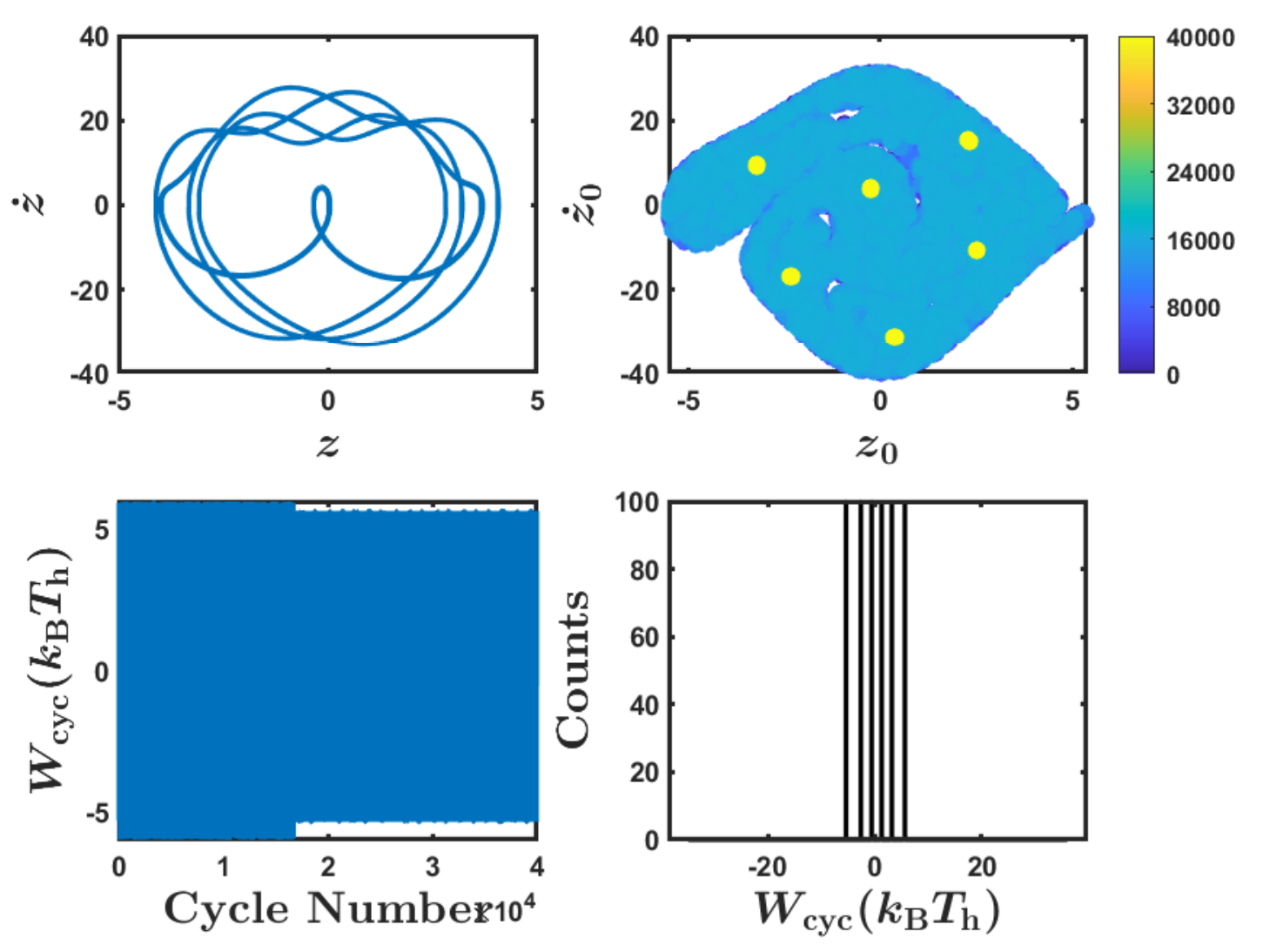}
 }
 \centerline{(a) $z(0)=4,\dot z(0)=-7$}
 \end{minipage}
 \begin{minipage}{0.329\textwidth}
 \centerline{
\includegraphics[width=\textwidth]{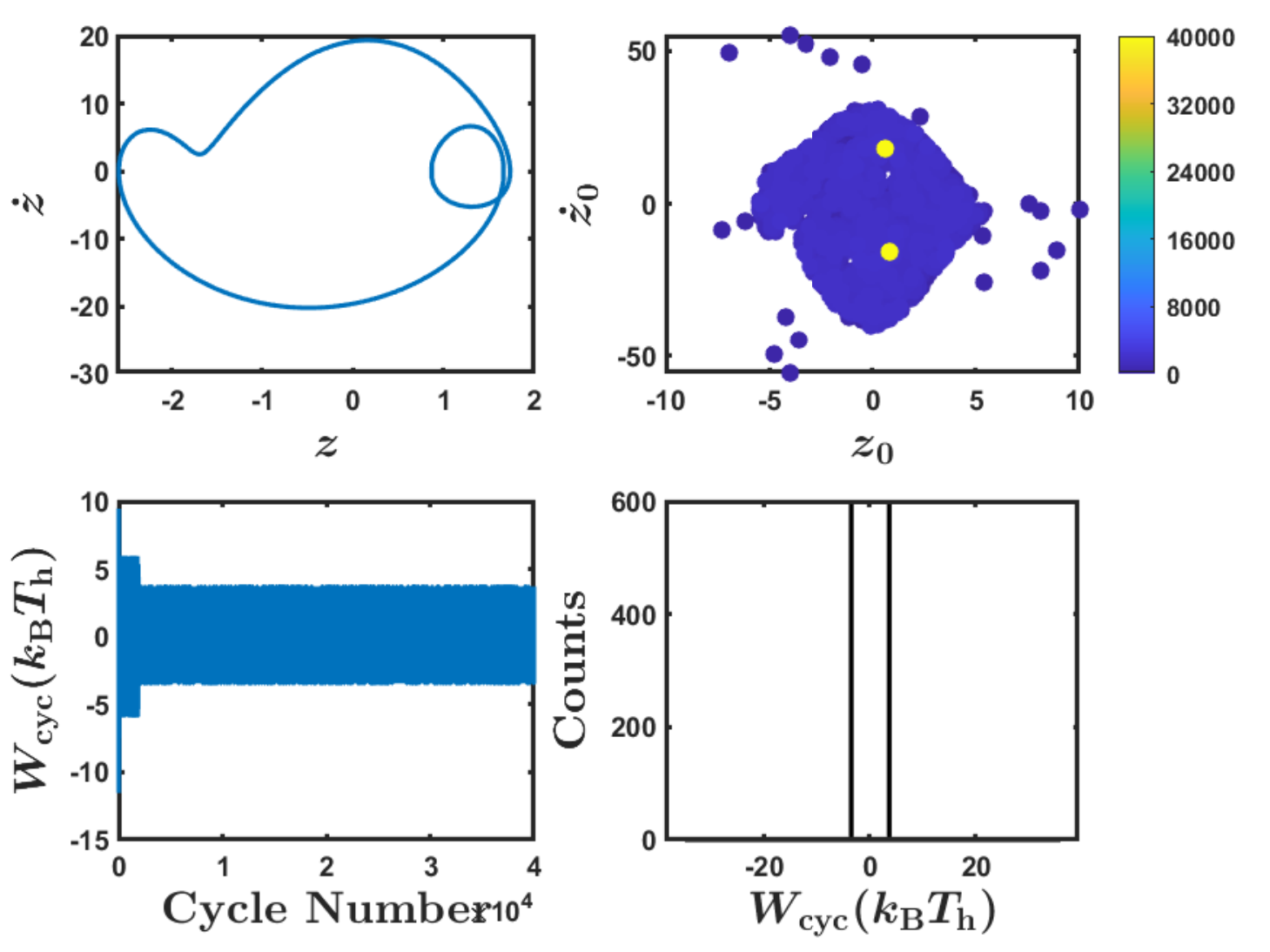}
 }
 \centerline{(b) $z(0)=10,\dot z(0)=-2$}
 \end{minipage}
 \begin{minipage}{0.329\textwidth}
 \centerline{
\includegraphics[width=\textwidth]{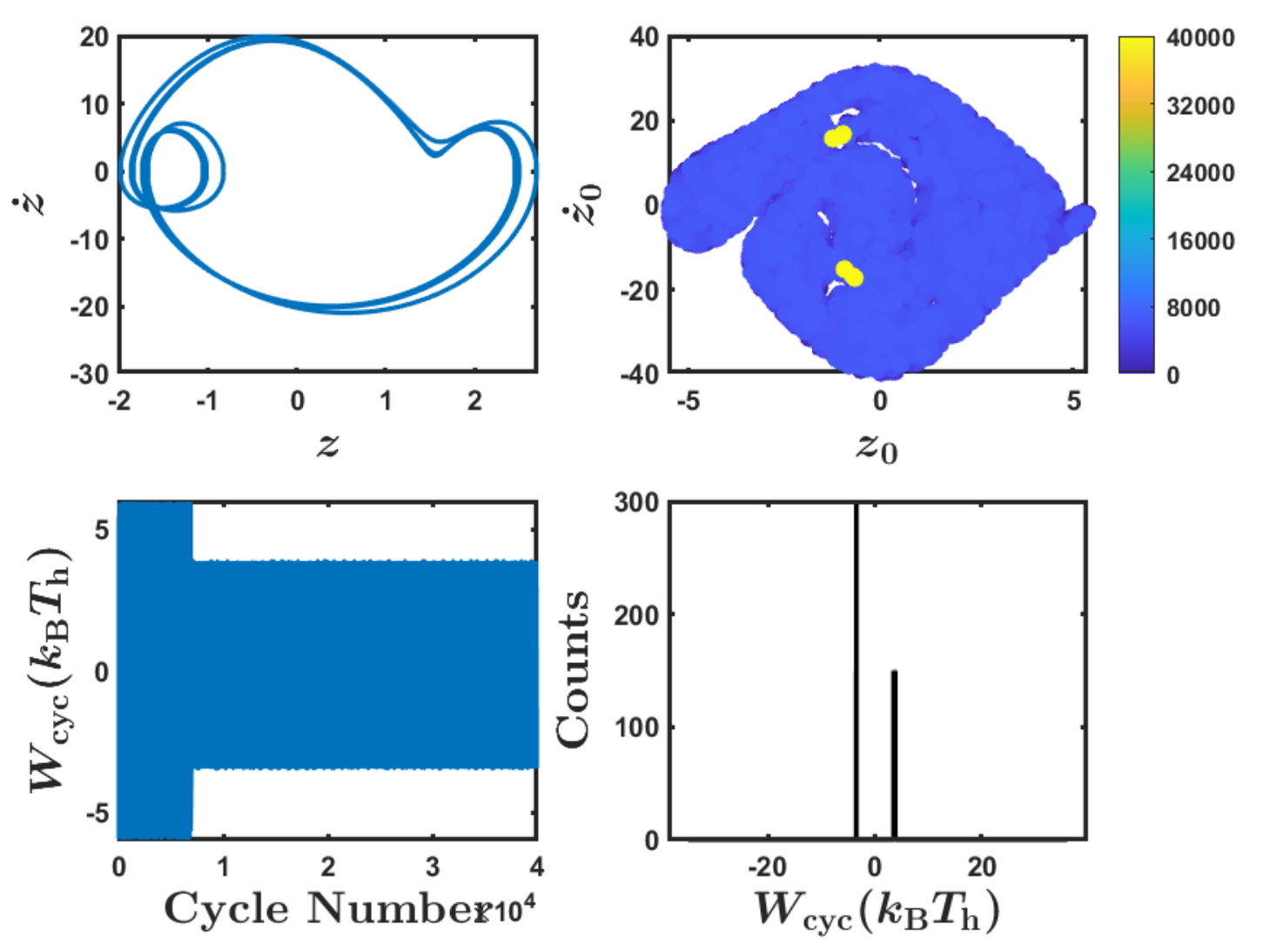}
 }
 \centerline{(c) $z(0)=4,\dot z(0)=3$}
 \end{minipage}\\
 \begin{minipage}{0.329\textwidth}
 \centerline{
\includegraphics[width=\textwidth]{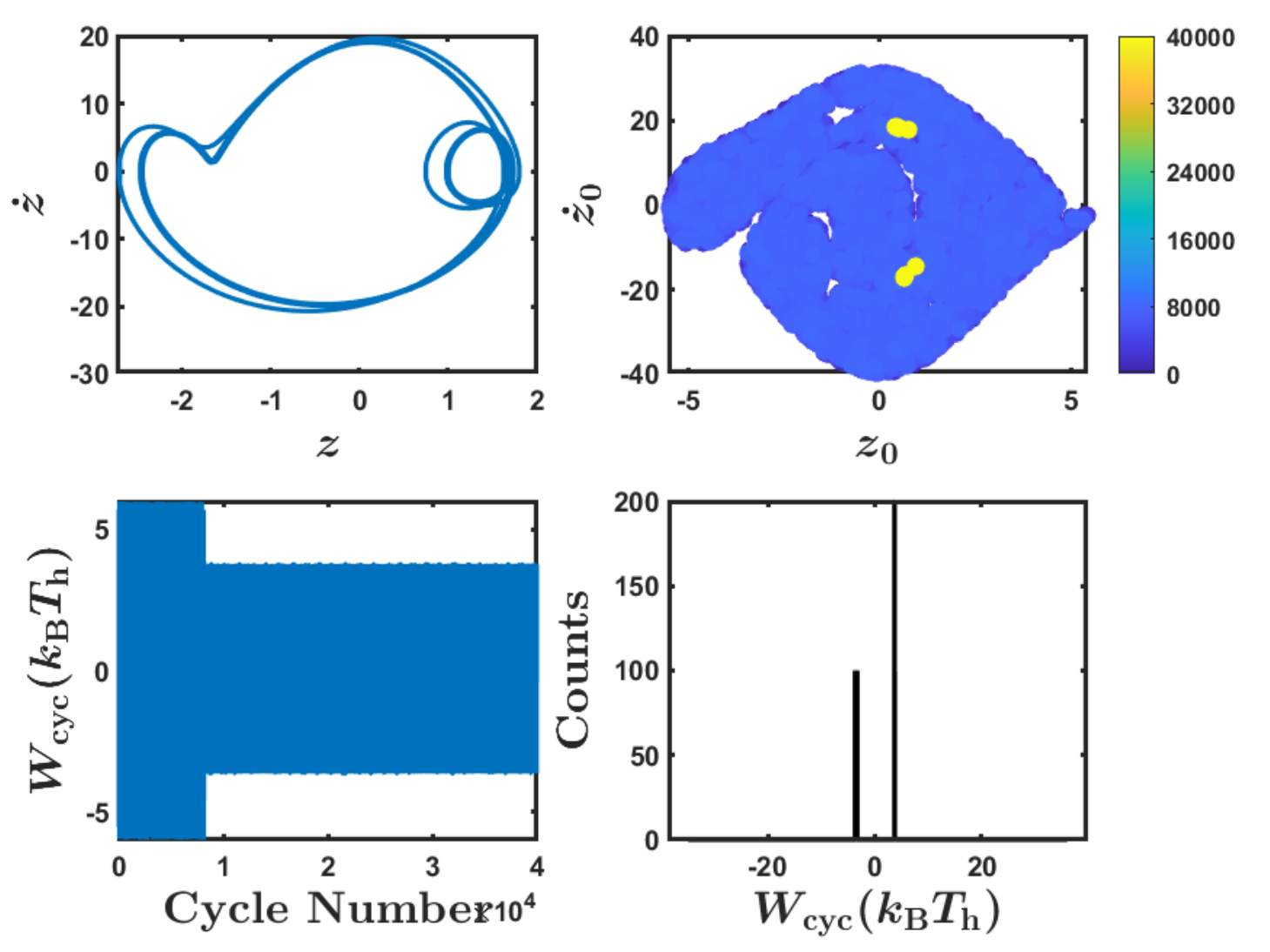}
 }
 \centerline{(d) $z(0)=-4,\dot z(0)=-5$}
 \end{minipage}
% \begin{minipage}{0.329\textwidth}
% \centerline{
%\includegraphics[width=\textwidth]{SFigures/MultipleSolutionsv01_m9_0_omgorgn.pdf}
% }
% \centerline{(e) $z(0)=-9,\dot z(0)=0$}
% \end{minipage}\\
 \begin{minipage}{0.329\textwidth}
 \centerline{
\includegraphics[width=\textwidth]{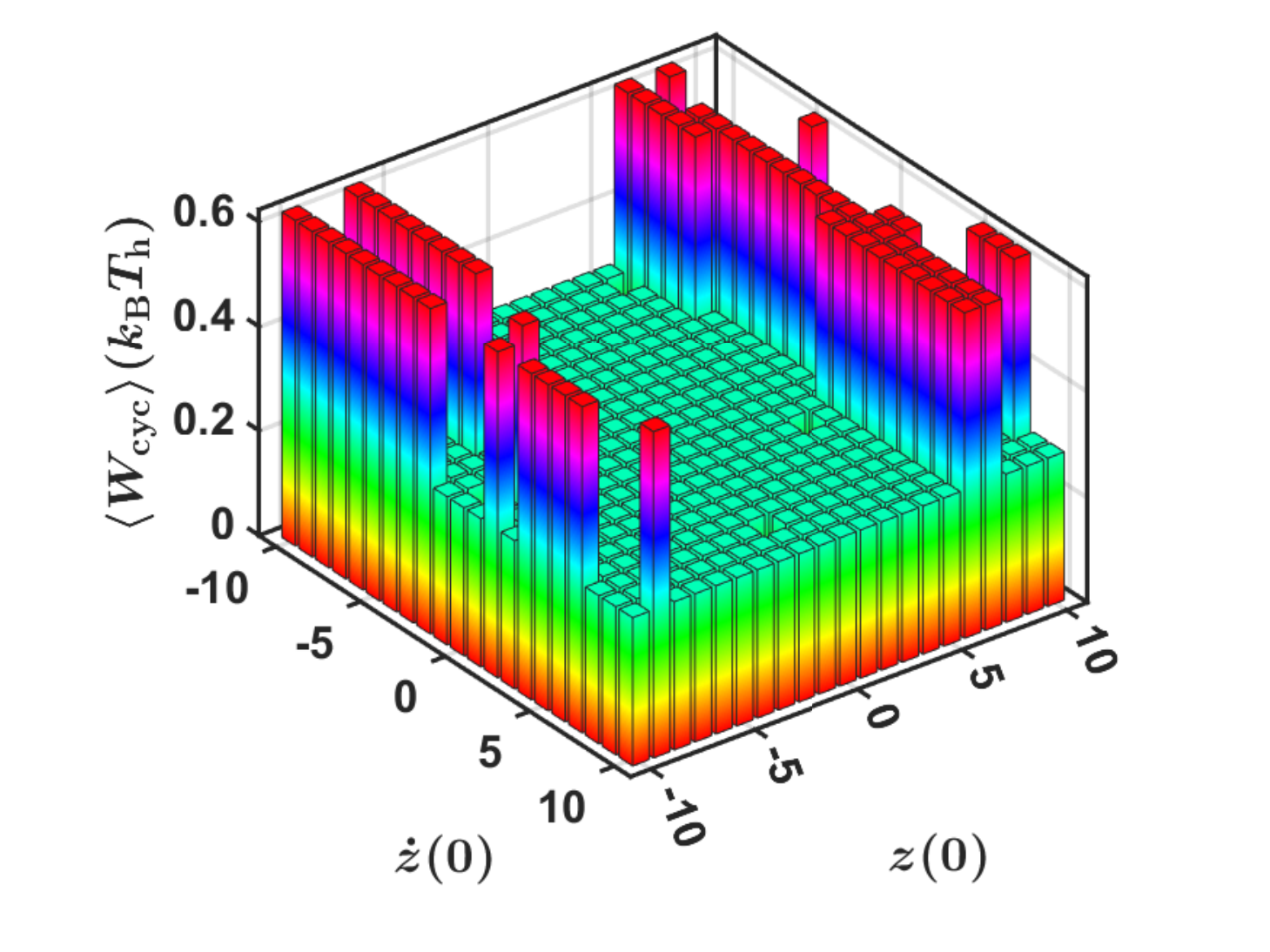}
 }
 \centerline{(e)}
 \end{minipage}
 \begin{minipage}{0.329\textwidth}
 \centerline{
\includegraphics[width=\textwidth]{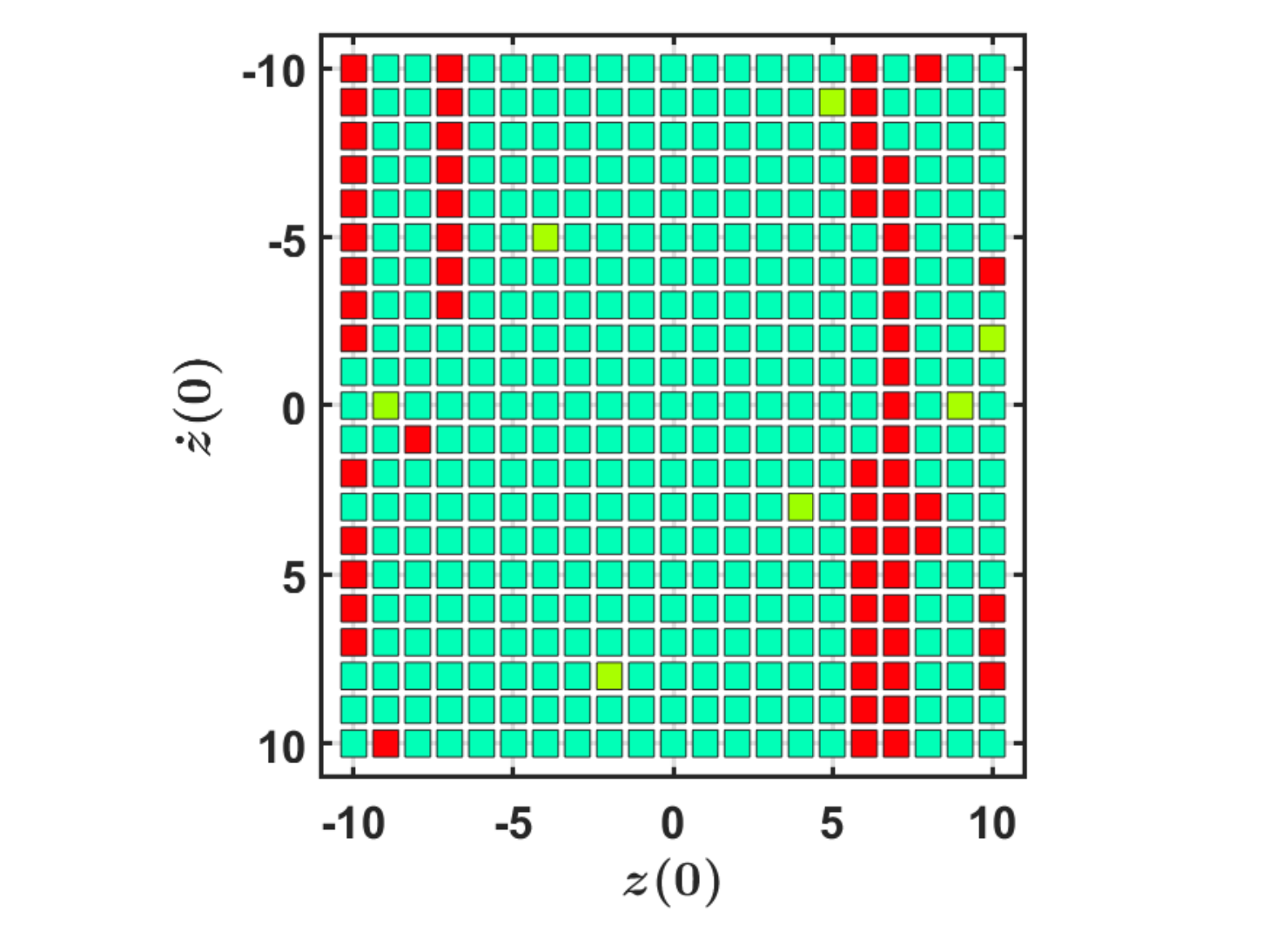}
 }
 \centerline{(f)}
 \end{minipage}

\caption{Different solutions of the zero temperature Langevin equation at $v_{\rm dr}=0.1\rm m/s$, $\eta=3.0$ and $\mu=4\times10^{4}\rm s^{-1}$. (a), (b), (c) and (d) Four different solutions of different initial conditions. In each of the four subfigures, the top left subgraph is the limit cycle on the $z-\dot z$ phase plane. The bottom left subgraph is the cycle work from the first to the last simulation cycle during the simulation time range. The bottom right subgraph is the count distribution of $W_{\rm cyc}$ of the last 1200 simulation cycles. The top right subgraph is the Poincare (or stroboscopic) map \cite{Seydel2010Ch7Stability} sampled at the starting point of each cycle, i.e. the set of the intial point $(z_0,\dot z_0)$ of each cycle with the cycle number indicated by the color of the points ranging from blue at the beginning to yellow at the end. (e) The mean cycle work $\langle W_{\rm cyc}\rangle$ corresponding to different initial conditions $(z(0),\dot z(0))$, which is different from the initial point $(z_0,\dot z_0)$ of each cycle. (f) The airview of (e). Other parameters are given in Sec. \ref{Langevindynamicssimulation}.\ref{ParametersUsed}.}
\label{multiplesolutions0p1}
\end{figure}

In Figure \ref{multiplesolutions0p1}, we plot the  solutions at $v_{\rm dr}=0.1\rm m/s$, $\eta=3.0$ and $\mu=4\times10^{4}\rm s^{-1}$ to give an impression of the peculiarity of the limit cycles and Poincare maps at some driving velocities. More limit cycles can be found in Figure \ref{fig:limitcyclesbackbone}, \ref{fig:limitcycleshigherperiods} and \ref{fig:limitcyclesnoperiod}.

\subsection{The nonlinear bifurcation of $\langle W_{\rm cyc}\rangle$ with respect to $v_{\rm dr}$}
\subsubsection{Continuation of the zero temperature $\langle W_{\rm cyc}\rangle-v_{\rm dr}$ curve}
We first need to transform the ordinary equation system Eq. \ref{ODE} into a form which is easy to continue. Substitute $y_1=z-\tilde v\tau$, $y_2=\dot z$, $y_3=\sin z$ and $ y_4=\cos z$ into Eq. \ref{ODE} we obtain
\begin{equation}
\begin{cases}
\dot y_1=y_2,\\
\dot y_2=-\beta\eta y_2-4\pi^2y_1+4\pi^2\eta y_3,\\
\dot y_3=(\dot y_1+\tilde v)\cos z,\\
\dot y_4=-(\dot y_1+\tilde v)\sin z,
\end{cases}
\end{equation}
which is equivalent to 
\begin{equation}
\label{eqn:ContEq}
\begin{cases}
\dot y_1=y_2,\\
\dot y_2=-\beta\eta y_2-4\pi^2y_1+4\pi^2\eta y_3,\\
\dot y_3=y_3+(y_2+\tilde v)y_4-y_3(y_3^2+y_4^2),\\
\dot y_4=y_4-(y_2+\tilde v)y_3-y_4(y_3^2+y_4^2).
\end{cases}
\end{equation}
This equation system can be used as a model system for the continuation software MatCont \cite{MatcontRef}.

We first solve the ordinary differential equation with the Matlab integrator ode113 to obtain several periodic solutions, i.e. limit cycles in the phase plane \cite{seydel2009practical}. Starting with one of the limit cycles, we can continue the branch containing this limit cycle.

The mean cycle work is calculated by
\begin{equation}
\label{eq:mcycworkode}
\begin{aligned}
\langle W_{\rm cyc}\rangle&=\frac1{Tv_{\rm dr}/a}\int_{0}^{T}\kappa(v_{\rm dr}t-x)v_{\rm dr}\mathrm dt=\frac a{Tv_{\rm dr}}\int_{0}^{T}\kappa(v_{\rm dr}t-x)\mathrm d(v_{\rm dr}t)\\
&=\frac a{\tilde T\tilde v\frac a{2\pi}}\int_{0}^{\tilde T}\kappa\frac a{2\pi}(\tilde v\tau-z)\mathrm d(\frac a{2\pi}\tilde v\tau)=\frac1{2\pi\tilde T}\int_0^{\tilde T}\kappa a^2(\tilde v\tau-z)\mathrm  d\tau,
\end{aligned}
\end{equation}
where $\tilde T=\frac{\omega_0T}{2\pi}$ is the nondimentional period of the limit cycle on the branch. The period $T$ of the limit cycle may equal to one or several $\frac a{v_{\rm dr}}$'s.

\subsubsection{Details of the $\langle W_{\rm cyc}\rangle-v_{\rm dr}$ bifurcation diagram}
Some details of the bifurcation diagram of $\langle W_{\rm cyc}\rangle$ with $v_{\rm dr}$ as the parameter, i.e. Figure 6 in the main text, are plotted in Figure \ref{BifurcationFigures}. 

The cycle number period 
\begin{equation}
\label{eqn:cyclenumperiod}
P_{\rm c.n.}=\frac{Tv_{\rm dr}}{a},
\end{equation}
is an invariant of each curve distinguished by the color. Here $T$ is the actual temporal period of the particle which should be one or more integer multiples of $\frac a{v_{\rm dr}}$. In the main text Figure 6, besides the first and higher order period doubling curves with $P_{\rm c.n.}=2^n,n=1,2,\cdots$ on the blue backbone curve ($P_{\rm c.n.}=1$) as shown in detail in the last subfigure in Figure \ref{BifurcationFigures}, other isolated loops have other values of $P_{\rm c.n.}$. For instance, on the black loops $P_{\rm c.n.}=3$; on the dark green loops $P_{\rm c.n.}=6$; on the dark red loops $P_{\rm c.n.}=12$; on the orange loop $P_{\rm c.n.}=9$; on the purple loop, which is a period doubling branch of the orange loop, $P_{\rm c.n.}=18$.
\begin{figure}[H]
\centering
\includegraphics[width=8.6cm]{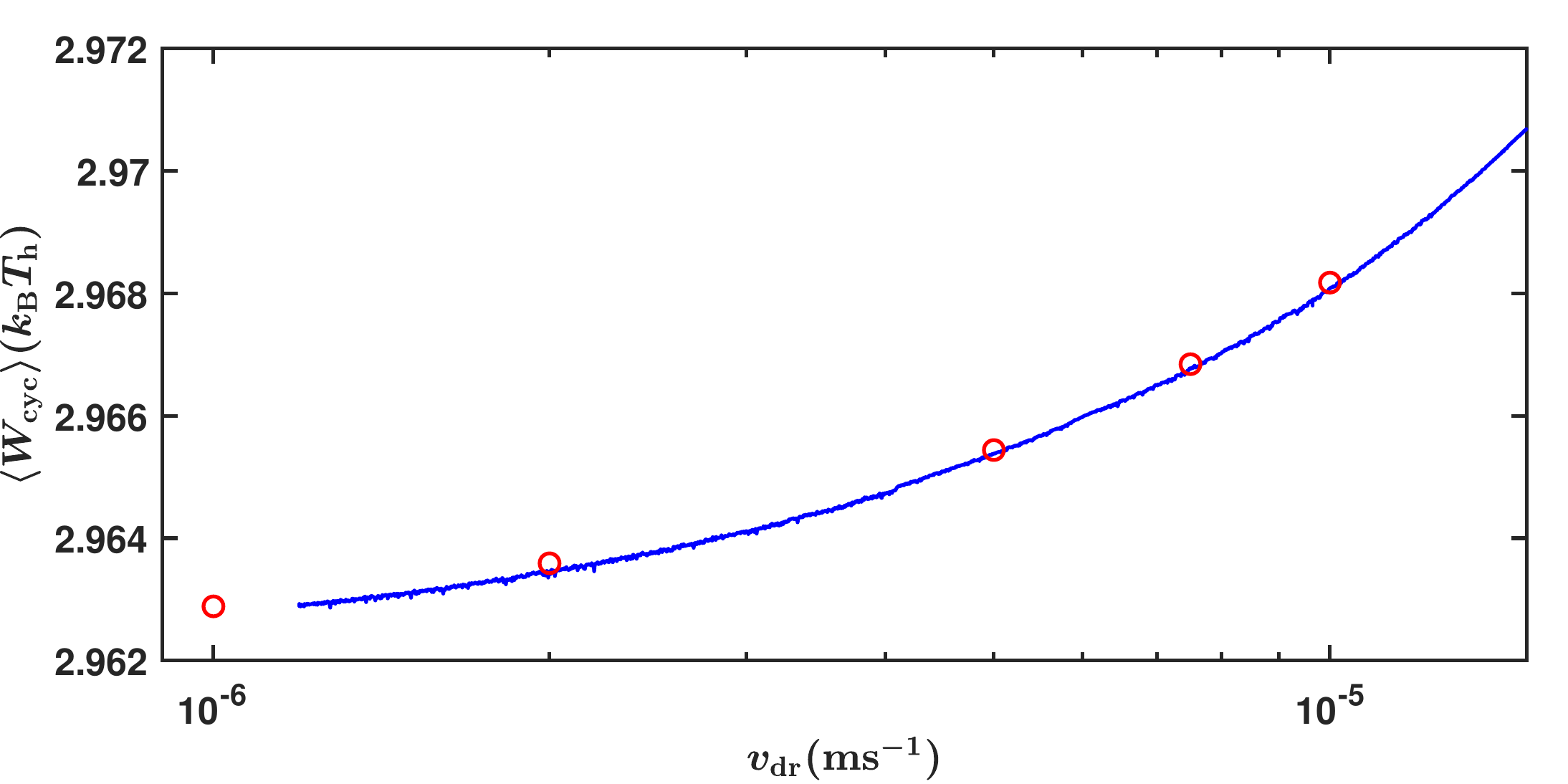}
\includegraphics[width=8.6cm]{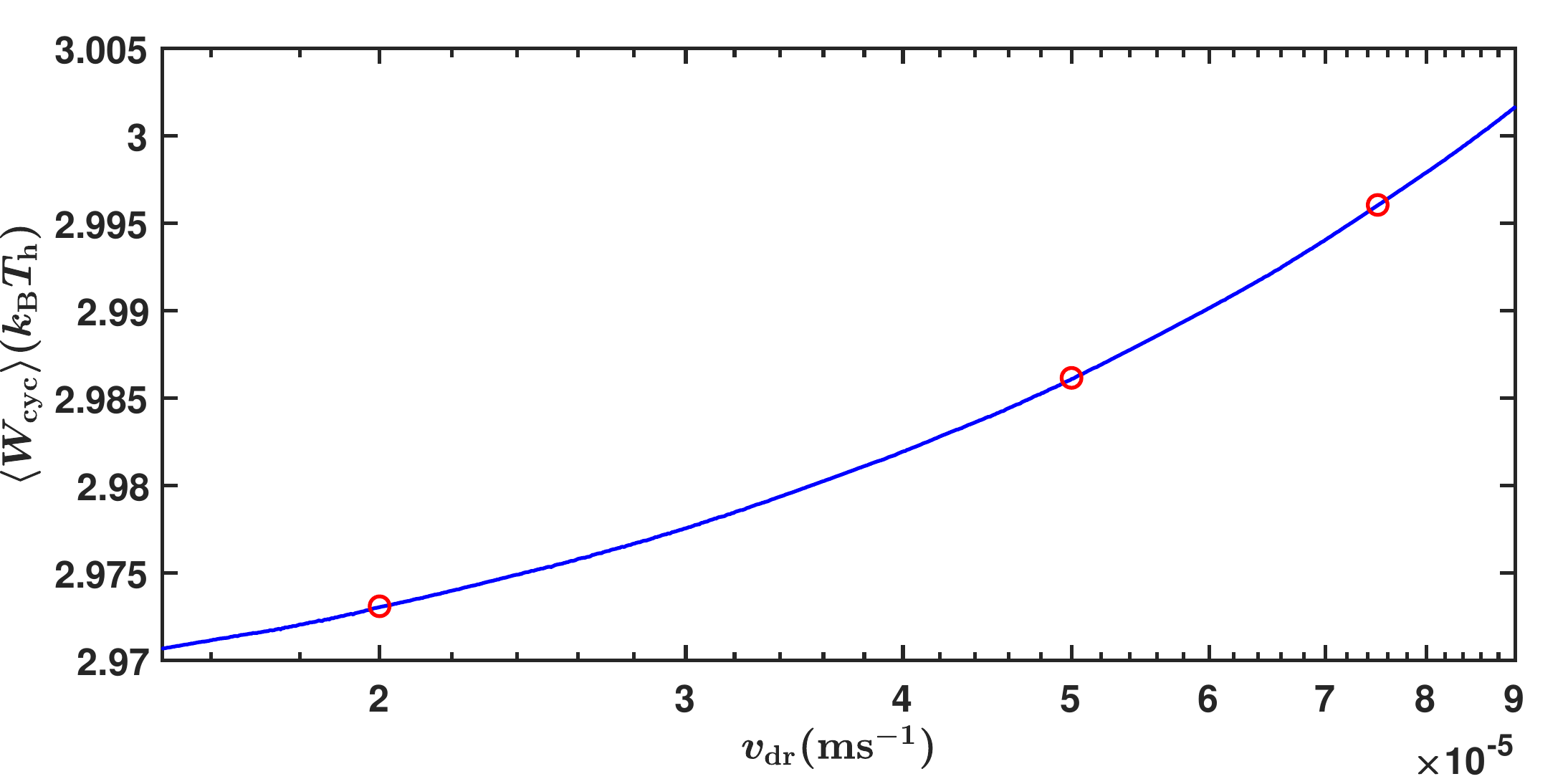}\\
\includegraphics[width=8.6cm]{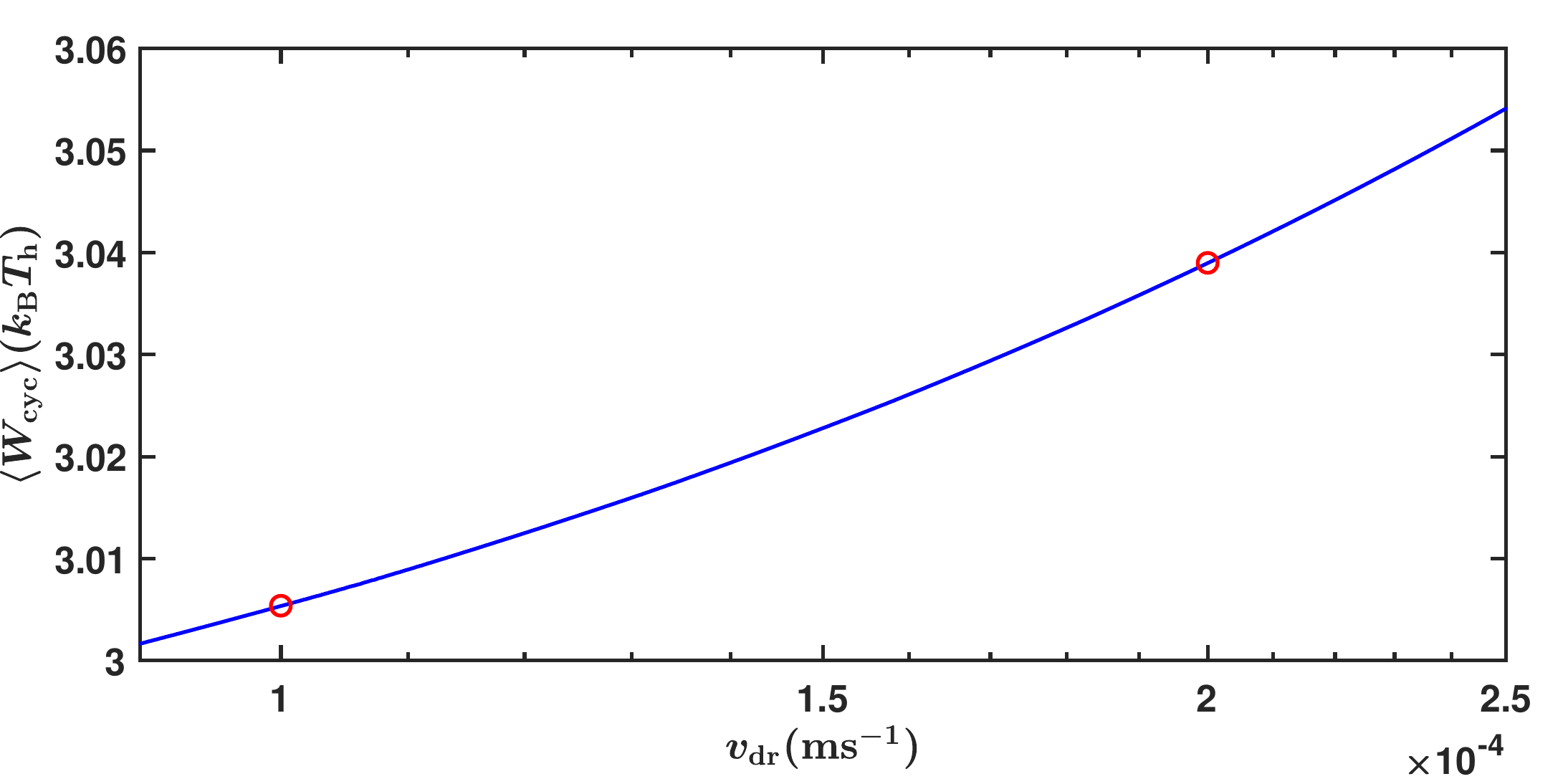}
\includegraphics[width=8.6cm]{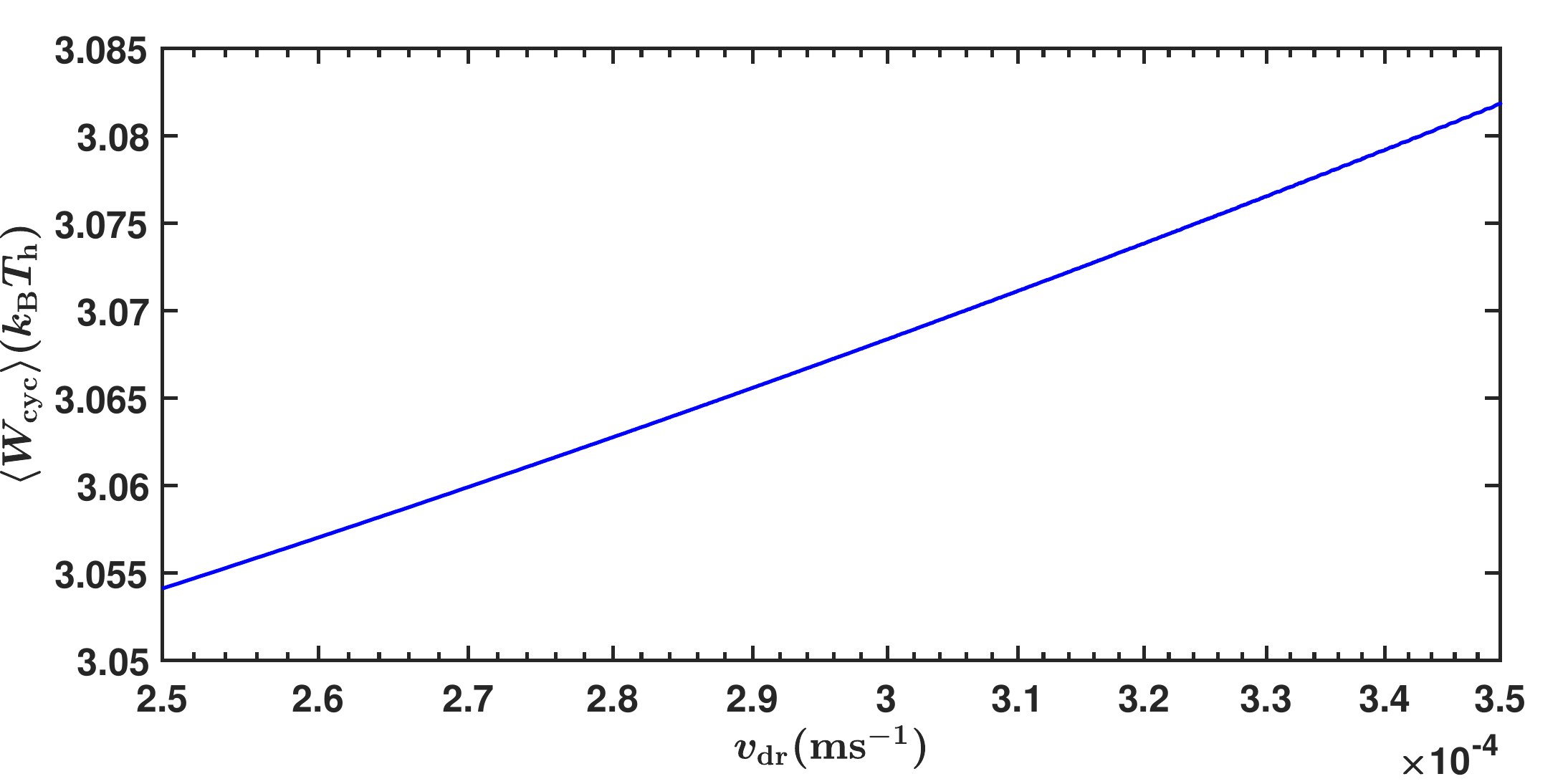}
\end{figure}
\begin{figure}[H]
\centering
\includegraphics[width=8.6cm]{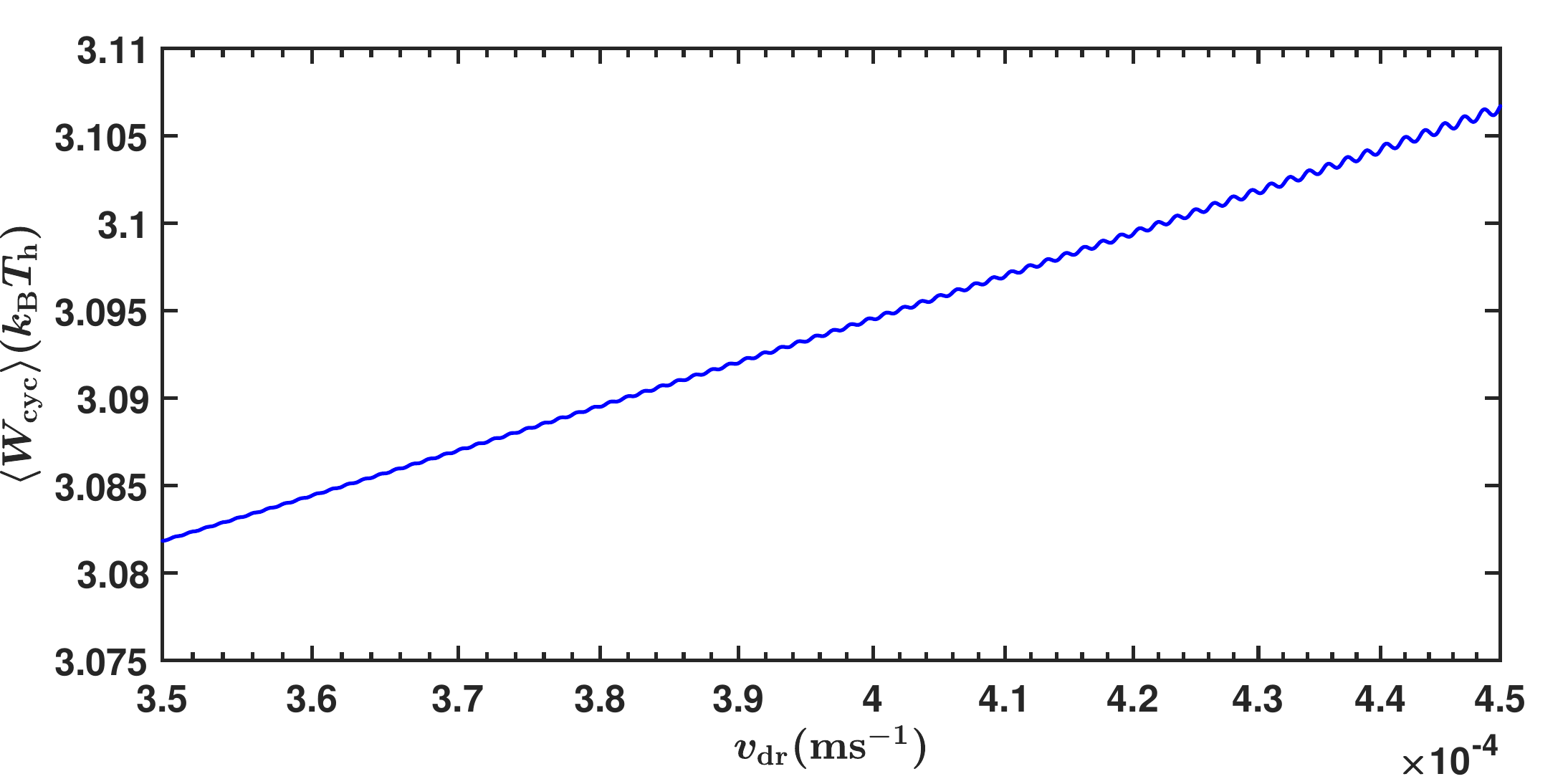}
\includegraphics[width=8.6cm]{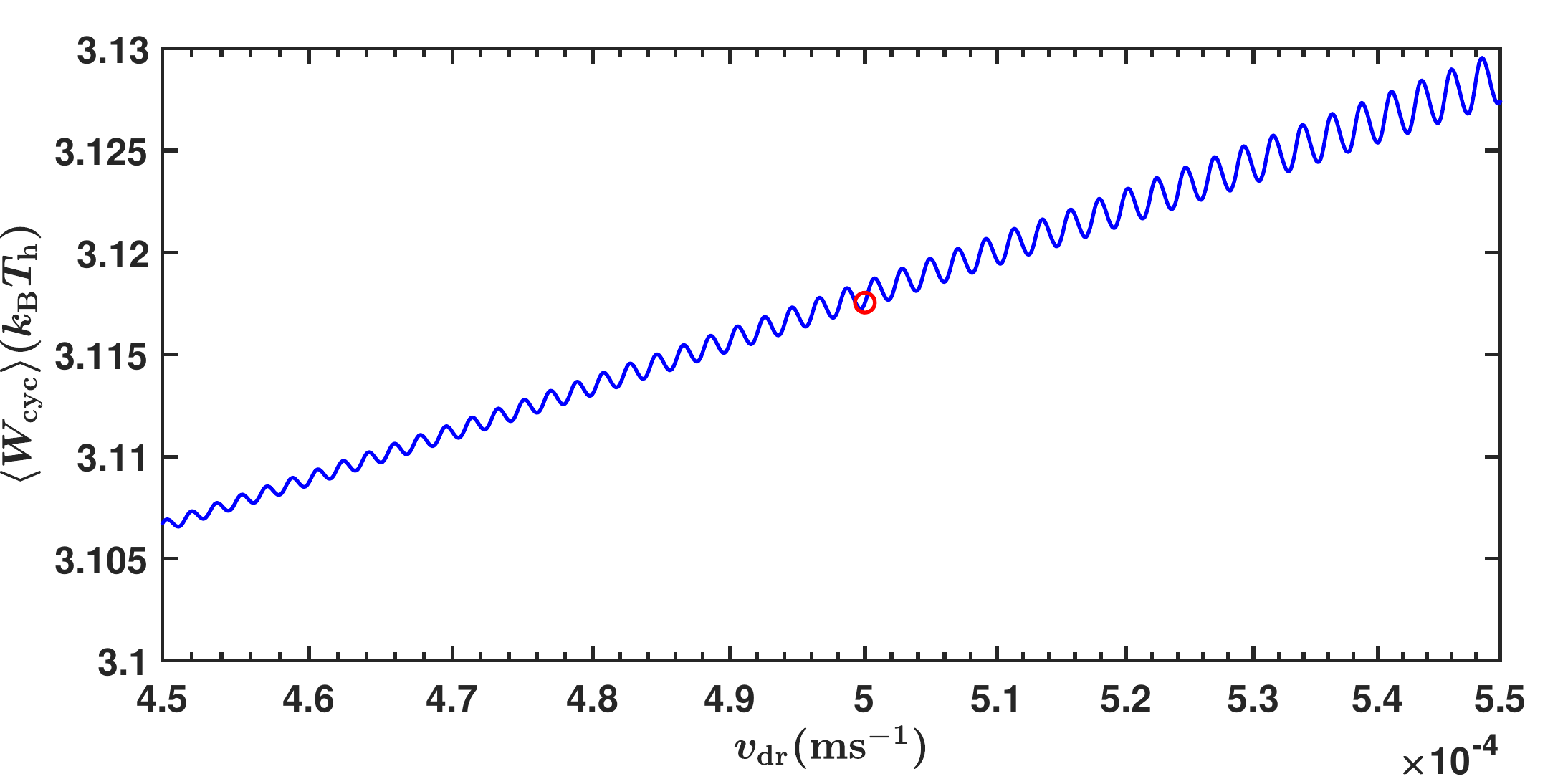}\\
\includegraphics[width=8.6cm]{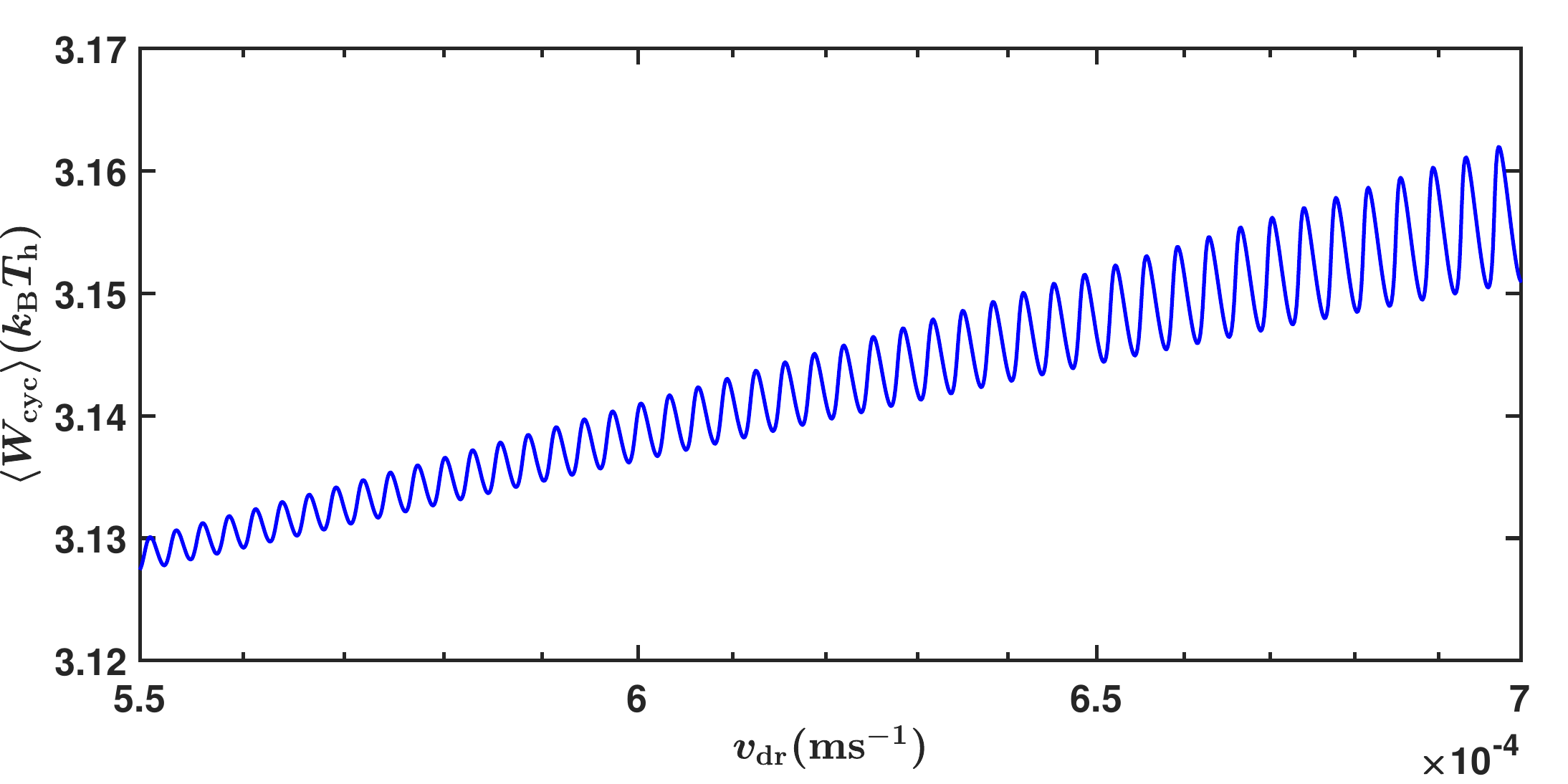}
\includegraphics[width=8.6cm]{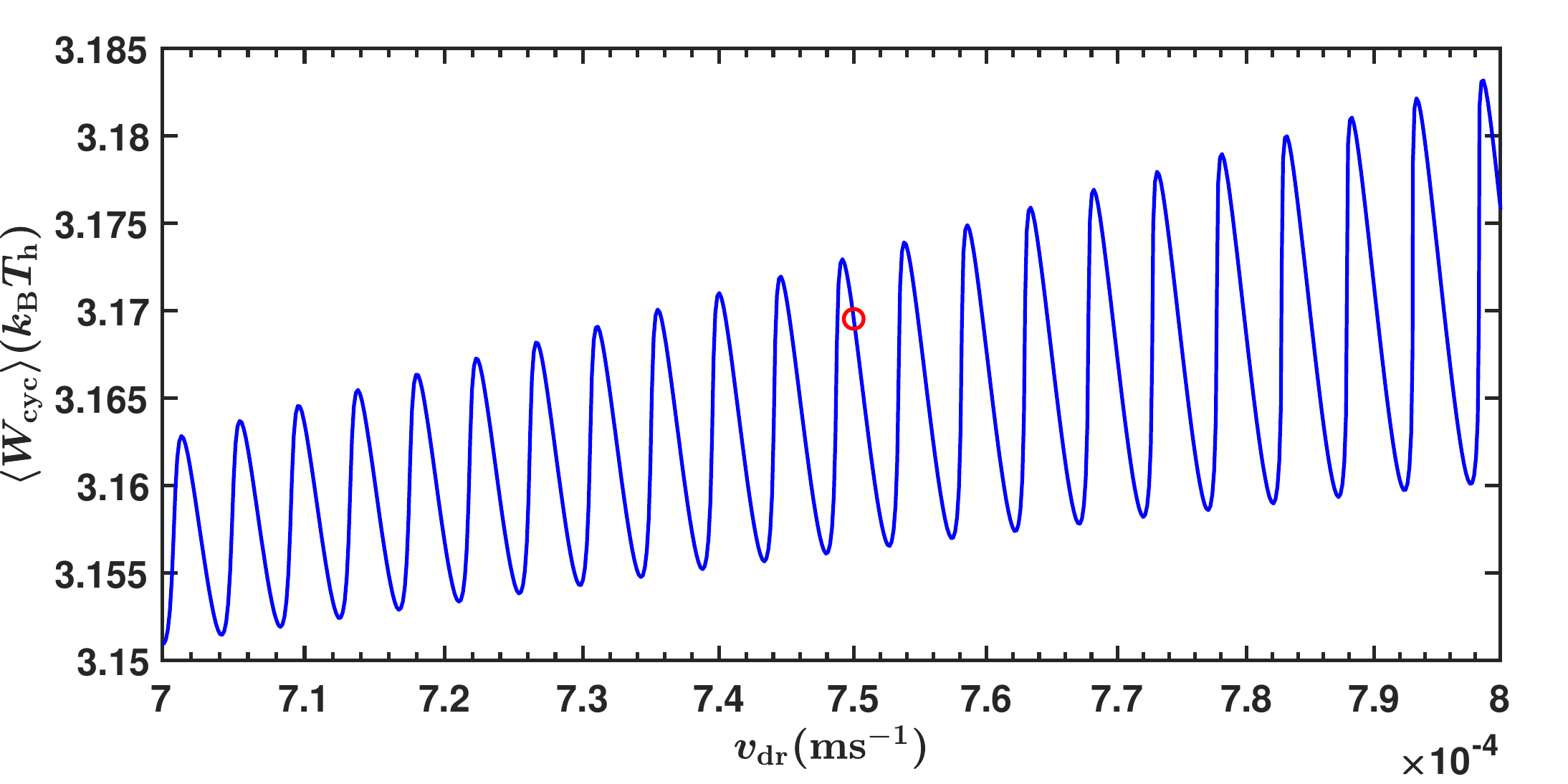}\\
\includegraphics[width=8.6cm]{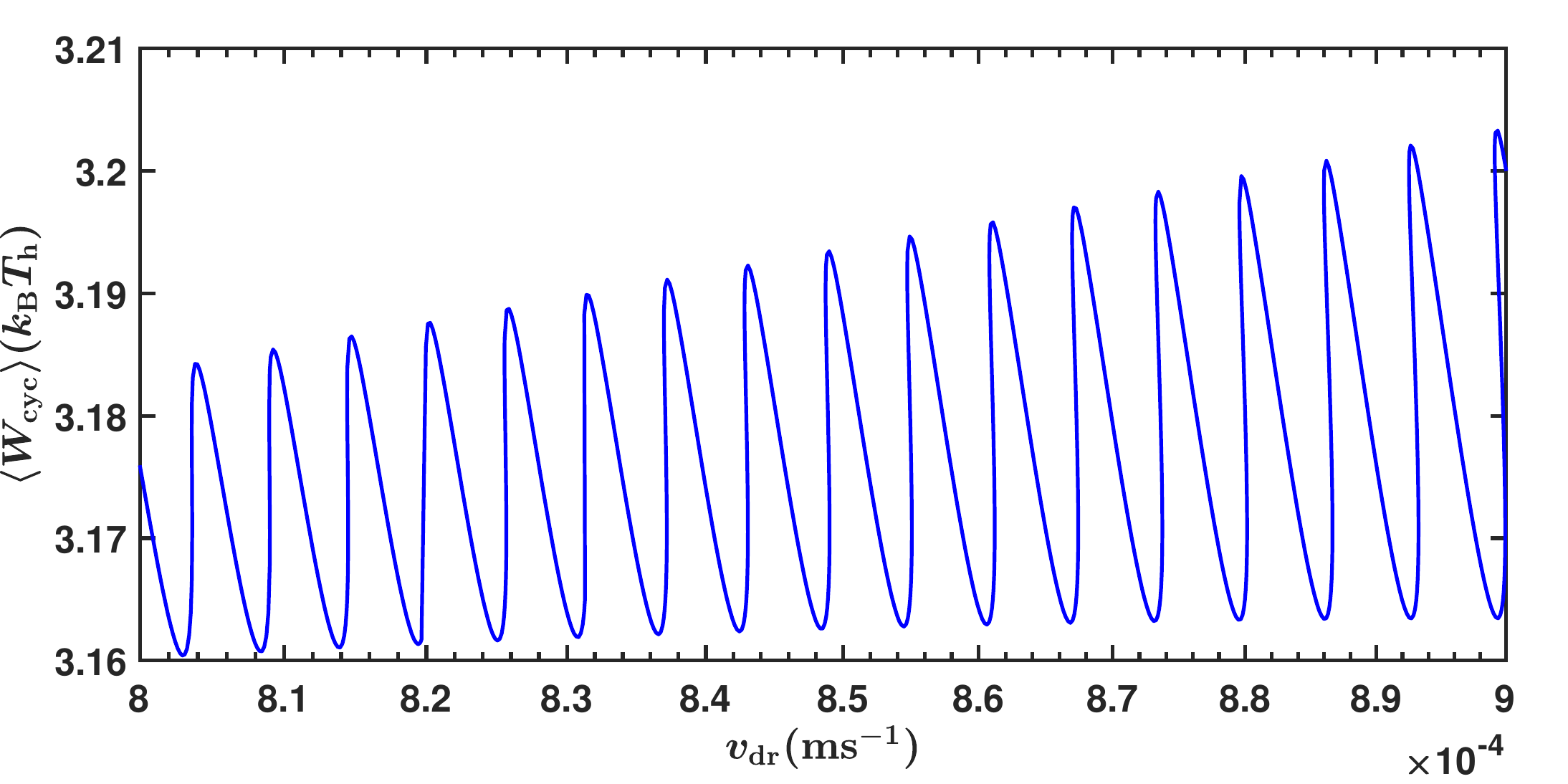}
\includegraphics[width=8.6cm]{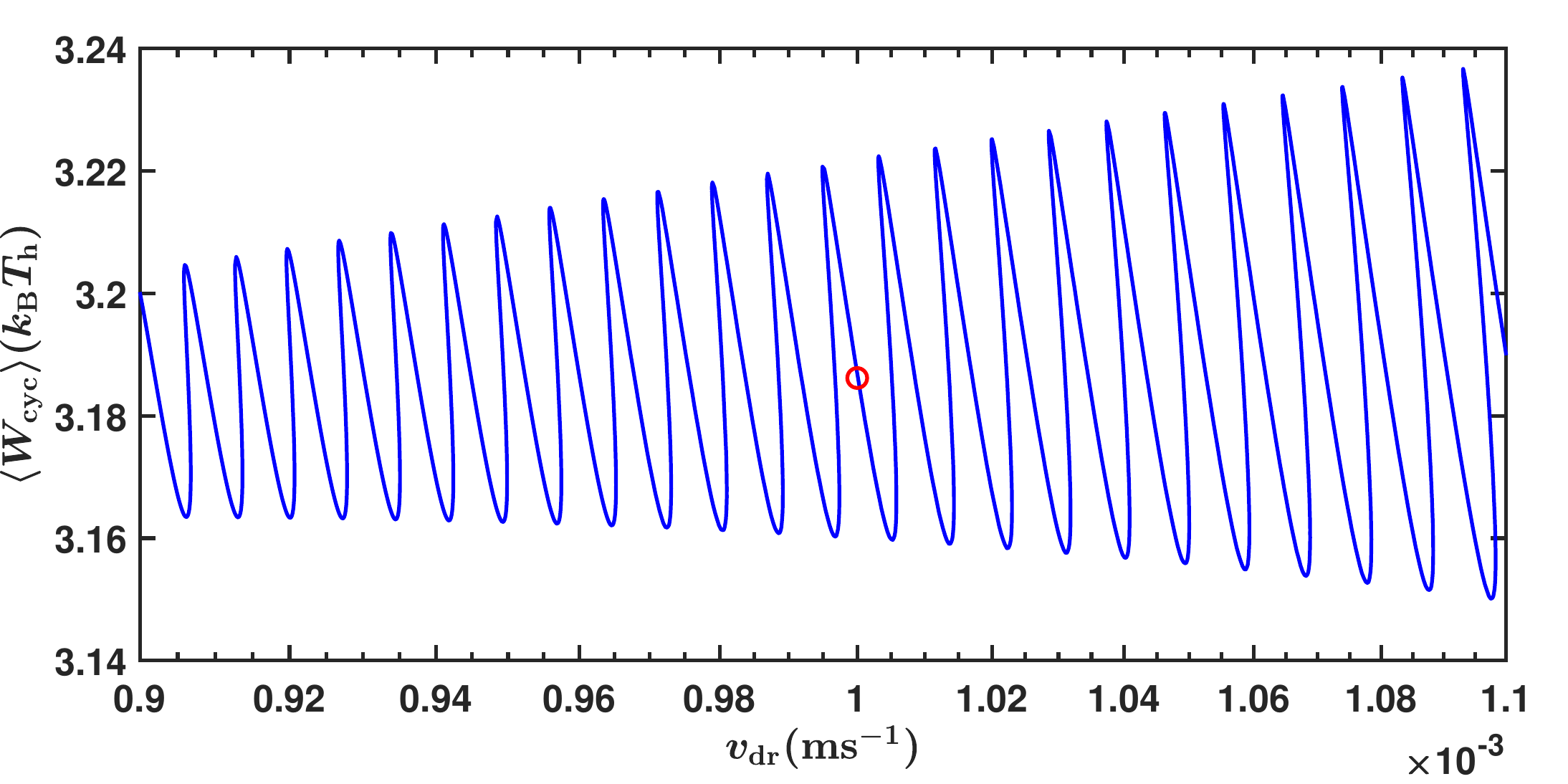}\\
\includegraphics[width=8.6cm]{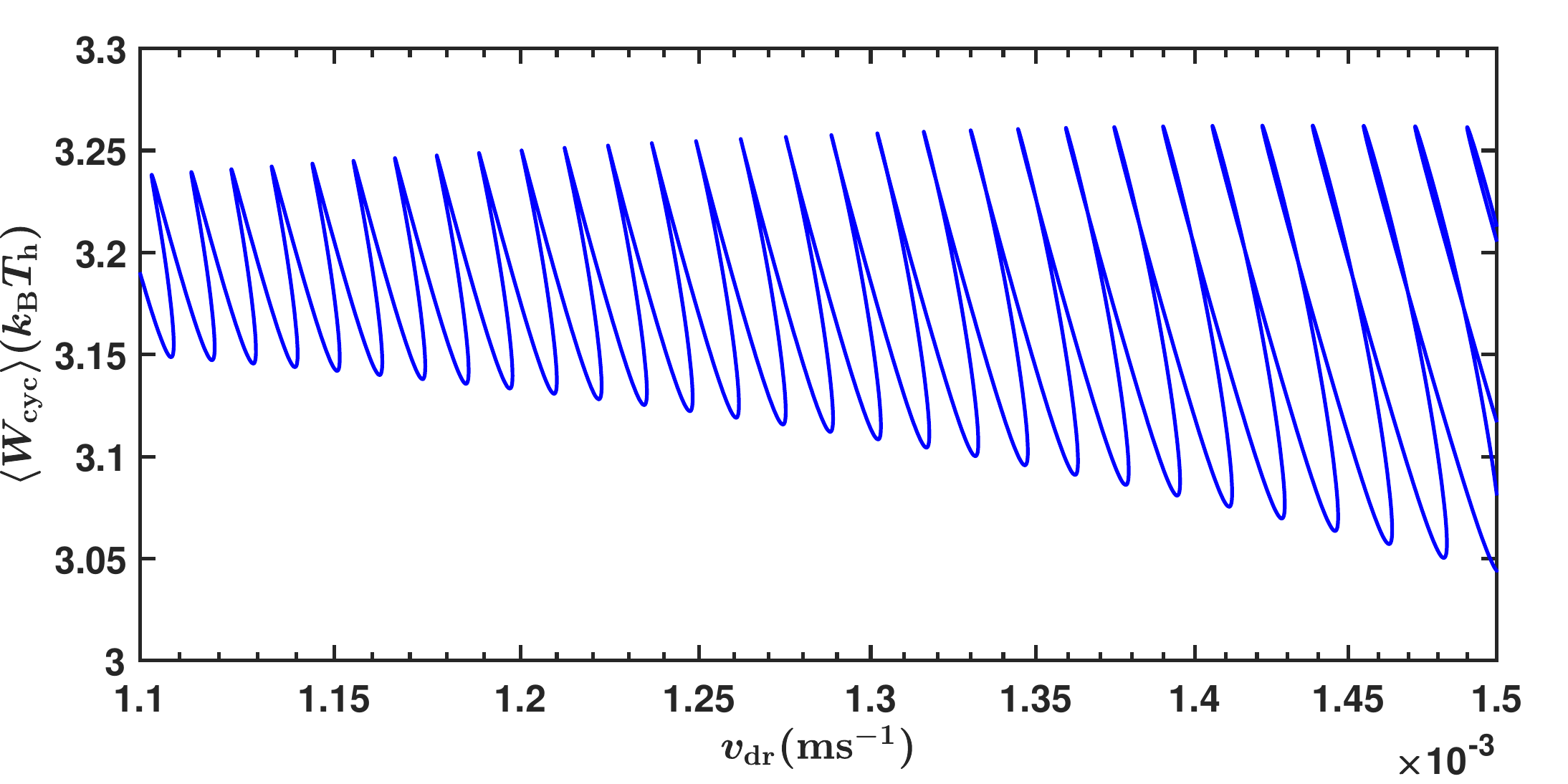}
\includegraphics[width=8.6cm]{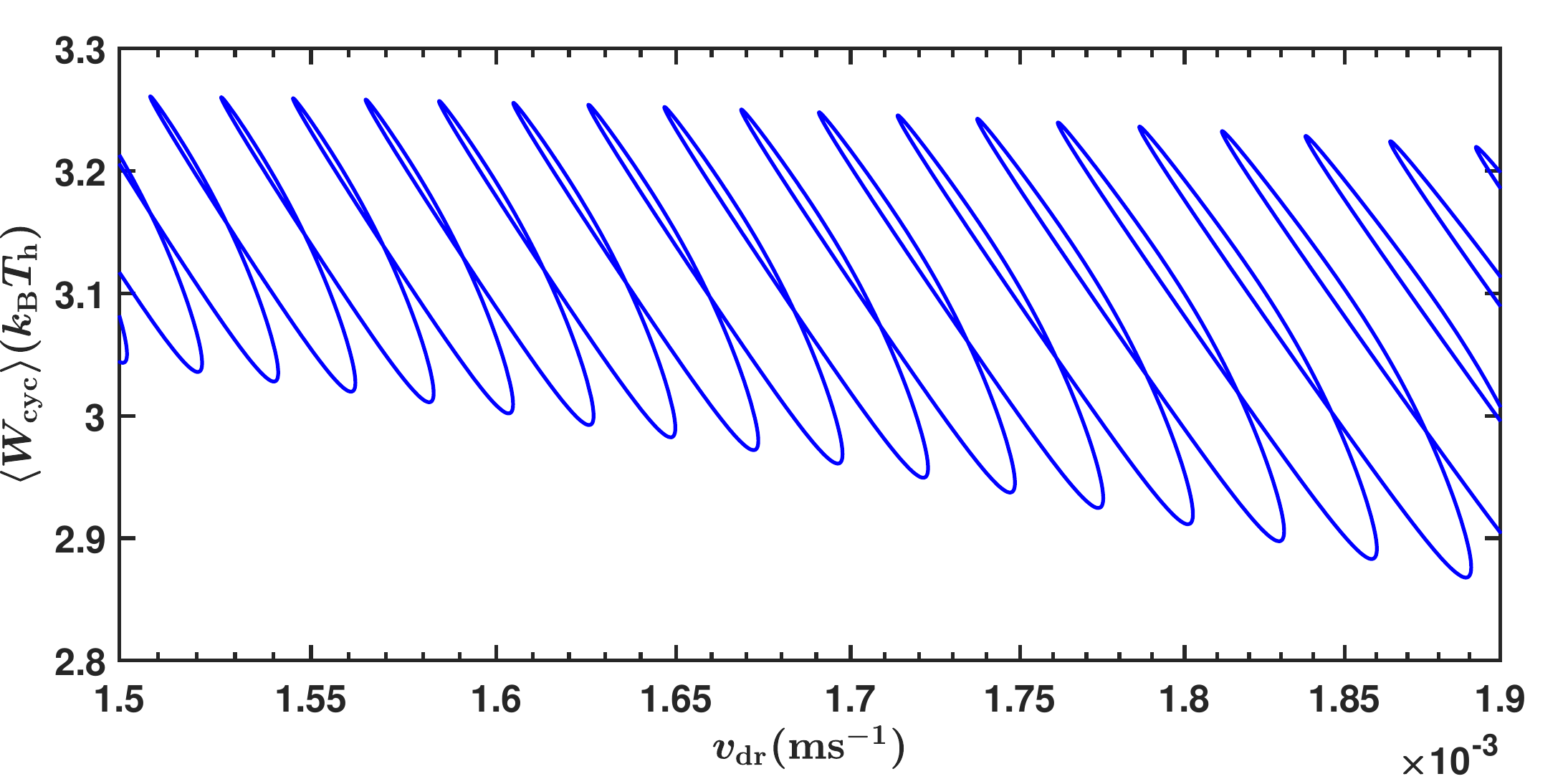}\\
\includegraphics[width=8.6cm]{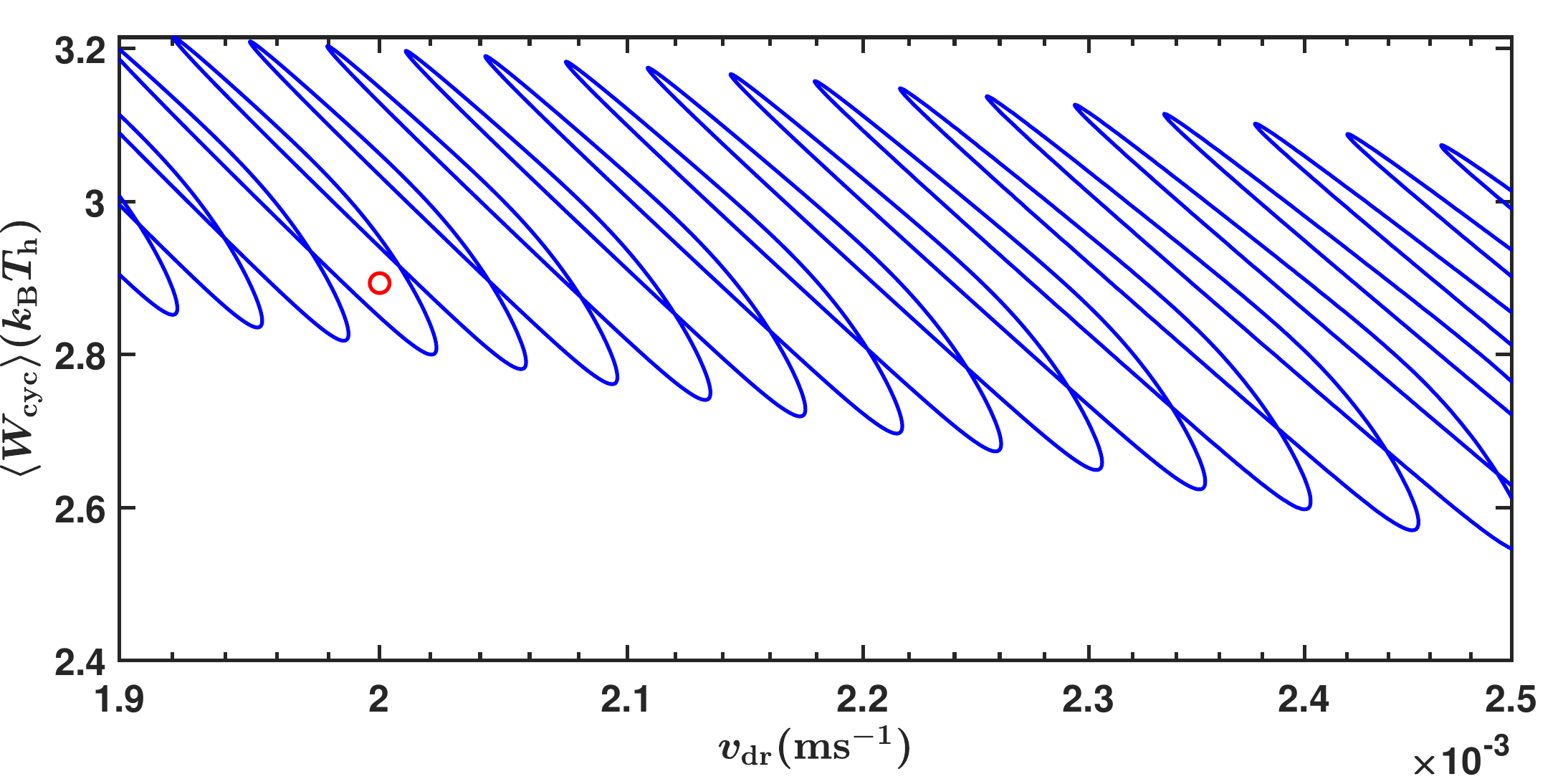}
\includegraphics[width=8.6cm]{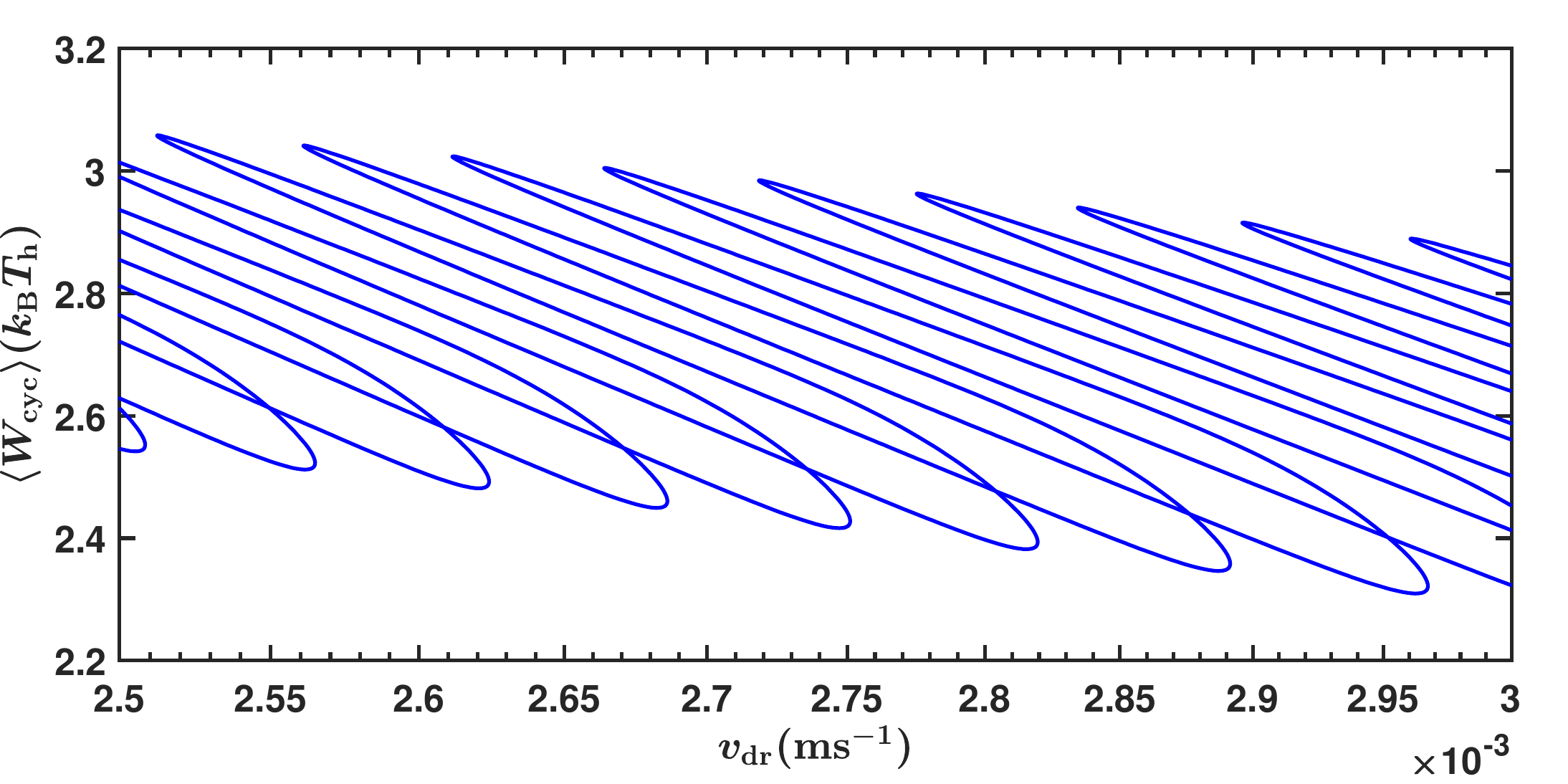}
\end{figure}
\begin{figure}[H]
\centering
\includegraphics[width=8.6cm]{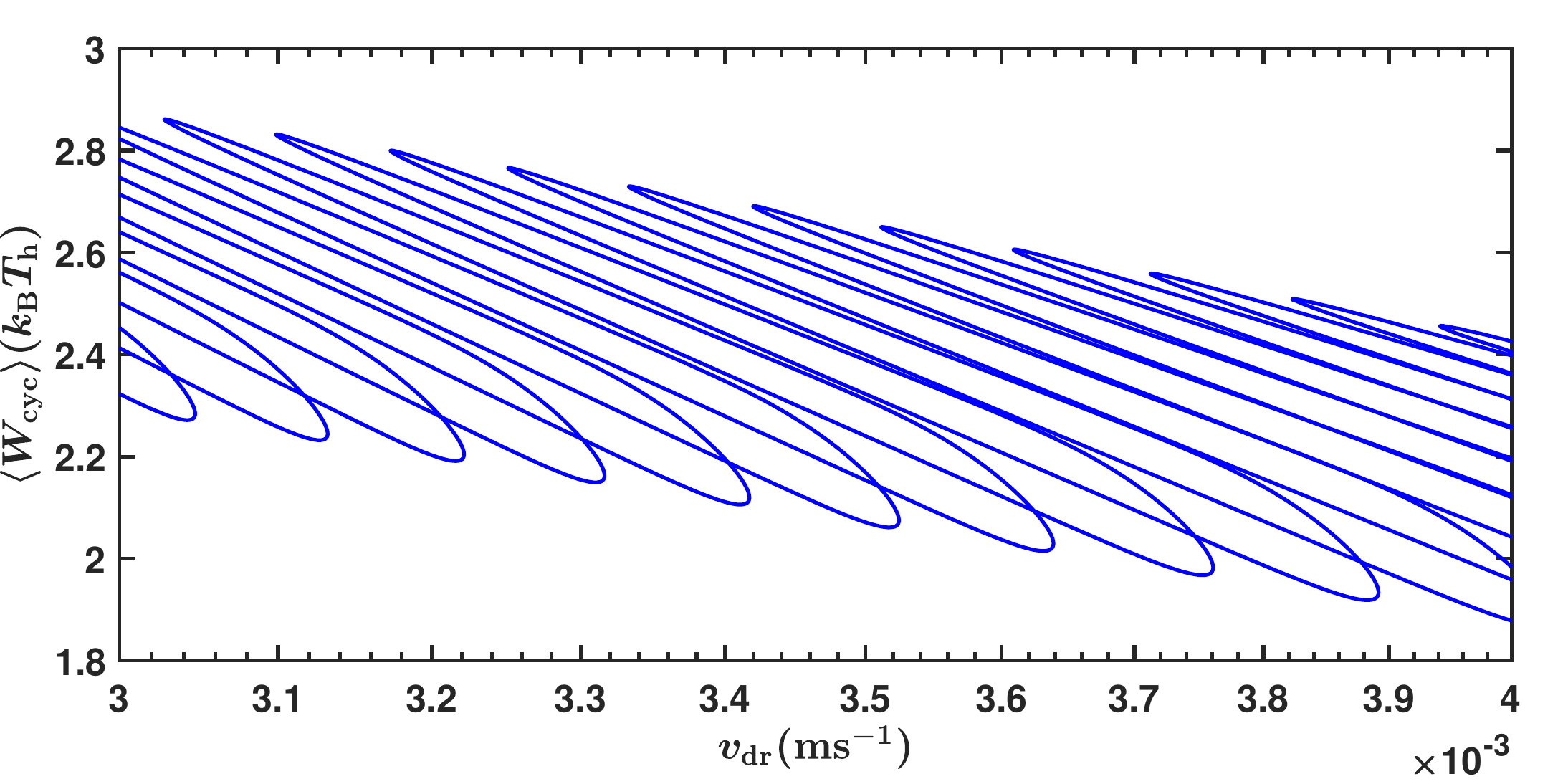}
\includegraphics[width=8.6cm]{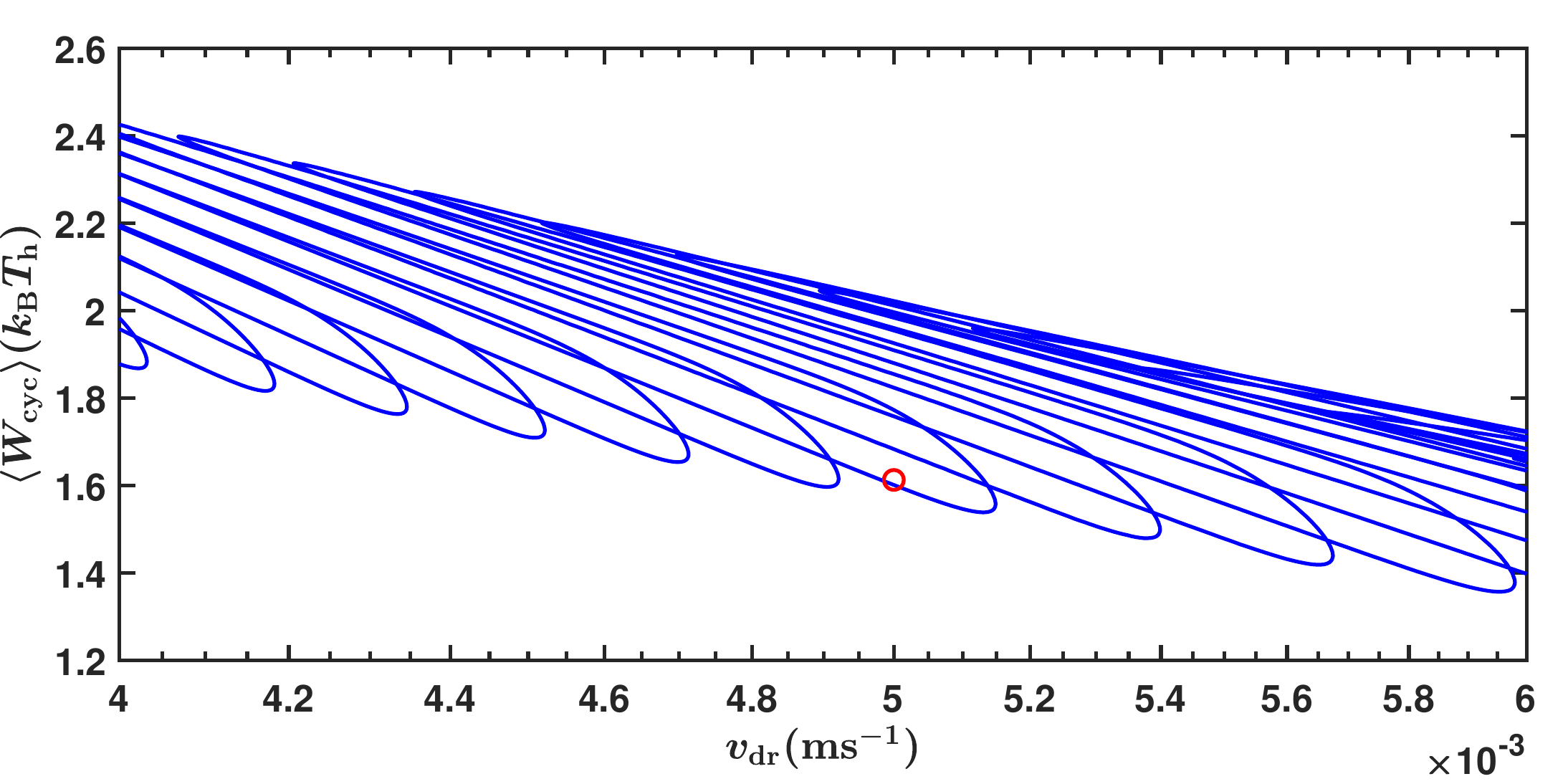}\\
\includegraphics[width=8.6cm]{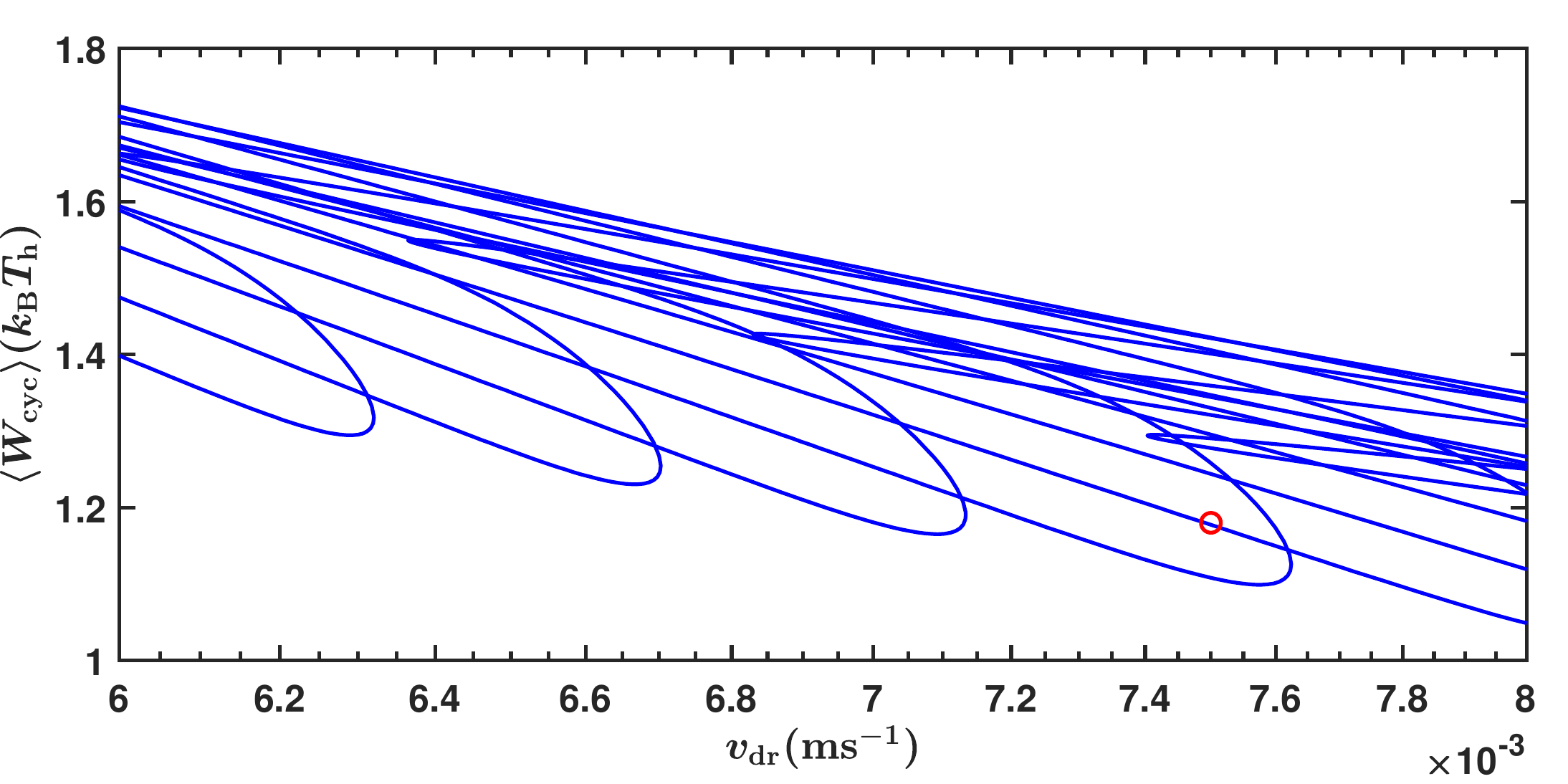}
\includegraphics[width=8.6cm]{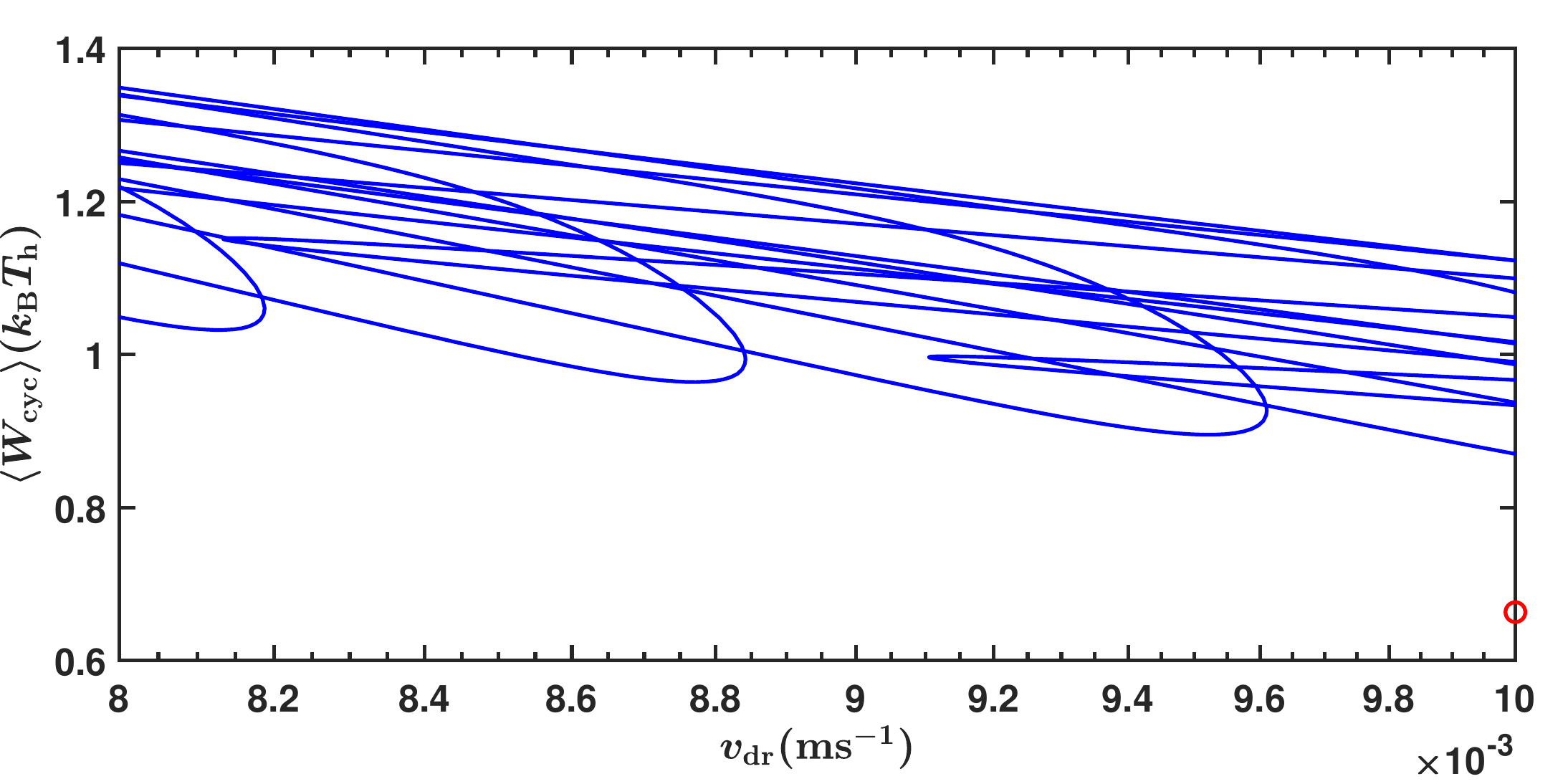}\\
\includegraphics[width=8.6cm]{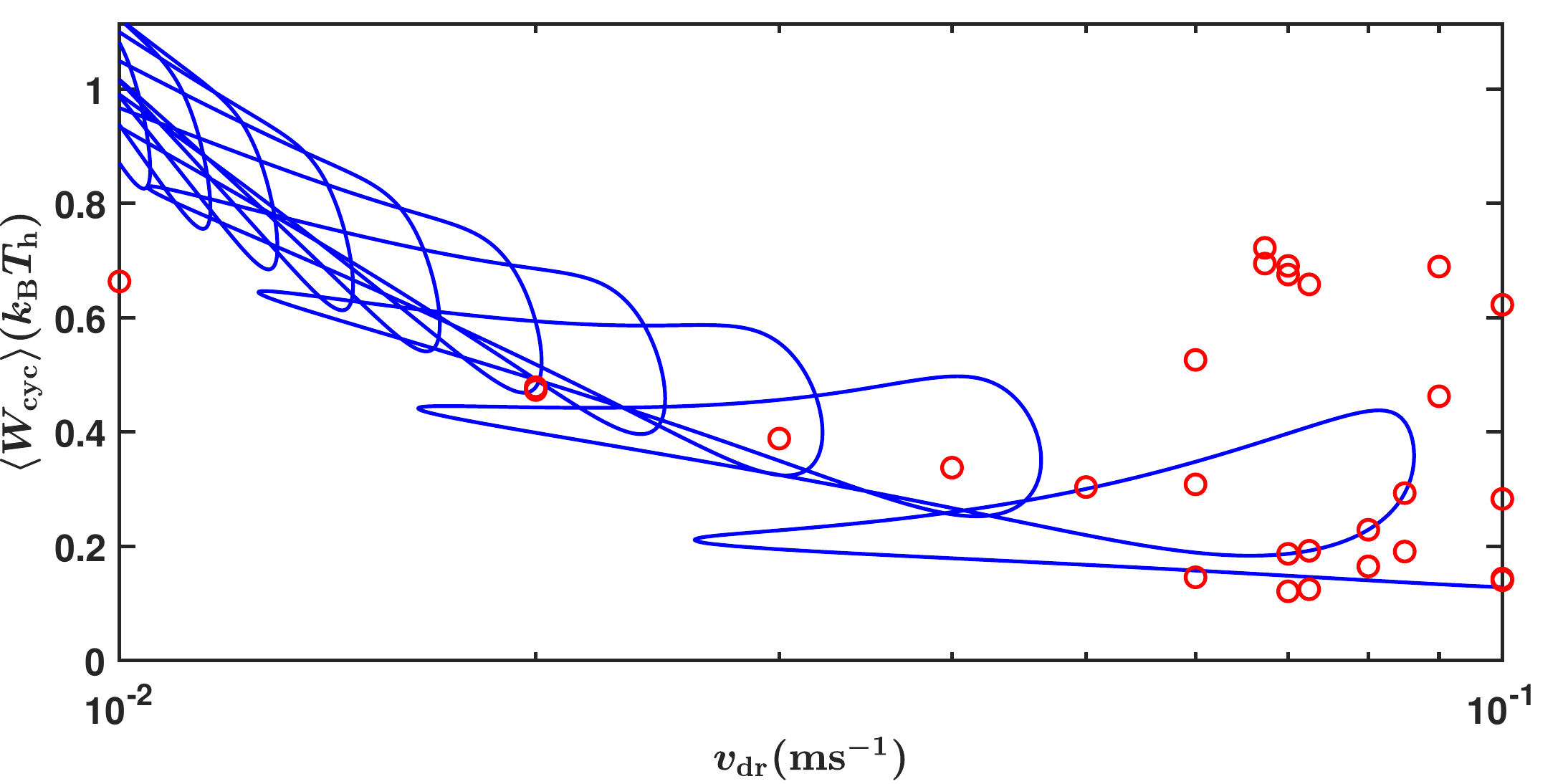}
\includegraphics[width=8.6cm]{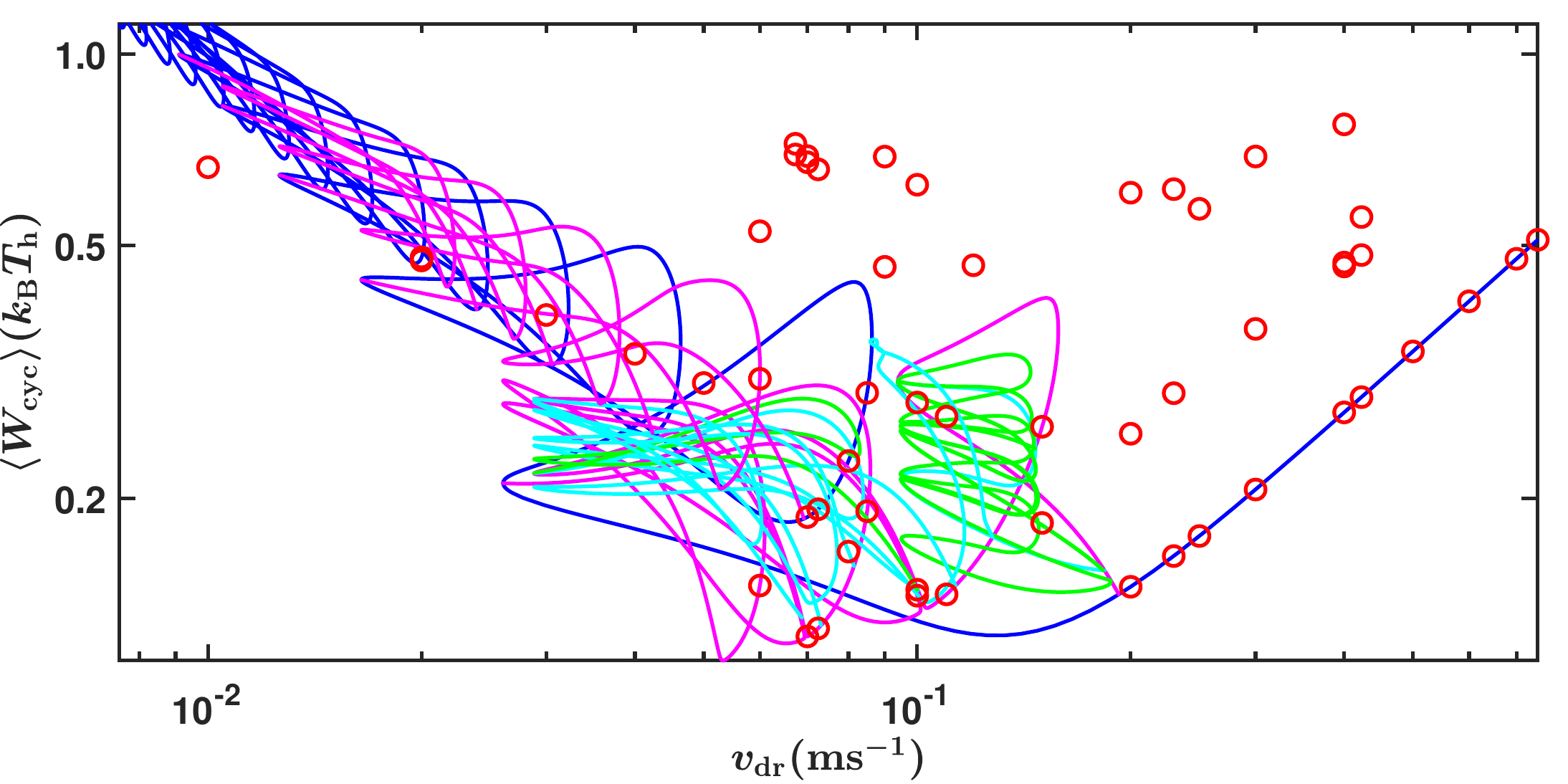}
\caption{Details of Figure 6 in the main text with some of its segments zoomed in. The last subfigure gives some of the first and higher order period doubling \cite{Seydel2010Ch7Stability} branches of the blue backbone curve on which the cycle number period $P_{\rm c.n.}=1$. The megenta curves have $P_{\rm c.n.}=2$; the cyan ones have $P_{\rm c.n.}=4$; and the green ones have $P_{\rm c.n.}=8$. The cycle number period $P_{\rm c.n.}$ is an invariant of each curve. Parameters: $\eta=3.0$, $\mu=4\times10^4\rm s^{-1}$, ${\it\Theta}=0$ and others are given in Sec. \ref{Langevindynamicssimulation}.\ref{ParametersUsed}.}
\label{BifurcationFigures}
 \end{figure}

\subsubsection{Some of the peculiar nonlinear features of the ordinary differential equation based on the PT model}
In this subsection, we discuss some of the peculiar nonlinear features of the ordinary differential equation based on the PT model, including the following three topics: symmetry, limit cycles and tori. We give a preliminary analysis of the three topics in this subsection to attract the readers', especially the nonlinear science community's, attention to the peculiar nonlinearity of the ordinary differential equation based on the PT model and and we hope more scholars can join us to analyze it further.

\begin{enumerate}
\item \textbf{Symmetry} 
When projected to the $\langle W_{\rm cyc}\rangle-v_{\rm dr}$ space, some unclosed $x(0)-v_{\rm dr}$ and $v(0)-v_{\rm dr}$ curves become closed as shown in Figure \ref{BifurcationUnclosed}. Here the $x(0)$ and $v(0)$ are the coordinates of the first point of the limit cycle at the specific $v_{\rm dr}$ calculated by continuation \cite{MatcontRef,seydel2009practical}. In Figure \ref{BifurcationClosed}, we plot a case of closed $x(0)-v_{\rm dr}$ and $v(0)-v_{\rm dr}$ curves remaining closed when projected to the $\langle W_{\rm cyc}\rangle-v_{\rm dr}$ space for comparison. 

In Figure \ref{BifurcationClosed}, we also plot the first order and and one of the second order period doubling branches. We can see that the closed period doubling loops in the $x(0)-v_{\rm dr}$ and $v(0)-v_{\rm dr}$ spaces become two unclosed curves when projected to the $\langle W_{\rm cyc}\rangle-v_{\rm dr}$ space, indicating some special symmetry \cite{Seydel2010Symmetry}.

\item \textbf{Limit cycles} In Figure \ref{fig:limitcyclesbackbone}, \ref{fig:limitcycleshigherperiods}, we plot the limit cycles corresponding to some of the red circles in Figure \ref{BifurcationFigures}. The rest red circles with no period or with a torus as the limit are given in Figure \ref{fig:limitcyclesnoperiod}. All the limit cycles are calculated with the method described in Sec. \ref{sec:apndRD}.\ref{sec:mulsolT0}. 

In Figure \ref{fig:limitcyclesbackbone}, all the limit cycles have the cycle number period $P_{\rm c.n.}=1$ and are on the blue backbone curve in the main text Figure 6, %(or equivalently Figure \ref{BifurcationFigures})
comparing with which we can see the corresponding features of the limit cycles at different driving velocity regimes, cf. the caption of Figure \ref{fig:limitcyclesbackbone}. 

The limit cycles of some of the red circles out of the blue backbone curve are plotted in Figure \ref{fig:limitcycleshigherperiods} to give the readers an impression of the exotic shapes of the limit cycles.

\item \textbf{Tori}
In Figure \ref{fig:limitcyclesnoperiod}, the six phase curves represent solutions with a torus as the limit or with no period. They are in the velocity weakening regime, or to be exact, in the transition regime between the stick-slip regime and the resonance regime, cf. Figure \ref{fig:StickSlipsSingleOnecycleXT0} and Sec. \ref{sec:apndRD}.\ref{sec:transitionEDFwdv}. 

If we set $y_5=\tilde v\tau$ and add $\dot y_5=\tilde v$ to Eq. \ref{eqn:ContEq}, the four dimensional phase space consisting of four variables $(y_1,y_2,y_3,y_4)$ becomes a five dimensional phase space of $(y_1,y_2,y_3,y_4,y_5)$. $y_5$ is a variable with radian as dimension and can be mapped onto a circle so that we can project the five dimensional phase space of $(y_1,y_2,y_3,y_4,y_5)$ in the three dimension cylindrical coordinate space $(y5,y1,y2)=(\theta,r,h)$. In the top left subgraph of each subfigure in Figure \ref{fig:ToriandNonPeriodSol}, we plot the projection results of the six cases in Figure \ref{fig:limitcyclesnoperiod}. To avoid the curves intersect themselves, we have add 10 to $y_1$ in these six subgraphs, i.e. the particle's trajectories are actually embedded in the cylindrical coordinate space $(\theta,r,h)=(\tilde v\tau,z-\tilde v\tau+10,\dot z)$ in the top left subgraph of each subfigure in Figure \ref{fig:ToriandNonPeriodSol}. The Poincare (or stroboscopic) map \cite{Seydel2010Ch7Stability} sampled at the intial instant of each cycle $(z_0,\dot z_0)$ are ploted in the top right subgraphs of each subfigure in Figure \ref{fig:ToriandNonPeriodSol} without shift of $y_1$. From the cylindrical coordinate embedded phase trajectories and the Poincare maps of the cases of $v_{\rm dr}=10^{-3}$ and $5\times10^{-3}\rm m/s$ we can recognize two clear tori, whose intersection with the $\tilde v\tau=0$ plane (the Poincare section) are on two ellipses after the steady state has been achieved. We can see that the two Poincare maps are continuous curves rather than discrete points, so the two trajectories have no period of rational multiples of $\frac a{v_{\rm dr}}$, i.e. the two solutions are qusiperiodic. So the angular frequency of the longitudinal motion along the axis of the torus $\omega_1=2\pi\frac{v_{\rm dr}}a$ and the angular frequency of the latitudinal motion surrounding the cross section of the torus $\omega_2$ are incommensurate, i.e. the ratio $\frac{\omega_1}{\omega_2}$ is irrational \cite{Seydel2010Ch7Stability}. For both of the two torus solutions, we can see that the mean cycle work $\langle W_{\rm cyc}\rangle$ denoted by the two red circles in Figure \ref{BifurcationFigures} are on the blue backbone curve at the corresponding $v_{\rm dr}$'s, indicating that at the two $v_{\rm dr}$'s the $P_{\rm c.n.}=1$ limit cycles are unstable and the torus bifurcation branches are stable with the two limit cycles as their axes \cite{Seydel2010Ch7Stability}. So after being averaged the torus branch will coincide with its axis limit cycle branch in the $\langle W_{\rm cyc}\rangle-v_{\rm dr}$ bifurcation diagram. This is a speculation and should be verified further.

At $v_{\rm dr}=2\times10^{-3}$ and $7.5\times10^{-3}\rm m/s$, there seems to be several tori for the particle to shift from one to another.
%, in which some are stable while others are unstable. 
The rest two cases in Figure \ref{fig:limitcyclesnoperiod} with no period indicating that there may be no stable solutions at these two driving velocities. We haven't considered the stability of the branches, which should be investigated further.

%After being averaged, the mean cycle work $\langle W_{\rm cyc}\rangle$ can on or out of the solution curves. The phase curves in 3-dimension and the Poincare sections of these solutions are plotted in Figure \ref{fig:ToriandNonPeriodSol}. We can see that the torus solution in 3-dimension phase space looks like enwinding a torus. Although having no period, the Poicare section of these torus solutions distribute on an ellipse. 

\begin{figure}[H]
\centering
\includegraphics[width=0.329\textwidth]{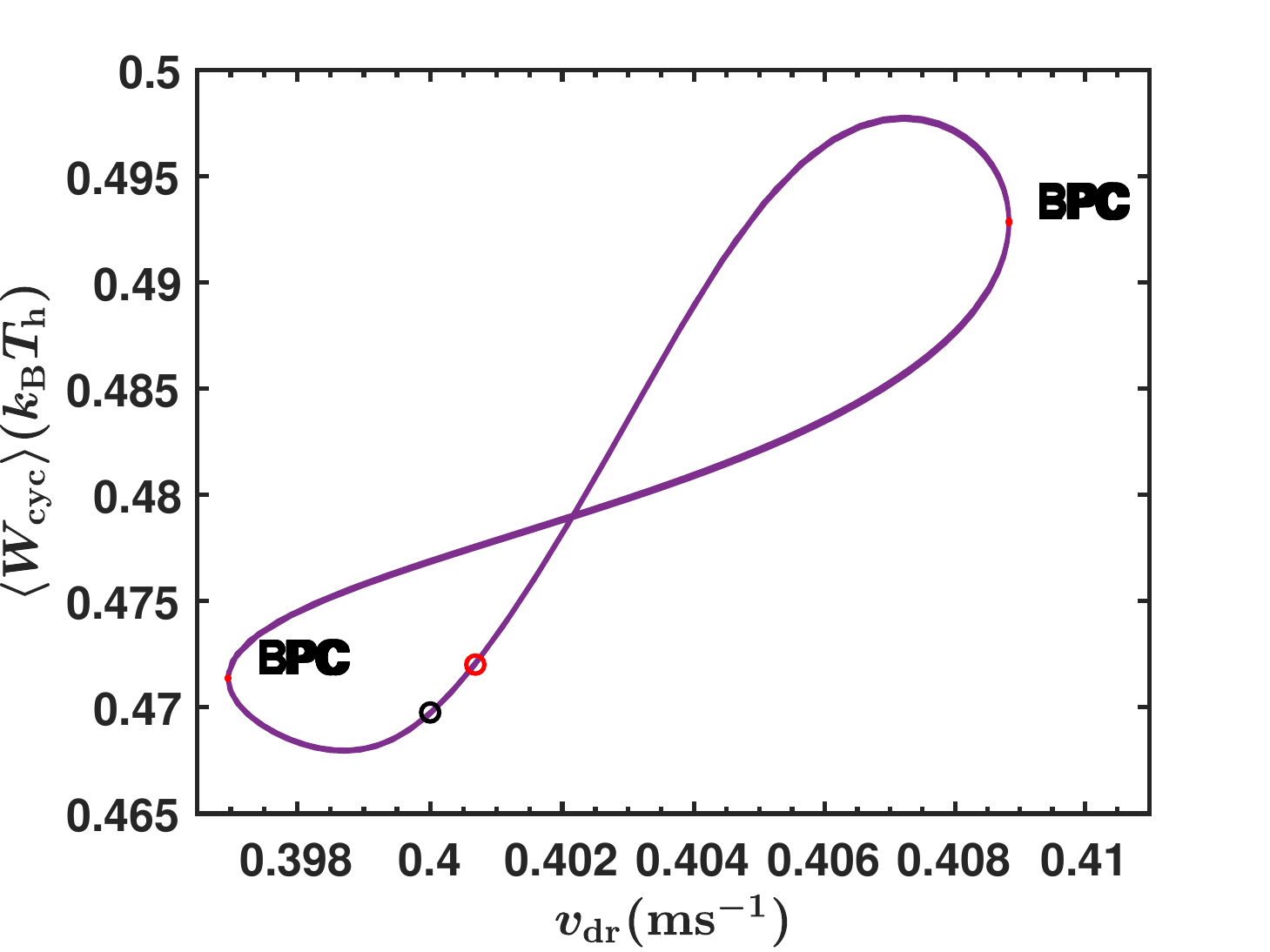}
\includegraphics[width=0.329\textwidth]{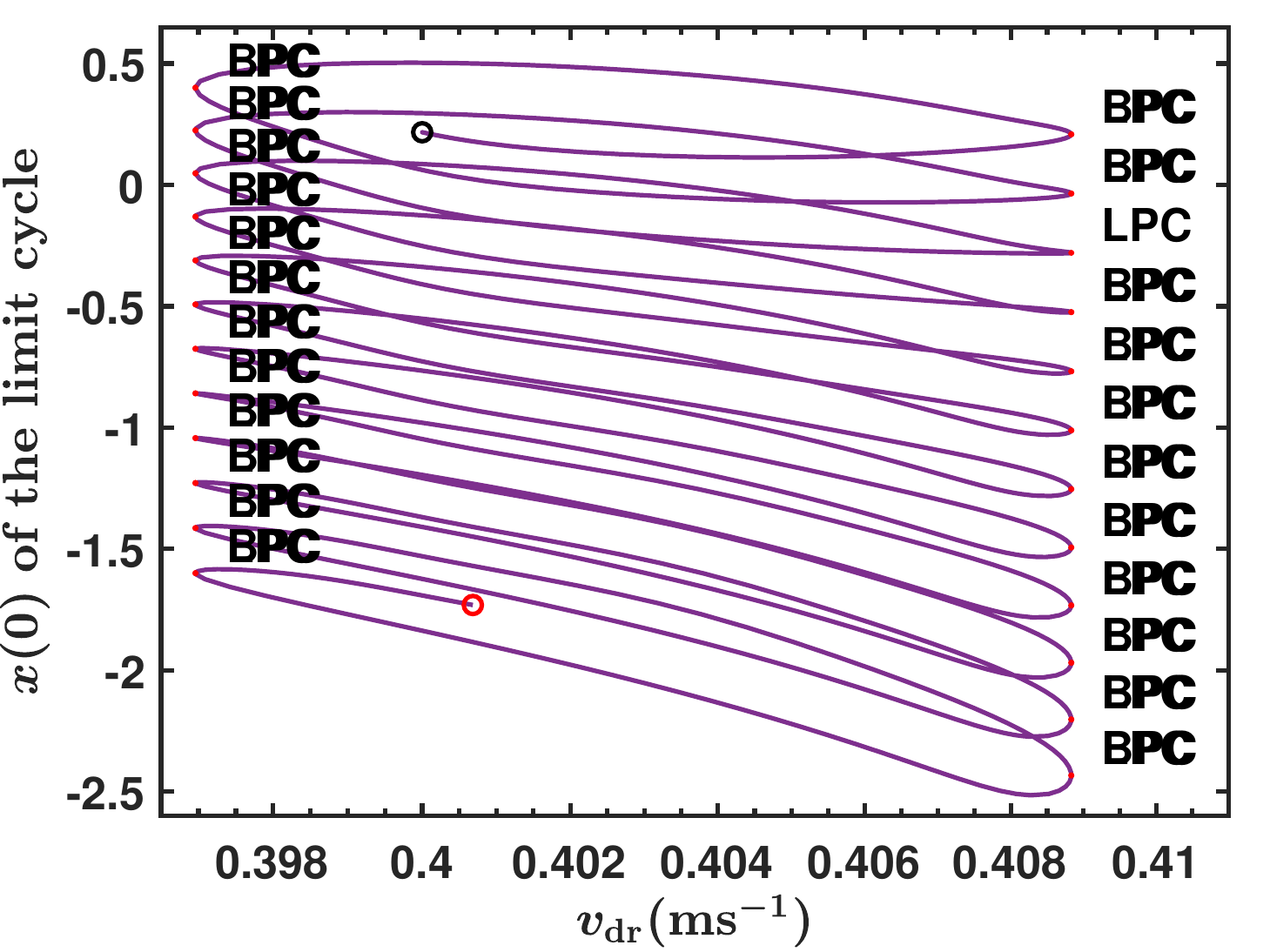}
\includegraphics[width=0.329\textwidth]{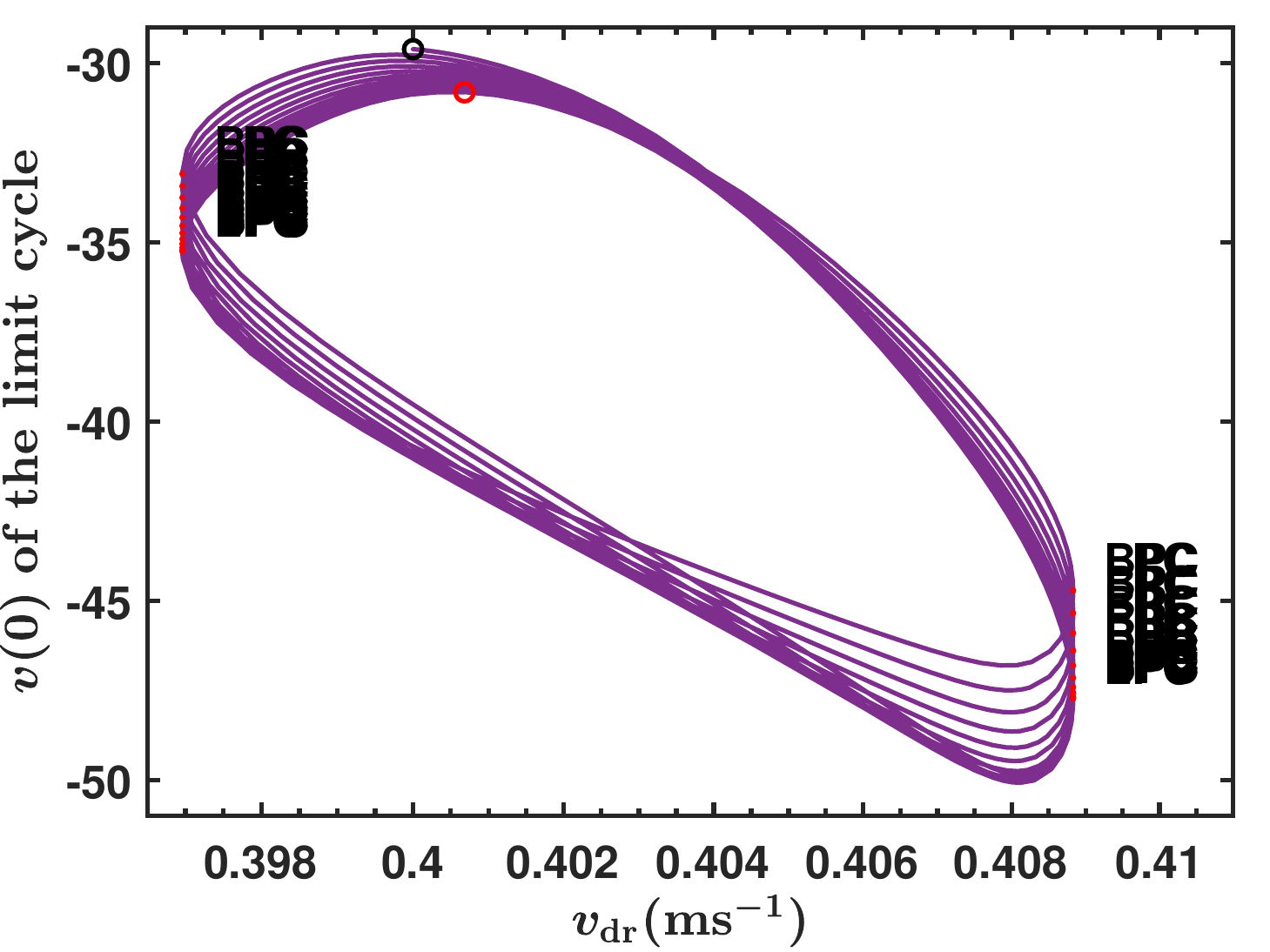}
\caption{Closed $\langle W_{\rm cyc}\rangle-v_{\rm dr}$ curve and unclosed $x(0)-v_{\rm dr}$ and $v(0)-v_{\rm dr}$ curves at zero temperature. The red circle at $v_{\rm dr}=0.4\rm m/s$ is the starting point of this batch of continuation and the black circle is the end point. For the $\langle W_{\rm cyc}\rangle-v_{\rm dr}$ curve, the starting and end points are on the same closed curve, while for the $x(0)-v_{\rm dr}$ and $v(0)-v_{\rm dr}$ curves, the starting and end points are at the two ends of an unclosed curve. Here BPC represents ``Branch Point of Cycles'' and LPC represent ``Limit Point bifurcation of Cycles'' \cite{MatcontRef}. Parameters: $\eta=3.0$, $\mu=4\times10^4\rm s^{-1}$, ${\it\Theta}=0$ and others are given in Sec. \ref{Langevindynamicssimulation}.\ref{ParametersUsed}.}
\label{BifurcationUnclosed}
\end{figure}

\begin{figure}[H]
 \centering
\includegraphics[width=0.329\textwidth]{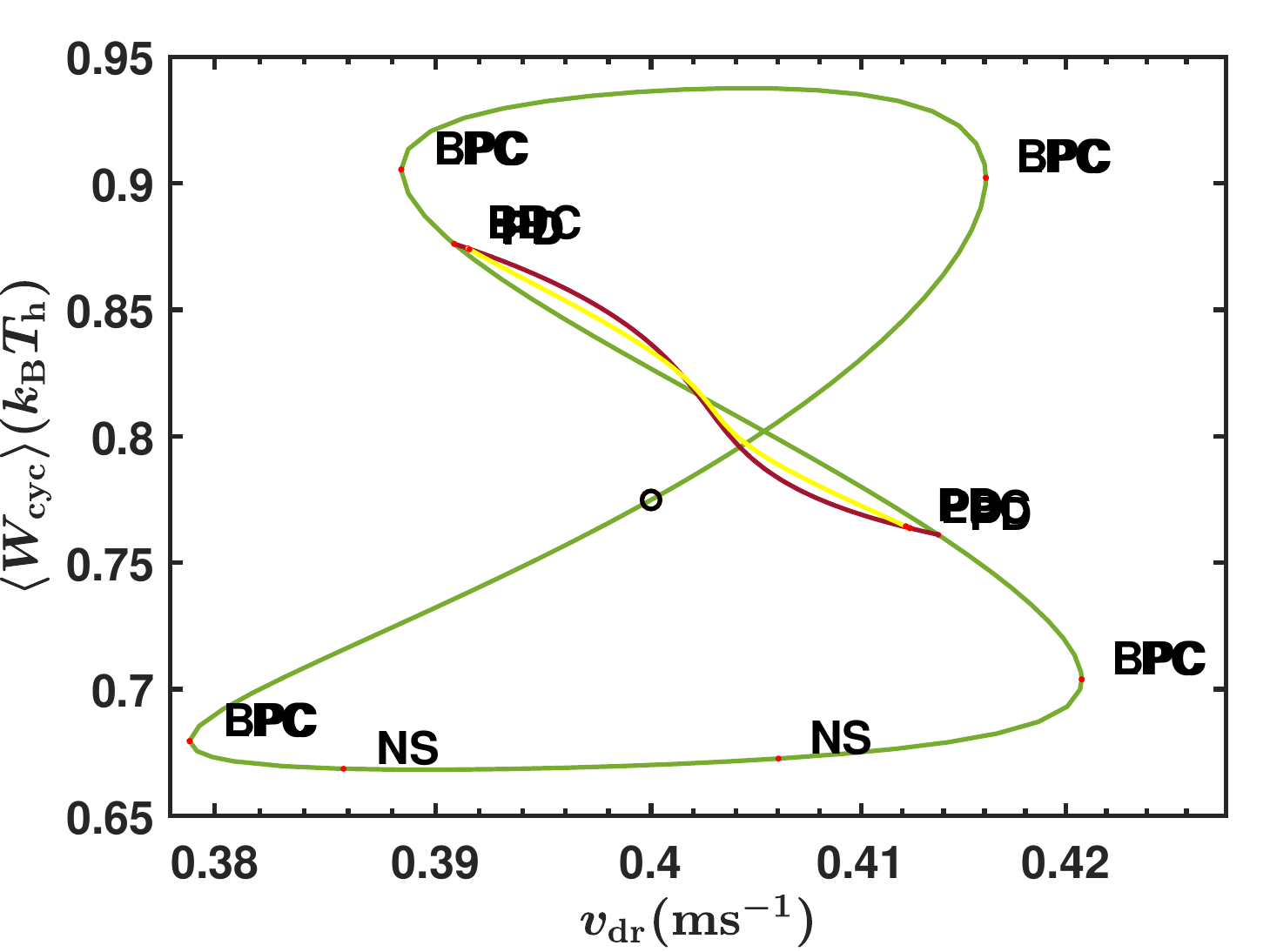}
\includegraphics[width=0.329\textwidth]{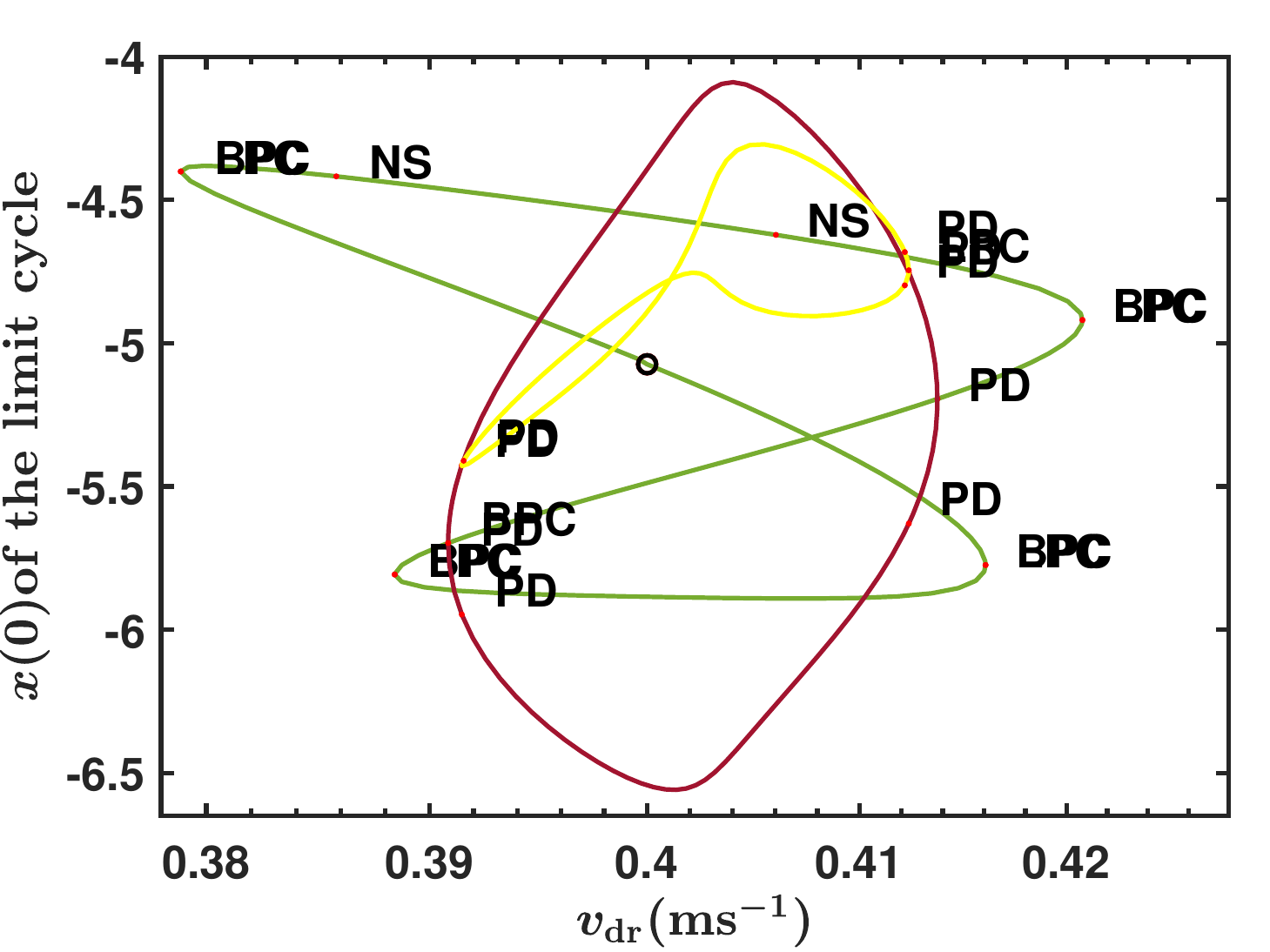}
\includegraphics[width=0.329\textwidth]{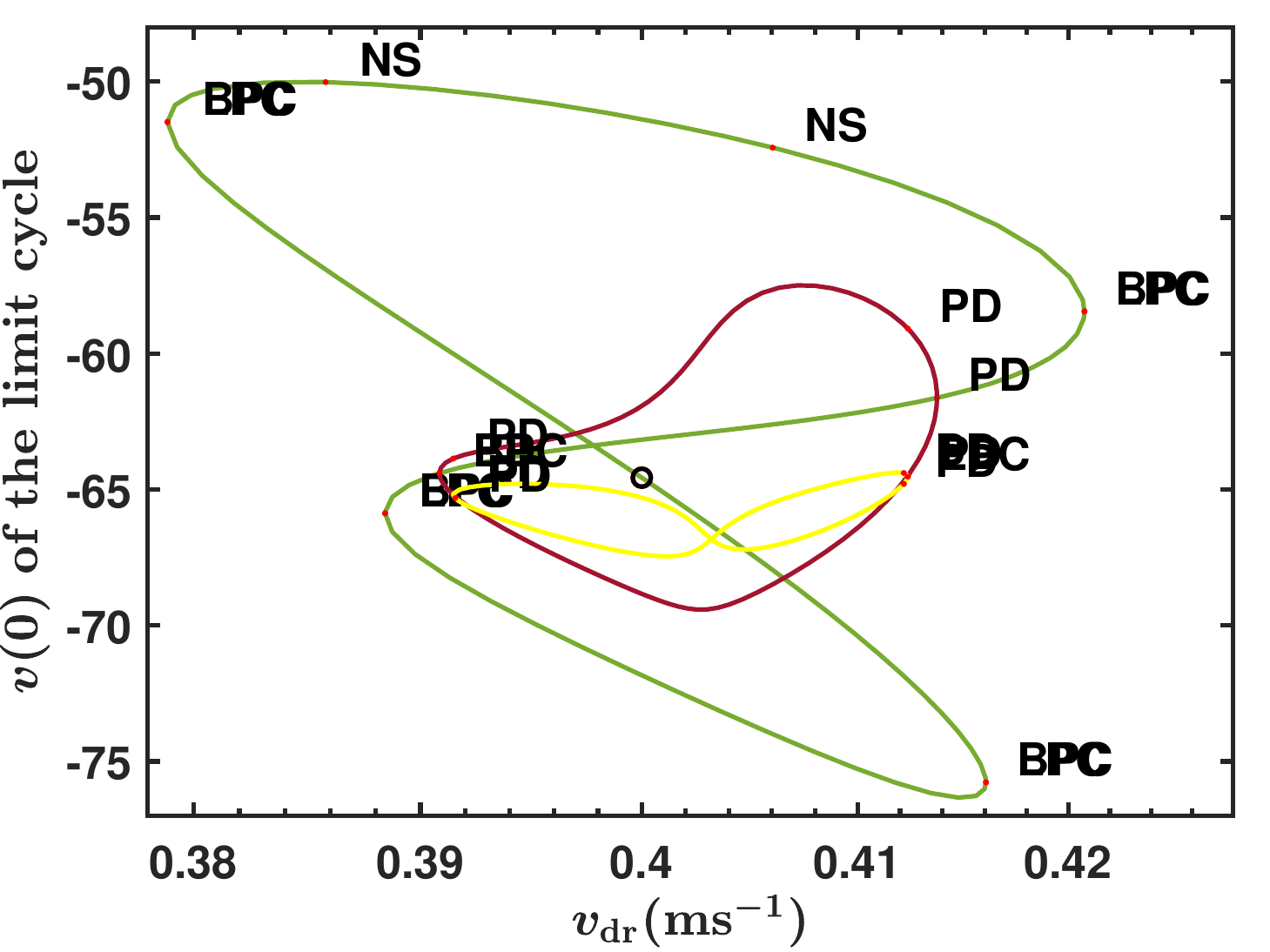}
\caption{Closed $\langle W_{\rm cyc}\rangle-v_{\rm dr}$, $x(0)-v_{\rm dr}$ and $v(0)-v_{\rm dr}$ curves at zero temperature. The red circle at $v_{\rm dr}=0.4\rm m/s$ is the starting point of this batch of continuation and the black circle is the end point. Because the two circles coincide, we can only see the black one. The dark green loop has a cycle number period $P_{\rm c.n.}=6$ (Eq. \ref{eqn:cyclenumperiod}). The dark red curve is a period doubling branch of the dark green one while the yellow curve is a period doubling branch of the dark red one. Here BPC represents ``Branch Point of Cycles''; LPC represents ``Limit Point bifurcation of Cycles'' and PD represents ``Period Doubling points'' \cite{MatcontRef}. Parameters: $\eta=3.0$, $\mu=4\times10^4\rm s^{-1}$, ${\it\Theta}=0$ and others are given in Sec. \ref{Langevindynamicssimulation}.\ref{ParametersUsed}.}
\label{BifurcationClosed}
\end{figure}

\begin{figure}[H]
 \centering
\includegraphics[width=\textwidth]{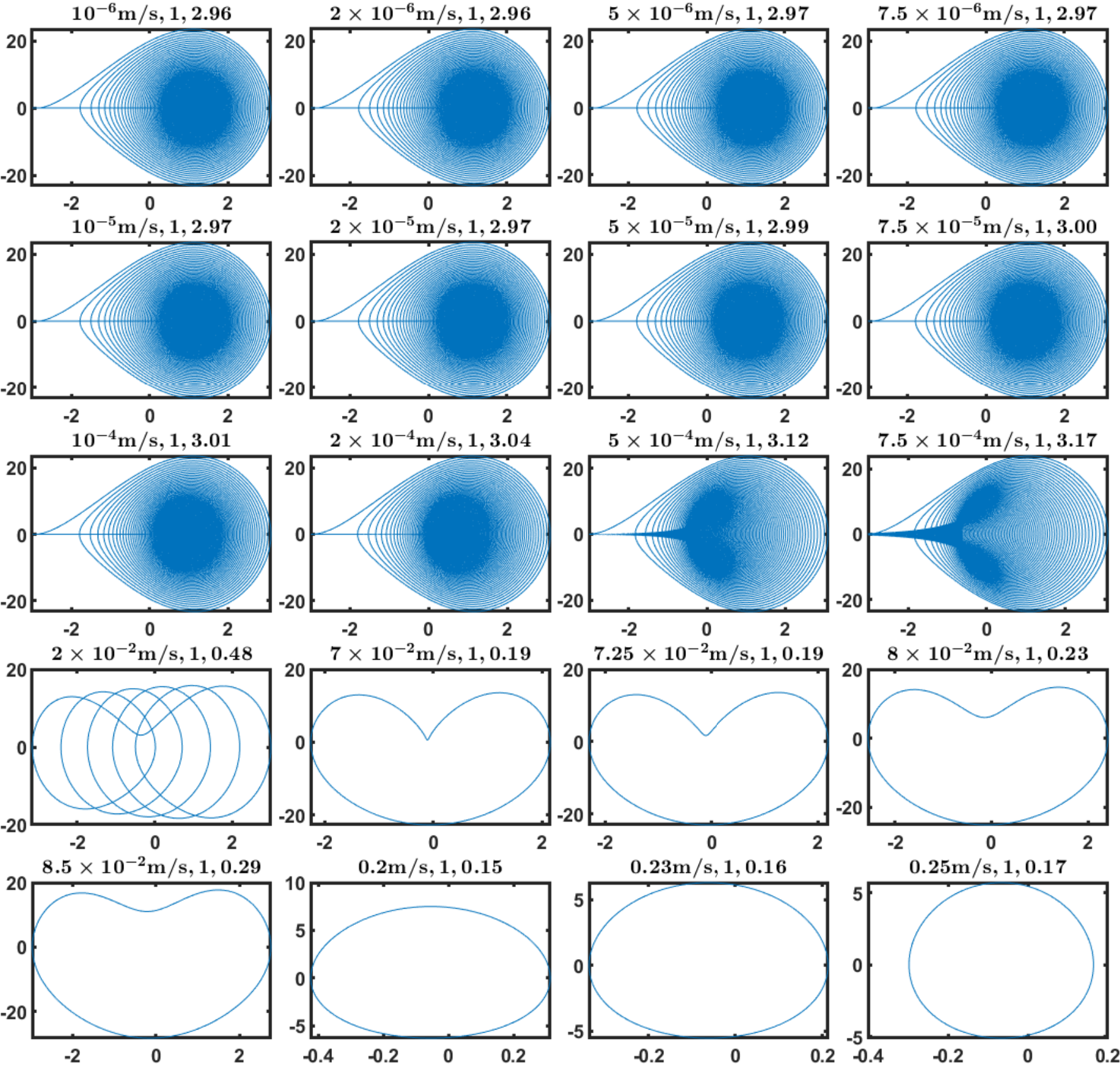}
\end{figure}

\begin{figure}[H]
 \centering
\includegraphics[width=\textwidth]{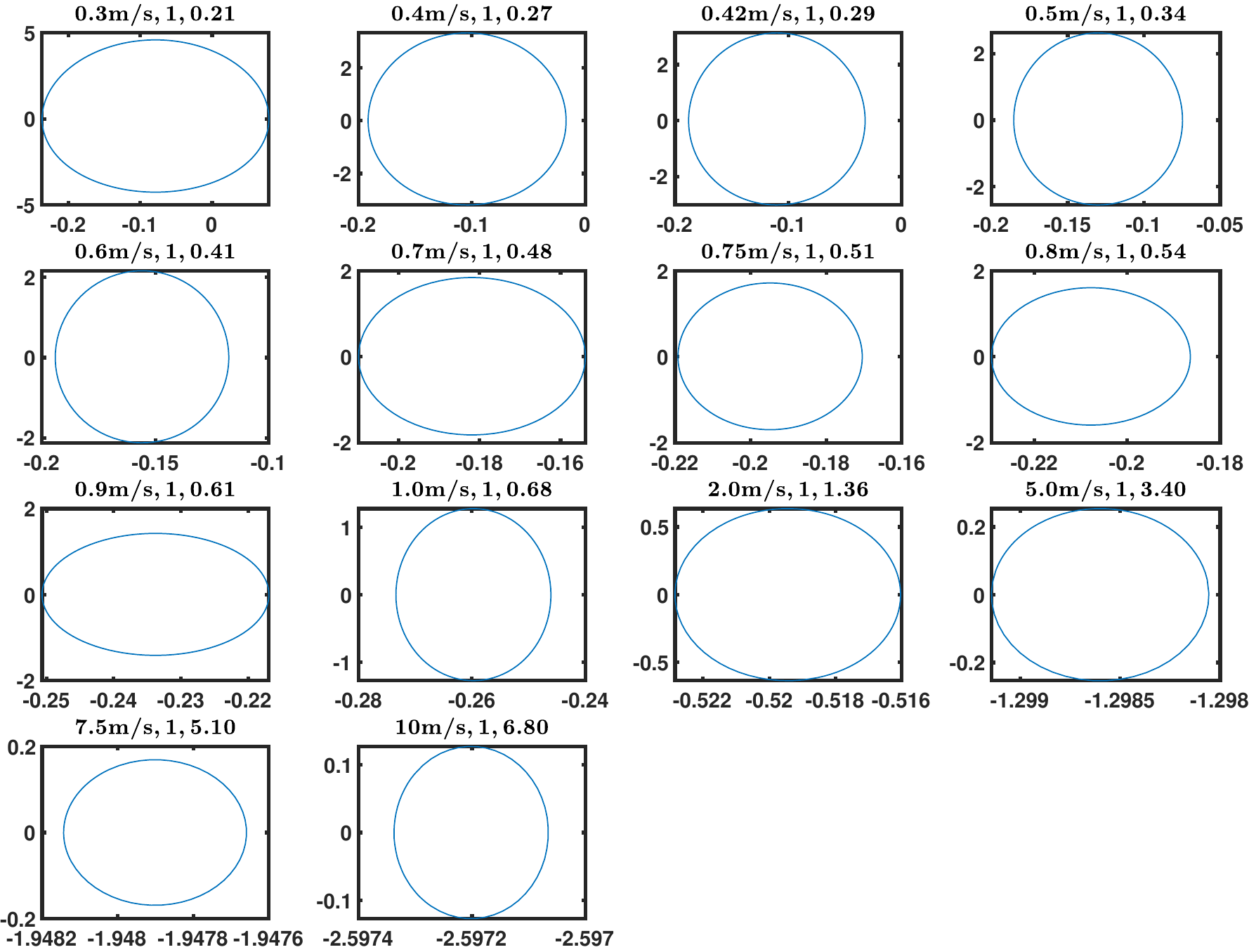}
\caption{Limit cycles with the cycle number period $P_{\rm c.n.}=1$ at different driving velocity $v_{\rm dr}$ corresponding to some red circles on the blue backbone curve in Figure \ref{BifurcationFigures} (or the main text Figure 6). The title of each limit cycle has three numbers: the driving velocity $v_{\rm dr}$, the cycle number period $P_{\rm c.n.}$ and the mean cycle work $\langle W_{\rm cyc}\rangle=W_{\rm cyc}(k_{\rm B}T_{\rm h})$. Here $k_{\rm B}T_{\rm h}$ is the same as that in Figure 6 in the main text. All the $x$-coordinates are the particle's nondimensional position $z$ and the $y$-coordinates are the the particle's nondimensional velocity $\dot z$. The first 9 limit cycles at $v_{\rm dr}\in[10^{-6},7.5\times10^{-4}]\rm m/s$ corresponding the stick-slip solutions, cf. Figure \ref{fig:StickSlipsSingleOnecycleXT0}. At and after $v_{\rm dr}=10^{-3}\rm m/s$, the interference from the last cycle occurs and there is a short transient stage of torus and nonperiodic solutions as shown in Figure \ref{fig:limitcyclesnoperiod}, with the limit cycle at $v_{\rm dr}=2\times10^{-2}\rm m/s$ as an exception, whose $P_{\rm c.n.}=2$ and whose shape is a little complicated. At $v_{\rm dr}\in[7\times10^{-2},8.5\times10^{-2}]\rm m/s$, the limit cycle becomes one simple loop with a notch. We note that in this velocity regime the blue backbone curve in Figure \ref{BifurcationFigures} has more than one solution at a specific $v_{\rm dr}$. At $v_{\rm dr}\geq0.2\rm m/s$ the blue backbone curve has only one solution and the notch disappears and the loop approaches to an ellipse.}
\label{fig:limitcyclesbackbone}
\end{figure}

\begin{figure}[H]
 \centering
\includegraphics[width=\textwidth]{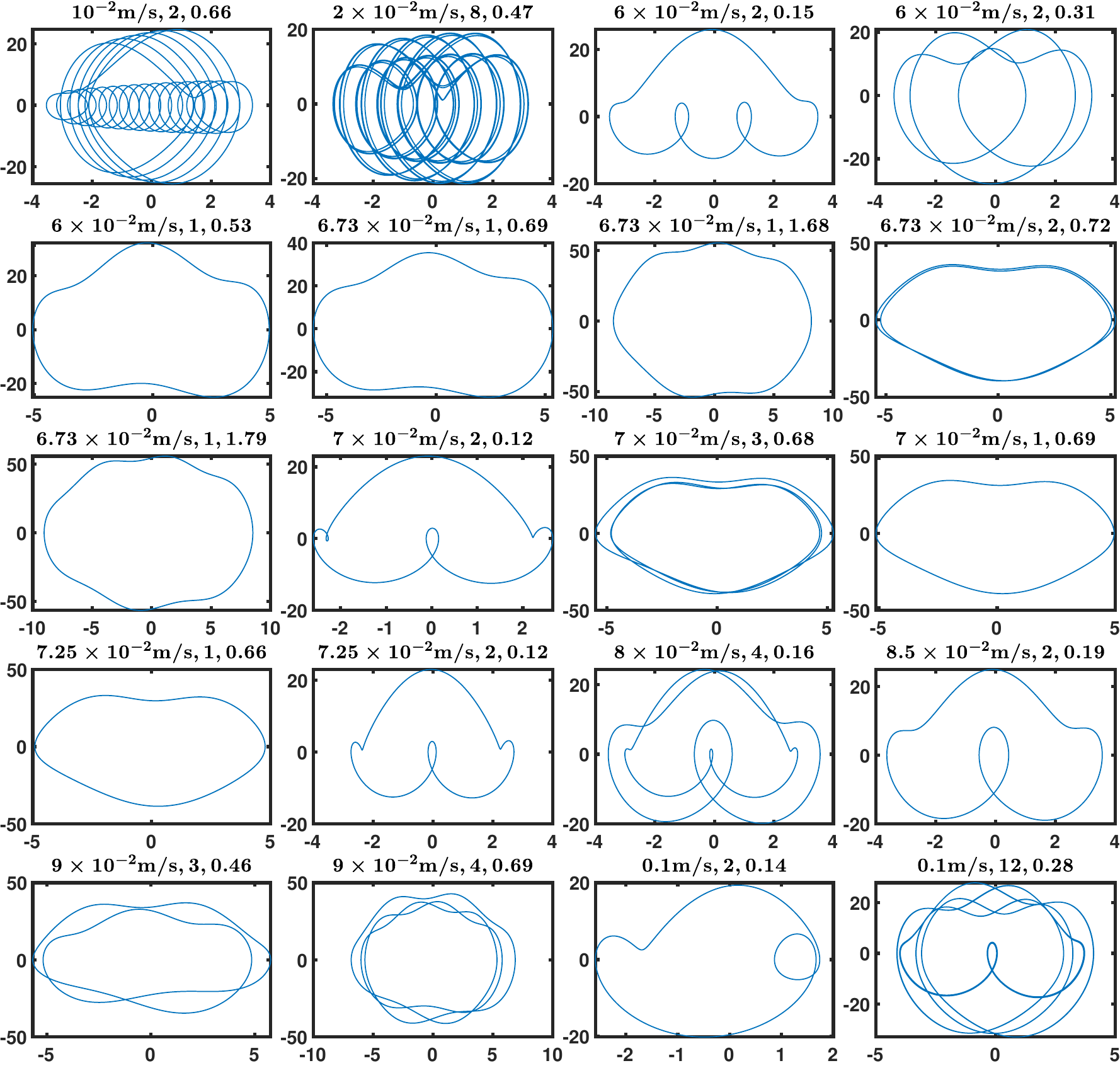}
\end{figure}

\begin{figure}[H]
 \centering
\includegraphics[width=\textwidth]{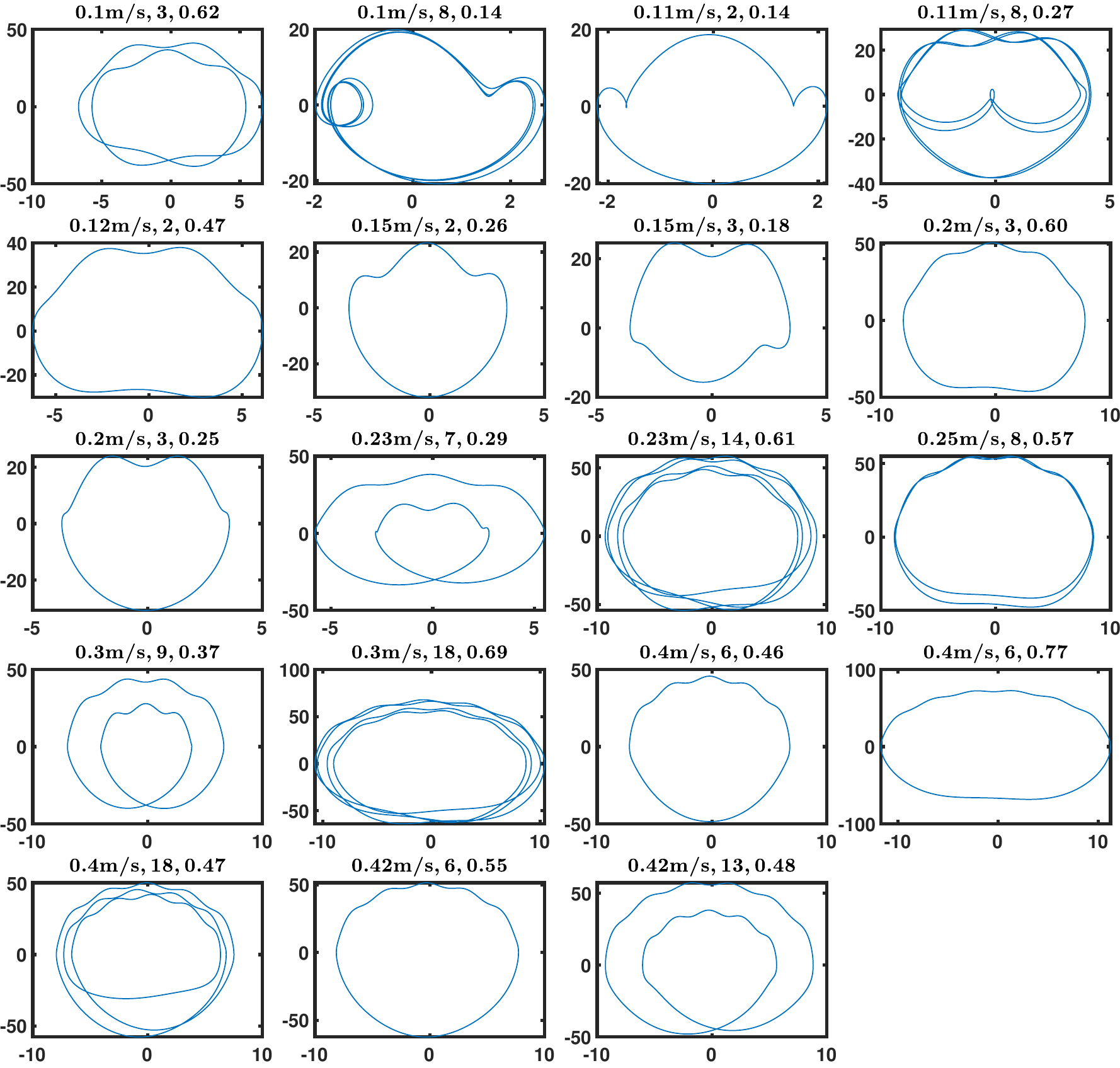}
\caption{Limit cycles of some of the red circles out of the blue backbone curve in Figure \ref{BifurcationFigures} (or the main text Figure 6). The title of each limit cycle has three numbers: the driving velocity $v_{\rm dr}$, the cycle number period $P_{\rm c.n.}$ and the mean cycle work $\langle W_{\rm cyc}\rangle(k_{\rm B}T_{\rm h})$. Here $k_{\rm B}T_{\rm h}$ is the same as that in Figure 6 in the main text. All the $x$-coordinates are the particle's nondimensional position $z$ and the $y$-coordinates are the the particle's nondimensional velocity $\dot z$. These limit cycles are solutions on the isolated loops or the period doubling branches in the main text Figure 6. We can see the shapes of these limit cycles are peculiar.}
\label{fig:limitcycleshigherperiods}
\end{figure}

\begin{figure}[H]
 \centering
\includegraphics[width=\textwidth]{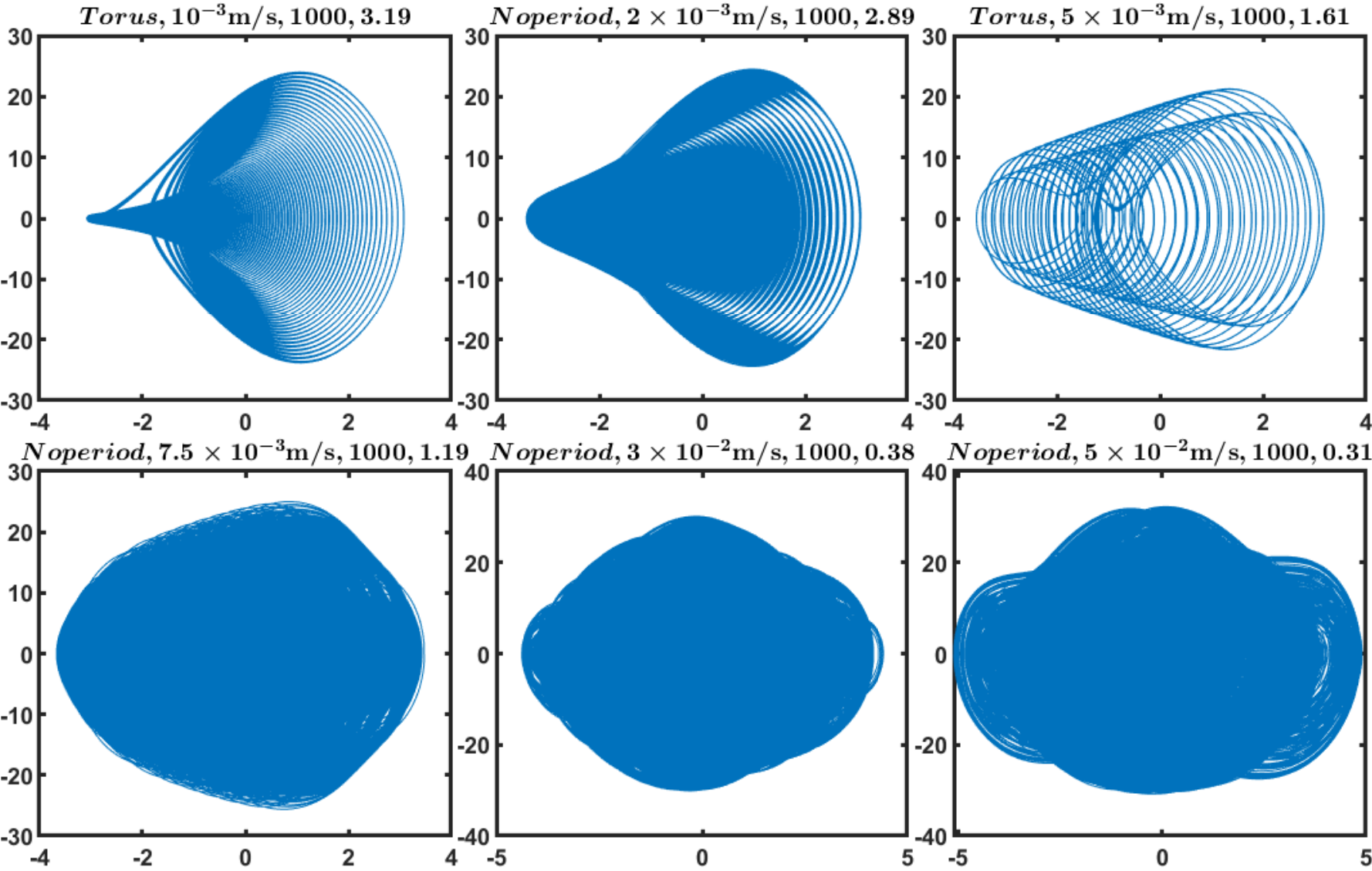}
\caption{Phase curves of solutions with a torus as the limit or with no period in Figure \ref{BifurcationFigures} (or the main text Figure 6). The title of each subfigure has four fields: Torus or Noperiod, the driving velocity $v_{\rm dr}$, the number of simulation cycles (different from Figure \ref{fig:limitcyclesbackbone} and \ref{fig:limitcycleshigherperiods}) and the mean cycle work $\langle W_{\rm cyc}\rangle(k_{\rm B}T_{\rm h})$ averaged over the simulation cycles. Here $k_{\rm B}T_{\rm h}$ is the same as that in Figure 6 in the main text. The initial values of the six cases are inherited from the end values of the last simulation cycle used to calculate the corresponding red circles in Figure \ref{BifurcationFigures} so that we can make sure that the steady state has already been achieved. All the $x$-coordinates are the particle's nondimensional position $z$ and the $y$-coordinates are the the particle's nondimensional velocity $\dot z$. These solutions are in the velocity weakening regime after the stick-slip regime and before the resonance regime (Figure \ref{fig:StickSlipsSingleOnecycleXT0} and Sec. \ref{sec:apndRD}.\ref{sec:transitionEDFwdv}). So the torus and nonperiodic solutions are a mark of transition between the stick-slip and the resonance regime. Other aspects of these six solutions are given in Figure \ref{fig:ToriandNonPeriodSol}.}
\label{fig:limitcyclesnoperiod}
\end{figure}

\begin{figure}[H]
\centering
 \begin{minipage}{0.49\textwidth}
 \centerline{
\includegraphics[width=\textwidth]{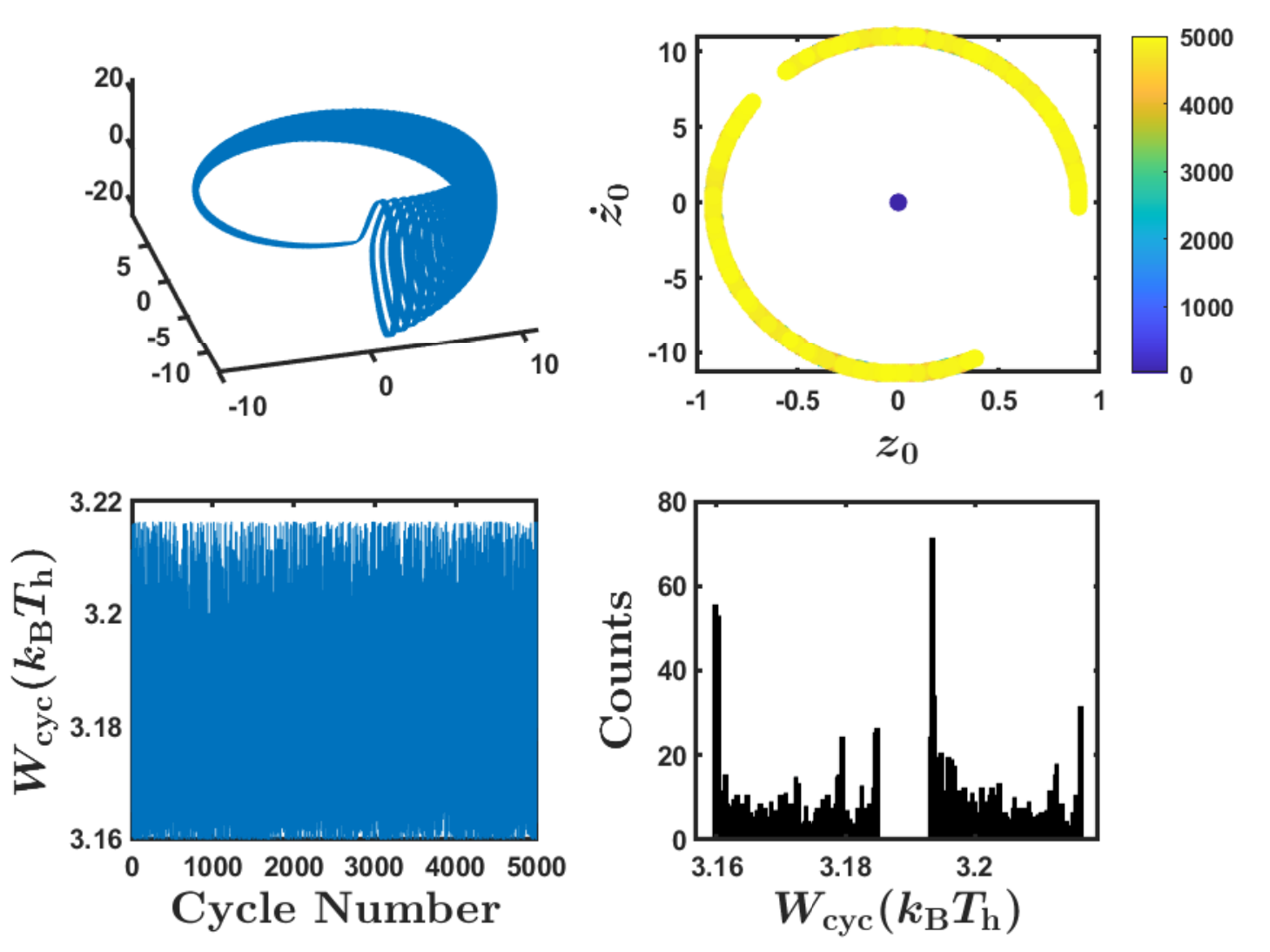}
 }
 \centerline{(a) $v_{\rm dr}=10^{-3}\rm m/s$}
 \end{minipage}
 \begin{minipage}{0.49\textwidth}
 \centerline{
\includegraphics[width=\textwidth]{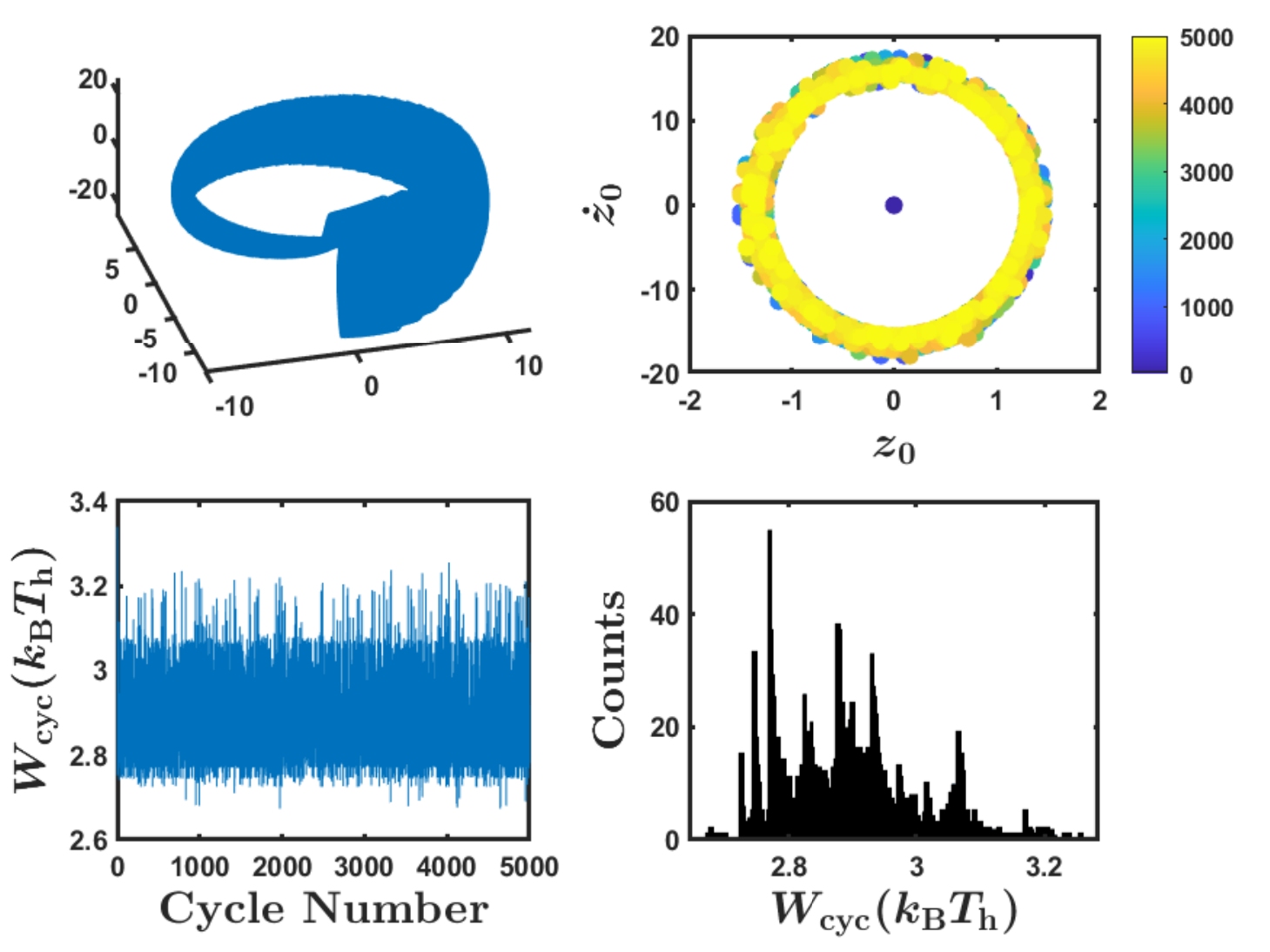}
 }
 \centerline{(b) $v_{\rm dr}=2\times10^{-3}\rm m/s$}
 \end{minipage}
 \begin{minipage}{0.49\textwidth}
 \centerline{
\includegraphics[width=\textwidth]{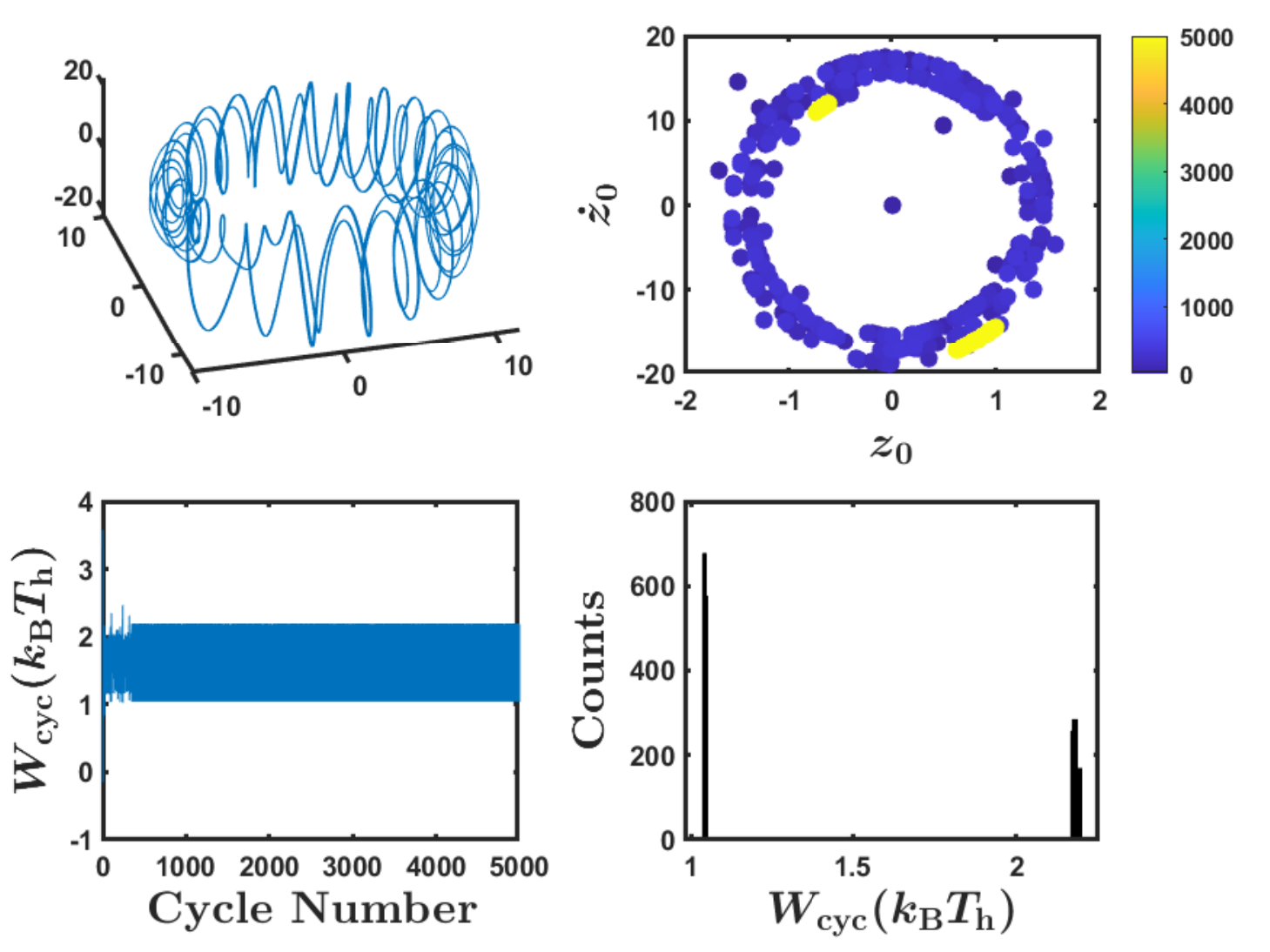}
 }
 \centerline{(c) $v_{\rm dr}=5\times10^{-3}\rm m/s$}
 \end{minipage}
 \begin{minipage}{0.49\textwidth}
 \centerline{
\includegraphics[width=\textwidth]{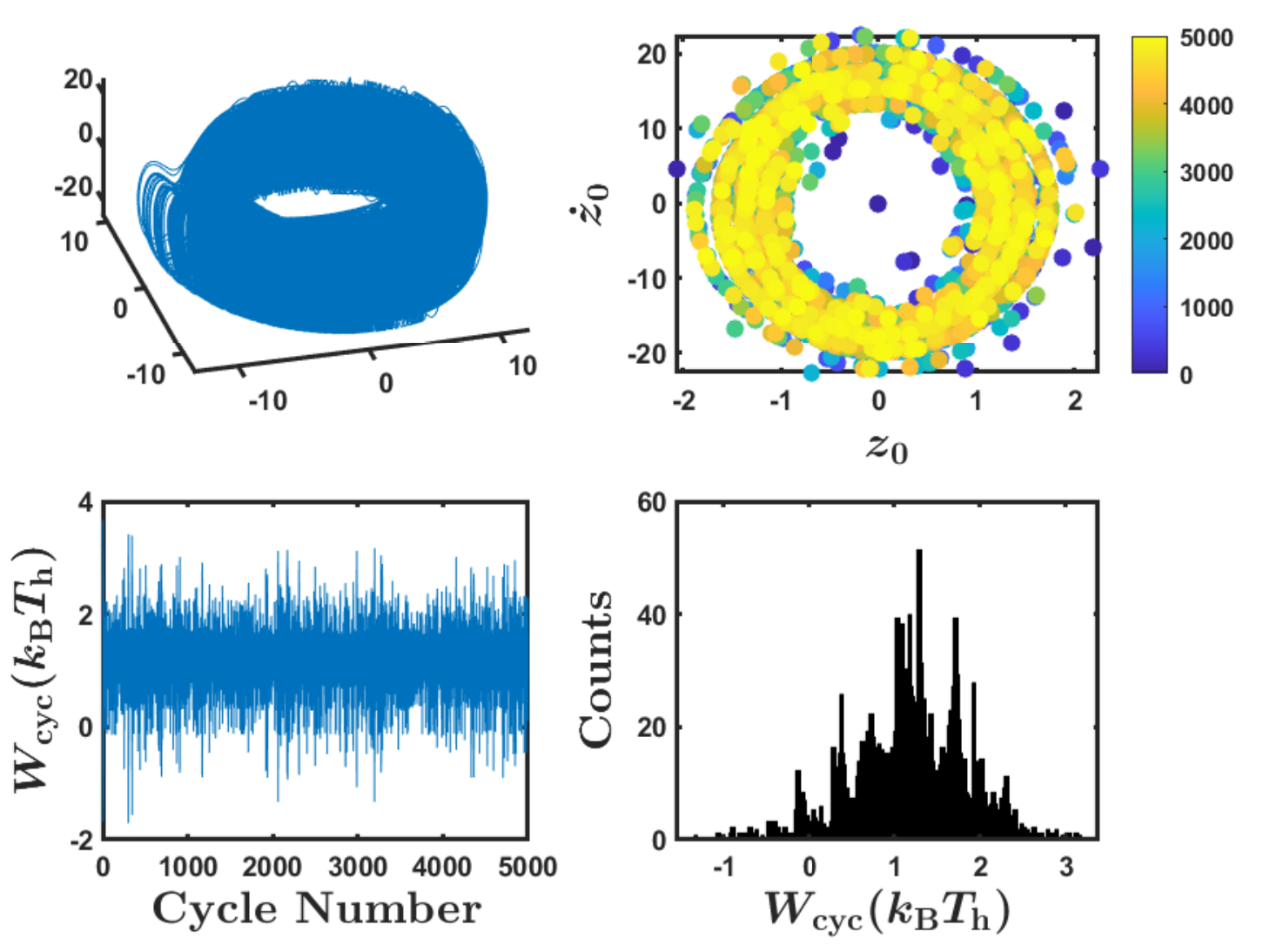}
 }
 \centerline{(d) $v_{\rm dr}=7.5\times10^{-3}\rm m/s$}
 \end{minipage}\\
 \begin{minipage}{0.49\textwidth}
 \centerline{
\includegraphics[width=\textwidth]{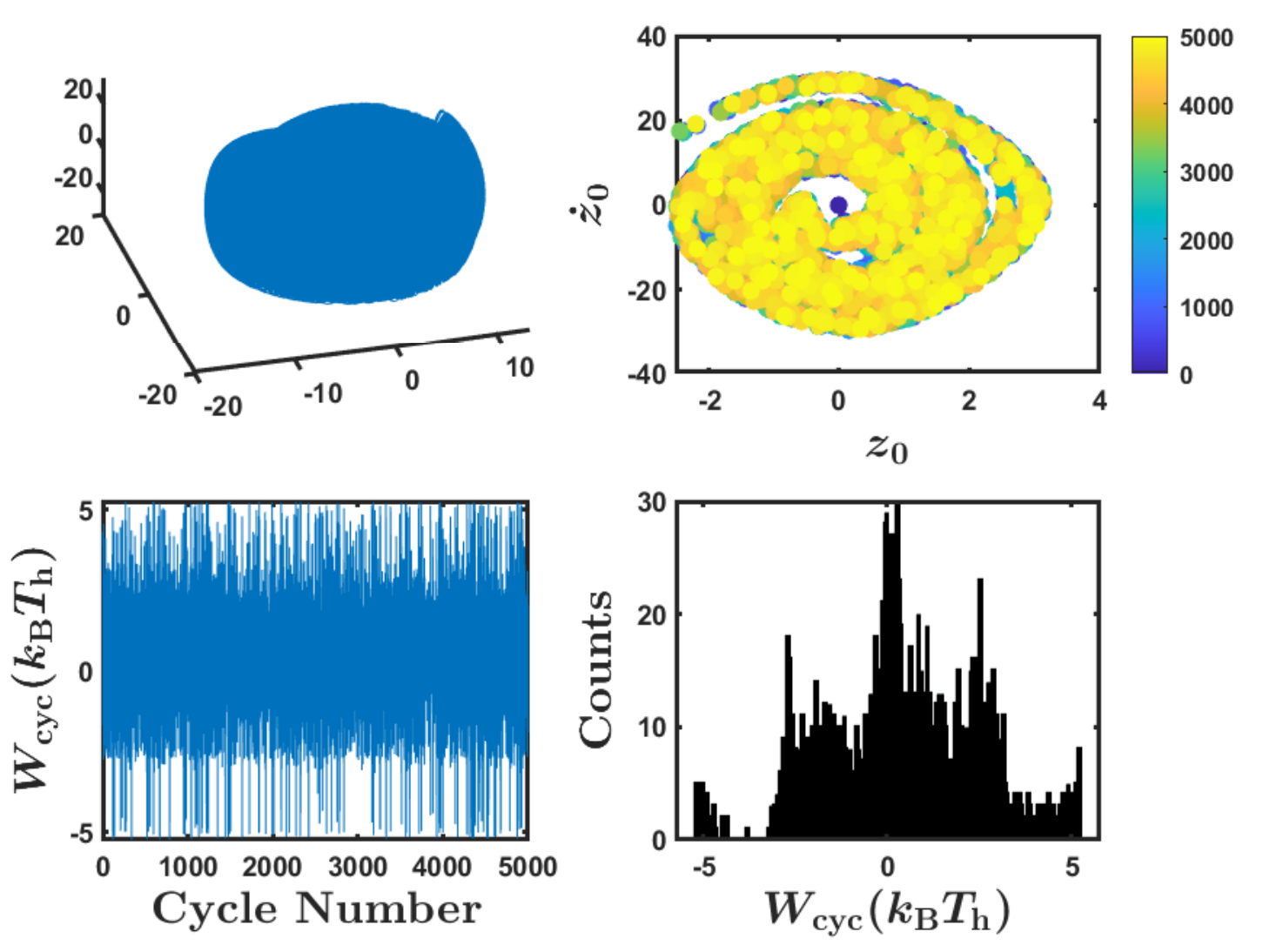}
 }
 \centerline{(e) $v_{\rm dr}=3\times10^{-2}\rm m/s$}
 \end{minipage}
 \begin{minipage}{0.49\textwidth}
 \centerline{
\includegraphics[width=\textwidth]{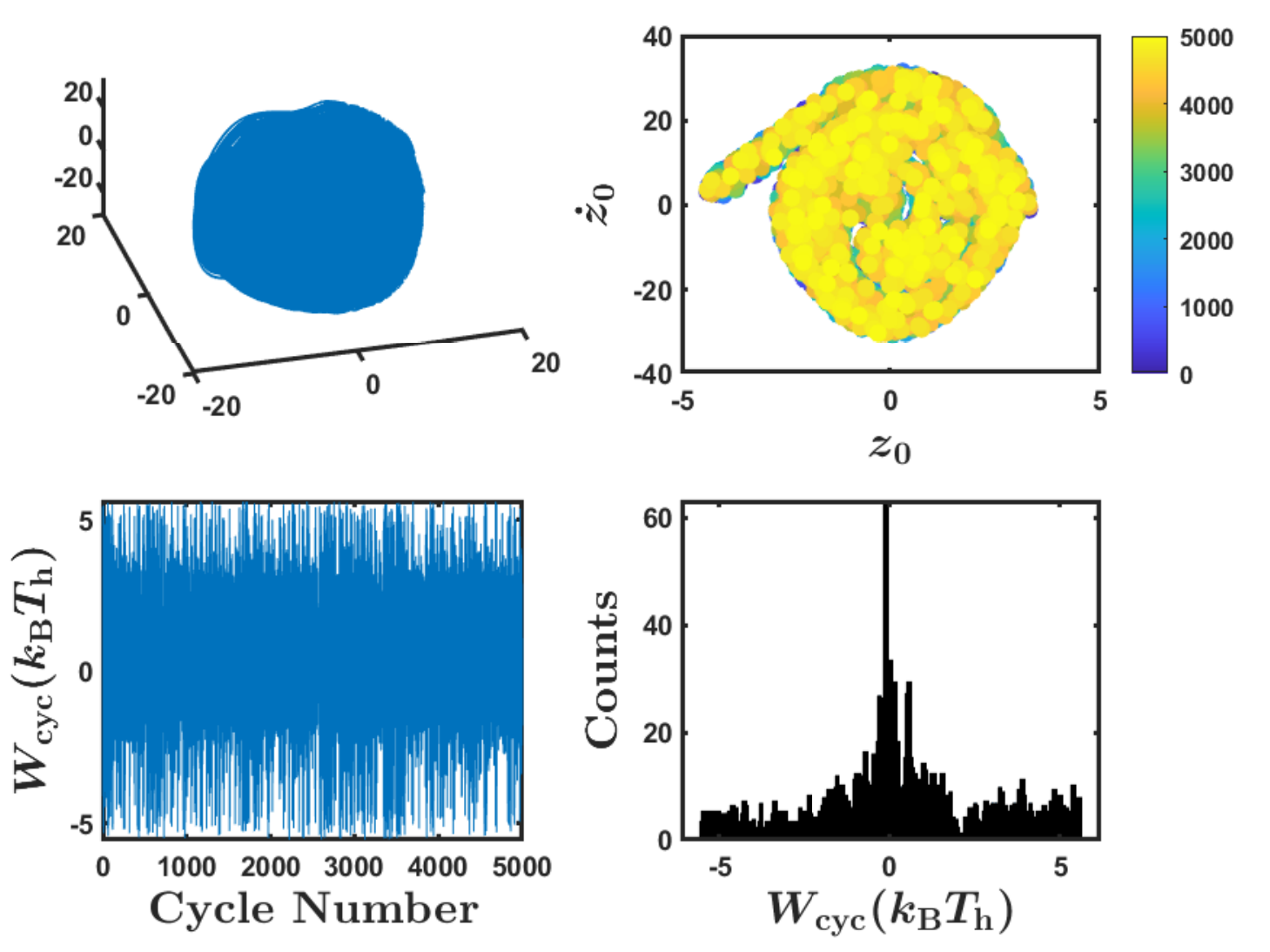}
 }
 \centerline{(f) $v_{\rm dr}=5\times10^{-2}\rm m/s$}
 \end{minipage}
\caption{Other aspects of the six solutions in Figure \ref{fig:limitcyclesnoperiod}. In each subfigure at different driving velocity $v_{\rm dr}$, the top left subgraph plots the phase trajectory embedded in the 3-dimensional cylindrical coordinate space $(\theta,r,h)=(\tilde v\tau,z-\tilde v\tau+10,\dot z)$. The top right subgraph plots the Poincare (or stroboscopic) map \cite{Seydel2010Ch7Stability} sampled at the starting point of each cycle, i.e. the set of the intial point $(z_0,\dot z_0)$ of each cycle with the cycle number indicated by the color of the points ranging from blue at the beginning to yellow at the end. The bottom left subgraph is the cycle work from the first to the last simulation cycle during the simulation time range. The bottom right subgraph is the count distribution of the cycle work $W_{\rm cyc}$ of the total 5000 simulation cycles. In (a) and (c), we can recognize two clear tori. In (b) and (d), there seems to be several tori for the particle to shift from one to another. In (e) and (f), there is no recognizable period.}
\label{fig:ToriandNonPeriodSol}
\end{figure}

\end{enumerate}
\subsection{More on the $\eta>4.6$ cases}

\begin{figure}[H]
 \centering
\includegraphics[width=0.49\textwidth]{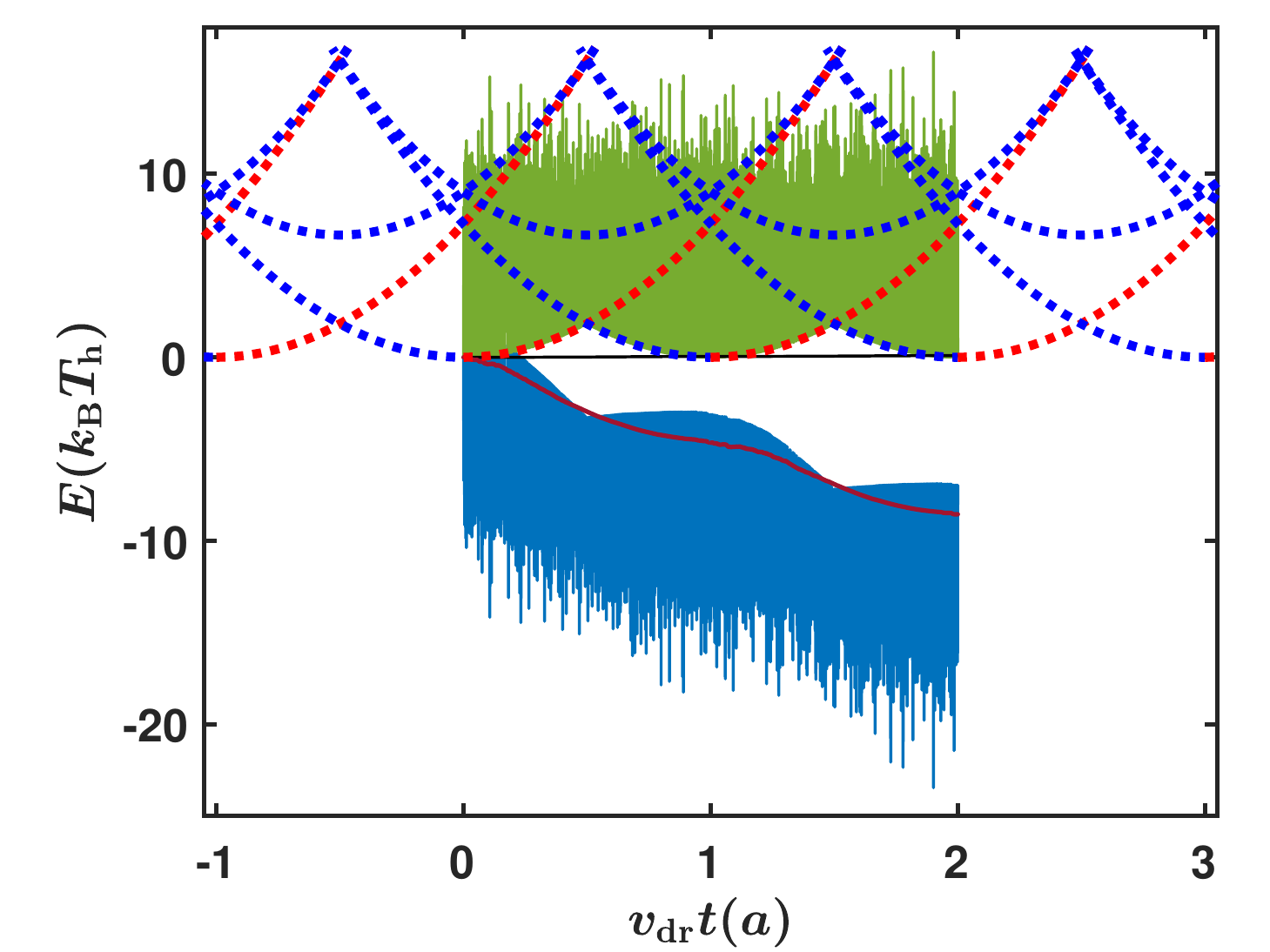}
\includegraphics[width=0.49\textwidth]{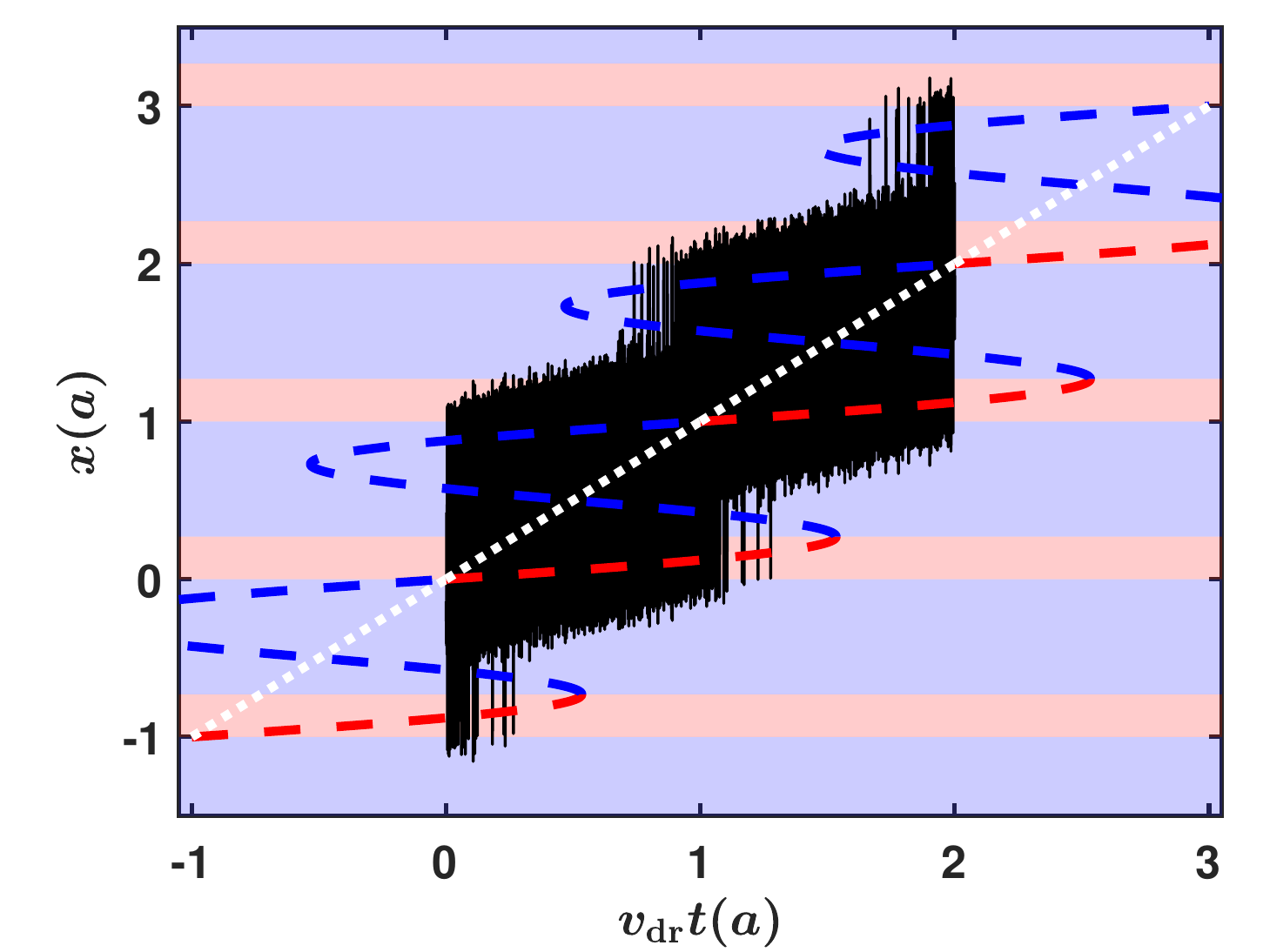}
\caption{Energy and displacement curves at $\eta=8$, ${\it\Theta}_{\rm h,c}=0.4,0.04$ and $\mu=4\times10^{4}\rm s^{-1}$. Curve colors and types are the same as those in the main text Figure 2(A) and (B) and $k_{\rm B}T_{\rm h}$ is of the same value as that in the main text Figure 2(A). The complete picture of the balanced resultant potential curves (the dotted red and blue curves in the left subfigure) and the loci of the balanced point (the dashed red and blue curves in the right subfigure) are given as a complement for the main text Figure 7(B) and (C). Comparing with the $\eta<4.6$ cases, we can see from the left subfigure that a dotted red local minimum branch goes through two engine cycles, in consistency with that it takes more than one cycle for the middle global minimum in the main text Figure 7(F1) to go leftward and upward through one of the hot zones and out of it at the instant of the BCP (F5) of the next cycle. Symmetrically, a dotted blue local minimum branch goes through two engine cycles from the instant of the FCP of the last cycle to the end of the current cycle. A dotted blue energy barrier peak branch even goes through three engine cycles from its appearance at the instant of the FCP of the last cycle to its disappearance at the instant of the BCP of the next cycle. The dashed red and blue curves in the right subfigure are of similar characteristic. So there are two hot and one cold local minima and two energy barriers at the beginning of one cycle while one hot and two cold local minima and two energy barriers at the end, cf. the main text Figure 7(F2) and (F6). In the middle of one cycle there are two hot and two cold local minima and three energy barriers, cf. the main text Figure 7(F4). Please pay attention to the representation of the local mimima and the energy barrier peaks by the dotted curves in the left subfigure and the dashed curves in the right subfigure.
%Comparing with the $\eta<4.6$ cases, we can see from the left figure that in one engine cycle, e.g. $v_{\rm dr}t/a\in[0,1]$, the hot local minimum branch intrudes into the next engine cycle and the cold local minimum branch intrudes into the last engine cycle while the energy barrier branch between them extrudes out of both ends of the current engine cycle. 
Simulation results in two consecutive steady state cycles are plotted here. In the middle of either engine cycle, the particle covers two local minima and most of the time stays around the cold local minimum point represented by the bottommost dotted blue branch [marked by the solid purple segment on the bottommost dotted blue curve in Figure \ref{fig:WorkLowerBoundHighEtas}(a)] in the left subfigure, so the middle stage of the work curve is parallel to the bottommost dotted blue branch in either cycle. However, at the beginning and end of either cycle, the particle covers all the three local minima and tends to neglect the two energy barriers (also see the right subfigure) so that its distribution center approaches the driver center and the work output at the beginning and end of one engine cycle are both reduced, leading to the beginning and end stages of the work curve not parallel to the bottommost dotted blue branch in either cycle in the left subfigure. Although very similar, the work curves during the two cycles are not identical, consistent with the nonzero standard deviation of the $\eta=8$ circle in the main text Figure 7(A1). %Please pay attention to the correspondance among the dotted blue and red curves in the left subfigure, the dashed blue and red curves in the right subfigure and the resultant potential energy curves in the main text Figure 7(F1)-(F7). For instance, the bottommost blue curve in the interval $[0,1]$ in the left subfigure represents the resultant potential energy transition from the right local minimum point in (F1) to the middle global mimimum point in (F7) in the main text Figure 7, and this very local minimum point's position is expressed by the upmost dashed blue curve in the zone $[0,1]\times[0,1]$ in the right subfigure. In one engine cycle, the right energy barrier in (F1) turns into the left energy barrier in (F7) of the main text Figure 7, and this transition process is expressed by the middle dashed blue curve in the zone $[0,1]\times[0,1]$ in the right subfigure and the resultant potential energy at this energy barrier is represented by the dotted blue curve above and neighbor to the bottommost dotted blue curve in the interval $[0,1]$ in the left subfigure.
Other parameters are given in Sec. \ref{Langevindynamicssimulation}.\ref{ParametersUsed}.
}
\label{fig:Eta8EnergyDisplacement}
\end{figure}

\begin{figure}[H]
 \centering
\includegraphics[width=0.49\textwidth]{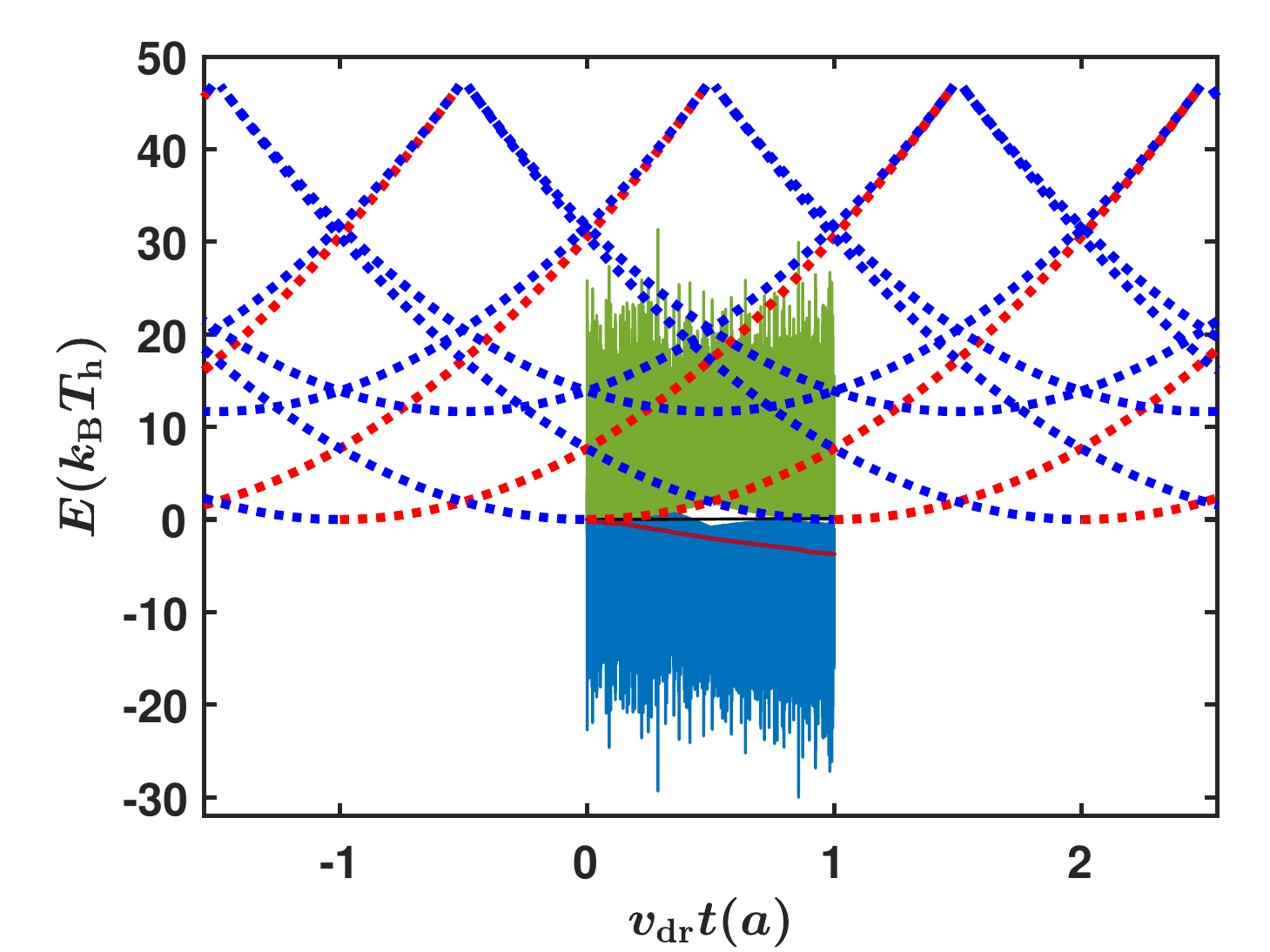}
\includegraphics[width=0.49\textwidth]{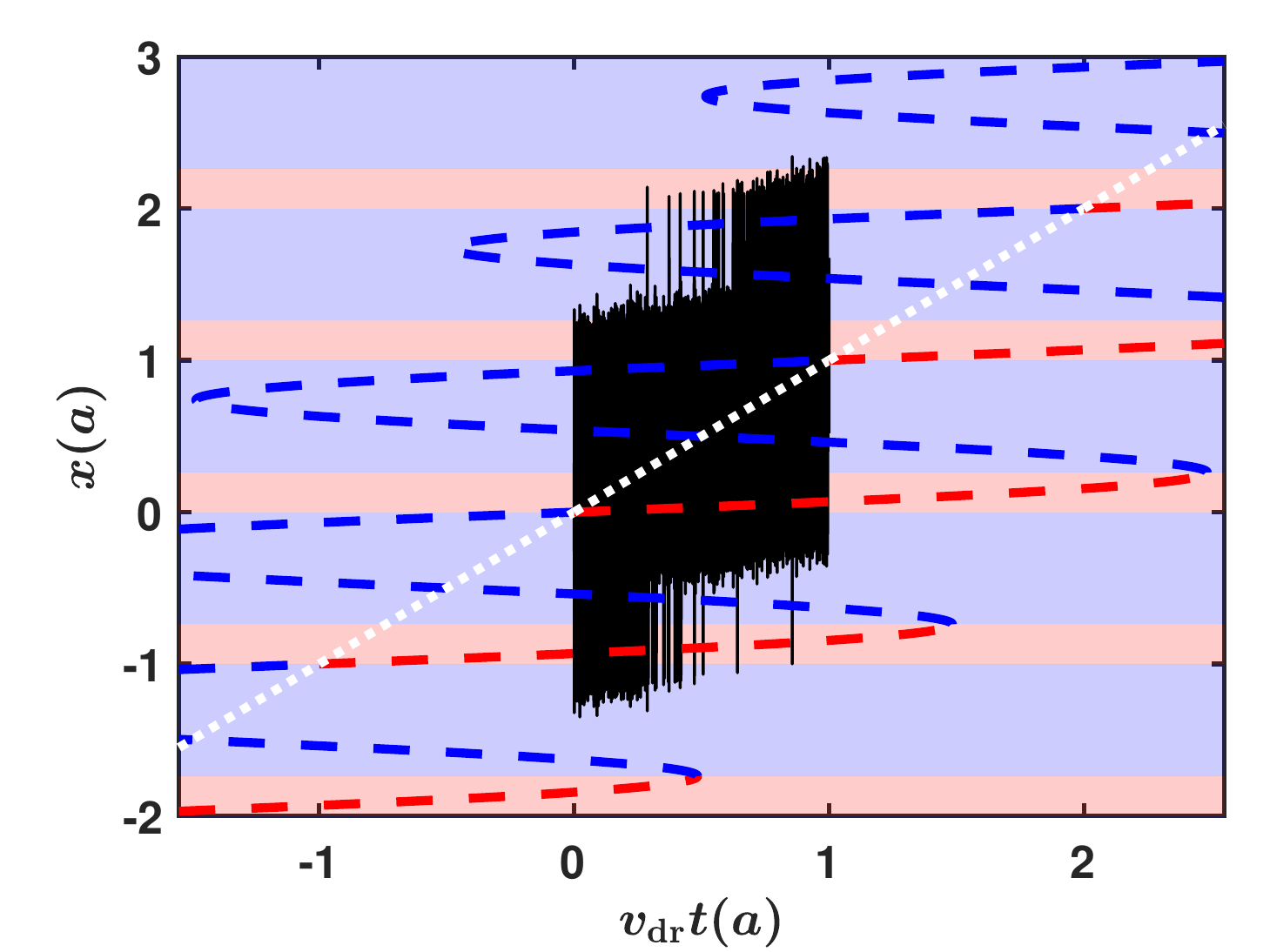}
\caption{Energy and displacement curves at $\eta=14$, ${\it\Theta}_{\rm h,c}=0.4,0.04$ and $\mu=4\times10^{4}\rm s^{-1}$. Curve colors and types are the same as those in the main text Figure 2(A) and (B) and $k_{\rm B}T_{\rm h}$ is of the same value as that in the main text Figure 2(A). The balanced resultant potential curves of the $\eta=14$ case is more complex than those of the $\eta=8$ and $\eta\leq4.6$ cases. We can see that, for instance, in the left subfigure a dotted red local minimum branch goes through 3 engine cycles and at the beginning of one engine cycle there are 3 hot and 2 cold local minima and 4 energy barrier peaks. 
Other parameters are given in Sec. \ref{Langevindynamicssimulation}.\ref{ParametersUsed}.}
\label{fig:Eta14EnergyDisplacement}
\end{figure}

In Figure \ref{fig:Eta8EnergyDisplacement} and \ref{fig:Eta14EnergyDisplacement}, we plot the energy and displacement curves at $\eta=8$ and $14$ for the subsection ``Mean Cycle Work at $\eta>4.6$'' in the main text to refer.

At the middle instant of one engine cycle, the bottommost energy barrier peak branch has the value of $V_0$ resulting from the superposition of the global minimum of the harmonic potential, which equals to zero, and a local maximum of the sinusoidal lattice potential, which equals to $V_0$. Due to the same ${\it\Theta}_{\rm h,c}=\frac{k_{\rm B}T_{\rm h}}{V_0}=0.4,0.04$ for both cases in Figure \ref{fig:Eta8EnergyDisplacement} and \ref{fig:Eta14EnergyDisplacement}, we can see from the two left subfigures that the dark green internal energy curve exceeds a similar proportion above the bottommost energy barrier peak branch at the middle instant of one engine cycle. However, for the $\eta=14$ case, the dotted red local minimum branch starting at the beginning of the last (not current) cycle, and the dotted blue local minimum branch ending at the end of the next (not current) cycle, intrude the internal energy curve more deeply than the $\eta=8$ case, %so the time of the particle's covering three wells and its position distribution center close to the driver center, is longer for the $\eta=14$ case 
so that the particle remains longer in the neighbor wells both ahead of and behind the global minimum, i.e. the particle covers three wells for a longer period (also cf. the two right subfigures in Figure \ref{fig:Eta8EnergyDisplacement} and \ref{fig:Eta14EnergyDisplacement}). Actually, from the right subfigure of Figure \ref{fig:Eta14EnergyDisplacement} we can see that the particle nearly always covers three wells during one cycle and sometimes even covers four wells in the cycle middle, rather than covering three wells only at the beginning and end of one cycle and two wells in the rest of one cycle as in the $\eta=8$ case. Thus the particle's position distribution center stays close to the driver center rather than nearby one of the local minimum points during the entire cycle, and so the work curve is no longer similar to not only the bottommost dotted blue branch [marked by the solid purple segment on the bottommost dotted blue curve in Figure \ref{fig:WorkLowerBoundHighEtas}(b)] but also any of the dotted branches in the left subfigure of Figure \ref{fig:Eta14EnergyDisplacement}, %Therefore, the work curve of the $\eta=14$ case
and it goes down much slower than the bottommost dotted blue local minimum branch. Comparing the two left subfigures of Figure \ref{fig:Eta8EnergyDisplacement} and \ref{fig:Eta14EnergyDisplacement}, we can see that the bottommost dotted blue local minimum branches are of similar height in one engine cycle, cf. the end of this subsection. So the work output of the $\eta=14$ case is declined more and smaller than that of the $\eta=8$ case [Figure 7(A1) in the main text]. Therefore for the $\eta=14$ case, the particle's %remaining longer in the side (especially backward) wells, i.e. its 
covering more (than two) wells causes the work output reduces a lot from the prediction of the potential mechanism. %, i.e. $W_{\rm cyc,e.p.}$, cf. Eq. \ref{worklimit_tight}.
%However, as we have mentioned in the main text, due to the high absolute temperatures $T_{\rm h,c}$ at high $\eta$ with the nondimensional temperatures ${\it\Theta}_{\rm h,c}$ keeping constant, another interpretation is that the particle tends to neglect the lattice potential, i.e. the particle is  (or exactly speaking, the energy barriers) rather than staying in the neighbor wells, i.e. the negative thermolubricity mechanism. 

\begin{figure}[H]
 \centering
\includegraphics[width=0.49\textwidth]{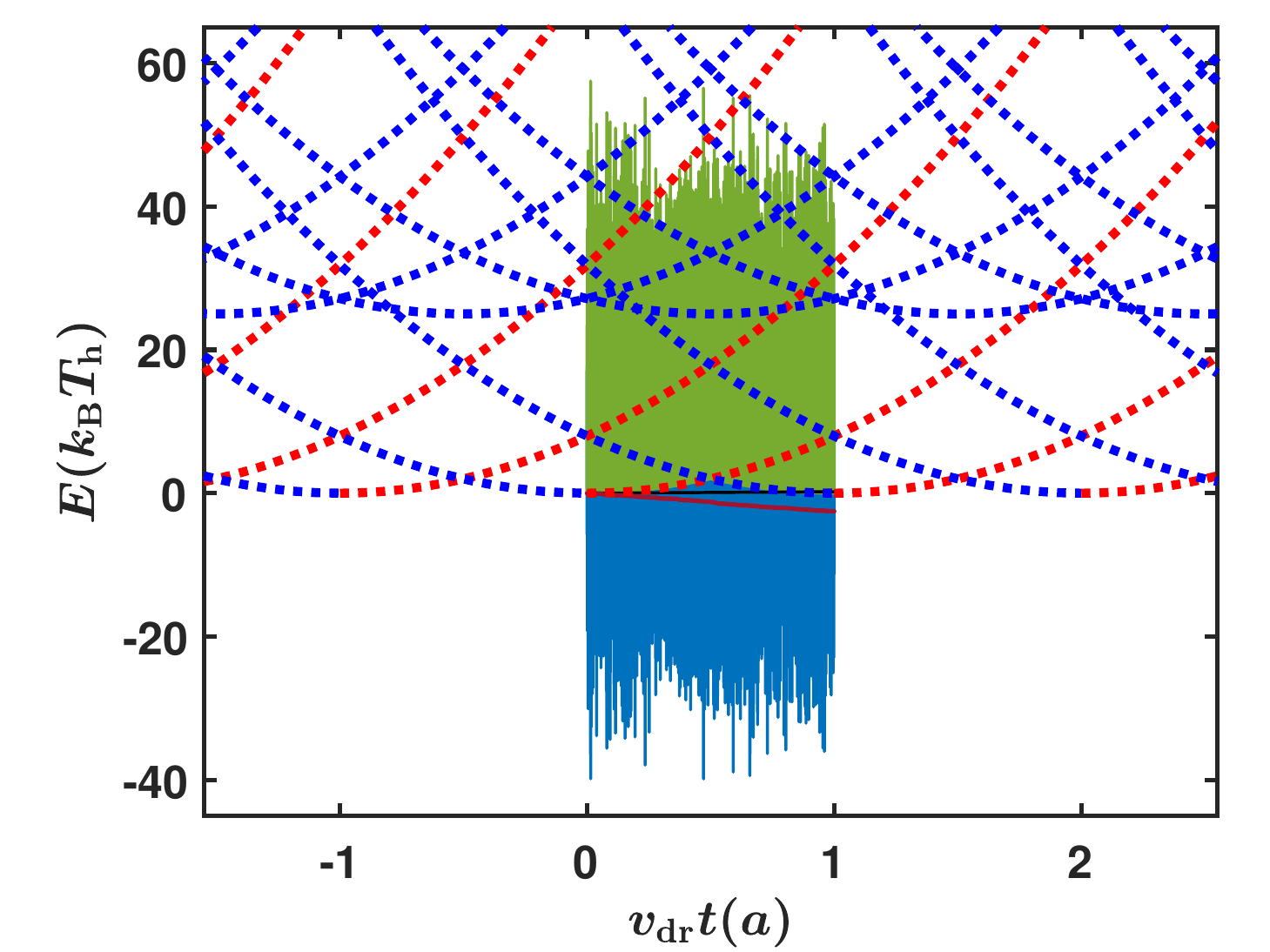}
\includegraphics[width=0.49\textwidth]{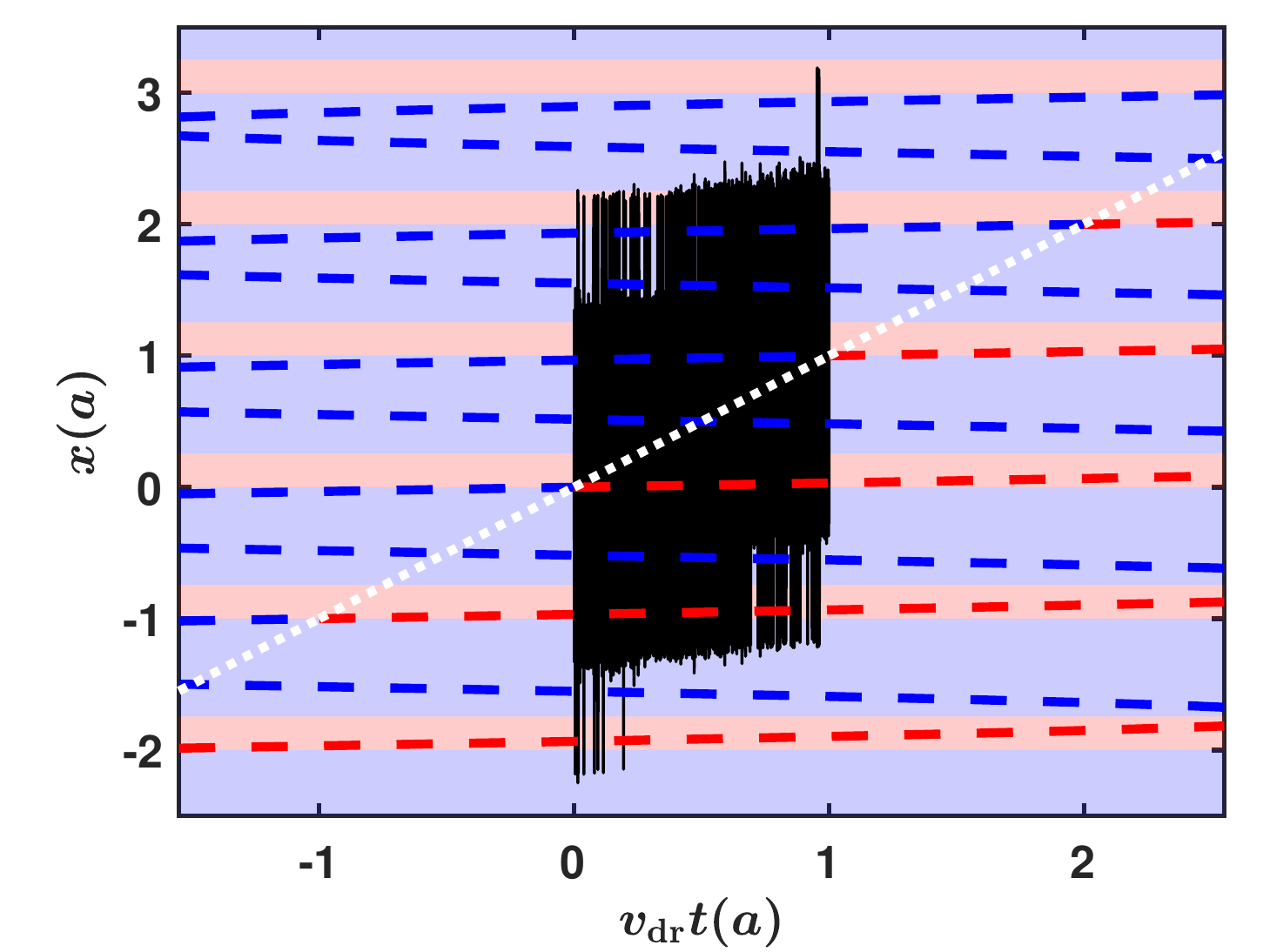}
\caption{Energy and displacement curves at $\eta=30$, ${\it\Theta}_{\rm h,c}=0.4,0.04$ and $\mu=4\times10^{4}\rm s^{-1}$. Curve colors and types are the same as those in the main text Figure 2(A) and (B) and $k_{\rm B}T_{\rm h}$ is of the same value as that in the main text Figure 2(A). The balanced resultant potential curves of the $\eta=30$ case are even more complex than those of the $\eta=14$ cases. In the right subfigure we can see that at the beginning of one engine cycle there are 3 hot and 2 cold local minima and 4 energy barrier peaks. In the left subfigure, the dotted red local minimum branches starting at the beginning of the two upstream cycles, and the dotted blue local minimum branches ending at the end of the two downstream cycles, intrude the dark green internal energy curve, so that the particle remains longer in more neighbor wells both ahead of and behind the global minimum. In fact, from the right subfigure we can see that the displacement curve covers two hot local minimum branches and two cold local minimum branches nearly all the way and a third hot branch or a third cold branch part of the way from the beginning to the end of one engine cycle,  
%In the middle of the engine cycle, we can see that the internal energy curve sometimes traverses through three hot and three cold branches. While for the $\eta=8$ and $14$ cases in Figure \ref{fig:Eta8EnergyDisplacement} and \ref{fig:Eta14EnergyDisplacement}, the internal energy curve covers one hot and one cold local minimum branches all the way and a second hot or a second cold local minimum branches part of the way from the cycle beginning to the end. 
i.e. the particle always covers four (lowest, of course) wells and sometimes covers five (lowest, of course) wells. 
%While for the $\eta=14$ case the particle always covers two wells and sometimes covers three to four wells and for $\eta=8$ always two and sometimes three. 
Thus the particle's position distribution center stays closer to the driver center and the work curve is very different from and becomes much flatter than the bottommost dotted blue branch in the left subfigure. %We can see from the left subfigure that the bottommost dotted blue branches are of similar height to the $\eta=8$ and $14$ cases in one engine cycle. %Therefore the mean cycle work output of the $\eta=30$ case is declined much more and thus smaller than the $\eta=14$ and $8$ cases. 
Other parameters are given in Sec. \ref{Langevindynamicssimulation}.\ref{ParametersUsed}.
}
\label{fig:Eta30EnergyDisplacement}
\end{figure}

In Figure \ref{fig:Eta30EnergyDisplacement}, we plot the energy and displacement curves of the case of $\eta=30$, ${\it\Theta}_{\rm h,c}=0.4,0.04$ and $\mu=4\times10^{4}\rm s^{-1}$. We can see that at the same ${\it\Theta}_{\rm h,c}$ as the $\eta=8$ and $14$ cases above, %the upper bound of 
the dark green internal energy curve exceeds a similar proportion above the bottommost dotted blue energy barrier peak branch at the middle point of one engine cycle. With $\eta$ increasing to $30$ there are more dotted red local minimum branches starting at the beginning of upstream cycles, and dotted blue local minimum branches ending at the end of downstream cycles, intrude the internal energy curve. We can see from the right subfigure of Figure \ref{fig:Eta30EnergyDisplacement} that the particle covers two hot and two cold wells all the way and a third hot (at the beginning) or a third cold (at the end) well part of the way from the beginning to the end of one cycle, and in total the particle always covers at least four wells, which is more than the $\eta=8$ and 14 cases. %while for the $\eta=8$ and $14$ case the particle covers one hot and one cold well all the way and a second hot or a second cold well part of the way from the beginning to the end of one cycle. 
So the particle's position distribution center is much closer to the driver center [Figure 7(E) in the main text] and the work curve's shape deviates from the bottommost dotted blue local minimum branch much more and is now very close to a horizontal line. At $\eta=30$ the bottomost dotted blue branch is still of similar height to those of the $\eta=8$ and $14$ cases in one engine cycle, cf. the end of this subsection. So the work output is reduced much more than the $\eta=14$ and also the $\eta=8$ case [Figure 7(A1) in the main text]. 

%We have seen that keeping ${\it\Theta}_{\rm h,c}$ constant makes the upper bound of the internal energy curve exceeds the bottommost energy barrier peak branch a similar proportion at the middle instant of one engine cycle, which makes it possible for us to distinguish the effect of the more teeth of the resultant potential curve at $\eta>4.6$. 

At $\eta=30$, the internal energy curve always covers five dotted local minimum branches (three red and two blue or two red and three blue) and sometimes covers six (three red and three blue) in the middle of one cycle. However, this doesn't mean that the particle always covers five wells. In fact, in the right subfigure of Figure \ref{fig:Eta30EnergyDisplacement}, we can see that the particle always covers four wells and five only at the beginning and end of one cycle. The reason is that to reach a remote outside well, the particle has to cross over an energy barrier. In the middle of the engine cycle, although the internal energy curve frequently covers the fifth blue or red local minimum branch, it cannot cover the neighbor energy barrier peak branch above it so that the particle cannot cross over the energy barrier in the way and reach the fifth well. Comparatively, in the $\eta=8$ case in Figure \ref{fig:Eta8EnergyDisplacement}, we can see that the third red or blue branch covered by the internal energy curve is close to the neighbor energy barrier peak branch above it and so the range of the third well covered by the particle is similar to the range of its near center neighbor energy barrier covered. In the $\eta=14$ case in Figure \ref{fig:Eta14EnergyDisplacement} the range of the third well covered by the particle is also approximate to the range of its near center neighbor energy barrier covered. Generally speaking, at higher $\eta$ the outward energy barriers outside become higher and harder to be crossed over. So our above conclusion that in the $\eta=8$ case in Figure \ref{fig:Eta8EnergyDisplacement}, the particle always covers two wells and three at the beginning and end of one cycle is correct and the conclusion that in the $\eta=14$ case in Figure \ref{fig:Eta14EnergyDisplacement}, the particle always covers three wells and four in the middle of one cycle is also roughly correct, which can be validated by the displacement curves in the two right subfigures of Figure \ref{fig:Eta8EnergyDisplacement} and \ref{fig:Eta14EnergyDisplacement}. To avoid making mistakes, we'd better get the information of the number of wells covered by the particle from the displacement curves.

We want to note that we mean the particle's covering two or more wells by that the particle can stay around one of or hop between or among these wells, which can be represented by the displacement curve (or the internal energy curve, although sometimes not rigorous as we have explained) covering two or more local minimum branches, as in the right subfigures in Figure  \ref{fig:Eta8EnergyDisplacement}, \ref{fig:Eta14EnergyDisplacement} and \ref{fig:Eta30EnergyDisplacement}. %For the $\eta=14$ case in Figure \ref{fig:Eta14EnergyDisplacement}, we can see that the particle always covers three wells during one cycle and sometimes four around the middle of one cycle, so the work curve is no longer similar to the bottommost branch due to the particle's covering more wells. 

\begin{figure}[H]
\centering
\includegraphics[width=0.75\textwidth]{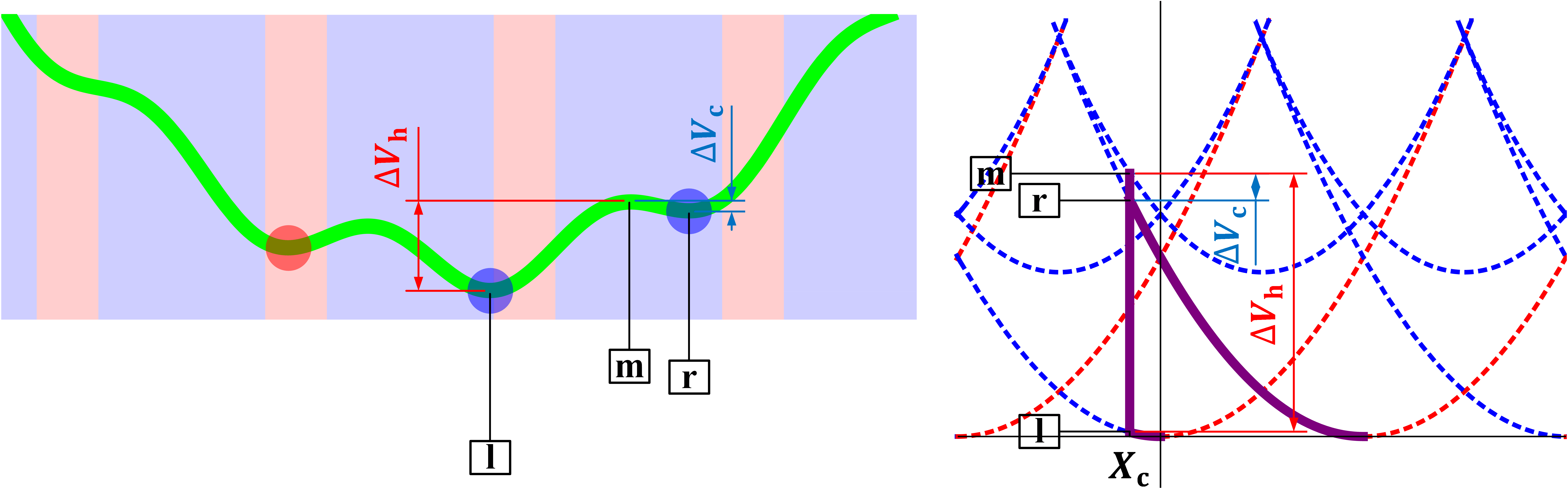}
\caption{The schematic of calculating the $W_{\rm cyc,e.p.}$ value at $\eta=8$ on the dot-dashed black curve in the main text Figure 7(A1) with the method described in Sec. \ref{sec:apndRD}.\ref{sec:wcyceploweta}. This schematic is the $\eta=8$ version of Figure \ref{Carnot} with the temperature background of the left subschematic and the balanced resultant potential curves of the right subschematic a little more delicate. On the right we can see that at $\eta=8>4.6$, the balanced resultant potential curve of the current engine cycle extends to the last and next cycles. At the ratio ${\it\Theta}_{\rm h}/{\it\Theta}_{\rm c}=10$, the jumping point of the particle $X_{\rm c}$ gets out of the current engine cycle to the last one, so the ascent stage of the resultant potential energy represented by the solid purple curve starting from the cycle starting point goes backward to the last cycle while the descent stage continues more than one cycle, which is not physical. Moreover, from the left subschematic we can see that the particle jumps from a cold well to another cold one which is different from the $\eta\leq4.6$ cases where the particle usually jumps from a hot well to a cold one. Therefore this method is only suitable for the $\eta\leq4.6$ cases and does not work for the $\eta>4.6$ cases, see the caption of Figure \ref{fig:WorkLowerBoundHighEtas}.}
\label{fig:CarnotHighEta}
\end{figure}

\begin{figure}[H]
\centering
\begin{minipage}{0.4\textwidth}
\centerline{
\includegraphics[width=\textwidth]{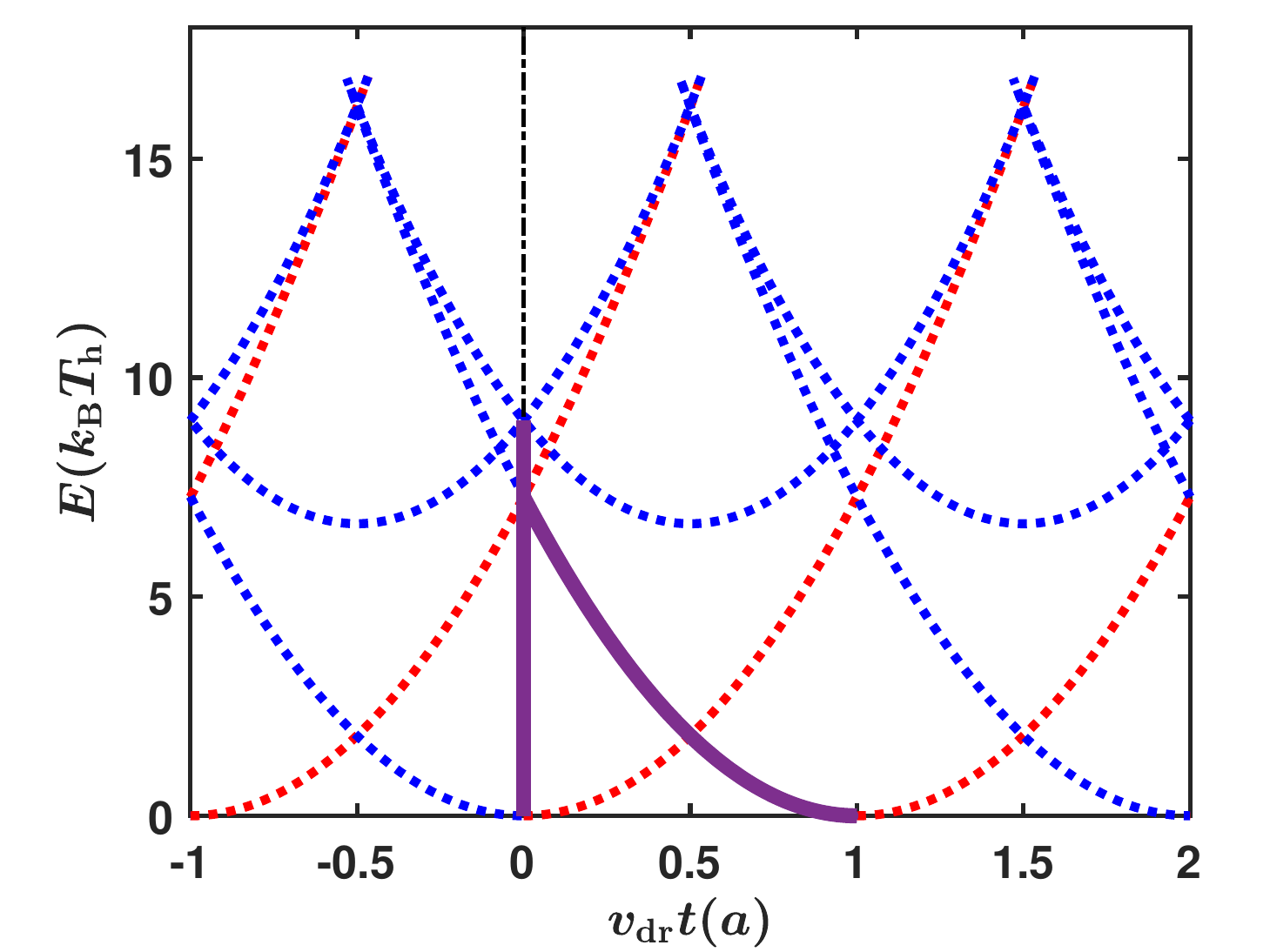}}
\centerline{(a)}
\end{minipage}
\begin{minipage}{0.4\textwidth}
\centerline{
\includegraphics[width=\textwidth]{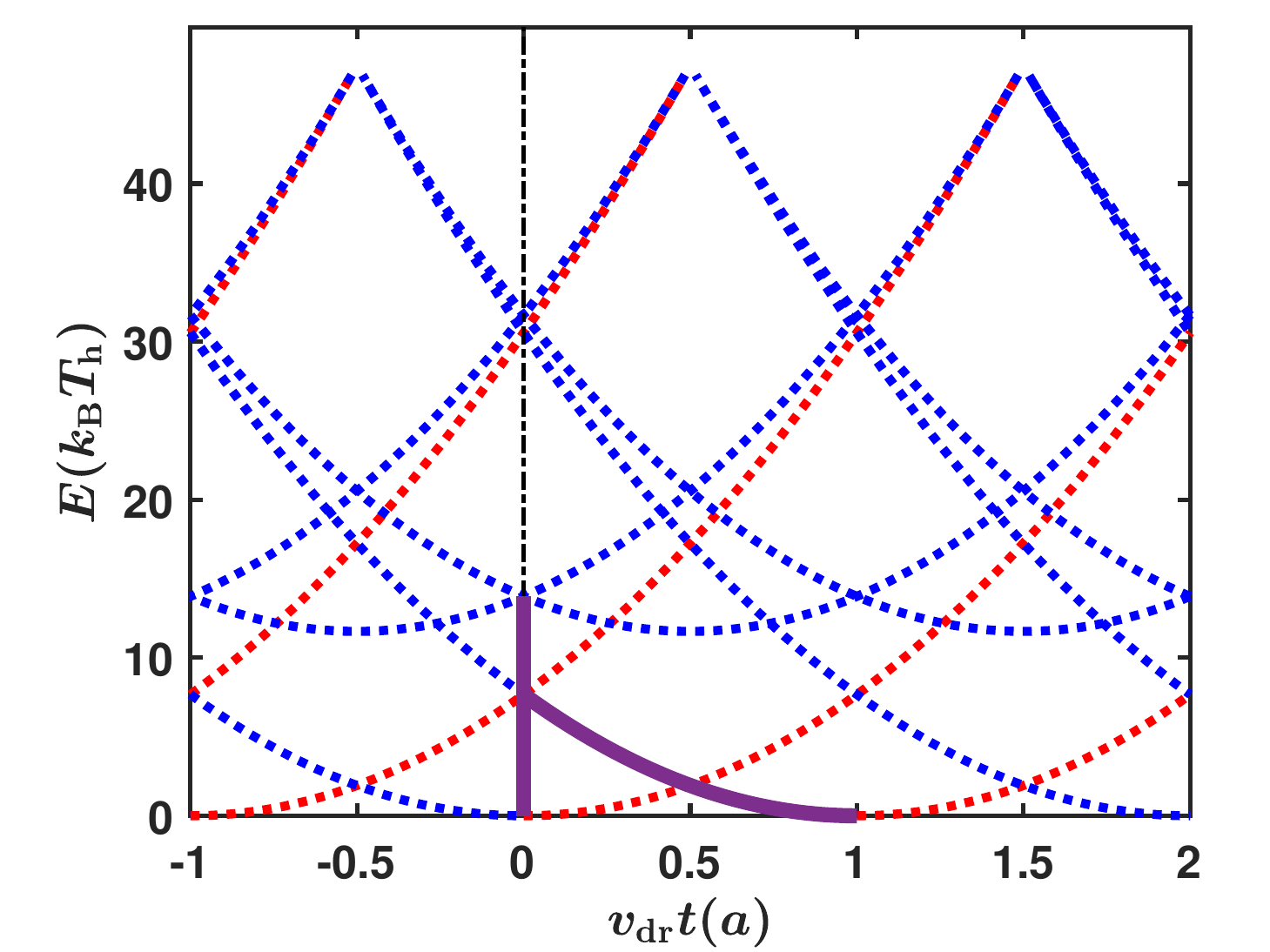}}
\centerline{(b)}
\end{minipage}\\
\begin{minipage}{0.4\textwidth}
\centerline{
\includegraphics[width=\textwidth]{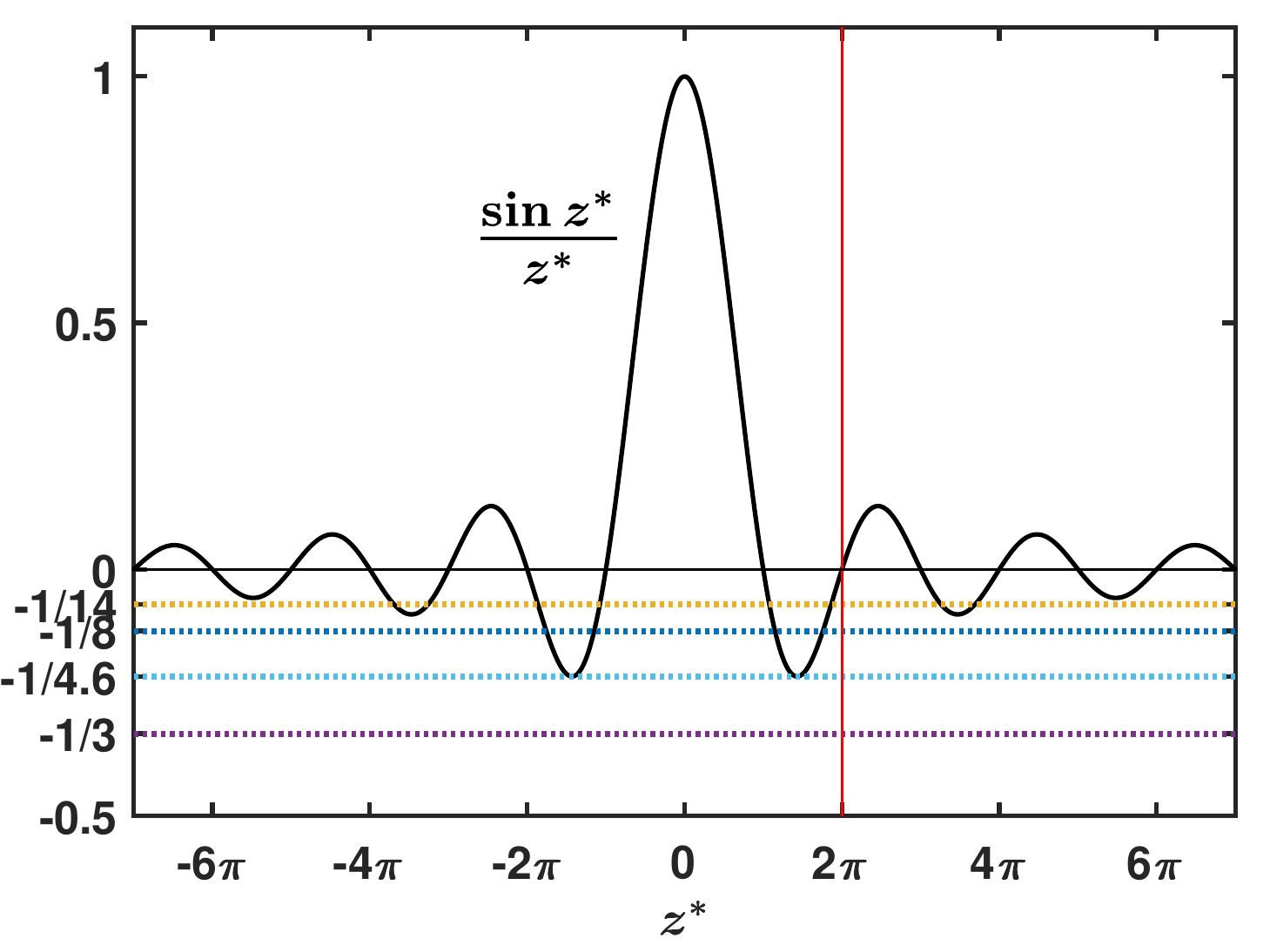}}
\centerline{(c)}
\end{minipage}
\begin{minipage}{0.4\textwidth}
\centerline{
\includegraphics[width=\textwidth]{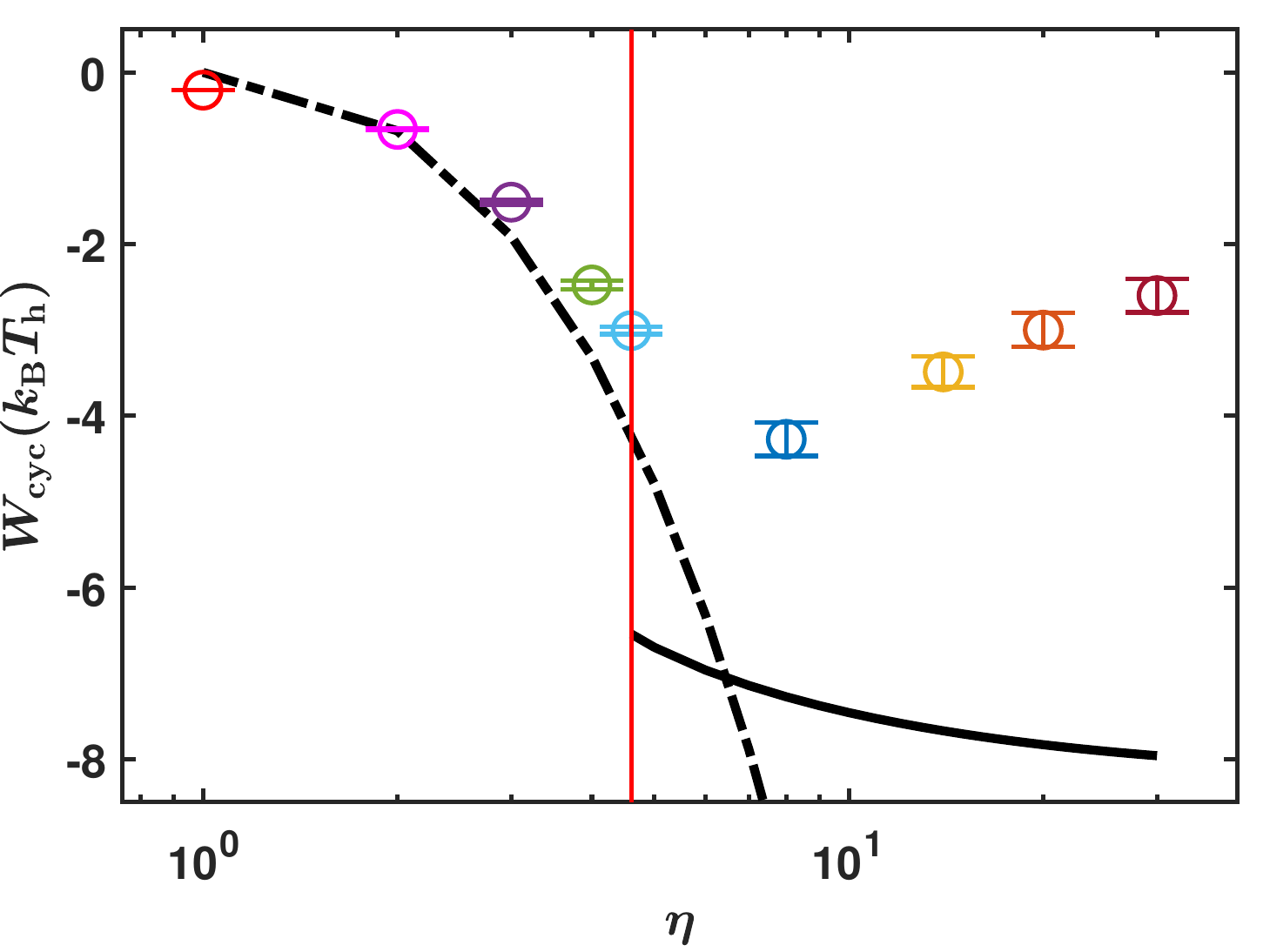}}
\centerline{(d)}
\end{minipage}
\caption{The approximation of the equilibrium limit cycle work output by the bottommost dotted blue local minimum resultant potential branch at $\eta>4.6$. In (a) and (b), the solid purple curve represents the particle's resultant potential energy jumps at the beginning of one engine cycle from the global minimum point over the right first energy barrier to the right neighbor local minimum point and then decreases along the bottommost dotted blue local minimum resultant potential branch to zero at $\eta=8$ and $14$ respectively. The dotted red and blue curves in (a) and (b) are inherited from Figure \ref{fig:Eta8EnergyDisplacement} and \ref{fig:Eta14EnergyDisplacement} respectively. In (c), we describe how to locate the nondimensional position $\hat{z^*}$ of the particle's landing point on the bottommost blue branch, which is represented by the intersection points of the horizontal line $-\frac1\eta$ with $\frac{\sin(z^*)}{z^*}$ nearby $z^*=2\pi$ indicated by the vertical solid red line. As $\eta\rightarrow+\infty$, $-\frac1\eta\rightarrow-0$ and $\hat{z^*}\rightarrow2\pi$. In (c), we can see that at $\eta<4.6$, the horizontal line $-\frac1\eta$ (e.g. the dotted purple $-\frac1\eta=-\frac13$ line) doesn't intersect $\frac{\sin(z^*)}{z^*}$ and at the first critical value of $\eta\approx4.6$, the $-\frac1\eta$ line touches $\frac{\sin(z^*)}{z^*}$ at its two lowest local minima and we can generalize that at the second critical value of $\eta$, the $-\frac1\eta$ line touches $\frac{\sin(z^*)}{z^*}$ at its two second lowest local minima and so on, which gives us another way to calculate the critical values of $\eta$ \cite{CorrugationNumber}, cf. Sec. \ref{sec:critcalvalueeta}. (d) The results of $W_{\rm cyc,e.p.\eta>4.6}$ at ${\it\Theta}_{\rm h,c}=0.4,0.04$ and $\mu=10^4\rm s^{-1}$ approximated by the height change of the bottommost dotted blue resultant potential curve during one engine cycle, represented by the solid black curve. The results are plotted in the main text Figure 7(A1) for comparison. The dot-dashed black curve representing $W_{\rm cyc,e.p.}$ approximated by the method introduced in Sec. \ref{sec:apndRD}.\ref{sec:wcyceploweta} and we have shown in Figure \ref{fig:CarnotHighEta} that this approximation is not physical at $\eta\gtrsim4.6$, which leads to a higher value than the approximation $W_{\rm cyc,e.p.\eta>4.6}\approx\frac{m\omega_0^2a^2}{4\pi^2}\tilde{V}(\hat{z^*})$ made here. Comparing Figure \ref{fig:CarnotHighEta} and (a), we can see that at $\eta\gtrsim4.6$, the approximation made with the method in Sec. \ref{sec:apndRD}.\ref{sec:wcyceploweta} is a loser lower bound, as can be validated by the lower value of the dot-dashed black curve than the solid black one at high $\eta$. Here ${\it\Theta}_{\rm h,c}=0.4,0.04$ and for this ratio $\frac{{\it\Theta}_{\rm h}}{{\it\Theta}_{\rm c}}=10$, the dot-dashed black curve still has a small physical range from $\eta=4.6$ to its intersection point with the solid black one, where although the balanced resultant potential curve of the current engine cycle extends out, the jumping point $X_{\rm c}$ is still in the current engine cycle. With $\eta$ increasing from $4.6$, $X_{\rm c}$ gradually approaches the cycle starting point until the intersection point where $X_{\rm c}$ is just at the cycle starting point, cf. Figure \ref{fig:CarnotHighEta}.}
\label{fig:WorkLowerBoundHighEtas}
\end{figure}

For the $\eta>4.6$ cases, due to the more local minima of the resultant potential, the approximation of the equilibrium limit of the cycle work is more difficult and the method we utilized in Sec. \ref{sec:apndRD}.\ref{sec:wcyceploweta} no longer works. For instance, in Figure \ref{fig:CarnotHighEta}, the schematic represents the $\eta=8$ case corresponding to Figure \ref{Carnot} which we have used to derive $W_{\rm cyc,e.p.}$ at $\eta\leq4.6$. We can see that the average jumping point gets out of the current engine cycle to last cycle's cold zone and the particle goes backward along the cold branch and then jumps to the next cold branch back to the current cycle over the energy barrier in-between. This is not physical but it is how we actually calculate the $W_{\rm cyc,e.p.}$ for the $\eta=8$ point on the black dot-dashed curve in the main text Figure 7(A1). So at $\eta>4.6$ the black dot-dashed curve in the main text Figure 7(A1) is incorrect or exactly speaking too lose as we will see next.

As we have explained above, at $\eta\approx8$ the work curve is similar to and at $\eta\gtrsim14$ the work curve gets flatter and flatter than the bottommost dotted blue local mimimum branch, so we can approximate the equilibrium cycle work output by the resultant potential energy change of the particle on this branch during one cycle, which is a lower bound and lose espacially at $\eta\gtrsim14$. In Figure \ref{fig:WorkLowerBoundHighEtas}(a) and (b), the solid purple vertical line indicates that the particle's resultant potential energy jumps at the beginning of one cycle from the global minimum over the right neighbor energy barrier to the right first local minimum. Then the resultant potential energy decreases to zero along the bottommost dotted blue branch indicated by the solid purple segment. (a) and (b) correspond to the cases of $\eta=8$ and $14$ respectively. The equilibrium work output $W_{\rm cyc,e.p.\eta>4.6}$ can be approximated by the height of the solid purple segment on the bottommost dotted blue branch, i.e. the value of the bottommost dottedd blue branch at $\frac{v_{\rm dr}t}{a}=0$. Note here we indeed approximate the work curve by the bottommost dotted blue branch, i.e. we assume the particle jumps over the right first energy barrier at the beginning of the cycle and then stays on the right first local minimum point driving the driver to transform the resultant potential energy into work output until the end of the cycle, cf. the main text Figure 7(F1)-(F7) where we can see the right first blue ball in (F1) goes backward to the end of the cycle approaching to the driver center, which is equivalent to that the blue ball goes forward to the end of the cycle and the driver center goes after and catches up with it at the end. So this approximation is still based on the potential mechanism, indicated by the subscript `p.'.

Set $\tilde X(z(\tau)^*)=0$ in Eq. \ref{eq:latticepotentialstickslip}, we will obtain
\begin{equation}\label{eqn:sincintersectVz}
\begin{cases}
0=z^*+\eta\sin(z^*),\\
\tilde{V}(z^*)=\frac12\eta^2\sin^2(z^*)+\eta[1-\cos(z^*)],
\end{cases}
\end{equation}
the first equation of which is equivalent to
\begin{equation}\label{eqn:sincintersect}
\frac{\sin(z^*)}{z^*}=-\frac1\eta,
\end{equation}
excluding the trivial solution $z^*=0$.

In Figure \ref{fig:WorkLowerBoundHighEtas}(c), we plot the function $\frac{\sin(z^*)}{z^*}$. The solution of Eq. \ref{eqn:sincintersect} can be obtained from the intersection points of $\frac{\sin(z^*)}{z^*}$ with the horizontal line $-\frac1\eta$. For instance, the dotted dark blue horizontal line representing $-\frac1\eta=-\frac18$ has four intersection points with $\frac{\sin(z^*)}{z^*}$ corresponding to the four nonzero, excluding the trivial zero, intersection points of the vertical dot-dashed line at $\frac{v_{\rm dr}t}{a}=0$ with the resultant potential curves in Figure \ref{fig:WorkLowerBoundHighEtas}(a). The middle two intersection points on the $-\frac1\eta=-\frac18$ line in Figure \ref{fig:WorkLowerBoundHighEtas}(c) corresponds to the upper two intersection points on the two dotted blue energy barrier peak branches of the current and last cycles in Figure \ref{fig:WorkLowerBoundHighEtas}(a). The intersection point $\hat{z^*}$ nearby $z^*=2\pi$ in Figure \ref{fig:WorkLowerBoundHighEtas}(c) corresponds to the intersection point of the vertical dot-dashed line with the bottommost blue branch in Figure \ref{fig:WorkLowerBoundHighEtas}(a), which is what we want. Solve the $\hat{z^*}$ out with a nonlinear solver and substitute it into the second equation of Eq. \ref{eqn:sincintersectVz}, we will get the height of the solid purple segment on the bottommost blue branch, $\tilde{V}(\hat{z^*})$, which is used to approximate $W_{\rm cyc,e.p.\eta>4.6}\approx\frac{m\omega_0^2a^2}{4\pi^2}\tilde{V}(\hat{z^*})$ after dimensionalizing back, the result of which is plotted in Figure \ref{fig:WorkLowerBoundHighEtas}(d) compared with the main text Figure 7(A1). In Figure \ref{fig:WorkLowerBoundHighEtas}(c), we can see that the dotted dark yellow horizontal line representing $-\frac1\eta=-\frac1{14}$ intersects $\frac{\sin(z^*)}{z^*}$ at $8$ points corresponding to the $8$ nonzero intersection points of the vertical dot-dashed line with the dotted resultant potential curves in Figure \ref{fig:WorkLowerBoundHighEtas}(b). And again the intersection point nearby $z^*=2\pi$ corresponds to the solid purple segment on the bottommost dotted blue branch at $\eta=14$.

%\begin{figure}[H]
%\centering
%\includegraphics[width=0.328\textwidth]{SFigures/WcycEta.pdf}
%\caption{The equilibrium limit of the cycle work at $\eta>4.6$, ${\it\Theta}_{\rm h,c}=0.4,0.04$ and $\mu=10^4\rm s^{-1}$ approximated by the height change of the bottommost dotted blue resultant potential curve during one engine cycle, represented by the solid black curve. The results in the main text Figure 7(A1) are plotted for comparison. The dot-dashed black curve representing $W_{\rm cyc,e.p.}$ approximated by the potential mechanism at $\eta<4.6$ introduced in Sec. \ref{sec:apndRD}.\ref{sec:wcyceploweta} and we have shown in Figure \ref{fig:CarnotHighEta} that this approximation is not physical at $\eta=8$, which leads to a higher value than the approximation $W_{\rm cyc,e.p.\eta>4.6}\approx\frac{m\omega_0^2a^2}{4\pi^2}\tilde{V}(\hat{z^*})$ made in Figure \ref{fig:WorkLowerBoundHighEtas}. Here ${\it\Theta}_{\rm h,c}=0.4,0.04$ and for this ratio $\frac{{\it\Theta}_{\rm h}}{{\it\Theta}_{\rm c}}=10$, the dot-dashed black curve still has a small physical range from $\eta=4.6$ to its intersection point with the solid black one, where although the resultant potential curve extends out of one engine cycle, the jumping point $X_{\rm c}$ is still in the current engine cycle. With $\eta$ increasing from $4.6$, $X_{\rm c}$ gradually approaches the cycle starting point until the intersection point where $X_{\rm c}$ is at the cycle starting point, cf. Figure \ref{fig:CarnotHighEta}.}
%\label{fig:WcycEtaHighBound}
%\end{figure}

From Figure \ref{fig:WorkLowerBoundHighEtas}(d), we can see that the above approximation $W_{\rm cyc,e.p.\eta>4.6}\approx\frac{m\omega_0^2a^2}{4\pi^2}\tilde{V}(\hat{z^*})$ is a lower bound and is lose especially at $\eta\gtrsim14$ when the particle always covers more than two wells and the work output is declined from $W_{\rm cyc,e.p.\eta>4.6}$ more. We must note that this curve is obtained at ${\it\Theta}_{\rm h,c}=0.4,0.04$, which is high at $\eta>4.6$ as can be seen in Figure \ref{fig:Eta8EnergyDisplacement}, \ref{fig:Eta14EnergyDisplacement} and \ref{fig:Eta30EnergyDisplacement} where the upper bound of the internal energy curve exceeds the bottommost dotted energy barrier peak branch by a large amount from the beginning to the end of one engine cycle, compared with the $\eta<4.6$ cases in the main text Figure 4. At other values of ${\it\Theta}_{\rm h,c}$, the results may be different especially when ${\it\Theta}_{\rm h,c}$ are low, as we will explain next.

We have seen that keeping ${\it\Theta}_{\rm h,c}$ constant makes the internal energy curve exceeds the bottommost energy barrier peak branch a similar proportion at the middle instant of one engine cycle and makes it possible for us to distinguish the effect of the more teeth of the resultant potential curve at $\eta>4.6$. Meanwhile, as we have chosen two relatively high values of ${\it\Theta}_{\rm h,c}=0.4,0.04$, we can achieve qusi-equilibrium at a not very low driving velocity $v_{\rm dr}=10^{-7}\rm m/s$. At lower $v_{\rm dr}$, more time steps are needed to simulate one engine cycle, causing high computational cost and long computing time, especially at high $\eta$ which leads to small time step (Eq. \ref{TimeStepSize}). We have mentioned in the main text that at higher $\eta$ the absolute temperatures $T_{\rm h,c}$ are higher so that the high absolute temperatures, or rather thermolubricity, play an important role in the reduction of the equilibrium cycle work output at high $\eta$, and to have a complete understanding about the effect of the increasing number of teeth of the resultant potential with increasing $\eta$ on the equilibrium limit of $\langle W_{\rm cyc}\rangle$, %more simulation should be done to 
we can keep the absolute temperatures $T_{\rm h,c}$ constant to see how $\langle W_{\rm cyc}\rangle $ varies with $\eta$ or keep $\eta$ constant at a high value to see how $\langle W_{\rm cyc}\rangle $ varies with ${\it\Theta}_{\rm h,c}$, at low driving velocity. Here we give a qualitative analysis.

At $8\lesssim\eta\lesssim14$ if we reduce $T_{\rm h,c}$, the period for the particle to cover more than two wells is shortened and the work curve may be more parallel to the bottommost curve so that the work output will increase, cf. Figure \ref{fig:Eta8EnergyDisplacement}. Nonetheless, the high temperature $T_{\rm h}$ cannot be too low for the particle to jump over the right first energy barrier early at the beginning of the cycle. At $\eta\gtrsim14$, even if we decrease $T_{\rm h,c}$, the work curve is not easy to parallel the bottommost blue branch. The reason is that $T_{\rm h}$ cannot be so low for the internal energy curve to get below the bottommost energy barrier peak branch, otherwise the work input in one cycle may get higher than the work output due to the jumping point moving to the second half of the cycle. Or the work output may be declined due to the jumping point moving to the next cycle or the cycle after the next and so on, which means that the particle starting at the cycle starting point of the current cycle won't slip until the next cycle or the cycle after the next and so on. In this case the engine cycle covers more than one lattice period and multislip may occur and the particle would cover more wells so that the work output would be declined, cf. Figure \ref{fig:Eta14EnergyDisplacement} and \ref{fig:Eta30EnergyDisplacement}. We speculate that although $T_{\rm h}$ is low, as long as $T_{\rm h}>T_{\rm c}$ there will still be work output even if declined, which is not easy to be validated by the Langevin dynamics simulation due to the low temperatures leading to the particle's difficulty to arrive at equilibrium.

Whereas, if $T_{\rm c}$ is still high and makes the internal energy curve exceed the bottommost energy barrier peak branch from the beginning to the end of one engine cycle, the particle will nearly always cover more than two wells so that the work curve cannot parallel the bottommost dotted blue branch, like the current cases of ${\it\Theta}_{\rm h,c}=0.4,0.04$ (at $\eta\gtrsim14$) we are considering. 

If $T_{\rm h}$ is high, $T_{\rm c}$ is low and the ratio $T_{\rm h}/T_{\rm c}$ is not too large, the particle could jump from the bottommost dotted red branch over the barrier in-between to the bottommost dotted blue branch in the first half of one engine cycle. For the $\eta=30$ case in Figure \ref{fig:Eta30EnergyDisplacement}, we can see that this situation occurs when $T_{\rm h}/T_{\rm c}$ is not greater than $2$. In this situation the work curve will similar to those in the main text Figure 4 with two segments and only the descent segment parallels the bottommost dotted blue branch. The high temperature $T_{\rm h}$ should not be too high or too low and the particle should just jump from the bottommost red branch to the bottommost blue branch over the energy barrier in-between and not jump too high to cross over the next energy barrier to the second bottommost red branch, which means the particle jumps to the backward well. Then multislip may occur and the particle may covers more wells and the work curve would deviate from the bottommost dotted blue branch and the cycle work output would be declined. With an appropriate $T_{\rm h}$, $T_{\rm c}$ could be chosen as low as possible for the particle to jump over the right first energy barrier at the beginning of the cycle and then to be immediately cooled down in the right first cold well, which will make the work curve parallel to the bottommost dotted blue branch and the cycle work output close to $W_{\rm cyc,e.p.\eta>4.6}$. However, the practical realization of $T_{\rm c}$ which is low enough is difficult and the validation by Langevin dynamics simulation of this situation is also of high computational cost and long computing time.

So we can see at $\eta>4.6$ the PTSHE is more complicated and the above analysis is preliminary. If the above explanation makes the readers feel confused, they don't need to worry. It's partly due to the author's weak expression skill. But it does output work and we can turn to the Fokker-Planck or Kramers equation methods \cite{KramersEqBookRisken} to get further understanding. %as we have mentioned in the main text.

We have presumed above that covering more wells makes the particle's position distribution center closer to the driver center and so the work output is much smaller, cf. the main text Figure 7(E). %And we can indeed see that the work curve of the $\eta=30$ case in Figure \ref{fig:Eta30EnergyDisplacement} is much flatter than the bottommost dotted blue local minimum branch. 
This presumption could be further validated by the Fokker-Planck or Kramers equation methods \cite{KramersEqBookRisken}.
%Compare the $\eta=8$ case in Figure \ref{fig:Eta8EnergyDisplacement}, we can see that the particle always covers two wells during one cycle and sometimes three at the beginning and end of one cycle and the work curve is parallel to the bottommost dotted blue local minimum branch in the middle of one cycle, indicating the particle stays around the local minimum represented by the bottommost branch most of the time in one cycle. 

We have mentioned above that at $\eta>4.6$ the bottommost dotted blue branches are of similar height in one engine cycle, which can be validated by the solid black curve in Figure \ref{fig:WorkLowerBoundHighEtas}(d), which changes mildly with $\eta$ and goes to a limit as $\eta\rightarrow+\infty$. The limit can be calculated. Substitute the first equation of Eq. \ref{eqn:sincintersectVz} into the second equation and eliminate $\eta$ we will obtain
\begin{equation}
\tilde{V}(\hat{z^*})=\frac12(\hat{z^*})^2+\frac{\hat{z^*}[\cos(\hat{z^*})-1]}{\sin(\hat{z^*})},
\end{equation}
in which we have replaced $z^*$ by the specific $\hat{z^*}$ nearby $z^*=2\pi$. We can see from Figure \ref{fig:WorkLowerBoundHighEtas} that as $\eta\rightarrow+\infty$, $-\frac1\eta\rightarrow-0$ and $\hat{z^*}\rightarrow2\pi-0$, so that both the numerator and denominator of the second term on the RHS goes to zero. Using L'H{\^o}pital's rule, the limit
\begin{equation}
\lim_{\hat{z^*}\rightarrow2\pi-0}\frac{\hat{z^*}[\cos(\hat{z^*})-1]}{\sin(\hat{z^*})}=\lim_{\hat{z^*}\rightarrow2\pi-0}\frac{[\cos(\hat{z^*})-1]-\hat{z^*}\sin(\hat{z^*})}{\cos(\hat{z^*})}=0.
\end{equation}
Therefore, as $\eta\rightarrow+\infty$,
\begin{equation}
\tilde{V}(\hat{z^*})=\frac12(\hat{z^*})^2+\frac{\hat{z^*}[\cos(\hat{z^*})-1]}{\sin(\hat{z^*})}\rightarrow\frac12(2\pi)^2=2\pi^2,
\end{equation}
and after dimensionalizing back, 
\begin{equation}
V(\hat{z^*})=\frac{m\omega_0^2a^2}{4\pi^2}\tilde{V}(\hat{z^*})\rightarrow2\pi^2\frac{m\omega_0^2a^2}{4\pi^2}=\frac12m\omega_0^2a^2=\frac12\frac{m\omega_0^2a^2}{0.4\frac{3m\omega_0^2a^2}{2\pi^2}}k_{\rm B}T_{\rm h}=\frac{\pi^2}{1.2}k_{\rm B}T_{\rm h}\approx8.22k_{\rm B}T_{\rm h},
\end{equation} 
where $k_{\rm B}T_{\rm h}=0.4\frac{3m\omega_0^2a^2}{2\pi^2}$ is the same as that of the main text Figure 7(A1).
\subsection{More remarks on the temperature}

Comparing the cases of $1\leq\eta\leq4.6$ in the same absolute temperatures in Figure 3(B) and in the same nondimensional temperatures in Figure 7(A), we can see that at the same $\eta$, in high absolute temperatures the probabilities of jumping over the barrier, $\exp(-\frac{\Delta V_{\rm h,c}}{k_{\rm B}T_{\rm h,c}})$, are higher so that the equilibrium state is easier to be achieved than the cases in lower absolute temperatures. In other words, in lower temperature equilibrium has to be achieved at lower driving velocity. For instance, the mean cycle work at the low driving velocity end of the $\eta=4.6$, ${\it\Theta}_{\rm h,c}=0.4,0.04$ case in Figure 7(A) has almost touches ground while that of the $\eta=4.6$, ${\it\Theta}_{\rm h,c}=0.26,0.026$ case in Figure 3(B) still has a trend to decrease if $v_{\rm dr}$ decreases further. For another instance, in Figure 3(B) we can see that in the same absolute temperatures, the velocity range of the work output shrinks to the left as $\eta$ increases. However, with the absolute temperatures increasing in Figure 7(A) the shrinked velocity ranges extend to the right again.

At nonequilibrium (or rather qusiequilibrium) such as when $v_{\rm dr}=10^{-5}\rm m/s$, the high and low temperature zone can be separately considered due to the rare occurance of the particle's crossing over the middle energy barrier. Then we can deduce that the hot temperature should make the particle easily cross over the barrier from the hot zone and the residual kinetic energy should not be too high for the particle to be cooled down on the right easily. In the cold zone, $k_{\rm B}T_{\rm c}$ should be adequately lower than the backward energy barrier $\Delta V_{\rm c}$ so that the particle can be cooled down and stay in the forward well and ahead of the driver center. On the other hand, at equilibrium, the particle crosses over the energy barrier frequently and the hot and cold zone on both sides of the energy barrier should be treated as a whole with the distribution of the particle at each instant calculated by considering the current shape of the resultant potential curve immersed in the alternative high and low temperature field.

In the main case of $\eta=3$, $\mu=4\times10^4\rm s^{-1}$, ${\it\Theta}_{\rm h,c}=0.4,0.04$, we can calculate that $\Delta V_{\rm h}\approx2.12k_{\rm B}T_{\rm h}$ and $\Delta V_{\rm c}\approx2.12k_{\rm B}T_{\rm c}$. At $v_{\rm dr}=10^{-5}\rm m/s$ (nonequilibrium) the two temperatures ${\it\Theta}_{\rm h,c}=0.4,0.04$ are a little higher and the driver center should go forward a little to make the forward barrier lower and the backward one higher for the particle to jump backward more rarely so that the mean cycle work output at $v_{\rm dr}=10^{-5}\rm m/s$ is smaller than $W_{\rm cyc,e.p.}$. 
\subsection{Topics about the practical implementation of the PTSHE}
\subsubsection{Parameter choice constrained by the obtainable values of and the mutual relations between temperature and other quantities}
\label{sec:parammutualconstraint}

In practice, to determine the nondimensional low temperature ${\it\Theta}_{\rm c}=\frac{k_{\rm B}T_{\rm c}}{V_0}$ we should first consider the obtainable range of $T_{\rm c}$ and $V_0$ and then choose an appropriate ${\it\Theta}_{\rm c}$ which is moderately low. The nondimensional high temperature ${\it\Theta}_{\rm h}=\frac{k_{\rm B}T_{\rm h}}{V_0}$ can then be chosen appropriately considering the obtainable $T_{\rm h}$.

Because the stiffness of the parabola harmonic potential $\kappa$ is restricted by $\eta=\frac{2\pi^2V_0}{\kappa a^2}$, the practically attainable range of $\kappa$ should also be taken into consideration to obtain an appropriate $\eta$. As $\kappa=m\omega_0^2$, the mass of the particle $m$, the intrinsic frequency of the parabola harmonic potential $\omega_0$ can be adjusted to achieve a suitable $\kappa$. The lattice period $a$ can also be chosen accordingly on condition that it is large enough relative to the particle's diameter. After choosing the parameters, we'd better do Langevin dynamics simulation to have a big picture of the performance of the PTSHE at these parameters, as different combination of parameters may lead to different behaviors of the PTSHE as we have seen in the main text Figure 3. In a word, the parameters are mutually constrained by each other and to achieve a good performance, we can solve a constraint optimization problem to get the optimum parameters of the PTSHE, in which the Fokker-Planck or Kramers equation method \cite{KramersEqBookRisken} might need to be turned to.

\subsubsection{Implementation of the spatially alternative high and low temperature field}

\begin{figure}[H]
\centering
\includegraphics[width=0.49\textwidth]{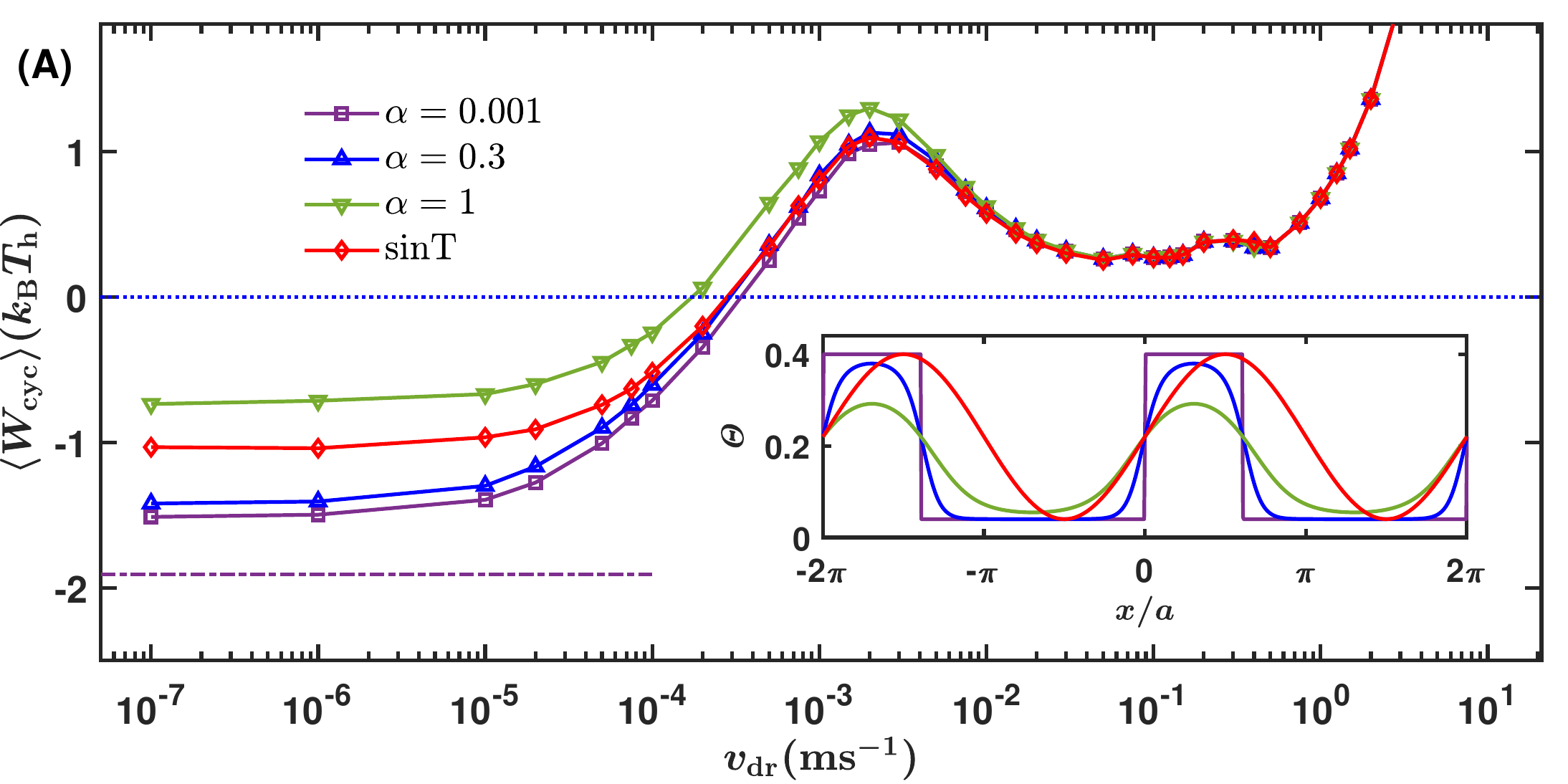}
\includegraphics[width=0.49\textwidth]{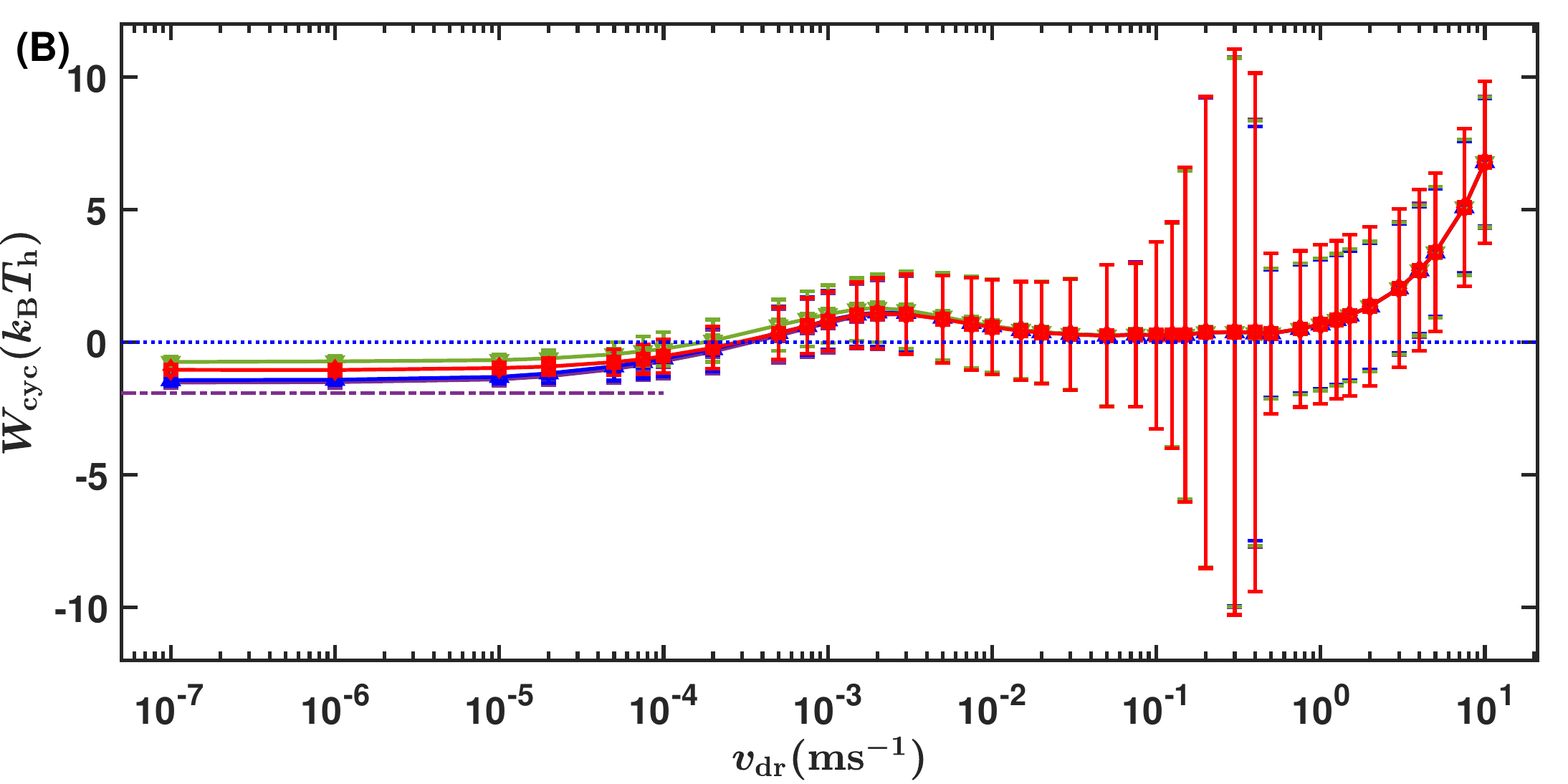}
\caption{Simulation results of mildly changing temperature fields compared with the sharply changing one. (A) The $\langle W_{\rm cyc}\rangle-v_{\rm dr}$ curves of different temperature field. The red curve with diamond denoted by sinT represents the case of a sinusoidal temperature field with the same wavelength and the same highest and lowest temperature values as the discontinuous one that we have adopted in the PTSHE. For the rest three curves, the temperature field ${\it\Theta}(z)$ is calculated by Eq. \ref{ContinuousTemp} with the same ${\it\Theta}_{\rm h,c}=0.4,0.04$ and different $\alpha$. The temperature profiles of the four curves are plotted in the inset with the purple one representing the $\alpha=0.001$ case as a reference, which we have utilized to approximate the discontinuous alternative high and low temperature field, cf. Sec. \ref{Langevindynamicssimulation}.\ref{sec:tempfunc}. The purple dot-dashed line on the bottom left is the equilibrium cycle work output approximated by the potential mechanism, $W_{\rm cyc,e.p.}$, cf. Eq. \ref{worklimit_tight}. We can see that there is still work output at low driving velocity for the case of $\alpha$ as large as $1.0$ as well as the case of the sinusoidal temperature profile, in both of which the temperature field changes very gently. So the demand for the temperature field is not very high and the experimental implementation is not unreachable. (B) The standard deviations of $W_{\rm cyc}$ with respect to $v_{\rm dr}$. Here $\eta=3.0,\ \mu=4\times10^{4}\rm s^{-1}$ and other parameters are given in Sec. \ref{Langevindynamicssimulation}.\ref{ParametersUsed}.}
\label{DifferentAlpha}
\end{figure}

It's not easy to implement a discontinuous temperature curve practically. In Figure \ref{DifferentAlpha} we simulate two $\langle W_{\rm cyc}\rangle -v_{\rm dr}$ curves with the change from the high temperature to the low temperature more gently and there is still work output at low driving velocity. Moreover, we can see that a sinusoidal temperature field with the same wavelength and the same highest and lowest temperature values as the discontinuous one can lead to considerable work output at low driving velocity too. So the temperature field is not that demanding. And a sinusoidal temperature field like that in the laser-induced transient thermal grating \cite{TTGPRL,TTGBook,TTGAR} may serve as a possible solution. To obtain a rectangular waveform temperature field, it's promising to superpose several sinusoidal waves with different amplitude, phase and wavelength. The optically levitated dielectric nanosphere system in cavity with two standing-wave optical modes, one for trapping and the other weaker one for cooling \cite{CavityOMOptLNPNAS,TwolasersPRA}, superposed with a uniformly moving Paul trap,
%(or equivalently with the Paul trap stationary and the two uniformly travelling-waves), 
may also work. And we can also speculate that the trapped ion system with two lasers, one for trappinng and the other for cooling, superposed with a uniformly moving Paul trap, may work too. 
%Nevertheless, if we only want to implement the PTSHE at the low velocity regime where it actually works and work output can be observed in one cycle, we can just setup four high and low temperature heat baths alternatively by e.g. four beams of lasers next to one another with alternative intensities.

\subsubsection{On the damping coefficient}
In the main text Figure 3(A), we can see that for the cases considered there the low driving velocity limit of the mean cycle work is nearly independent on $\mu$, i.e. the overdamped and underdamped cases have similar work output. Therefore we can implement the PTSHE both with microscale colloidal particle in liquid medium at the overdamped regime \cite{Stirling,BrownianCarnot} and with optically levitated nanoparticle \cite{UnderdampedEngine} or trapped ion \cite{ScienceTIFE} at the underdamped regime.

\begin{figure}[H]
 \centering
\includegraphics[width=0.49\textwidth]{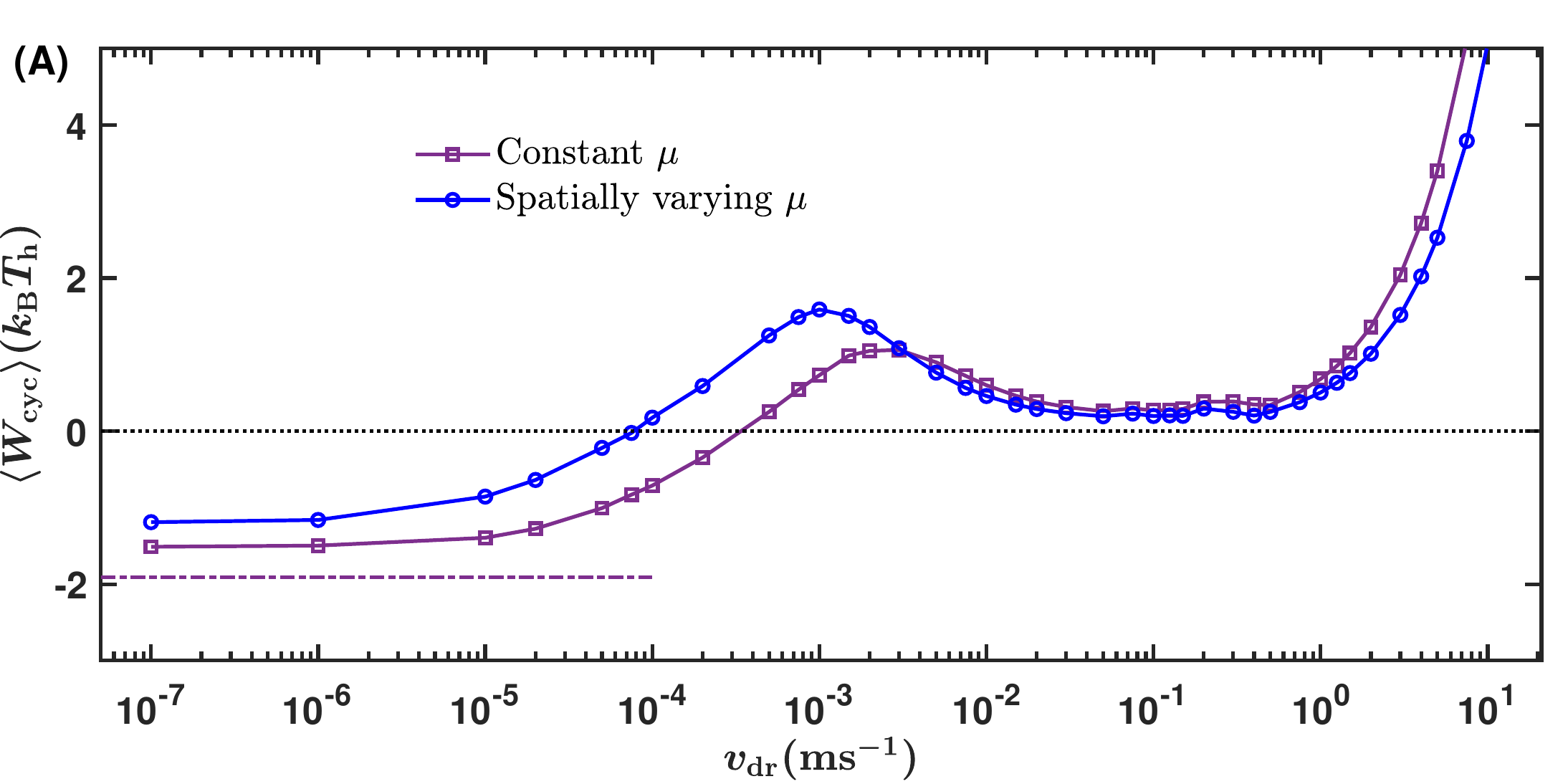}
\includegraphics[width=0.49\textwidth]{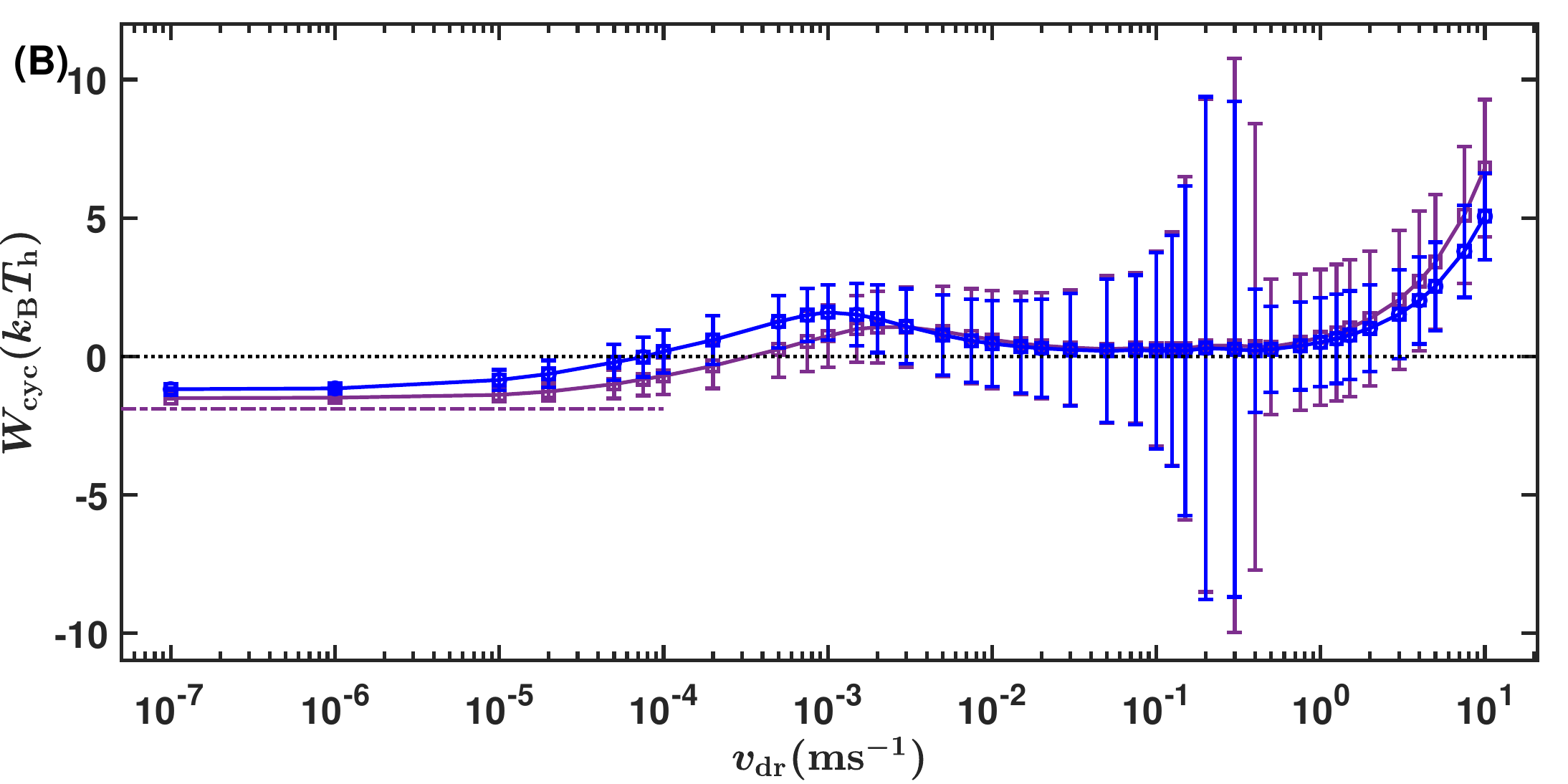}
\caption{The simulation results at spatially varying damping coefficient compared with the constant damping coefficient case. The $\langle W_{\rm cyc}\rangle-v_{\rm dr}$ curves are plotted in (A) and the standard deviations of $W_{\rm cyc}$ are given in (B). Here $\eta=3.0,\ {\it\Theta}_{\rm h,c}=0.4,0.04$ and other parameters except $\mu(z)$ are given in Sec. \ref{Langevindynamicssimulation}.\ref{ParametersUsed}.}
\label{fig:Wcyc_Vdr_viscosvar}
 \end{figure}

To evaluate the temperature dependence of the damping coefficient, we simulate a case in which the damping coefficient field $\mu(z)$ is proportional to the temperature field ${\it\Theta}(z)$ (Eq. \ref{ContinuousTemp}):
\begin{equation}\label{varyingmu}
\mu(z)=\frac12\left\{\mu_{\rm h}+\mu_{\rm c}+(\mu_{\rm h}-\mu_{\rm c})\tanh(\frac1\alpha[\sin(z+\arctan\sqrt{\frac{\eta-1}{\eta+1}})-\sqrt{\frac{\eta-1}{2\eta}}])\right\},
\end{equation}
where we choose $\mu_{\rm h}\approx6.32\times10^3\rm s^{-1}$ and $\mu_{\rm c}=4\times10^4\rm s^{-1}$. The value of $\mu_{\rm h}$ is calculated from $\mu_{\rm c}=4\times10^4\rm s^{-1}$ and the assumption that ${\it\Theta}\propto\frac1{\mu^2}$, based on which we will actually obtain $\mu_{\rm h}=\frac2{50\times\sqrt{10}\times10^{-6}}\rm s^{-1}\approx1.26\times10^4\rm s^{-1}$. To make the damping coefficient change more noteworthy, we halved this value to $\mu_{\rm h}=\frac12\frac2{50\times\sqrt{10}\times10^{-6}}\rm s^{-1}\approx6.32\times10^3\rm s^{-1}$. Indeed, Eq. \ref{varyingmu} would be very close to $\mu(z)=\mu_{\rm c}\sqrt{\frac{{\it\Theta}_c}{{\it\Theta}(z)}}$ if we chose $\mu_{\rm h}\approx1.26\times10^4\rm s^{-1}$, because ${\it\Theta}(z)$ is very close to a picewise function composed of two constant functions. The simulation results are plotted in Figure \ref{fig:Wcyc_Vdr_viscosvar}. We can see that there is still distinct work output at low driving velocity.

In the main text, we have mentioned that it seems that the low velocity limit of the mean cycle work is nearly independent on the damping coefficient for the parameter ranges we considered there. However, from Figure \ref{fig:Wcyc_Vdr_viscosvar}, we can see the damping coefficient actually affects the equilibrium cycle work. If we observe carefully the low driving velocity end of the main text Figure 3(A), we can see that the equilibrium cycle work output seems to first increase and then decrease with the damping coefficient increasing. It's valuable to investigate that how 
%in what degree 
the equilibrium crossover probability is dependent on the damping coefficient and what's the effect the damping coefficient has on the equilibrium cycle work output, for which we may need to turn to the Fokker-Planck or Kramers equation method \cite{KramersEqBookRisken} for help.

Qualitatively speaking, the damping coefficient affects the relaxation time from out of equilibrium to equilibrium as well as the damping force $m\mu \dot x$ which is proportional to the velocity $\dot x$ of the particle. In the underdamped regime, the particle will relax to equilibrium more quickly when the damping coefficient increases, i.e. equilibrium is arrived at more quickly relative to the driving velocity. So the driving velocity range of work output extends to the right [Figure 3(A) in the main text], until the damping coefficient is so large that the damping force becomes very high at low driving velocity, which leads to the driving velocity range of work output shrinks to the left again. Therefore, to avoid too small driving velocity a moderate $\mu$ should be chosen.
 
% As $\mu$ increases from $0$ to $4\times10^7\rm s^{-1}$, the velocity range of work output first expands to the right and then shrinks back, so to avoid too small driving velocity a moderate $\mu$ should be chosen.

%\subsection{The $\langle W_{\rm cyc}\rangle-v_{\rm dr}$ curves at spatially varying damping coefficient}
%To evaluate the effect of the temperature dependence of the damping coefficient, we consider the damping coefficient proportional to the temperature ${\it\Theta}$ in Eq. (\ref{ContinuousTemp})
%\begin{equation}\label{varyingmu}
%\mu(z)=\frac12\left\{\mu_{\rm h}+\mu_{\rm c}+(\mu_{\rm h}-\mu_{\rm c})\tanh(\frac1\alpha[\sin(z+\arctan\sqrt{\frac{\eta-1}{\eta+1}})-\sqrt{\frac{\eta-1}{2\eta}}])\right\},
%\end{equation}
%where we choose $\mu_{\rm h}=\frac1{5\times\sqrt{10}\times10^{-6}}\rm s^{-1}\approx6.32\times10^3\rm s^{-1}$ and $\mu_{\rm c}=4\times10^4\rm s^{-1}$. The simulation results is plotted in Figure \ref{fig:Wcyc_Vdr_viscosvar}. We can see that there is still work output at low velocity.
% \begin{figure}[H]
% \centering
%\includegraphics[width=0.49\textwidth]{SFigures/Wcyc_viscosvar.pdf}
%\includegraphics[width=0.49\textwidth]{SFigures/Wcyc_viscosvar_Error.pdf}
%\caption{The $\langle W_{\rm cyc}\rangle-v_{\rm dr}$ curves at spatially varying damping coefficient compared with the constant damping coefficient case. Here $\eta=3,\ {\it\Theta}_{\rm h,c}=0.4,0.04$ and other parameters are given in Sec.~\ref{ParametersUsed}.}
%\label{fig:Wcyc_Vdr_viscosvar}
% \end{figure}
 
\section{Fluctuation-dissipation theorem}
\label{FluctDissTheoSec}
In Langevin Eq. 4 in the main text, we assume that the intensity $\Gamma$ of the stochastic force $\xi(t)=\Gamma\zeta(t)$ satisfies the fluctuation-dissipation theorem. Here we give a brief derivation.
The Langevin equation can be linearized into a nonhomogeneous ordinary differential equation discribing a Brownian motion of a harmonic oscillator \cite{PathriaChapter15}
\begin{equation}\label{LinearLangevinEq}
\ddot{x}(t)+\mu\dot{x}(t)+\omega_0^2(1+\eta)x(t)=\frac\Gamma m\zeta(t),
\end{equation}
which is equivalent to a first order differential equation system
%\begin{equation}
%\begin{cases}
%\frac{\mathrm dx}{\mathrm dt}=\dot x(t),\\
%\frac{\mathrm d\dot x}{\mathrm dt}=-\mu\dot x-\omega_0^2(1+\eta)x+\frac\Gamma m\zeta(t),
%\end{cases}
%\end{equation}
%i.e.
\begin{equation}
\frac{\mathrm d}{\mathrm dt}\left(\begin{matrix}x\\
\dot x\end{matrix}\right)=\begin{bmatrix}0&1\\
-\omega_0^2(1+\eta)&-\mu
\end{bmatrix}\left(\begin{matrix}x\\
\dot x\end{matrix}\right)+\left(\begin{matrix}0\\
\frac\Gamma m\zeta(t)\end{matrix}\right).
\end{equation}
It can be represented compactly as 
\begin{equation}\label{LinearVectorEq}
\dot{\bm x}=\bm A\bm x+\bm b.
\end{equation}

We first consider the case in which $\mu\neq2\omega_0\sqrt{1+\eta}$ and $\bm A$ has two eigenvalues 
\begin{equation}\label{eq:lamda12muneqomega0}
\begin{aligned}
\lambda_1=\frac{-\mu-\sqrt{\mu^2-4\omega_0^2(1+\eta)^2}}2,\\\lambda_2=\frac{-\mu+\sqrt{\mu^2-4\omega_0^2(1+\eta)^2}}2,
\end{aligned}
\end{equation}
with two eigenvectors 
\begin{equation}
\bm v_1=\left(\begin{matrix}\frac{\lambda_2}{\omega_0^2(1+\eta)^2}\\1\end{matrix}\right),\bm v_2=\left(\begin{matrix}\frac{\lambda_1}{\omega_0^2(1+\eta)^2}\\1\end{matrix}\right),
\end{equation}
respectively. The corresponding homogeneous equation $\dot{\bm x}=\bm A\bm x$ has general solutions of the form $\bm x=\left(\begin{matrix}\mathrm e^{\lambda_1 t}\bm v_1&\mathrm e^{\lambda_2 t}\bm v_2\end{matrix}\right)\left(\begin{matrix}c_1\\c_2\end{matrix}\right)$ with two constants $c_1$ and $c_2$. By the method of variation of constant, the solution of Eq. \ref{LinearVectorEq} can be obtained as
\begin{equation}\label{SolLinearLangevinEq}
\begin{aligned}
\bm x=\left(\begin{matrix}x\\
\dot x\end{matrix}\right)&=c_1(0)\mathrm e^{\lambda_1 t}\bm v_1+c_2(0)\mathrm e^{\lambda_2 t}\bm v_2\\
&\qquad\qquad+\frac{\Gamma}{m}\frac1{\sqrt{\mu^2-4\omega_0^2(1+\eta)^2}}\left[\left(\begin{matrix}1\\ \lambda_1\end{matrix}\right)\mathrm e^{\lambda_1 t}\int_0^t-\mathrm e^{-\lambda_1 t'}\zeta(t')\mathrm dt'+\left(\begin{matrix}1\\ \lambda_2\end{matrix}\right)\mathrm e^{\lambda_2 t}\int_0^t\mathrm e^{-\lambda_2 t'}\zeta(t')\mathrm dt'\right]\\
&=c_1(0)\mathrm e^{\lambda_1 t}\bm v_1+c_2(0)\mathrm e^{\lambda_2 t}\bm v_2\\
&\qquad\qquad+\frac{\Gamma}{m}\frac1{\sqrt{\mu^2-4\omega_0^2(1+\eta)^2}}\int_0^t\left[-\left(\begin{matrix}1\\ \lambda_1\end{matrix}\right)\mathrm e^{\lambda_1(t-t')}+\left(\begin{matrix}1\\ \lambda_2\end{matrix}\right)\mathrm e^{\lambda_2(t-t')}\right]\zeta(t')\mathrm dt',
\end{aligned}
\end{equation}
with two constants $c_{1,2}(0)$ determined by the initial conditions.
The mean squared value is calculated by

\begin{equation}
\begin{aligned}
\langle\bm x^T\bm x\rangle&=\langle x^2\rangle+\langle\dot x^2\rangle=[c_1(0)\mathrm e^{\lambda_1 t}\bm v_1+c_2(0)\mathrm e^{\lambda_2 t}\bm v_2]^2\\
&\qquad+2[c_1(0)\mathrm e^{\lambda_1 t}\bm v_1+c_2(0)\mathrm e^{\lambda_2 t}\bm v_2]^T\frac{\Gamma}{m}\frac1{\sqrt{\mu^2-4\omega_0^2(1+\eta)^2}}\int_0^t\left[-\left(\begin{matrix}1\\ \lambda_1\end{matrix}\right)\mathrm e^{\lambda_1(t-t')}+\left(\begin{matrix}1\\ \lambda_2\end{matrix}\right)\mathrm e^{\lambda_2(t-t')}\right]\langle\zeta(t')\rangle\mathrm dt'\\
&\qquad+\frac{\Gamma^2}{m^2}\frac1{\mu^2-4\omega_0^2(1+\eta)^2}\left\langle\int_0^t\left[-\left(\begin{matrix}1\\ \lambda_1\end{matrix}\right)\mathrm e^{\lambda_1(t-t')}+\left(\begin{matrix}1\\ \lambda_2\end{matrix}\right)\mathrm e^{\lambda_2(t-t')}\right]\zeta(t')\mathrm dt'\right\rangle^2.
\end{aligned}
\end{equation}
As $t\to\infty$, when the particle arrives at equilibrium, the first term on the RHS goes to zero due to the real parts of $\lambda_1$ and $\lambda_2$ are negative, cf. Eq. \ref{eq:lamda12muneqomega0}. The second one is also zero in that $\langle\zeta(t)\rangle=0$. Because $\langle\zeta(t)\zeta(t')\rangle=\delta(t-t')$, in the third term we have

\begin{equation}
\begin{aligned}
&\left\langle\int_0^t\left[-\left(\begin{matrix}1\\ \lambda_1\end{matrix}\right)\mathrm e^{\lambda_1(t-t')}+\left(\begin{matrix}1\\ \lambda_2\end{matrix}\right)\mathrm e^{\lambda_2(t-t')}\right]\zeta(t')\mathrm dt'\right\rangle^2\\
=&\int_0^t\int_0^t\left[-\left(\begin{matrix}1\\ \lambda_1\end{matrix}\right)\mathrm e^{\lambda_1(t-t')}+\left(\begin{matrix}1\\ \lambda_2\end{matrix}\right)\mathrm e^{\lambda_2(t-t')}\right]^T\left[-\left(\begin{matrix}1\\ \lambda_1\end{matrix}\right)\mathrm e^{\lambda_1(t-t'')}+\left(\begin{matrix}1\\ \lambda_2\end{matrix}\right)\mathrm e^{\lambda_2(t-t'')}\right]\langle\zeta(t')\zeta(t'')\rangle\mathrm dt'\mathrm dt''\\
=&\int_0^t\int_0^t\left[-\left(\begin{matrix}1\\ \lambda_1\end{matrix}\right)\mathrm e^{\lambda_1(t-t')}+\left(\begin{matrix}1\\ \lambda_2\end{matrix}\right)\mathrm e^{\lambda_2(t-t')}\right]^T\left[-\left(\begin{matrix}1\\ \lambda_1\end{matrix}\right)\mathrm e^{\lambda_1(t-t'')}+\left(\begin{matrix}1\\ \lambda_2\end{matrix}\right)\mathrm e^{\lambda_2(t-t'')}\right]\delta(t'-t'')\mathrm dt'\mathrm dt''\\
=&\int_0^t\left[-\left(\begin{matrix}1\\ \lambda_1\end{matrix}\right)\mathrm e^{\lambda_1(t-t')}+\left(\begin{matrix}1\\ \lambda_2\end{matrix}\right)\mathrm e^{\lambda_2(t-t')}\right]^T\left[-\left(\begin{matrix}1\\ \lambda_1\end{matrix}\right)\mathrm e^{\lambda_1(t-t')}+\left(\begin{matrix}1\\ \lambda_2\end{matrix}\right)\mathrm e^{\lambda_2(t-t')}\right]\mathrm dt'\\
=&\int_0^t\left[-\mathrm e^{\lambda_1(t-t')}+\mathrm e^{\lambda_2(t-t')}\right]^2\mathrm dt'+\int_0^t\left[-\lambda_1\mathrm e^{\lambda_1(t-t')}+\lambda_2\mathrm e^{\lambda_2(t-t')}\right]^2\mathrm dt'\\
%=&\left[\mathrm e^{2\lambda_1t}(-\frac1{2\lambda_1})\left.\mathrm e^{-2\lambda_1t'}\right|_0^t-2\mathrm e^{(\lambda_1+\lambda_2)t}(-\frac1{\lambda_1+\lambda_2})\left.\mathrm e^{-(\lambda_1+\lambda_2)t'}\right|_0^t+\mathrm e^{2\lambda_2t}(-\frac1{2\lambda_2})\left.\mathrm e^{-2\lambda_2t'}\right|_0^t\right]\\
%&+\left[\lambda_1^2\mathrm e^{2\lambda_1t}(-\frac1{2\lambda_1})\left.\mathrm e^{-2\lambda_1t'}\right|_0^t-2\lambda_1\lambda_2\mathrm e^{(\lambda_1+\lambda_2)t}(-\frac1{\lambda_1+\lambda_2})\left.\mathrm e^{-(\lambda_1+\lambda_2)t'}\right|_0^t+\lambda_2^2\mathrm e^{2\lambda_2t}(-\frac1{2\lambda_2})\left.\mathrm e^{-2\lambda_2t'}\right|_0^t\right]\\
=&\left[(-\frac1{2\lambda_1})(1-\mathrm e^{2\lambda_1t})-2(-\frac1{\lambda_1+\lambda_2})(1-\mathrm e^{(\lambda_1+\lambda_2)t})+(-\frac1{2\lambda_2})(1-\mathrm e^{2\lambda_2t})\right]\\
&+\left[\lambda_1^2(-\frac1{2\lambda_1})(1-\mathrm e^{2\lambda_1t})-2\lambda_1\lambda_2(-\frac1{\lambda_1+\lambda_2})(1-\mathrm e^{(\lambda_1+\lambda_2)t})+\lambda_2^2(-\frac1{2\lambda_2})(1-\mathrm e^{2\lambda_2t})\right]\\
\xrightarrow{t\to\infty}&\left[(-\frac1{2\lambda_1})-2(-\frac1{\lambda_1+\lambda_2})+(-\frac1{2\lambda_2})\right]+\left[\lambda_1^2(-\frac1{2\lambda_1})-2\lambda_1\lambda_2(-\frac1{\lambda_1+\lambda_2})+\lambda_2^2(-\frac1{2\lambda_2})\right]\\
%=&\left(-\frac{\lambda_1+\lambda_2}{2\lambda_1\lambda2}+\frac2{\lambda_1+\lambda_2}\right)+\left(-\frac12(\lambda_1+\lambda_2)+\frac{2\lambda_1\lambda_2}{\lambda_1+\lambda_2}\right)\\
=&\frac1{\lambda_1\lambda_2}\left(-\frac{\lambda_1+\lambda_2}2+\frac{2\lambda_1\lambda_2}{\lambda_1+\lambda_2}\right)+\left(-\frac{\lambda_1+\lambda_2}2+\frac{2\lambda_1\lambda_2}{\lambda_1+\lambda_2}\right)\\
%=&\left[-\frac{\lambda_1+\lambda_2}{2\lambda_1\lambda2}+\frac2{\lambda_1+\lambda_2}\right]+\lambda_1\lambda_2\left[-\frac{\lambda_1+\lambda_2}{2\lambda_1\lambda_2}+\frac2{\lambda_1+\lambda_2}\right]\\
%=&\frac1{\omega_0^2(1+\eta)^2}\left[\frac\mu2+\frac{2\omega_0^2(1+\eta)^2}{-\mu}\right]+\left[\frac\mu2+\frac{2\omega_0^2(1+\eta)^2}{-\mu}\right]\\
=&\frac1{\omega_0^2(1+\eta)^2}\frac{\mu^2-4\omega_0^2(1+\eta)^2}{2\mu}+\frac{\mu^2-4\omega_0^2(1+\eta)^2}{2\mu}.
\end{aligned}
\end{equation}

The limit step above roots in that the real parts of $2\lambda_1,(\lambda_1+\lambda_2),2\lambda_2$ are all negative, whether $\mu>2\omega_0\sqrt{1+\eta}$ or $\mu<2\omega_0\sqrt{1+\eta}$, cf. Eq. \ref{eq:lamda12muneqomega0}. 

Therefore, at the equilibrium state,

\begin{equation}
\begin{aligned}\label{EquilibriumXV}
\langle x(t)^2\rangle&=\frac{\Gamma^2}{m^2}\frac1{\mu^2-4\omega_0^2(1+\eta)^2}\frac1{\omega_0^2(1+\eta)^2}\frac{\mu^2-4\omega_0^2(1+\eta)^2}{2\mu}=\frac{\Gamma^2}{2\mu m^2\omega_0^2(1+\eta)^2},\\
\langle v(t)^2\rangle&=\frac{\Gamma^2}{m^2}\frac1{\mu^2-4\omega_0^2(1+\eta)^2}\frac{\mu^2-4\omega_0^2(1+\eta)^2}{2\mu}=\frac{\Gamma^2}{2\mu m^2}.
\end{aligned}
\end{equation}

Substitute the equipartition theorem
\begin{equation}\label{EquipartitionTheorem}
\begin{aligned}
\frac12m\omega_0^2(1+\eta)\langle x(t)^2\rangle&=\frac12k_{\rm B}T,\\
\frac12m\langle v(t)^2\rangle&=\frac12k_{\rm B}T.
\end{aligned}
\end{equation}
into Eq. \ref{EquilibriumXV} we will obtain
\begin{equation}\label{FlucDiss}
\Gamma=\sqrt{2m\mu k_{\rm B}T}
\end{equation}
from both equations. This is a form of fluctuation-dissipation theorem for the Brownian motion of a linear harmonic oscillator \cite{PathriaChapter15}. 

For the case in which $\mu=2\omega_0\sqrt{1+\eta}$, $\bm A$ has two degenerate eigenvalues 
\begin{equation}
\lambda_1=\lambda_2=\frac{-\mu}2=-\omega_0\sqrt{1+\eta}=\lambda,
\end{equation}
with one eigenvector 
\begin{equation}
\bm v=\left(\begin{matrix}\frac{-\mu}{2\omega_0^2(1+\eta)^2}\\1\end{matrix}\right)=\left(\begin{matrix}-\frac1{\omega_0(1+\eta)}\\1\end{matrix}\right).
\end{equation}
The corresponding homogeneous equation $\dot{\bm x}=\bm A\bm x$ has general solutions of the form $\bm x=\mathrm e^{\lambda t}\left(\begin{matrix}\bm v&\bm v_0+\bm vt\end{matrix}\right)\left(\begin{matrix}c_1\\c_2\end{matrix}\right)$, with $\bm v_0$ satisfying $\begin{cases}(\bm A-\lambda\bm I)^2\bm v_0=\bm 0,\\
(\bm A-\lambda\bm I)\bm v_0=\bm 0,\end{cases}$ and two cosntants $c_1$ and $c_2$. Again by the method of variation of constant, the solution of Eq. \ref{LinearVectorEq} can be obtained. We will achieve the same fluctuation-dissipation theorem as Eq. \ref{FlucDiss} with a similar procedure as above.

For our nonlinear Langevin equation, we assume the noise intensity $\Gamma$ is related to the friction coefficient $\mu$ by the same fluctuation-dissipation theorem as the linear case, which is customary \cite{zwanzig}.

We can see that for a harmonic oscillator, the fluctuation-dissipation theorem Eq. \ref{FlucDiss} is equivalent to the equipartition theorem Eq. \ref{EquipartitionTheorem}. In the next section the Langevin dynamics simulation method will be evaluated basing on this point to guarantee that it satisfys the fluctuation-dissipation theorem and that the time stepsize we choose is reasonable.

\section{Langevin dynamics simulation}
\label{Langevindynamicssimulation}
\subsection{Nodimensionalization of the Langevin equation}
The Langevin Eq. 4 in the main text is numerically solved by the 4th order Stochastic Runge-Kutta (SRK4) method \cite{DongAnalytical,KasdinSRK4}. We first nondimensionalize it to reduce the floating point computation error \cite{gangloffphdthesis}:
\begin{equation}\label{NLEqn}
\frac{\mathrm d^2z}{\mathrm d\tau^2}+\beta\eta\frac{\mathrm dz}{\mathrm d\tau}+4\pi^2(z-\tilde{v}\tau)+4\pi^2\eta\sin z=4\pi^2\eta\Xi(\tau),
\end{equation}
with $z=\frac{2\pi}ax,\ \tau=\frac{\omega_0}{2\pi}t,\  \beta=\frac{m\mu\omega_0a^2}{\pi V_0}$, and $\eta=\frac{2\pi^2V_0}{m\omega_0^2a^2}$ being the nondimentional corrugation number, $\tilde v=\frac{4\pi^2}{a\omega_0}v_{\rm dr}$ and 

\begin{equation}\label{NondimXi}
\begin{aligned}
\Xi(\tau)\mathrm d\tau&=\frac a{\pi V_0}\xi(\frac{2\pi}{\omega_0}\tau)\mathrm d\tau=\frac a{\pi V_0}\Gamma(\frac a{2\pi}z)\zeta(\frac{2\pi}{\omega_0}\tau)\mathrm d\tau=\frac{a\omega_0}{2\pi^2 V_0}\Gamma(\frac a{2\pi}z)\zeta(\frac{2\pi}{\omega_0}\tau)\mathrm d(\frac{2\pi}{\omega_0}\tau)\\
&=\frac{a\omega_0}{2\pi^2 V_0}\Gamma(\frac a{2\pi}z)\zeta(t)\mathrm dt=\frac{a\omega_0}{2\pi^2 V_0}\Gamma(\frac a{2\pi}z)\mathrm d\mathcal{W}_t=\frac{a\omega_0}{2\pi^2 V_0}\Gamma(\frac a{2\pi}z)\sqrt{\frac{2\pi}{\omega_0}}\mathrm d\mathcal{W}_\tau\\
&=\frac{a\omega_0}{2\pi^2 V_0}\sqrt{2m\mu k_BT(\frac a{2\pi}z)}\sqrt{\frac{2\pi}{\omega_0}}\mathrm d\mathcal{W}_\tau=\sqrt{\frac{1}{\pi^2}\frac{m\mu\omega_0 a^2}{\pi V_0} \frac{k_BT(\frac a{2\pi}z)}{V_0}}\mathrm d\mathcal{W}_\tau\\
&=\sqrt{\frac{\beta}{\pi^2}\frac{k_BT(\frac a{2\pi}z)}{V_0}}\mathrm d\mathcal{W}_\tau\\
&=\sqrt{\frac{\beta}{\pi^2}{\it\Theta}(z)}\mathrm d\mathcal{W}_\tau.
\end{aligned}
\end{equation}
Here, ${\it\Theta(z)}=\frac{k_{\rm B}T(\frac a{2\pi}z)}{V_0}$ is the nondimensional temperature. $\mathcal{W}_t=\mathcal{W}(t)$ is a Weiner process with independent increment $\mathrm d\mathcal{W}_t=\zeta(t)\mathrm dt$ satisfying
\begin{equation}\label{dWdt}
\begin{aligned}
\langle \mathrm d\mathcal{W}_t\rangle&=0,\\
\langle\mathrm d\mathcal{W}_t^2\rangle&=\mathrm dt,
\end{aligned}
\end{equation}
i.e. $\mathrm d\mathcal{W}_t\sim N(0,\mathrm dt)$ \cite{Gardinerhandbook}. After nondimensionalizing $\mathrm d\mathcal{W}_t$ to $\mathrm d\mathcal{W}_\tau$, $\mathrm d\mathcal{W}_\tau$ should satisfy $\mathrm d\mathcal{W}_\tau\sim N(0,\mathrm d\tau)$ and
\begin{equation}
\frac{\mathrm d\mathcal{W}_\tau-0}{\sqrt{\mathrm d\tau}}=\frac{\mathrm d\mathcal{W}_t-0}{\sqrt{\mathrm dt}}\sim N(0,1),
\end{equation}
i.e.
\begin{equation}
\mathrm d\mathcal{W}_t=\sqrt{\frac{\mathrm d t}{\mathrm d\tau}}\mathrm d\mathcal{W}_\tau=\sqrt{\frac{2\pi}{\omega_0}}\mathrm d\mathcal{W}_\tau,
\end{equation}
which leads to the last equality in the second line of Eq. \ref{NondimXi}.
\subsection{Equivalant form of the Stratonovich and the Ito stochastic differential equation}
Eq. \ref{NLEqn} can be transformed to the first-order stochastic differential equation system

\begin{equation}\label{StraSDE}
\begin{cases}
\mathrm dz=\dot z\mathrm d\tau,\\
\mathrm d\dot z=\mathrm d\frac{\mathrm dz}{\mathrm d\tau}=[-\beta\eta\dot z-4\pi^2(z-\tilde{v}\tau)-4\pi^2\eta\sin z]\mathrm d\tau+4\pi^2\eta\sqrt{\frac{\beta}{\pi^2}{\it\Theta}(z)}\mathrm d\mathcal{W}_\tau=a(z,\dot z,\tau)\mathrm d\tau+b(z)\mathrm d\mathcal{W}_\tau.
\end{cases}
\end{equation}

Here the symbol $\mathrm d\mathcal{W}_\tau$ should be interpreted as obeying the Stratonovich's mid-point rule when integrated, i.e. this is a Stratonovich stochastic differential equation \cite{StratonovichSDE,Gardinerhandbook}. 
It's easier to implement numerical discretization on the Ito stochastic differential equation, which is explicit \cite{Gardinerhandbook}. And the stochastic Runge-Kutta method used in this paper is based on the discretization of the Ito stochastic differential equation \cite{KasdinSRK4}. So we'd better transform the Stratonovich stochastic Eq. \ref{StraSDE} into Ito's form. Fortunately, because the temperature in this paper is constant or varies with the nondimential spatial coordinate $z$, not with the nondimential velocity $\dot z$, the Stratonovich and the Ito form are the same \cite{Gardinerhandbook}. Here we give a brief and not rigorous derivation.

Consider discretization of the Stratonovich stochastic integral
\begin{equation}\label{DiscrtStratStoInt}
\int_{\tau_0}^\tau b(z)\circ\mathrm d\mathcal W_\tau=\sum_{i=1}^{n}b(\frac{z_{i-1}+z_{i}}2)\Delta\mathcal W_i,
\end{equation}
in which $\Delta\mathcal W_i=\mathcal W_{i}-\mathcal W_{i-1}$, and the $\circ$ represents the Stratonovich's mid-point rule (see the next subsection) for the moment. The mid-point value of $b(z)$ can be Taylor expanded to 

\begin{equation}\label{MidPt_b}
\begin{aligned}
b(\frac{z_{i-1}+z_{i}}2)&=b(z_{i-1}+\frac{z_{i}-z_{i-1}}2)\\
&=b(z_{i-1})+\frac{\mathrm d}{\mathrm dz} b(z_{i-1})\frac{z_{i}-z_{i-1}}2+\frac12\frac{\mathrm d^2}{\mathrm dz^2}b(z_{i-1})(\frac{z_{i}-z_{i-1}}2)^2+\cdots\\
&=b(z_{i-1})+\frac{\mathrm d}{\mathrm dz} b(z_{i-1})\frac{\dot z_{i-1}\Delta\tau_i}2+\frac12\frac{\mathrm d^2}{\mathrm dz^2}b(z_{i-1})(\frac{\dot z_{i-1}\Delta\tau_i}2)^2+\cdots,
\end{aligned}
\end{equation}
in which $\Delta\tau_i=\tau_i-\tau_{i-1}$ and $z_{i}-z_{i-1}=\dot z_{i-1}\Delta\tau_i$ because of the discretization of the first equation of Eq. \ref{StraSDE} (we only need to keep the first order term here because the higher-order terms will vanish, cf. Eq. \ref{nonlinearTaylor2}). Substitude Eq. \ref{MidPt_b} into Eq. \ref{DiscrtStratStoInt} and take the mean square limit \cite{Gardinerhandbook}, which can be regarded as omitting the terms of order higher than 1 ($\Delta\mathcal W_i$ can be regarded as of order $0.5$), we obtain
 
\begin{equation}\label{StraequIto}
\int_{\tau_0}^\tau b(z)\circ\mathrm d\mathcal W_\tau=\sum_{i=1}^{n}\left[ b(z_{i-1})+\frac{\mathrm d}{\mathrm dz} b(z_{i-1})\frac{\dot z_{i-1}\Delta\tau_i}2+\frac12\frac{\mathrm d^2}{\mathrm dz^2}b(z_{i-1})(\frac{\dot z_{i-1}\Delta\tau_i}2)^2+\cdots\right]\Delta\mathcal W_i=\int_{\tau_0}^\tau b(z)\mathrm d\mathcal W_\tau,
\end{equation}
where the last integral is Ito stochastic integral. Hence the two forms of stochastic integrals (and also the stochastic differential equations) are the same and we can omit the $\circ$ in Eq. \ref{StraSDE}.

Therefore we can apply the numerical methods below directly to Eq. \ref{StraSDE}.

\subsection{The mid-point rule of discretizing the Stratonovich stochastic integral}
\label{midpointrule}
Here we give a brief and not rigorous derivation of the mid-point rule, i.e. the Stratonovich rule. The necessity of the mid-point rule results from the form invariance of the first order differential of the Stratonovich interpretation of the stochastic Eq. \ref{StraSDE}, i.e. for a smooth enough function $f(z(\tau),\dot z(\tau),\tau)$
\begin{equation}\label{StraInvariance}
\mathrm d f(z(\tau),\dot z(\tau),\tau)=\frac{\partial f}{\partial z}\circ\mathrm dz+\frac{\partial f}{\partial\dot z}\circ\mathrm d\dot z+\frac{\partial f}{\partial \tau}\mathrm d\tau,
\end{equation}
which is also known as the chain rule. To obtain this form, we first Taylor expand $f(z(\tau),\dot z(\tau),\tau)$ into
\begin{equation}\label{ItoFormula}
\begin{aligned}
\mathrm df(z(\tau),\dot z(\tau),\tau)&=f(z(\tau)+\mathrm dz(\tau),\dot z(\tau)+\mathrm d\dot z(\tau),\tau+\mathrm d\tau)-f(z(\tau),\dot z(\tau),\tau)\\
&=\left[\frac{\partial}{\partial z}\mathrm dz(\tau)+\frac{\partial}{\partial \dot z}\mathrm d\dot z(\tau)+\frac{\partial}{\partial \tau}\mathrm d\tau\right]f+\frac12\left[\frac{\partial}{\partial z}\mathrm dz(\tau)+\frac{\partial}{\partial \dot z}\mathrm d\dot z(\tau)+\frac{\partial}{\partial \tau}\mathrm d\tau\right]^2f+\cdots\\
&=\left[\frac{\partial}{\partial z}\mathrm dz(\tau)+\frac{\partial}{\partial \dot z}\mathrm d\dot z(\tau)+\frac{\partial}{\partial \tau}\mathrm d\tau\right]f+\frac12\frac{\partial^2f}{\partial\dot z^2}b(z)^2\mathrm d\tau+o(\mathrm d\tau),
\end{aligned}
\end{equation}
in which the third equality results from substituting Eq. \ref{StraSDE} into the second term on the RHS of the second equality, expanding this term, replacing $\mathrm d\mathcal W_\tau^2$ with $\mathrm d\tau$ [here we have used Eq. \ref{dWdt} and omitted the process of taking mean square limit \cite{Gardinerhandbook}] and absorbing the terms of higher than first order into $o(\mathrm d\tau)$. Omitting $o(\mathrm d\tau)$, the last equality is a form of Ito's formula \cite{Gardinerhandbook}, in which an extra term $\frac12\frac{\partial^2f}{\partial\dot z^2}b(z)^2\mathrm d\tau$ is added to Eq. \ref{StraInvariance} if we neglect the $\circ$. 
%The Ito's formula is resulted from the left-point rule, utlizing which Eq. \ref{StraInvariance} is no longer valid. 

Next we will show that utlizing the mid-point rule, we can transform Eq. \ref{StraInvariance} into Ito's formula. In the integral form, Eq. \ref{StraInvariance} can be represented by
\begin{equation}\label{IntegralInvariance}
f(z(\tau),\dot z(\tau),\tau)=f(z(\tau_0),\dot z(\tau_0),\tau_0)+\int_{\tau_0}^\tau\frac{\partial f(z(s),\dot z(s),s)}{\partial z}\circ\mathrm dz(s)+\int_{\tau_0}^\tau\frac{\partial f(z(s),\dot z(s),s)}{\partial\dot z(s)}\circ\mathrm d\dot z(s)+\int_{\tau_0}^\tau\frac{\partial f(z(s),\dot z(s),s)}{\partial \tau}\mathrm ds.
\end{equation}
Substitute Eq. \ref{StraSDE} into Eq. \ref{IntegralInvariance}, we will obtain 
\begin{equation}\label{IntFormIvarSubs}
\begin{aligned}
f(z(\tau),\dot z(\tau),\tau)&=f(z(\tau_0),\dot z(\tau_0),\tau_0)+\int_{\tau_0}^\tau\frac{\partial f(z(s),\dot z(s),s)}{\partial z}\dot z(s)\mathrm ds\\
&\qquad\qquad+\int_{\tau_0}^\tau\frac{\partial f(z(s),\dot z(s),s)}{\partial\dot z}\left[a(z(s),\dot z(s),s)\mathrm ds+b(z(s))\circ\mathrm d\mathcal W_s\right]+\int_{\tau_0}^\tau\frac{\partial f(z(s),\dot z(s),s)}{\partial \tau}\mathrm ds.
\end{aligned}
\end{equation}
The stochastic integral on the RHS can be discretized using mid-point rule:
\begin{equation}\label{DisStraIntegralMP}
\begin{aligned}
\int_{\tau_0}^\tau\frac{\partial f(z(s),\dot z(s),s)}{\partial\dot z}b(z(s))\circ\mathrm d\mathcal W_s&=\sum_{i=1}^{n}\frac{\partial f(\frac{z(\tau_i)+z(\tau_{i-1})}2,\frac{\dot z(\tau_i)+\dot z(\tau_{i-1})}2,\tau_{i-1})}{\partial\dot z}b(\frac{z(\tau_i)+z(\tau_{i-1})}2)(\mathcal W_{\tau_i}-\mathcal W_{\tau_{i-1}})\\
&=\sum_{i=1}^{n}g(\frac{z(\tau_i)+z(\tau_{i-1})}2,\frac{\dot z(\tau_i)+\dot z(\tau_{i-1})}2,\tau_{i-1})(\mathcal W_{\tau_i}-\mathcal W_{\tau_{i-1}}),
\end{aligned}
\end{equation}
where we have replaced $\frac{\partial f(z(\tau),\dot z(\tau),\tau)}{\partial\dot z}b(z(\tau))$ by $g(z(\tau),\dot z(\tau),\tau)$ for compactness. As in Eq. \ref{MidPt_b} we can expand the mid-point value of $g(z(\tau),\dot z(\tau),\tau)$ into
\begin{equation}\label{eqn:midpointexpg}
\begin{aligned}
&g(\frac{z(\tau_i)+z(\tau_{i-1})}2,\frac{\dot z(\tau_i)+\dot z(\tau_{i-1})}2,\tau_{i-1})\\
=&g(z_{i-1}+\frac{z_i-z_{i-1}}2,\dot z_{i-1}+\frac{\dot z_i-\dot z_{i-1}}2,\tau_{i-1})\\
=&g(z_{i-1},\dot z_{i-1},\tau_{i-1})+\frac{\partial g(z_{i-1},\dot z_{i-1},\tau_{i-1})}{\partial z}\frac{z_i-z_{i-1}}2+\frac{\partial g(z_{i-1},\dot z_{i-1},\tau_{i-1})}{\partial\dot z}\frac{\dot z_i-\dot z_{i-1}}2+\cdots\\
=&g(z_{i-1},\dot z_{i-1},\tau_{i-1})+\frac{\partial g(z_{i-1},\dot z_{i-1},\tau_{i-1})}{\partial z}\frac{\dot z_{i-1}(\tau_i-\tau_{i-1})}2\\
&\qquad\qquad\qquad+\frac{\partial g(z_{i-1},\dot z_{i-1},\tau_{i-1})}{\partial\dot z}\frac{a(z_{i-1},\dot z_{i-1},\tau_{i-1})(\tau_i-\tau_{i-1})+b(z_{i-1})(\mathcal W_{\tau_i}-\mathcal W_{\tau_{i-1}})}2+\cdots,
\end{aligned}
\end{equation}
substitute which into Eq. \ref{DisStraIntegralMP} and discard the terms of higher than first order of $\tau_i-\tau_{i-1}$, we will get
\begin{equation}\label{TransformStraIto}
\begin{aligned}
\int_{\tau_0}^\tau\frac{\partial f(z(s),\dot z(s),s)}{\partial\dot z}b(z(s))\circ\mathrm d\mathcal W_s&=\sum_{i=1}^{n}\left[g(z_{i-1},\dot z_{i-1},\tau_{i-1})(\mathcal W_{\tau_i}-\mathcal W_{\tau_{i-1}})+\frac12\frac{\partial g(z_{i-1},\dot z_{i-1},\tau_{i-1})}{\partial\dot z}b(z(\tau_{i-1}))(\mathcal W_{\tau_i}-\mathcal W_{\tau_{i-1}})^2\right]\\
&=\sum_{i=1}^{n}\left[g(z_{i-1},\dot z_{i-1},\tau_{i-1})(\mathcal W_{\tau_i}-\mathcal W_{\tau_{i-1}})+\frac12\frac{\partial g(z_{i-1},\dot z_{i-1},\tau_{i-1})}{\partial\dot z}b(z(\tau_{i-1}))(\tau_i-\tau_{i-1})\right]\\
&=\int_{\tau_0}^\tau\frac{\partial f(z(s),\dot z(s),s)}{\partial\dot z}b(z(s))\mathrm d\mathcal W_s+\frac12\int_{\tau_0}^\tau\frac{\partial g(z(s),\dot z(s),s)}{\partial\dot z}b(z(s))\mathrm ds\\
&=\int_{\tau_0}^\tau\frac{\partial f(z(s),\dot z(s),s)}{\partial\dot z}b(z(s))\mathrm d\mathcal W_s+\frac12\int_{\tau_0}^\tau\frac{\partial^2 f(z(s),\dot z(s),s)}{\partial\dot z^2}b(z(s))^2\mathrm ds,
\end{aligned}
\end{equation}
where on the RHS of the second equality we replace $(\mathcal W_{\tau_i}-\mathcal W_{\tau_{i-1}})^2$ by $(\tau_i-\tau_{i-1})$ [here we have used $\langle (\mathcal W_{\tau_i}-\mathcal W_{\tau_{i-1}})^2\rangle=\tau_i-\tau_{i-1}$ and omitted the process of taking mean square limit \cite{Gardinerhandbook}] and the two stochastic integrals on the RHS of the last two equalities are of Ito's form. 

In the last equality of Eq. \ref{eqn:midpointexpg}, we have substituted $\dot z_i-\dot z_{i-1}$ with $a(z_{i-1},\dot z_{i-1},\tau_{i-1})(\tau_i-\tau_{i-1})+b(z_{i-1})(\mathcal W_{\tau_i}-\mathcal W_{\tau_{i-1}})$ utilizing the left-point rule for compactness. We can also substitute $\dot z_i-\dot z_{i-1}$ with $a(z_{i-1},\dot z_{i-1},\tau_{i-1})(\tau_i-\tau_{i-1})+b(\frac{z_{i-1}+z_i}2)(\mathcal W_{\tau_i}-\mathcal W_{\tau_{i-1}})$ and then Taylor expand $b(\frac{z_{i-1}+z_i}2)$ as in Eq. \ref{MidPt_b}. After substituting what we get of Eq. \ref{eqn:midpointexpg} in this way into Eq. \ref{DisStraIntegralMP} and discard the terms of higher than first order of $\tau_i-\tau_{i-1}$, we will obtain Eq. \ref{TransformStraIto} again. 

Substitute Eq. \ref{TransformStraIto} into Eq. \ref{IntFormIvarSubs} and convert it back into differential form we will obtain Ito's formula Eq. \ref{ItoFormula}. From the derivation process in Eq. \ref{ItoFormula}, we can see that the extra term $\frac12\frac{\partial^2f}{\partial\dot z^2}b(z)^2\mathrm d\tau$ of Ito's formula is resulted from the left-point rule, which makes Eq. \ref{StraInvariance} no longer valid. To apply the form variance of the first order differential, i.e.  Eq. \ref{StraInvariance}, as in the ordinary calculus, the mid-point rule is necessary. (Actually here we can only prove that the mid-point rule is sufficient, there may be another rule which can also lead to Eq. \ref{StraInvariance}, so our derivation is not rigorous.)

In Eq. \ref{StraequIto}, $\frac{\partial f(z(\tau),\dot z(\tau),\tau)}{\partial\dot z}=1$ and the second term on the RHS of Eq. \ref{TransformStraIto} vanishes, so that the Stratonovich and the Ito stochastic integral as well as the Stratonovich and the Ito stochastic differential equation are the same and we can integrate the Stratonovich stochastic differential Eq. \ref{StraSDE} using methods applicable to the Ito stochastic differential equation. However for other stochastic integrals, the function $f(z(\tau),\dot z(\tau),\tau)$ may not be that trivial. So we think that it's necessary to use the mid-point rule when integrate stochastic integrals with stochastic quantities like $z(\tau)$ and $\dot z(\tau)$ as integral variables, e.g. the one in Eq. \ref{QInt}. Although in the stochastic integral in Eq. \ref{QInt}, the integral variable is $x(t)$ and we can change $\mathrm dx(t)$ into $\dot x(\tau)\mathrm d\tau$ to avoid the mid-point rule, we find that it's not as precise as we use $\mathrm dx(t)$ directly with the mid-point rule and the absolute value of the energy difference $\Delta=\Delta U+Q-W$, which should be as close to zero as possible to obey the first law of thermodynamics (cf. Sec. \ref{Langevindynamicssimulation}.\ref{sec:furthercheckfirstlaw}), is about one order of magnitude higher than using the mid-point rule. Maybe we should change $\mathrm dx(t)$ into a higher order approximation as in Eq. \ref{nonlinearTaylor2}. In our opinion, because $\mathrm d\dot z(\tau)$ includes $\mathrm d\mathcal W_{\tau}$, the Stratonovich stochastic integral with respect to $\dot z(\tau)$ should be discretized using the mid-point rule. And because $\mathrm dz(\tau)$ can be represented by $\dot z(\tau)$ and $\mathrm d\mathcal W_{\tau}$ (cf. Eq. \ref{StraSDE}), the Stratonovich stochastic integral in Eq. \ref{QInt} should be discretized by the mid-point rule too. Therefore we should add a  $\circ$ in front of $\mathrm dx(t)$ of this stochastic integral. We omit it to avoid some unnecessary confusion.

There is another form of mid-point rule like that in \cite{BrownianCarnot}:
\begin{equation}\label{DisStraIntegralAnotherMP}
\begin{aligned}
\int_{\tau_0}^\tau\frac{\partial f(z(s),\dot z(s),s)}{\partial\dot z}b(z(s))\circ\mathrm d\mathcal W_s&=\sum_{i=1}^{n}\frac12\left[\frac{\partial f(z(\tau_{i}),\dot z(\tau_i),\tau_i)}{\partial\dot z}b(z(\tau_i))+\frac{\partial f(z(\tau_{i-1}),\dot z(\tau_{i-1}),\tau_{i-1})}{\partial\dot z}b(z(\tau_{i-1}))\right](\mathcal W_{\tau_i}-\mathcal W_{\tau_{i-1}})\\
&=\sum_{i=1}^{n}\frac12\left[g(z(\tau_i),\dot z(\tau_i),\tau_i)+g(z(\tau_{i-1}),\dot z(\tau_{i-1}),\tau_{i-1})\right](\mathcal W_{\tau_i}-\mathcal W_{\tau_{i-1}}),
\end{aligned}
\end{equation}
which we actually use in our simulation.
Because
\begin{equation}
\begin{aligned}
&\frac12\left[g(z(\tau_i),\dot z(\tau_i),\tau_i)+g(z(\tau_{i-1}),\dot z(\tau_{i-1}),\tau_{i-1})\right]\\
=&g(z(\tau_{i-1}),\dot z(\tau_{i-1}),\tau_{i-1})+\frac12\left[g(z(\tau_i),\dot z(\tau_i),\tau_i)-g(z(\tau_{i-1}),\dot z(\tau_{i-1}),\tau_{i-1})\right]\\
=&g(z(\tau_{i-1}),\dot z(\tau_{i-1}),\tau_{i-1})+\frac12\left[\frac{\partial}{\partial z}(z(\tau_i)-z(\tau_{i-1}))+\frac{\partial}{\partial\dot z}(\dot z(\tau_i)-\dot z(\tau_{i-1}))+\frac{\partial}{\partial \tau}(\tau_i-\tau_{i-1})\right]g(z(\tau_{i-1}),\dot z(\tau_{i-1}),\tau_{i-1})+\cdots\\
=&g(z(\tau_{i-1}),\dot z(\tau_{i-1}),\tau_{i-1})\\
&\qquad+\frac12\left\{\frac{\partial}{\partial z}\dot z(\tau_{i-1})(\tau_i-\tau_{i-1})+\frac{\partial}{\partial\dot z}\left[a(z(\tau_i),\dot z(\tau_{i-1}),\tau_{i-1})(\tau_i-\tau_{i-1})+b(z(\tau_{i-1}))(\mathcal W_{i}-\mathcal W_{i-1})\right]+\frac{\partial}{\partial \tau}(\tau_i-\tau_{i-1})\right\}\\
&\qquad\qquad\times g(z(\tau_{i-1}),\dot z(\tau_{i-1}),\tau_{i-1})+\cdots,
\end{aligned}
\end{equation}
substitute which into Eq. \ref{DisStraIntegralAnotherMP} and discard the terms of higher than first order of $\tau_i-\tau_{i-1}$, we will get Eq. \ref{TransformStraIto} again.

Therefore the two forms of mid-point rule are equivalent. And we find that using the two methods to calculate $\Delta=\Delta U+Q-W$ leads to similar results and they work equally well.
\subsection{The continuous temperature function}
\label{sec:tempfunc}
\begin{figure}[H]
 \centering
 \begin{minipage}{0.329\textwidth}
 \centerline{
 \includegraphics[width=\textwidth]{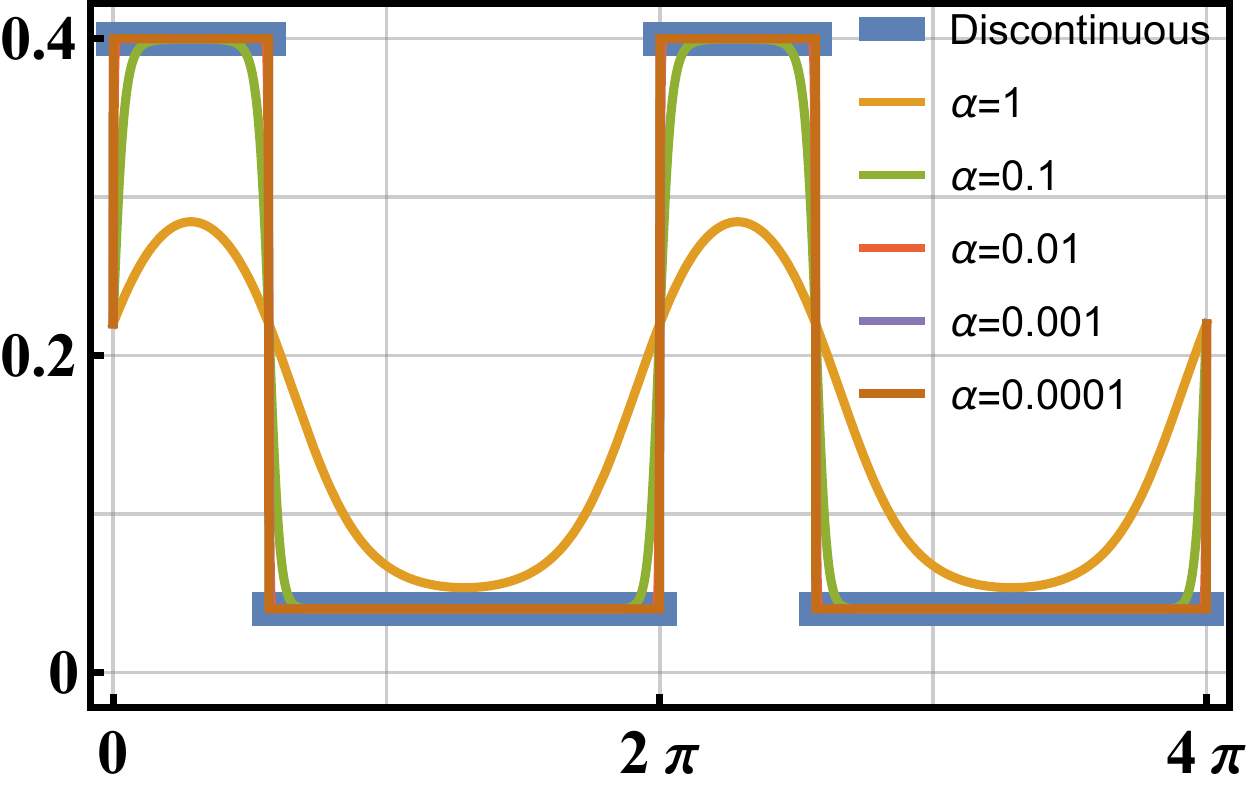}}
 \centerline{(a)\ $\eta=4.60$}
 \end{minipage}
 \begin{minipage}{0.329\textwidth}
 \centerline{\includegraphics[width=\textwidth]{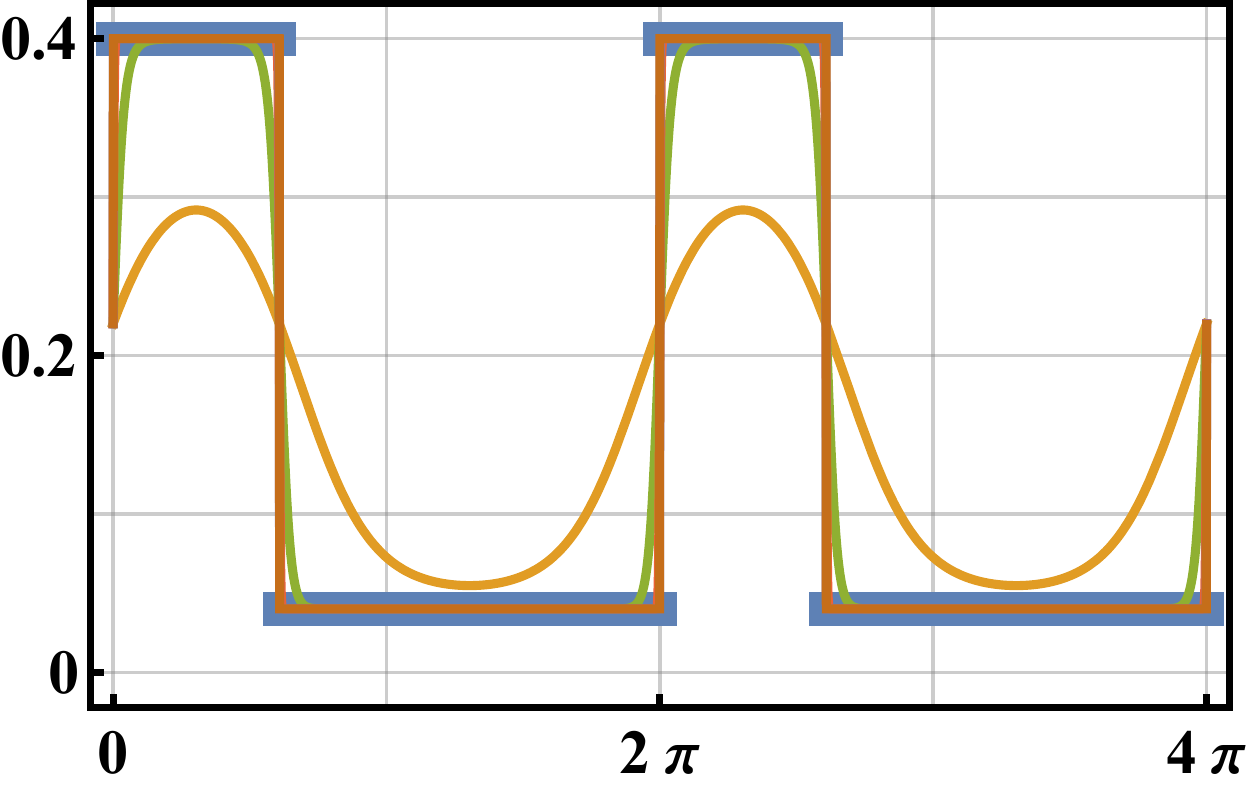}}
 \centerline{(b)\ $\eta=3.00$}
 \end{minipage}
 \begin{minipage}{0.329\textwidth}
 \centerline{\includegraphics[width=\textwidth]{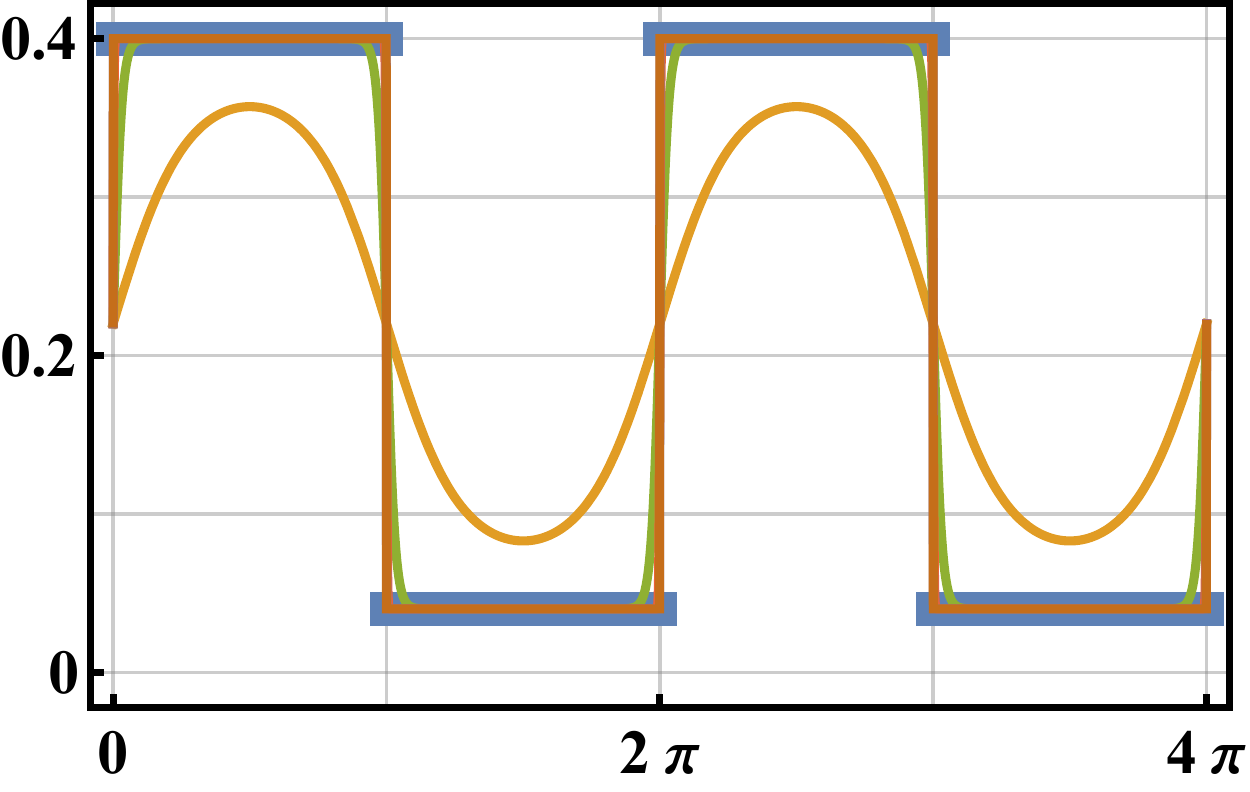}}
 \centerline{(c)\ $\eta=1.00$}
 \end{minipage}
 \caption{
 The function ${\it\Theta}(z)$ at different $\alpha$ with $\eta=4.60,3.00,1.00$ from left to right respectively. The corresponding discontinuous temperature function ${\it\Theta}^*(z)$ is plotted for comparison. Here ${\it\Theta}_{\rm h}=0.4,{\it\Theta}_{\rm c}=0.04$. The three curves of $\alpha\leq0.01$ almost overlap so we can only see the curve of $\alpha=0.0001$.
 }
 \label{TempDist}
 \end{figure}

In the derivation of Eq. \ref{StraequIto} and the mid-point rule, the function $b(z)$ should be continuous. To avoid the not well-defined derivation resulting from the discontinuity of the temperature field in our PTSHE, we construct a continuous function to approximate the nondimensional temperature \cite{ContinuousTemp}

\begin{equation}\label{ContinuousTemp}
{\it\Theta}(z)=\frac12\left\{{\it\Theta}_{\rm h}+{\it\Theta}_{\rm c}+({\it\Theta}_{\rm h}-{\it\Theta}_{\rm c})\tanh(\frac1\alpha[\sin(z+\arctan\sqrt{\frac{\eta-1}{\eta+1}})-\sqrt{\frac{\eta-1}{2\eta}}])\right\},
\end{equation}
where ${\it\Theta}_{\rm h,c}$ represents the temperature of the hot and the cold zones in our PTSHE.
In Figure \ref{TempDist}, we compare ${\it\Theta}(z)$ with its discontinuous version
\begin{equation}\label{DiscontinuousTemp}
{\it\Theta}^*(z)=
\begin{cases}
{\it\Theta}_{\rm h},&2n\pi\leq z<2n\pi+\arccos(-\frac1\eta),\\
{\it\Theta}_{\rm c},&2n\pi+\arccos(-\frac1\eta)\leq z<2(n+1)\pi,
\end{cases}
n\in\mathbb Z,
\end{equation}
 at different $\alpha$ with $\eta=4.60,3.00,1.00$. When $\alpha$ is small enough, i.e. $\alpha\leq0.01$, ${\it\Theta}(z)$ is nearly the same as the discontinuous one. We choose $\alpha=0.001$ in all our simulations unless otherwise noted.

\subsection{Stochastic Runge-Kutta method to integrate the Langevin equation}
We use the 4-step Stochastic Runge-Kutta (SRK4) method in \cite{DongAnalytical,KasdinSRK4} to numerically discretize the stochastic Eq. \ref{StraSDE} into

\begin{equation}\label{SRK4}
\begin{aligned}
z_1&=\dot z_k\Delta\tau,\\
\dot z_1&=a(z_k,\dot z_k,\tau_k)\Delta\tau\\
&\qquad\qquad\qquad\qquad+b(z_k)\sqrt{q_1\Delta\tau}w_{k,1};\\
z_2&=(\dot z_k+a_{21}\dot z_1)\Delta\tau,\\
\dot z_2&=a(z_k+a_{21}z_1,\dot z_k+a_{21}\dot z_1,\tau_k+a_{21}\tau_1)\Delta\tau\\
&\qquad\qquad\qquad\qquad+b(z_k+a_{21}z_1)\sqrt{q_2\Delta\tau}w_{k,2};\\
z_3&=(\dot z_k+a_{31}\dot z_1+a_{32}\dot z_2)\Delta\tau,\\
\dot z_3&=a(z_k+a_{31}z_1+a_{32}z_2,\dot z_k+a_{31}\dot z_1+a_{32}\dot z_2,\tau_k+a_{31}\tau_1+a_{32}\tau_2)\\
&\qquad\qquad\qquad\qquad+b(z_k+a_{31}z_1+a_{32}z_2)\sqrt{q_3\Delta\tau}w_{k,3};\\
z_4&=(\dot z_k+a_{41}\dot z_1+a_{42}\dot z_2+a_{43}\dot z_3)\Delta\tau,\\
\dot z_4&=a(z_k+a_{41}z_1+a_{42}z_2+a_{43}z_3,\dot z_k+a_{41}\dot z_1+a_{42}\dot z_2+a_{43}\dot z_3,\tau_k+a_{41}\tau_1+a_{42}\tau_2+a_{43}\tau_3)\\&\qquad\qquad\qquad\qquad+b(z_k+a_{41}z_1+a_{42}z_2+a_{43}z_3)\sqrt{q_4\Delta\tau}w_{k,4};\\
z_{k+1}&=z_k+\alpha_1z_1+\alpha_2z_2+\alpha_3z_3+\alpha_4z_4,\\
\dot z_{k+1}&=\dot z_k+\alpha_1\dot z_1+\alpha_2\dot z_2+\alpha_3\dot z_3+\alpha_4\dot z_4.
\end{aligned}
\end{equation}

The random numbers are independent and obey identical standard normal distribution with zero-mean and one-variance: $w_{k,i}\sim N(0,1),i=1,2,3,4$ .
The parameters are 

\noindent
\begin{center}
\begin{tabular}{|cccc|c|}
\hline
 & & & & $q_1$\\ 
$a_{21}$ & & & & $q_2$\\
$a_{31}$ & $a_{32}$ & & & $q_3$\\
$a_{41}$ & $a_{42}$ & $a_{43}$ & & $q_4$\\
\hline
$\alpha_1$ & $\alpha_2$ & $\alpha_3$ & $\alpha_4$ & \\
\hline
\end{tabular}=\begin{tabular}{|cccc|c|}
\hline
 & & & & $3.99956364361748$\\ 
$0.66667754298442$ & & & & $1.64524970733585$\\
$0.63493935027993$ & $0.00342761715422$ & & & $1.59330355118722$\\
$-2.32428921184321$ & $2.69723745129487$ & $0.29093673271592$ & & $0.26330006501868$\\
\hline
$0.25001351164789$ & $0.67428574806272$ & $-0.00831795169360$ & $0.08401868181222$ & \\
\hline
\end{tabular}.
\end{center}

\subsection{Choice of the time stepsize}
In order to aviod loss of precision or waste of time, the time stepsize $\Delta\tau$ cannot be too large or too small. It is not only constrained by the original differential equation for the specific practical problem, but also determined by the prescion of the numerical method we used. We calculate the time stepsize from approximation of the Langevin equation first. The disscusion of the feasibility of the time stepsize we will choose is given in the next subsection.

The second derivative of the resultant potential energy Eq. \ref{ResPotential} with respect to $x$ is 
\begin{equation}
\begin{aligned}
\frac{\partial^2V}{\partial x^2}=\kappa+\kappa\eta\cos(\frac{2\pi}{a}x(t))&\leq\kappa(1+\eta)\\
&=m\omega_0^2(1+\eta).
\end{aligned}
\end{equation}
We can approximate the resultant potential energy as $V\approx\frac12m\omega_0^2(1+\eta)x(t)^2$ at the points where its stiffness as well as intrinsic oscillating frequency are both maximum: $\omega_0\sqrt{1+\eta}$. 
At these points neglecting the stochastic force $\xi(t)$, the Langevin Eq. 4 in the main text can be approximated by
\begin{equation}
m\ddot{x}(t)+m\mu\dot{x}(t)+m\omega_0^2(1+\eta)x(t)=0.
\end{equation}
Let's change the sign before the third term on the LHS, 
\begin{equation}
m\ddot{x}(t)+m\mu\dot{x}(t)-m\omega_0^2(1+\eta)x(t)=0,
\end{equation}
so that the  characteristic equation 
\begin{equation}
r^2+\mu r-\omega_0^2(1+\eta)=0
\end{equation}
with two characteristic roots $r_{1,2}=\frac{-\mu\pm\sqrt{\mu^2+4\omega_0^2(1+\eta)}}2$,
achieves a larger ``characteristic frequency'':
\begin{equation}
\frac1{t^*}=\frac{\mu+\sqrt{\mu^2+4\omega_0^2(1+\eta)}}2,
\end{equation}
which is an upper bound of the frequency scale of
%with respect to 
both the oscillation and the damping relaxation of the particle.

Another frequency scale affecting the disctete time step is that of the driver center, and also the particle, passing over the periodic lattice:
\begin{equation}
\frac1{t^{**}}=\frac{v_{\rm dr}}{a}.
\end{equation}
When $v_{\rm dr}$ is small, the particle oscillates many cycles during it passing over one lattice period, so the former frequency scale contributes mainly. On the other hand, when $v_{\rm dr}$ is large, the mass point can pass over many lattice period during one oscillating cycle, and the latter frequency scale contributes more. Therefore, the discrete time stepsize can be computed from 
\begin{equation}
\Delta t=\frac1{100}\frac1{\frac1{t^*}+\frac1{t^{**}}}=\frac1{100}\frac1{\frac{\mu+\sqrt{\mu^2+4\omega_0^2(1+\eta)}}2+\frac{v_{\rm dr}}{a}},
\end{equation}
which is a lower bound. The choice of the factor $\frac1{100}$ is described in the next subsection.

After being nondimensionalized, the simulation discrete time stepsize can be chosen as
\begin{equation}\label{TimeStepSize}
\begin{aligned}
\Delta\tau=\frac{\omega_0}{2\pi}\Delta t
%&=\frac1{50}\frac1{\frac{\mu+\sqrt{\mu^2+4\omega_0^2(1+\eta)}}{2\frac{\omega_0}{2\pi}}+\frac{v_{\rm dr}}{a\frac{\omega_0}{2\pi}}}\\&
=\frac1{100}\frac1{\frac{\beta\eta+\sqrt{(\beta\eta)^2+16\pi^2(1+\eta)}}{2}+\frac{\tilde v}{2\pi}}.
\end{aligned}
\end{equation}

\subsection{Evaluation of the SRK4 method}\label{SRK4Eval}
We now evaluate the SRK4 method introduced above. Kasdin, J \cite{KasdinSRK4} made all the four substeps stochastic with four independent random numbers respectly, which results in more degrees of freedom so that the coefficients of his SRK4 method can be derived by %matching the a Taylor series expansion of the covariance of the discrete approximation of SRK4 with that of the continuous covariance of the solution to a specific order.
matching the covariance of the Taylor expansion of the discrete solution with the Taylor expansion of the covariance of the continuous solution to a specific order. 
Because it's difficult to obtain the covariance of the solution of a nonlinear stochastic differential equation, his derivation is based on linear stochastic differential equation with zero mean. He applied this method to a nonlinear stochastic differential equation and proposed that the same set of coefficients derived for the time-varying linear case are appropriate for the nonlinear case.

In Sec. \ref{FluctDissTheoSec}, we derived the fluctuation-dissipation theorem of the linearized Langevin Eq. \ref{LinearLangevinEq}, the solution of which, Eq. \ref{SolLinearLangevinEq}, is zero-mean with the covariances satisfying the equipartition theorem Eq. \ref{EquipartitionTheorem}, as $t\to\infty$. Thus, at equilibrium the simulation results of the linear Eq. \ref{LinearLangevinEq} with this SRK4 method, Eq. \ref{SRK4}, should satisfy equipartition theorem in high precision. 

Eq. \ref{LinearLangevinEq} can be transformed into a nondimensional stochastic differential equation system
%\begin{equation}\label{NondimLinearLangevinEq}
%\frac{\mathrm d^2z}{\mathrm d\tau^2}+\beta\eta\frac{\mathrm dz}{\mathrm d\tau}+4\pi^2(1+\eta)z=4\pi^2\eta\Xi(\tau),
%\end{equation}

\begin{equation}\label{NondimLinearLangevinEq}
\begin{cases}
\mathrm dz=\dot z\mathrm d\tau,\\
\mathrm d\dot z=[-\beta\eta\dot z-4\pi^2(1+\eta)z]\mathrm d\tau+4\pi^2\eta\sqrt{\frac{\beta}{\pi^2}{\it\Theta}}\mathrm d\mathcal{W}_\tau=\bar a(z,\dot z)\mathrm d\tau+\bar b\mathrm d\mathcal{W}_\tau.
\end{cases}
\end{equation}

\begin{figure}[H]
 \centering
\begin{minipage}{0.95\textwidth}
 \centerline{
 \includegraphics[width=0.5\textwidth]{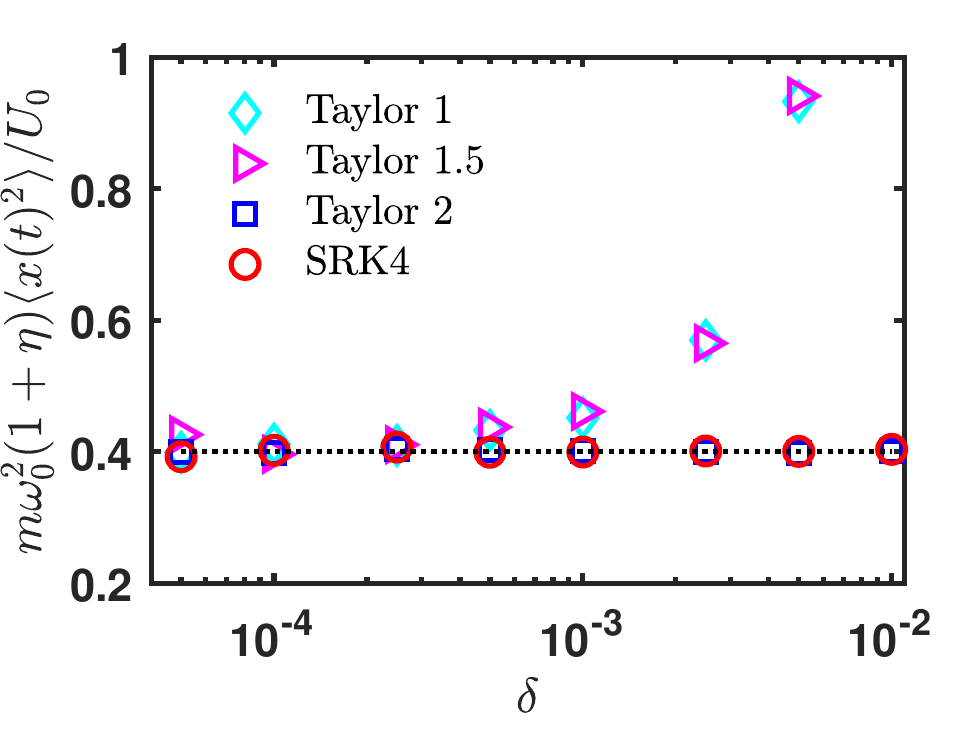}
 \includegraphics[width=0.5\textwidth]{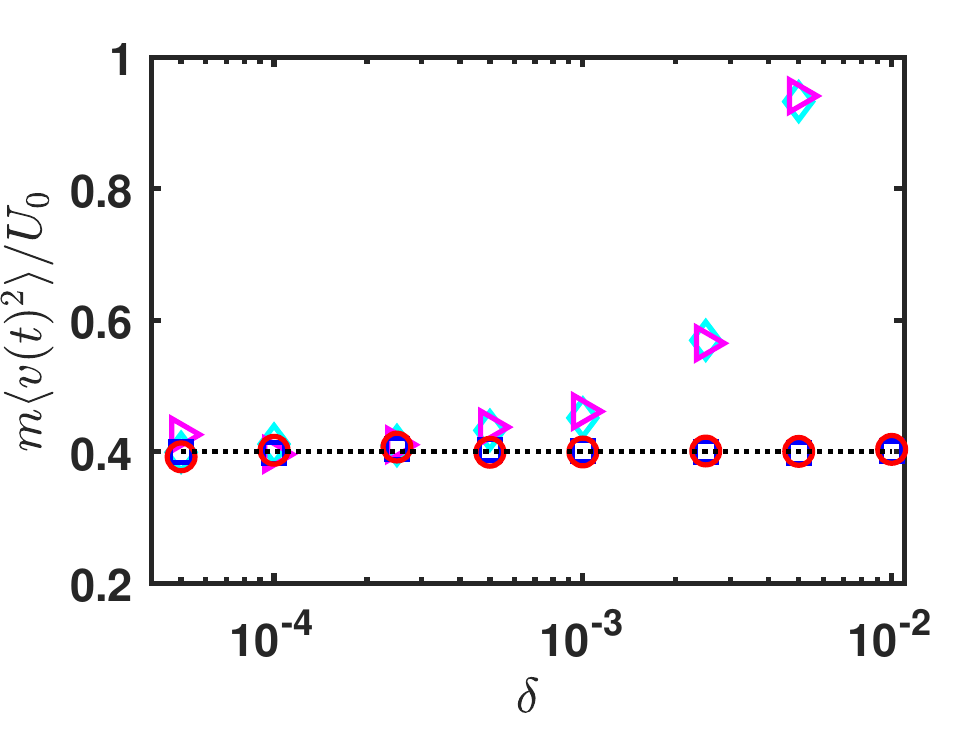}}
 \centerline{(a)\ ${\it\Theta}=0.4$}
 \end{minipage}
\begin{minipage}{0.95\textwidth}
 \centerline{
 \includegraphics[width=0.5\textwidth]{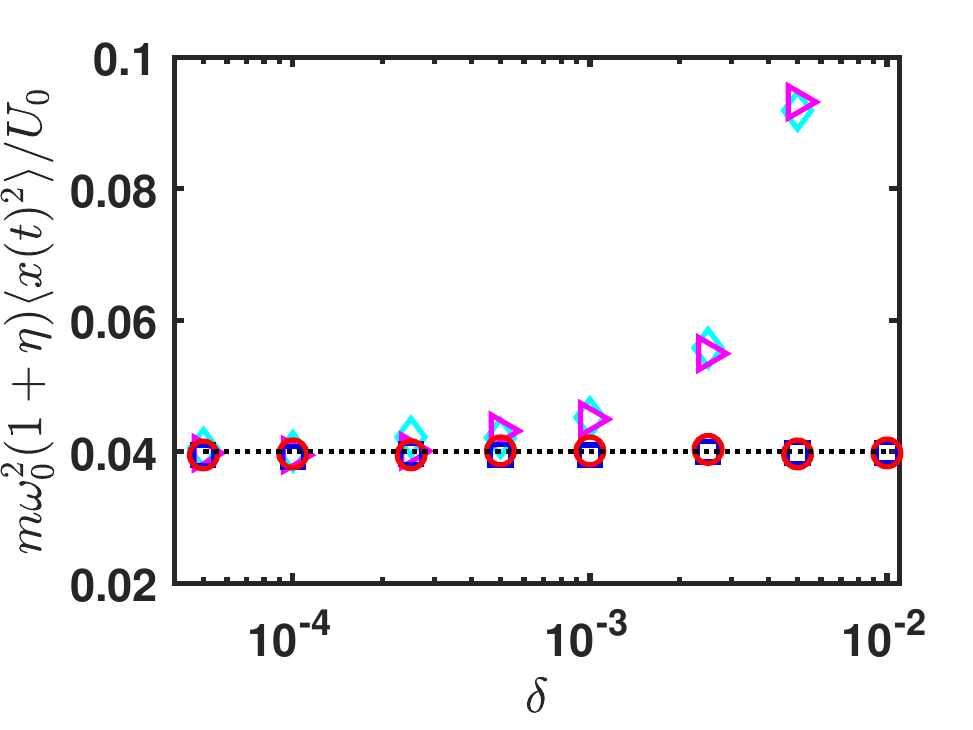}
 \includegraphics[width=0.5\textwidth]{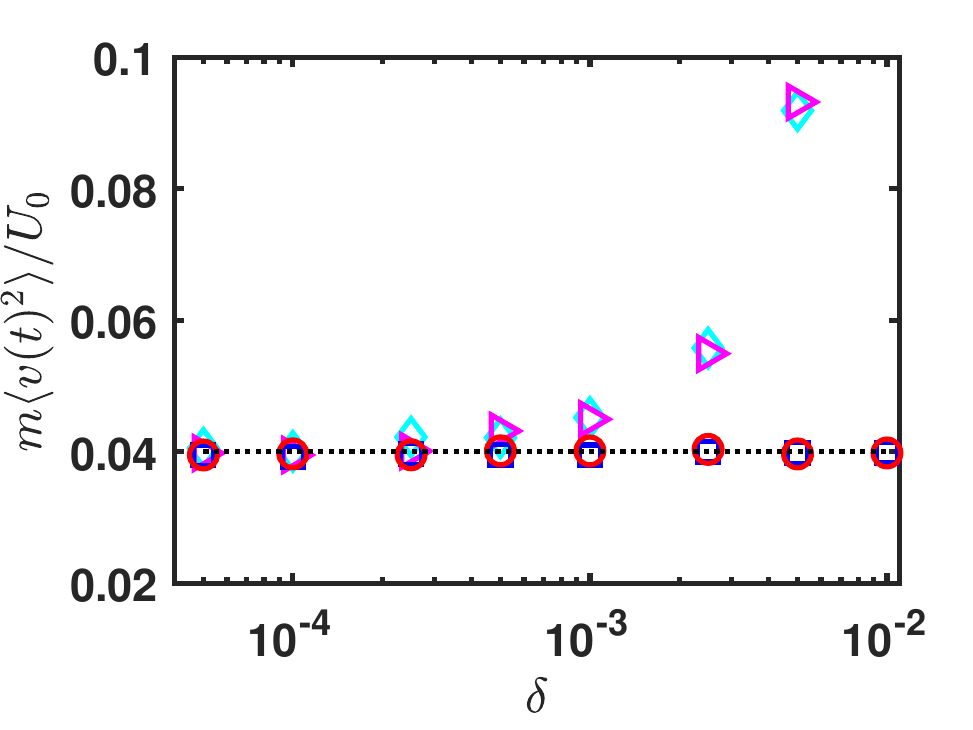}}
 \centerline{(b)\ ${\it\Theta}=0.04$}
 \end{minipage}
 \caption{Comparison between different simulation methods. The dotted black lines represent the values of $m\omega_0^2(1+\eta)\langle x(t)^2\rangle/V_0$ and $m\langle v(t)^2\rangle/V_0$ at equilibrium approximated by the equipartition theorem, which should be equal to the corresponding nondimensional temperature $\it\Theta$. The $x$-coordinates of all the four subfigures are the time stepsize coefficient $\delta$, see the text below. The SRK4 method performs well for all the $\delta$ values considered. The order 2 strong Taylor scheme works equally well. Whereas, the performance of the order 1 and 1.5 strong Taylor schemes depend much on the time stepsize. Note that when $\delta=0.01$, the time step is too large for the order 1 and 1.5 strong Taylor schemes to converge.}
\label{EvalSRK4}
\end{figure}

We integrate it using the SRK4 method described in Eq. \ref{SRK4}. The results are given in Figure \ref{EvalSRK4} represented by the red circles. We can see that for all the time stepsizes considered, the simulation results of Eq. \ref{NondimLinearLangevinEq} at equilibrium with the SRK4 method obey equipartition theorem (represented by the black dotted lines) in high prescision.

Here the time stepsizes are calculated by Eq. \ref{TimeStepSize} with $\frac1{100}$ replaced by a coefficient $\delta$ and $\tilde v=0$. We start the particle at $(x=0,v=0)$ and iterate an enough number of time steps to ensure equilibrium has been arrived at so that we can utilize the time average to approximate the ensemble average.
%to ensure equilibrium has been arrived at and that at equilibrium the time average is equal to the ensemble average. 
The $\delta$'s and the corresponding number $N$'s of time steps which we finally chose in simulation of Eq. \ref{NondimLinearLangevinEq} after some trial and error are listed in Table \ref{EvalSRK4TSteps}. Each ensemble average is calculated by the time average from the $(N/2+1)$st time step, at which we presume that the equilibrium has been arrived at, to the end.

\begin{table}[H]
\centering
\caption{\label{EvalSRK4TSteps}The $\delta$'s and the corresponding $N$'s in simulation of Eq. \ref{NondimLinearLangevinEq} with different methods}

 \begin{tabular}{ccccccccc}
 \toprule
 $\delta$ & $5\times10^{-5}$ & $1\times10^{-4}$ & $2.5\times10^{-4}$ & $5\times10^{-4}$ & $1\times10^{-3}$ & $2.5\times10^{-3}$ & $5\times10^{-3}$ & $1\times10^{-2}$\\
\midrule
 $N(2\times10^8)$ & $200$ & $100$ & $40$ & $20$ & $10$ & $10$ & $10$ & $10$\\
\bottomrule
\end{tabular}
\end{table}

We compare this SRK4 method to the stochastic Taylor approximations of strong convergence order 1.0, 1.5 and 2.0 \cite{NumericalSolSDEStrong}, which are explicit iterative method utilizing the stochastic Taylor expansion \cite{StochasticTaylorExpansion} to different orders. We give a brief derivation of the stochastic Taylor expansion at the end of this subsection. The order 1, 1.5 and 2.0 Taylor schemes to integrate the Langevin Eq. \ref{NondimLinearLangevinEq} are listed here.
 
 Order 1 strong Taylor scheme \cite{NumericalSolSDEStrong} (Taylor 1):
\begin{equation}\label{LinearTaylor1}
\begin{aligned}
z_{k+1}&=z_k+\dot z_k\Delta \tau,\\
\dot z_{k+1}&=\dot z_k+\bar b\sqrt{\Delta\tau}w_k+\bar a(z_k,\dot z_k)\Delta\tau,
\end{aligned}
\end{equation}
with $w_k$ independently normally distributed with zero-mean and one-variance: $w_k\sim N(0,1)$. This scheme is also called Milstein scheme \cite{NumericalSolSDEStrong} and is derived from truncating Eq. \ref{eqn:linearlangevinz} and Eq. \ref{eqn:linearlangevindotz} to the first order. We can obtain the same scheme using (3.2) in \cite{NumericalSolSDEStrong} with $b^1$ and $b^2$ substituted by $0$ and $\bar b$ respectively and because the function $\bar b$ is independent of $\dot z$, the third term on the RHS of (3.2) in \cite{NumericalSolSDEStrong} vanishes. Therefore for Eq. \ref{NondimLinearLangevinEq} this scheme is the same as the Euler scheme, i.e. (2.4) in \cite{NumericalSolSDEStrong}, which is of strong order 0.5.
 
 Order 1.5 strong Taylor scheme \cite{NumericalSolSDEStrong} (Taylor 1.5):
\begin{equation}\label{LinearTaylor1p5}
\begin{aligned}
z_{k+1}&=z_k+\dot z_k\Delta \tau+\bar b\Delta Z_k,\\
\dot z_{k+1}&=\dot z_k+\bar b \Delta W_k+\bar a(z_k,\dot z_k)\Delta\tau-\beta\eta\bar b\Delta Z_k,
\end{aligned}
\end{equation}
where $\Delta W_k=w_{k,1}\sqrt{\Delta\tau},\ \Delta Z_k=\frac12\Delta\tau^{\frac32}(w_{k,1}+\frac1{\sqrt3}w_{k,2})$ with $w_{k,i},\ i=1,2$ independently normally distributed with zero-mean and one-variance: $w_{k,i}\sim N(0,1),\ i=1,2$. This scheme is derived from truncating Eq. \ref{eqn:linearlangevinz} and Eq. \ref{eqn:linearlangevindotz} to order 1.5. And we can also obtain the same scheme using (4.5) in \cite{NumericalSolSDEStrong} with $b^1$ and $b^2$ substituted by $0$ and $\bar b$ respectively. Here we haven't included the second order term on the RHS of (4.5) in \cite{NumericalSolSDEStrong}, whose order is higher than 1.5.

 Order 2 strong Taylor scheme \cite{NumericalSolSDEStrong} (Taylor 2):
\begin{equation}\label{LinearTaylor2}
\begin{aligned}
z_{k+1}&=z_k+\dot z_k\Delta \tau+\bar b\Delta Z_k+\frac12\bar a(z_k,\dot z_k)\Delta\tau^2,\\
\dot z_{k+1}&=\dot z_k+\bar a(z_k,\dot z_k)\Delta\tau+\bar b\Delta W_k-\beta\eta\bar b\Delta Z_k+\frac12[-4\pi^2(1+\eta)\dot z_k-\beta\eta\bar a(z_k,\dot z_k)]\Delta\tau^2,
\end{aligned}
\end{equation}
where $\Delta W_k=w_{k,1}\sqrt{\Delta\tau},\ \Delta Z_k=\frac12\Delta\tau^{\frac32}(w_{k,1}+\frac1{\sqrt3}w_{k,2})$ with $w_{k,i},i=1,2$ independently normally distributed with zero-mean and one-variance: $w_{k,i}\sim N(0,1),\ i=1,2$. The scheme is derived from truncating Eq. \ref{eqn:linearlangevinz} and Eq. \ref{eqn:linearlangevindotz} to order 2. 
%And we can also obtain the same scheme using (5.2) in \cite{NumericalSolSDEStrong}, although it is of the Stratonovich form. 
%Because $\bar a(z(\tau),\dot z(\tau))=-\beta\eta\dot z-4\pi^2(1+\eta)z$ and $\bar b$ is constant, 
%As $\bar b$ is constant, the $\underline{a}^2$ in (5.2) of \cite{NumericalSolSDEStrong} can be substituted by $\underline{\bar a}(z_k,\dot z_k)=\bar a(z_k,\dot z_k)-\frac12L^1\bar b=\bar a(z_k,\dot z_k)$. By the way, as $\bar b$ is constant, i.e. the noise here is additive, the Ito and Stratonovich multiple stochastic integrals are the same, so we can treat the Stratonovich mutiple integrals in (5.2) of \cite{NumericalSolSDEStrong} as Ito multiple integrals, although it's unnecessary, in that all the higher order multiple integrals diminish because of the simple form of Eq. \ref{NondimLinearLangevinEq}.

The simulation results are given in Figure \ref{EvalSRK4}. We can see that the order 1 and 1.5 strong Taylor schemes are not precise when the stepsize is large. The computation even diverges at $\delta=0.01$. The Taylor 1.5 scheme (Eq. \ref{LinearTaylor1p5}) appends the 1.5 order term containing $\Delta Z_k$ to the Taylor 1 scheme (Eq. \ref{LinearTaylor1}), which improves little and the two schemes performs almost equally.

In constast, appended by the second order term, the Taylor 2 scheme improves a lot and performs equally to the SRK4 method. It's appeal to use Taylor 2 scheme instead of SRK4, because the Taylor 2 scheme needs less random numbers generated and less function evaluations, so it runs fast. However, our nonlinear Langevin Eq. \ref{StraSDE} is not so trival as the linear one.

\begin{figure}[H]
 \centering
 \includegraphics[width=0.5\textwidth]{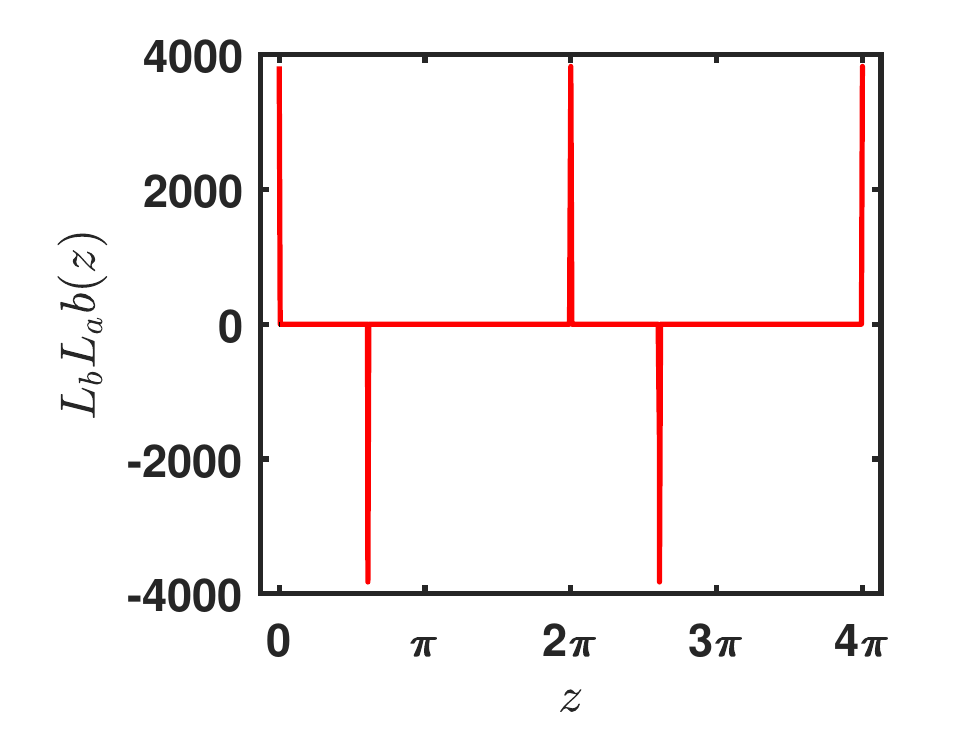}
 \caption{The function $L_bL_ab(z)$ is nearly a train of periodic pulses.}
\label{figLb_La_b_func_eval}
 \end{figure}

The order 2 strong Taylor scheme for nonlinear Langevin Eq. \ref{StraSDE} is

\begin{equation}\label{nonlinearTaylor2}
\begin{aligned}
z_{k+1}&=z_k+b(z_k)\Delta Z_k+\dot z_k\Delta \tau+\frac12 a(\tau_k,z_k,\dot z_k)\Delta\tau^2,\\
\dot z_{k+1}&=\dot z_k+a(\tau_k,z_k,\dot z_k)\Delta\tau+b(z_k)\Delta W_k\\
&\qquad+L_ba(\tau_k,z_k,\dot z_k)\Delta Z_k+L_ab(z_k)(\Delta\tau\Delta W_k-\Delta Z_k)+\frac12L_a a(\tau_k,z_k,\dot z_k)\Delta\tau^2+L_bL_ab(z_k)I_{(1,0,1)},
\end{aligned}
\end{equation}
%where 
%
%\begin{equation}\notag
%\begin{aligned}
%L_ba(\tau,z(\tau),\dot z(\tau))&=b(z(\tau))\frac{\partial}{\partial\dot z}a(\tau,z(\tau),\dot z(\tau))=b(z(\tau))(-\beta\eta),\\
%L_ab(z(\tau))&=(\frac{\partial}{\partial t}+\dot z\frac\partial{\partial z}+a(\tau,z(\tau),\dot z(\tau))\frac{\partial}{\partial\dot z})b(z(\tau))\\
%&=\dot z\pi^2\eta\sqrt{\frac\beta{\pi^2}}(\Theta_h-\Theta_c)\frac1{\sqrt{\Theta(x_1)}}\frac1{\cosh^2(\frac1\alpha[\sin(z+\arctan\sqrt{\frac{\eta-1}{\eta+1}})-\sqrt{\frac{\eta-1}{2\eta}}]))}\\
%&\qquad\qquad\qquad\qquad\qquad\qquad\times\frac1\alpha\cos(z+\arctan\sqrt{\frac{\eta-1}{\eta+1}}),\\
%L_{a}a\left( \tau,z\left( \tau \right),\dot z\left( \tau \right) \right)&=(\frac{\partial}{\partial t}+\dot z\frac\partial{\partial z}+a(\tau,z(\tau),\dot z(\tau))\frac{\partial}{\partial\dot z})a\left( \tau,z\left( \tau \right),\dot z\left( \tau \right) \right)\\
%&=4\pi^2\tilde v+\dot z(-4\pi^2-4\pi^2\eta\cos z)+a(\tau,z,\dot z)(-\beta\eta),
%\end{aligned}
%\end{equation}
%
%
%
%\begin{equation}
%\begin{aligned}
%L_bL_ab(z(\tau))&=b(z(\tau))\frac{\partial}{\partial\dot z}L_ab(z(\tau))\\
%&=b(z(\tau))\pi^2\eta\sqrt{\frac\beta{\pi^2}}(\Theta_h-\Theta_c)\frac1{\sqrt{\Theta(x_1)}}\frac1{\cosh^2(\frac1\alpha[\sin(z+\arctan\sqrt{\frac{\eta-1}{\eta+1}})-\sqrt{\frac{\eta-1}{2\eta}}]))}\\
%&\qquad\qquad\qquad\qquad\qquad\qquad\times\frac1\alpha\cos(z+\arctan\sqrt{\frac{\eta-1}{\eta+1}}),
%\end{aligned}
%\end{equation}
%
%and
%\begin{equation}
%I_{(1,0,1)}=\int_{\tau_0}^\tau\int_{\tau_0}^s\int_{\tau_0}^u\mathrm d\mathcal W_v\mathrm d u\mathrm d\mathcal W_s. 
%\end{equation}
which is derived from Eq. \ref{eqn:STEDz} and Eq. \ref{eqn:STEDdotz}.
%This scheme is derived from (5.2) in \cite{NumericalSolSDEStrong}. Here for our special form of Eq.~(\ref{NondimLinearLangevinEq}), the Strotonovich multiple integral is equivalent to the Ito one, so we can substitute the $J_{(1,0,1)}$ there with our $I_{(1,0,1)}$. 
We can see that for the discretization of the nonlinear Langevin equation to the second order, we need to evaluate more functions. What's more, the multiple stochastic integral $I_{(1,0,1)}$ in the last term needs to be simulate, which is complicated. To estimate this term we plot the function $L_bL_ab(z)$ in Figure \ref{figLb_La_b_func_eval}. We can see that it's almost zero everywhere except at the boundaries of the hot and the cold heat baths. In view of this we as a compromise neglect the term $L_bL_ab(z_k)I_{(1,0,1)}$, which is of second order. We can regard it as an earlier trancation of the stochastic Taylor expansion to less than 2nd order. And we will call this scheme Taylor 2-.
 
We have seen that the appending the terms of order 1.5, which is stochastic, to the Taylor 1 scheme doesn't improve the performance at least in the sense of mean square values. Therefore we can presume that neglecting the stochastic term 
$L_bL_ab(z_k)I_{(1,0,1)}$ won't affect the performance of the Taylor 2 scheme much either. If the temperature of the heat bath is constant, i.e. $b(z_k)=\text{constant}$, this term will vanish, and the Taylor 2- method can achieve higher precision than 2nd order. Moreover, by choosing small enough time step size, the Taylor 2- scheme can achieve the accuracy we need.

\begin{figure}[H]
\centering
\includegraphics[width=0.49\textwidth]{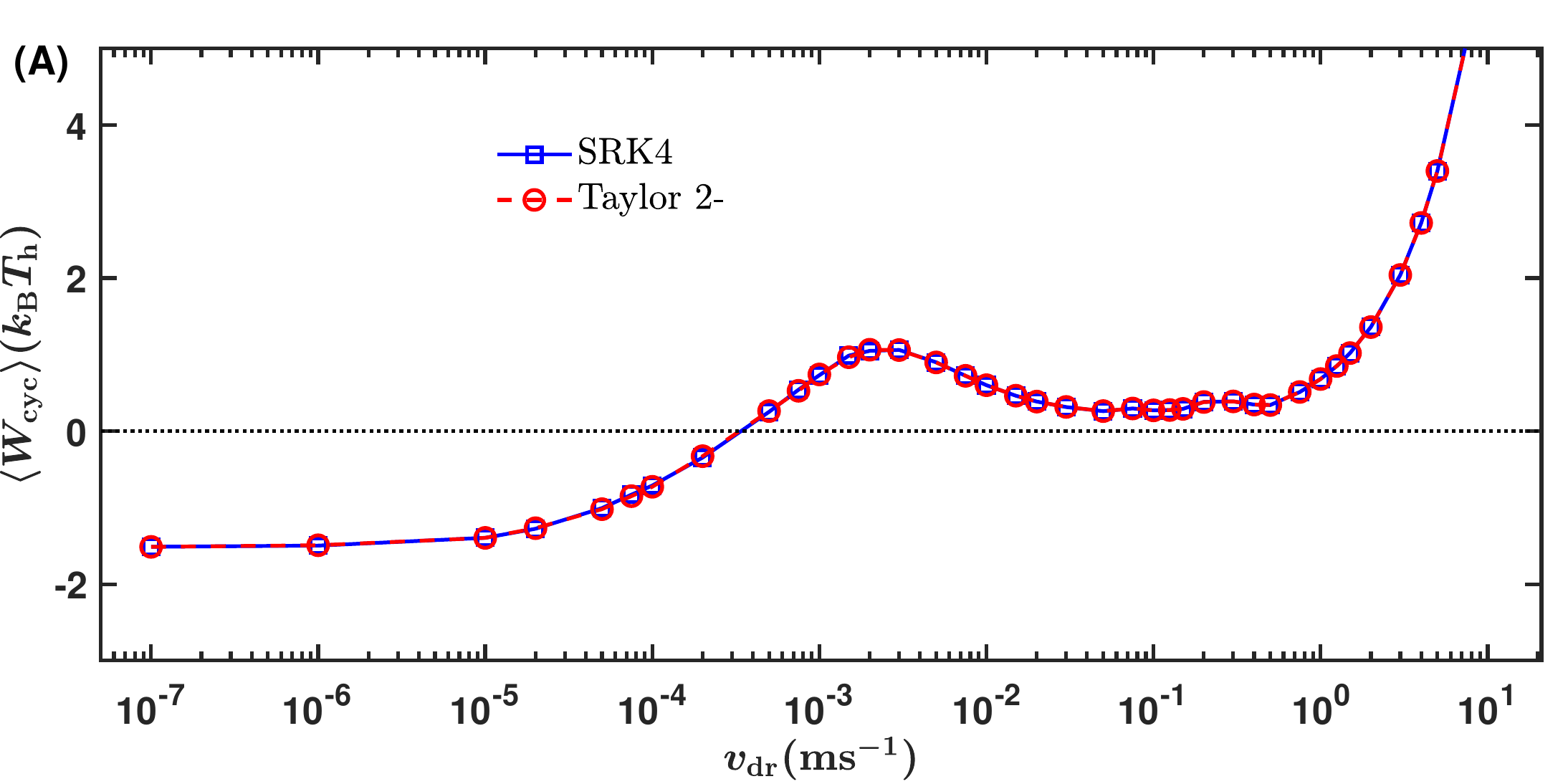}
\includegraphics[width=0.49\textwidth]{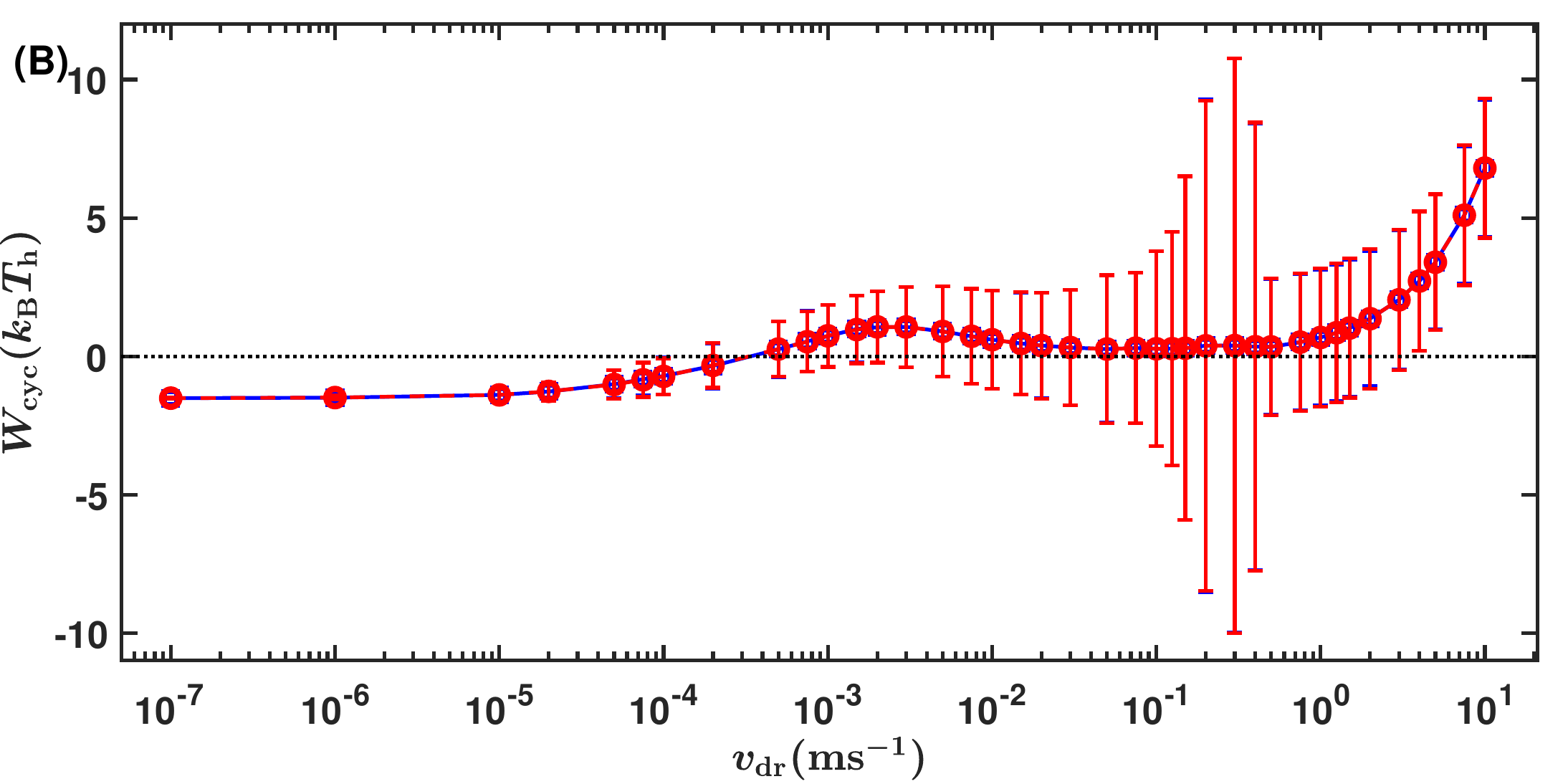}
\caption{The cycle work $W_{\rm cyc}$ with respect to the driving velocity $v_{\rm dr}$ simulated by Taylor 2- and SRK4. The left plot gives the mean values and the right one is added by the standard deviations to the left one. Both the mean values and the standard deviations simulated by the two methods coincide. The time stepsize is calculated by Eq. \ref{TimeStepSize}. The parameters are $\eta=3.0,\mu=4\times10^4\rm s^{-1},{\it\Theta}_h=0.4,{\it\Theta}_c=0.04$, and other parameters are given in Table \ref{parameters}.}
\label{figTaylor2_SRK4}
\end{figure}

In Figure \ref{figTaylor2_SRK4}, we compare the Taylor 2- scheme with SRK4 in calculating the cycle work $W_{\rm cyc}$. We can see that both the mean value and the standard deviation values are nearly the same using the two methods at each driving velocity $v_{\rm dr}$. Therefore we can conclude that the SRK4 and the Taylor 2- method are both appropriate for the simulation of our nonlinear Langevin equation, although the SRK4 scheme is derived from the linear stochastic differential equation \cite{KasdinSRK4}. And we can also conclude that the time stepsize Eq. \ref{TimeStepSize} we have chosen is also appropriate. 
 
 The Taylor 2- scheme runs a little faster than the SRK4, and is more comprehensible. In spite of this, we still choose the SRK4 method in all our simulations of the nonlinear Langevin equation in the main text. It is easy to implement and there is ready-made codes \cite{DongAnalytical}. 

Another advantage of the SRK4 is that it does not need the evaluations of the partial derivatives as in the case of Taylor 2-. At our first simulations of the PTSHE, we treat the temperature field with the discontinuous function Eq. \ref{DiscontinuousTemp}, which is not differentiable at the boundaries of the hot and the cold heat baths. With the SRK4, we can still process our simulations without dealing with the derivatives, although it's not very rigorous. We have compared the results of the discontinuous temperature with that of the continuous one, and we didn't find any significant differences. So it is possible that the equivalence of the Stratonovich and the Ito form of the Langevin Eq. \ref{StraSDE} is still correct with discontinuous spatio temperature function. Maybe there are relevant theorems which have been proved. We hope we can find or prove this in the future.

\subsubsection*{The stochastic Taylor expansion}
Here we give a brief derivation of the Ito stochastic Taylor expansion \cite{StochasticTaylorExpansion} based on the form
\begin{equation}\label{eqn:StochasticTaylorExpan}
\begin{cases}
\mathrm dz=\dot z(\tau)\mathrm d\tau,\\
\mathrm d\dot z=a(z(\tau),\dot z(\tau),\tau)\mathrm d\tau+b(z(\tau))\mathrm d\mathcal W_\tau,
\end{cases}
\end{equation}
by which both the nonlinear and the linear Langevin Eq. \ref{StraSDE} and Eq. \ref{NondimLinearLangevinEq} can be represented. Integrate Eq. \ref{eqn:StochasticTaylorExpan} from $\tau_0$ to $\tau$ we obtain
\begin{equation}\label{eqn:StochasticTaylorExpanIntegral}
\begin{cases}
\Delta z=\int_{\tau_0}^\tau\dot z(s)\mathrm ds,\\
\Delta\dot z=\int_{\tau_0}^\tau a(z(s),\dot z(s),s)\mathrm ds+\int_{\tau_0}^\tau b(z(s))\mathrm d\mathcal W_s,
\end{cases}
\end{equation}
in which $\Delta z=z(\tau)-z(\tau_0)$. The Ito formula Eq. \ref{ItoFormula} will be used iteratively next, so we write it down here again in the integral form
\begin{equation}\label{ItoformulaIntegral}
\begin{aligned}
f(z(\tau),\dot z(\tau), \tau)&=f(z(\tau_0),\dot z(\tau_0), \tau_0)+\int_{\tau_0}^\tau\left[\frac{\partial f}{\partial z}\mathrm dz(s)+\frac{\partial f}{\partial \dot z}\mathrm d\dot z(s)+\frac{\partial f}{\partial \tau}\mathrm ds\right]+\int_{\tau_0}^\tau\frac12\frac{\partial^2f}{\partial\dot z^2}b(z(s))^2\mathrm ds\\
&=f(z_0,\dot z_0, \tau_0)+\int_{\tau_0}^\tau\left\{\frac{\partial f}{\partial z}\dot z\mathrm ds+\frac{\partial f}{\partial \dot z}\left[a(z,\dot z,s)\mathrm ds+b(z)\mathrm d\mathcal W_s\right]+\frac{\partial f}{\partial \tau}\mathrm ds\right\}+\int_{\tau_0}^\tau\frac12\frac{\partial^2f}{\partial\dot z^2}b(z)^2\mathrm ds\\
&=f(z_0,\dot z_0, \tau_0)+\int_{\tau_0}^\tau\left[\frac{\partial f}{\partial \tau}+\dot z\frac{\partial f}{\partial z}+a(z,\dot z,s)\frac{\partial f}{\partial\dot z}+\frac12\frac{\partial^2f}{\partial\dot z^2}b(z)^2\right]\mathrm ds+\int_{\tau_0}^\tau b(z)\frac{\partial f}{\partial \dot z}\mathrm d\mathcal W_s\\
&=f(z_0,\dot z_0, \tau_0)+\int_{\tau_0}^\tau L_af(z(s),\dot z(s),s)\mathrm ds+\int_{\tau_0}^\tau L_bf(z(s),\dot z(s),s)\mathrm d\mathcal W_s,
\end{aligned}
\end{equation}
in which the second equality results from substituting Eq. \ref{eqn:StochasticTaylorExpan} and we use the operators
\begin{equation}
\begin{aligned}
L_a&=\frac{\partial}{\partial \tau}+\dot z\frac{\partial}{\partial z}+a(z(\tau),\dot z(\tau),\tau)\frac{\partial}{\partial\dot z}+\frac12b(z(\tau))^2\frac{\partial^2}{\partial\dot z^2},\\
L_b&=b(z(\tau))\frac{\partial}{\partial \dot z},
\end{aligned}
\end{equation}
and $z_0=z(\tau_0)$, $\dot z_0=\dot z(\tau_0)$ for compactness.

We first expand $\Delta z$. Apply Ito formula Eq. \ref{ItoformulaIntegral} to $\dot z(s)$ and substitute it into the first equation of Eq. \ref{eqn:StochasticTaylorExpanIntegral} we will obtain
\begin{equation}
\begin{aligned}
\Delta z&=\int_{\tau_0}^\tau\left[\dot z_0+\int_{\tau_0}^s a(z(u),\dot z(u),u)\mathrm du+\int_{\tau_0}^s b(z(u))\mathrm d\mathcal W_u\right]\mathrm ds\\
&=\dot z_0\Delta\tau+\int_{\tau_0}^\tau\int_{\tau_0}^sa(z(u),\dot z(u),u)\mathrm du\mathrm ds+\int_{\tau_0}^\tau\int_{\tau_0}^sb(z(u))\mathrm d\mathcal W_u\mathrm ds,
\end{aligned}
\end{equation}
in which $\Delta\tau=\int_{\tau_0}^\tau\mathrm ds$.
Apply Ito formula again to $a(z(u),\dot z(u),u)$ and $b(z(u))$ and substitute them to this equation, we will obtain
\begin{equation}\label{eq:Deltazbeforelast}
\begin{aligned}
\Delta z&=\dot z_0\Delta\tau+\int_{\tau_0}^\tau\int_{\tau_0}^s\left[a(z_0,\dot z_0,\tau_0)+\int_{\tau_0}^uL_aa(z(v),\dot z(v),v)\mathrm dv+\int_{\tau_0}^uL_bz(z(v),\dot z(v),v)\mathrm d\mathcal W_v\right]\mathrm du\mathrm ds\\
&\qquad\quad\ \ +\int_{\tau_0}^\tau\int_{\tau_0}^s\left[b(z_0)+\int_{\tau_0}^uL_ab(z(v))\mathrm dv+\int_{\tau_0}^uL_bb(z(v))\mathrm d\mathcal W_v\right]\mathrm d\mathcal W_u\mathrm ds\\
&=\dot z_0\Delta\tau+a(z_0,\dot z_0,\tau_0)\frac12\Delta\tau^2+\int_{\tau_0}^\tau\int_{\tau_0}^s\int_{\tau_0}^uL_aa(z(v),\dot z(v),v)\mathrm dv\mathrm du\mathrm ds+\int_{\tau_0}^\tau\int_{\tau_0}^s\int_{\tau_0}^uL_bz(z(v),\dot z(v),v)\mathrm d\mathcal W_v\mathrm du\mathrm ds\\
&\qquad\quad\ \ +b(z_0)I_{(1,0)}+\int_{\tau_0}^\tau\int_{\tau_0}^s\int_{\tau_0}^uL_ab(z(v))\mathrm dv\mathrm d\mathcal W_u\mathrm ds+\int_{\tau_0}^\tau\int_{\tau_0}^s\int_{\tau_0}^uL_bb(z(v))\mathrm d\mathcal W_v\mathrm d\mathcal W_u\mathrm ds\\
&=\dot z_0\Delta\tau+a(z_0,\dot z_0,\tau_0)\frac12\Delta\tau^2+b(z_0)I_{(1,0)}+\int_{\tau_0}^\tau\int_{\tau_0}^s\int_{\tau_0}^uL_bb(z(v))\mathrm d\mathcal W_v\mathrm d\mathcal W_u\mathrm ds+o(\Delta\tau^2),
\end{aligned}
\end{equation}
in which $I_{(1,0)}=\int_{\tau_0}^\tau\int_{\tau_0}^s\mathrm d\mathcal W_u\mathrm ds$, which is of order 1.5 because $\mathrm d\mathcal W_u$ is of order 0.5. In this equation, all the multiple stochastic integrals of order higher than 2 are absorbed in $o(\Delta\tau^2)$, except the remaining one $\int_{\tau_0}^\tau\int_{\tau_0}^s\int_{\tau_0}^uL_bb(z(v))\mathrm d\mathcal W_v\mathrm d\mathcal W_u\mathrm ds$, which is order 2. 

Apply the Ito formula to $L_bb(z(v))$, substitute it into Eq. \ref{eq:Deltazbeforelast} and absorb the stochastic multiple integrals of order higher than 2 into $o(\Delta\tau^2)$, we will obtain
\begin{equation}\label{eqn:STEDzGeF}
\Delta z=\dot z_0\Delta\tau+b(z_0)I_{(1,0)}+a(z_0,\dot z_0,\tau_0)\frac12\Delta\tau^2+L_bb(z_0)I_{(1,1,0)}+o(\Delta\tau^2),
\end{equation}
where $I_{(1,1,0)}=\int_{\tau_0}^\tau\int_{\tau_0}^s\int_{\tau_0}^u\mathrm d\mathcal W_v\mathrm d\mathcal W_u\mathrm ds$ is of order 2.

Then we expand $\Delta\dot z$. Apply Ito formula to $a(z(s),\dot z(s),s)$ and $b(z(s))$, and substitute them into the second equation of Eq. \ref{eqn:StochasticTaylorExpanIntegral}, we will obtain
\begin{equation}
\begin{aligned}
\Delta\dot z&=\int_{\tau_0}^\tau\left[a(z_0,\dot z_0,\tau_0)+\int_{\tau_0}^sL_aa(z(u),\dot z(u),u)\mathrm du+\int_{\tau_0}^sL_ba(z(u),\dot z(u),u)\mathrm d\mathcal W_u\right]\mathrm ds\\
&\qquad+\int_{\tau_0}^\tau\left[b(z_0)+\int_{\tau_0}^sL_ab(z(u))\mathrm du+\int_{\tau_0}^sL_bb(z(u))\mathrm d\mathcal W_u\right]\mathrm d\mathcal W_s\\
&=a(z_0,\dot z_0,\tau_0)\Delta\tau+\int_{\tau_0}^\tau\int_{\tau_0}^sL_aa(z(u),\dot z(u),u)\mathrm du\mathrm ds+\int_{\tau_0}^\tau\int_{\tau_0}^sL_ba(z(u),\dot z(u),u)\mathrm d\mathcal W_u\mathrm ds\\
&\qquad+b(z_0)\Delta\mathcal W_\tau+\int_{\tau_0}^\tau\int_{\tau_0}^sL_ab(z(u))\mathrm du\mathrm d\mathcal W_s+\int_{\tau_0}^\tau\int_{\tau_0}^sL_bb(z(u))\mathrm d\mathcal W_u\mathrm d\mathcal W_s,
\end{aligned}
\end{equation}
where $\Delta\mathcal W_\tau=\int_{\tau_0}^\tau\mathrm d\mathcal W_s$.% and the stochastic mutiple integral of order higher than 2 is absorbed into $o(\Delta\tau^2)$.

Apply the Ito formula to $L_aa(z(u),\dot z(u),u)$, $L_ba(z(u),\dot z(u),u)$, $L_ab(z(u))$ and $L_bb(z(u))$ again and substitute them into the above equation, we will obtain
\begin{equation}
\begin{aligned}
\Delta\dot z&=a(z_0,\dot z_0,\tau_0)\Delta\tau+\int_{\tau_0}^\tau\int_{\tau_0}^s\left[L_aa(z_0,\dot z_0,\tau_0)+\int_{\tau_0}^uL_aL_aa(z(v),\dot z(v),v)\mathrm dv+\int_{\tau_0}^uL_bL_aa(z(v),\dot z(v),v)\mathrm d\mathcal W_v\right]\mathrm du\mathrm ds\\
&\qquad+\int_{\tau_0}^\tau\int_{\tau_0}^s\left[L_ba(z_0,\dot z_0,\tau_0)+\int_{\tau_0}^uL_aL_ba(z(v),\dot z(v),v)\mathrm dv+\int_{\tau_0}^uL_bL_ba(z(v),\dot z(v),v)\mathrm d\mathcal W_v\right]\mathrm d\mathcal W_u\mathrm ds\\
&\qquad+b(z_0)\Delta\mathcal W_\tau+\int_{\tau_0}^\tau\int_{\tau_0}^s\left[L_ab(z_0)+\int_{\tau_0}^uL_aL_ab(z(v))\mathrm dv+\int_{\tau_0}^uL_bL_ab(z(v))\mathrm d\mathcal W_v\right]\mathrm du\mathrm d\mathcal W_s\\
&\qquad+\int_{\tau_0}^\tau\int_{\tau_0}^s\left[L_bb(z_0)+\int_{\tau_0}^uL_aL_bb(z(v))\mathrm dv+\int_{\tau_0}^uL_bL_bb(z(v))\mathrm d\mathcal W_v\right]\mathrm d\mathcal W_u\mathrm d\mathcal W_s\\
&=a(z_0,\dot z_0,\tau_0)\Delta\tau+L_aa(z_0,\dot z_0,\tau_0)\frac12\Delta\tau^2+L_ba(z_0,\dot z_0,\tau_0)I_{(1,0)}+\int_{\tau_0}^\tau\int_{\tau_0}^s\int_{\tau_0}^uL_bL_ba(z(v),\dot z(v),v)\mathrm d\mathcal W_v\mathrm d\mathcal W_u\mathrm ds\\
&\qquad+b(z_0)\Delta\mathcal W_\tau+L_ab(z_0)I_{(0,1)}+\int_{\tau_0}^\tau\int_{\tau_0}^s\int_{\tau_0}^uL_bL_ab(z(v))\mathrm d\mathcal W_v\mathrm du\mathrm d\mathcal W_s\\
&\qquad+L_bb(z_0)I_{(1,1)}+\int_{\tau_0}^\tau\int_{\tau_0}^s\int_{\tau_0}^uL_aL_bb(z(v))\mathrm dv\mathrm d\mathcal W_u\mathrm d\mathcal W_s+\int_{\tau_0}^\tau\int_{\tau_0}^s\int_{\tau_0}^uL_bL_bb(z(v))\mathrm d\mathcal W_v\mathrm d\mathcal W_u\mathrm d\mathcal W_s+o(\Delta\tau^2),
\end{aligned}
\end{equation}
where $I_{(0,1)}=\int_{\tau_0}^\tau\int_{\tau_0}^s\mathrm du\mathrm d\mathcal W_s$ and $I_{(1,1)}=\int_{\tau_0}^\tau\int_{\tau_0}^s\mathrm d\mathcal W_u\mathrm d\mathcal W_s$ and the stochastic mutiple integrals of order higher than 2 are absorbed into $o(\Delta\tau^2)$.

Apply the Ito formula iteratively, absorb the stochastic mutiple integrals of order higher than 2, we can obtain the stochastic Taylor expansion of $\Delta\dot z$ up to the second order
\begin{equation}\label{eqn:STEDdotzGeF}
\begin{aligned}
\Delta\dot z&=a(z_0,\dot z_0,\tau_0)\Delta\tau+L_aa(z_0,\dot z_0,\tau_0)\frac12\Delta\tau^2+L_ba(z_0,\dot z_0,\tau_0)I_{(1,0)}+L_bL_ba(z_0,\dot z_0,\tau_0)I_{(1,1,0)}\\
&\qquad\ \ +b(z_0)\Delta\mathcal W_\tau+L_ab(z_0)I_{(0,1)}+L_bL_ab(z_0)I_{(1,0,1)}\\
&\qquad\qquad\qquad\quad\ +L_bb(z_0)I_{(1,1)}+L_aL_bb(z_0)I_{(0,1,1)}+L_bL_bb(z(v))I_{(1,1,1)}+L_bL_bL_bb(z_0)I_{(1,1,1,1)}+o(\Delta\tau^2)\\
&=b(z_0)\Delta\mathcal W_\tau\\
&\quad+a(z_0,\dot z_0,\tau_0)\Delta\tau+L_bb(z_0)I_{(1,1)}\\
&\quad+L_ba(z_0,\dot z_0,\tau_0)I_{(1,0)}+L_ab(z_0)I_{(0,1)}+L_bL_bb(z_0)I_{(1,1,1)}\\
&\quad+L_aa(z_0,\dot z_0,\tau_0)\frac12\Delta\tau^2+L_bL_ba(z_0,\dot z_0,\tau_0)I_{(1,1,0)}+L_bL_ab(z_0)I_{(1,0,1)}+L_aL_bb(z_0)I_{(0,1,1)}\\
&\quad+L_bL_bL_bb(z_0)I_{(1,1,1,1)}+o(\Delta\tau^2),
\end{aligned}
\end{equation}
in which $I_{(1,1,0)}=\int_{\tau_0}^\tau\int_{\tau_0}^s\int_{\tau_0}^u\mathrm d\mathcal W_v\mathrm d\mathcal W_u\mathrm ds$, $I_{(1,0,1)}=\int_{\tau_0}^\tau\int_{\tau_0}^s\int_{\tau_0}^u\mathrm d\mathcal W_v\mathrm du\mathrm d\mathcal W_s$, $I_{(0,1,1)}=\int_{\tau_0}^\tau\int_{\tau_0}^s\int_{\tau_0}^u\mathrm dv\mathrm d\mathcal W_u\mathrm d\mathcal W_s$, $I_{(1,1,1)}=\int_{\tau_0}^\tau\int_{\tau_0}^s\int_{\tau_0}^u\mathrm d\mathcal W_v\mathrm d\mathcal W_u\mathrm d\mathcal W_s$ and $I_{(1,1,1,1)}=\int_{\tau_0}^\tau\int_{\tau_0}^s\int_{\tau_0}^u\int_{\tau_0}^v\mathrm d\mathcal W_p\mathrm d\mathcal W_v\mathrm d\mathcal W_u\mathrm d\mathcal W_s$.

The stochastic Taylor expansions of $\Delta z$ (Eq. \ref{eqn:STEDzGeF}) and $\Delta\dot z$ (Eq. \ref{eqn:STEDdotzGeF}) are a little complicated in the general form. However, we can simplify them utilizing the relatively simple form of our linear and nonlinear Langevin equations.

In the linear case, $\bar a(z(\tau),\dot z(\tau))=-\beta\eta\dot z-4\pi^2(1+\eta)z$ and $\bar b$ is constant, then
\begin{equation}\label{eqn:linearlangevinz}
\Delta z=\dot z_0\Delta\tau+\bar bI_{(1,0)}+\frac12\bar a(z_0,\dot z_0)\Delta\tau^2+o(\Delta\tau^2)
\end{equation} 
and
\begin{equation}\label{eqn:linearlangevindotz}
\begin{aligned}
\Delta\dot z&=\bar b\Delta\mathcal W_\tau+\bar a(z_0,\dot z_0)\Delta\tau-\beta\eta\bar bI_{(1,0)}+\frac12\left\{[-4\pi^2(1+\eta)]\dot z_0+(-\beta\eta)\bar a(z_0,\dot z_0)\right\}\Delta\tau^2+o(\Delta\tau^2).
\end{aligned}
\end{equation}

In the nonlinear case, $a(z(\tau),\dot z(\tau),\tau)=-\beta\eta\dot z-4\pi^2(z-\tilde{v}\tau)-4\pi^2\eta\sin z$ and $b(z(\tau))=4\pi^2\eta\sqrt{\frac{\beta}{\pi^2}{\it\Theta}(z(\tau))}$, which is only dependent on $z(\tau)$, then
\begin{equation}\label{eqn:STEDz}
\Delta z=\dot z_0\Delta\tau+b(z_0)I_{(1,0)}+\frac12a(z_0,\dot z_0,\tau_0)\Delta\tau^2+o(\Delta\tau^2),
\end{equation}
and
\begin{equation}\label{eqn:STEDdotz}
\begin{aligned}
\Delta\dot z&=b(z_0)\Delta\mathcal W_\tau+a(z_0,\dot z_0,\tau_0)\Delta\tau+L_ba(z_0,\dot z_0,\tau_0)I_{(1,0)}+L_ab(z_0)I_{(0,1)}+\frac12L_aa(z_0,\dot z_0,\tau_0)\Delta\tau^2+L_bL_ab(z_0)I_{(1,0,1)}+o(\Delta\tau^2),
\end{aligned}
\end{equation}
in which 
\begin{equation}\notag
\begin{aligned}
L_ba(z(\tau),\dot z(\tau),\tau)&=b(z(\tau))\frac{\partial}{\partial\dot z}a(z(\tau),\dot z(\tau),\tau)=b(z(\tau))(-\beta\eta),\\
L_ab(z(\tau))&=\left[\frac{\partial}{\partial\tau}+\dot z(\tau)\frac\partial{\partial z}+a(z(\tau),\dot z(\tau),\tau)\frac{\partial}{\partial\dot z}+\frac12b(z(\tau))^2\frac{\partial^2}{\partial\dot z^2}\right]b(z(\tau))\\
&=\dot z(\tau)\pi^2\eta\sqrt{\frac\beta{\pi^2}}({\it\Theta}_{\rm h}-{\it\Theta}_{\rm c})\frac1{\sqrt{{\it\Theta}(z(\tau))}}\frac1{\cosh^2(\frac1\alpha[\sin(z(\tau)+\arctan\sqrt{\frac{\eta-1}{\eta+1}})-\sqrt{\frac{\eta-1}{2\eta}}])}\\
&\qquad\qquad\qquad\qquad\qquad\qquad\times\frac1\alpha\cos(z(\tau)+\arctan\sqrt{\frac{\eta-1}{\eta+1}}),\\
L_{a}a\left(z\left( \tau \right),\dot z\left( \tau \right),  \tau\right)&=\left[\frac{\partial}{\partial\tau}+\dot z(\tau)\frac\partial{\partial z}+a(z(\tau),\dot z(\tau),\tau)\frac{\partial}{\partial\dot z}+\frac12b(z(\tau))^2\frac{\partial^2}{\partial\dot z^2}\right]a\left( z\left( \tau \right),\dot z\left( \tau \right),\tau \right)\\
&=4\pi^2\tilde v+\dot z(\tau)(-4\pi^2-4\pi^2\eta\cos z)+a(z(\tau),\dot z(\tau),\tau)(-\beta\eta),\\
%\end{aligned}
%\end{equation}
%\begin{equation}
%\begin{aligned}
L_bL_ab(z(\tau))&=b(z(\tau))\frac{\partial}{\partial\dot z}L_ab(z(\tau))\\
&=b(z(\tau))\pi^2\eta\sqrt{\frac\beta{\pi^2}}({\it\Theta}_{\rm h}-{\it\Theta}_{\rm c})\frac1{\sqrt{\Theta(z(\tau))}}\frac1{\cosh^2(\frac1\alpha[\sin(z(\tau)+\arctan\sqrt{\frac{\eta-1}{\eta+1}})-\sqrt{\frac{\eta-1}{2\eta}}])}\\
&\qquad\qquad\qquad\qquad\qquad\qquad\times\frac1\alpha\cos(z(\tau)+\arctan\sqrt{\frac{\eta-1}{\eta+1}}).
\end{aligned}
\end{equation}
We now calculate the mean values and variances of $I_{(1,0)}$ and $I_{(0,1)}$, both of which are normal.
\begin{equation}\label{I01mean}
\langle I_{(0,1)}\rangle=\langle \int_{\tau_0}^\tau\int_{\tau_0}^s\mathrm du\mathrm d\mathcal W_s\rangle=\int_{\tau_0}^\tau(s-\tau_0)\langle\mathrm d\mathcal W_s\rangle=0.
\end{equation}
%\begin{equation}
%\langle I_{(1,0)}\rangle=\langle \int_{\tau_0}^\tau\int_{\tau_0}^s\mathrm d\mathcal W_u\mathrm ds\rangle=\int_{\tau_0}^\tau\int_{\tau_0}^s\langle \mathrm d\mathcal W_u\rangle\mathrm ds=0.
%\end{equation}
\begin{equation}\label{I01var}
\begin{aligned}
\langle I_{(0,1)}^2\rangle&=\langle\int_{\tau_0}^\tau\int_{\tau_0}^s\mathrm du\mathrm d\mathcal W_s\int_{\tau_0}^{\tau}\int_{\tau_0}^{s'}\mathrm du'\mathrm d\mathcal W_{s'}\rangle=\langle\int_{\tau_0}^\tau(s-\tau_0)\mathrm d\mathcal W_s\int_{\tau_0}^{\tau}(s'-\tau_0)\mathrm d\mathcal W_{s'}\rangle\\
&=\langle\int_{\tau_0}^\tau\int_{\tau_0}^{\tau}(s-\tau_0)(s'-\tau_0)\mathrm d\mathcal W_s\mathrm d\mathcal W_{s'}\rangle=\int_{\tau_0}^\tau\int_{\tau_0}^{\tau}(s-\tau_0)(s'-\tau_0)\langle\mathrm d\mathcal W_s\mathrm d\mathcal W_{s'}\rangle\\
&=\int_{\tau_0}^\tau\int_{\tau_0}^{\tau}(s-\tau_0)(s'-\tau_0)\delta(s-s')\mathrm ds\mathrm ds'\\
&=\int_{\tau_0}^\tau(s-\tau_0)^2\mathrm ds=\frac13\Delta\tau^3,
\end{aligned}
\end{equation}
where we have used the independence of the increments at different instants of the Gausian process, i.e. $\langle\mathrm d\mathcal W_s\mathrm d\mathcal W_{s'}\rangle=\delta(s-s')\mathrm ds\mathrm ds'$. 
%$\langle\mathrm d\mathcal W_s\mathrm d\mathcal W_{s'}\rangle=\langle\mathrm d\mathcal W_s\rangle\langle\mathrm d\mathcal W_{s'}\rangle$, and $\langle \mathrm d\mathcal W_{s'}\rangle=\langle \mathrm d\mathcal W_s\rangle=0$.

Because $I_{(0)}=\int_{\tau_0}^\tau\mathrm ds$, $I_{(1)}=\int_{\tau_0}^\tau\mathrm d\mathcal W_{s}=\Delta\mathcal W_\tau$, we can constitute a stochastic differential equation system
\begin{equation}
\begin{cases}
\mathrm dI_{(0)}=\mathrm d\tau,\\
\mathrm dI_{(1)}=\mathrm d\mathcal W_\tau,
\end{cases}
\end{equation}
which is of the same form as Eq. \ref{eqn:StochasticTaylorExpan}. Utilizing the Ito formula Eq. \ref{ItoformulaIntegral} leads to
\begin{equation}
\begin{aligned}
I_{(0)}I_{(1)}&=\int_{\tau_0}^\tau\left[\frac{\partial I_{(0)}I_{(1)}}{\partial I_{(0)}}\mathrm dI_{(0)}+\frac{\partial I_{(0)}I_{(1)}}{\partial I_{(1)}}\mathrm dI_{(1)}+\frac{\partial I_{(0)}I_{(1)}}{\partial\tau}\mathrm ds\right]+\int_{\tau_0}^\tau \frac12\frac{\partial^2I_{(0)}I_{(1)}}{\partial I_{(1)}^2}\times1^2\mathrm ds\\
&=\int_{\tau_0}^\tau\left[\frac{\partial I_{(0)}I_{(1)}}{\partial I_{(0)}}\mathrm ds+\frac{\partial I_{(0)}I_{(1)}}{\partial I_{(1)}}\mathrm d\mathcal W_s+\frac{\partial I_{(0)}I_{(1)}}{\partial\tau}\mathrm ds\right]+\int_{\tau_0}^\tau \frac12\frac{\partial^2I_{(0)}I_{(1)}}{\partial I_{(1)}^2}\times1^2\mathrm ds\\
&=\int_{\tau_0}^\tau I_{(1),s}\mathrm ds+\int_{\tau_0}^\tau I_{(0),s}\mathrm d\mathcal W_s=\int_{\tau_0}^\tau \int_{\tau_0}^s\mathrm d\mathcal W_u\mathrm ds+\int_{\tau_0}^\tau \int_{\tau_0}^s\mathrm du\mathrm d\mathcal W_s\\
&=I_{(1,0)}+I_{(0,1)}.
\end{aligned}
\end{equation}
Then $I_{(1,0)}=I_{(1)}I_{(0)}-I_{(0,1)}$.
%Because $I_{(0,1)}=\int_{\tau_0}^\tau\int_{\tau_0}^s\mathrm du\mathrm d\mathcal W_s=\int_{\tau_0}^\tau I_{(0),s}\mathrm d\mathcal W_s$, we can constitute another stochastic differential equation
%\begin{equation}
%\begin{cases}
%\mathrm dI_{(0,1)}=I_{(0)}\mathrm d\mathcal W_\tau,\\
%\mathrm dI_{(1)}=\mathrm d\mathcal W_\tau,
%\end{cases}
%\end{equation}
%and utilizing the Ito formula Eq. \ref{ItoformulaIntegral} leads to
%\begin{equation}
%\begin{aligned}
%I_{(1)}I_{(0,1)}&=\int_{\tau_0}^\tau\left[\frac{\partial I_{(1)}I_{(0,1)}}{\partial I_{(0,1)}}\mathrm dI_{(0,1)}+\frac{\partial I_{(1)}I_{(0,1)}}{\partial I_{(1)}}\mathrm dI_{(1)}+\frac{\partial I_{(1)}I_{(0,1)}}{\partial\tau}\mathrm ds\right]+\int_{\tau_0}^\tau\frac12\frac{\partial^2I_{(1)}I_{(0,1)}}{\partial I_{(1)}^2}\times 1^2\mathrm ds\\
%&=\int_{\tau_0}^\tau\left[\frac{\partial I_{(1)}I_{(0,1)}}{\partial I_{(0,1)}}I_{(0)}\mathrm ds+\frac{\partial I_{(1)}I_{(0,1)}}{\partial I_{(1)}}\mathrm d\mathcal W_s+\frac{\partial I_{(1)}I_{(0,1)}}{\partial\tau}\mathrm ds\right]+\int_{\tau_0}^\tau\frac12\frac{\partial^2I_{(1)}I_{(0,1)}}{\partial I_{(1)}^2}\times 1^2\mathrm ds\\
%&=\int_{\tau_0}^\tau I_{(1),s}I_{(0),s}\mathrm ds+\int_{\tau_0}^\tau I_{(0,1)}\mathrm d\mathcal W_s\\
%&=\int_{\tau_0}^\tau \left[I_{(1,0),s}+I_{(0,1),s}\right]\mathrm ds+\int_{\tau_0}^\tau I_{(0,1)}\mathrm d\mathcal W_s.
%\end{aligned}
%\end{equation}
Therefore we can treat $I_{(1)}$ and $I_{(0,1)}$ as two normal random variables: $I_{(1)}=\Delta\mathcal W_\tau\sim N(0,\Delta\tau)$ and $I_{(0,1)}\sim N(0,\frac13\Delta\tau^3)$, and then $I_{(1,0)}$ can be obtained. The covariance of $I_{(1)}$ and $I_{(0,1)}$ is
\begin{equation}\label{I1I01covar}
\langle I_{(1)}I_{(0,1)}\rangle=\langle\int_{\tau_0}^\tau\mathrm d\mathcal W_s\int_{\tau_0}^\tau\int_{\tau_0}^{s'}\mathrm du\mathrm d\mathcal W_{s'}\rangle=\int_{\tau_0}^\tau\int_{\tau_0}^\tau(s'-\tau_0)\langle\mathrm d\mathcal W_s\mathrm d\mathcal W_{s'}\rangle=\int_{\tau_0}^\tau(s-\tau_0)\mathrm ds=\frac12\Delta\tau^2.
\end{equation}
To simulate $I_{(1)}$ and $I_{(0,1)}$ with two independent standard normal random variables $w_{1,2}\sim N(0,1)$, we can institute
\begin{equation}
I_{(1)}=Aw_1+Bw_2,\quad
I_{(0,1)}=Cw_1+Dw_2,
\end{equation}
into $\langle\Delta\mathcal W_\tau^2\rangle=\Delta\tau$, Eq. \ref{I01mean}, Eq. \ref{I01var} and Eq. \ref{I1I01covar}. Solve the equation system with respect to $A$, $B$, $C$ and $D$, we will achieve
\begin{equation}
I_{(1)}=\sqrt{\Delta\tau}w_1,\quad
I_{(0,1)}=\frac12\Delta\tau^{\frac32}w_1+\frac1{2\sqrt{3}}\Delta\tau^{\frac32}w_2.
\end{equation}

There is another way around, which is used in \cite{NumericalSolSDEStrong} and which we actually adopt in this paper. We can represent $I_{(0,1)}$ with $I_{(1)}$ and $I_{(1,0)}$: $I_{(0,1)}=I_{(1)}I_{(0)}-I_{(1,0)}$, and the mean and variance of $I_{(1,0)}$ can be calculated by
\begin{equation}\label{I10mean}
\langle I_{(1,0)}\rangle=\langle \int_{\tau_0}^\tau\int_{\tau_0}^s\mathrm d\mathcal W_u\mathrm ds\rangle=\int_{\tau_0}^\tau\int_{\tau_0}^s\langle \mathrm d\mathcal W_u\rangle\mathrm ds=0,
\end{equation}
\begin{equation}
\begin{aligned}
\langle I_{(1,0)}^2\rangle&=\langle\left[I_{(0)}I_{(1)}-I_{(0,1)}\right]^2\rangle=\langle I_{(0)}^2I_{(1)}^2-2I_{(0)}I_{(1)}I_{(0,1)}+I_{(0,1)}^2\rangle=\Delta\tau^2\langle\Delta\mathcal W_\tau^2\rangle-2\Delta\tau\langle I_{(1)}I_{(0,1)}\rangle+\langle I_{(0,1)}^2\rangle\\
&=\Delta\tau^2\Delta\tau-2\Delta\tau\frac12\Delta\tau^2+\frac13\Delta\tau^3=\frac13\Delta\tau^3,
\end{aligned}
\end{equation}
and we can also calculate the covariance of $I_{1}$ and $I_{(1,0)}$:
\begin{equation}\label{I1I10covar}
\begin{aligned}
\langle I_{(1)}I_{(1,0)}\rangle&=\langle\int_{\tau_0}^\tau\mathrm d\mathcal W_s\int_{\tau_0}^\tau\int_{\tau_0}^{s'}\mathrm d\mathcal W_u\mathrm ds'\rangle=\langle\int_{\tau_0}^\tau\left[\int_{\tau_0}^\tau\mathrm d\mathcal W_s\int_{\tau_0}^{s'}\mathrm d\mathcal W_u\right]\mathrm ds'\rangle=\int_{\tau_0}^\tau\left[\int_{\tau_0}^{s'}\int_{\tau_0}^\tau\langle\mathrm d\mathcal W_s\mathrm d\mathcal W_u\rangle\right]\mathrm ds'\\
&=\int_{\tau_0}^\tau\left[\int_{\tau_0}^{s'}\int_{\tau_0}^\tau\delta(s-u)\mathrm ds\mathrm du\right]\mathrm ds'=\int_{\tau_0}^\tau\left[\int_{\tau_0}^{s'}\mathrm du\right]\mathrm ds'=\frac12\Delta\tau^2.
\end{aligned}
\end{equation}
Because $\langle I_{(1,0)}\rangle$, $\langle I_{(1,0)}^2\rangle$ and $\langle I_{(1)}I_{(1,0)}\rangle$ are equal to $\langle I_{(0,1)}\rangle$, $\langle I_{(0,1)}^2\rangle$ and $\langle I_{(1)}I_{(0,1)}\rangle$ respectively, we can simulate $I_{(1)}$ and $I_{(1,0)}$ with two independent standard normal random variables: $w_{1,2}\sim N(0,1)$ in the same way as $I_{(1)}$ and $I_{(0,1)}$:
\begin{equation}
I_{(1)}=\sqrt{\Delta\tau}w_1,\quad
I_{(1,0)}=\frac12\Delta\tau^{\frac32}w_1+\frac1{2\sqrt{3}}\Delta\tau^{\frac32}w_2.
\end{equation}
\subsection{Parameters used in the simulation}\label{ParametersUsed}
Although we have nondimensionalized the Langevin equation, we will use the parameters in the trapped ion friction emulating experiment implemented by Gangloff et al. \cite{NPVelocityTuning,ScienceTIFE,MultislipTIFE} so that our results will be more realistic and we can validate our numerical model through comparing with the experimental results. We must emphasize that because our simulation results are nondimensional, our simulation results may not only apply to the trapped ion system but may also apply to other systems such as a levitated nanosphere in optomechanical cavity \cite{UnderdampedEngine,CavityCoolNanospherePRL} and a microparticle suspended in liquid \cite{Stirling}. The implementation of the PTSHE should depend on the practical obtainable ranges of all the parameters, which are mutually constrained as discussed in Sec. \ref{sec:apndRD}.\ref{sec:parammutualconstraint}.

In \cite{NPVelocityTuning,ScienceTIFE,MultislipTIFE}, $\rm{Yb^+}$ ion is chosen as the trapped ion whose mass $m=2.8887\times10^{-25}\rm{kg}$ and diameter $a=185\times10^{-9}\rm m$.
%Throughout the paper, our discussion will focus on the interval $\eta\in(1,4.6)$, in which the resultant potential energy curve has at most two local minima. 
The corrugation number $\eta$ is determined by the amplitute of the lattice potential $V_0/2$ and the Paul trap longitudinal vibrational frequency $\omega_0/(2\pi)=364\rm kHz$, both of which can be tuned to obtain a specific value of $\eta$. If we fix $\omega_0$ and choose a specific $\eta$, then $V_0$ will be determined by $\eta=\frac{2\pi^2V_0}{m\omega_0^2a^2}$.
% the typical value of which is $\eta=3$. 
The range of temperature is constrained by the experimental condition which is several to dozens of $\rm\mu K$. Temperature $T$ and $V_0$ determine the nondimentional temperature ${\it\Theta}=\frac{k_BT}{V_0}$. 
%which is calculated in the experiment as the temperature parameter. 
%In this paper, unless otherwise specified, we fix $\omega_0$ and then $V_0$ is determined by $\eta=\frac{2\pi^2V_0}{m\omega_0^2a^2}$, the typical value of which is $\eta=3$. 
We choose ${\it\Theta}=0.4$ and $0.04$ as the high and the low temperature respectively in the main simulation case in the main text and the corresponding absolute temperatures are $22.77\rm\mu K$ and $2.277\rm\mu K$. In the parameter analysis, we have to change $\eta$, ${\it\Theta}$ and $\mu$. Attention has to be payed to the mutual constraints between these paremeters. Ranges of all the parameters used in this paper are listed in Table \ref{parameters} unless otherwise specified.

\begin{table}[H]
\centering
\caption{Parameters used in the simulation}
\label{parameters}
\begin{tabular}{ccccccccc}
\toprule
 Parameters & $m(10^{-25}\rm{kg})$ & $a(10^{-9}\rm{m})$ & $\omega_0/(2\pi)(\rm{kHz})$ & $\eta$ & $\mu(\rm{s^{-1}})$ & ${\it\Theta}$ & $V_0$ & $\beta$\\
\midrule
 Values & $2.8887$ & $185$ & $364$ & $[1,30]$ & $[0, 4\times10^7]$ & $[0,400]$ & $\eta\frac{m\omega_0^2a^2}{2\pi^2}$ & $\frac{m\mu\omega_0a^2}{\pi V_0}$\\
\bottomrule
\end{tabular}
\end{table}
The numbers of simulation cycles should be large enough for the particle to arrive at steady state. However, we cannot simulate too many cycles at low driving velocity due to the small time stepsize and the large number of time steps in one cycle leading to the long simulation time. We chose the number of simulation cycles at each driving velocity after some trial and error and they are listed in Table \ref{tab:Numberofcycles}. 

At $v_{\rm dr}=0.4\rm m/s$ and $0.5\rm m/s$, the numbers of simulation cycles are larger than their neighbors, because we find that for the main simulation case with parameters $\eta=3.0$, $\mu=4\times10^4\rm s^{-1}$ and ${\it\Theta}_{\rm h,c}=0.4,0.04$, at these driving velocities
%, stochastic bifurcation are distinct (Figure \ref{}), and 
the stable mean values and standard deviations of $W_{\rm cyc}$ cannot be achieved until the number of cycles are large enough. The two numbers are chosen after some trial and error when the mean values and standard deviations of $W_{\rm cyc}$ no longer change obviously with the number of simulation cycles increasing. 

For the main simulation case of $\eta=3.0$, $\mu=4\times10^4\rm s^{-1}$ and ${\it\Theta}_{\rm h,c}=0.4,0.04$ computed in a Dell Inc. PowerEdge T630 Server with two Intel(R) Xeon(R) CPU's (E5-2680 v4 @ 2.40GHz) and 192GB RAM with Matlab R2021a on Windows Server 2019 Standard Evaluation, the time to simulate all the cycles at all the driving velocities using SRK4 is about 20 hours.

\begin{table}[H]
\centering
\caption{The number of simulation cycles at different driving velocities}
\label{tab:Numberofcycles}
\begin{tabular}{cccccccccccc}
\toprule
$v_{\rm dr}(\rm m/s)$ & $10^{-7}$ & $10^{-6}$ & $10^{-5}$ & $2\times10^{-5}$ & $5\times10^{-5}$ & $7.5\times10^{-5}$ & $10^{-4}$ & $2\times10^{-4}$ & $5\times10^{-4}$ & $7.5\times10^{-4}$\\
\midrule
Number of cycles & 20 & 200 & 2000 & 2000 & 2000 & 2000 & 5000 & 5000 & 8000 & 8000\\
\midrule
$v_{\rm dr}(\rm m/s)$ & $10^{-3}$ & $1.5\times10^{-3}$ & $2\times10^{-3}$ & $3\times10^{-3}$ & $5\times10^{-3}$ & $7.5\times10^{-3}$ & $10^{-2}$ & $1.5\times10^{-2}$ & $2\times10^{-2}$ & $3\times10^{-2}$\\
\midrule
Number of cycles & 20000 & 30000 & 40000 & 60000 & 100000 & 150000 & 200000 & 300000 & 400000 & 600000\\
\midrule
$v_{\rm dr}(\rm m/s)$ & $5\times10^{-2}$ & $7.5\times10^{-2}$ & $0.1$ & $0.125$ & $0.15$ & $0.2$ & $0.3$ & $0.4$ & $0.5$ & $0.75$\\
\midrule
Number of cycles & 1000000 & 1500000 & 3000000 & 3000000 & 3000000 & 3000000 & 3000000 & 22000000 & 12000000 & 3000000\\
\midrule
$v_{\rm dr}(\rm m/s)$ & $1$ & $1.25$ & $1.5$ & $2$ & $3$ & $4$ & $5$ & $7.5$ & $10$\\
\midrule
Number of cycles & 3000000 & 3000000 & 3000000 & 3000000 & 3000000 & 3000000 & 3000000 & 3000000 & 3000000\\
\bottomrule
\end{tabular}
\addtabletext{For the case of $\mu=4\times10^{-1}\rm s^{-1}$ in the main text Figure 3(A), six more driving velocities are simulated at the low driving velocity end to make the curve smoother. The simulation cycles at these driving velocities are: 40 at $2\times10^{-7}\rm m/s$, 100 at $5\times10^{-7}\rm m/s$, 150 at $7.5\times10^{-7}\rm m/s$, 200 at $2\times10^{-6}\rm m/s$, 200 at $5\times10^{-6}\rm m/s$ and 200 at $7.5\times10^{-6}$\rm m/s.}
\end{table}

At each driving velocity $v_{\rm dr}$, the initial conditions are $(z(0),\dot z(0))=(0,0)$. Incidentally, we now realize that if we chose $(z(0),\dot z(0))=(0,\tilde v)$, the transient stage before steady state being arrived at might be shortened at the beginning of the simulation time range, especially at the high driving velocity regime where $\tilde v$ is large.

The mean values and standard deviations of the cycle work $W_{\rm cyc}$ are averaged from the last $2000000$ simulation cycles when the cycle number is $3000000$, from the last $8000000$ simulation cycles at $v_{\rm dr}=0.4\rm m/s$, from the last $5000000$ simulation cycles at $v_{\rm dr}=0.5\rm m/s$, and from the last $90\%$ of the simulation cycles at the rest driving velocities. These sample numbers are chosen after some trial and error to eliminate the influence from the beginning transient stage and to make sure that the sample numbers are large enough to reflect the true first and second order moments of the stochastic variable $W_{\rm cyc}$. 

\begin{figure}[H]
\centering
\begin{minipage}{0.49\textwidth}
\centerline{
\includegraphics[width=\textwidth]{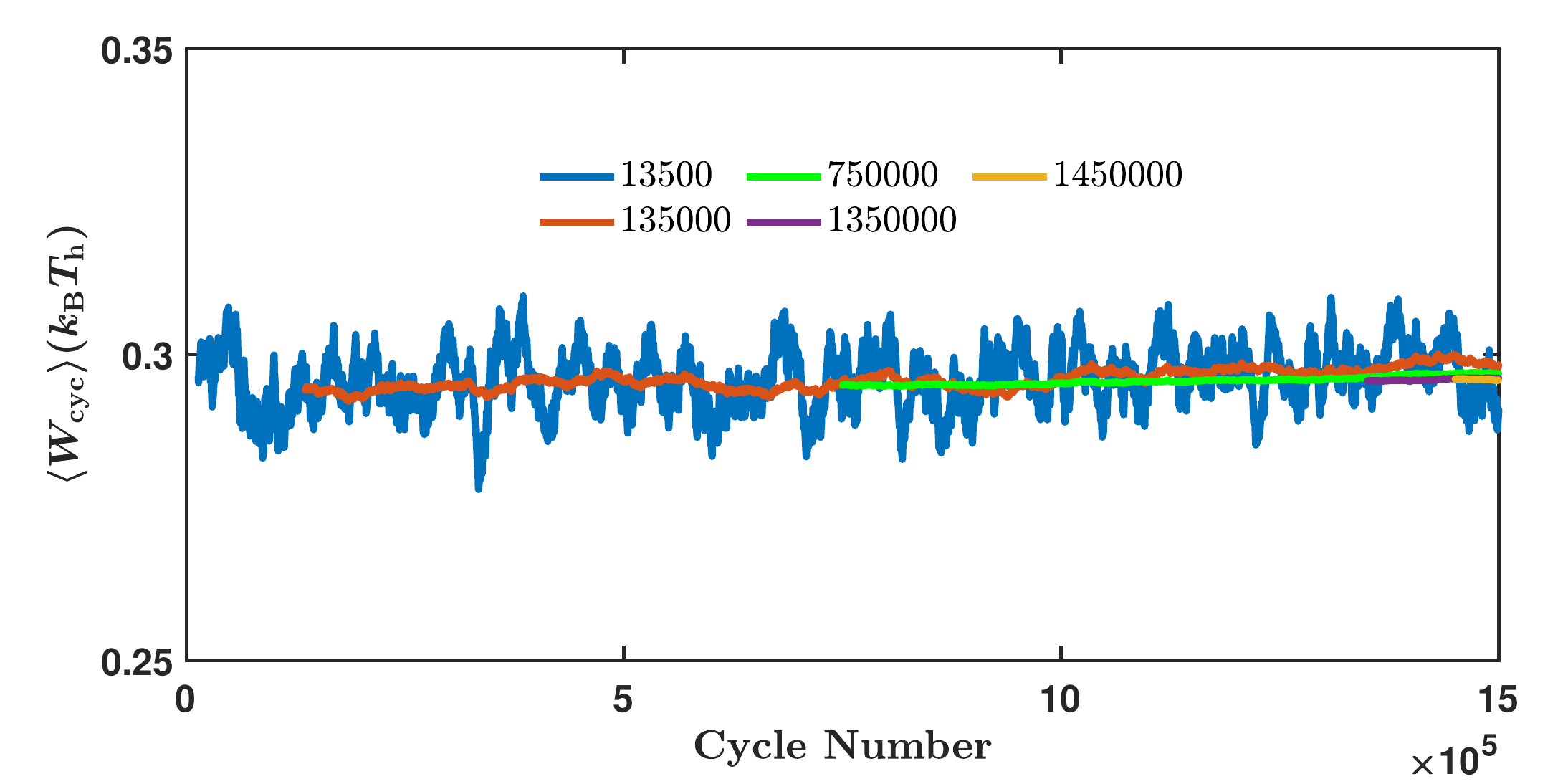}}
\centerline{(a)\ $v_{\rm dr}=7.5\times10^{-2}\rm m/s$}
\end{minipage}
\begin{minipage}{0.49\textwidth}
\centerline{
\includegraphics[width=\textwidth]{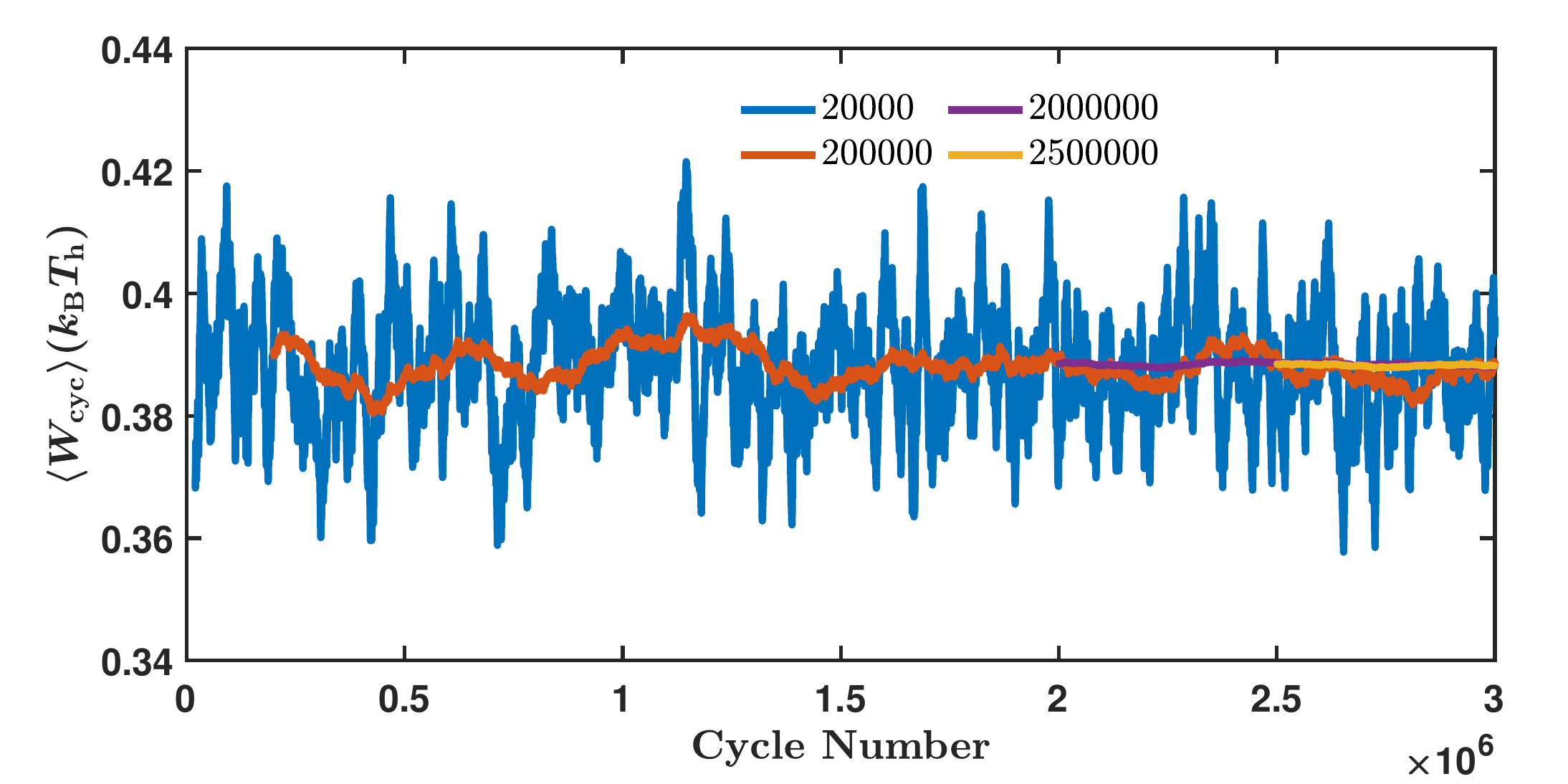}}
\centerline{(b)\ $v_{\rm dr}=0.3\rm m/s$}
\end{minipage}\\
\begin{minipage}{0.49\textwidth}
\centerline{
\includegraphics[width=\textwidth]{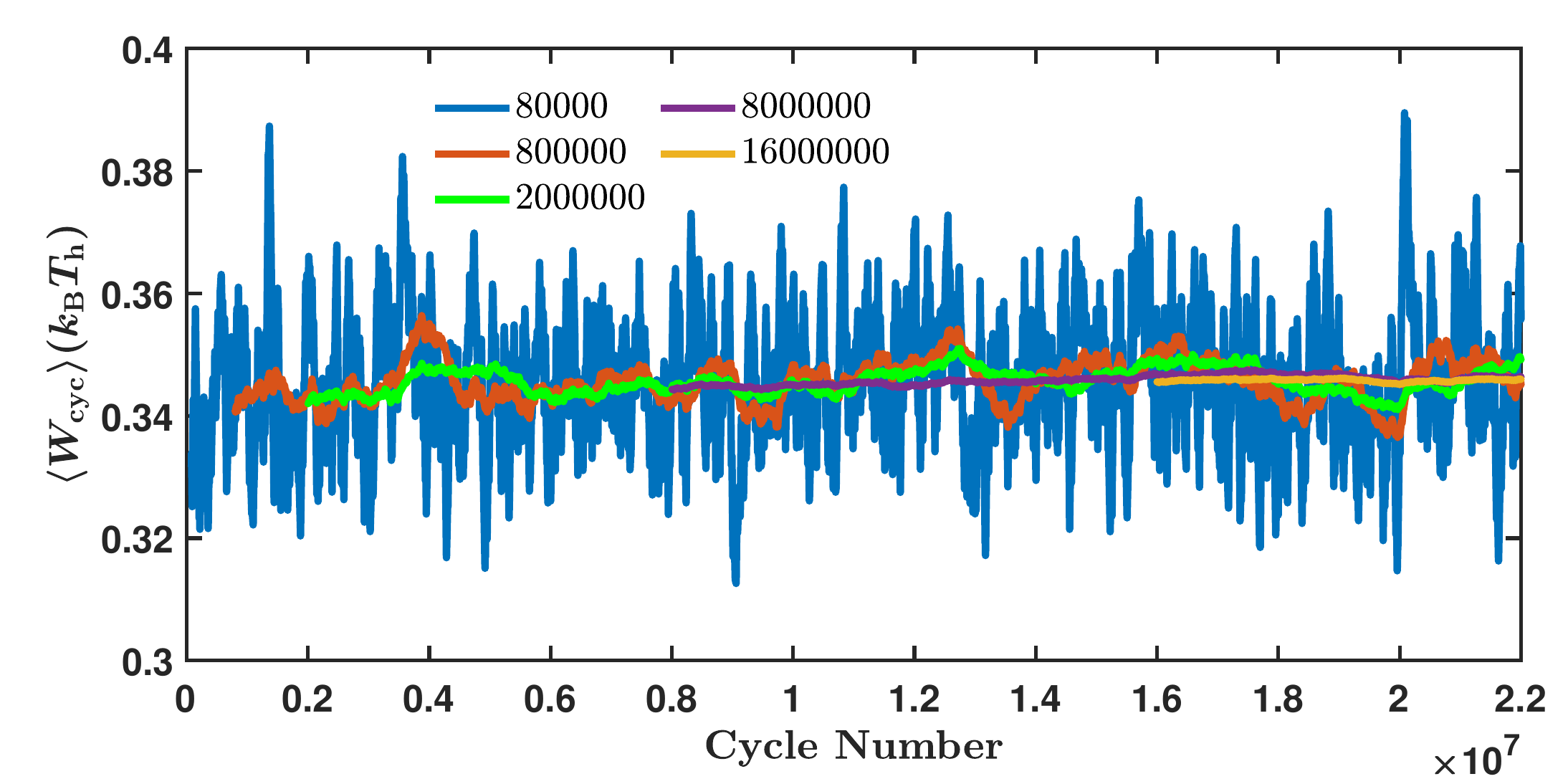}}
\centerline{(c)\ $v_{\rm dr}=0.4\rm m/s$}
\end{minipage}
\begin{minipage}{0.49\textwidth}
\centerline{
\includegraphics[width=\textwidth]{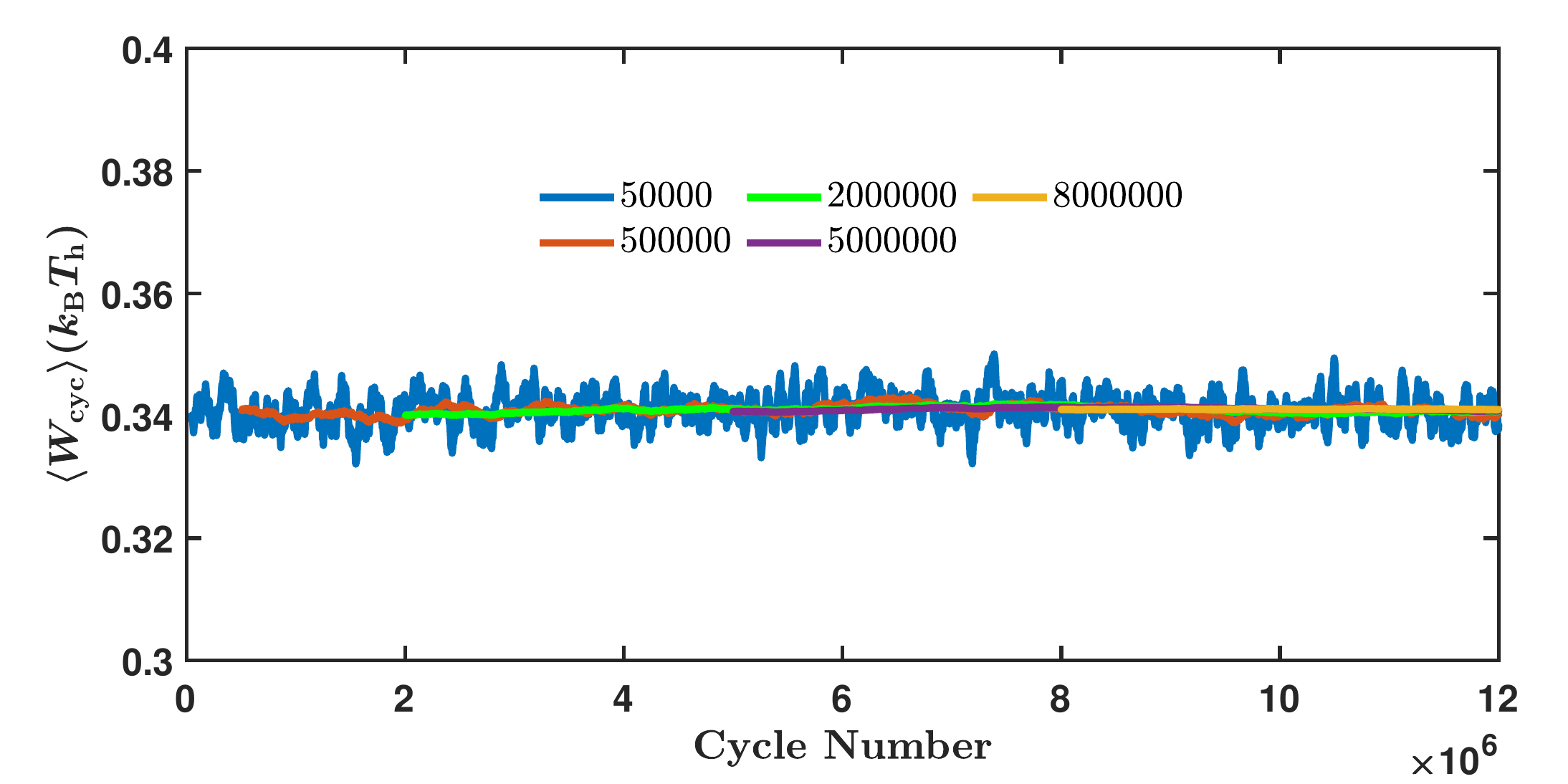}}
\centerline{(d)\ $v_{\rm dr}=0.5\rm m/s$}
\end{minipage}
\caption{The mean cycle work calculated from different sample number of simulation cycles at four typical driving velocities. The $y$-coordinate is the moving mean cycle work $\langle W_{\rm cyc}\rangle$ averaged over a sliding window of different sample number of simulation cycles. The $x$-coordinate is the simulation cycle number from the beginning to the end of the simulation time range. Different color indicates different sample number to calculate the mean value $\langle W_{\rm cyc}\rangle$. We take the case of $\eta=3.0$, $\mu=4\times10^4\rm s^{-1}$ and ${\it\Theta}_{\rm h,c}=0.4,0.04$ as an example. Each curve begins after the cycle number is greater than the corresponding sample number. We can see that at all the four driving velocities, our choice of the sample numbers (represented by the purple curves) are suitable in that the mean value $\langle W_{\rm cyc}\rangle$ remains nearly unchanged with the sample number increased further. The range of the $y$-axis is 1 $k_{\rm B}T_{\rm h}$ for all the four subfigures for comparison. In (c), we can see that at $v_{\rm dr}=0.4\rm m/s$, the amplitude of $\langle W_{\rm cyc}\rangle$ averaged from 80000 cycles is large and similar to that averaged from 20000 cycles at $v_{\rm dr}=0.3\rm m/s$ in (b). At $v_{\rm dr}=0.4\rm m/s$, the moving mean $\langle W_{\rm cyc}\rangle$ averaged from 2000000 cycles still varies a little violently compared with the purple curve in (b) at $v_{\rm dr}=0.3\rm m/s$, indicating that at $v_{\rm dr}=0.4\rm m/s$, 2000000 cycles are not enough for the particle to traverse all the possible states sufficiently and the frequency obtained deviates from the steady-state probability to a certain extent, while at $v_{\rm dr}=0.3\rm m/s$, the sample number of 2000000 is enough for the particle to traverse all the possible states with the frequency obtained close to the steady-state probability. 
%Therefore we have to choose a larger simulation cycle number and a larger sample number to calculate the mean and standard deviations at $v_{\rm dr}=0.4\rm m/s$. 
The starting transient stage is obsevable when the sample number is small [$\leq$800000 in (c) and $\leq$200000 in (b)] and becomes less observable when the sample number increases. At $v_{\rm dr}=0.5\rm m/s$ in (d), we can see that the simulation number of 12000000 and the sample number of 5000000 are chosen excessively and 3000000 and 2000000 for the simulation and the sample number respectively should be enough. In (a) at $v_{\rm dr}=7.5\times10^{-2}\rm m/s$, we choose the last $90\%$ simulation cycles as the sample, i.e. a sample number of 1350000. We presume that $10\%$ simulation cycles at the beginning are enough for the particle to achieve steady state and it seems reasonable. There should be a critical sample number of simulation cycles at steady state for the particle's frequency distribution to reflect the steady-state probability distribution. This number should depend on the driving velocity, which is different from the dependence of the standard deviation of $W_{\rm cyc}$ on the driving velocity in that the standard deviation of $W_{\rm cyc}$ at $v_{\rm dr}=0.3\rm m/s$ is greater than that at $v_{\rm dr}=0.4\rm m/s$. This number should reflect the complexity of the steady-state probability distribution of $W_{\rm cyc}$ and may be related to the entropy of the particle. We will consider the entropy of the particle in the PTSHE in the future.
%For the case of $v_{\rm dr}=0.3\rm m/s$, the distribution is not as complicated as the two cases in (c) and (d), cf. 
}
\label{fig:CheckNumforMeanStd}
\end{figure}

We can give a check for the choice of the sample numbers. In Figure \ref{fig:CheckNumforMeanStd}, we calculate the moving mean of the cycle work $W_{\rm cyc}$ with a sliding window of different number of simulation cycles. We can see that when the sample number is large enough the mean cycle work $\langle W_{\rm cyc}\rangle$ is nearly constant with influence from the starting transient stage nearly unobservable. The purple curves represent our final choice of the sample numbers to calculate $\langle W_{\rm cyc}\rangle$, which remains nearly unchanged when we increase the sample number further. Actually, we can see that at $v_{\rm dr}=0.5\rm m/s$, the simulation cycle number we chose is excessive and a simulation cycle number of 3000000 and a sample number of 2000000 should be enough. Thus our choice of the sample numbers as well as the simulation cycle numbers is still rough. 
%Although the choice of two numbers of cycles is still rough,
Even so, considering that our choice of the simulation cycle numbers is not very small and that the change of the mean values and the standard deviations calculated from different numbers of simulation cycles is not very large and doesn't affect our main conclusions, we didn't choose the number of simulation cycles for each simulation case specificly and used this set of cycle numbers for all of our simulation cases.

The count distribution of the cycle work at each driving velocity in Figure \ref{WcycDist} is obtained from the same sample of simulation cycles as those used to calculate the mean values and the standard deviations at the same driving velocities.

\subsection{Further check of the simulation method: the first law of thermodynamics}
\label{sec:furthercheckfirstlaw}
In the above, we have evaluated the SRK4 method by using it to validate the equipartition theorem for the linear Langevin equation and comparing its performance with a nonlinear scheme from the simulation of our nonlinear Langevin equation. Through both of the two aspects we are able to choose an appropriate time stepsize and confirm that the SRK4 is suitable for solving our nonlinear Langevin equation.
% Because the fluction-dissipation theorem, which is equivalent to the equipartition theorem in the linear case, is the core assumption of our nonlinear Langevin Eq. 4 in the main text, we can conclude that the SRK4 is suitable in the sense of satisfying our core assumption. 
 In this and next subsections, we will further check the SRK4 method, in the sense of low cumulative errors and the parameters chosen leading to results consistent with experiments.
%\subsubsubsection{The first law of thermodynamics}

We have derived the first law of thermodynamics from the Langevin equation in Sec. \ref{sec:firstlaw}. The error 
\begin{equation}
\delta_{\rm err}=\mathrm dU+\text{\dj} Q-\text{\dj} W,
\end{equation}
should keep small throughout the time range of each simulation case \cite{FirstLawPRL}. To measure this, we consider its integral over one cycle in which the driver goes through one lattice period
\begin{equation}
\Delta_{\rm err,cyc}=\Delta U_{\rm cyc}+Q_{\rm cyc}-W_{\rm cyc}.
\end{equation}
 
\begin{figure}[H]
 \centering
 \begin{minipage}{\textwidth}
 \centerline{
 \includegraphics[width=0.5\textwidth]{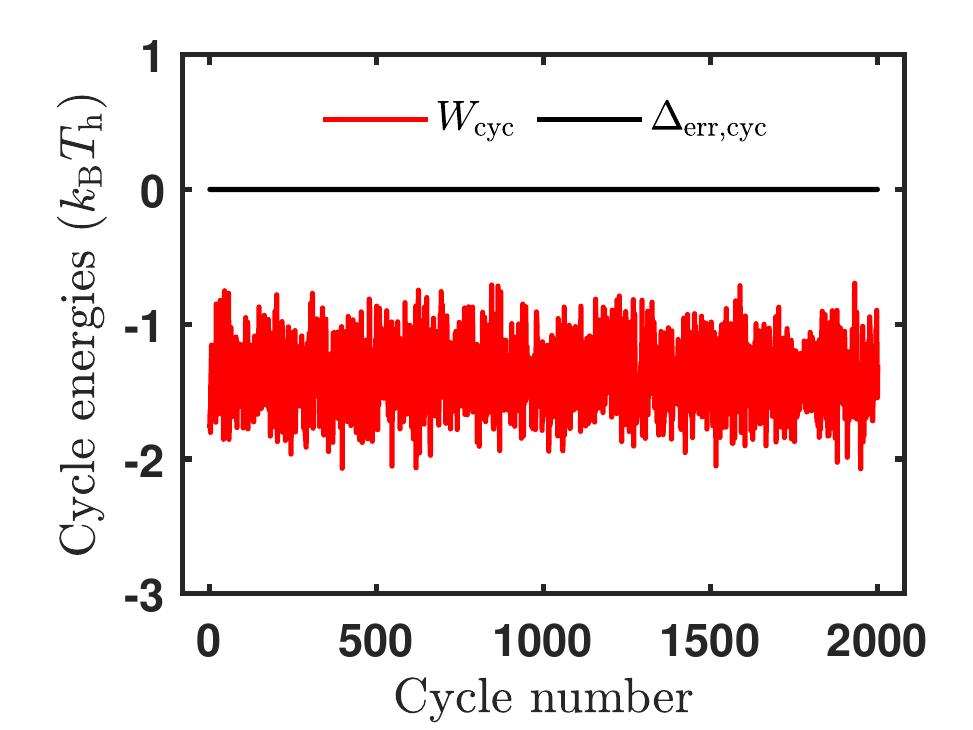}
 \includegraphics[width=0.5\textwidth]{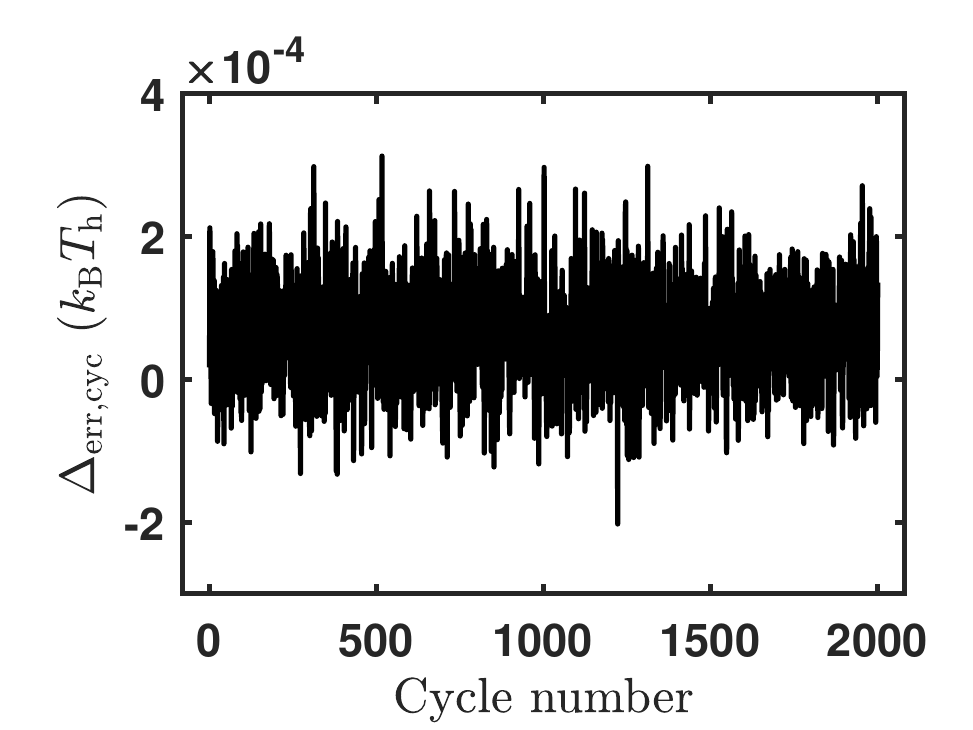}}
 \centerline{(a)\ $v_{\rm dr}=10^{-5}\rm m/s$}
 \end{minipage}
 \begin{minipage}{\textwidth}
 \centerline{
 \includegraphics[width=0.5\textwidth]{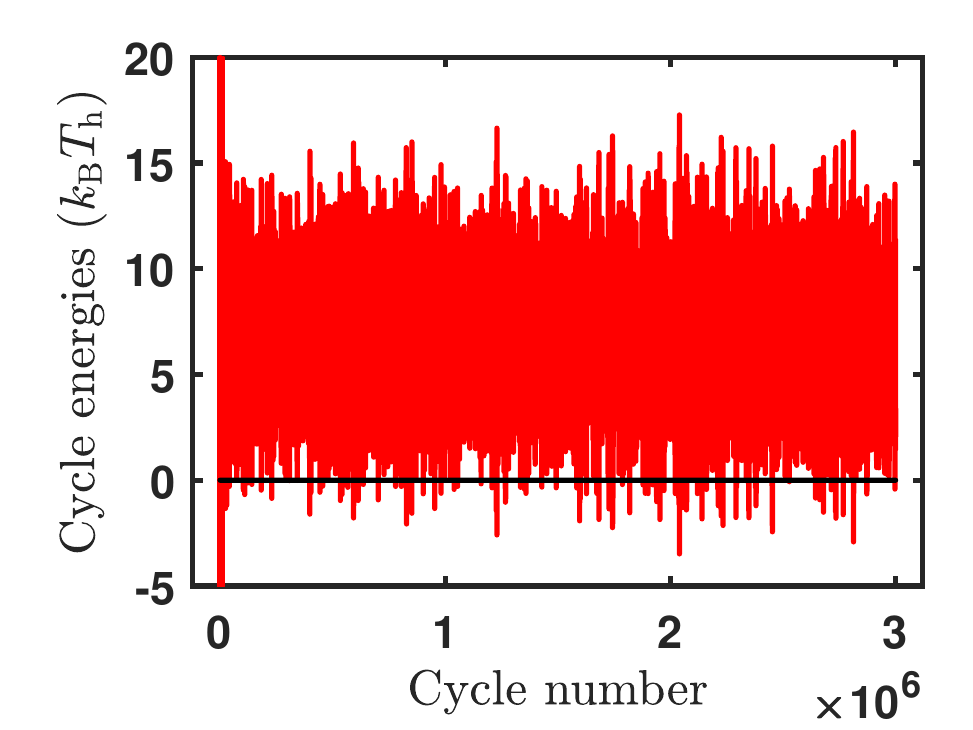}
 \includegraphics[width=0.5\textwidth]{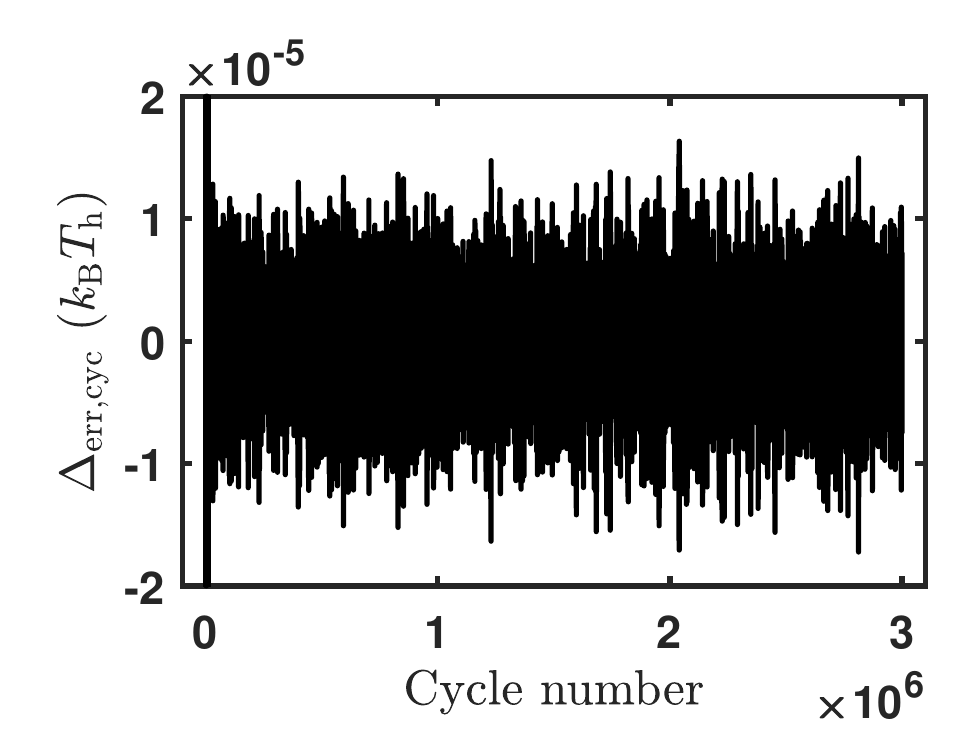}}
 \centerline{(b)\ $v_{\rm dr}=10\rm m/s$}
 \end{minipage}
 \caption{The cycle error $\Delta_{\rm err,cyc}$ compared with the cycle work $W_{\rm cyc}$. The upper two figures are simulation results at $v_{\rm dr}=10^{-5}\rm m/s$, in the case of which we simulate $2000$ cycles. The lower two figures correspond to $v_{\rm dr}=10\rm m/s$, in the case of which we simulate $3\times10^6$ cycles. When compared with the cycle work, the cycle errors keep small throughout the simulating time range, as shown in the left two figures. The right two figures are the zoom-in of the cycle errors on the left, indicating that the cycle errors are about four orders of magnitude lower than the cycle work values.}
 \label{FigFirstLaw}
 \end{figure}

 In Figure \ref{FigFirstLaw}, the cycle work $W_{\rm cyc}$ is compared with $\Delta{\rm err,cyc}$. We can see that for both the low driving velocity $v_{\rm dr}=10^{-5}\rm m/s$ and the high driving velocity $v_{\rm dr}=10\rm m/s$, the cycle error $\Delta{\rm err,cyc}$ keeps small and is about four orders of magnitude lower than $W_{\rm cyc}$ from the beginning to the end of the time range of each simulation case, which is true even in the violent transient process at the beginning of the $v_{\rm dr}=10\rm m/s$ case.  

\subsection{Further check of the simulation method: comparison with the experimental results}
\label{SecExperimentalDataVerify}
% \begin{figure*}
% \includegraphics[width=8.6cm]{SFigures/eta22kbToUl015.pdf}%
% \caption{\label{eta22kbToUl015}}
% \end{figure*}
% \begin{figure*}
% \includegraphics[width=8.6cm]{SFigures/eta46kbToUl004.pdf}%
% \caption{\label{eta46kbToUl004}}
% \end{figure*}
% \begin{figure*}
% \includegraphics[width=8.6cm]{SFigures/eta46kbToUl017.pdf}%
% \caption{\label{eta46kbToUl017}}
% \end{figure*}
To confirm that our mathematical model is predictable for practical problems, we do more numerical experiments to simulate the trapped ion friction simulator in \cite{ScienceTIFE,NPVelocityTuning} and compare our results with the experiments. 

In \cite{NPVelocityTuning,ScienceTIFE}, the fluorescence of the ion, which is proportional to the lattice potential energy $\tilde V_{\rm l}=\eta(1-\cos z)$, is detected to measure the friction force. The green dashed curves in Figure \ref{ExperimentalDataVerify_eta2p2}, \ref{ExperimentalDataVerify_eta4p6_T004} and \ref{ExperimentalDataVerify_eta4p6_T017} represent the lattice potential energy $\tilde V_{\rm l}$ at the resultant potential energy extrema $z^*$ with respect to $\tilde X(t)=\frac{v_{\rm dr}t}a$, which can be represented by Eq. \ref{eq:latticepotentialstickslip} and has been obtained in Figure \ref{fig:StickSlips}. As we have shown in Figure \ref{fig:StickSlips}, at zero temperature and infinite damping coefficient, i.e. qusis-tatic state, the particle slips at the instant of the BCP [i.e. $\tilde X(z_1^{**})$, cf. Sec. \ref{CriticalPtsofTempZoneSec} and Sec. \ref{sec:georoot}] of one cycle if the driver moves forward. If the driver moves backward, the particle will slip at the instant of the FCP [i.e. $\tilde X(z_2^{**})$, cf. Sec. \ref{CriticalPtsofTempZoneSec} and Sec. \ref{sec:georoot}]. So the forward and backward sweeping of the driver causes hysterises, and half of the separation between the two critical points is proportional to the maximum friction force defined in \cite{NPVelocityTuning,ScienceTIFE}:
\begin{equation}
\begin{aligned}
F_{\eta}&=\frac12m\omega_0^2a\left[\tilde X(z_1^{**})-\tilde X(z_2^{**})\right]\\
&=\frac12m\omega_0^2a\left\{z_1^{**}+\eta\sin(z_1^{**})-\left[z_2^{**}+\eta\sin(z_2^{**})\right]\right\}\\
&=\frac12m\omega_0^2a\left\{\arccos(-\frac1\eta)+\eta\sin(\arccos(-\frac1\eta))-\left[2\pi-\arccos(-\frac1\eta)+\eta\sin(2\pi-\arccos(-\frac1\eta))\right]\right\}\\
&=m\omega_0^2a\left[-\pi+\arccos(-\frac1\eta)+\eta\sin(\arccos(-\frac1\eta))\right],
\end{aligned}
\end{equation}
which is only determined by $\eta$. When the temperature is finite, the particle tends to slips early and both of the two slipping points approach to the middle and the friction force $F$ will be smaller than the maximum friction force. In the experiment the driver moves forward for several lattice periods and then moves backward and this loop is repeated many times. The fluoescence counts are detected at different time bins and then averaged over the all the repeated loops. To simulate the experiment as closely as possible, we integrate the Langevin equation forward three (or six at some very high velocities, cf. Figure \ref{ExperimentalDataVerify_eta2p2}, \ref{ExperimentalDataVerify_eta4p6_T004} and \ref{ExperimentalDataVerify_eta4p6_T017}) lattice periods and then backward three (or six at some very high velocities, cf. Figure \ref{ExperimentalDataVerify_eta2p2}, \ref{ExperimentalDataVerify_eta4p6_T004} and \ref{ExperimentalDataVerify_eta4p6_T017}) lattice periods, and this loop is repeated for a large number of times which we presume are enough to reflect the true statisics. The number of simulation loops are given in Table \ref{tab:ExperimentalDataVerify_eta2p2}, \ref{tab:ExperimentalDataVerify_eta4p6_T004} and \ref{tab:ExperimentalDataVerify_eta4p6_T017} before Figure \ref{ExperimentalDataVerify_eta2p2}, \ref{ExperimentalDataVerify_eta4p6_T004} and \ref{ExperimentalDataVerify_eta4p6_T017} respectively. During each loop, the forward and backward three (six) lattice periods passed by the driver center are both divided into 50 (100) bins. In each bin, the value of $\cos^2(z/2)$ at each discrete time step is summed and averaged to achieve the $\overline{\cos^2(z/2)}$ at the middle of each bin. Then we average $\overline{\cos^2(z/2)}$ at each specific bin through all the repeated simulation loops and obtain $\langle\overline{\cos^2(z/2)}\rangle$ at the middle of each bin, which is proportional to the fluorescence counts of the ion in the experiments. The results are given in Figure \ref{ExperimentalDataVerify_eta2p2}, \ref{ExperimentalDataVerify_eta4p6_T004} and \ref{ExperimentalDataVerify_eta4p6_T017}.

All the x-coordinates in the three figures are applied force $F/(m\omega_0^2a)$ \cite{ScienceTIFE} which is equal to the nondimensional driver center position $\tilde X(\tau)=X(t)/a$ relative to the starting point $0$. The error bars represent the standard deviations calculated from the ensemble of the repeated simulation loops of each simulation case at each specific bin. In each of the three cases, the separations of the peaks (at which slips occur) of the hysteresis loops first increase (and then descrease in two of the three cases), i.e. twofold of the so-defined friction forces \cite{NPVelocityTuning,ScienceTIFE} first increase (and then decrease in two of the three cases). The peaks of the hysteresis loops are each fitted to the quadratic polynomial $a(x-b)^2+c$ from 7 points (3 points on either side of the local maximum point) by quadratic nonlinear least square method. Half of the separation of two $b$'s in one hysteresis loop is equal to the friction force $F/(m\omega_0^2a)$. The error of $F$ is calculated from the $67\%$ confidence intervals of the two $b$'s. At each driving velocity $v_{\rm dr}$, we can get two or five friction forces and their errors. We then average these values to obtain the final friction forces and errors.

\begin{figure}[H]
 \centering
 \includegraphics[width=0.329\textwidth]{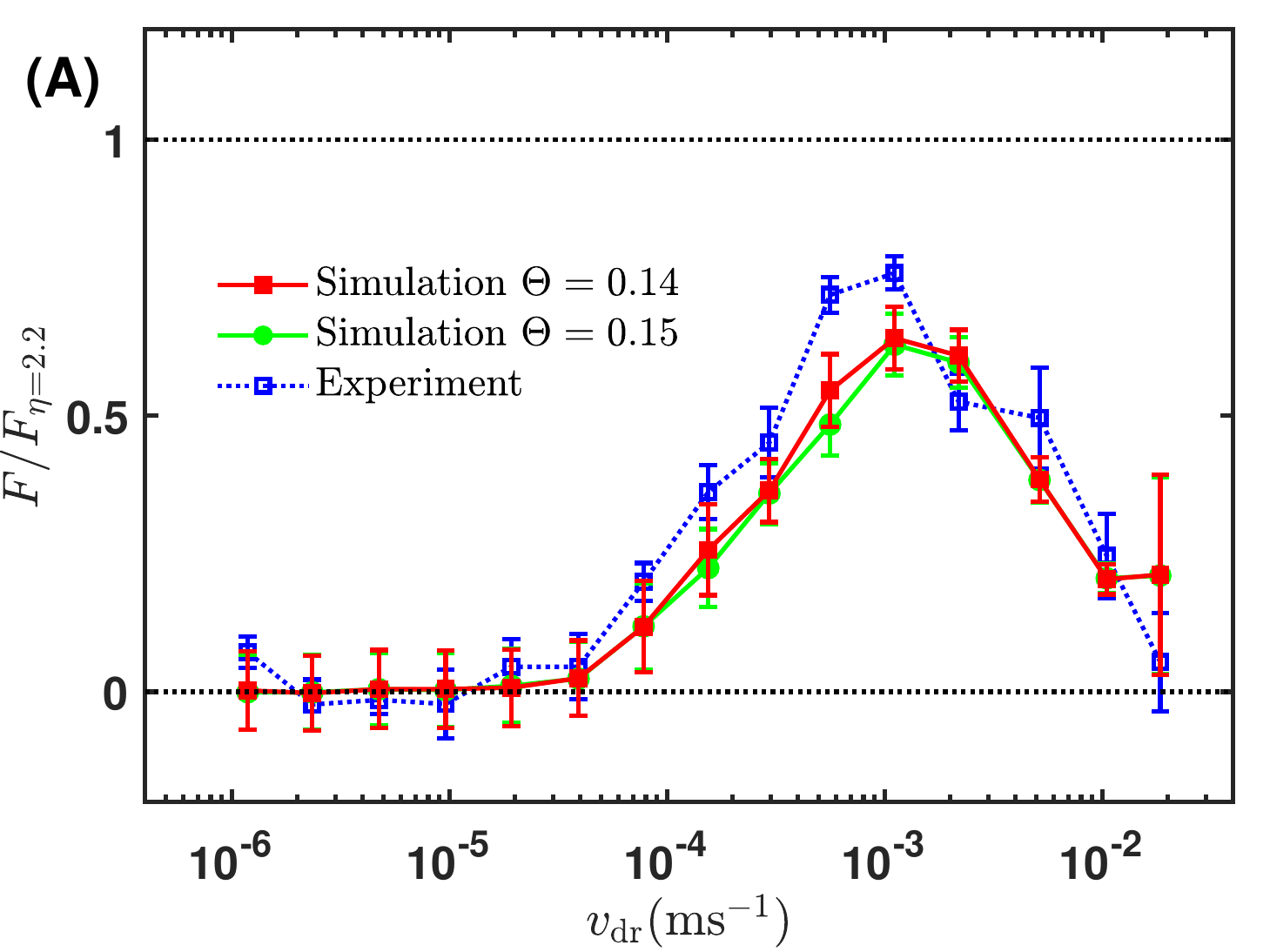}
 \includegraphics[width=0.329\textwidth]{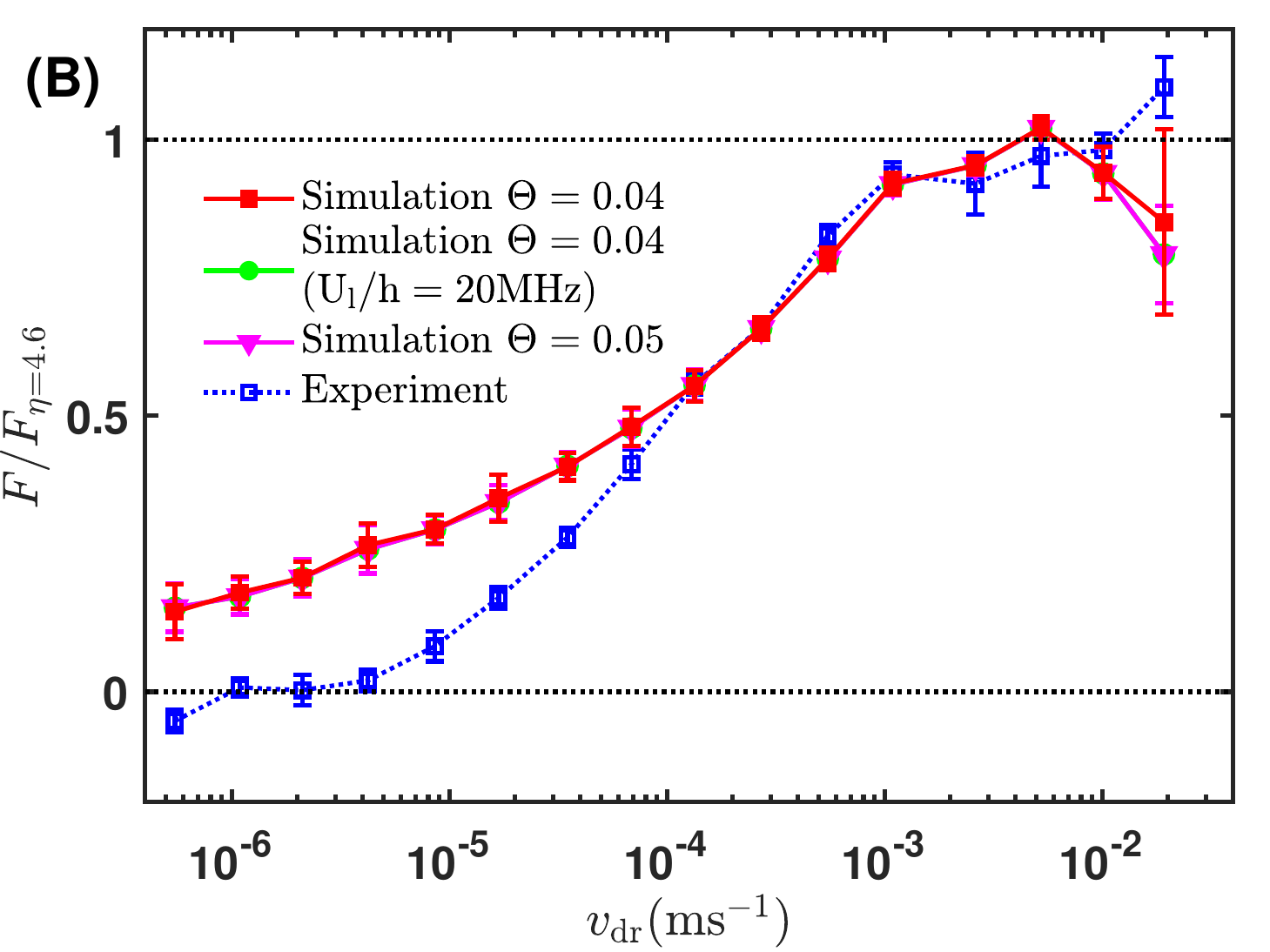}
 \includegraphics[width=0.329\textwidth]{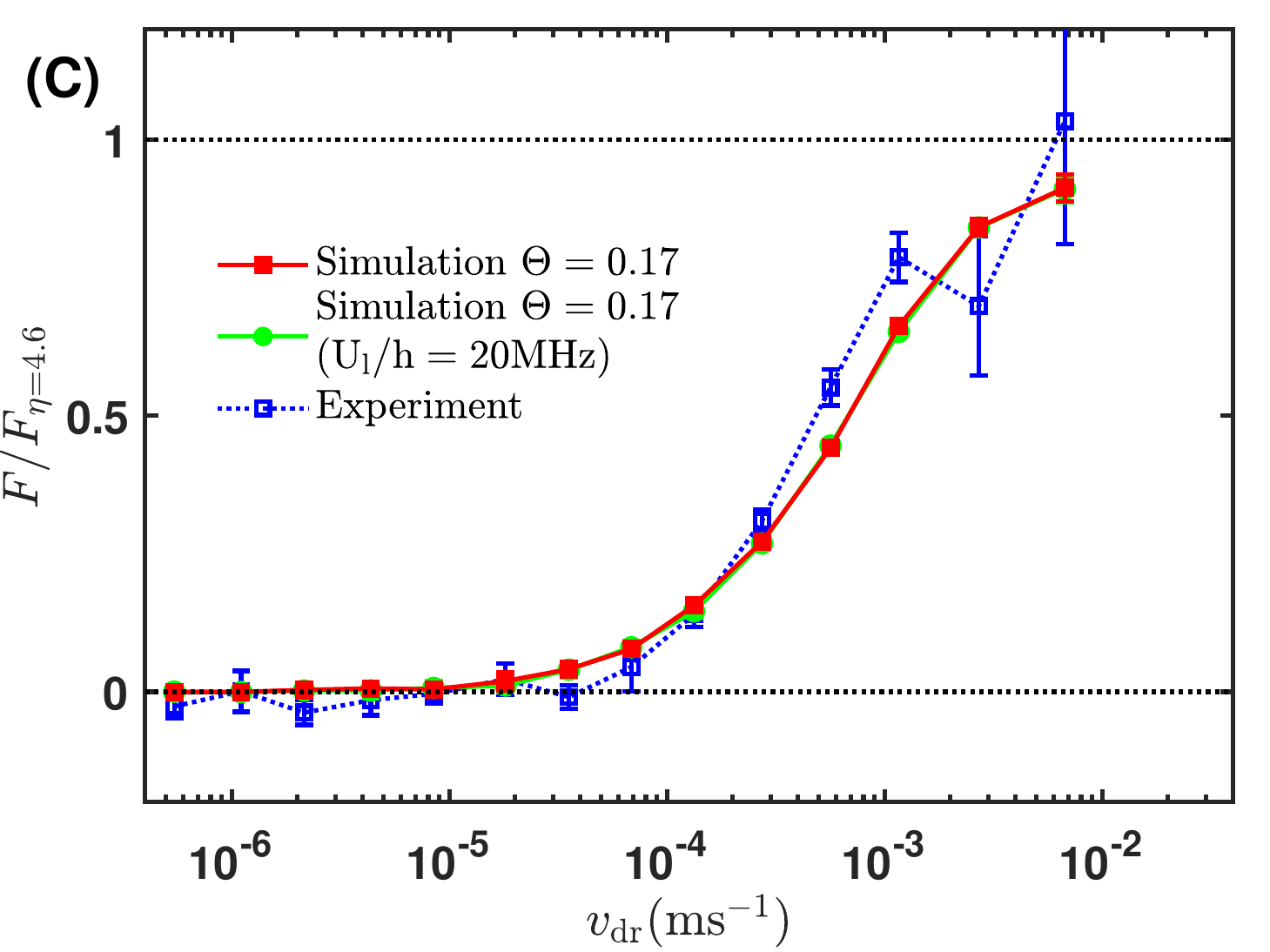}
 \caption{The friction force with respect to the driving velocity: comparison between the simulation and the experiment. (A) The results of the case of $\eta=2.2,\ \mu=2\times10^{4}\rm s^{-1},\ V_0=2\pi\hbar\times9.5\rm MHz$ and ${\it\Theta}=0.14(0.15)$. (B) The results of the case of $\eta=4.6,\ \mu=4\times10^{4}\rm s^{-1},\ V_0=2\pi\hbar\times18(20)\rm MHz$ and ${\it\Theta}=0.04(0.05)$. (C) The results of the case of $\eta=4.6,\ \mu=4\times10^{4}\rm s^{-1},\ V_0=2\pi\hbar\times18(20)\rm MHz$ and ${\it\Theta}=0.17$. The experimental data are from Figure 2 of \cite{NPVelocityTuning} courtesy of the authors. In each subfigure, we plot two or three sets of simulation results at parameters a little different from each other [corresponding to the tiny difference between the parameters in Figure 2 of the authors' paper \cite{NPVelocityTuning} and those in Figure 15-5 of one of the authors' theses \cite{bylinskiiphdthesis} respectively] and we can see that there is no significant deviation.}
 \label{ExperimentalDataVerify}
 \end{figure}

\begin{figure}[H]
\centering
 \includegraphics[width=5.73cm]{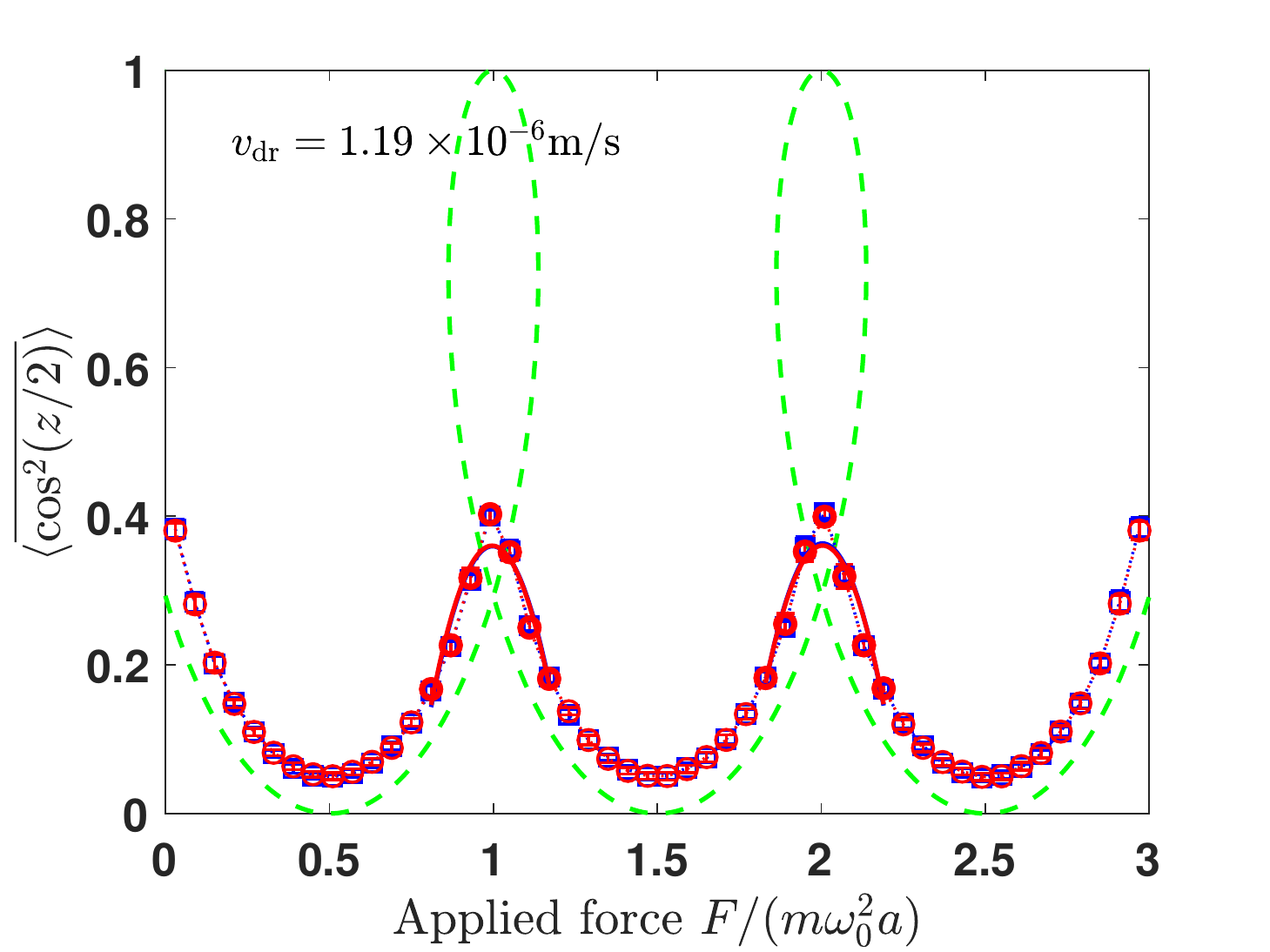}
 \includegraphics[width=5.73cm]{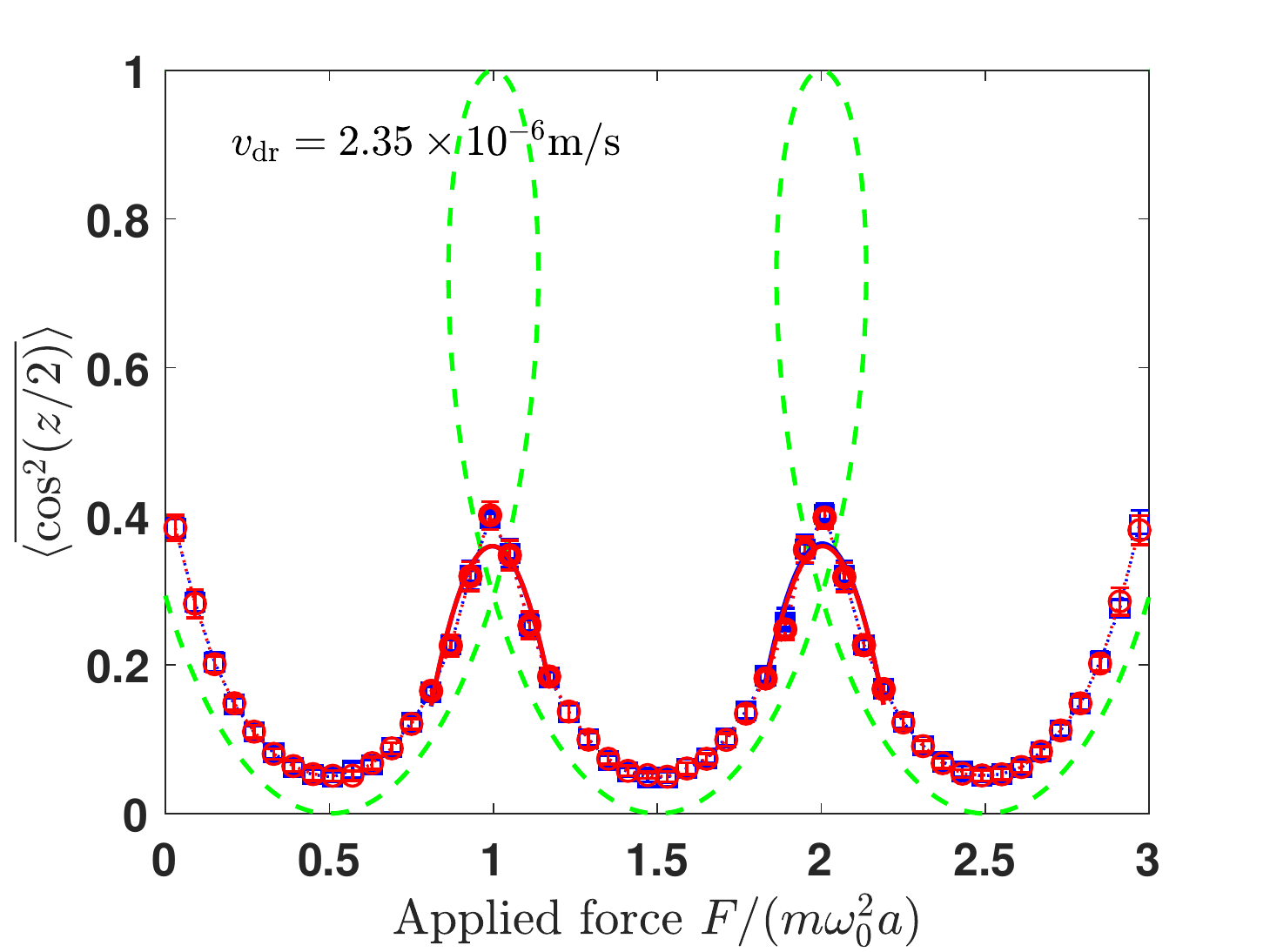}
 \includegraphics[width=5.73cm]{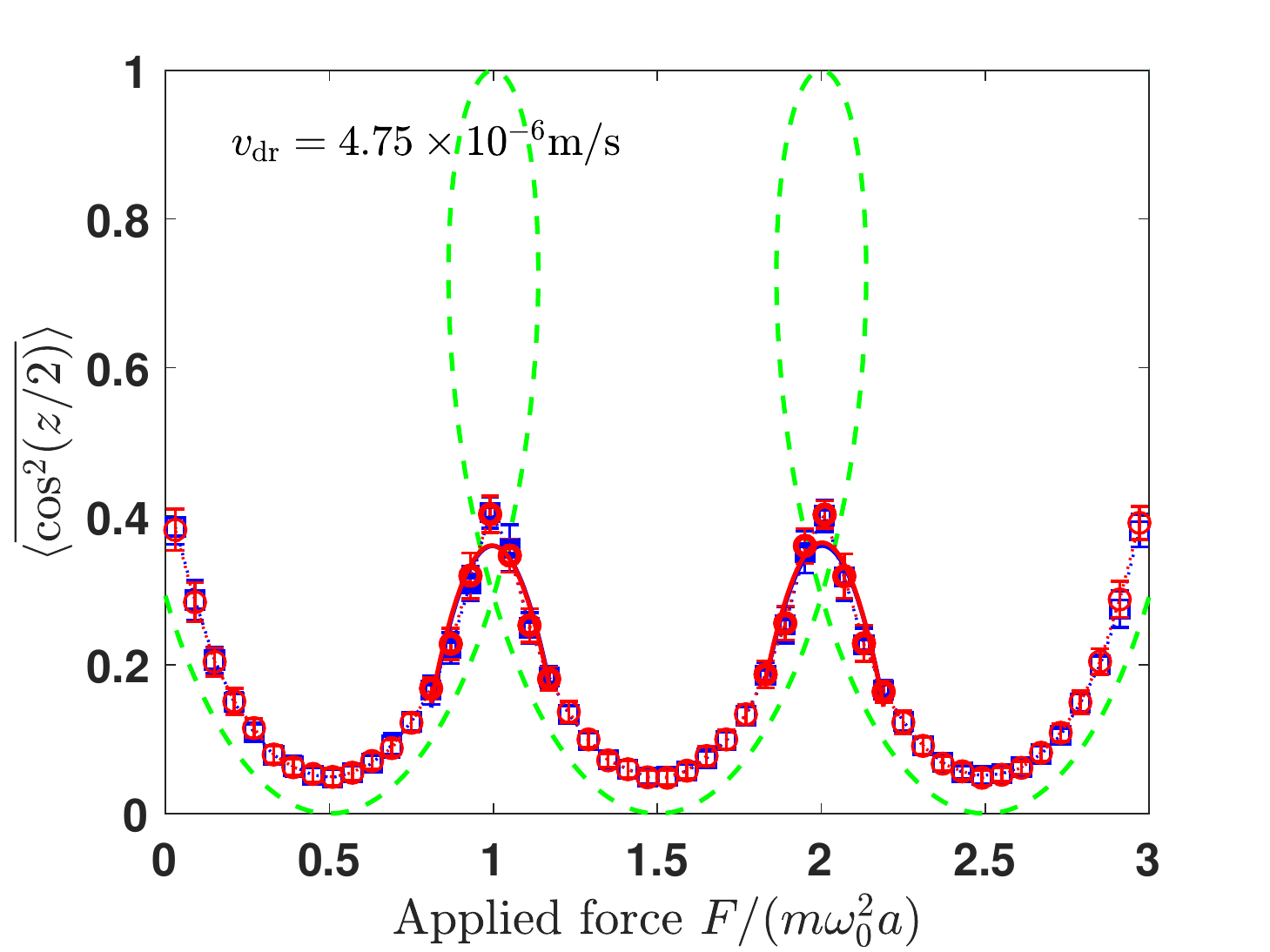}\\
 \includegraphics[width=5.73cm]{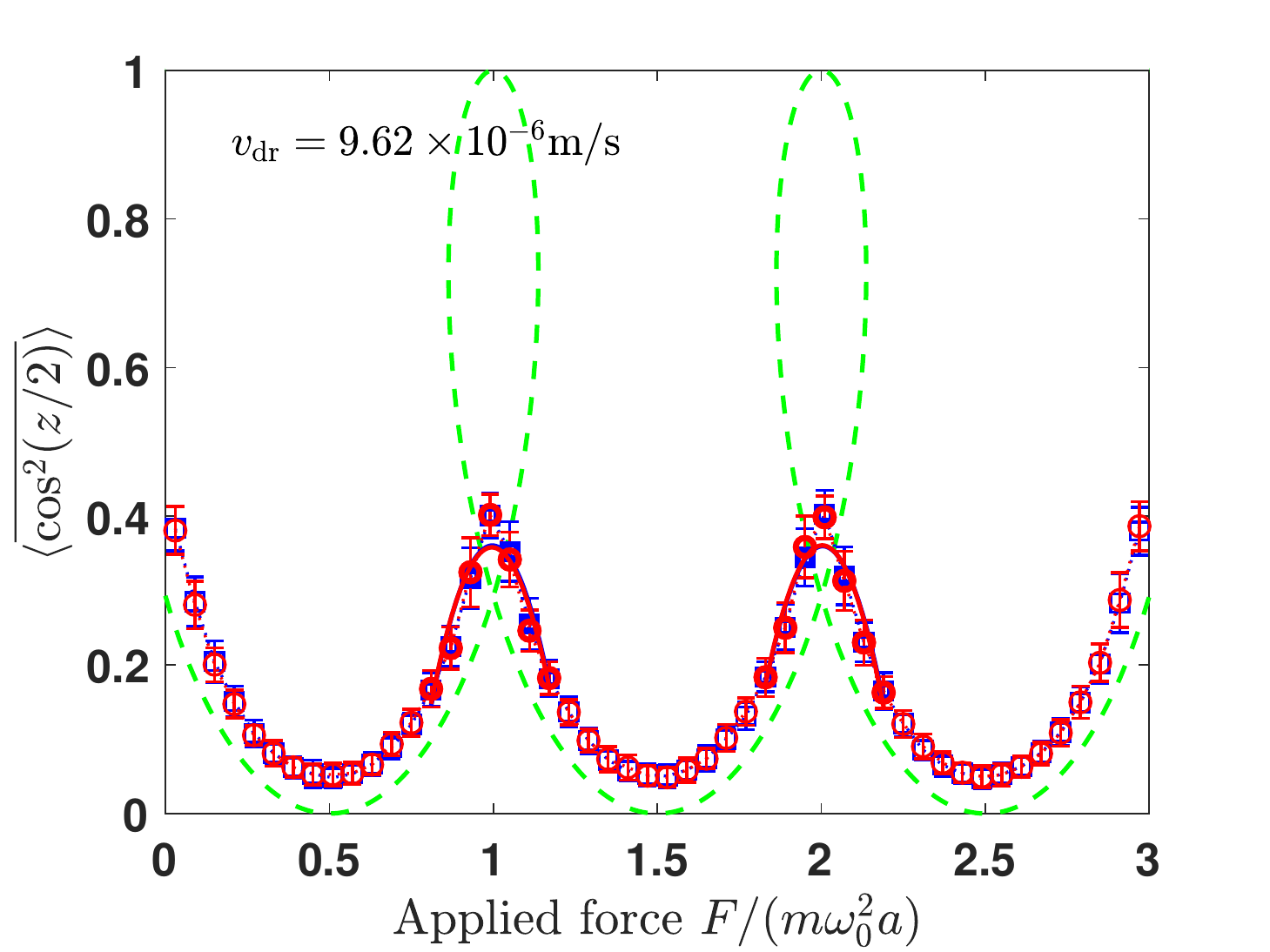}
 \includegraphics[width=5.73cm]{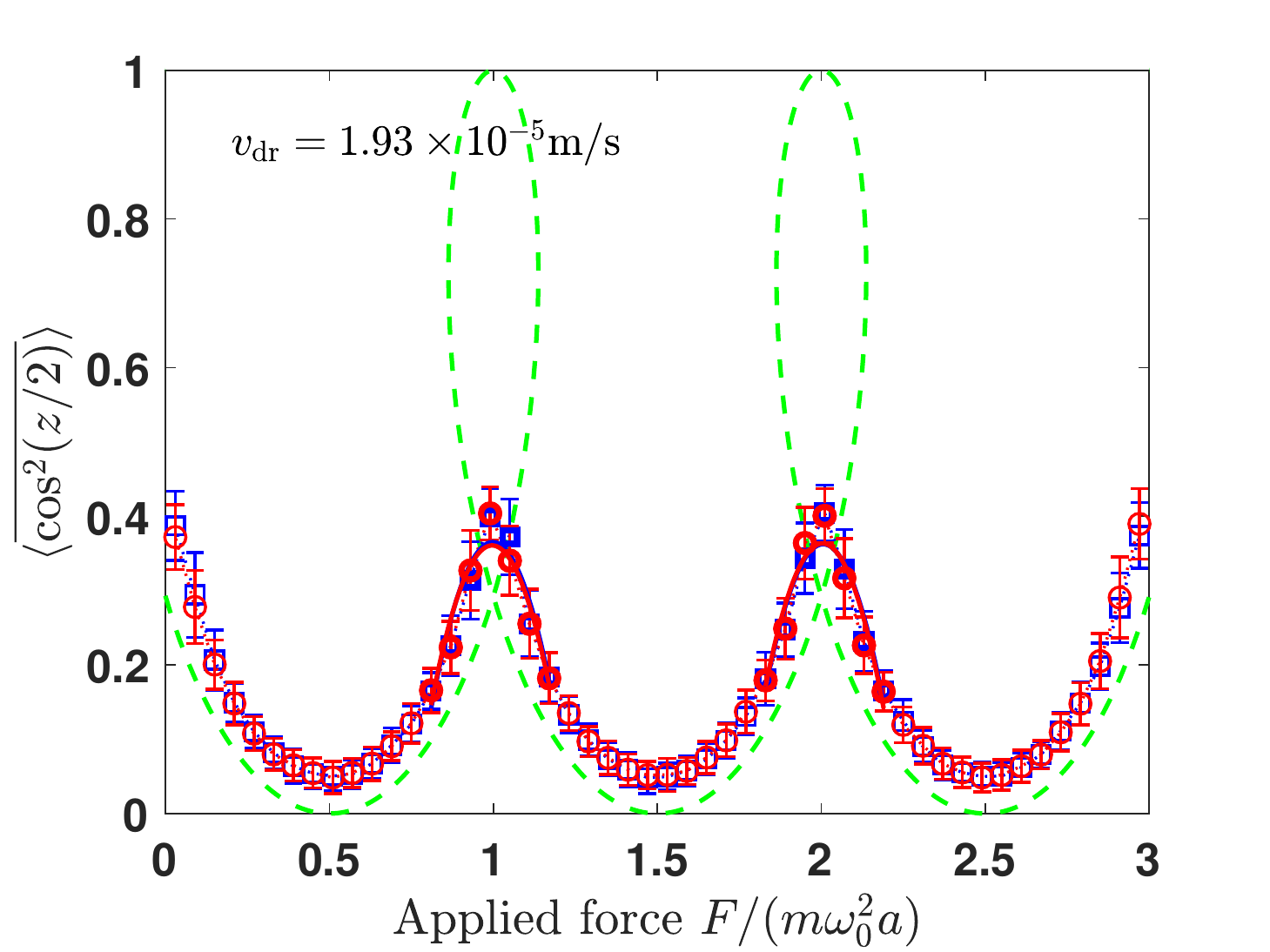}
 \includegraphics[width=5.73cm]{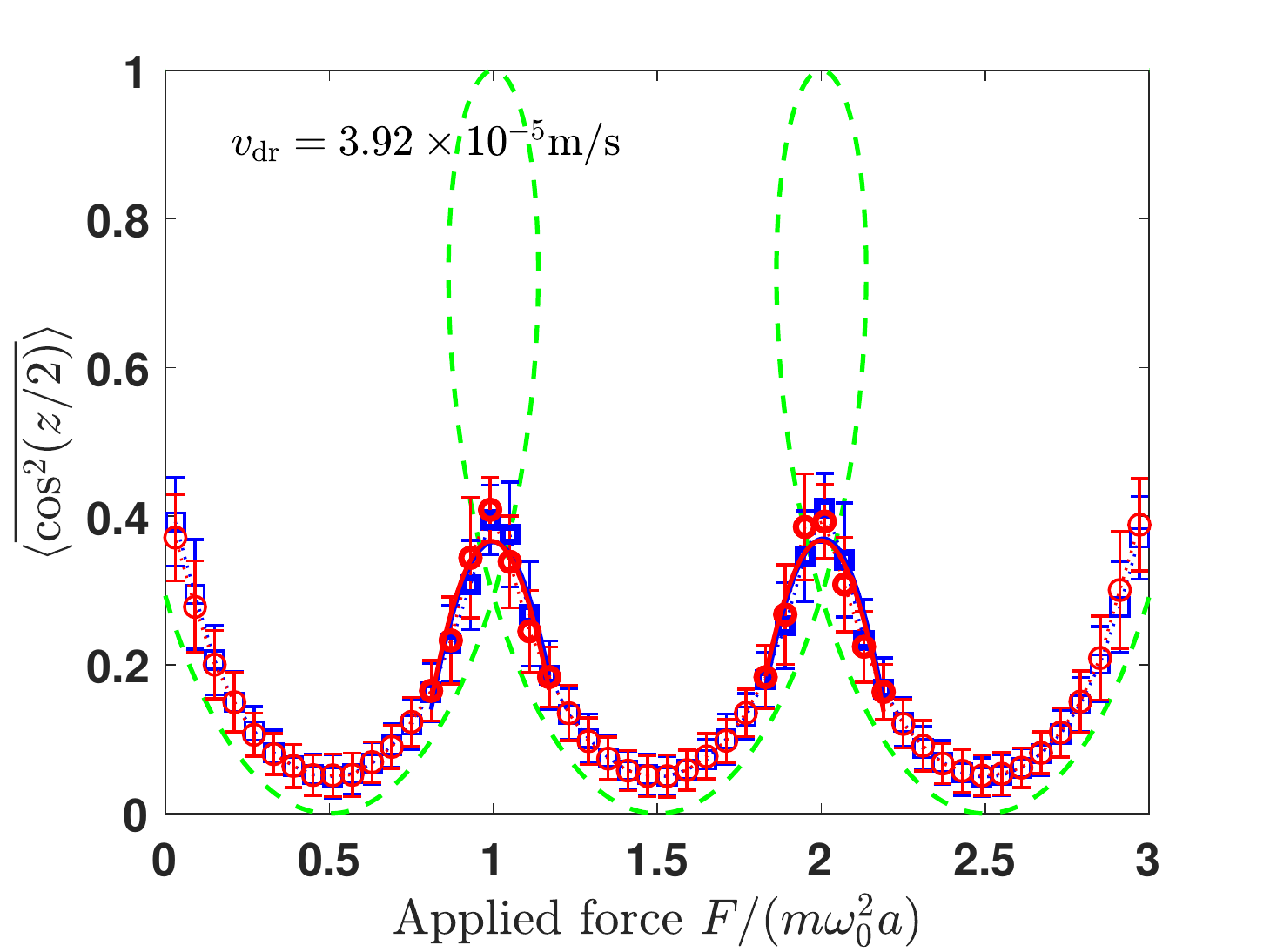}\\
\includegraphics[width=5.73cm]{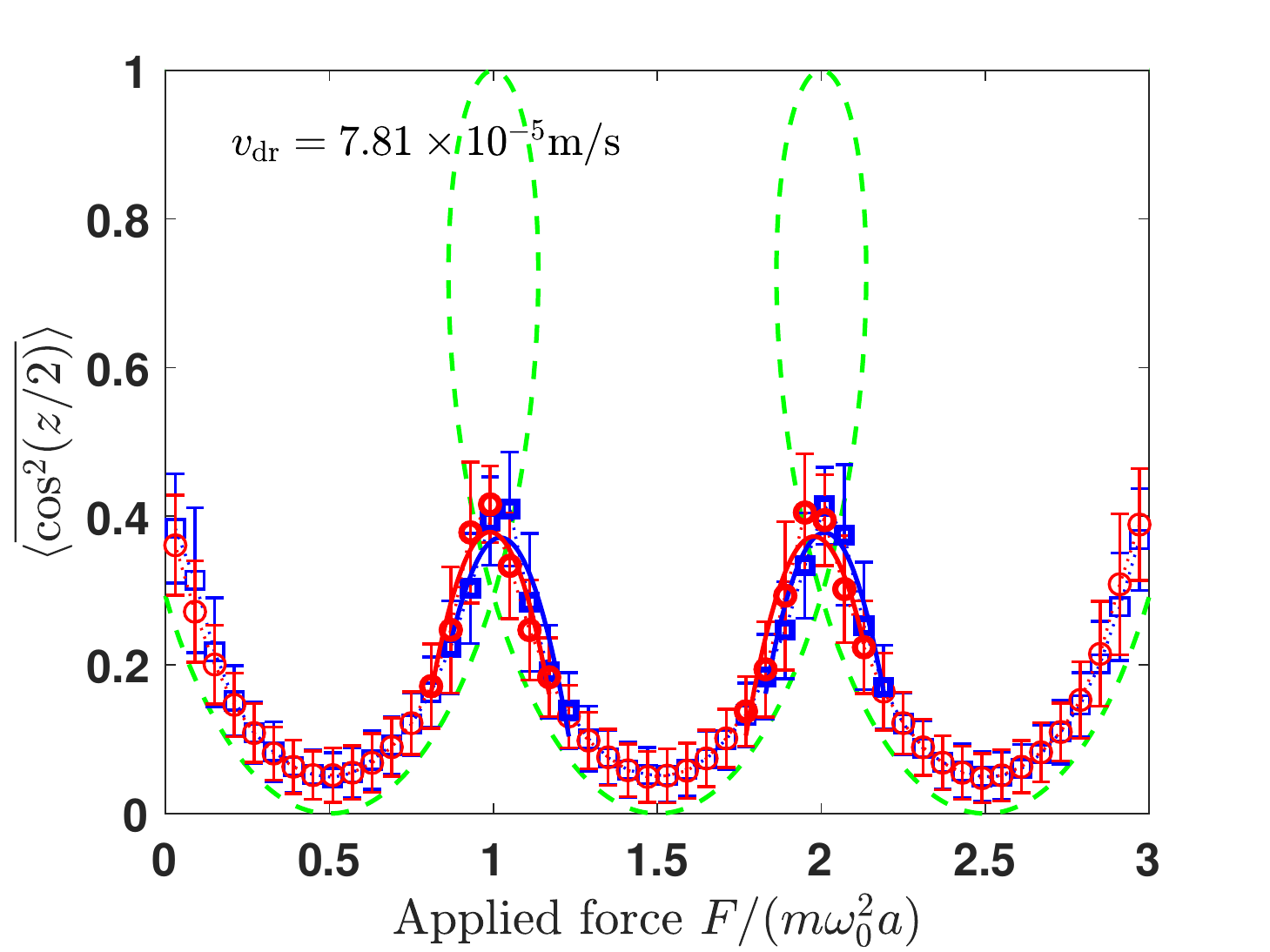}
\includegraphics[width=5.73cm]{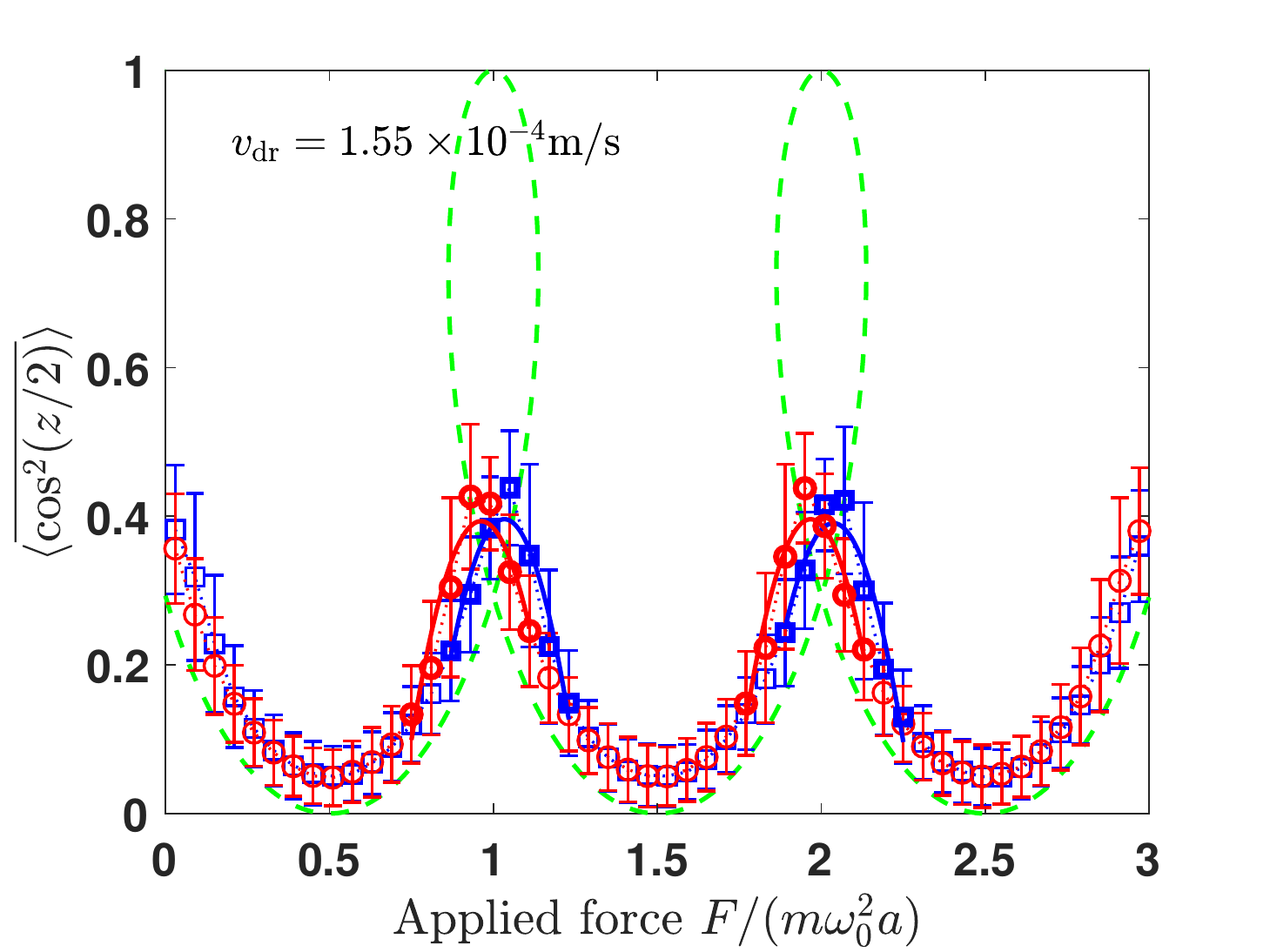}
\includegraphics[width=5.73cm]{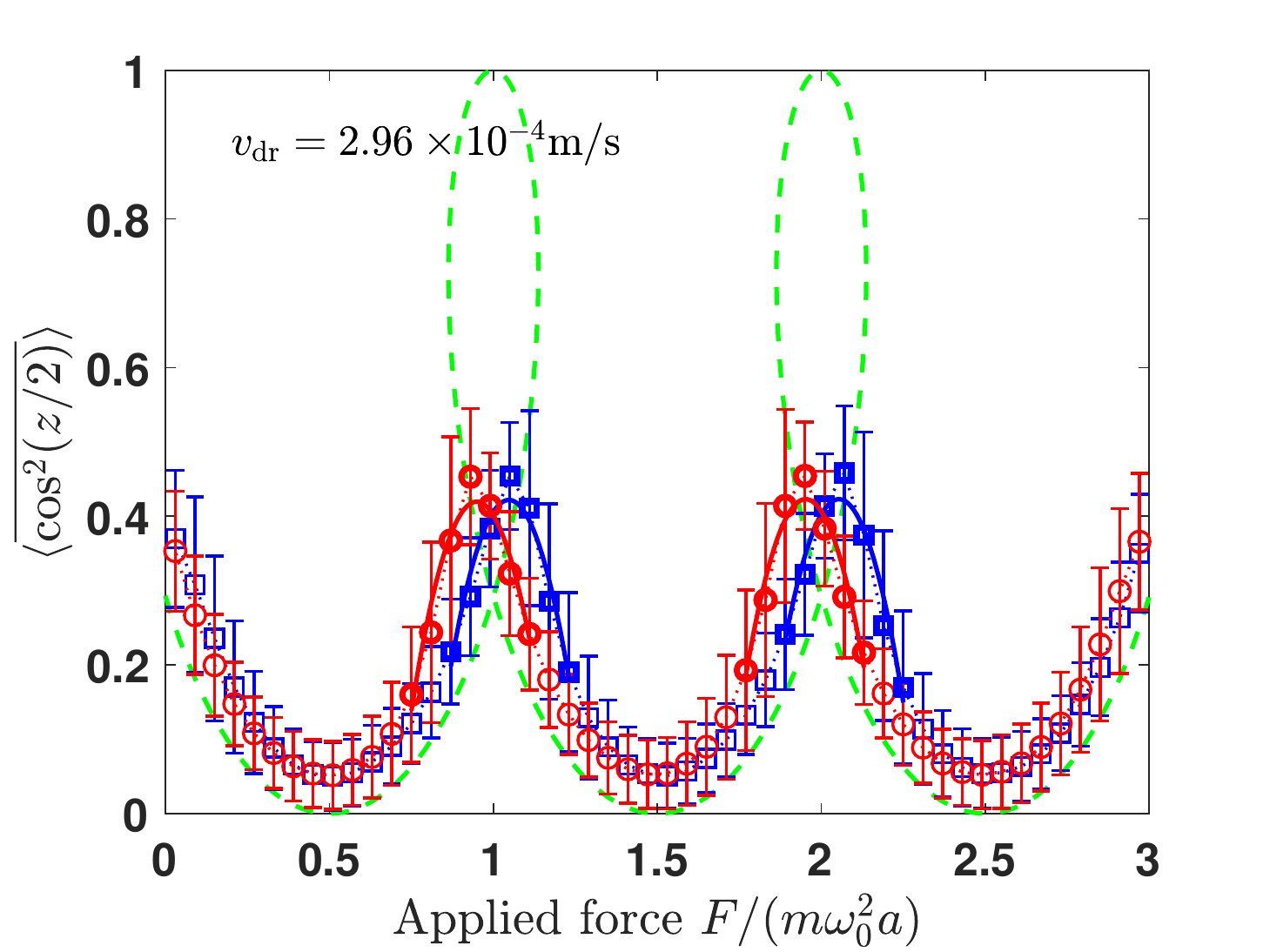}\\
\includegraphics[width=5.73cm]{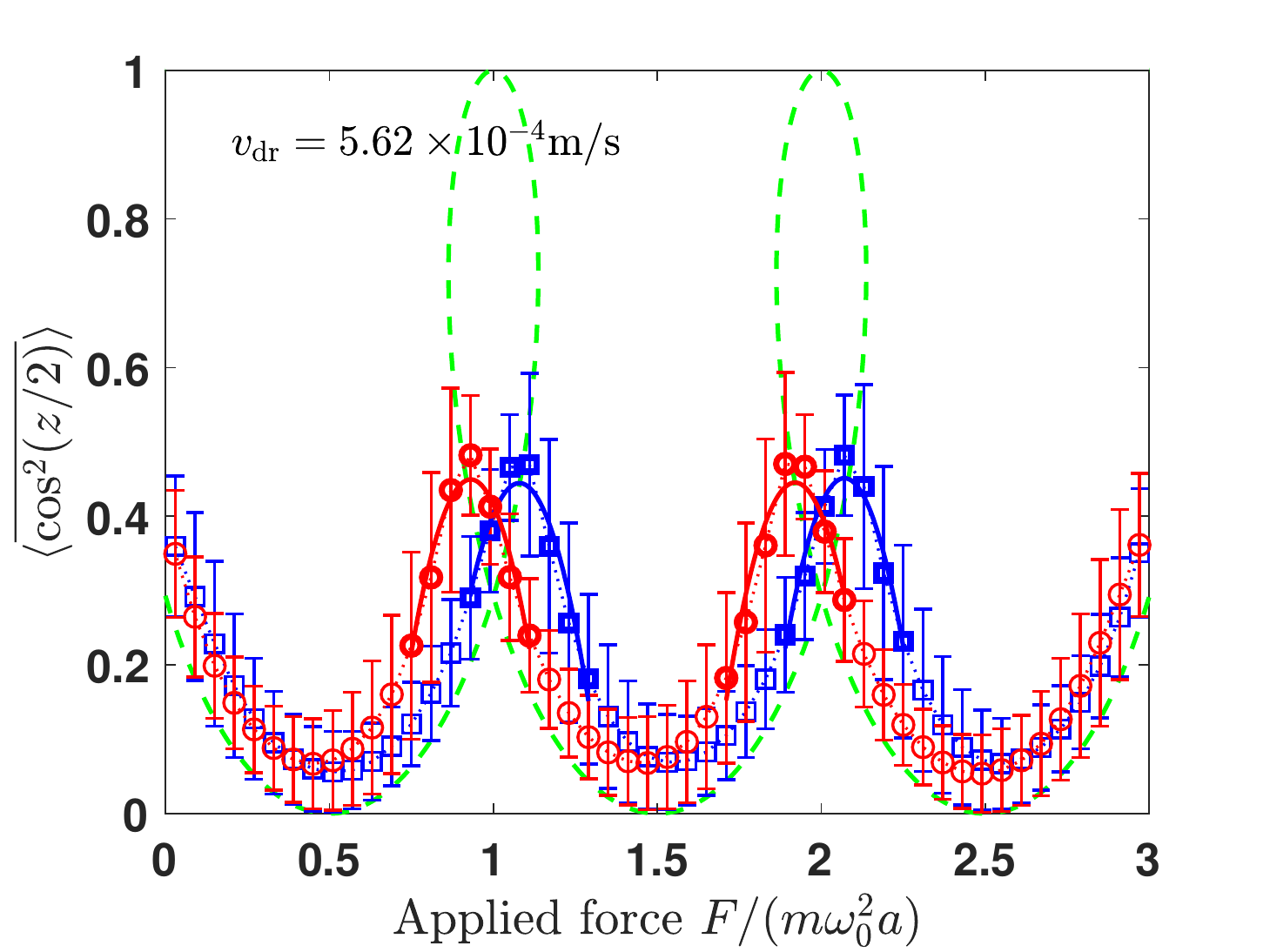}
\includegraphics[width=5.73cm]{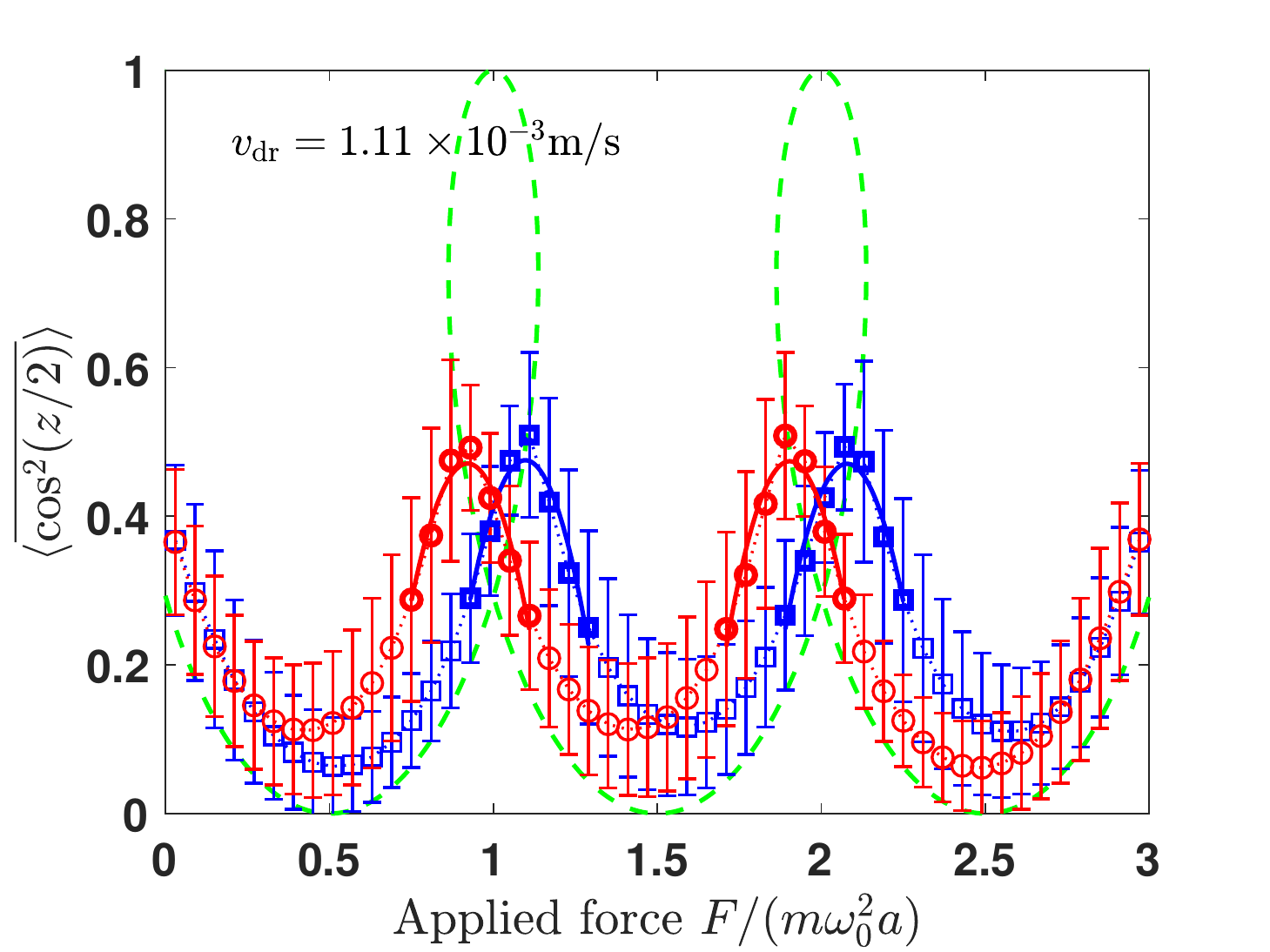}
\includegraphics[width=5.73cm]{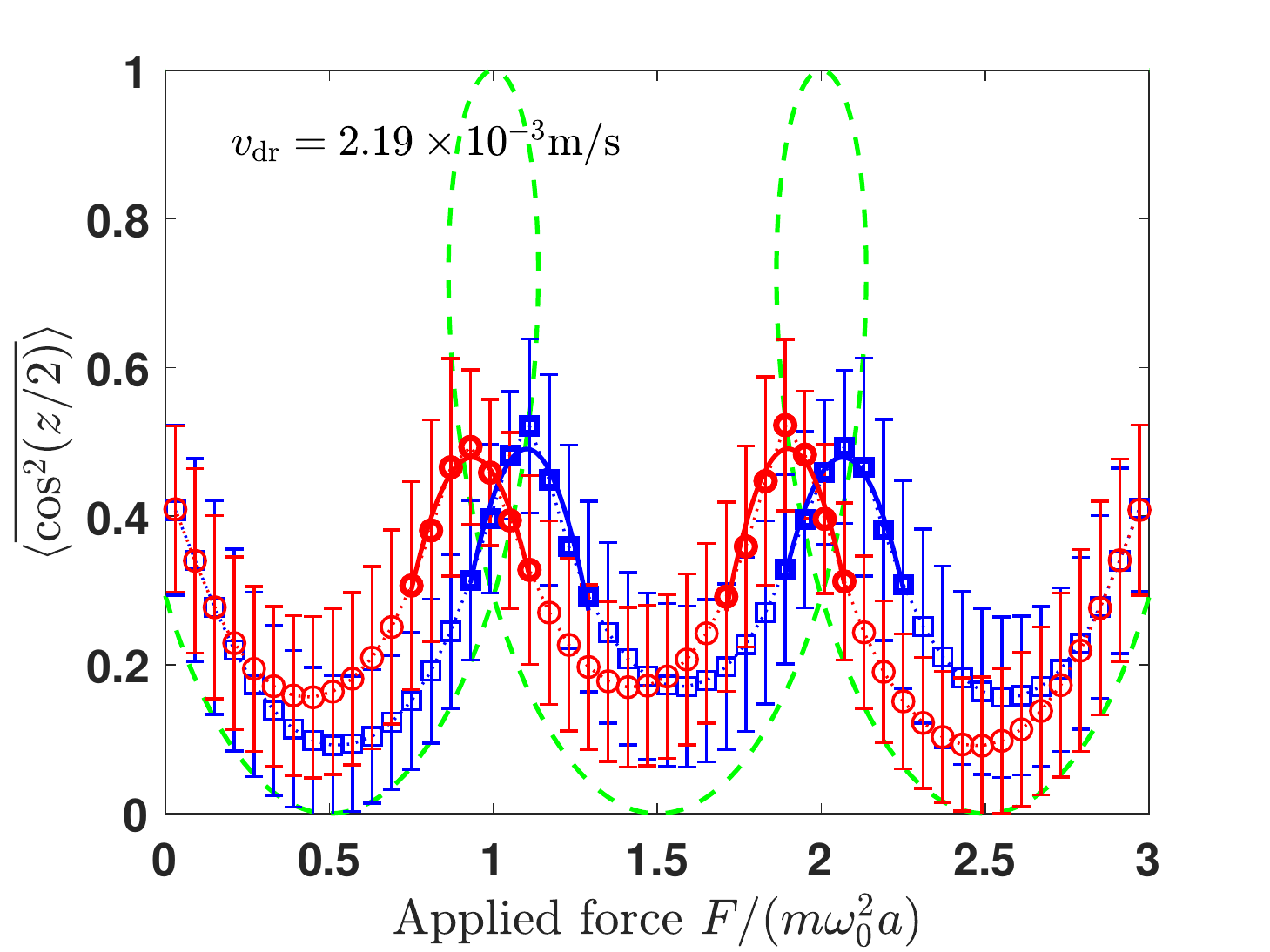}\\
\includegraphics[width=5.73cm]{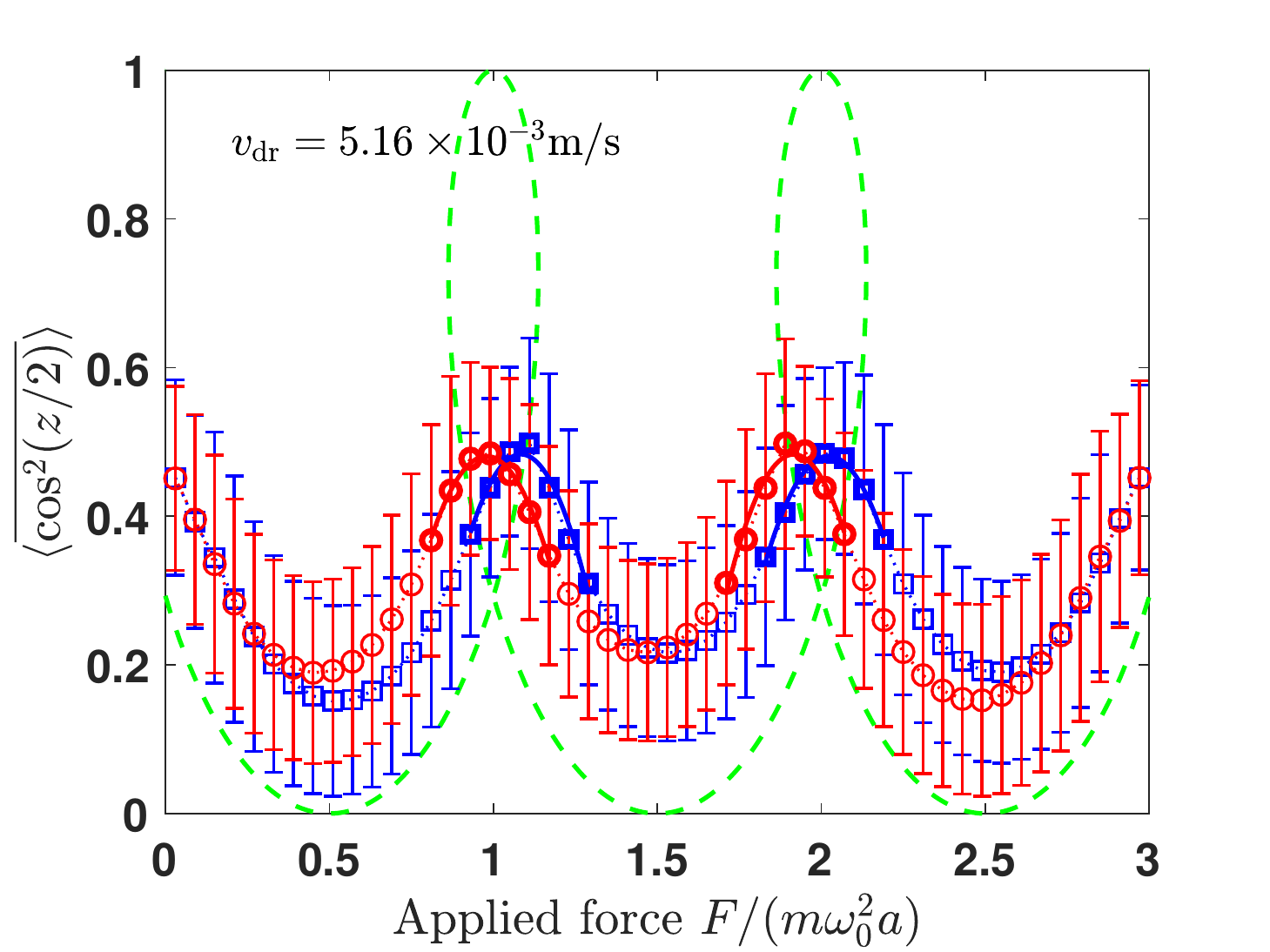}
\includegraphics[width=5.73cm]{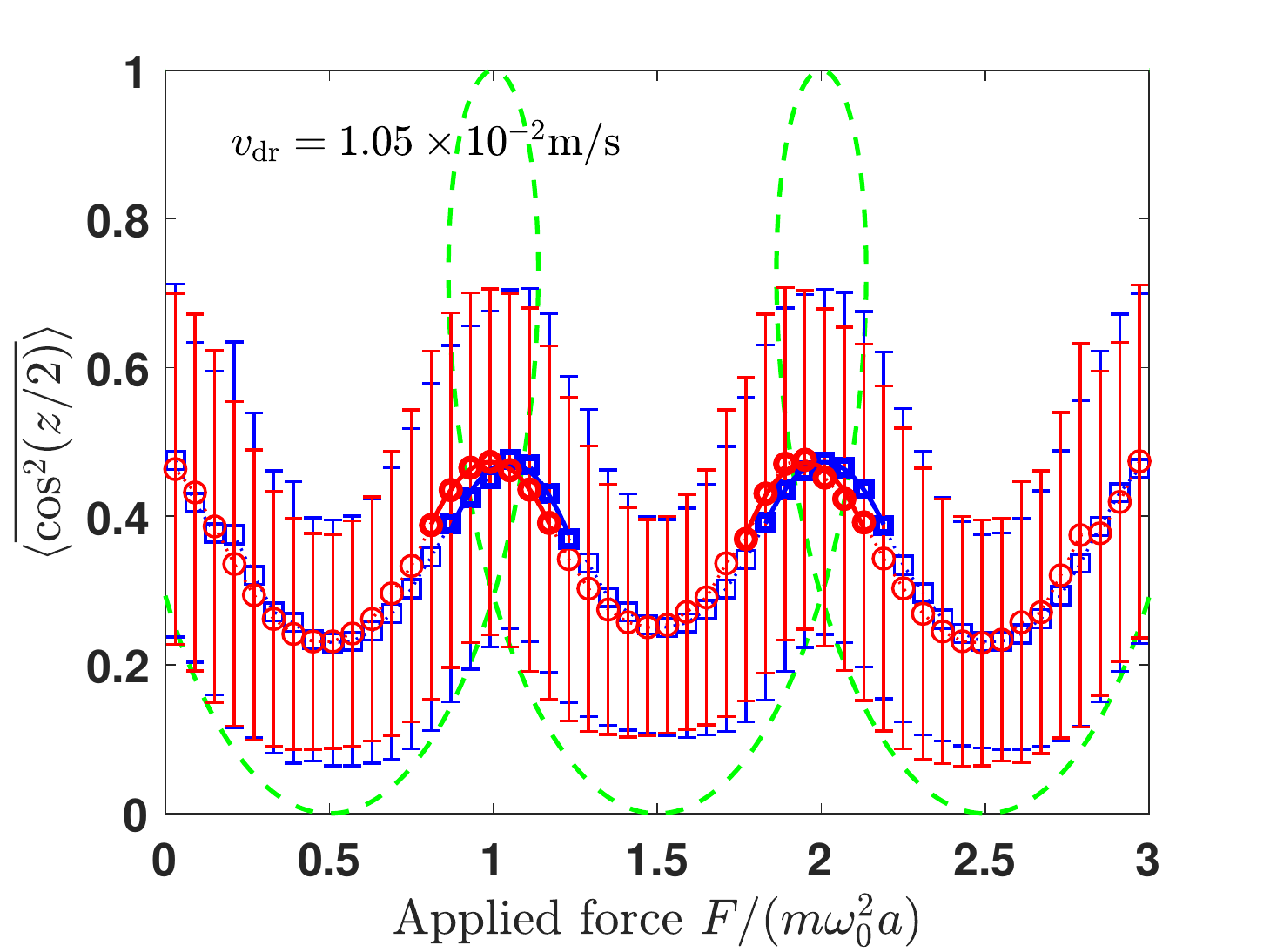}
\includegraphics[width=5.73cm]{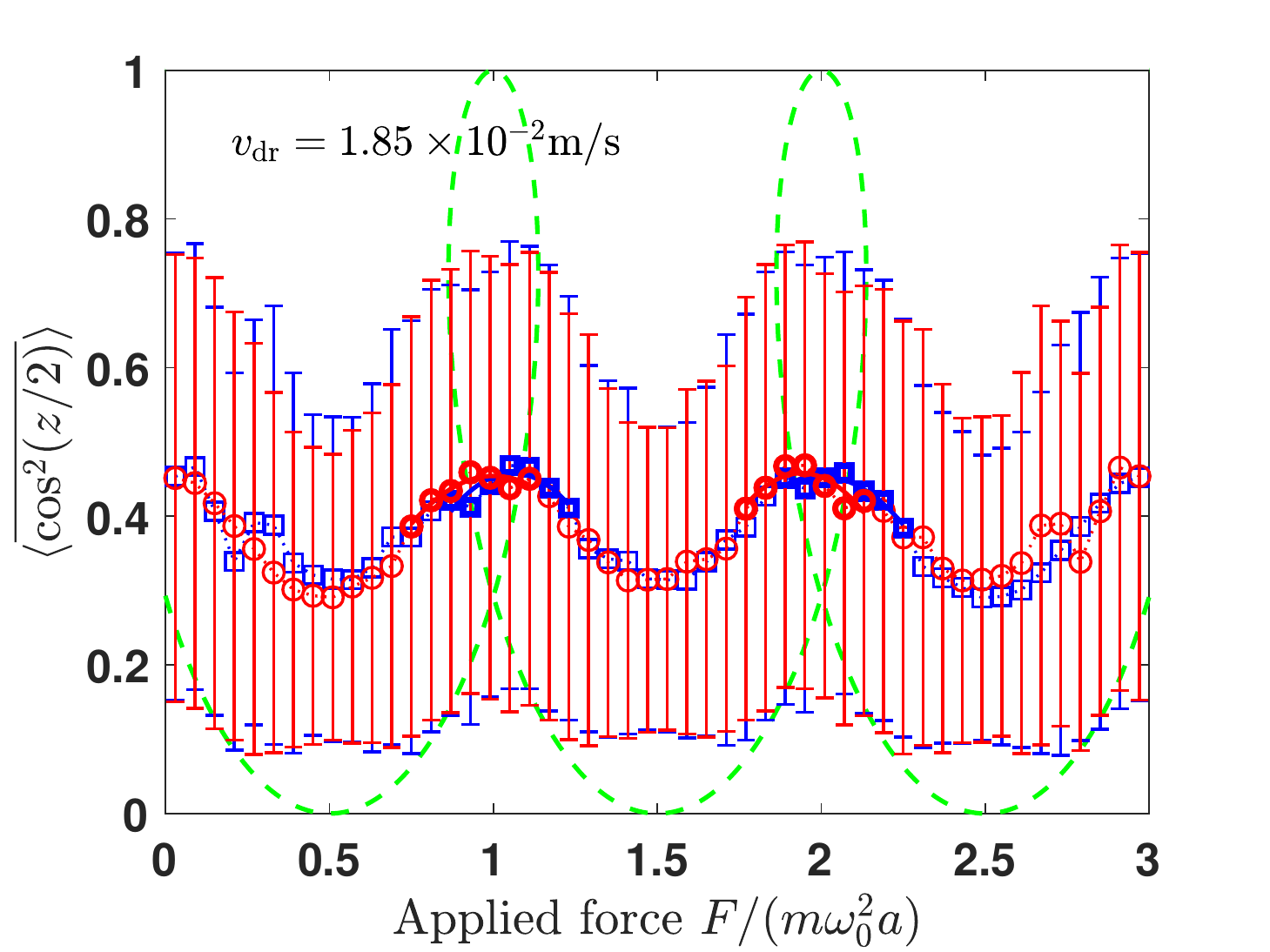}
 \caption{The simulation fluorescence counts at different driving velocity $v_{\rm dr}$ in the case of $\eta=2.2,\ \mu=2\times10^{4}\rm s^{-1},\ V_0=2\pi\hbar\times9.5\rm MHz$ and ${\it\Theta}=0.14$. The parameters are inherited from
%Figure 2 in \cite{NPVelocityTuning} and 
Figure 15-5 in \cite{bylinskiiphdthesis}. $\omega_0$ is determined by $\eta$ and $V_0$ with $\eta=\frac{2\pi^2V_0}{m\omega_0^2a^2}$ in which $m$ and $a$ are given in Table \ref{parameters}. The green dashed curves are $\cos^2(z^*/2)$ at the balanced points $z^*$ of the resultant potential energy with respect to the driver center's position $\tilde X(z^*)$, which is equal to the applied force $F/(m\omega_0^2a)$, cf. Figure \ref{fig:StickSlips}. The friction forces measured from different hysteresis loops in each subfigure are averaged to obtain the final friction force at each driving velocity.}
\label{ExperimentalDataVerify_eta2p2}
\end{figure}

In Figure \ref{ExperimentalDataVerify}, we plot the simulation friction force with respect to the driving velocity compared with the experimental results for the three cases in Figure 2 of \cite{NPVelocityTuning}. We can see that for the first and the third case in (A) and (C) respectively, the simulation results and the experimental results are close. For the second case, the simulation results are a little larger than the experimental results at low driving velocities. From Figure \ref{ExperimentalDataVerify_eta4p6_T004} we can see that at low driving velocities the measuring of the friction force is conservative, while in spite of this the simulation results are still larger. We note that in \cite{NPVelocityTuning} the authors correct the temperature in their simulation and anlytical models [see the supplementary information of \cite{NPVelocityTuning}], while we  introduce no correction. Therefore our simulation results agree quantitatively or at least semi-quantitatively with the experimental results and we can conclude that our numerical method is predictable for practical problems. 

\begin{table}[H]
\centering
\caption{Number of simulation loops  for Figure \ref{ExperimentalDataVerify_eta2p2}}
\label{tab:ExperimentalDataVerify_eta2p2}
\begin{tabular}{lrlrlr}
$v_{\rm dr}(\rm m/s)$ & Number of loops & $v_{\rm dr}(\rm m/s)$ & Number of loops & $v_{\rm dr}(\rm m/s)$ & Number of loops \\
\midrule
1.18549201434396e-06 & 30 & 2.34735773684298e-06 & 30 & 4.75054129690064e-06 & 48\\
9.61785513478092e-06 & 100 & 1.92947430887720e-05 & 200 & 3.92472869445721e-05 & 400\\
7.81396573846716e-05 & 790 & 0.000154898829332650 & 1200 & 0.000295777367511797 & 3000\\
0.000562210180376583 & 5700 & 0.00111396132026529 & 12000 & 0.00218933180613166 & 22000\\
0.00515587947242487 & 57000 & 0.0105075865958613 & 110000 & 0.0185006051024171 & 190000\\
\bottomrule
\end{tabular}
\end{table}

\begin{table}[H]
\centering
\caption{Number of simulation loops  for Figure \ref{ExperimentalDataVerify_eta4p6_T004}}
\label{tab:ExperimentalDataVerify_eta4p6_T004}
\begin{tabular}{lrlrlr}
$v_{\rm dr}(\rm m/s)$ & Number of loops & $v_{\rm dr}(\rm m/s)$ & Number of loops & $v_{\rm dr}(\rm m/s)$ & Number of loops \\
\midrule
5.48330003252868e-07 & 15 & 1.09148805346082e-06 & 30 & 2.10838483259046e-06 & 30\\
4.23341275737471e-06 & 43 & 8.60114892152780e-06 & 87 & 1.68661013327135e-05 & 170\\
3.49433799327154e-05 & 350 & 6.86477082436799e-05 & 690 & 0.000134028433498140 & 1400\\
0.000271253909318031 & 2800 & 0.000548809185400129 & 5500 & 0.00109233256749753 & 11000\\
0.00262136175653125 & 27000 & 0.00524259937197039 & 53000 & 0.0101243891976080 & 110000\\
0.0192306183601816 & 200000 & & & & \\
\bottomrule
\end{tabular}
\end{table}

\begin{figure}[H]
 \centering
 \includegraphics[width=5.73cm]{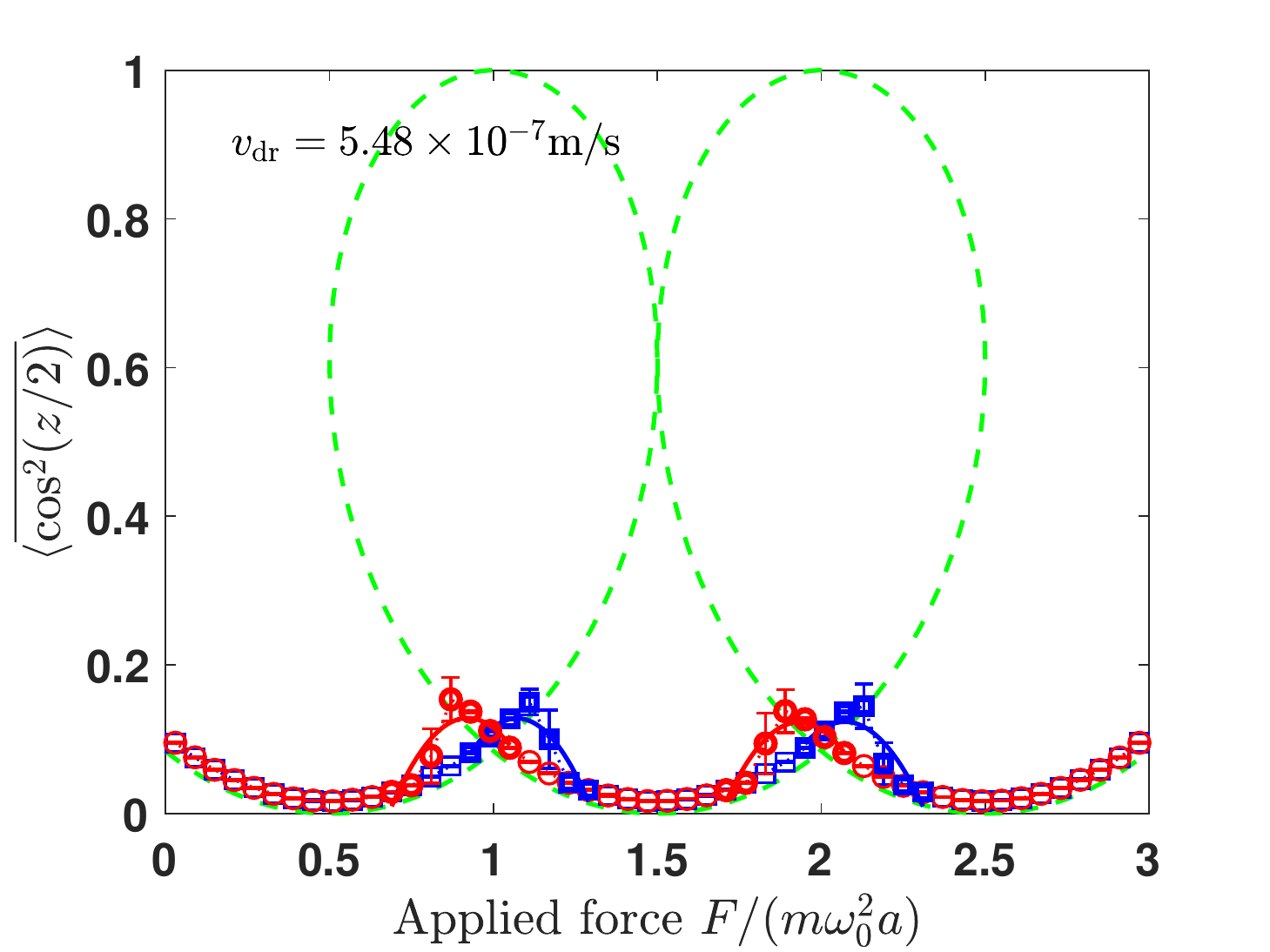}
 \includegraphics[width=5.73cm]{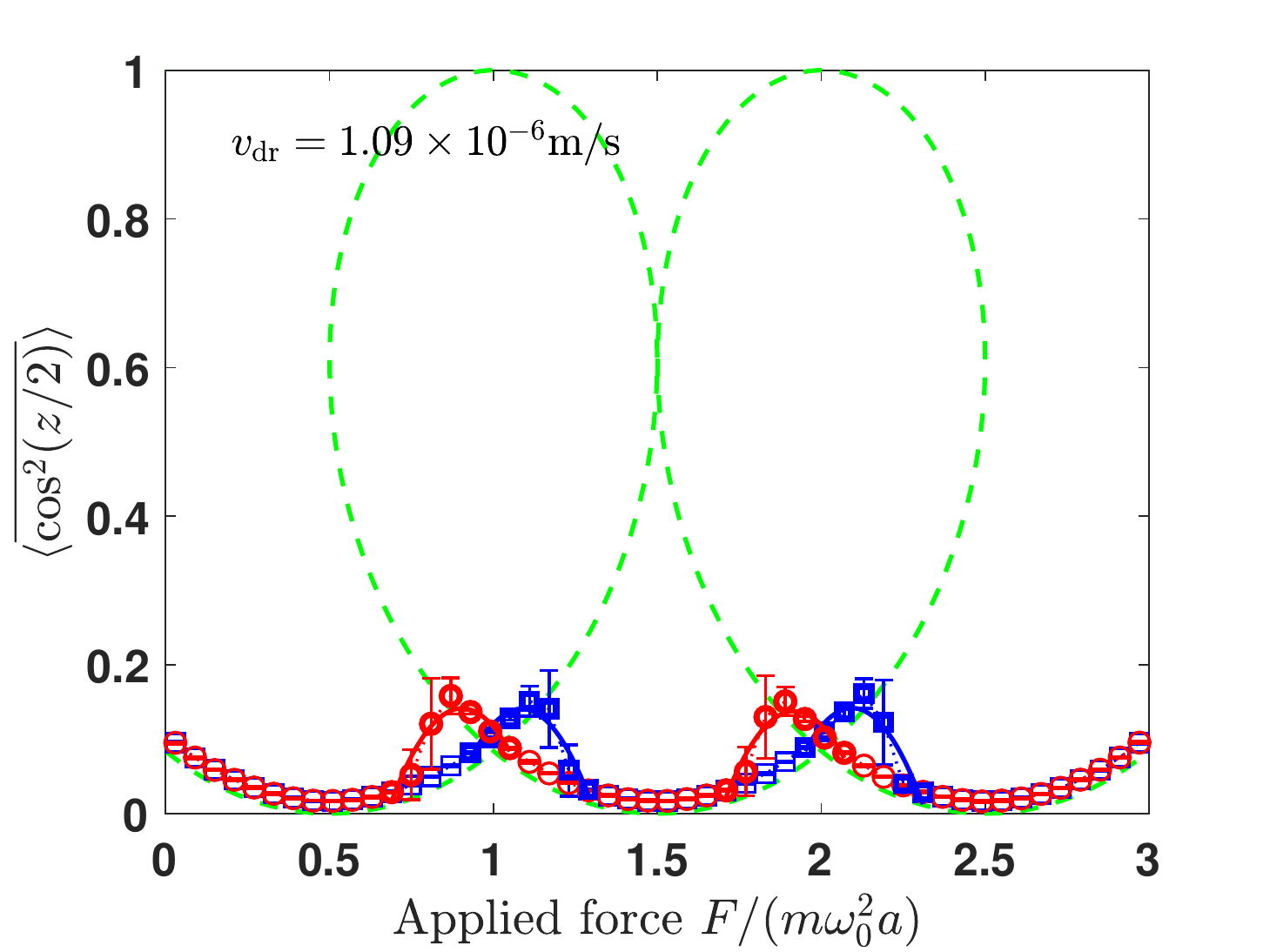}
 \includegraphics[width=5.73cm]{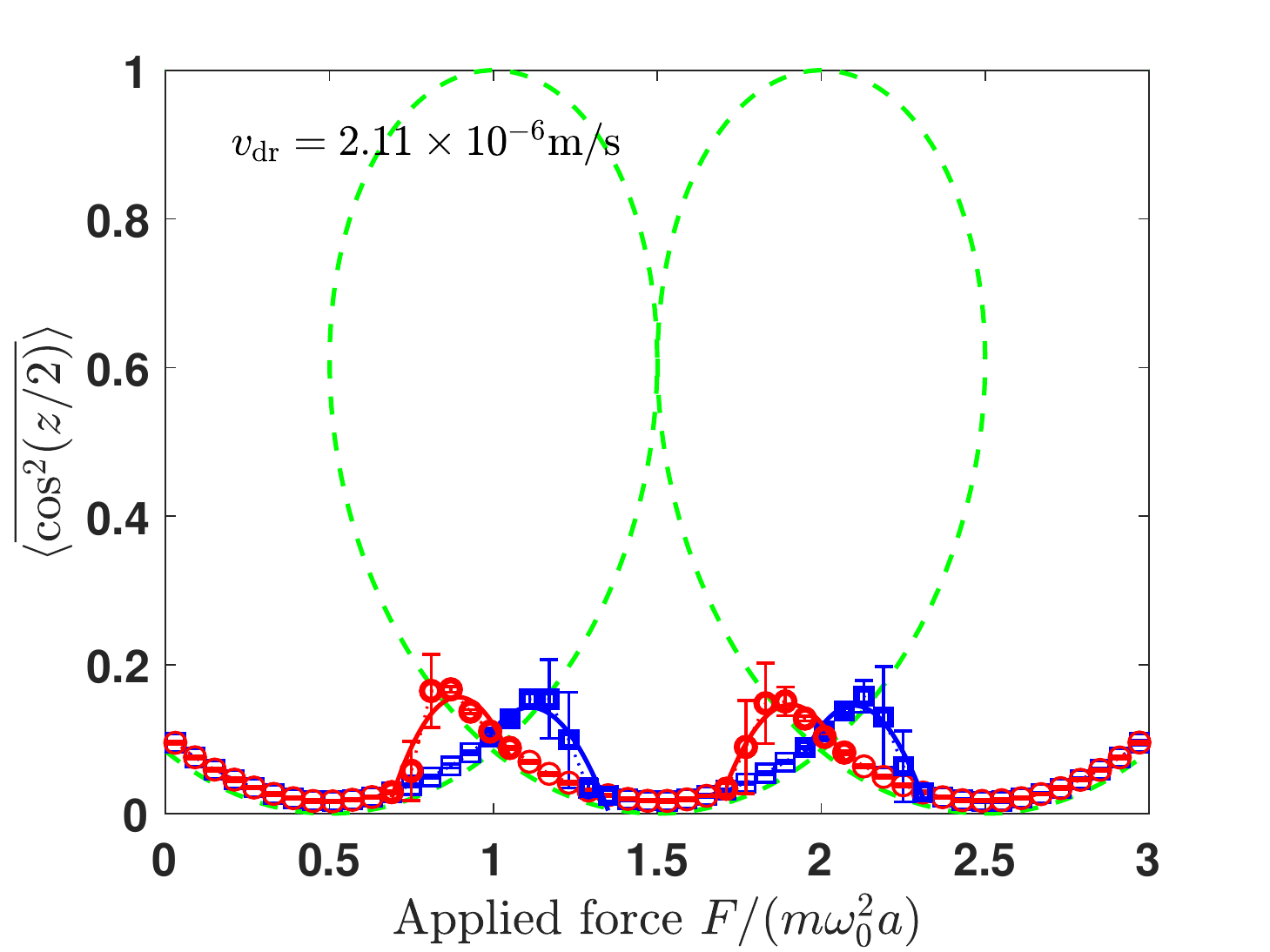}\\
 \includegraphics[width=5.73cm]{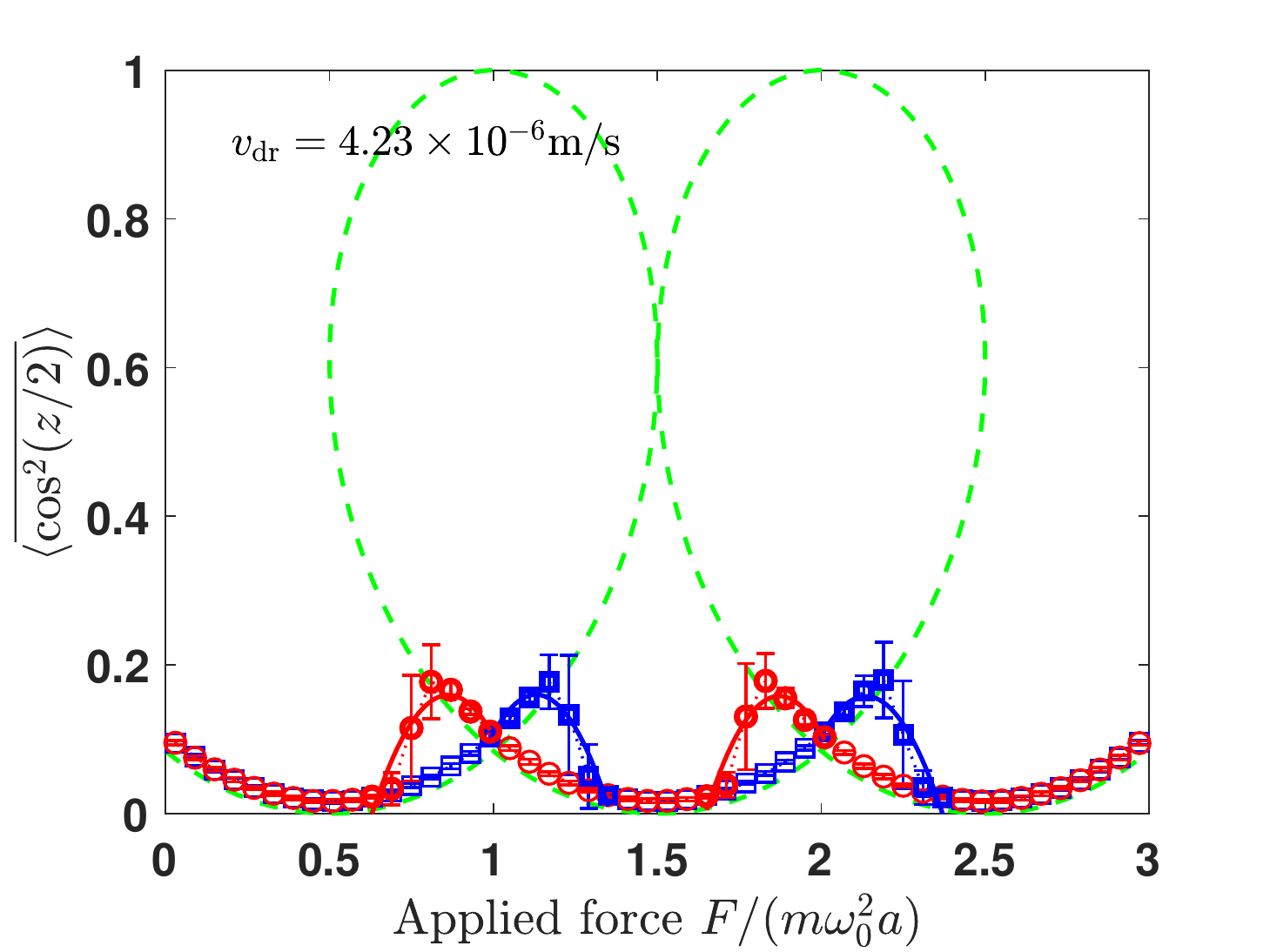}
 \includegraphics[width=5.73cm]{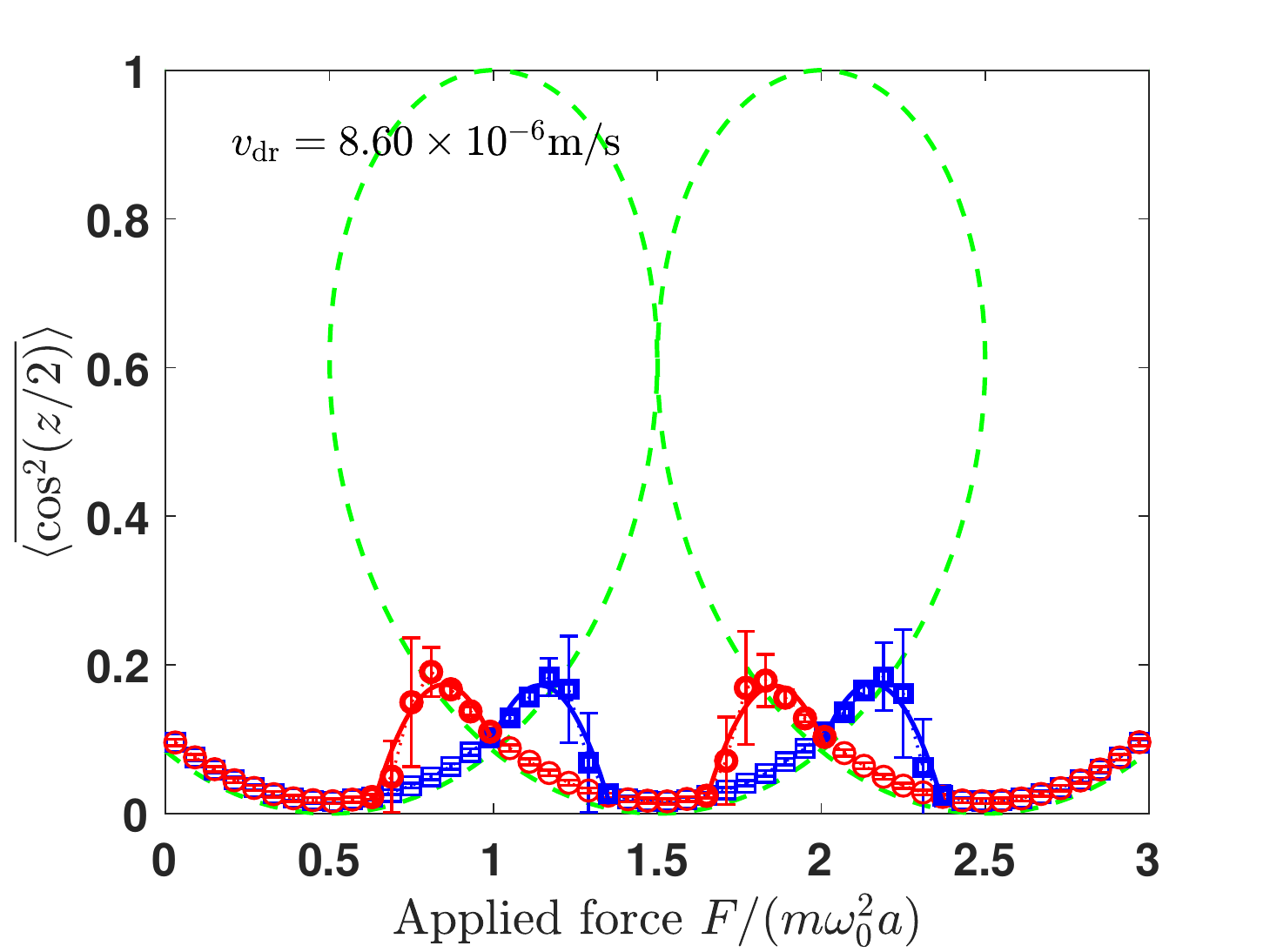}
 \includegraphics[width=5.73cm]{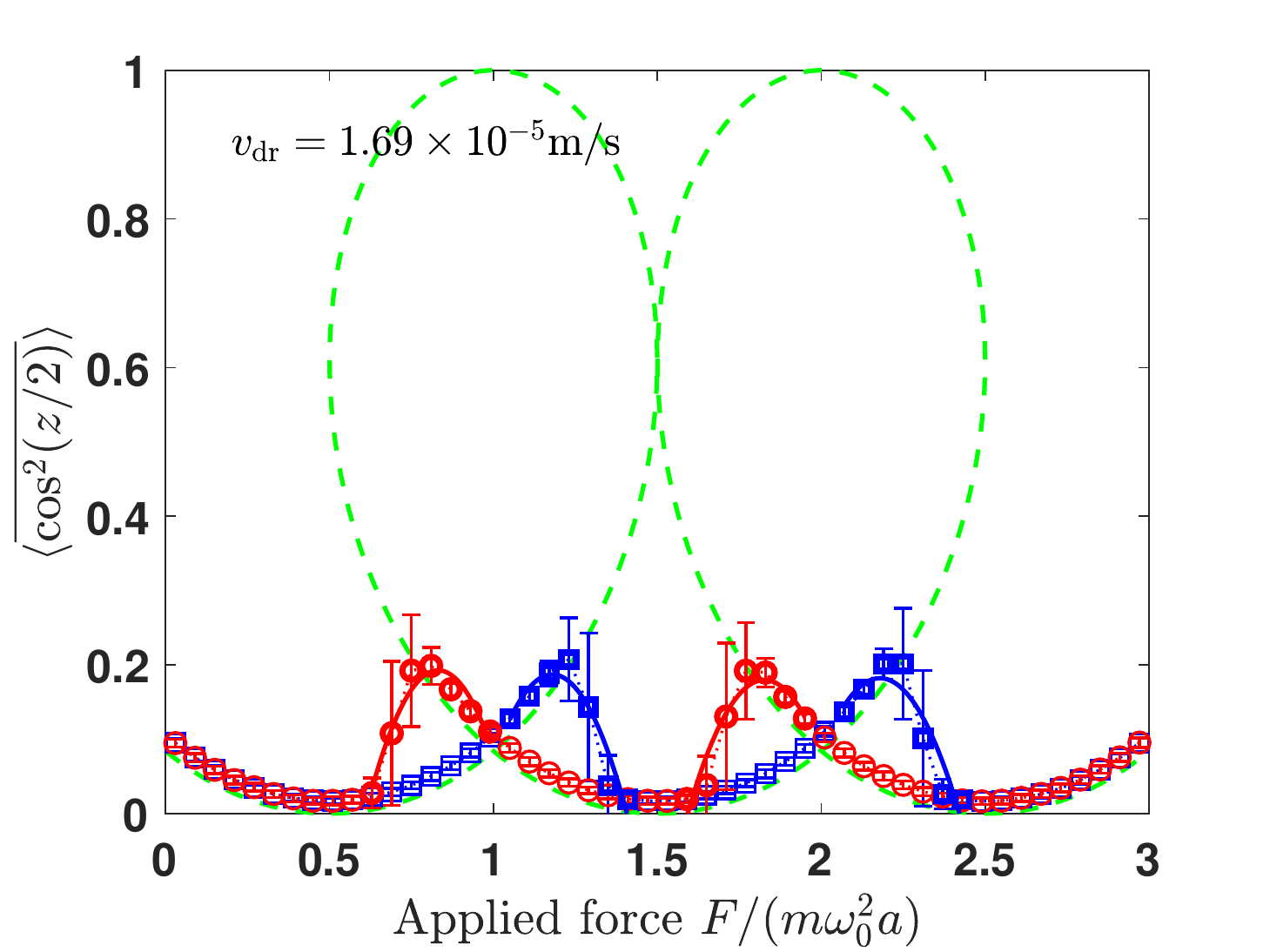}\\
\includegraphics[width=5.73cm]{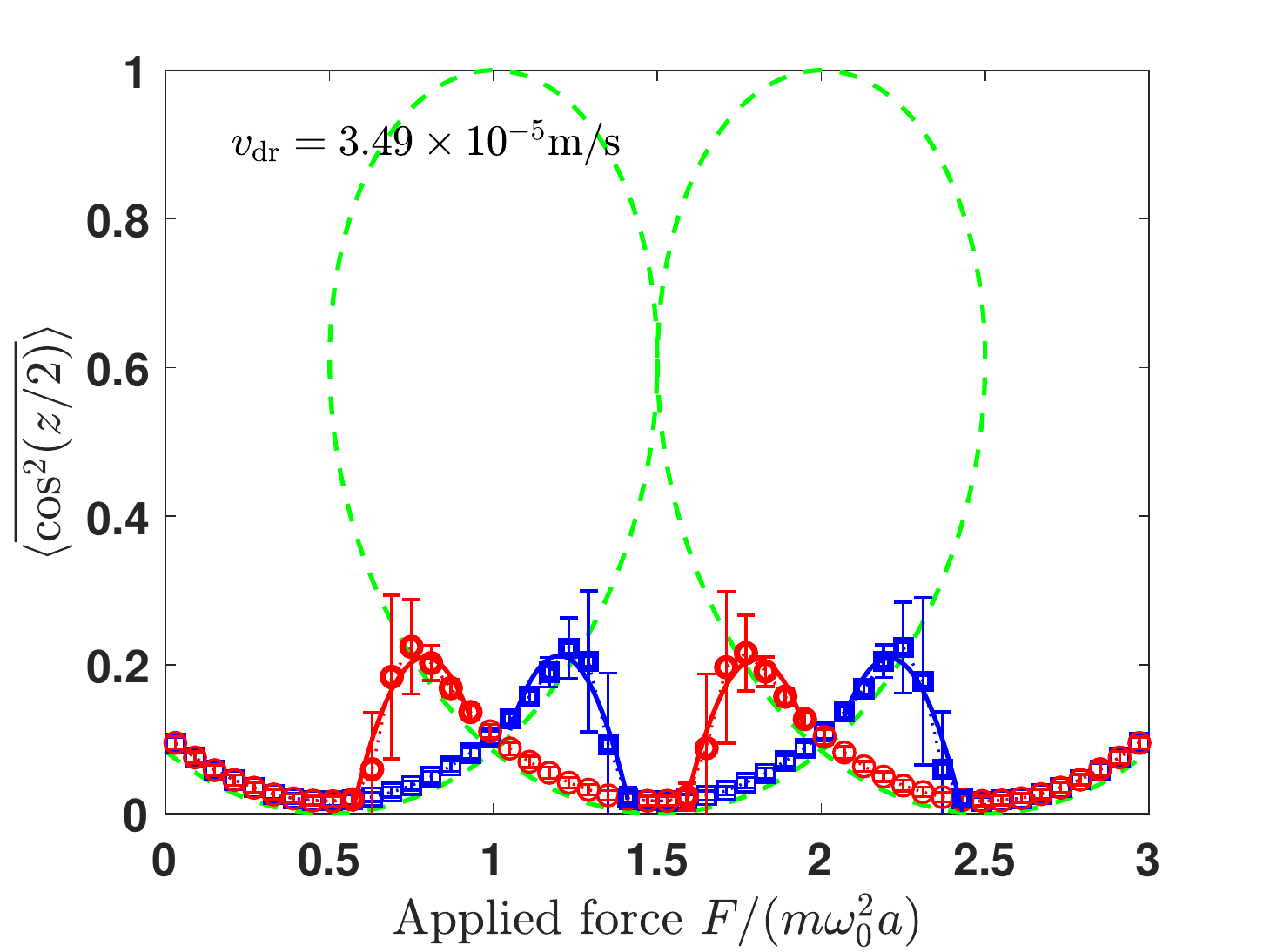}
\includegraphics[width=5.73cm]{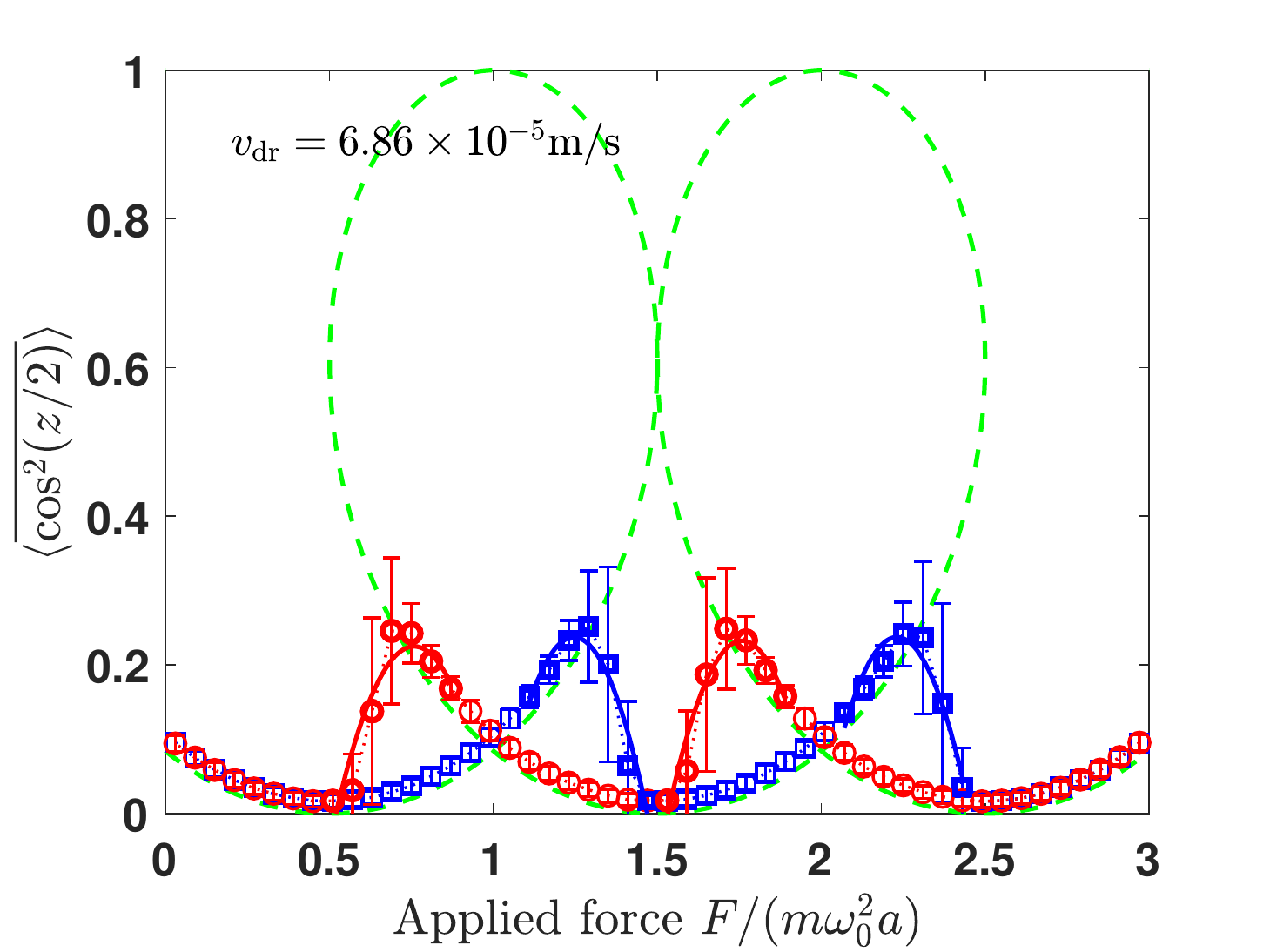}
\includegraphics[width=5.73cm]{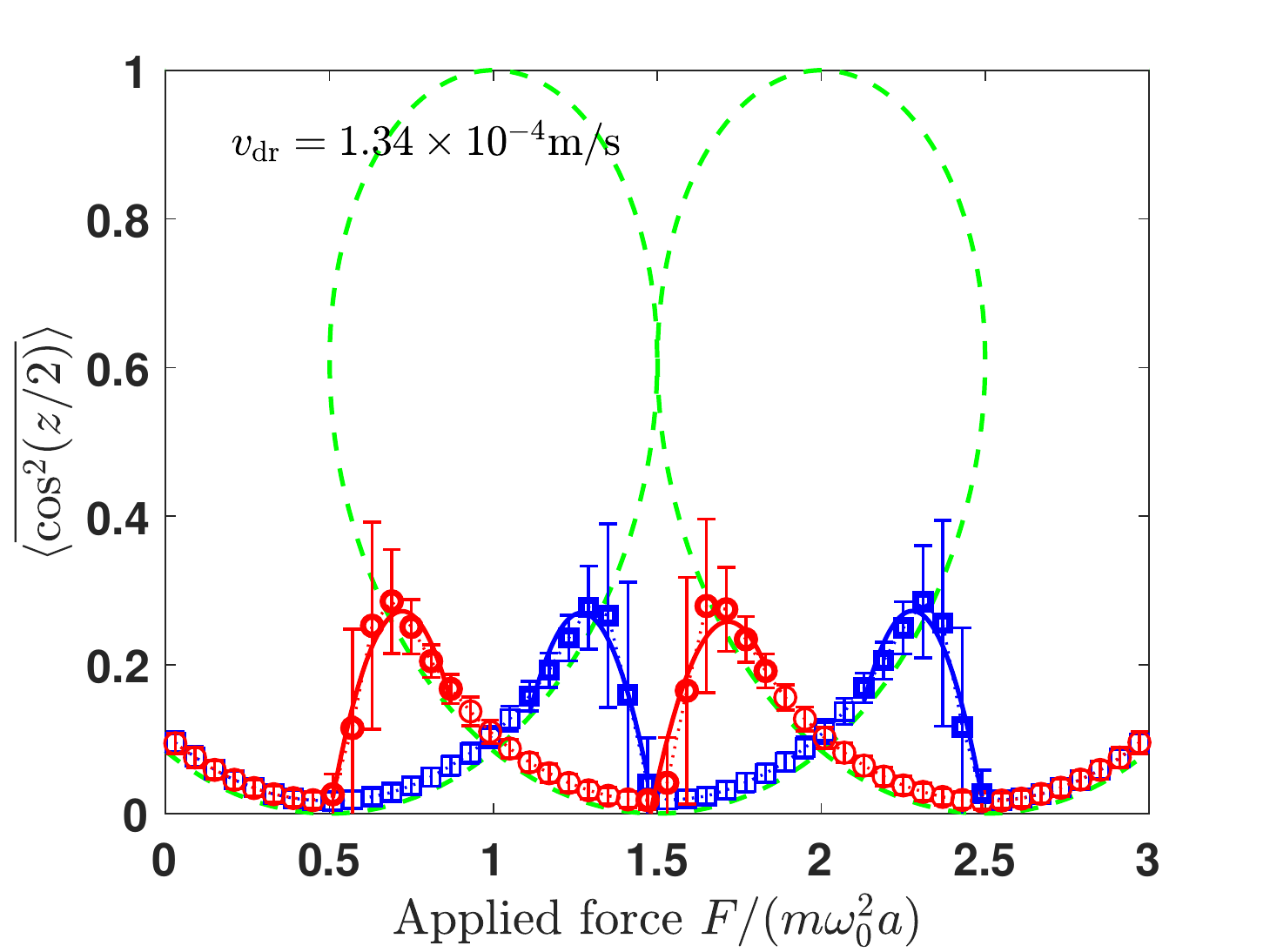}\\
\includegraphics[width=5.73cm]{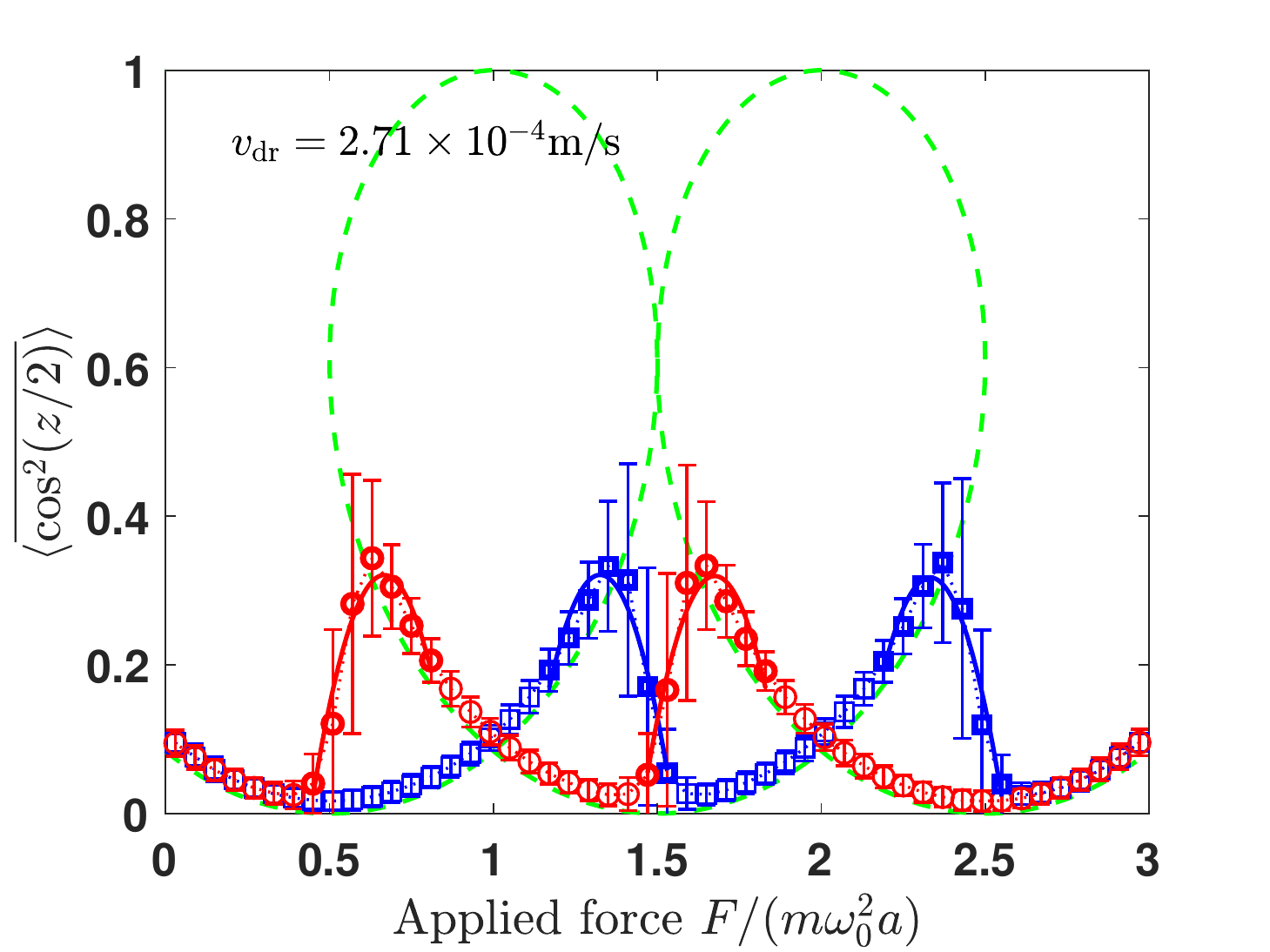}
\includegraphics[width=5.73cm]{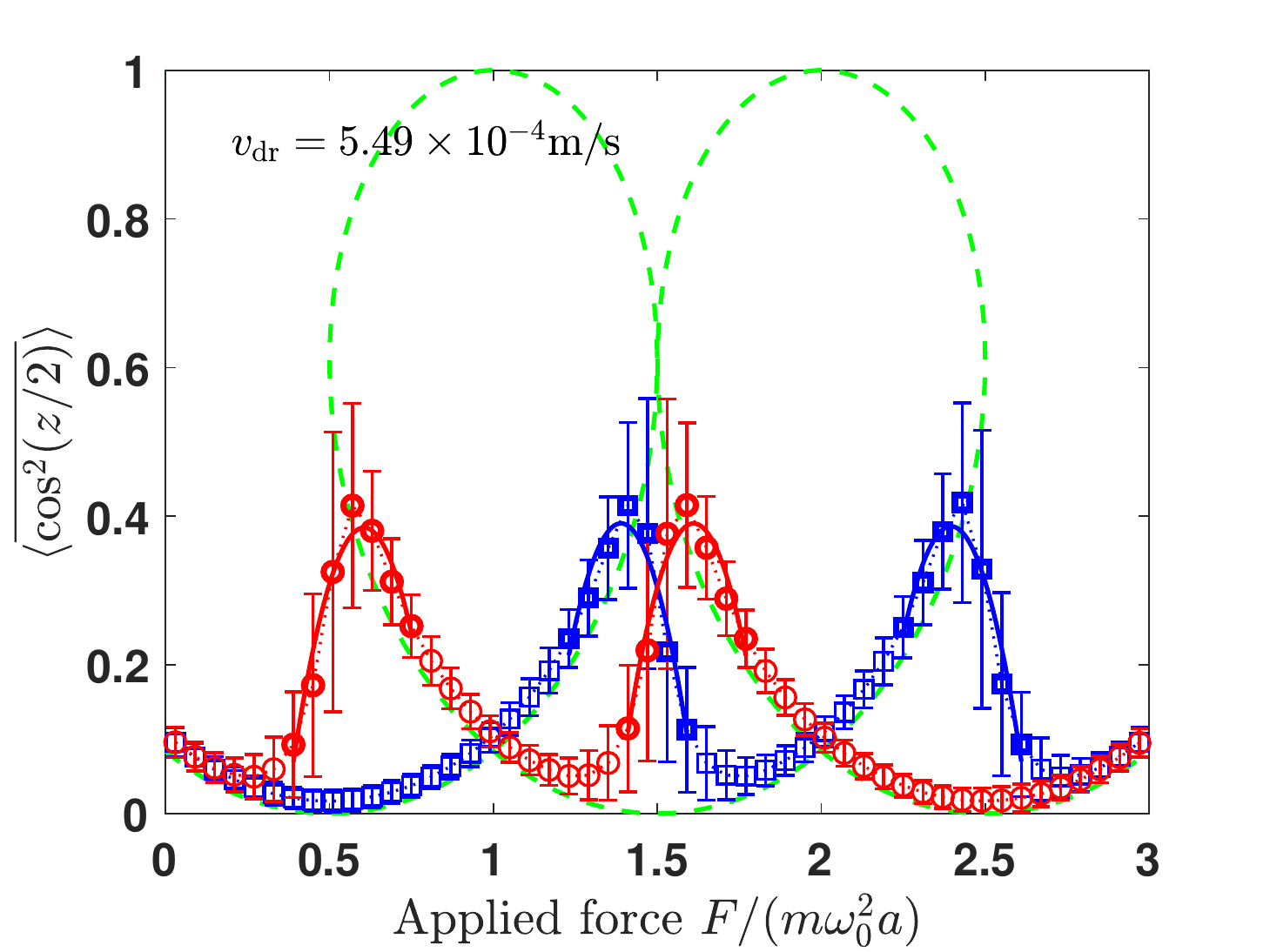}
\includegraphics[width=5.73cm]{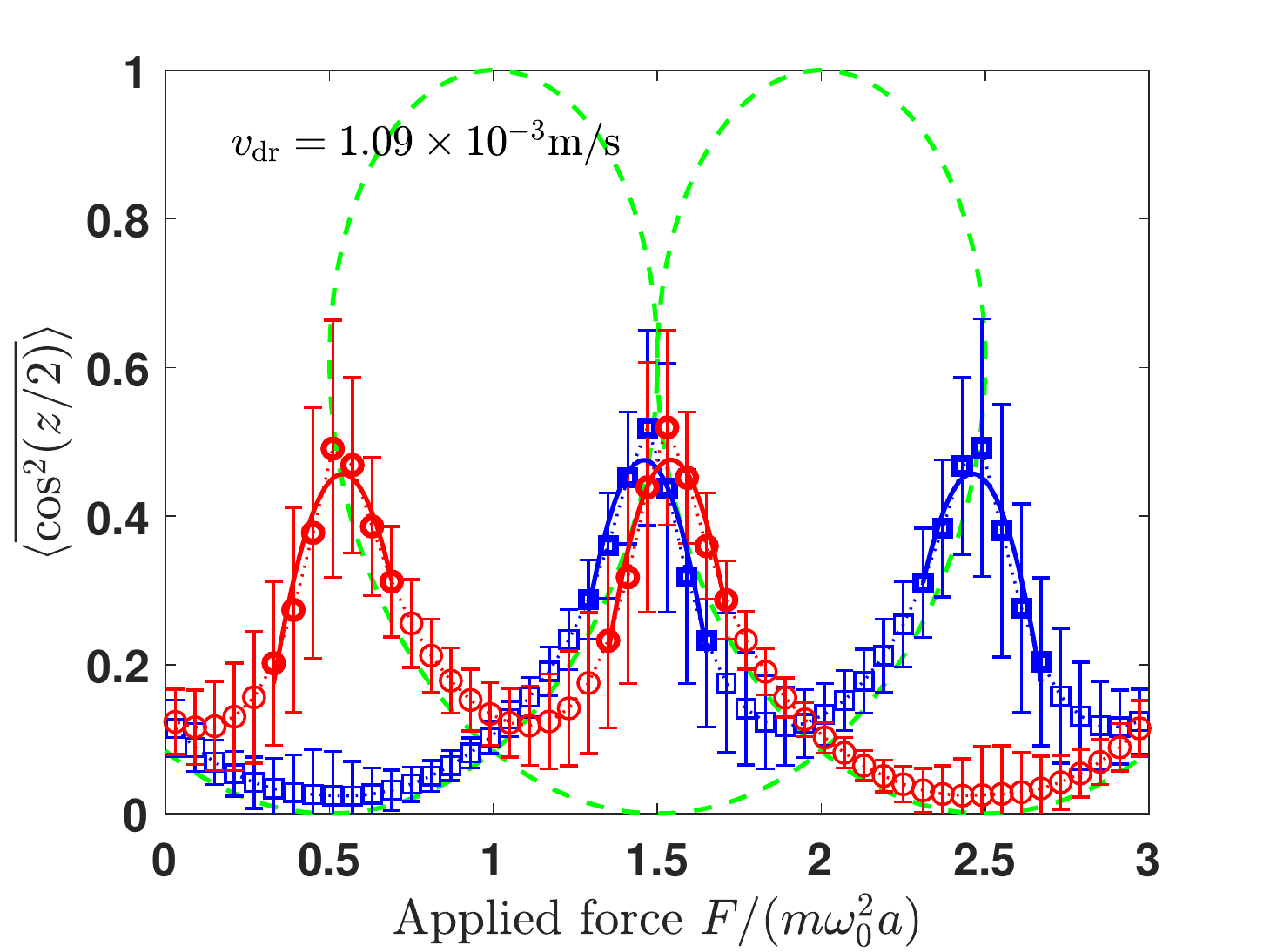}\\
\end{figure}
\begin{figure}[H]
 \centering
\includegraphics[width=5.73cm]{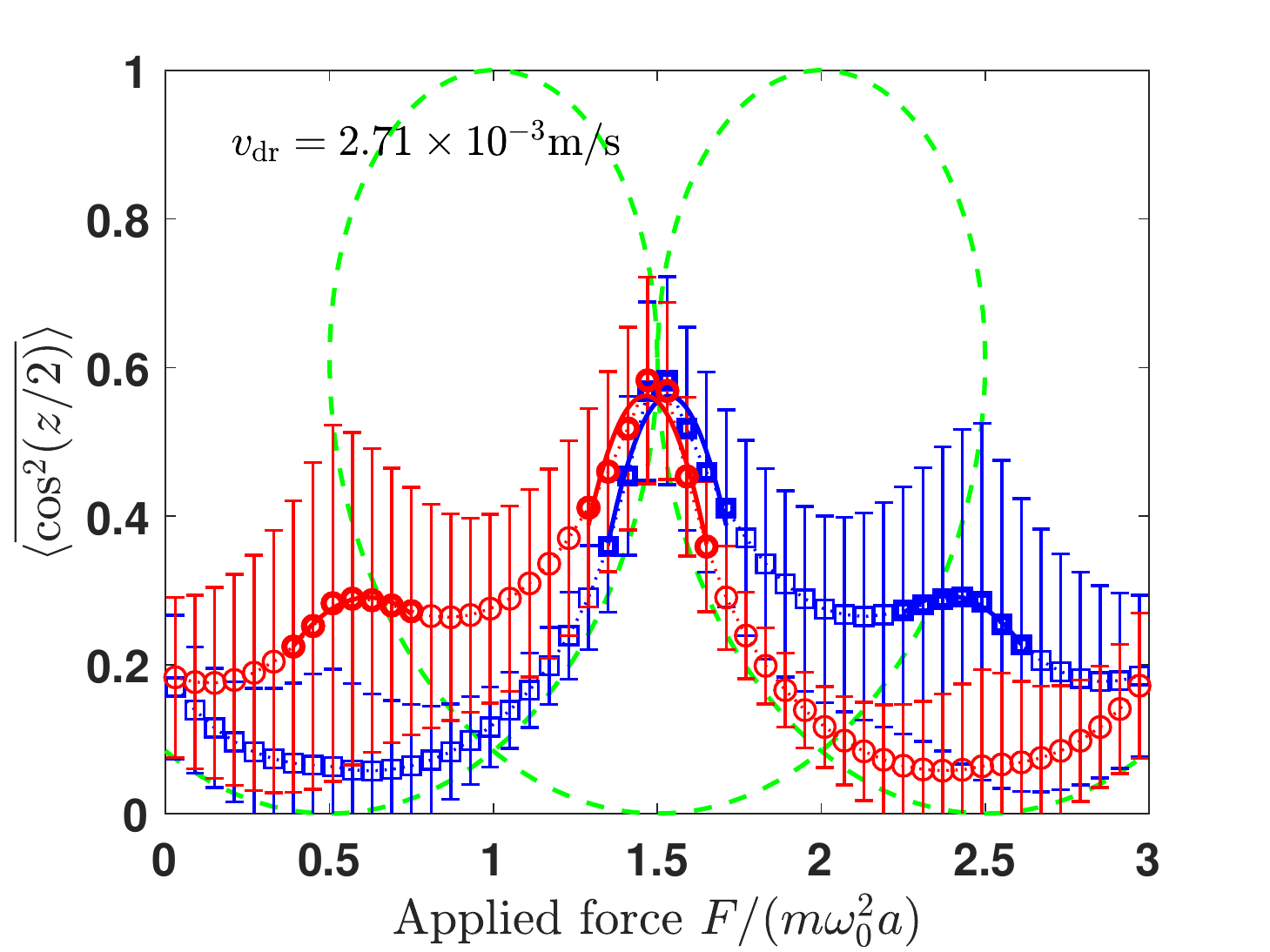}\includegraphics[width=11.46cm]{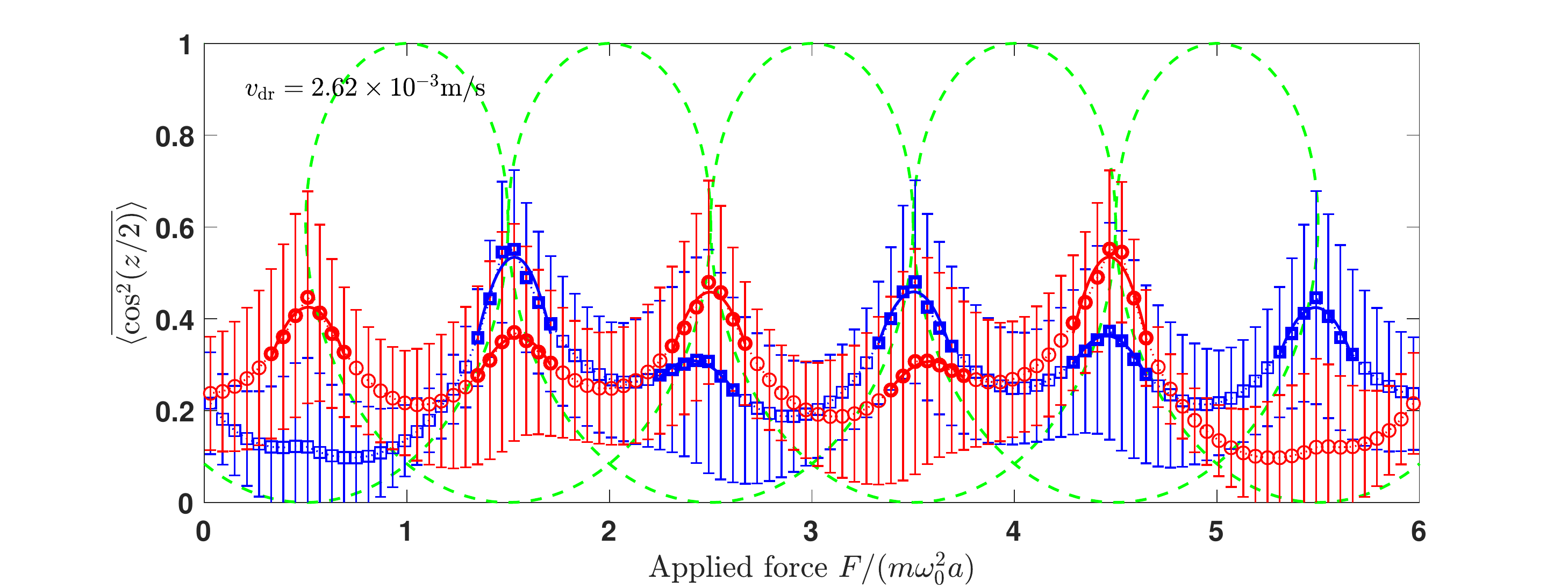}\\
\includegraphics[width=5.73cm]{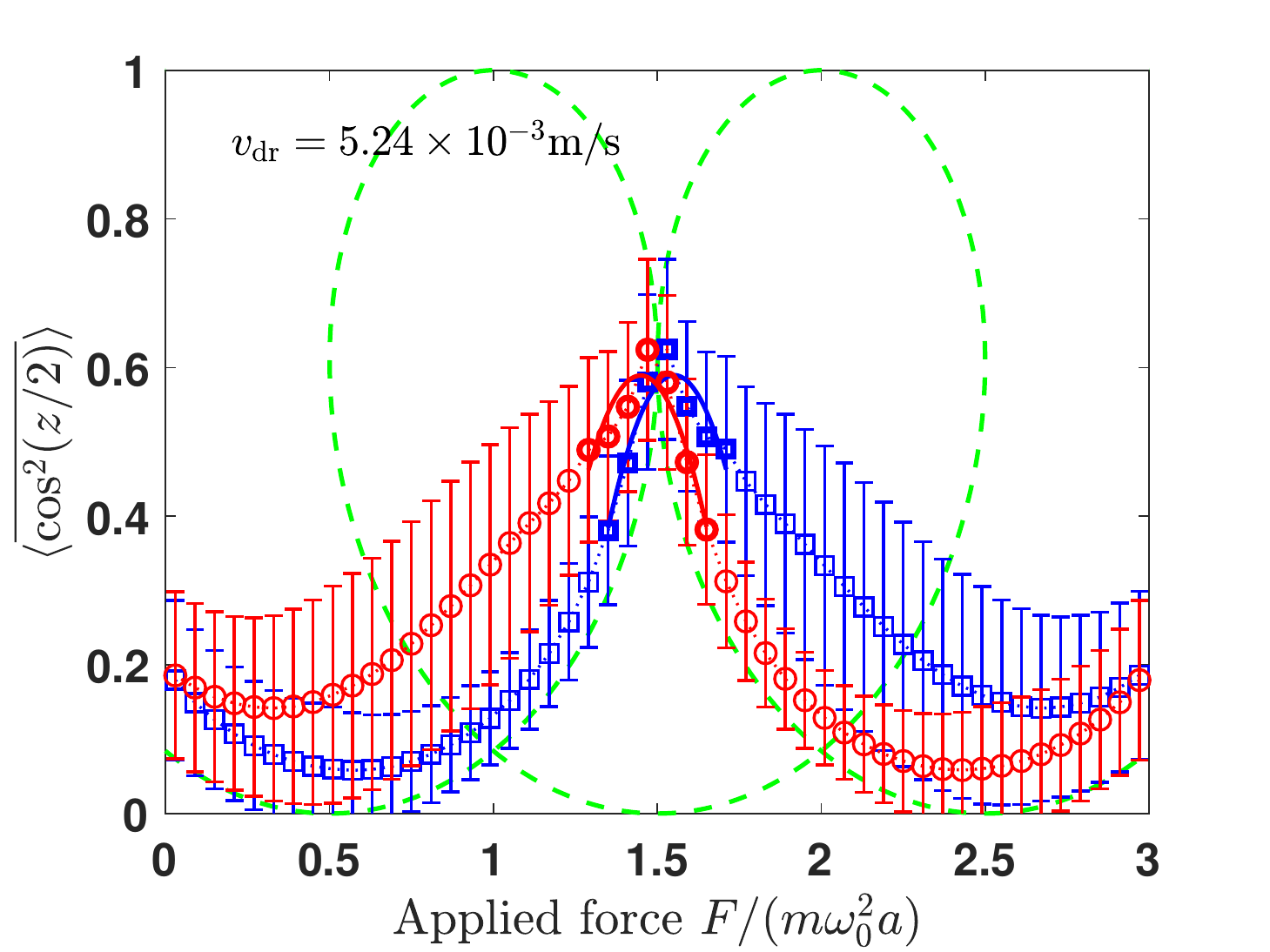}\includegraphics[width=11.46cm]{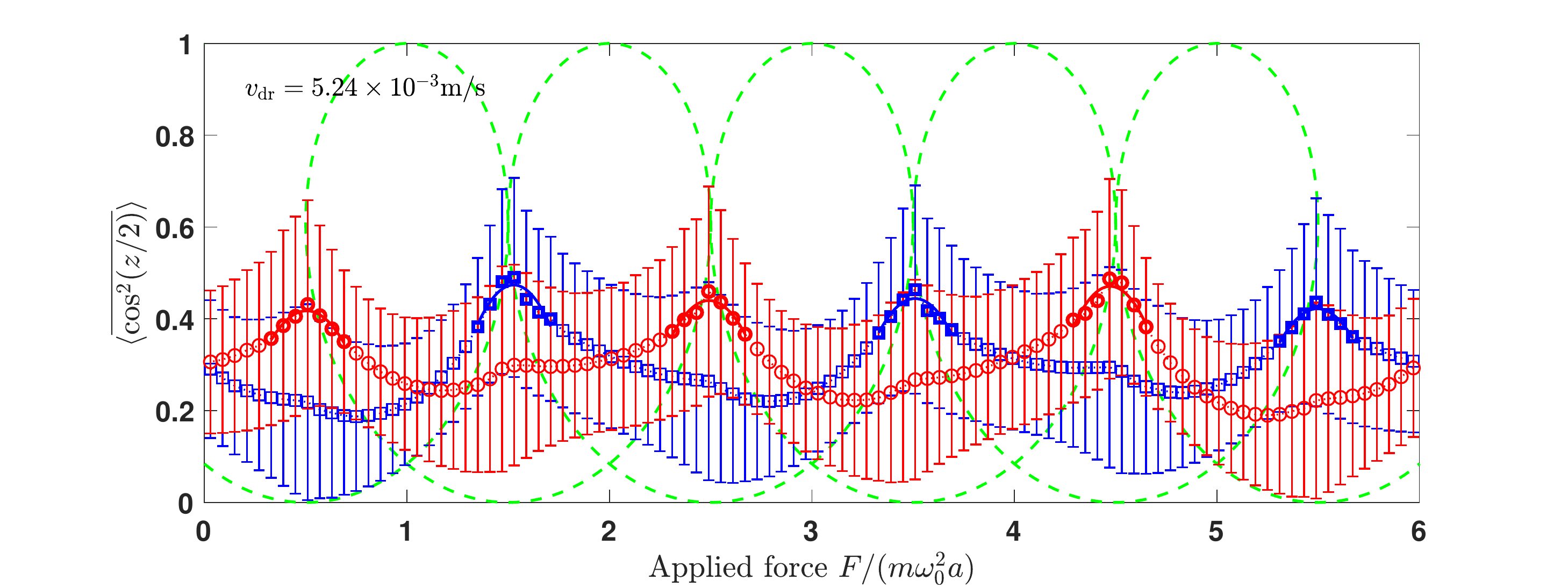}\\
\includegraphics[width=5.73cm]{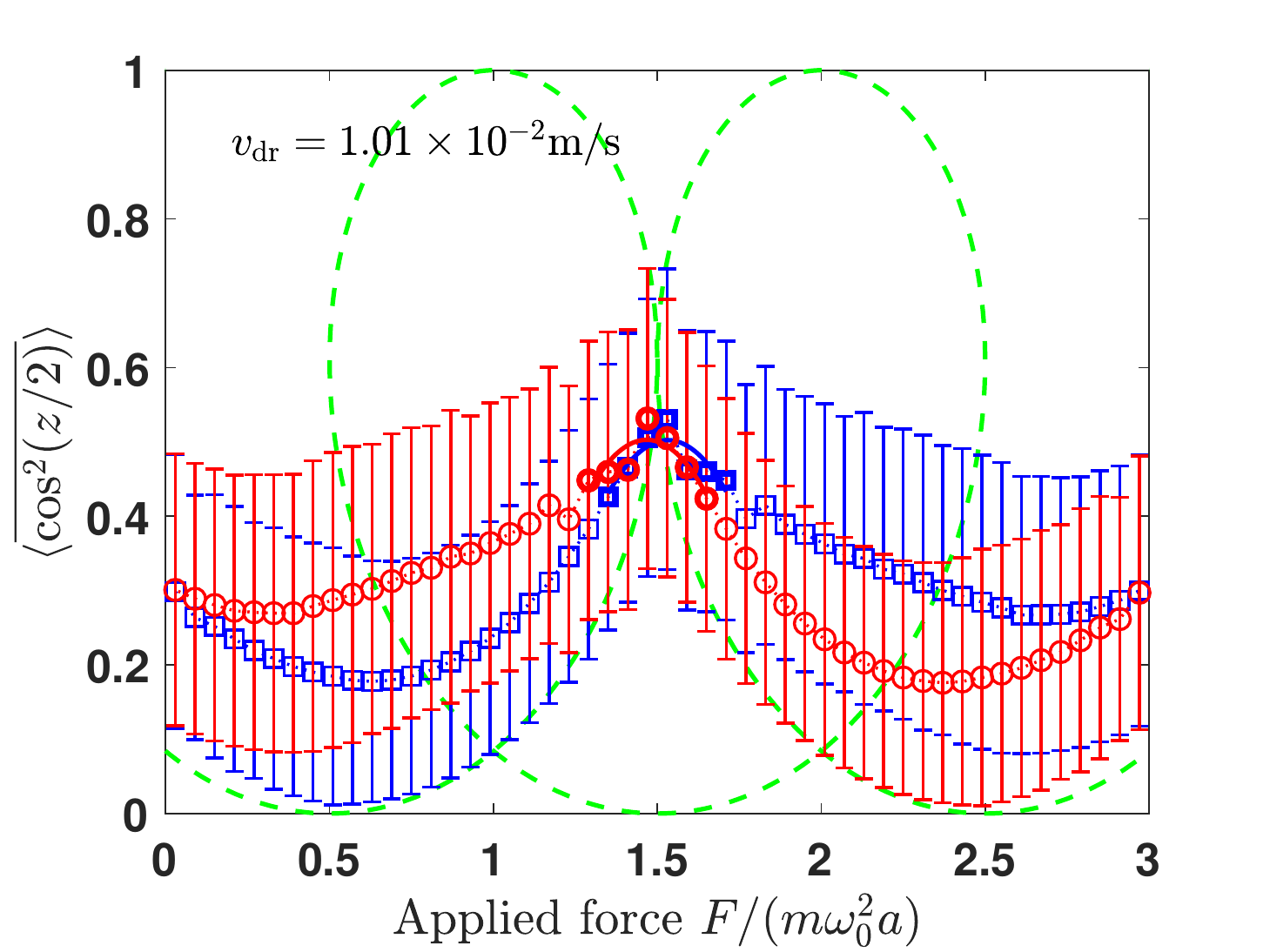}\includegraphics[width=11.46cm]{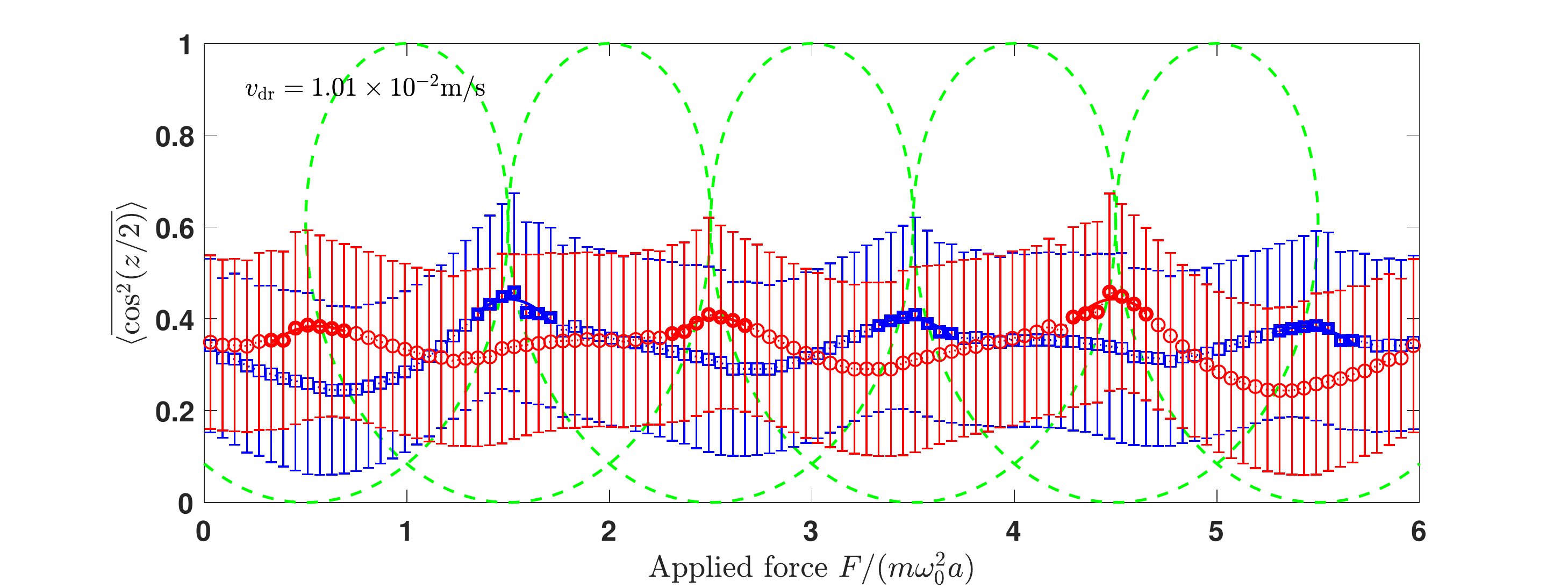}\\
\includegraphics[width=5.73cm]{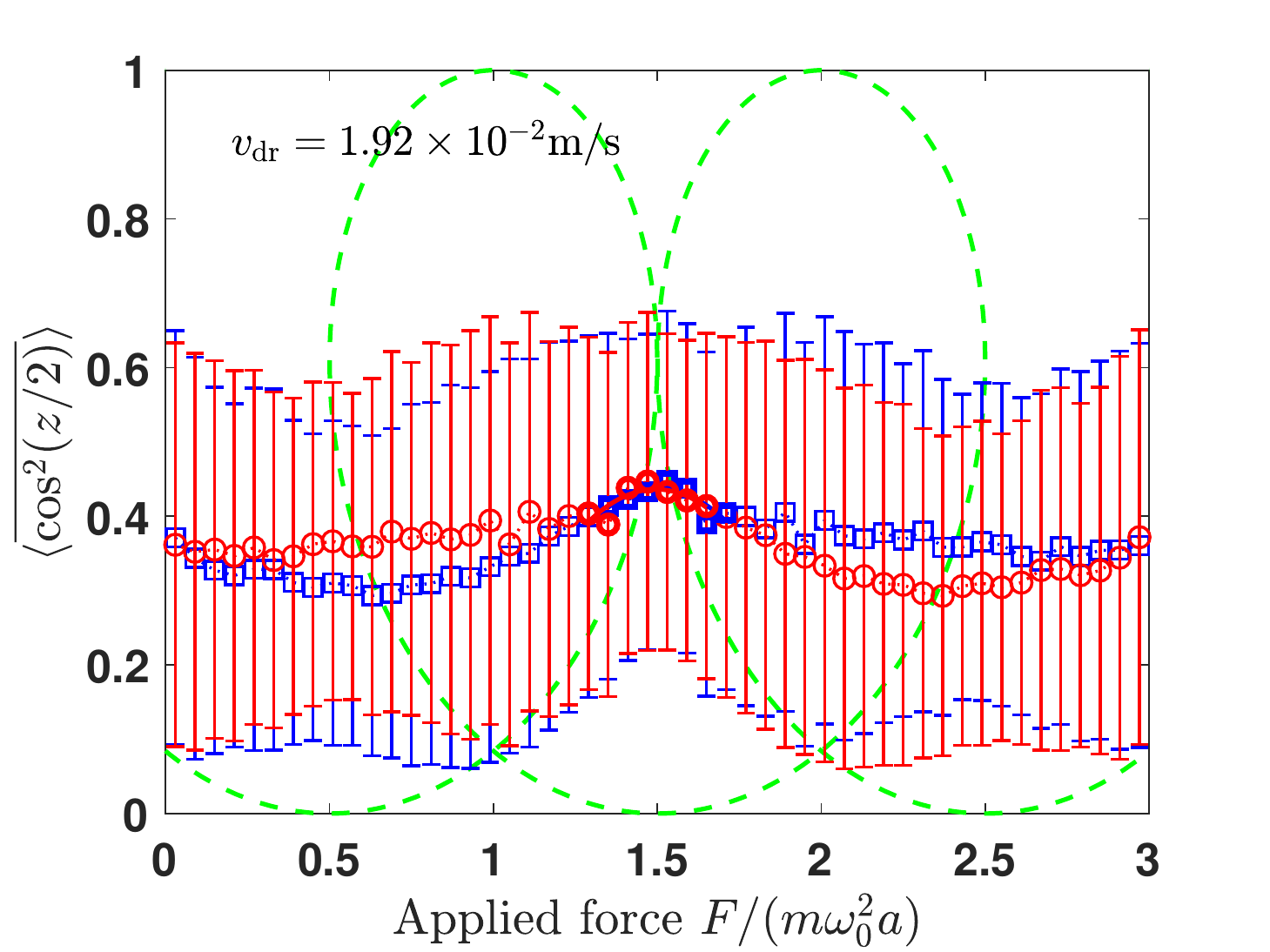}\includegraphics[width=11.46cm]{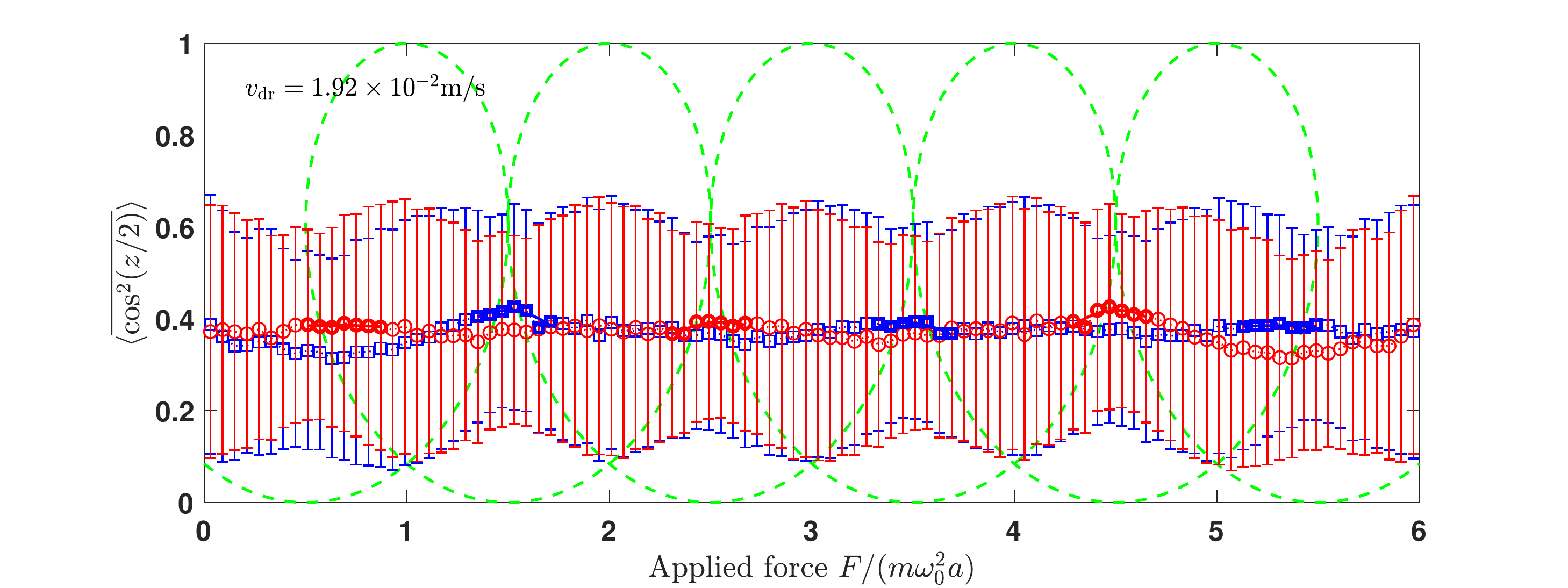}
 \caption{The simulation fluorescence counts at different driving velocity $v_{\rm dr}$ in the case of $\eta=4.6,\ \mu=4\times10^{4}\rm s^{-1},\ V_0=2\pi\hbar\times18\rm MHz$ and ${\it\Theta}=0.04$. The parameters are inherited from
%Figure 2 in \cite{NPVelocityTuning} and 
Figure 15-5 in \cite{bylinskiiphdthesis}. $\omega_0$ is determined by $\eta$ and $V_0$ with $\eta=\frac{2\pi^2V_0}{m\omega_0^2a^2}$ in which $m$ and $a$ are given in Table \ref{parameters}. The green dashed curves are $\cos^2(z^*/2)$ at the balanced points $z^*$ of the resultant potential energy with respect to the driver center's position $\tilde X(z^*)$, which is equal to the applied force $F/(m\omega_0^2a)$, cf. Figure \ref{fig:StickSlips}. At the last four driving velocities, the adjacent hysteresis loops interfere with each other and 3 periods are not adequate for recognizable hysteresis loops, so we simulate 6 periods in either direction and the 3-period results are given on the left of them for comparison. At the last three driving velocities, there are three recognizable hysteresis loops in the 6-period subfigures; at the fourth from the last one, there are five and at the others, there are two. The friction forces measured from different hysteresis loops in each subfigure are averaged to obtain the final friction force at each driving velocity. Incidentally, at the fourth from the last driving velocity, there are still two recognizable hysteresis loops in the 3-period subfigure and the friction force averaged from measuring the two hysteresis loops is close to the result obtained from the 6-period subfigure. Nevertheless, we get the friction force from the 6-period subfigure if there is one.}
\label{ExperimentalDataVerify_eta4p6_T004}
\end{figure}

\begin{table}[H]
\centering
\caption{Number of simulation loops  for Figure \ref{ExperimentalDataVerify_eta4p6_T017}}
\label{tab:ExperimentalDataVerify_eta4p6_T017}
\begin{tabular}{lrlrlr}
$v_{\rm dr}(\rm m/s)$ & Number of loops & $v_{\rm dr}(\rm m/s)$ & Number of loops & $v_{\rm dr}(\rm m/s)$ & Number of loops \\
\midrule
5.44388025066915e-07 & 15 & 1.10564067717483e-06 & 30 & 2.15141515728251e-06 & 30\\
4.35733373350793e-06 & 44 & 8.46740770553410e-06 & 85 & 1.81424066649033e-05 & 190\\
3.54768157470305e-05 & 360 & 6.86415767660501e-05 & 690 & 0.000133298297451480 & 1400\\
0.000273712287989294 & 2800 & 0.000564691945089056 & 5700 & 0.00116318457296758 & 12000\\
0.00271481432041413 & 28000 & 0.00674279208568709 & 68000 &  & \\
\bottomrule
\end{tabular}
\end{table}

\begin{figure}[H]
\centering
 \includegraphics[width=5.73cm]{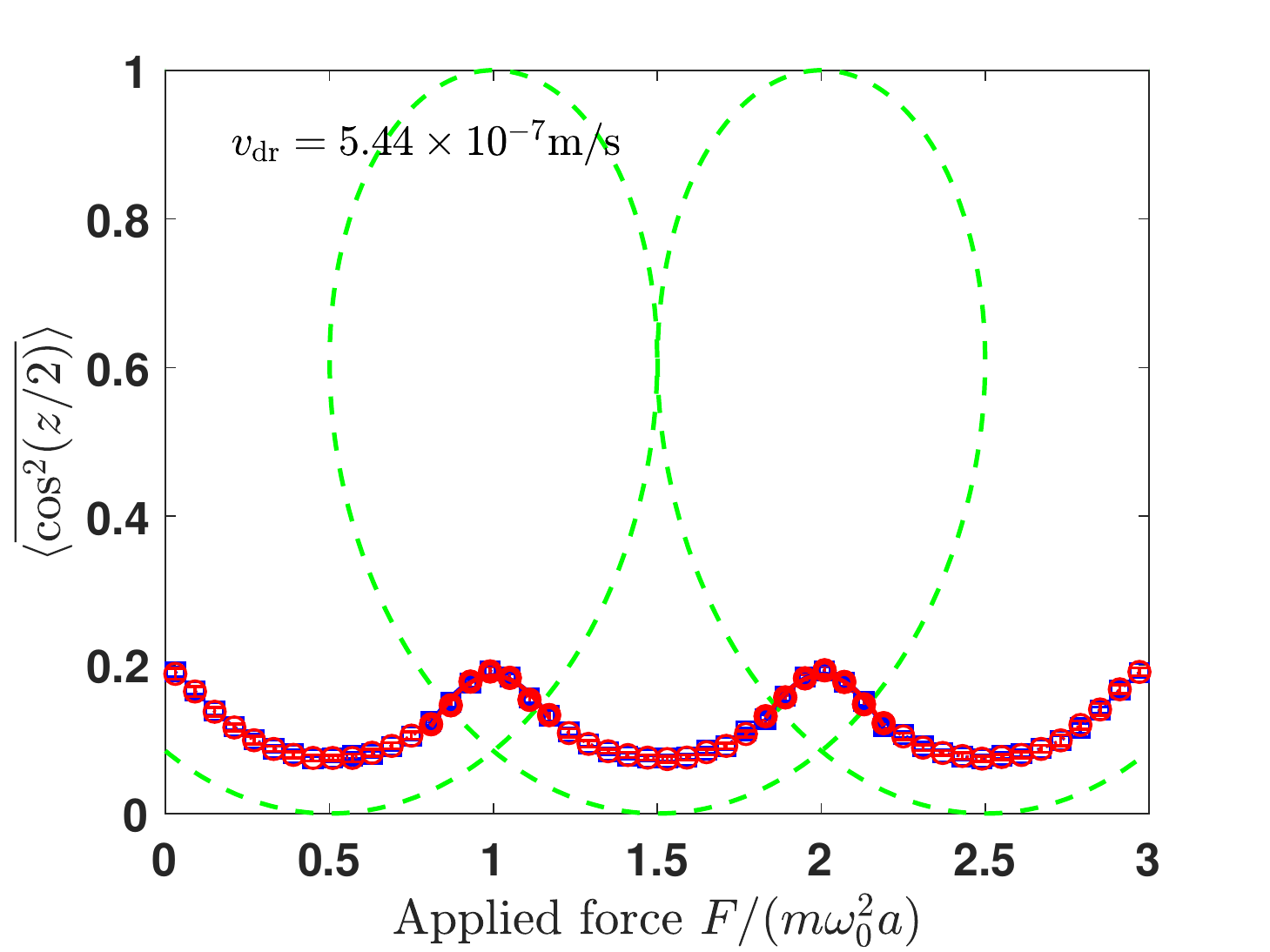}
 \includegraphics[width=5.73cm]{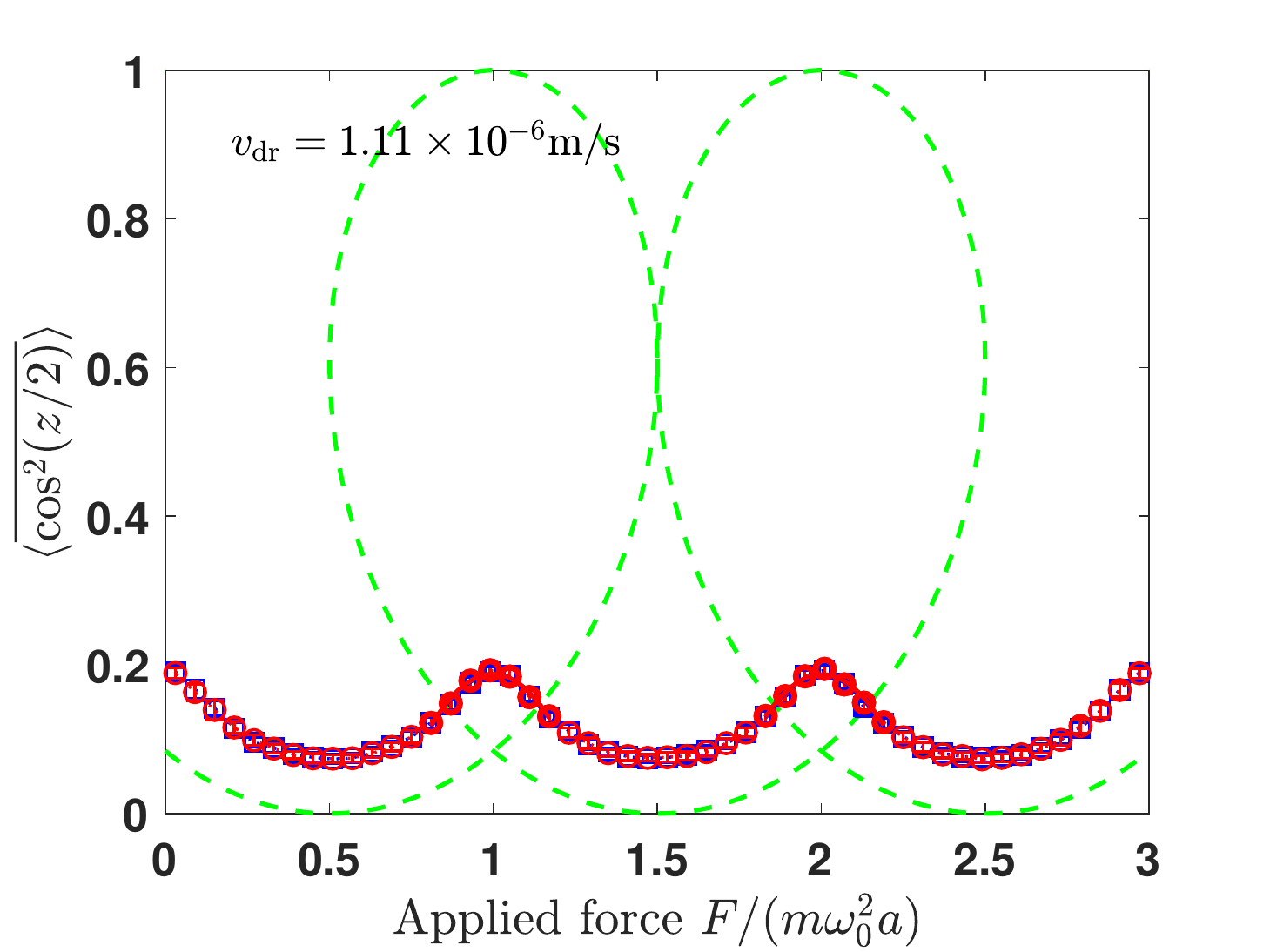}
 \includegraphics[width=5.73cm]{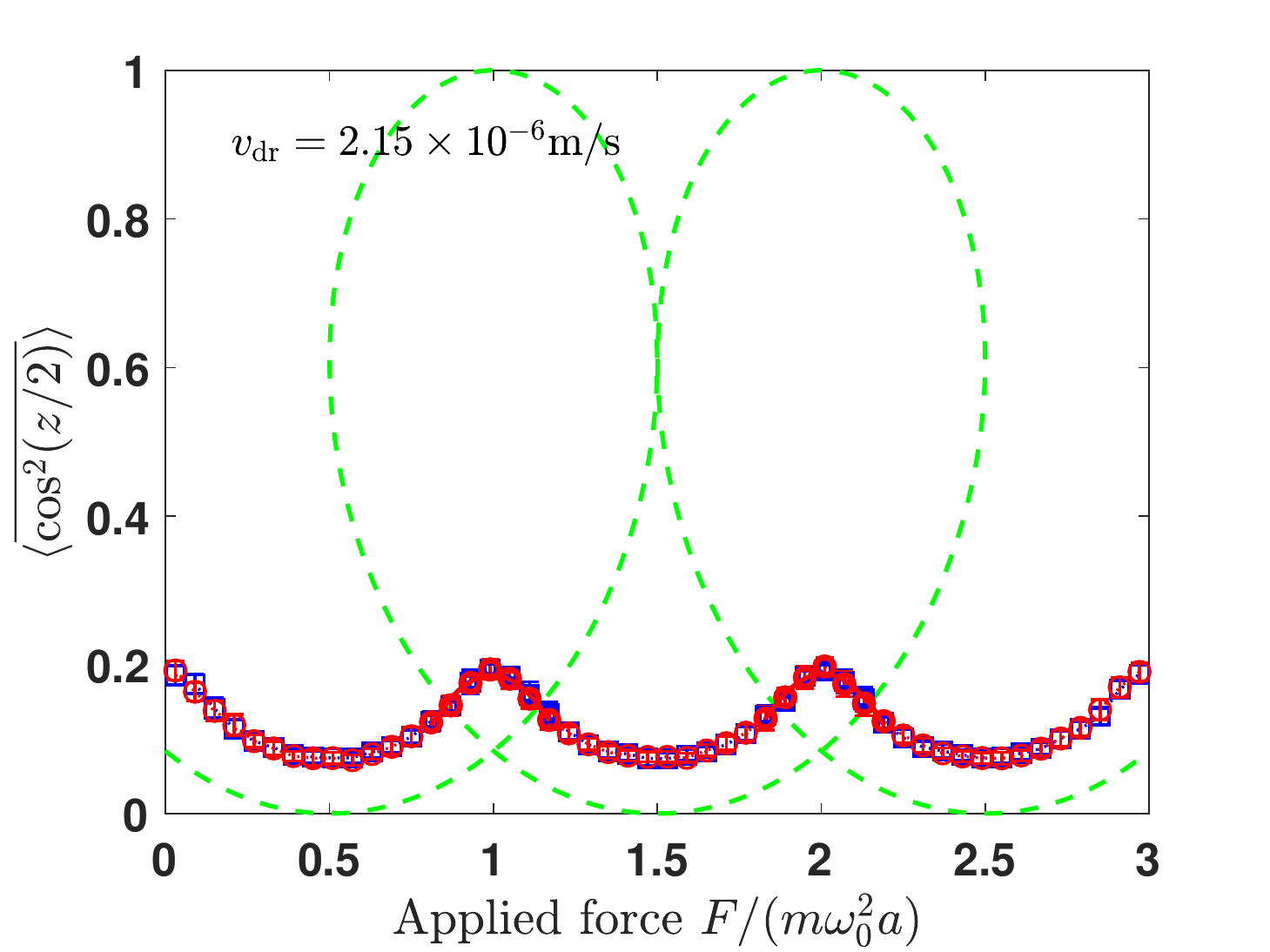}\\
 \includegraphics[width=5.73cm]{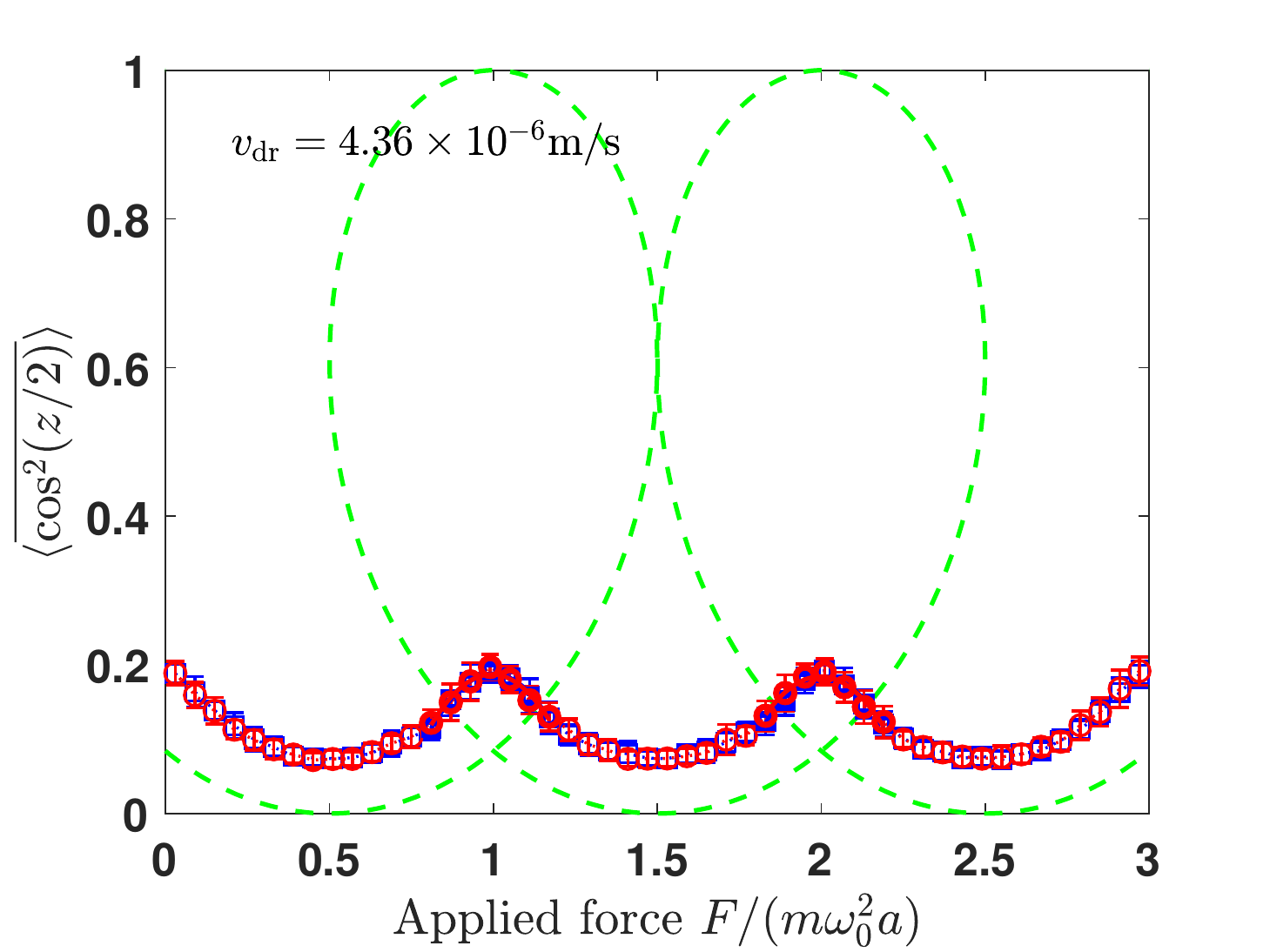}
 \includegraphics[width=5.73cm]{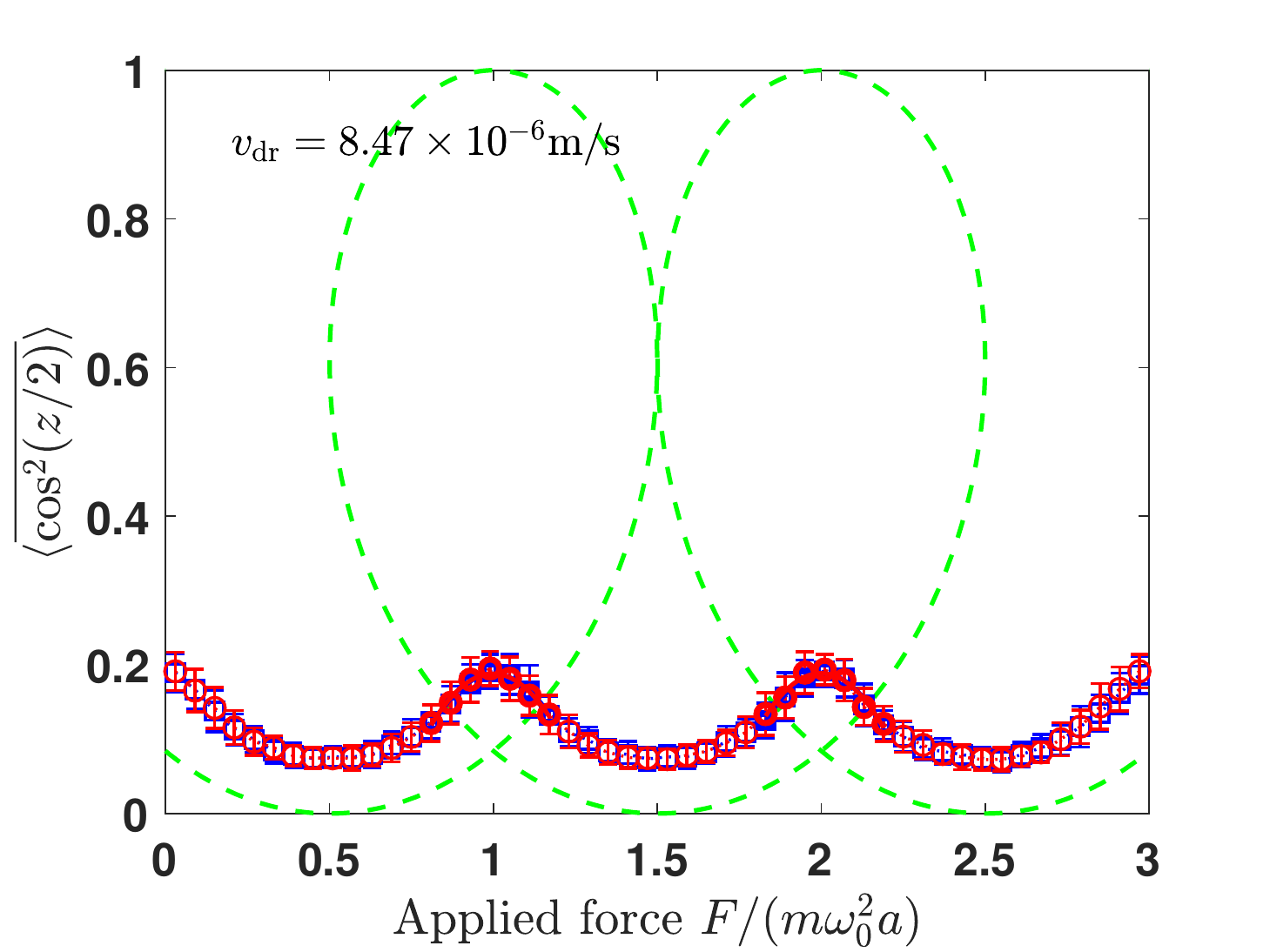}
 \includegraphics[width=5.73cm]{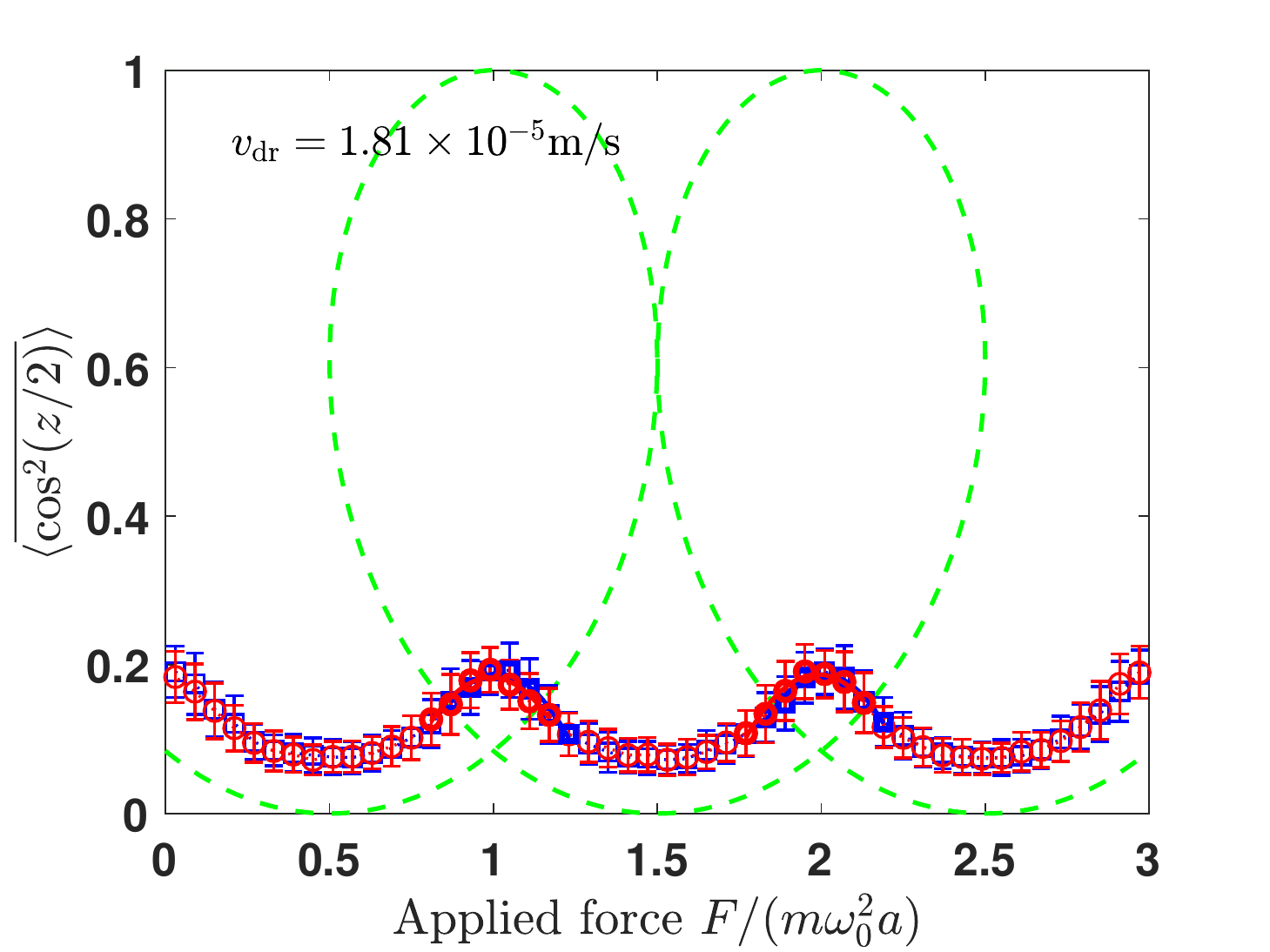}\\
\includegraphics[width=5.73cm]{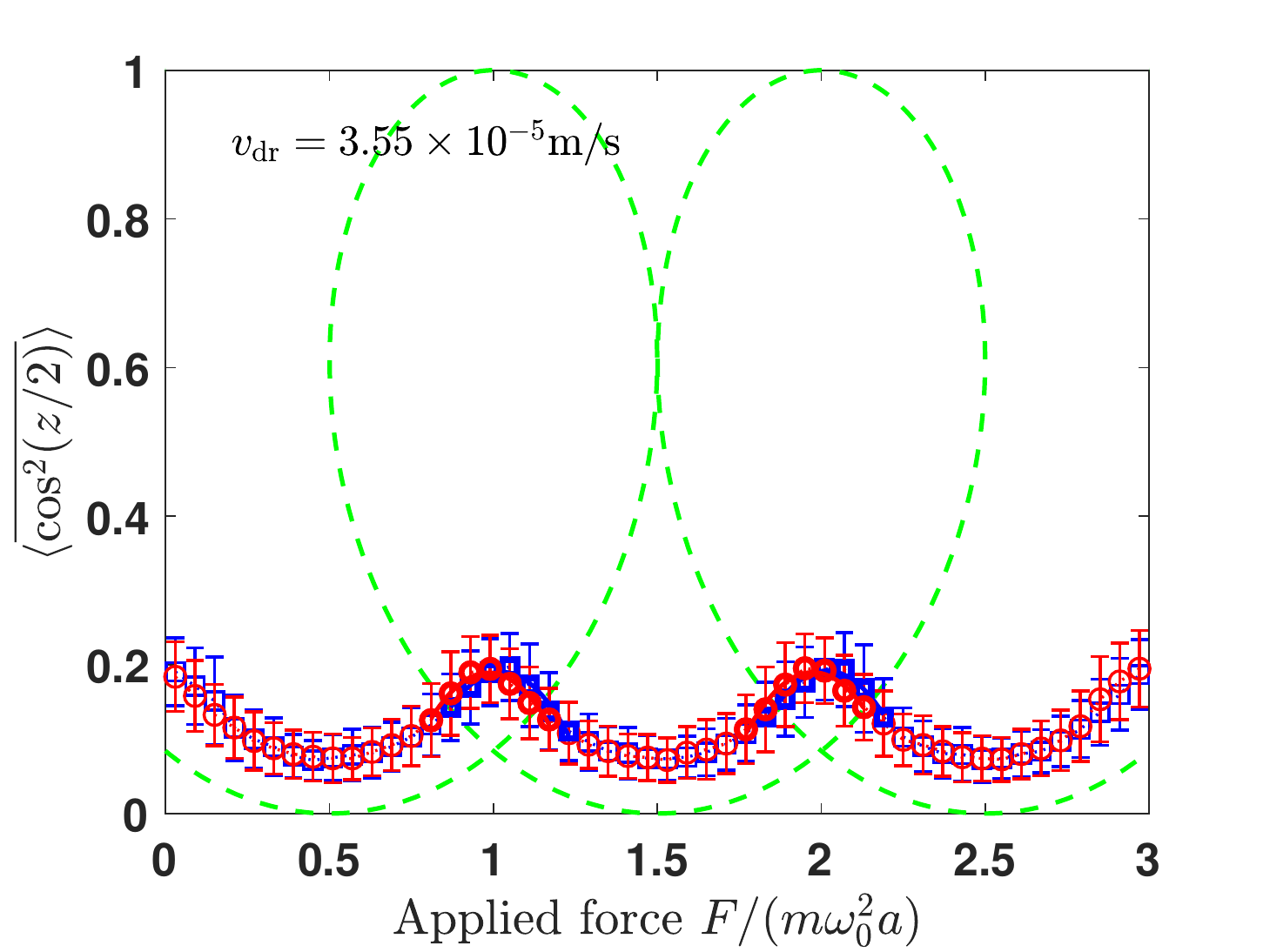}
\includegraphics[width=5.73cm]{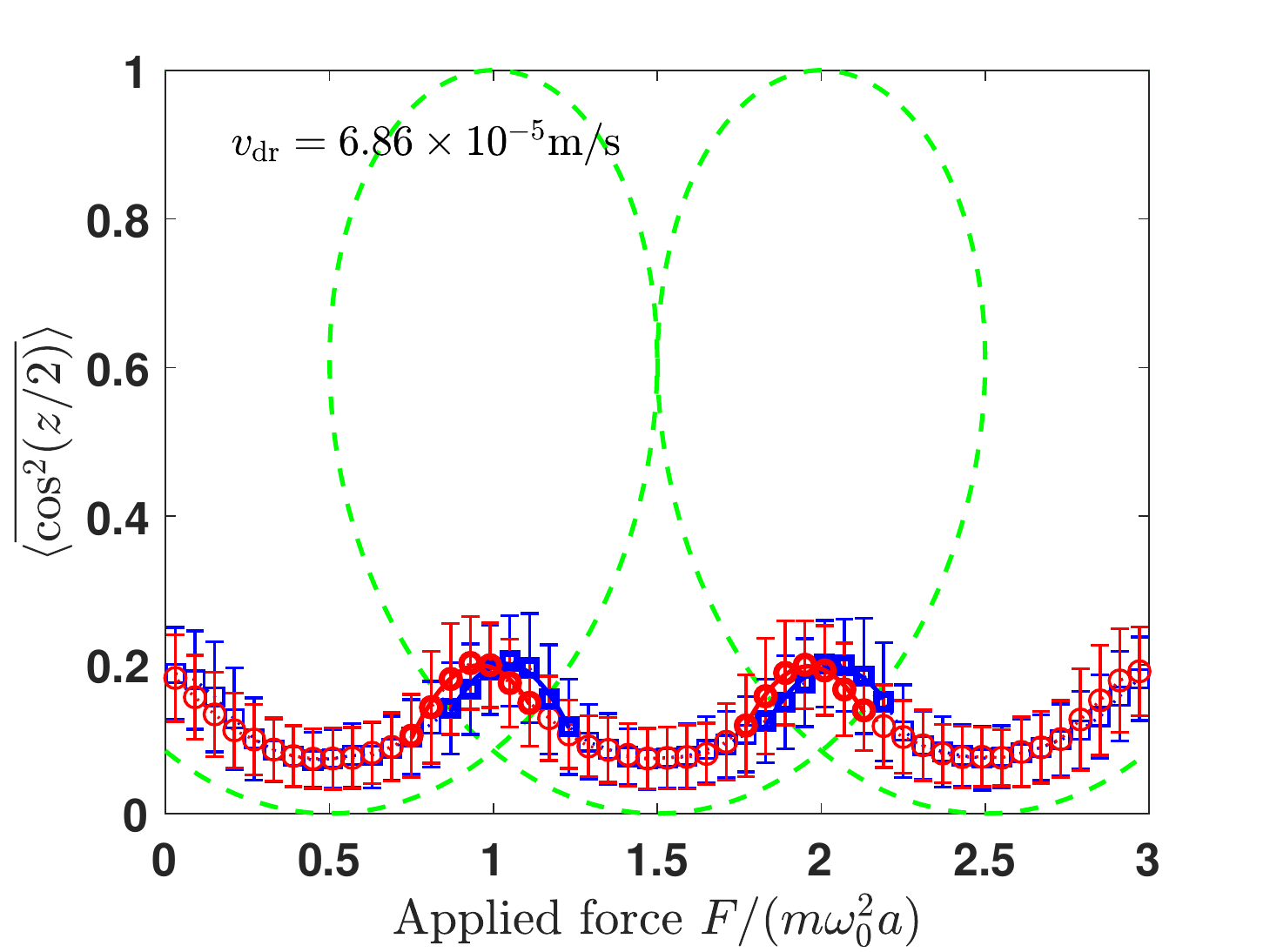}
\includegraphics[width=5.73cm]{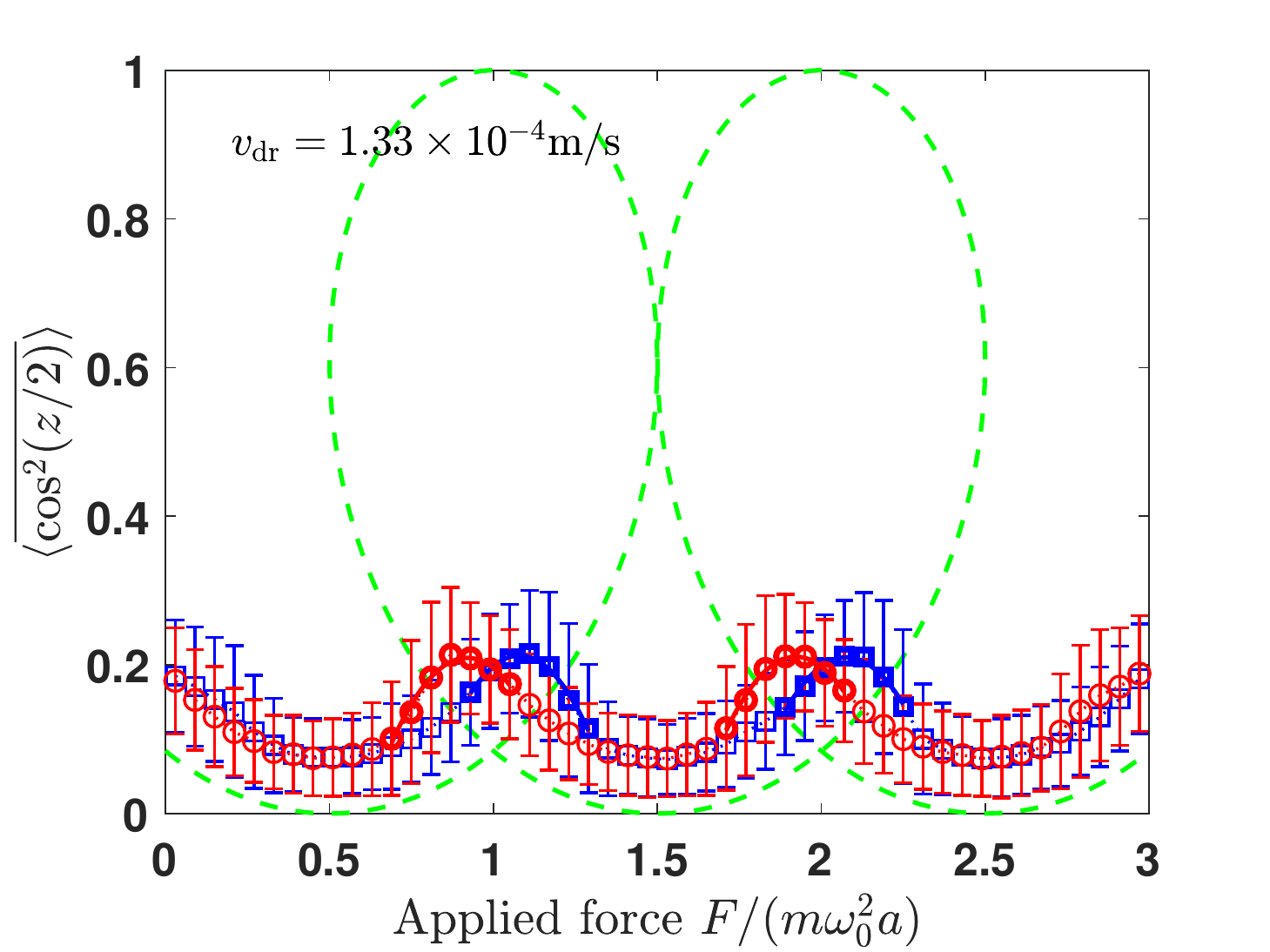}\\
\includegraphics[width=5.73cm]{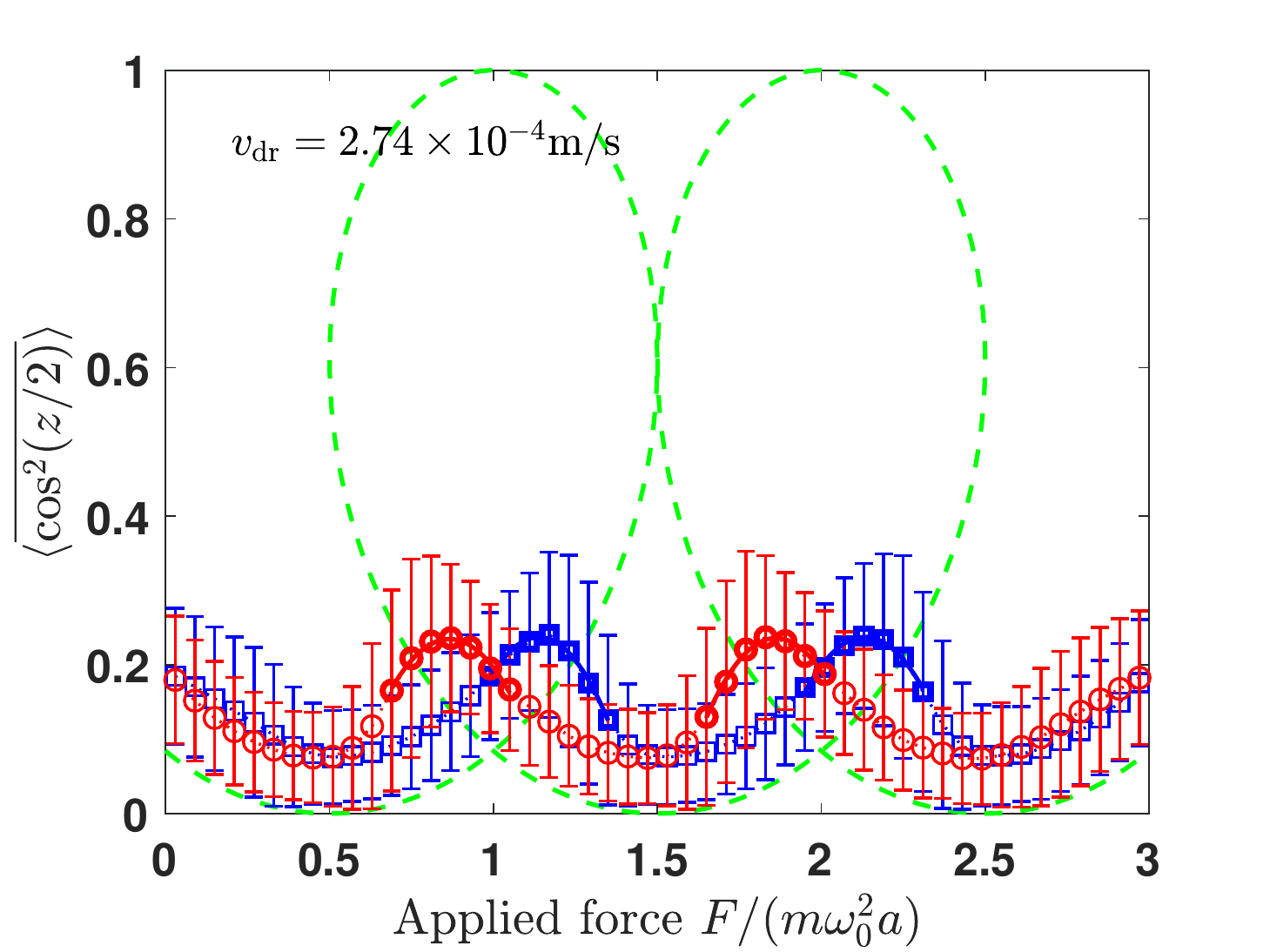}
\includegraphics[width=5.73cm]{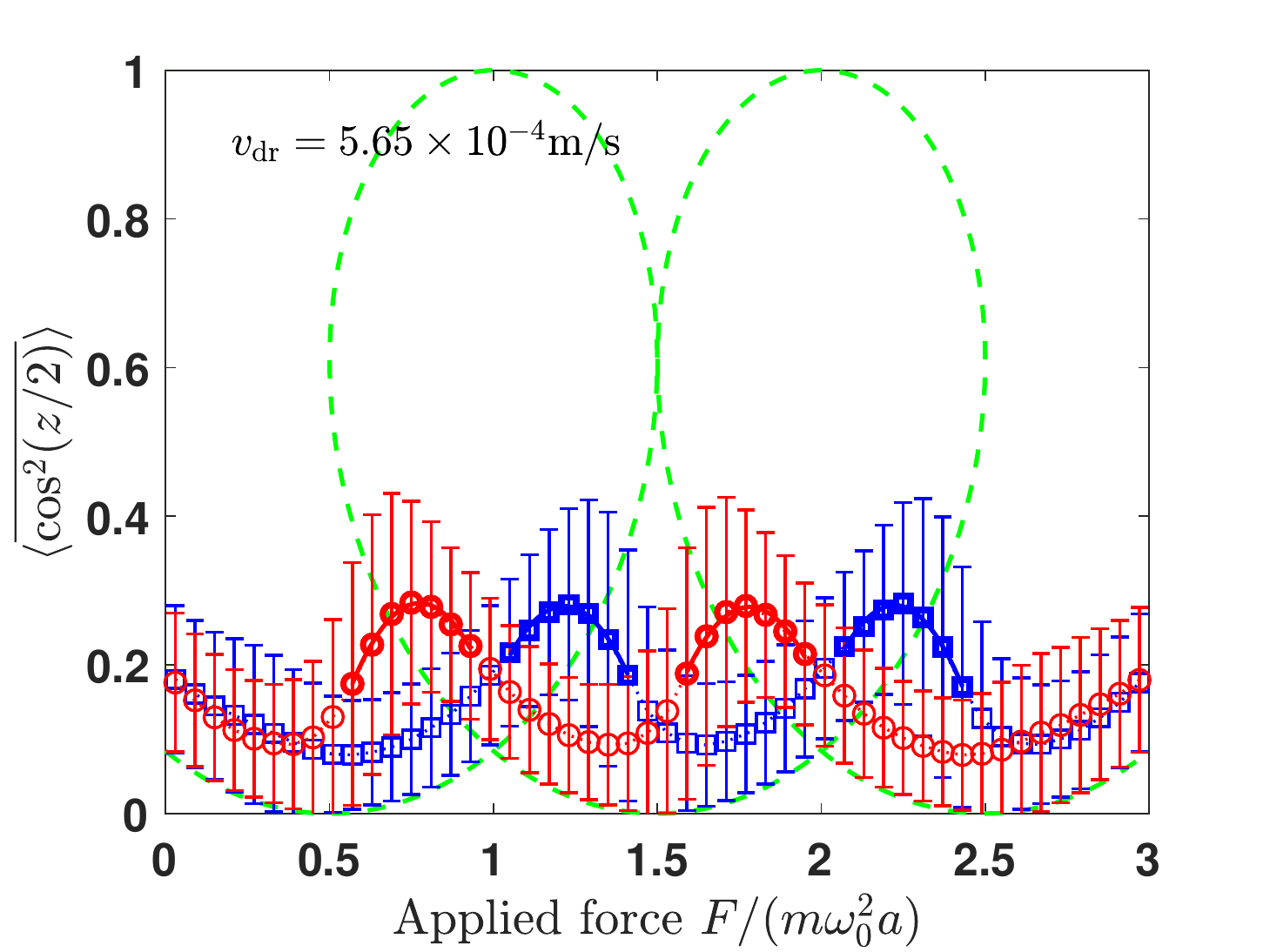}
\includegraphics[width=5.73cm]{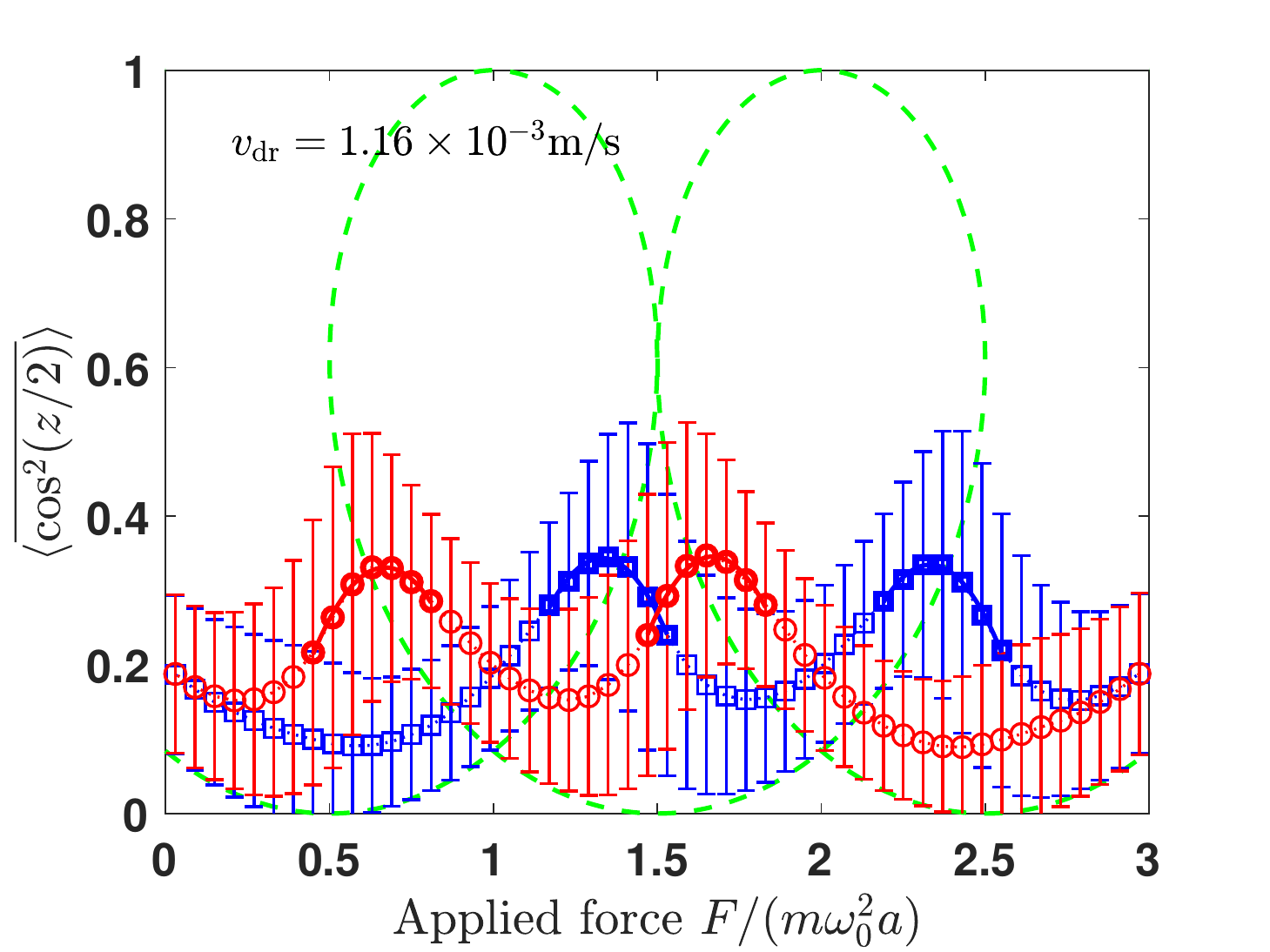}
\end{figure}
\begin{figure}[H]
\centering
\includegraphics[width=5.73cm]{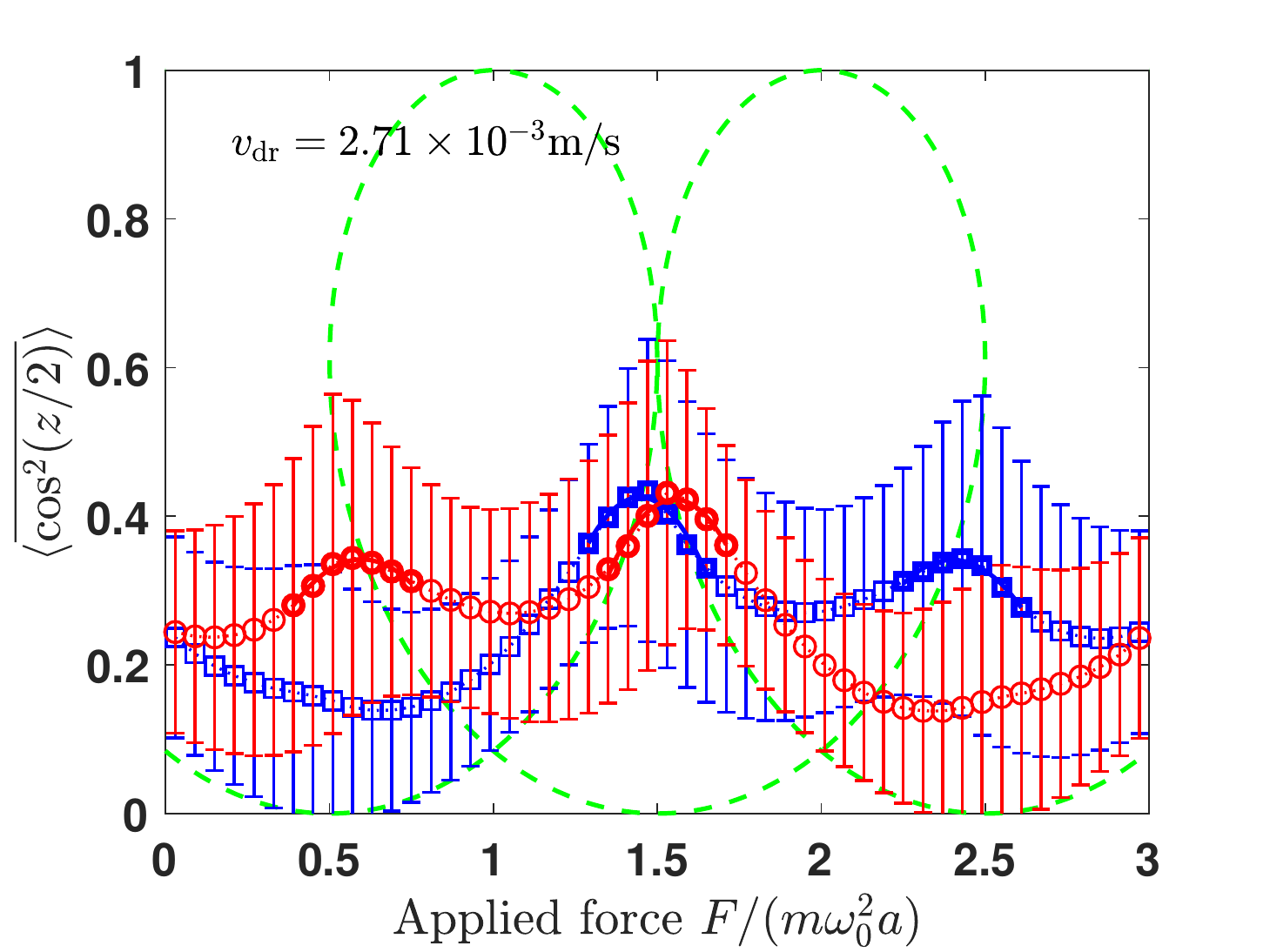}\includegraphics[width=11.46cm]{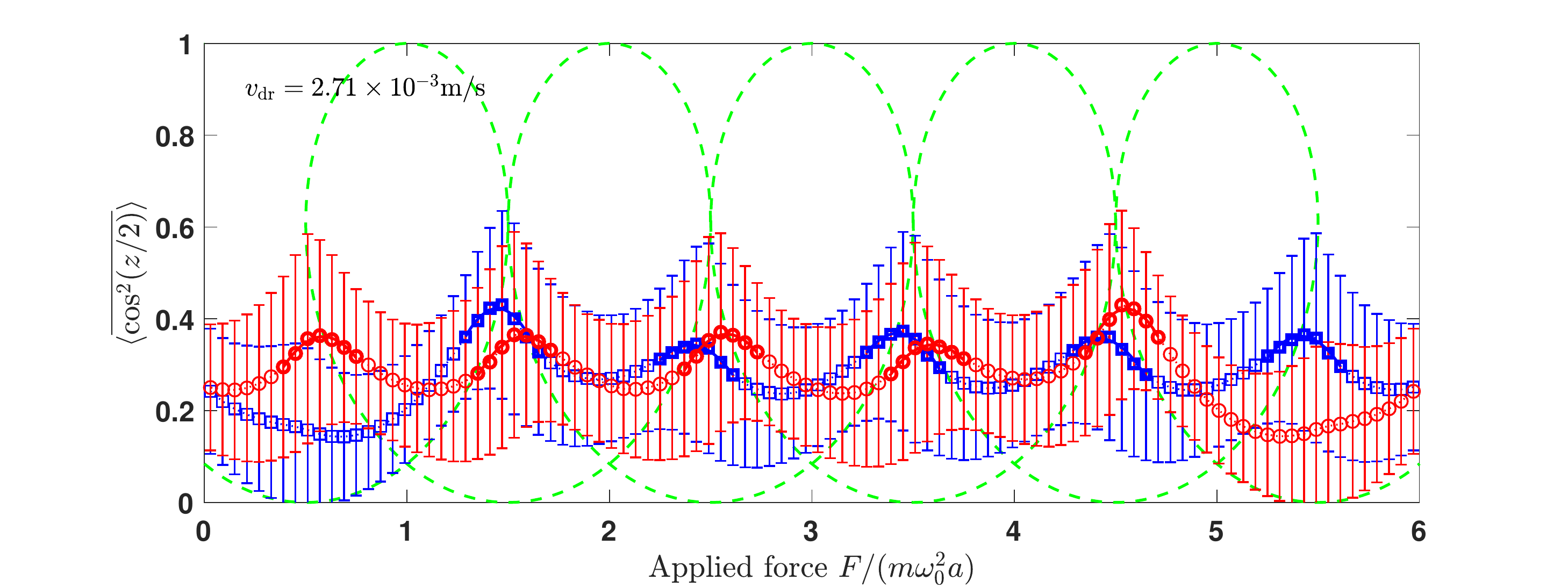}\\
\includegraphics[width=5.73cm]{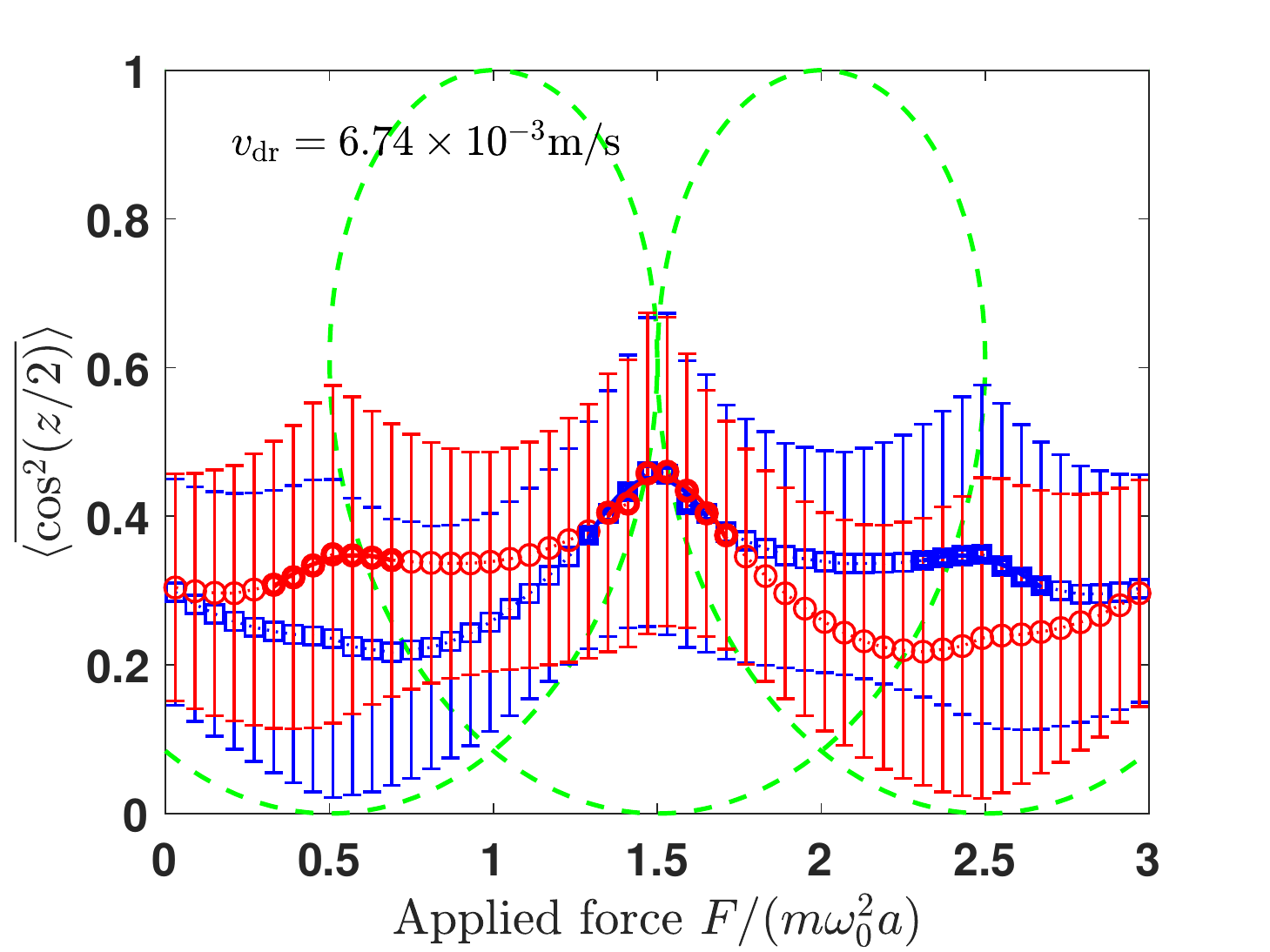}\includegraphics[width=11.46cm]{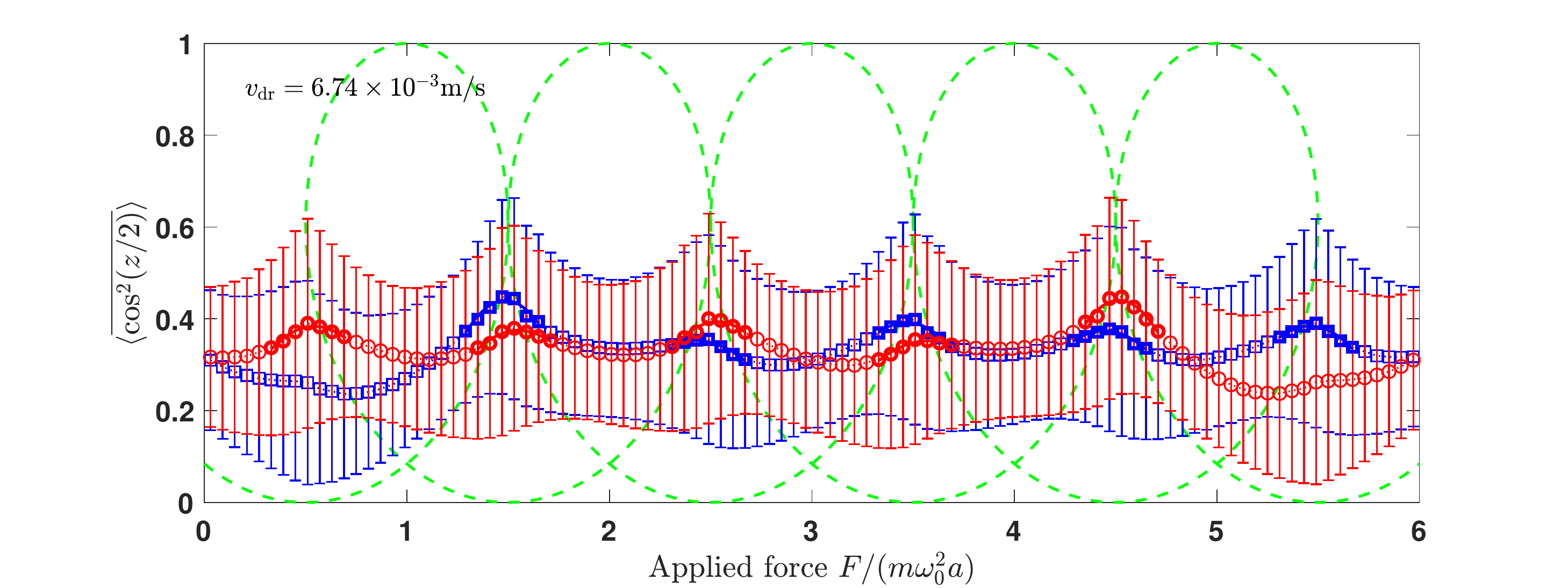}\\
\caption{The simulation fluorescence counts at different driving velocities in the case of $\eta=4.6,\ \mu=4\times10^{4}\rm s^{-1},\ V_0=2\pi\hbar\times18\rm MHz$ and ${\it\Theta}=0.17$. These parameters are inherited from
%Figure 2 in \cite{NPVelocityTuning} and 
Figure 15-5 in \cite{bylinskiiphdthesis}. $\omega_0$ is determined by $\eta$ and $V_0$ with $\eta=\frac{2\pi^2V_0}{m\omega_0^2a^2}$ in which $m$ and $a$ are given in Table \ref{parameters}. The green dashed curves are $\cos^2(z^*/2)$ at the balanced points $z^*$ of the resultant potential energy with respect to the driver center's position $\tilde X(z^*)$, which is equal to the applied force $F/(m\omega_0^2a)$, cf. Figure \ref{fig:StickSlips}. At the last two driving velocities, the adjacent hysteresis loops tend to interfere with each other and 3 periods seem not adequate for recognizable hysteresis loops, so we simulate 6 periods in either direction with the 3-period results given on the left of them for comparison. At the last two driving velocities, there are five recognizable hysteresis loops in the 6-period subfigures and at the others, there are two. The friction forces measured from different hysteresis loops in each subfigure are averaged to obtain the final friction force at each driving velocity. Actually, at the last two driving velocities, there are still two recognizable hysteresis loops in the two 3-period subfigures and the friction forces averaged from measuring the two hysteresis loops in both of them are close to the results obtained from the corresponding 6-period subfigures. Nevertheless, we get the friction forces from the two 6-period subfigures at the last two driving velocities.}
\label{ExperimentalDataVerify_eta4p6_T017}
\end{figure}

%%% Add this line AFTER all your figures and tables
\FloatBarrier

%\movie{Type legend for the movie here.}
%
%\movie{Type legend for the other movie here. Adding longer text to show what happens, to decide on alignment and/or indentations.}
%
%\movie{A third movie, just for kicks.}
%
%\dataset{dataset_one.txt}{Type or paste legend here.}
%
%\dataset{dataset_two.txt}{Type or paste legend here. Adding longer text to show what happens, to decide on alignment and/or indentations for multi-line or paragraph captions.}

\clearpage
\phantomsection
\addcontentsline{toc}{section}{Reference}
\bibliography{supplement}